\documentclass[]{aa} 

\usepackage{graphicx}
\usepackage{txfonts}
\usepackage{multirow}
\usepackage[final]{hyperref}
\hypersetup{colorlinks=true, linkcolor=red,citecolor=blue,urlcolor=magenta} 

\usepackage{comment}
\usepackage{longtable}
\usepackage{rotating}
\usepackage{placeins}
\usepackage{tikz}
\usepackage{booktabs}
\usepackage{nicefrac}
\usepackage{xcolor}

\newcommand{\met}{CH$_3$OH\xspace}
\newcommand{\et}{C$_2$H$_5$OH\xspace}
\newcommand{\etc}{C$_2$H$_5$CN\xspace}
\newcommand{\vc}{C$_2$H$_3$CN\xspace}
\newcommand{\dme}{CH$_3$OCH$_3$\xspace}
\newcommand{\mf}{CH$_3$OCHO\xspace}
\newcommand{\mic}{CH$_3$NCO\xspace}
\newcommand{\ad}{CH$_3$CHO\xspace}
\newcommand{\aan}{NH$_2$CH$_2$CN\xspace}
\newcommand{\fmm}{NH$_2$CHO\xspace}
\newcommand{\cyano}{HC$^{13}$CCN\xspace}
\newcommand{\isoc}{\textit{i}-C$_3$H$_7$CN\xspace}
\newcommand{\nsoc}{\textit{n}-C$_3$H$_7$CN\xspace}
\newcommand{\2}{$_2$}
\newcommand{\3}{$_3$}
\newcommand{\4}{$_4$}
\newcommand{\5}{$_5$}
\newcommand{\kms}{\,km\,s$^{-1}$\xspace}
\newcommand{\sdu}{S$_{\rm D}$\xspace}
\newcommand{\wdu}{W$_{\rm D}$\xspace}
\newcommand{\sco}{S$_{\rm CO}$\xspace}
\newcommand{\wco}{W$_{\rm CO}$\xspace}

\begin{document}

   
   \title{Resolving desorption of complex organic molecules in a hot core:}
   \subtitle{Transition from non-thermal to thermal desorption or two-step thermal desorption?}

   \author{Laura\,A.\,Busch
          \inst{1}\thanks{Member of the International Max\,Planck Research School\,(IMPRS) for Astronomy and Astrophysics at the Universities of Bonn and Cologne.}
          \and
          Arnaud\,Belloche\inst{1}
          \and
          Robin\,T.\,Garrod\inst{2}
          \and
          Holger\,S.\,P.\,M\"uller\inst{3}
          \and
          Karl\,M.\,Menten\inst{1}
          }

   \institute{Max-Planck-Institut f\"ur Radioastronomie, Auf dem H\"ugel 69, 53121 Bonn, Germany\\
              \email{labusch@mpifr-bonn.mpg.de}
              \and 
              Departments of Chemistry and Astronomy, University of Virginia, Charlottesville, VA 22904, USA
              \and
              I. Physikalisches Institut, Universit\"at zu K\"oln, Z\"ulpicher Str. 77, 50937 K\"oln, Germany
             }

   \date{Received ; accepted }

  \abstract  
   {The presence of many interstellar complex organic molecules (COMs) in the gas phase in the vicinity of protostars has long been associated with their formation on icy dust grain surfaces before the onset of protostellar activity, and their subsequent thermal co-desorption with water, the main constituent of the grains' ice mantles, as the protostar heats its environment to $\sim$100\,K. } 
   {Using the high angular resolution provided by the Atacama Large Millimetre/submillimetre Array (ALMA) we want to resolve the COM emission in the hot molecular core Sagittarius\,B2\,(N1) and thereby shed light on the desorption process of COMs in hot cores.} 
   {We use data taken as part of the 3\,mm spectral line survey Re-exploring Molecular Complexity with ALMA (ReMoCA) to investigate the morphology of COM emission in Sagittarius\,B2\,(N1). We also use ALMA continuum data at 1\,mm taken from the literature. Spectra of ten COMs (including one isotopologue) are modelled under the assumption of Local Thermodynamic Equilibrium (LTE) and population diagrams are created for these COMs for positions at various distances to the south and west from the continuum peak. Based on this analysis, we produce resolved COM rotation temperature and column density profiles. H\2 column density profiles are derived from dust continuum emission and C$^{18}$O 1--0 emission and used to derive COM abundance profiles as a function of distance and temperature. These profiles are compared to astrochemical models.} 
   {Based on the morphology, a rough separation into O- and N-bearing COMs can be done. The temperature profiles span a range of 80--300\,K with power-law indices from $-0.4$ to $-0.8$, in agreement with expectations of protostellar heating of an envelope with optically thick dust.  
   Column density and abundance profiles reflect a similar trend as seen in the morphology. While abundances of N-bearing COMs peak only at highest temperatures, those of most O-bearing COMs peak at lower temperatures and remain constant or decrease towards higher temperatures. Many abundance profiles show a steep increase at $\sim$100\,K. To a great extent, the observed results agree with results of astrochemical models that, besides the
   co-desorption with water, predict that O-bearing COMs are mainly formed on dust grain surfaces at low temperatures while at least some N-bearing COMs and CH\3CHO are substantially formed in the gas phase at higher temperatures.} 
   {Our observational results, in comparison with model predictions, suggest that COMs that are exclusively or to a great extent formed on dust grains desorb thermally at $\sim$100\,K from the grain surface likely alongside water. A dependence on the COM binding energy is not evident from our observations. Non-zero abundance values below $\sim$100\,K suggest that another desorption process of COMs is at work at these low temperatures: either non-thermal desorption or partial thermal desorption related to the lower binding energies experienced by COMs in the outer, water-poor ice layers. In either case, this is the first time that the transition between two regimes of COM desorption has been resolved in a hot core. } 
   
   \keywords{astrochemistry -- ISM: molecules -- stars: formation -- Galaxy: centre -- individual objects: Sagittarius B2}
    \authorrunning{L.A.\,Busch et al.}
    \titlerunning{COM chemistry in Sgr\,B2\,(N1)}
   \maketitle
   
\section{Introduction}
To date, more than 270 molecules have been detected in space\footnote{see \url{https://cdms.astro.uni-koeln.de/classic/molecules}} and the detection rate is speeding up thanks in particular to the perpetual advancement of broadband receivers on single-dish telescopes like the Yebes\,40m, IRAM\,30m telescopes, or the Green Bank Telescope allowing for observations of sensitive spectral-line surveys that cover a wide range of frequencies \citep[][]{McGuire22}.
Observations of molecules help us understand the structure and evolution of the interstellar medium (ISM) and shed light on the various phases of the star formation process from diffuse clouds ($n_{\rm H} \sim 10^2-10^3$\,cm$^{-3}$), to dense clouds and cores with densities beyond $10^8$\,cm$^{-3}$, to protostellar and protoplanetary disks \citep[for a recent review see][]{Jorgensen20}.
Amongst these, observations of complex organic molecules (COMs), which are carbon-bearing molecules that consist, per definition, of at least six atoms \citep[][]{Herbst09}, are of special interest as they are potential precursors of even more complex and possibly prebiotic species. Therefore, they may play a role in the build-up of the complex organic and prebiotic inventory seen in meteorites and comets in the solar system and, going one step further, in the emergence of life on Earth. 

Many COMs have first been detected in the close vicinity of high- and low-mass protostars. These regions are referred to as hot cores and hot corinos, respectively. The spatial extent of these regions is usually associated with a gas temperature of $\sim$100\,K \citep{Jorgensen20}. The characteristic temperature of $\sim$100\,K denotes the point when water desorbs from the dust grain surfaces, along with many other species, perhaps including the COMs themselves. At the end of the cold collapse phase of a dense cloud core, an ice mantle consisting of mostly water \citep[with traces of, e.g., CO, CO\2, and NH\3,][]{Boogert15} covers the grain. A widely accepted paradigm for the production of COMs was the photodissociation of molecules in this ice mantle, notably methanol, to produce radicals that would become mobile, and thus reactive, during the gradual warm-up of dust grains that occurs once a protostar is born and starts heating up its collapsing envelope \citep[][]{Garrod06}. At high enough temperatures, both the simple and complex molecules would thermally desorb from the grains into the gas phase. The release of the simple species in particular could also induce gas-phase chemistry. Recent models by \citet{Garrod22}, which include non-diffusive surface and bulk-ice chemistry, indicate that significant COM production on dust grains may actually occur at low temperatures when the simple ices themselves are beginning to form, in agreement with recent experimental findings involving the co-deposition of atomic H and simple molecules such as CO \citep[e.g.,][]{Fedoseev15}.

Assuming that COMs are indeed formed on the dust grains (which does not rule out complementary gas-phase formation in some cases), their temperature-dependent desorption behaviour will be important to the observed spatial morphologies of their emission regions.
The desorption of COMs from the grains has been studied in the laboratory using interstellar ice analogues \citep[e.g.,][]{Collings04,Martin-Domenech14}, but the precise temperature dependence is complex. In particular, for a COM more volatile than water, the desorption of that molecule from a water ice mixture may occur partially at a relatively low temperature, corresponding to direct thermal desorption, perhaps mediated by some diffusive transport of molecules through surface pore structures. Another component would be trapped within the water-ice structure, eventually co-desorbing with the water, and therefore, exhibiting a similar desorption temperature as water itself.

Astrochemical models have typically used simple treatments for the desorption of COMs; so-called two-phase models \citep[e.g.,][]{Garrod08}, which consider the grain-surface ices to be a single physical phase, and three-phase models  \citep[e.g.,][]{Garrod13}, which distinguish the bulk ice from the ice-mantle surface, usually allow molecules to desorb according to their own binding energies, meaning that there is no trapping of any kind. The more recent three-phase model of \citet{Garrod22} adopts the other extreme; material in the bulk ice is trapped beneath the upper layer, meaning that only a small fraction of COMs in the bulk are able to desorb until water rapidly begins to desorb at its own characteristic desorption temperature \citep[][used an empirical treatment to model temperature-dependent molecular desorption, but did not include actual grain-surface chemistry, only desorption]{Viti04}.
There has so far been no direct observational confirmation of the co-desorption of COMs with water, mainly because the angular resolution of observations has been insufficient to resolve the temperature gradient as traced by the COM emission, which is needed to pinpoint their desorption temperatures. The alternative picture, in which COMs do not desorb alongside water but according to their own independent binding energies, would lead to the appearance of COMs at different temperatures even below 100\,K. This dependence on binding energy has neither been confirmed nor refuted observationally, either, again due to observational limitations.

Moreover, observations of COMs in the absence of protostars in cold ($T\lesssim20$\,K) regions, for example, the dark cloud TMC--1 \citep[e.g.,][and references therein]{Agundez21} or prestellar cores \citep[][]{Bacmann12,Jimenez-Serra16,Jimenez-Serra21,Scibelli20} but also in protostellar outflows such as L1157 \citep{Codella15,Codella17} and, for example, the shocked region (while quiescent in star formation) G+0.693$-$0.027 located in the Galactic centre region \citep[e.g.,][]{Requena-Torres06,Zeng18}, point to additional or purely non-thermal desorption processes. 
These processes may include grain-sputtering and subsequent release of COMs into the gas phase due to shock passing induced by outflows, accretion shocks \citep[e.g.,][]{Csengeri19}, or cloud interactions, desorption after interaction of molecules on dust grain surfaces with cosmic rays \citep[][]{Shingledecker18,Paulive21}, or other diffusive and non-diffusive reactions taking place on the dust surface that lead to the chemical desorption of COMs \citep[e.g,][]{Ruaud15,Jin20}.

The investigation of COMs and the many possible ways for them to arrive in the gas phase is one of the objectives of the ReMoCA (Re-exploring Molecular Complexity with ALMA) survey \citep[][]{Belloche19}, an imaging spectral line survey performed at 3\,mm with the Atacama Large
Millimetre/submillimetre Array (ALMA) towards the giant molecular cloud Sagittarius\,B2 (hereafter Sgr\,B2), which is located at a distance of 8.2\,kpc from the sun \citep{Reid19} in the Galactic centre region. Sgr\,B2 hosts two sites of active high-mass star formation.
Amongst them, Sagittarius\,B2\,North (hereafter Sgr\,B2\,(N)) has a high record of first detections of interstellar molecules turning it into an excellent target to look out for COMs.
Observations towards this region revealed several high-mass protostars with hot cores \citep{Bonfand17}, H{\small II} regions \citep{Gaume95,DePree15}, filaments through which accretion probably happens \citep[][]{Schwoerer19}, and outflows \citep[][]{Higuchi15,Bonfand17}. Therefore, it fulfils the requirements to study all thermal and non-thermal desorption processes mentioned above.

Before ReMoCA, its predecessor, the EMoCA survey \citep[Exploring Molecular Complexity with ALMA,][]{Belloche16}, which itself succeeded a spectral line survey of Sgr\,B2\,(N) and (M) with the IRAM\,30\,m single-dish telescope \citep[][]{Belloche13}, already allowed to study Sgr\,B2\,(N) in the 3\,mm spectral window at a high angular resolution of 1.6\arcsec. Insights that were obtained from the data of this survey included first detections of new COMs such as \textit{i}-C$_3$H$_7$CN \citep{Belloche14} and CH$_3$NHCHO \citep[][]{Belloche17,Belloche19}, a detailed study on the complex chemistry in translucent clouds along the line of sight \citep[][]{Thiel17}, the discovery of three new hot cores and their outflows, if present \citep{Bonfand17}, and an inventory of COMs in the hot cores Sgr\,B2\,(N2--N5) \citep[][]{Bonfand19}.
The latter analysis does not include the main and most massive hot core Sgr\,B2\,(N1) because the observed line forest at the confusion limit, spectral lines that have full-width at half-maximum (FWHM) of $\sim$7\,\kms and additional line wings, and the high column density and optical depth of the dust continuum made it challenging to probe the inner part of this hot core at this wavelength. Now, with the ReMoCA survey and its sub-arcsecond angular resolution and four-times better sensitivity it becomes possible to resolve the COM emission in Sgr\,B2\,(N1).

In this work, we aim to shed light on the thermal desorption process, in particular on the question whether COMs co-desorb with water or rather depending on their binding energies.
To do so we study the morphology of the emission of 16 organic molecules and
choose ten of them, all COMs, for further analysis. The spectra of these ten COMs (including one isotopologue) are modelled under the assumption of Local Thermodynamic Equilibrium (LTE) and population diagrams are derived. Based on this analysis, we derive resolved profiles of rotation temperature and COM abundance starting from $\sim$0.5\arcsec distance from the continuum peak and proceeding to the point of non-detection for each COM. By comparing these profiles with results from state-of-the-art astrochemical models, we want to trace (back) each species' evolution in the solid and gas phases, and the transition from one phase to the other in order to answer the question above. 
The article is structured as follows: Section\,\ref{s:obs} provides details on the used data and the LTE modelling of the spectra. In Sect.\,\ref{s:results} we present our results including an analysis of the continuum data, a description of the morphology of the COM emission, and the path towards derivation of the COM abundance profiles. The discussion on the results is done in Sect.\,\ref{s:discussion} including the comparison to chemical models and a conclusion is provided in Sect.\,\ref{s:conclusion}.

\section{Observations and method of analysis}\label{s:obs}
\subsection{The ReMoCA survey}
The observations of the ReMoCA line survey were conducted with the ALMA interferometre and cover the complete frequency range from 84.1 to 114.4\,GHz. 
The phase centre is located at $(\alpha,\delta)_\mathrm{J2000}=(17^\mathrm{h}47^\mathrm{m}19\overset{s}{.}87,-28^\circ22^\prime16\overset{\arcsec}{.}00)$ halfway between Sgr\,B2\,(N1) and (N2). 
The survey is split into five observational setups, called S1 to S5, each of which delivers four spectral windows. The observations were performed with different antenna configurations, which yield mean synthesised beam sizes that vary from $\sim$0.75\arcsec in setup 1 to 0.3\arcsec in setup 5. The detailed list of covered frequency ranges, beam sizes, and average rms noise levels for all spectral windows can be found in Table 2 of \citet[][]{Belloche19}. 
We refer to the latter article for more details on the observations and the data reduction. The FWHM of the primary beam varies between 69\arcsec at 84\,GHz and 51\arcsec at 114\,GHz \citep[][]{ALMAc4}. The reduced spectra have a spectral resolution of 488\,kHz, which translates to 1.7--1.3\kms.

\subsection{1.3\,mm ALMA data}
\citet{Sanchez-Monge17} observed Sgr\,B2\,(N) and (M) with ALMA using an extended configuration. The continuum map used here was obtained at 242\,GHz with a synthesised beam of 0.4\arcsec. The average rms noise level in the map is 12.7\,mK. The phase centre is located at  $(\alpha,\delta)_\mathrm{J2000}=(17^\mathrm{h}47^\mathrm{m}19\overset{s}{.}887,-28^\circ22^\prime15\overset{\arcsec}{.}76)$ but is shifted to the phase centre of the ReMoCA data in this work. More details on these observations are presented in \citet{Sanchez-Monge17}.

\subsection{LTE modelling}

To analyse the spectral line data we perform radiative transfer modelling under conditions of LTE, which is appropriate given the high densities found towards Sgr\,B2\,(N) \citep[$\sim10^7$\,cm$^{-3}$,][]{Bonfand17}. 
The spectral line emission has already been split from the continuum emission as reported by \citet{Belloche19} (see their paper for details), who performed a preliminary baseline subtraction, which has been improved subsequently \citep{Melosso20}. Another subtraction is performed in this work, when residual baseline issues are identified, however, only for spectra at distances larger than 2\arcsec from Sgr\,B2\,(N1) because of the difficulty of distinguishing between the continuum and the pervasive line emission.

The line identification and modelling of the spectra are done with Weeds \citep[][]{Maret11}, which is an extension of the GILDAS/CLASS software\footnote{\url{https://www.iram.fr/IRAMFR/GILDAS/}}.
To produce a synthetic spectrum, Weeds requires five input parameters, which are: total column density, rotational temperature, source size, velocity offset with respect to the systemic
velocity of the source, and line width (FWHM).
We perform 1D Gaussian fits to optically thin and unblended transitions, from which we obtain the velocity offset and the line width.  
Column density and rotational temperature are determined in an iterative way. From a first model, in which these parameters are selected by eye, we derive population diagrams (see Sect.\,\ref{ss:profiles}), which also yield column density and rotational temperature. The Weeds parameters are adjusted until the model and the results of the population diagram are in good agreement.
The molecular emission is assumed to follow a 2D Gaussian distribution in Weeds. Because the COM emissions show extended but also small-scale structures, the determination of the source size is challenging and will be discussed in more detail in Sect.\,\,\ref{sss:COMmorph}. 

\begin{figure*}
\centering
    \hspace{-0.5cm}
    \includegraphics[width=\textwidth]{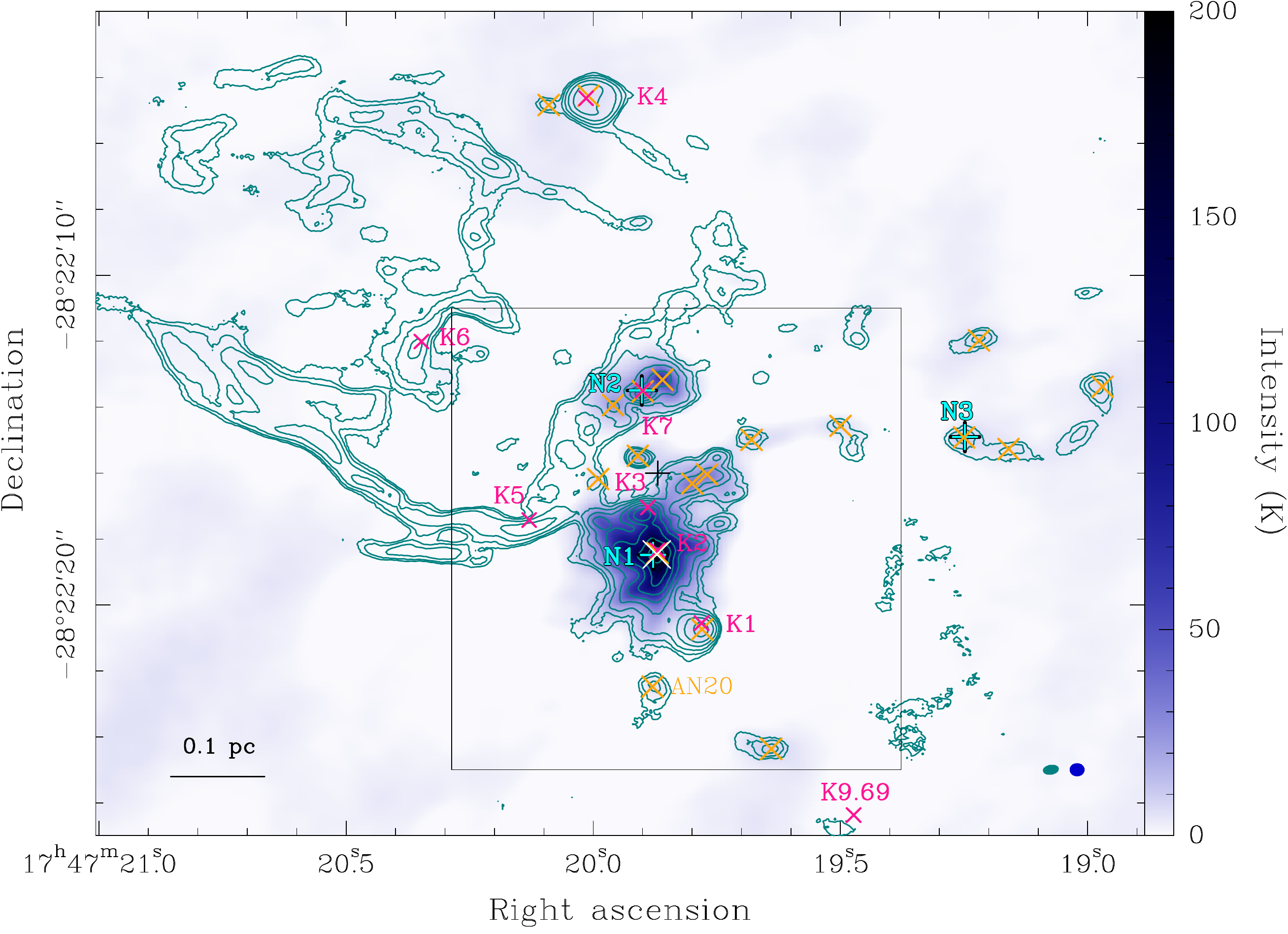}
    \caption{Continuum maps at 242\,GHz \citep{Sanchez-Monge17} shown in colour scale and at 99.2\,GHz (ReMoCA) shown with teal contours. The contour steps start at 4$\sigma$ and then increase by a factor of 2 with $\sigma=0.4$\,K the rms noise level of the continuum map at 99.2\,GHz. The black cross shows the phase centre of the ReMoCA observations. The white cross shows the intensity-weighted 3\,mm continuum peak determined in this work. Orange crosses indicate 1.3\,mm continuum sources from \citet{Sanchez-Monge17}, cyan crosses the continuum peaks of the hot cores N1 and N2 as well as the line density peak of N3 reported by \citet{Bonfand17} on the basis of the EMoCA survey
    at 1.6\arcsec resolution, and pink crosses H\small{II} regions reported by \citet{Gaume95} (K1--K6, K9.69) and \citet{DePree15} (K7) with the VLA. The black box indicates the region for which spectral indices are determined as shown in Fig.\,\ref{fig:spectral-index}. The respective synthesised beams for both maps are shown in the lower right corner. }
    \label{fig:continuum-maps}
\end{figure*}  
\begin{figure}
    \centering
    \includegraphics[width=0.45\textwidth]{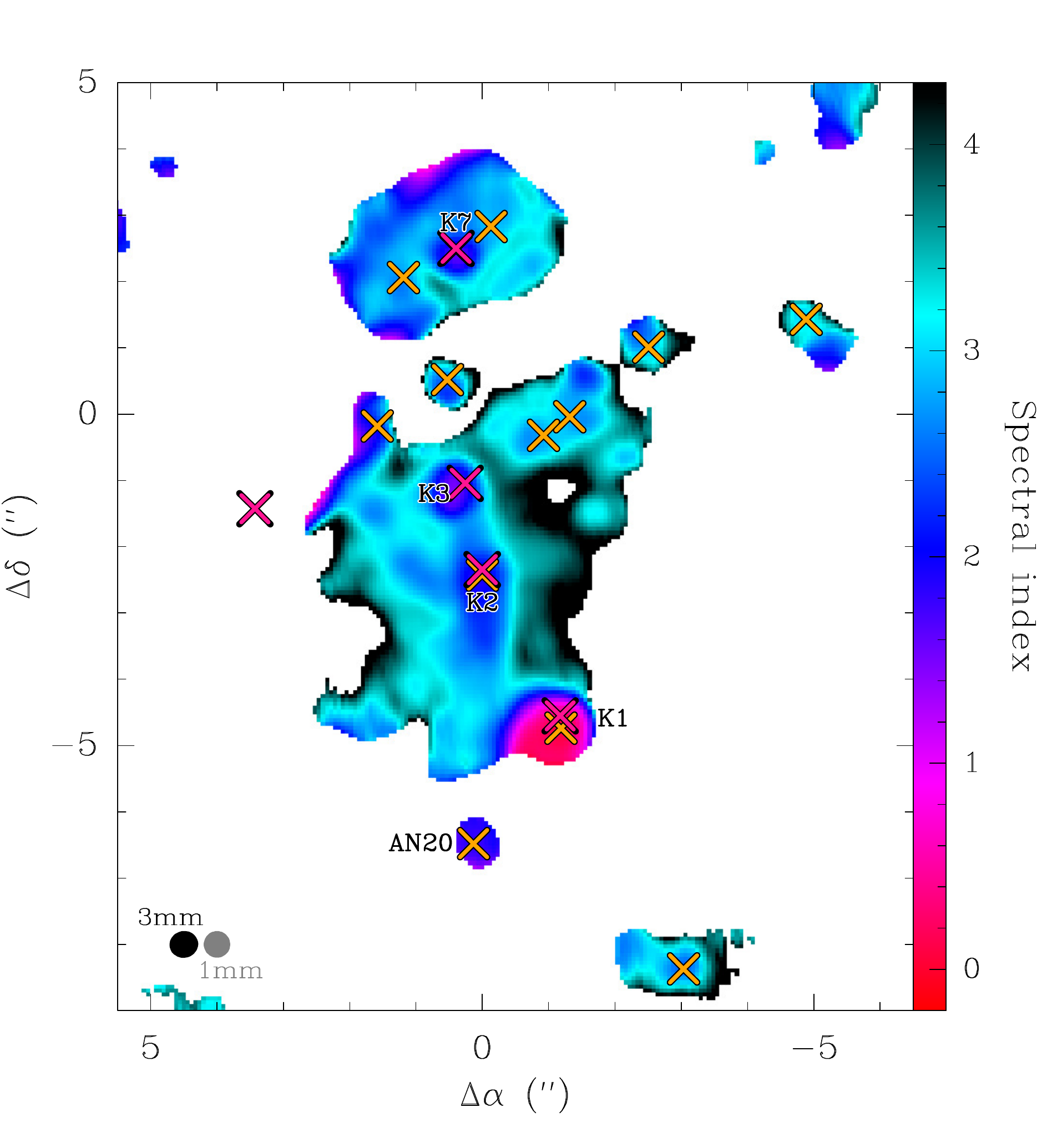}
    \caption{Spectral index map derived from the 1.3 and 3\,mm continuum emission for the region around N1 and N2 indicated with the black box in Fig.\,\ref{fig:continuum-maps}. The crosses indicate the same sources of respective colour as in Fig.\,\ref{fig:continuum-maps}. The synthesised beams (after smoothing for the 3\,mm map) are shown in the lower left corner. The position offsets are given with respect to the ReMoCA phase centre.} 
    \label{fig:spectral-index}
\end{figure}

A species is identified when the synthetic spectrum correctly predicts each observed transition. Models of all analysed species are subsequently combined to a complete synthetic spectrum, in which blending of lines is treated by adding up the model predictions of individual species. More details on the modelling procedure can be found in \citet{Belloche16}.

The spectroscopic calculations used to create the models are based on the CDMS \citep[Cologne Database for Molecular Spectroscopy,][]{CDMS} or JPL \citep[Jet Propulsion Laboratory,][]{JPL} catalogues in most instances. We provide details on the laboratory background and on the vibrational spectroscopy in Appendix\,\ref{app:specs}. The former is necessary to give credit to at least the most important contributors who often carried out the respective measurements to support the radio astronomical community. The latter is important to account for the molecules in excited vibrational states which are populated at excited temperatures such as in our work but may be characterised too little or not at all or may be too weak to be detected securely.

\section{Astronomical results}\label{s:results}
\subsection{Continuum emission at 1.3\,mm and 3\,mm} \label{sss:continuum}

Figure\,\ref{fig:continuum-maps} shows the continuum map at 1.3\,mm \citep[242\,GHz,][]{Sanchez-Monge17} in colour scale and a ReMoCA continuum map at 3\,mm (99.2\,GHz) with contours. The continuum at 1.3\,mm is dominated by dust emission and peaks towards the two main hot cores N1 and N2. 
We show continuum peak positions of Sgr\,B2\,(N1) derived by various authors along with the position determined by us in the following way: using continuum maps obtained from setup 5, which is the ReMoCA setup of highest angular resolution, and computing the peak position by intensity-weighting the four brightest pixels, we locate the 3\,mm continuum peak at (0.02\,$\pm$\,0.03,$-$2.47\,$\pm$\,0.04)\arcsec from the phase centre (see the white cross in Fig.\,\ref{fig:continuum-maps}).

Besides N1 and N2, hot core N3 is indicated, which has only been recently discovered \citep[together with N4 and N5, which are located outside the field shown in Fig.\,\ref{fig:continuum-maps},][]{Bonfand17}.
Further continuum sources at 1.3\,mm have been identified by \citet{Sanchez-Monge17}, those of which located within the shown field are indicated with orange crosses in Fig\,\ref{fig:continuum-maps}. 
Each of these sources has a counterpart in the 3\,mm continuum map as can be seen from the contours suggesting that a substantial part of the continuum emission traces dust. 
One exception is H{\small II} region K1 which is seen at both wavelengths but whose emission seems to be dominated by free-free emission even at 1.3\,mm \citep[see Fig.\,\ref{fig:spectral-index} and][]{Sanchez-Monge17}.
Moreover, there are features that are not seen at 1.3\,mm, for example, the large-scale structure to the north-east. It resembles the one reported by \citet{Gaume95}, who performed observations of the continuum towards Sgr\,B2\,(N) at 1.3\,cm with the VLA. 
At this wavelength, free-free emission originating from H{\small II} regions dominates over dust. Therefore, the continuum at 3\,mm is not exclusively tracing the dust but also free-free emission to a substantial extent. The positions of the H{\small II} regions reported by \citet{Gaume95} and \citet{DePree15} are indicated with pink crosses in Fig.\,\ref{fig:continuum-maps}, and all (except K9.69) are associated with intensity peaks in the 3\,mm continuum map. 

We created a spectral index map using the 1.3 and 3\,mm continuum maps and show the result in Fig.\,\ref{fig:spectral-index}. The spectral index $\alpha$ is computed by  
\begin{align}
    \alpha = \ln\left(\frac{S_\mathrm{242}}{S_\mathrm{99.2}}\right)\times\frac{1}{\ln\left(\frac{242\,\mathrm{GHz}}{99.2\mathrm{GHz}}\right)},
\end{align}
where $S_\mathrm{242}$ and $S_\mathrm{99.2}$ are the flux densities measured at each position in the continuum map of the respective frequency. 
We smoothed the 3\,mm data to the resolution of the 1.3\,mm data for a more precise analysis. 
The spectral index is only computed above a flux density threshold of 1$\sigma$, where $\sigma$ is the average noise level in the respective continuum map. 
For most parts in Fig.\,\ref{fig:spectral-index}, the spectral index has values ranging from 2.5 to $\gtrsim$4 suggesting prevalence of optically thin dust emission. In some regions the spectral index is smaller than 2--2.5 including the H{\small II} regions K1, K3, and K7 \citep[][]{Gaume95}, but also AN\,20 \citep[][]{Sanchez-Monge17}, whose nature is not certainly known. 
Besides at 99.2\,GHz and 242\,GHz, the latter source was detected at 40\,GHz \citep{Rolffs11a,Sanchez-Monge17} suggesting a possible contribution of free-free emission to the observed flux.  K1 even shows values $<$1, which is most likely the result of a mixture of dust and free-free emission \citep[see also][]{Sanchez-Monge17}.

We also expect a contribution of free-free emission originating from K2 to the continuum at the position of Sgr\,B2\,(N1) as they are nearly co-located. Because we want to derive H\2 column densities from the 3\,mm dust emission at positions close to Sgr\,B2\,(N1) it is important to disentangle the contributions of free-free and dust emission, particularly at this position. \citet{DePree15} reported a flux density of 80\,mJy for K2 at 44.2\,GHz for a size of the emitting region of 0.22$^{\prime\prime}$\,$\times$\,0.1\arcsec. 
Given a total peak continuum flux of 190\,mJy/beam at the position of Sgr\,B2\,(N1) at 99.2\,GHz for a beam size of 0.4\arcsec, free-free emission contributes $\sim$40\% to this flux assuming that the free-free emission is optically thin ($\alpha_{\rm ff,thin}=-0.1$) above 44.2\,GHz. 

\begin{figure}
    \hspace{-0.7cm}
    \includegraphics[width=0.52\textwidth]{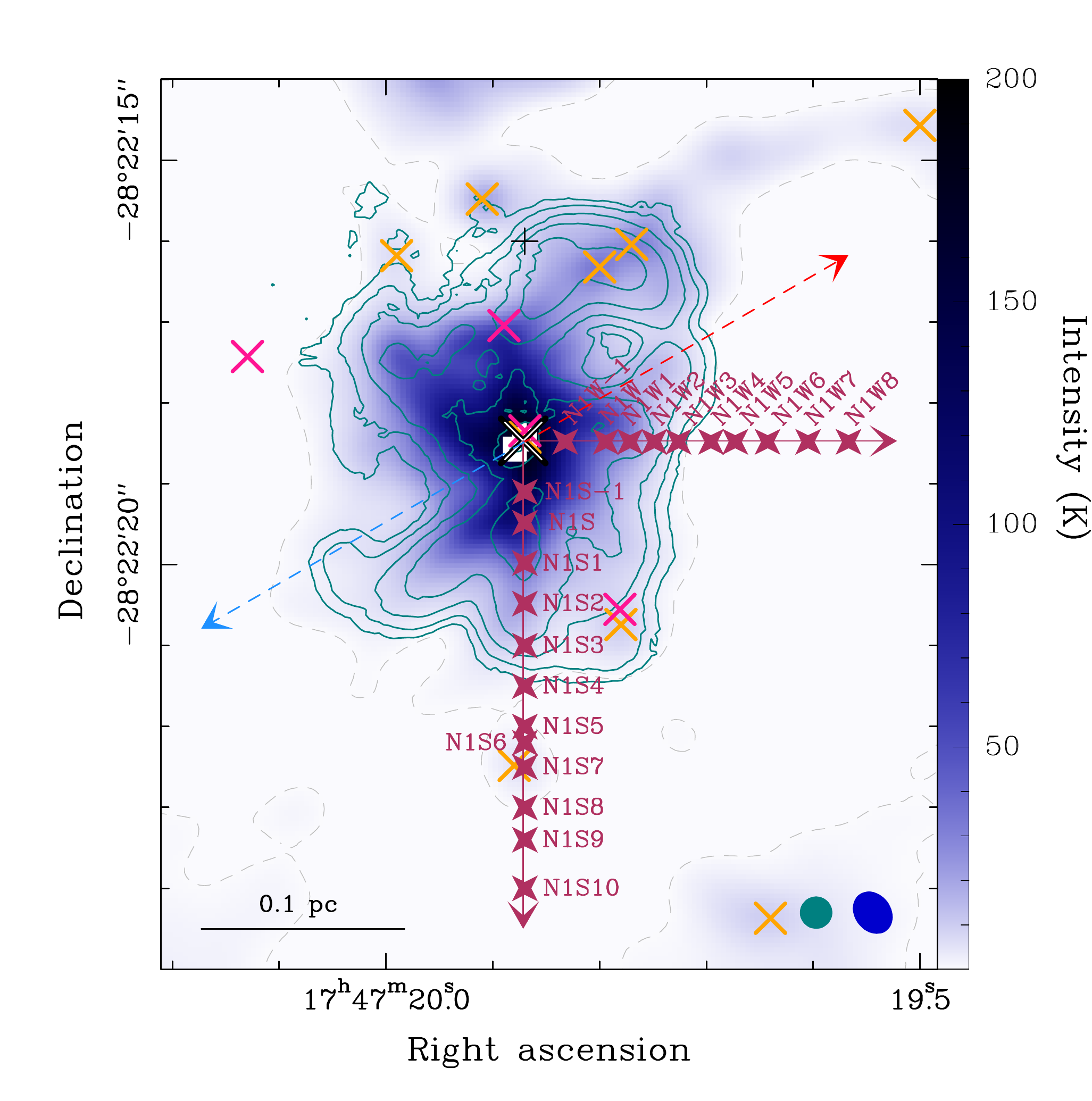}
    \caption{Continuum at 242\,GHz \citep{Sanchez-Monge17} in colour scale overlaid by a Line-width- and Velocity-corrected INtegrated Emission (LVINE) map of \et (97.139 GHz, $E_u=264$\,K) shown with teal contours. The contour steps start at 3$\sigma$ and then increase by a factor of 2 with $\sigma=8.3$\,K\,\kms. The grey dashed contour indicates the 3$\sigma$ level of the continuum with $\sigma=12.7$\,mK. The closest region around Sgr\,B2\,(N1) is masked out due to high continuum optical depth (see Appendix\,\ref{app:Cmask}).
    Coloured crosses indicate continuum sources introduced in Fig.\,\ref{fig:continuum-maps}. The black cross shows the phase centre of the ReMoCA observations. Red and blue dashed arrows indicate the outflow axis reported by \citet{Higuchi15}. The beams of the maps are shown in the respective colour in the lower right corner. Maroon arrows indicate the directions along which positions (maroon star symbols) are chosen for the subsequent analysis (see Sect.\,\ref{sss:positions}).}
    \label{fig:1mm+et}
\end{figure}

An advantage of the 3\,mm continuum is that dust opacities are smaller at this wavelength than at 1.3\,mm. 
To confirm this, we estimate the optical depth $\tau$ at the position of Sgr\,B2\,(N1) and the position N1S (0.00,$-$3.48)\arcsec (see Fig.\,\ref{fig:1mm+et}), which was used for analysis in \citet{Belloche19}. The flux density at the position of Sgr\,B2\,(N1) is corrected for the contribution of free-free emission as derived above, while for N1S the contribution is negligible.
We use the equation presented in \citet{Bonfand17} which is
\begin{align}\label{eq:tau}
   \tau = -\ln\left(1-\frac{S_\nu^\mathrm{beam}}{\Omega_\mathrm{beam}B_\nu(T_d)}\right),
\end{align}
where $\Omega_\mathrm{beam} = \frac{\pi}{4\ln 2}\times HPBW_\mathrm{min}\times HPBW_\mathrm{maj}$ is the beam solid angle, $S_\nu^\mathrm{beam}$ the peak flux density, and $B_\nu(T_d)$ the Planck function of the dust emission evaluated for the dust temperature $T_d$. 
Because the optical depth is sensitive to the dust temperature, we assume two values, which are $T_d = 200$\,K and 250\,K. We use the smoothed map for a direct comparison of the 1.3 and 3\,mm continuum and therefore, a beam size of $HPBW_\mathrm{min} = HPBW_\mathrm{maj}= 0.4$\arcsec. 
\begin{table}[]
    \caption{Optical depth at 1.3 and 3\,mm assuming two different dust temperatures. }
    \centering
    \begin{tabular}{lccccc}
       \hline\hline\\[-0.2cm]
        Position &  $\Delta\alpha$ (\arcsec) & $\Delta\delta$ (\arcsec) & $T_d$ (K) & $\tau_\mathrm{1.3mm}$ & $\tau_\mathrm{3mm}$ \\[0.1cm]\hline\\[-0.3cm]
        N1 & 0.02 & $-$2.47 & 200 & 2.58 & 0.83 \\ 
        & & & 250 & 1.32 & 0.60 \\ 
        N1S & 0.00 & $-$3.48 & 200 & 0.67 & 0.40 \\
        & & & 250 & 0.49 & 0.30 \\
        \hline\hline
    \end{tabular}
    \label{tab:tau}
\end{table}
The results are summarised in Table\,\ref{tab:tau} and show that opacities are lower for the observations at 3\,mm at both positions. We have a closer look at the optical depth in Sect.\,\ref{ss:NH2} when we derive H\2 column densities.

\begin{figure*}
    \centering
    \includegraphics[width=\textwidth]{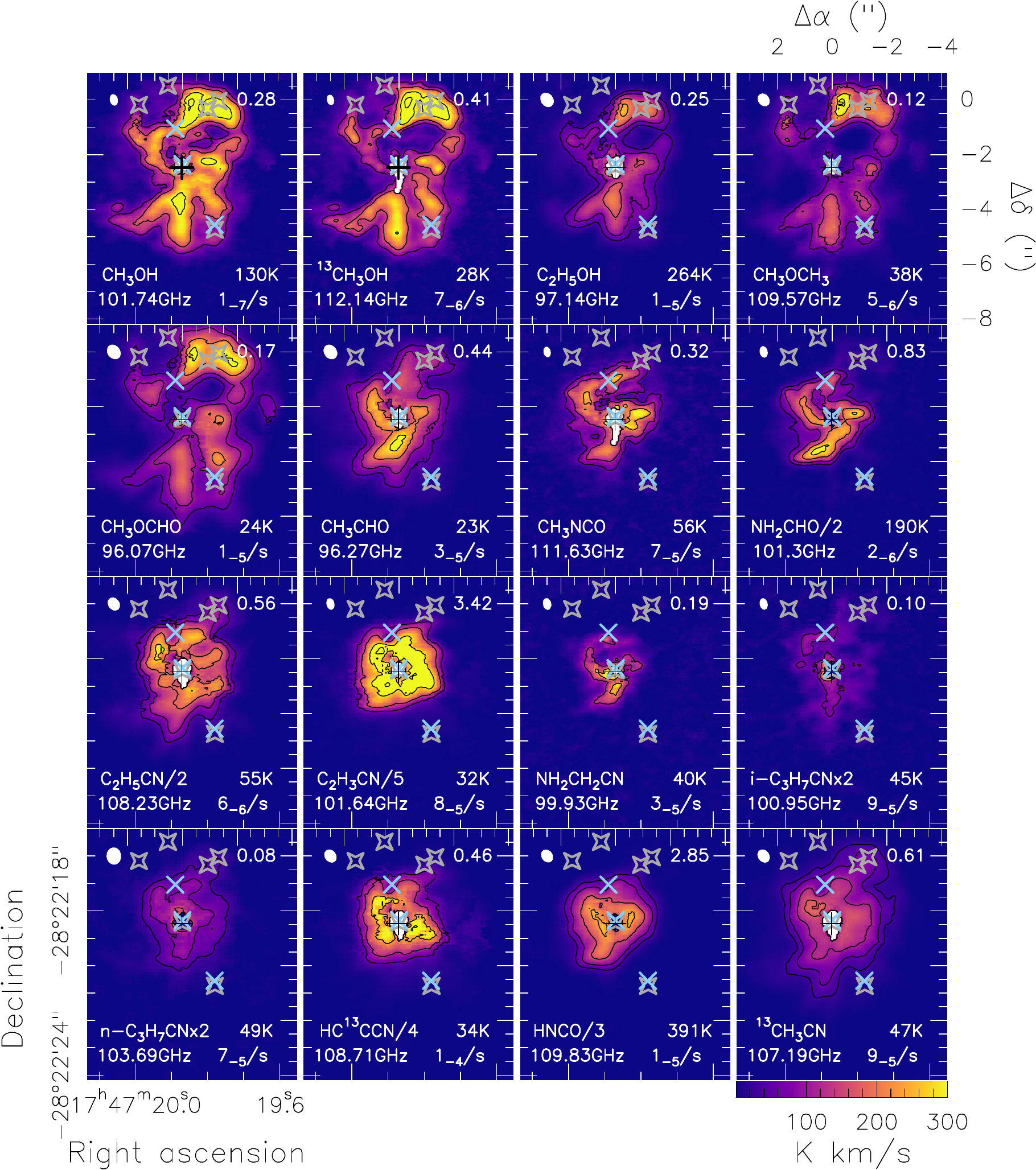}
    \caption{LVINE maps of various COMs. Integrated intensities of \etc are scaled down by a factor of 2, those of \vc by a factor 5, \cyano by a factor 4, HNCO by a factor 3, and NH\2CHO by a factor 2. Integrated intensities of \isoc and \nsoc are scaled up by a factor 2, respectively. Because the noise level within each map is not uniform due to the pixel-dependent integration limits, contours are at 15\%, 30\%, 60\%, and 90\% of the respective peak intensity and additional signal-to-noise maps are shown in Fig.\,\ref{fig:snr}. The name of the COM and the frequency of the used transition are shown in the lower left corner. Upper level energies and Einstein A coefficients (where $x_{-y}/{\rm s} = x\times 10^{-y}$\,s$^{-1}$) are shown in the lower right, optical depth at position N1S in the upper right, and the synthesised beam in the upper left corner. The closest region around Sgr\,B2\,(N1) is masked out due to high frequency- and beam-size-dependent continuum optical depth (see Appendix\,\ref{app:Cmask}). The black cross indicates the position of Sgr\,B2\,(N1) determined in this work. H{\small II} regions reported by \citet{Gaume95} are shown with blue crosses, continuum sources reported by \citet{Sanchez-Monge17} with grey polygons.}
    \label{fig:COM-Maps}
\end{figure*}%

An effect biasing the comparison between the continuum levels may be spatial filtering by the ALMA interferometre, which occurs as a consequence of limited $(u,v)$ coverage at shortest baselines. This leads to a lower sensitivity to larger-scale emission, that is, emission that is extended beyond a certain spatial scale might be filtered out, which eventually leads to a decrease of the observed flux density. This maximum recoverable scale depends on the length of the shortest baseline \citep{ALMAc4}.
For the observation at 1.3\,mm, 34--36 antennae were used covering baselines from 30\,m to 650\,m and hence, ensuring sensitivity to structures on scales of 0.4--5\arcsec at an angular resolution of 0.4\arcsec  \citep[][]{Sanchez-Monge17}. 
The ReMoCA survey has been observed with different configurations using 36--46 antennae that cover baselines from 15\,m to 3600\,m \citep[cf. Table\,2 in][]{Belloche19}. Therefore, maximum scales that can be recovered at 100\,GHz are between $\sim$3\arcsec and $\sim$8\arcsec for the highest and lowest angular resolutions, respectively \citep[see Table\,7.1 in][]{ALMAc4}.  
Because the two datasets have a similar maximum recoverable scale the spectral index map in Fig.\,\ref{fig:spectral-index} should be reliable for the scales traced by both. 

\subsection{Spatial distribution of COMs}\label{sss:COMmorph}

To explore the morphology of the COM emission we created Line-width- and Velocity-corrected INtegrated Emission maps (LVINE). The LVINE method is an extension of the VINE method, which was introduced by \citet{Calcutt18}, and takes into account gradients of velocity and spectral line width in the observed region. The LVINE method is explained in more detail in Appendix\,\ref{app:LVINE}.

Figure\,\ref{fig:1mm+et} shows the 1.3\,mm continuum map in colour scale, which is overlaid by contours of an LVINE map of a \et transition at 97.139\,GHz ($E_u=264$\,K). The continuum exhibits a lot of structure which was associated with filaments by \citet{Schwoerer19}, who described the continuum emission at 1.3\,mm in detail. 
The authors reported on the differences in the morphology of O- and S-bearing COM emission that follows the structured continuum emission, which can also be seen for \et in Fig.\,\ref{fig:1mm+et}, and N-bearing species, which reveal a more spherical morphology. In order to investigate these reported differences we show LVINE maps of emission from various COMs in Fig.\,\ref{fig:COM-Maps}.

Similarly to \et, the emissions of \met, its $^{13}$C isotopologue, \dme, and \mf, all of which are O-bearing COMs, closely follow the morphology of the continuum. The distribution of \ad emission is slightly different, however, it shows the tendency of the morphology seen for the other O-bearing COMs. The emissions of the N-bearing COMs such as $^{13}$CH\3CN, \etc, \vc, and \cyano share a similar morphology that is less structured and more compact around the central region of Sgr\,B2\,(N1), with the latter two showing the smallest spatial extent. 
The morphology of HNCO emission is similar to the N-bearing COMs.
The spatial distributions of \mic, NH\2CHO, and \aan emissions are similar, with the latter being more compact, and follow a slightly different morphology as it shows the compactness of the N-bearing COMs with a pattern similar to what was seen for the O-bearing COMs. \isoc and \nsoc emissions are extremely compact and weak and cannot be compared to the other COMs.
In summary, we observe a spatial segregation of species, approximately between O- and N-bearing COMs. Moreover, there may even be indications of a third group.  

The determination of the physical size of the emission region is challenging because the morphology for all COMs does not simply follow a 2D Gaussian profile as assumed by Weeds, instead it shows extended emission as well as structures on smaller scales. In the following we assume that we always resolve the emission region. In the Weeds model we set the size (FWHM) of the emission region to 2\arcsec, which yields a beam filling factor of approximately 1 and thus, satisfies the requirement. 

In the region closest to Sgr\,B2\,(N1) we find that the continuum becomes optically thick and obscures the COM emission. Therefore, the true intensity of the COM emission is unknown and this inner region is masked in the maps shown in Figs.\,\ref{fig:1mm+et} and \ref{fig:COM-Maps}. The determination of the mask size is explained in Appendix\,\ref{app:Cmask}. It varies between the observational setups and spectral windows because the opacity effect on the continuum is higher for higher frequency and higher angular resolution.

\subsection{Line and position selection}\label{sss:positions}
We focus our analysis on commonly abundant O- and N-bearing COMs, that show numerous strong lines and have enough lines that are not optically thick. This selection comprises \met (methanol), \et (ethanol), \dme (dimethyl ether), \mf (methyl formate), \ad (acetaldehyde), \mic (methyl isocyanate), \etc (ethyl cyanide), \vc (vinyl cyanide), and \fmm (formamide). 

As discussed above, optical depth of the continuum at its peak, and therefore obscuration of the COM emission, prevents us from starting our analysis at this position. 
Therefore, \citet{Belloche19} selected a position that is 1\arcsec to the south of Sgr\,B2\,(N1) for their analysis and called it N1S. Spectral lines at this position suffer less from masking by the continuum. Moreover, they have moderate average line widths of $\sim$\,5\kms and do not show wings, which would be indicative of emission originating from the outflow of Sgr\,B2\,(N1) \citep[][]{Higuchi15}. 
Based on this reasoning we decided to start from this position and go further south in order to determine the COM rotational temperature and column density profiles along this direction. The positions are chosen so that they are approximately one beam size apart (0.5\arcsec). The transition to source AN\,20 is analysed with an additional position (N1S6). The analysis is continued up to the distance where none of the selected COMs is detected any longer. This maximum distance varies for different COMs and can either result from the limited sensitivity of the observations or is due to the prevailing chemical or physical conditions. 
The used positions to the south are indicated with maroon star symbols in Fig.\,\ref{fig:1mm+et} and listed in Table\,\ref{tab:positions}, where we adopt the naming from \citet{Belloche19} and term positions N1Si (i=1,2,...) with increasing distance starting from N1S. Additionally, we study the position N1S-1, which is at a distance of 0.6\arcsec from Sgr\,B2\,(N1). During the analysis we find that this position is already affected by the high continuum optical depth, severely in spectral windows 2 and 3 of observational setup 5, which cover the highest frequencies and have the highest angular resolution. Therefore, transitions covered by these spectral windows at N1S-1 are not used in the following (see also Appendix\,\ref{app:Cmask}).

The results obtained southwards are compared to another direction. Because in this study we focus on the COMs' behaviour under the influence of heating by the protostar, that is the process of thermal desorption, we do not use directions towards the south-east and north-west as they correspond to the axis of the Sgr\,B2\,(N1) outflow \citep{Higuchi15}. 
The outflow induces shocks in the ambient gas which may trigger non-thermal desorption of COMs. The impact of the outflow on the COM emission will be topic of a forthcoming paper. 
To the north, there are multiple continuum sources that may bias the results of COM abundances and to the north-east, we face uncertain contributions of free-free emission (see Fig.\,\ref{fig:continuum-maps}) to the continuum. 
Therefore, these directions are not used, instead we decided to go west as indicated with maroon symbols in Fig.\,\ref{fig:1mm+et}. 
The positions are listed in Table\,\ref{tab:positions}. Because the molecular emission seemed slightly less spatially extended than to the south (cf. Fig.\,\ref{fig:COM-Maps}), the positions where the emission is more intense are chosen in a somewhat smaller distance from each other.
Similar to N1S, we refer the distance at 1\arcsec from Sgr\,B2\,(N1) to as N1W and then increase numbers with increasing distance. Additionally, we look at position N1W-1 which is at a distance of 0.5\arcsec from Sgr\,B2\,(N1). Positions closer to Sgr\,B2\,(N1) in either direction are not used because the continuum becomes optically thick.

In Appendix\,\ref{app:tables} we provide the parameters used for the LTE modelling of the selected COMs at each position. Additionally, we show observed spectra for each COM at position N1S (or N1S1 for \dme) together with the synthetic spectra in Appendix\,\ref{app:spectra}. These figures only show transitions that have been used to create the respective population diagram (see Sect.\,\ref{ss:profiles}).

\subsection{COM rotational temperature and column density profiles}\label{ss:profiles}

\begin{figure}
    \hspace{-.3cm}
    \includegraphics[width=0.51\textwidth]{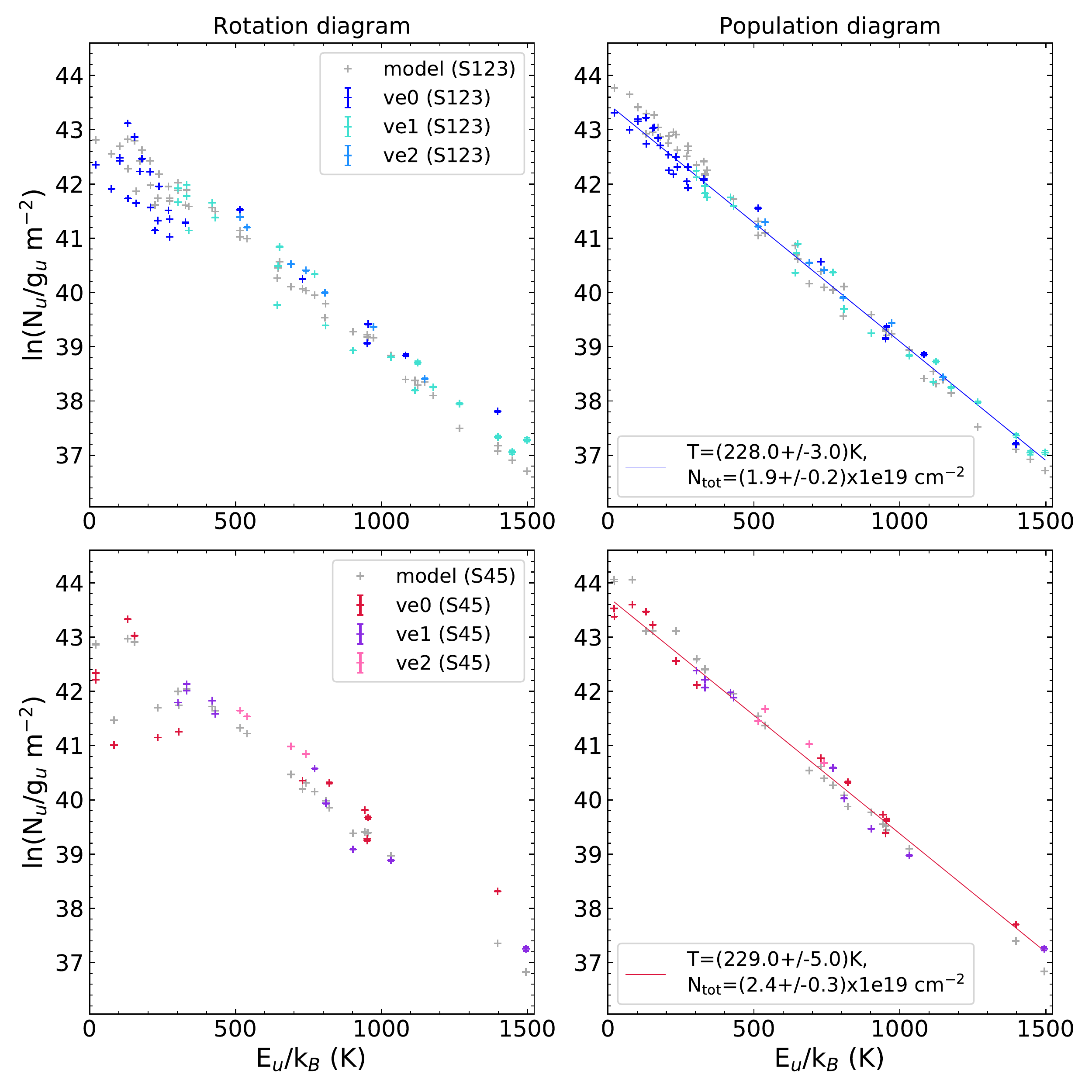}
    \caption{Population diagram for \met towards Sgr\,B2\,(N1S), where setups 1--3 and 4--5 are considered separately in the upper and lower row, respectively. Observed data points are shown in colours as indicated in the upper right corner in the left panel while the synthetic data points are shown in grey. No corrections are applied in the left panel while in the right panel corrections for opacity and contamination by other molecules have been considered for both the observed and synthetic populations. The blue and red lines are linear fits to the observed data points (in linear-logarithmic space) obtained with setups 1--3 and 4--5, respectively. The results of the fit are shown in the right panels. }
    \label{fig:PD-ch3oh-N1S}
\end{figure}

\begin{table}[]
    \caption{Slopes of the rotational temperature profiles shown in Fig.\,\ref{fig:Tprofiles}.}
    \centering
    \small
    \begin{tabular}{lcccc}
    \hline\hline\\[-0.3cm]
        COM & \multicolumn{2}{c}{South} & \multicolumn{2}{c}{West} \\ 
         & All\tablefootmark{a} & Far\tablefootmark{b} & All\tablefootmark{c} & Far\tablefootmark{d} \\[0.05cm] \hline\\[-0.3cm]
         \met & $-$0.67(4) & -- & $-$0.56(7) & $-$0.80(6) \\
         $^{13}$\met & $-$0.67(4) & -- & $-$0.68(5) & -- \\
         \et & $-$0.74(7) & -- & $-$0.73(5) & -- \\
         \dme & $-$0.69(5) & -- & $-$0.94(9) & --\\
         \mf & $-$0.84(8) & -- & $-$0.52(6) & -- \\
         \ad & $-$0.70(6) & -- & $-$0.82(13) & $-$1.16(21) \\
         \mic & $-$0.29(6) & $-$0.46(17) & $-$0.43(8) & $-$0.66(17) \\
         \etc & $-$0.52(8) & $-$0.74(3) & $-$0.48(6) & $-$0.59(9) \\
         \vc & $-$0.59(9) & -- & $-$0.48(2) & -- \\
         \fmm & $-$0.30(14) & $-$0.51(19) & $-$0.30(5) & $-$0.39(8)\\
    \hline\hline
    \end{tabular}
    \tablefoot{The values in parentheses show the uncertainties in units of the last digit. \\ \tablefoottext{a}{Linear fit is done using temperatures at all positions up to N1S4 (except for \vc (N1S3), \mic (N1S3) and \fmm (N1S2)).} \tablefoottext{b}{Linear fit is done for positions beyond 1\arcsec for \mic and \fmm or 1.5\arcsec for \etc (S) and up to maximum distances from (a).}\tablefoottext{c}{Linear fit is performed for all positions up to N1W4 for \met and $^{13}$\met or N1W3 for the rest, except for \dme (N1S--N1S2).}\tablefoottext{d}{Linear fit is done for positions beyond 1\arcsec and up to maximum distances from (c).}} 
    \label{tab:Tprofiles}
\end{table}

\begin{figure*}
    \centering
    \includegraphics[width=1.0\textwidth]{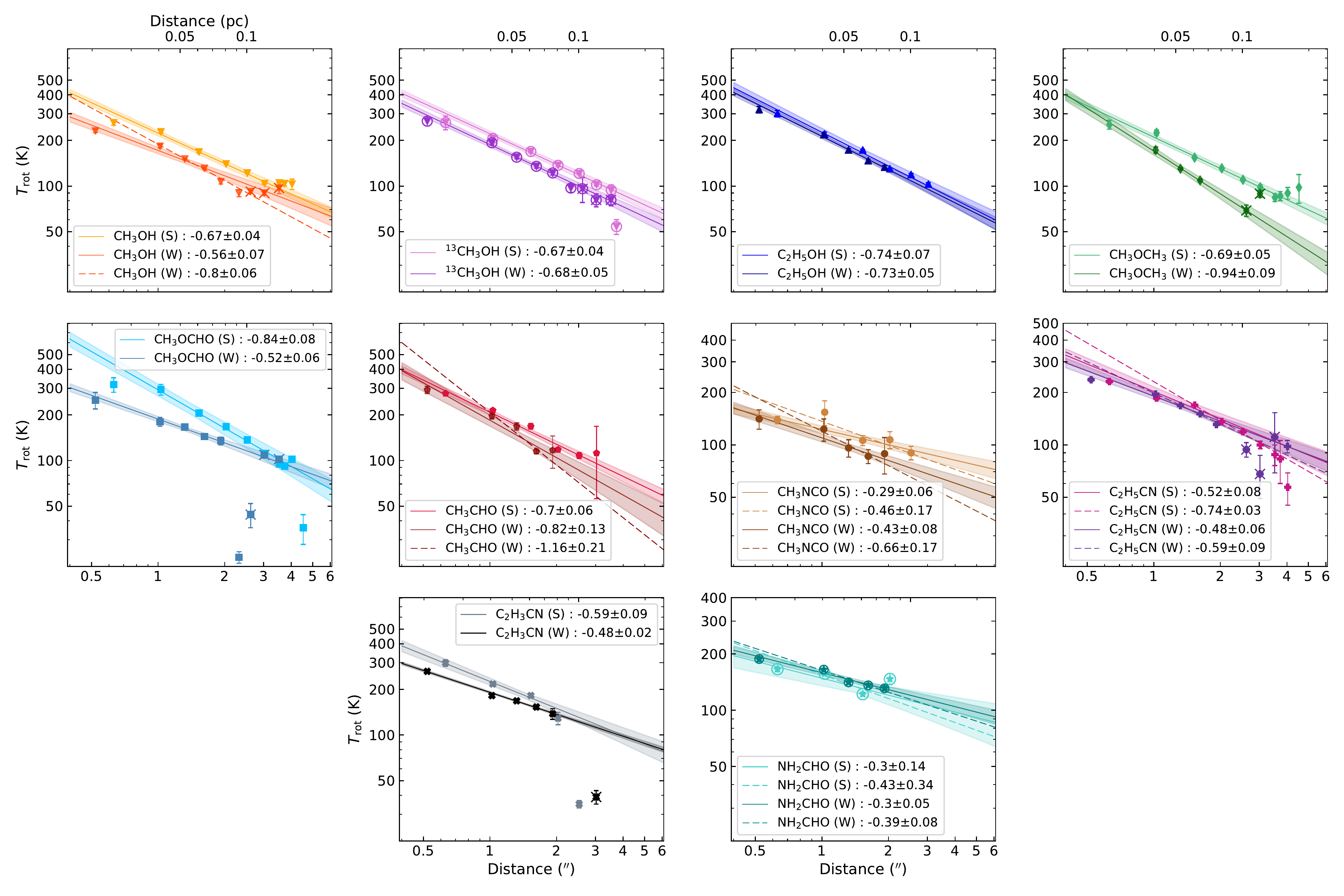}
    \caption{Rotational temperature profiles to the south (S) and west (W), where temperatures for all COMs are taken from the linear fit in the population diagrams using setups 4--5, except for $^{13}$\met and \fmm, for which setups 1--3 have to be used (encircled symbols). Solid lines show the fit including all positions up to N1W4 (except for \mf and \etc) to the west and up to N1S4 to the south, while dashed lines indicate the fit using only positions beyond 1\arcsec (\met (W), \ad (W), \mic (W), \etc (W), \fmm) or 1.5\arcsec (\etc (S)) and up to N1S4 and N1W4, respectively. Additional crosses on the markers indicate positions for which the velocity offset changed from $\lesssim$\,2 to $\sim$7\,\kms. The shaded areas indicate the uncertainty on the fit using all positions. The fit results are shown in the respective legend and are summarised in Table\,\ref{tab:Tprofiles}.}
    \label{fig:Tprofiles}
\end{figure*}

At each position selected in the previous section we derive population diagrams in order to determine the rotational temperature $T_\mathrm{rot}$ and column density $N_\mathrm{col}$ profiles for each COM. In the population diagrams we plot the natural logarithm of the upper level column density $N_u$ divided by the level degeneracy $g_u$ against the upper level energy $E_u$ \citep[][]{Mangum15} as they are related by 
\begin{align}
    \ln\left(\frac{N_u}{g_u}\right) = \ln\left(\frac{8\,\pi\,k_B\,\nu^2 \int J(T_B) \mathrm{d}\varv}{c^3\,h\,A_\mathrm{ul}\,g_u\,B}\right) = \ln\left(\frac{N_\mathrm{tot}}{Q(T_\mathrm{rot})}\right) - \frac{E_u}{k_B T_\mathrm{rot}},
\end{align}
where $k_B$ is the Boltzmann constant, $c$ the speed of light, $h$ the Planck constant,  $B=\frac{\mathrm{source\,size}^2}{\mathrm{source\,size}^2\,+\,\mathrm{beam\,size}^2}$ the beam filling factor, $A_\mathrm{ul}$ the Einstein A coefficient, $N_\mathrm{tot}$ the total column density, and $Q$ the partition function. Intensities in brightness temperature scale $J(T_B)$ are integrated over a manually selected velocity range d$\varv$. 

In Fig.\,\ref{fig:PD-ch3oh-N1S} the population diagram for \met towards N1S is shown as an example including data points for the observed as well as the modelled transitions. We use the vibrational ground state $v=0$ and the first and second torsionally excited states $v=(1,2)$ of \met, all of which are modelled with the same Weeds input parameters. 
We consider setups 1--3 and 4--5 separately, that is, we use different input parameters for both, if necessary, because of their different angular resolution that can vary by more than a factor of 2 (see Fig.\,C.1 of \citet{Motiyenko20} for another version of this population diagram of methanol that includes all setups). 
Without this separation, we might introduce larger scatter in the population diagram because different beams may see different portions of the gas. 
Because rotational temperatures derived from the population diagrams for either setups 1--3 or 4--5 differ by at most 3$\sigma$, where $\sigma$ is the error on the linear fit, and in roughly two thirds of the cases the difference is even $\lesssim$1$\sigma$ we only use the higher-resolution setups 4--5 to determine the temperature and column density profiles in the following. 
Except for $^{13}$\met and \fmm, for which we have to use setups 1--3, there are enough transitions of each COM detected in setups 4--5 to make reliable linear fits to the data in the population diagrams. Population diagrams for all analysed COMs and for all possible positions can be found in Appendix\,\ref{app:popdiagrams}. All vibrational states of a COM that are used during the analyses are indicated in these population diagrams and are summarised in Table\,\ref{tab:vibstates}.

We apply two corrections to the data points, which are only considered in the right panels of Fig.\,\ref{fig:PD-ch3oh-N1S} and Figs.\,\ref{fig:PD_met}--\ref{fig:wPD_fmm}. First, although we select only transitions which are not too contaminated some contamination is inevitable due to high spectral line intensity. 
Therefore, we create a complete Weeds model that includes all transitions of all species used for the analysis. Based on this model, the contamination of neighbouring transitions of at least other identified species is estimated and subtracted from the integrated intensities.

Second, to consider possible effects of line optical depth we multiply the integrated intensities of both the observed and modelled transitions, with a correction factor $\frac{\tau}{1-\mathrm{e}^{-\tau}}$ \citep[see][]{Goldsmith99,Mangum15}, where we use the opacities obtained from our Weeds model for the respective transition. Depending on the species, the upper level energy of a transition, and the distance to Sgr\,B2\,(N1), opacities can be extremely high, which cannot be properly accounted for in our simple Weeds model. Hence, the correction factor might not be able to counteract the effect on the intensity. This may be seen for the lowest upper-level energy transitions of \met from setups 4--5 in Fig.\,\ref{fig:PD-ch3oh-N1S}, for which we obtain an optical depth of $\sim$3 from the model. When corrected for it, the data points (red) remain slightly below the fitted line while the model (grey) overestimates the observation for these two transitions. Therefore, we only consider transitions in the population diagram that have opacities of $\lesssim$\,2--3 for all molecules.

After applying these two corrections, deviations between the observed and synthesised data points are generally small indicating that our chosen input parameters for the Weeds modelling are reliable. However, scatter amongst the observed data points can still occur due to a remaining contamination by unidentified lines, line opacity effects, or inaccurate baseline subtraction due to line confusion. 
Furthermore, the Weeds model is produced by taking into account the background continuum temperature, which is given by the baseline level initially subtracted from the spectra. Due to the different angular resolutions of the observational setups the continuum level varies, which can introduce scatter amongst both the observed and synthesised data points.
Moreover, at positions close to Sgr\,B2\,(N1), we face the problem that the continuum starts becoming optically thick and therefore, hiding the line emission from our view. 
Because in Weeds the continuum is only considered as another background source, the model does not perform properly in this situation. Accordingly, intensities of some transitions may be overestimated by the model resulting in larger scatter in the population diagrams of observed data points. 
This is indeed evident in populations diagrams at position N1S-1, where the continuum can be expected to become optically thick. Because transitions covered by spectral windows 2 and 3 of observational setup 5 suffer most from this effect they are not included for the analysis at this position as mentioned in Sect.\,\ref{sss:positions}.  

At larger distances where the COM emission becomes weak, scatter amongst the observed data points may be introduced when the larger beams of setup 4 capture material closer to the protostar that is no longer seen with the smaller beams of setup 5. For COMs and positions for which this effect is severe, we use spectra from setup 5 that are smoothed with a 2D Gaussian kernel to the angular resolution of spectral window 3 of setup 4.

For \mf, and to a lesser extent for \mic, we notice a systematic deviation of observed points from setups 4 and 5. We do not know the reason for this for sure. This is not caused by the different angular resolutions of the setups as smoothing the spectra of setup 5 to the resolution of setup 4 does not yield improvement. 
A possible explanation may however be filtering of extended emission by ALMA (see Sect.\,\ref{sss:continuum}) which would affect setup 5 more than setup 4 due to the average shortest baselines that are longer for setup 5 \citep[see Table\,1 in ][]{Belloche19}. 
Data points observed with setup 5 would be expected to lie below those of setup 4, if emission was filtered, which is what we see in the population diagrams of these two molecules. However, this effect cannot be confirmed quantitatively and its impact remains questionable as it is not clear why the two mentioned COMs should be more affected than others.

Many of the above-mentioned uncertainties cannot be meaningfully quantified. Therefore, the error bars shown in the population diagrams only reflect the standard deviation from fitting and an additional factor of 1$\sigma$, where $\sigma$ is the the median noise level measured in channel maps of the continuum-removed data cubes taken from Table\,2 in \citet{Belloche19}, 
to account for the uncertainty in the continuum level. However, the fit is performed by not weighting the data points by the errors in order to avoid giving too much weight to extremely strong lines and/or contaminated lines.

The data points in all population diagrams follow a linear trend meaning a single temperature can explain the level distribution for each molecule at each position. Therefore, we fit a straight line to the observed data points in all cases and obtain the desired temperature profiles for each COM. We perform the fit only as far as positions N1S4 and N1W4, respectively, for reasons explained below. 
The profiles are shown as solid lines in Fig.\,\ref{fig:Tprofiles} and the slopes of the linear fits to these profiles are summarised in Table\,\ref{tab:Tprofiles}. The errors on the data points correspond to values obtained from the population diagrams and determine the uncertainty on the temperature profiles. 
Few profiles seem to deviate from a straight line at shorter distances to Sgr\,B2\,(N1). Therefore, we apply an additional fit to the data that includes only those positions at larger distances from Sgr\,B2\,(N1) that seemingly lie on a straight line (dashed lines in Fig.\,\ref{fig:Tprofiles}). We keep the second fit only when the slopes of both fits differ by more than 15\%.

To the south, \met, \dme, and \mf are the COMs detected farthest from Sgr\,B2\,(N1) with a reliable temperature determination beyond a distance of 3.5\arcsec. Especially for \met and \dme it is evident in Fig.\,\ref{fig:Tprofiles} that the rotational temperatures beyond this distance (corresponding to positions N1S5--N1S7) lie above the fit. 
These slightly higher temperatures can possibly be associated with the continuum source AN\,20 (see Figs.\,\ref{fig:continuum-maps} and \ref{fig:1mm+et}) that heats the gas in addition to Sgr\,B2\,(N1). Therefore, we exclude these positions from the fit of the temperature profiles. 
To the west, there seems to be a drop in temperature at positions N1W4 and N1W5 for \mf and less pronounced for \etc and a subsequent rise to $\gtrsim$100\,K at larger distances. Therefore, position N1W4 is not included in the fit for these two molecules. A same trend may be expected for \dme and \vc, however, these molecules are not detected at these two positions. They reappear at larger distances. 
Remarkably, there is a change in velocity and line width starting from N1W5. Up to this position, spectral lines are detected at $\varv_{\rm lsr}\sim65$\,\kms with $FWHM\sim5$\,\kms. This component is undetected at larger distances, instead the line appears at $\varv_{\rm lsr}\sim69$\,\kms and is extremely narrow with $FWHM\sim2-3$\,\kms (indicated with crosses in all figures where necessary). This is also evident in the peak-velocity map in Fig.\,\ref{fig:LVINE_O}. 
We do not include these positions to the fit of the temperature profile. 

\begin{figure*}[]
    \hspace{-0.2cm}
    \includegraphics[width=\textwidth]{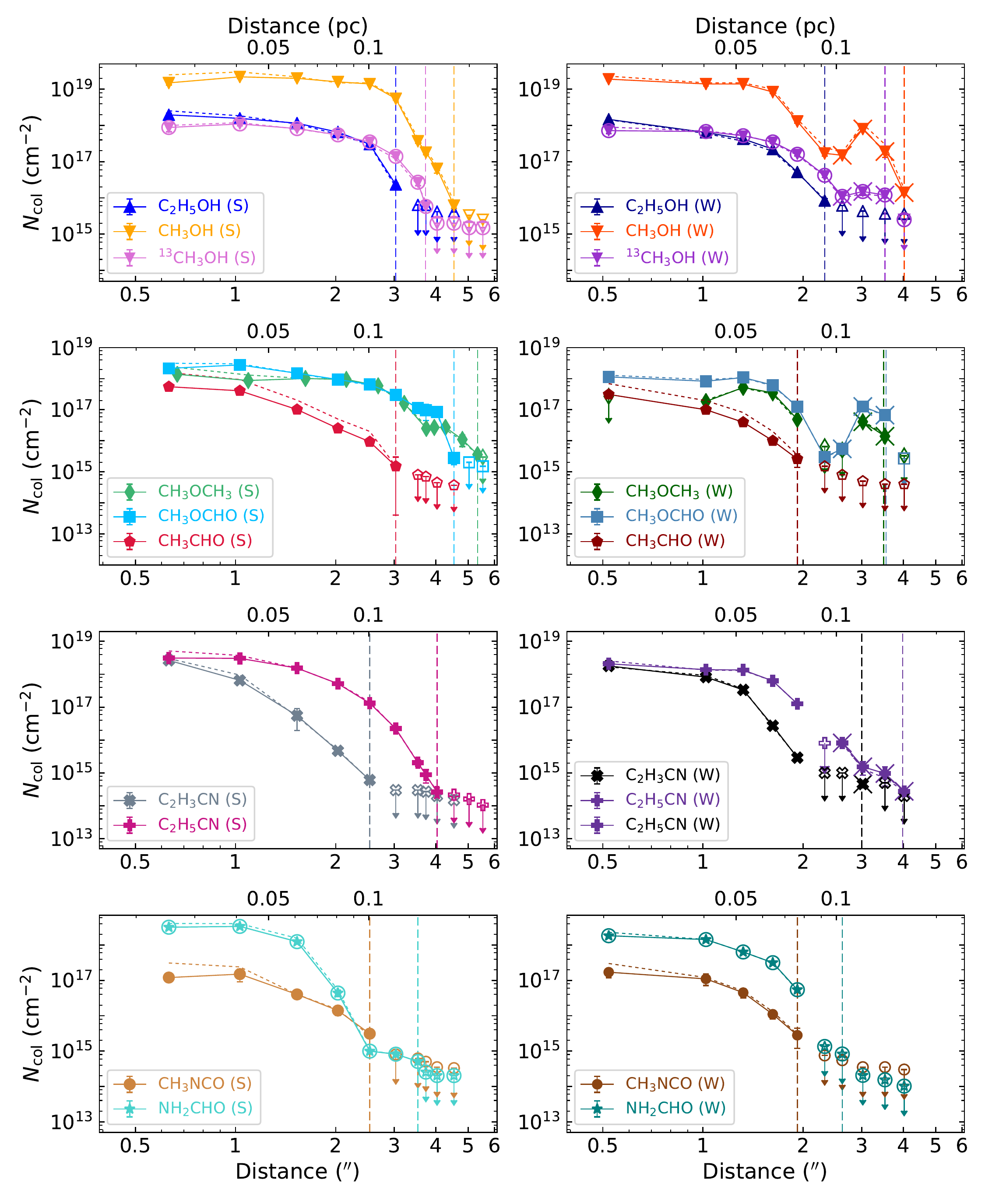}
    \caption{COM column density profiles to the south (S) and west (W), where solid curves show the total column densities derived from the linear fit in the population diagrams using setups 1--3 for $^{13}$\met and \fmm (encircled symbols) and setups 4--5 for the rest of the COMs. Dotted curves indicate column densities used during the radiative transfer modelling with Weeds. Additional crosses on the markers indicate positions for which the velocity offset changed from $\lesssim$\,2 to $\sim$7\,\kms. Isolated unfilled markers with arrows indicate upper limits. Vertical lines mark the distance from Sgr\,B2\,(N1) beyond which a COM is no longer detected.}
    \label{fig:Ncolprofiles}
\end{figure*}

The slopes of the southbound profiles for the O-bearing COMs vary in a range from $-$0.6 to $-$0.8 when including all positions up to N1S4, except \mf, whose profile is slightly steeper. However, given the apparent separation of data points from setups 4 and 5 at some positions for this molecule, we may expect a greater overall uncertainty for its profile.  
The westbound profiles of \met, its $^{13}$C isotopologue, \et, and \ad show similar slopes as their southbound counterparts. The westbound profile of \dme has a much steeper slope of $-1$ while that of \mf has a much shallower slope ($\sim-0.5$) than to the south. When including only points farther away from Sgr\,B2\,(N1), \met shows a slightly steeper slope of $-$0.81, which is still in the above-mentioned range. Also, \ad has a steeper slope of $-$1.15. 

The slopes of the N-bearing COMs, except \fmm, are slightly shallower ($-$0.4 to $-$0.6, \fmm: $-$0.3) than those of the O-bearing COMs when including all points. Only when the closest positions to Sgr\,B2\,(N1) are excluded, \etc and \fmm show a steeper slope that is comparable to those of the O-bearing COMs. 
The slopes of the westbound profiles of the N-bearing COMs are similar to the southbound ones. When considering only larger distances the slopes of \etc and \mic again become similar to those of the O-bearing COMs. The slope of \fmm remains shallow. 
Up to now, it is unclear what may cause a deviation from a linear fit with all points included and the difference between the COMs. Aside from an underestimation of uncertainties, possible reasons may be presented in an underestimation of line optical depth or the absence of a species in the closest vicinity of Sgr\,B2\,(N1). In the latter case, we may then observe a cloud layer where this species is still abundant and which has a lower temperature. In the case of \fmm (also $^{13}$\met) the larger beams of setups 1--3 may influence the result, especially to the west, since the distances between positions are smaller than the average beam size and hence, adjacent positions are not independent of each other.

The temperature profile ($T\propto D^\gamma$) is derived from the diffusion equation \citep[e.g.,][]{Kenyon93}.
The slope $\gamma$ varies depending on whether the dust emission is optically thin $\gamma=-\frac{2}{4+\beta}$ or thick $\gamma=-\frac{1-n}{4-\beta}$, where $\beta$ is the dust emissivity spectral index, which likely varies in a range from 0.5 to 2 at 3\,mm in dense cores \citep[e.g.,][]{Kwon09,Li17}, and $n$ is the density power law index, which typically is $n=-1.5$ for a free-falling envelope \citep{Shu77}. Assuming these values the slope $\gamma$ varies from $\sim$\,$[-0.3,-0.4]$ for optically thin and $\sim$\,$[-0.7,-1.3]$ for optically thick dust emission. 
Therefore, the slopes derived above, that include all positions, generally lie in between these ranges with the tendency towards the range for optically thick continuum. 

Total column density profiles for each COM are obtained from the linear fit performed in the population diagrams and are shown in Fig.\,\ref{fig:Ncolprofiles}, where the errors on the data points again correspond to those obtained from the population diagrams. In addition, column densities used for the modelling with Weeds are shown as dotted lines. Usually, the difference is less than a factor 2, except for \ad and \mic, for which the difference can be up to a factor 3--4 for some positions. This is mainly caused by the fact that the Weeds model considers the background continuum while it is ignored (and cannot be properly accounted for) when fitting the population diagram. For this reason, we will only use column densities from Weeds in the following. To all values we apply a vibrational correction factor that accounts for contributions to the column density by vibrational states, in the cases where the partition function provided by the various databases does not include this contribution that cannot be neglected at the high temperatures measured here. 
Southbound profiles of almost all O-bearing species resemble a plateau between 0.5\arcsec and 2.5\arcsec and then start to decrease at larger distances from Sgr\,B2\,(N1), with those of \met and \et dropping more abruptly than \dme and \mf. The column density of \ad rather continuously decreases with increasing distance. 
To the south, column densities of N-bearing COMs decrease similarly to \ad, with \vc and \fmm showing a steeper decrease of column density with increasing distance from Sgr\,B2\,(N1). 

At close distances, the westbound profiles of all COMs, except \dme, behave similarly to their southbound counterparts with values that are equal or slightly smaller. The column density of \dme decreases towards Sgr\,B2\,(N1) such that at N1S-1, we are not able to derive a population diagram and hence, only an upper limit from the Weeds model is determined by extrapolating its temperature profile. 
At large distances the west- and southbound profiles start deviating from each other. On the one hand, \et, \ad, and \mic are only detected to shorter westward distances than to the south, while on the other hand, \met  and \mf show a cavity to the west, where column densities drop by up to two orders of magnitude reaching a minimum at N1W4--N1W5 only to increase again beyond these distances. This coincides with the drop in temperature seen in Fig.\,\ref{fig:Tprofiles}. This cavity is also observed for \dme, \vc, and \etc, because although they are not even detected at positions N1W4 and, for the former two, N1W5, their column densities can be determined at the larger distances. For $^{13}$\met, this drop in column density is less pronounced if evident at all. This may be because it has been smeared out due to the larger beam sizes of setups 1--3 that are used for this COM. \fmm is detected up to position N1W5, however only upper limits (see below) can be determined. According to these, \fmm may show a trend similar $^{13}$\met concerning the cavity.

At some position at farther distance from Sgr\,B2\,(N1) some COMs may still be detected, however, the linear fit in the population diagram does not yield a reliable result. In theses cases we fix the rotational temperature by extrapolating the derived temperature profile of the respective species shown in Fig.\,\ref{fig:Tprofiles} in order to determine column densities.
Vertical lines in Fig.\,\ref{fig:Ncolprofiles} indicate the largest projected distance from Sgr\,B2\,(N1) up to which the COM is still detected and its column density can be derived. Accordingly, the COMs \textit{disappearing} southwards first are \vc and \mic, followed by \ad and \et, \fmm, \etc, and at last \met, \mf, and \dme. 
Due to the cavity in column density, such a statement is difficult to make to the west. However, all COMs that are detected farthest from Sgr\,B2\,(N1) to the south are also the ones detected at largest distances to the west, which is also seen in Fig.\,\ref{fig:Dmax-COM}, where these maximum distances are summarised. 
At distances beyond the shown vertical lines and at positions of the cavity, if necessary, upper limits for the column densities are determined, where the rotational temperature is again obtained from the extrapolation of the temperature profile. These upper limits are computed assuming an average velocity offset and line width determined from detected molecules at a specific position and making sure 
that the Weeds model for the undetected species does not overestimate the intensity over a 3$\sigma$ threshold, where $\sigma$ is taken from Table\,2 in \citet{Belloche19}. 

\begin{figure}\hspace{-0.5cm}
    \includegraphics[width=.5\textwidth]{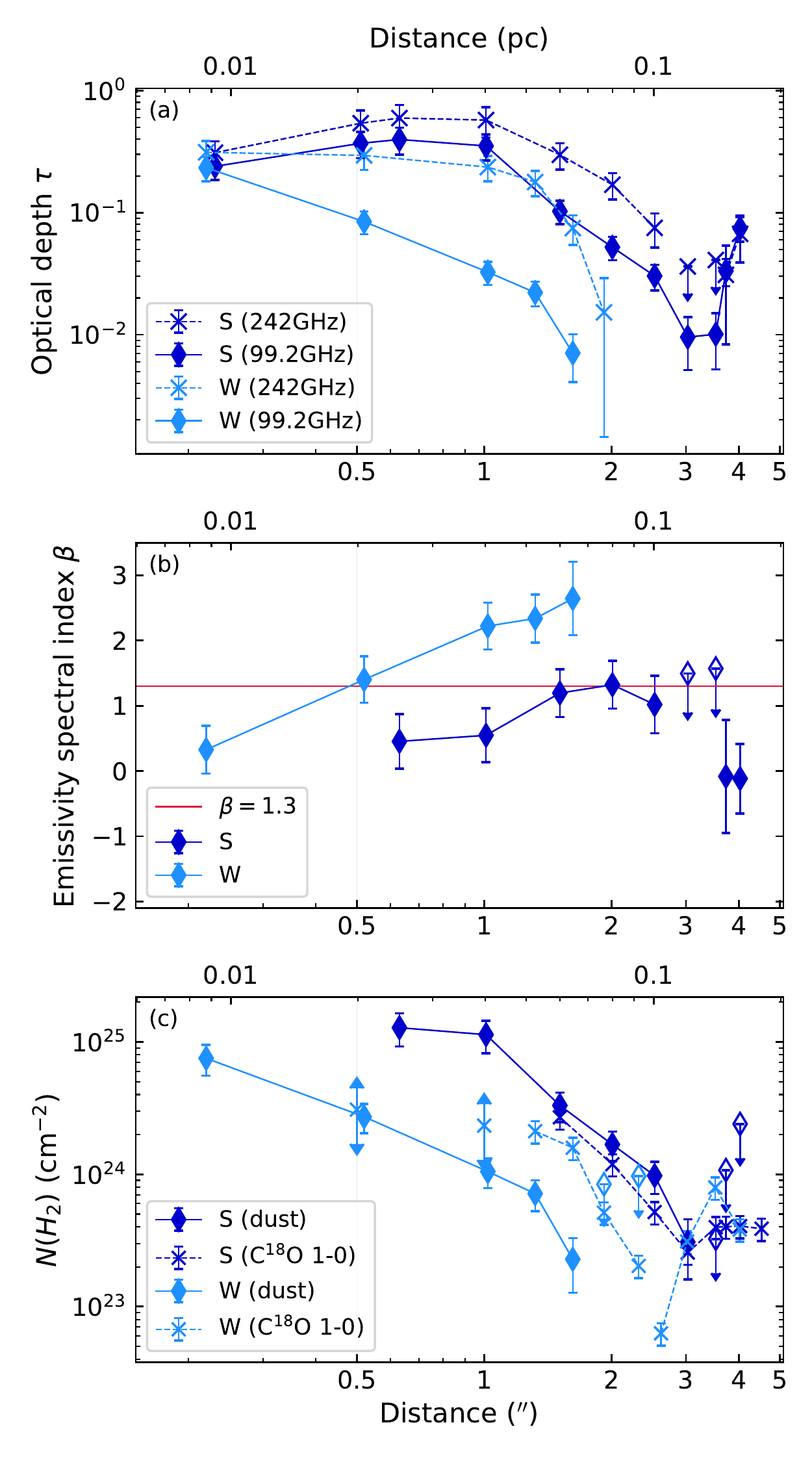}
    \caption{\textbf{(a)} Optical depth $\tau$ as a function of distance from Sgr\,B2\,(N1) in south (S) and west (W) direction and for the 1.3\,mm and 3\,mm continuum, respectively. The 3\,mm data has been corrected for free-free emission. \textbf{(b)} Emissivity spectral index $\beta$ as a function of distance from Sgr\,B2\,(N1) derived from the optical depth ratio of the 1.3\,mm and 3\,mm data shown in (a). \textbf{(c)} H\2 column densities derived from the 3\,mm continuum emission, using $\beta=1.3$ \citep[adopted from][see text]{Bonfand19}, and from C$^{18}$O 1--0 emission using conversion factors of ${\rm C}^{16}{\rm O}/{\rm C}^{18}{\rm O}=250\pm 30$ \citep[][]{Henkel94} and ${\rm H}_2/{\rm C}^{16}{\rm O}=10^4$ \citep[][]{Rodriguez-Fernandez01}. The break in the (dashed) westbound profile indicates the observed velocity shift of the line from $\sim$64\kms to 69\kms. In all panels upper and lower limits are indicated with arrows.}
    \label{fig:NH2}
\end{figure}
\begin{figure*}
    \hspace{-0.3cm}
    \includegraphics[width=\textwidth]{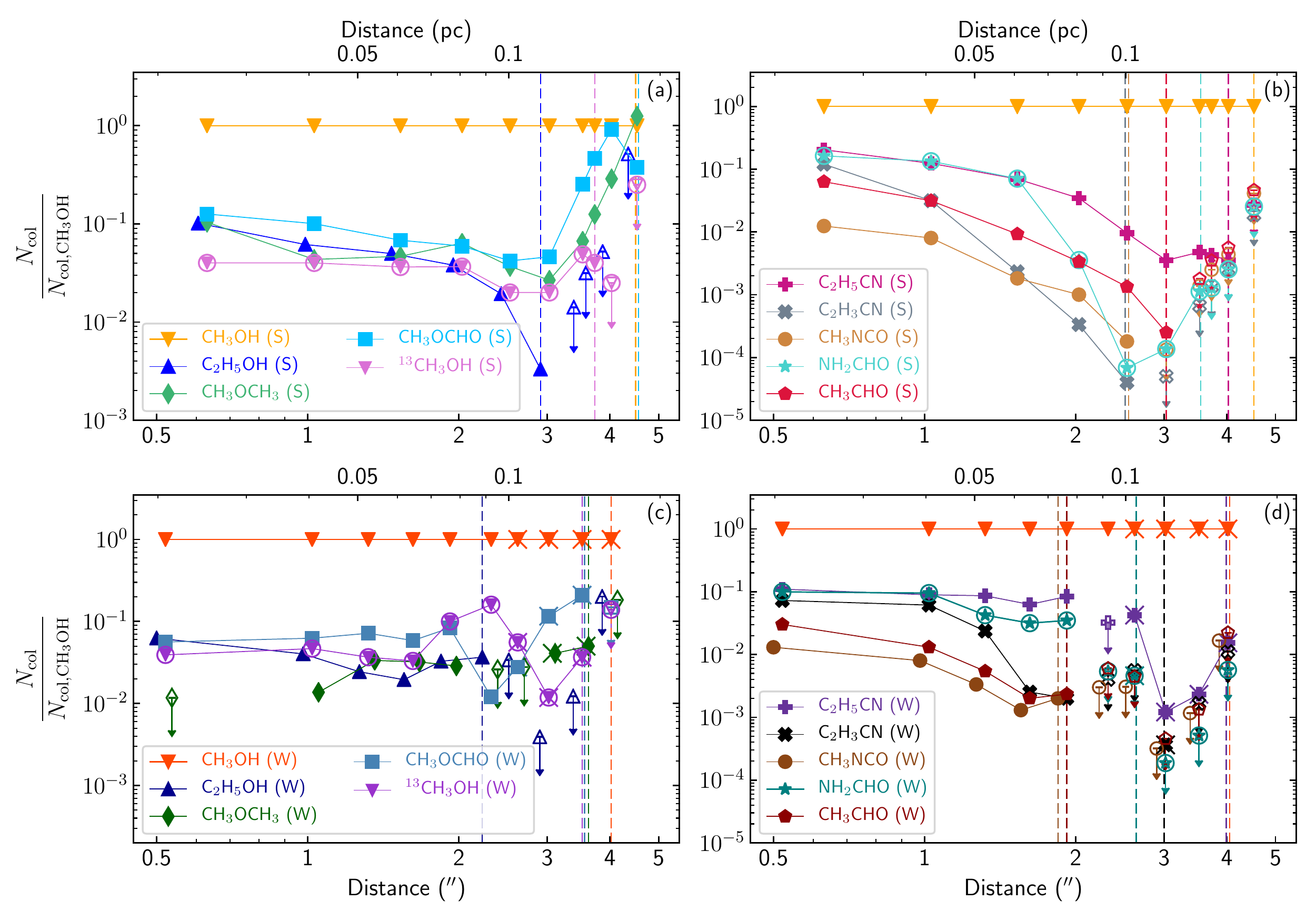}
    \caption{COM abundance profiles with respect to \met to the south (S, \textbf{(a) -- (b)}) and to the west (W, \textbf{(c) -- (d)}) based on data from observational setups 1--3 (encircled symbols) and 4--5. Additional crosses on the markers indicate positions for which the velocity offset changed from $\lesssim$\,2 to $\sim$7\,\kms. Unfilled symbols with arrows indicate upper limits. Vertical dashed lines mark the distance from Sgr\,B2\,(N1) beyond which the respective COM is no longer detected.}
    \label{fig:DvsXratio}
\end{figure*}

\subsection{H\2 column densities from dust}\label{ss:NH2}

Assuming that LTE conditions apply to the COMs, the rotational temperatures derived in Sect.\,\ref{ss:profiles} correspond to the kinetic temperature of the gas. Moreover, assuming that the gas is well coupled to the dust, the kinetic temperature of the gas also corresponds to that of the dust. 
Based on this, we compute again optical depth using Eq.\,\ref{eq:tau}, however, this time, we make use of the rotational temperatures derived in Sect.\,\ref{ss:profiles} and determine optical depth as a function of distance from Sgr\,B2\,(N1) for both the continua at 1.3 and 3\,mm. The profiles are shown in Fig.\,\ref{fig:NH2}a.  We use the temperature profile of \et 
because its population diagrams together with the temperature profile seem to be the most reliable amongst the COMs analysed here. The error bars correspond to an uncertainty that allows the dust temperature to differ by 20\% from the gas temperature, which dominates over other uncertainties. 

Based on the contribution of free-free emission to the 3\,mm continuum at the position of Sgr\,B2\,(N1) derived in Sect.\,\ref{sss:continuum}, we estimate the remaining contribution at closest distance to Sgr\,B2\,(N1) considered here ($\sim$0.2\arcsec, i.e, roughly half the average beam size of spectral window 0 of setup 5) to be $\sim$16\% and negligible at larger distances. The continuum at 1.3\,mm has no significant contribution of free-free emission.

The results shown in Fig.\,\ref{fig:NH2}a demonstrate again that the 3\,mm continuum is less optically thick than the 1.3\,mm continuum. Opacities are generally lower and the difference between the two continua is more pronounced to the west. 
The maximum distance from Sgr\,B2\,(N1) denotes where the continuum is detected above 2$\sigma$. At positions N1S4 and N1S5, the continuum at 242\,GHz drops below 1$\sigma$, hence we use an upper limit of 2$\sigma$ at these positions in the following.
At distances beyond 3.5\arcsec (corresponding to positions N1S5--N1S7) the optical depth increases again, which can be associated with continuum source AN\,20 (see Figs.\,\ref{fig:continuum-maps} and \ref{fig:1mm+et}), whose location coincides with these positions.

To the south, the profile for both frequencies slightly decreases at shorter distance to Sgr\,B2\,(N1) between 0.6\arcsec and 0.2\arcsec which may be the result of an underestimation of the continuum level. The continuum at both frequencies corresponds to the baseline, which is subtracted from the spectra. 
At these short distances, line emission is detected in almost every channel making the identification of the true baseline level more difficult. We checked the already baseline-subtracted spectra \citep[done by][]{Belloche19} at the closest distance to Sgr\,B2\,(N1) and estimated a possibly remaining continuum level of at most 20\,K that would lead to an increase of the current level by about 10\%. 
Therefore, missing continuum emission cannot fully account for the lower optical depth at short distances at least at 99.2\,GHz. 
Though we account for a possible difference between gas and dust temperature in the error bars we may still underestimate the optical depth, that is, the dust temperature that is assumed to be the gas temperature may be overestimated. The underestimation of optical depth would in turn lead to an underestimation of H\2 column densities to an unknown extent. Therefore, southbound distances to Sgr\,B2\,(N1) smaller than 0.6\arcsec are not considered in the following.

From the optical depth H\2 column densities $N(\mathrm{H}_2)$ can be derived using
\begin{align}\label{Eq:NH2}
    N(\mathrm{H}_2) = \frac{\tau}{\mu_{H_2}\,m_H\,\kappa_\nu},
\end{align}
where $\mu_{H_2}= 2.8$ is the mean molecular weight per hydrogen molecule, $m_H$ the mass of an hydrogen atom, and $\kappa_\nu$ the dust mass opacity. The latter follows a power law given by:
\begin{align}\label{eq:kappa}
    \kappa_\nu = \frac{\kappa_0}{\chi}\left(\frac{\nu}{\nu_0}\right)^\beta,
\end{align}
where $\chi = 100$ is the standard gas-to-dust ratio and $\kappa_0$ is the dust mass absorption coefficient at frequency $\nu_0$.

Assuming that H\2 column densities derived from the 1.3\,mm and the 3\,mm continuum yield the same result, the dust emissivity spectral index can be derived using the formalism introduced by \citet{Bonfand17}, which in our case leads up to
\begin{align}
    \beta = \ln\left(\frac{\tau_\mathrm{99.2}}{\tau_\mathrm{242}}\right)\times\frac{1}{\ln\left(\frac{99.2\,\mathrm{GHz}}{242\,\mathrm{GHz}}\right)},
\end{align}
where $\tau_\mathrm{99.2}$ and $\tau_\mathrm{242}$ are continuum optical depths at the respective frequency.
We show $\beta$ as a function of distance from Sgr\,B2\,(N1) to the south and west in Fig.\,\ref{fig:NH2}b. 
The points at largest distance to the south corresponding to N1S5--N1S7 are not reliable as these positions are attributed to AN\,20 and the contribution of free-free emission there is unknown. To the south, $\beta$ has a value 0.4 at closest distances to Sgr\,B2\,(N1) and increases to 1.3 at larger distances, while to the west, $\beta$ increases from roughly 0 up to a value of $\sim$2.5. Based on these profiles, it is therefore difficult to assign a single $\beta$ value that would be able to describe the dust in both directions. 
Furthermore, $\beta$ as well as $\kappa$ rely on the dust properties, which are generally highly uncertain.
Therefore, we decide to follow the approach described in \citet{Bonfand19}, who derived $\kappa$ as a function of wavelength for $\beta=1.2$ (see their Fig.\,F.1) and compared their results to models by \citet{Ossenkopf94}. 
Ultimately, they used $\beta=1.3$, which fits best the model that includes no ice mantles and assumes a gas density of 10$^6$\,cm$^{-3}$. Accordingly, they used $\kappa_0=1.99$\,cm$^{2}$\,g$^{-1}$ at $\nu_0=230$\,GHz (that is $\lambda_0=1.3$\,mm). In the following we will adopt these values for $\kappa_0$ and $\beta$ and assume an uncertainty of $\Delta\beta=0.5$.

With optical depth and the dust mass opacity at hand, H\2 column densities can be derived using Eq.\,\ref{Eq:NH2}. The results are shown in Fig.\,\ref{fig:NH2}c.
H\2 column densities are listed in Table\,\ref{tab:H2}, together with the continuum intensity and optical depth. In both directions we determine a maximum H\2 column density of $\sim$\,10$^{25}$\,cm$^{-2}$ which agrees well with the value of $(1.3\pm0.2)\times10^{25}$\,cm$^{-2}$ derived by \citet{Bonfand17} for Sgr\,B2\,(N1) based on the data of the EMoCA survey, and a minimum value of $\sim$2$\,\times$\,10$^{23}$\,cm$^{-2}$. However, the westbound profile decreases faster and at already shorter distances than the southbound one. %
At positions where H\2 column densities are determined along both directions, the westbound values are lower than the southbound ones by a factor of a few or even by roughly one order of magnitude. If free-free emission was indeed contributing to the continuum emission of AN\,20, H\2 column densities would be lower. Therefore, we consider the values at positions N1S5--N1S7 as upper limits. To the west, we add upper limits of $2\sigma$. 

\subsection{H\2 column densities from C$^{18}$O}\label{ss:c18o}
Additionally, H\2 column densities are derived from CO, where we use the $J=1-0$ rotational transition of its isotopologue $^{12}$C$^{18}$O. Upper level column densities of this isotopologue are converted to those of the main one $^{12}$C$^{16}$O by multiplying by the C$^{16}$O/C$^{18}$O isotopic ratio of 250$\pm$30 \citep[][]{Henkel94} and subsequently, to H\2 column densities by multiplying with $10^4$ \citep[][]{Rodriguez-Fernandez01}.
We use the opacity-corrected column densities of the Weeds model, for which we fix the temperature and the velocity offset to that of ethanol. 
The resulting H\2 column density profiles are shown in Fig.\,\ref{fig:NH2}c. Column densities of H\2 and C$^{18}$O, and Weeds parameters are summarised in Table\,\ref{tab:H2}.
At distances $\leq$1\arcsec only upper limits can be determined to the west as the line is too contaminated by emission from other species at the two closest positions.\footnote{One of the contaminating species at lower frequencies is \\ HNCO,$v=0$. Emission coming from higher frequencies could not be assigned to a molecule. However, in comparison with other molecules, there may be a blue-shifted component of C$^{18}$O originating from the outflow, which may then also apply to HNCO. At distances $>$1\arcsec, the remaining contamination to the line core of C$^{18}$O seems vanishingly low.} However, at these positions, the baseline subtraction may be overestimated due to spectral confusion, such that the upper limit could well be higher. Therefore we also show lower limits. At positions N1S-1 and N1S column densities cannot be computed because there is an absorption component absorbing the emission at the respective velocity. 
Between distances of 1.5\arcsec and 3\arcsec the southbound profiles of H\2 column densities derived from dust and C$^{18}$O follow the same trend with values from C$^{18}$O being only slightly smaller. 
The westbound profiles show greater difference. The value estimated at N1W-1 is in agreement with the one derived from dust while at N1W, the H\2 column density estimated from C$^{18}$O may be larger. At even larger distances from Sgr\,B2\,(N1), the column densities derived from the dust are surprisingly smaller than the column densities derived from C$^{18}$O by a factor 4--5 or even more, which cannot be explained by contamination alone.
In both directions, we are able to go to farther distances with the profile from C$^{18}$O emission. 

\subsection{COM abundances as a function of distance and temperature}\label{Rss:Xprofiles}
\begin{figure*}[h!]
    \centering
    \includegraphics[width=\textwidth]{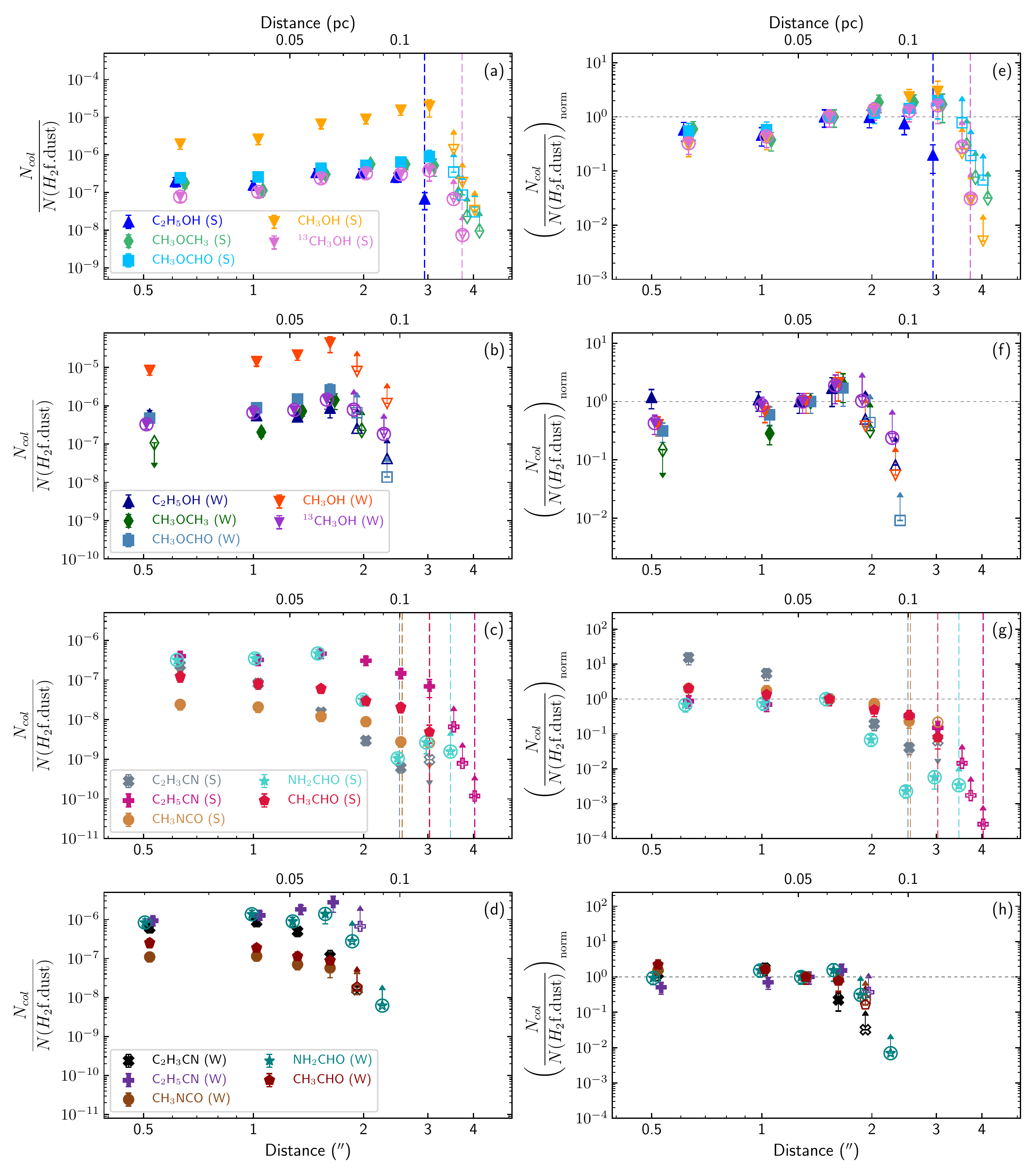}
    \caption{\textbf{(a)} -- \textbf{(d)} COM abundance profiles with respect to H\2 (\textit{f. dust} -- derived from dust emission) to the south (S) and to the west (W) based on data from observational setups 1--3 (encircled symbols) and 4--5. \textbf{(e)} -- \textbf{(h)} Same as in (a) -- (b) but with COM abundances normalised to the value at 1.5\arcsec distance to Sgr\,B2\,(N1). Unfilled symbols with arrows pointing downwards or upwards indicate upper
    or lower limits, respectively. Vertical dashed lines mark the distance from Sgr\,B2\,(N1) beyond which the respective COM is no longer detected.} 
    \label{fig:DvsX}
\end{figure*}

\begin{figure*}[h!]
    \centering
    \includegraphics[width=\textwidth]{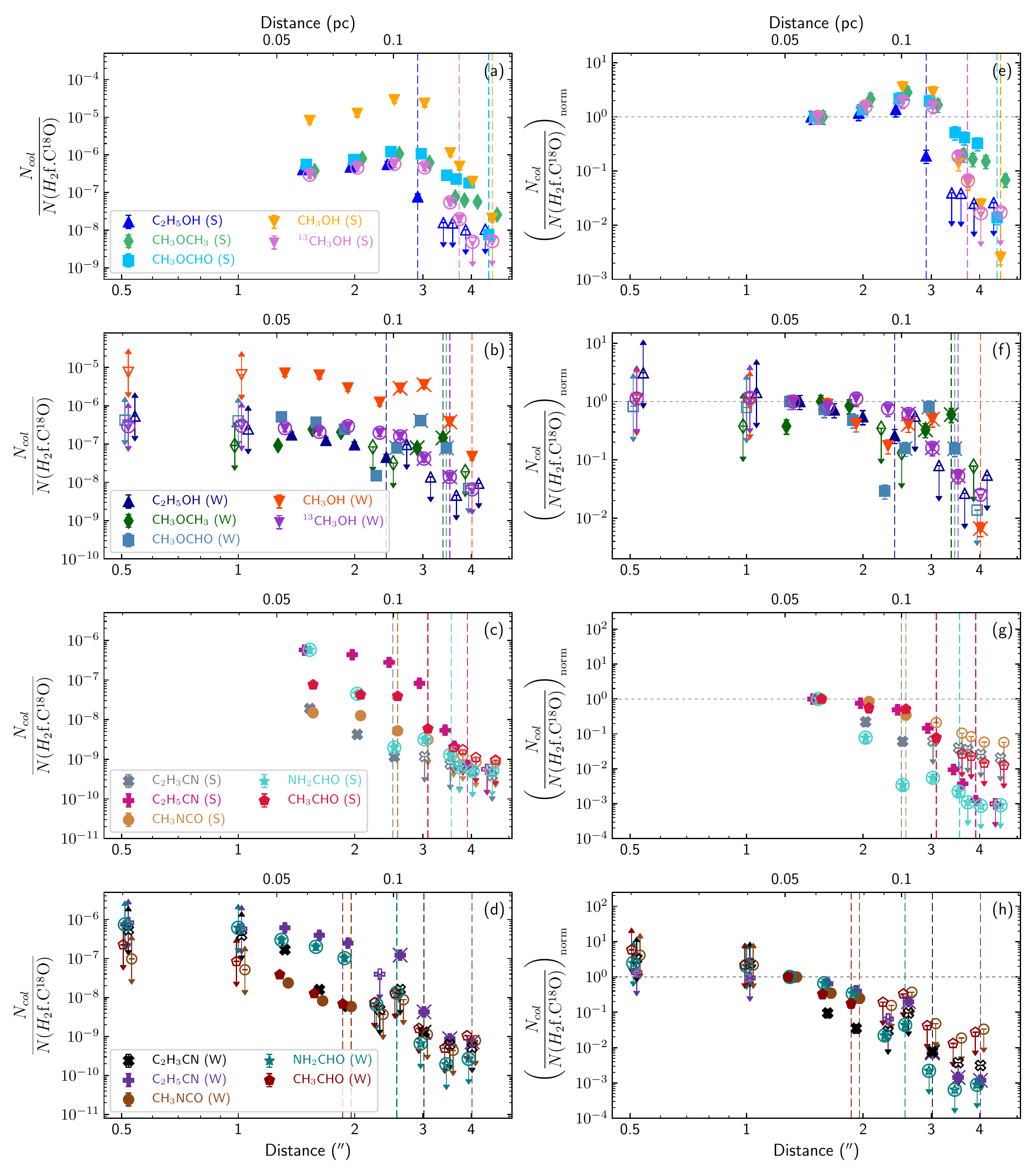}
    \caption{\textbf{(a)} -- \textbf{(d)} COM abundance profiles with respect to H\2 (\textit{f. C$^{18}$O} -- derived from C$^{18}$O 1--0 emission) to the south (S) and to the west (W) based on data from observational setups 1--3 (encircled symbols) and 4--5. \textbf{(e)} -- \textbf{(h)} Same as in (a) -- (b) but with COM abundances normalised to the value at 1.5\arcsec distance to Sgr\,B2\,(N1). Unfilled symbols with arrows pointing downwards or upwards indicate upper or lower limits, respectively. Vertical dashed lines mark the distance from Sgr\,B2\,(N1) beyond which the respective COM is no longer detected.} 
    \label{fig:DvsX_co}
\end{figure*}

Figure\,\ref{fig:DvsXratio} shows COM abundances relative to \met as a function of distance from Sgr\,B2\,(N1) to the south (panels a--b) and to the west (panels c--d). 
Figures\,\ref{fig:DvsX}\,a--d and \ref{fig:DvsX_co}\,a--d show COM abundances relative to H\2 derived from dust and C$^{18}$O 1--0 emissions, respectively, as a function of distance from Sgr\,B2\,(N1) in both directions. For a better comparison, the abundance profiles are normalised to the respective value at 1.5\arcsec distance to Sgr\,B2\,(N1) as is shown in panels e--h, respectively. 
Given the resemblance of the H\2 column density profiles to the south, also the abundance profiles are similar where possible to compare. 
Abundance profiles that are derived from dust continuum emission cover a distance range of 0.6\arcsec to 3\arcsec to the south (we refer to these profiles as \sdu hereafter) and 0.5\arcsec to $\sim$2\arcsec to the west (\wdu hereafter). The range is shorter for \wdu because of the lack of H\2 column density at farther distances.
In contrast, abundance profiles that are derived by using C$^{18}$O reach farther out to maximum distances of 4.5\arcsec to the south (\sco hereafter) and 3.5\arcsec to the west (\wco hereafter). 
However, at distances $\leq$1\arcsec, there are only lower/upper limits for \wco or no values at all for \sco. Moreover, with the cavity observed for \wco peak abundances for the COMs are more difficult to locate.

\begin{figure*}[]
    \includegraphics[width=\textwidth]{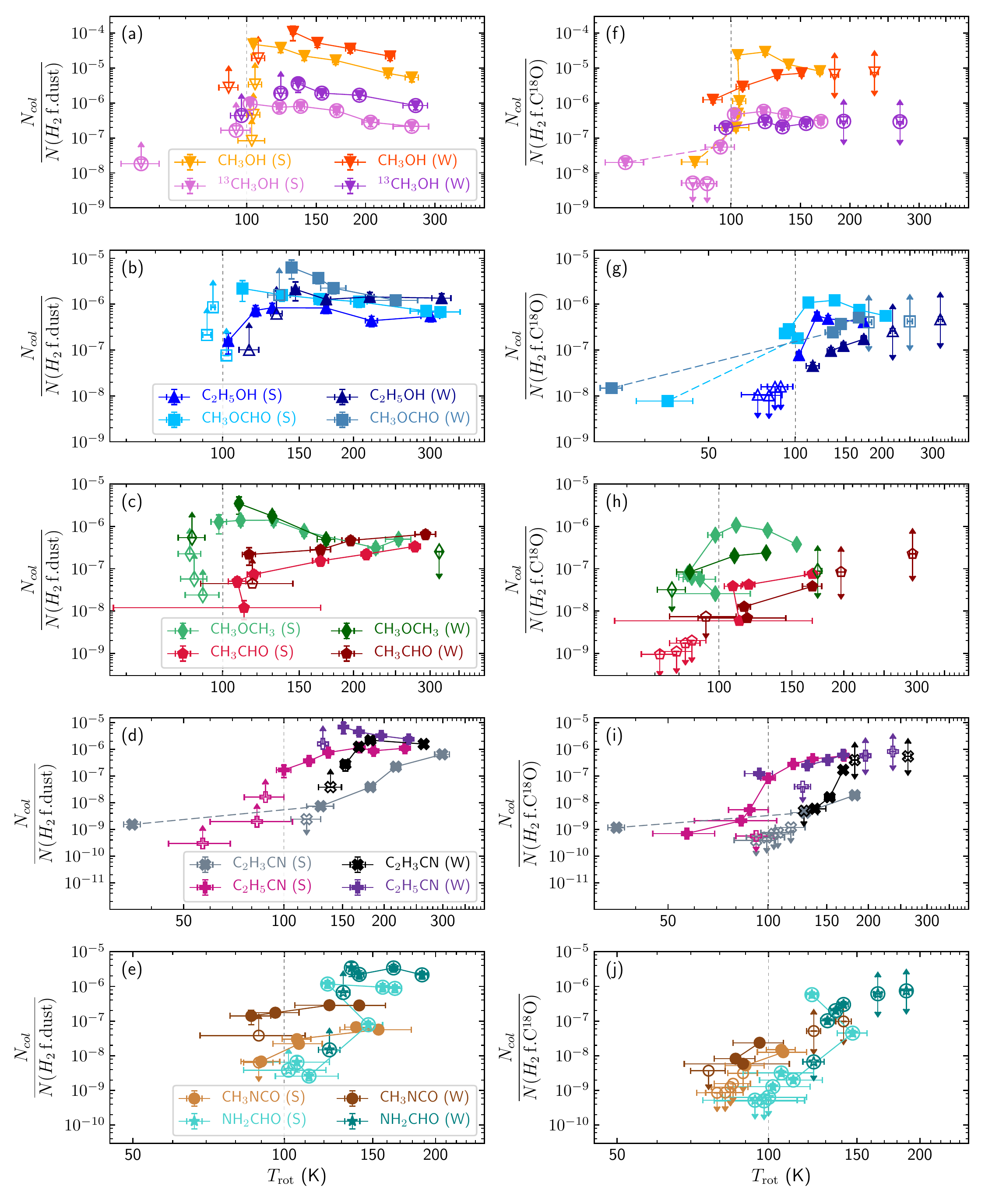}
    \caption{\textbf{(a)} -- \textbf{(e)} COM abundance profiles with respect to H$_2$ derived from dust emission (\textit{f. dust}) to the south (S) and west (W) as a function of rotational temperature taken from Fig.\,\ref{fig:Tprofiles}. \textbf{(f)} -- \textbf{(j)} Same as in (a) -- (e) but with H\2 column densities derived from C$^{18}$O emission (\textit{f. C$^{18}$O}). Coloured dashed lines indicate the connection to points for which the temperature was either fixed in the population diagram, where a value was obtained from extrapolating the respective temperature profile, or derived from the population diagram but not considered in the fit to the temperature profiles shown in Fig.\,\ref{fig:Tprofiles}. Unfilled symbols with arrows pointing downwards or upwards indicate upper or lower limits, respectively. Encircled symbols indicate the species for which data from observational setups 1--3 were used instead of 4--5.} 
    \label{fig:TvsX}
\end{figure*}

Figure\,\ref{fig:TvsX} also shows COM abundances relative to H\2 derived from dust (panels a--e) and C$^{18}$O 1--0 emissions (panels f--j), but as a function of rotation temperature using the temperatures derived in Sect.\,\ref{ss:profiles}. The temperatures covered by the abundance profiles are in a range of $\sim$100--300\,K for \sdu, well-above 100\,K for \wdu, and between $\sim$80 and 170--200\,K for \sco and \wco. Therefore, with the additional positions at larger distances for \sco and \wco, also abundances at lower temperatures can be analysed.

\subsubsection{Methanol \met}\label{sss:met}

Methanol abundances increase steeply at around 100\,K, that is between 3\arcsec and 4\arcsec distance from Sgr\,B2\,(N1), to peak at 100--130\,K, that is between $\sim$1.5\arcsec and 3\arcsec. Then, the profiles decrease steadily towards higher temperatures and shorter distances for directions \sdu, \wdu, and \sco. For \wco, the profile remains constant or possibly shows a slight increase towards higher temperatures with a peak that could be above 150\,K given the limits on abundance. With little variation most probably due to the different observational setups used for the analysis, that is different angular resolution, $^{13}$\met follows the same trend as the main isotopologue. 


\subsubsection{Ethanol \et}\label{sss:et}

Ethanol and methanol show similar behaviour as can be seen from Fig.\,\ref{fig:DvsXratio} where the ethanol abundance profile relative to methanol is relatively flat. However, there is a slight decrease with increasing distance to the south, also indicated to the west. This reflects the fact that the abundance of ethanol relative to H2 stays approximately constant with increasing temperature while that of methanol decreases, as can be seen in Fig.\,\ref{fig:TvsX}. 
Abundances with respect to H\2 peak at 120--150\,K for all directions but \wco, which corresponds to distances of 1.5--2\arcsec for \sdu and \wdu and 2.3\arcsec for \sco, which is closer to Sgr\,B2\,(N1) and at slightly higher temperatures on average than for methanol. Abundances start decreasing at shorter distances from Sgr\,B2\,(N1) and higher temperatures than for \met.

\subsubsection{Dimethyl ether \dme}\label{sss:dme}

The observed abundance profile as function of both distance and temperature shows similar behaviour compared to ethanol and methanol including a steep increase of abundance at 3--4\arcsec distance to Sgr\,B2\,(N1), corresponding to temperatures at $\sim$100\,K, followed by an abundance peak at 110--130\,K and a subsequent decrease with increasing temperature or shorter distance. However, while the increase of abundance at $\sim$100\,K is slightly less steep than for methanol as can be seen in the profiles for \sco and \wco, the decrease towards higher temperatures along the west direction is steeper (cf. Fig.\,\ref{fig:DvsXratio}). The COM is not detected in the cavity of \wco, however, it reappears at larger distances from Sgr\,B2\,(N1).

\subsubsection{Methyl formate \mf}\label{sss:mf}

The abundance profile of \mf closely resembles that of \dme, maybe slightly shifted by about $+$10\,K. Abundances increase to peak at temperature between 110\,K and 160\,K corresponding to distances of 2.5--3\arcsec for all but \wco, for which it is $\sim$1.5\arcsec. Similar to all other O-bearing COMs, abundance profiles then decrease towards higher temperatures. 

\subsubsection{Acetaldehyde \ad}\label{sss:ad}

In contrast to the other O-bearing COMs discussed above, the observed abundance profile of \ad shows a different behaviour as it continuously increases to a peak just below 300\,K, which is the maximum temperature measured at closest distance to Sgr\,B2\,(N1) for \sdu, \wdu, and \wco (cf. Figs.\,\ref{fig:DvsX}--\ref{fig:TvsX}). For \sco, the maximum temperature is only 170\,K due to missing data at positions closest to Sgr\,B2\,(N1). We do not trace temperatures lower than 100\,K  therefore, a conclusion on a possibly steeper increase similar to other O-bearing COMs cannot be drawn from the current data.

\subsubsection{Vinyl cyanide \vc}\label{sss:vc}

The observed abundance profile of \vc shows an increasing behaviour towards shorter distance and higher temperature, similar to \ad but steeper. Moreover, a difference is evident between the south- and westbound profiles because to the south, the profile increases continuously while to the west the increase is steeper between $\sim$120--170\,K or 1--2\arcsec and flattens for higher temperatures and shorter distances. 
The location of the abundance peak shows some variance but may well be at highest temperatures and 
closest to Sgr\,B2\,(N1) for all cases when considering all uncertainties. The different behaviours of the profiles could possibly be explained by an under- and overestimation of the line opacity to the west and south, respectively, since more transitions throughout all vibrational states used to derive the population diagram of \vc are optically thick at closest distances to Sgr\,B2\,(N1) compared to the other COMs, which can have an influence on the results of the population diagram. Similar to \ad, the abundance profiles only reliably go down to temperatures of $>$100\,K and therefore, we are not able to observe a possible steep increase at these temperatures.
And, similar to \dme, vinyl cyanide is not detected in the cavity of \wco, however, it is observed at one more position at larger distances.

\subsubsection{Ethyl cyanide \etc}\label{sss:etc}

The abundance profiles of \etc show a similarly steep increase as the aforementioned O-bearing COMs at distances of 3--4\arcsec from Sgr\,B2\,(N1), which, for \sdu and \sco, corresponds to a temperature of around 100\,K. For \wdu and \wco, the lack of H\2 column density and the cavity, respectively, prevent us from making a statement on the temperature to west. Abundances then keep gently increasing to a peak at 150--170\,K (1.5--2\arcsec distance). Towards higher temperatures and shorter distances, the profile slightly decreases for \wdu, while for the others the profile seems to remain flat or slightly increase.  
With this behaviour, \etc is placed somewhere in between the results seen for \vc and \ad and the other O-bearing COMs. 

\subsubsection{Methyl isocyanate \mic}\label{sss:mic}
Similar to \ad, we do not observe a steep increase of abundance for \mic because the COM is not detected at distances where the increase for other molecules is identified. The profiles do show a continuous increase to a peak at closest distances to Sgr\,B2\,(N1). The same trend is indicated in the profile as function of temperature, however the error bars on the measured temperatures are large. The temperature of the abundance peak is located at $\sim$110--150\,K.

\subsubsection{Formamide \fmm}\label{sss:fmm}
Abundance profiles of \fmm show a similarly steep increase as methanol, however, starting only at a distance of 2.5\arcsec to Sgr\,B2\,(N1), that is, at much shorter distances than for other COMs. The temperature of this increase is uncertain, however, it seems to be above 100\,K. 
The southbound abundance profile of formamide shows a plateau at the largest distances, that is at roughly 100\,K. The profiles peak at 120--140\,K or $\sim$1.5\arcsec for \sdu, \sco, and \wdu, and show a subsequent plateau or slight decrease. 

\section{Discussion}\label{s:discussion}

\subsection{H\2 column densities}\label{Dss:H2}

H\2 column densities were derived in two ways: from dust emission as well as from C$^{18}$O 1--0 emission.
While the H\2 column density profiles derived from both show only small differences to the south, the westbound profiles show larger differences at least at distances $>$1\arcsec, where H\2 column densities derived from dust are smaller by factors of a few than those from C$^{18}$O. In Sect.\,\ref{ss:c18o} we concluded, that contamination of the C$^{18}$O intensity by other species cannot fully explain this difference.
The conversion factors from CO to H\2 may have some greater uncertainties than considered here. 
Chemical models by \citet{Garrod22} (hereafter G22) show that the typical CO abundance with respect to H\2 of $10^{-4}$ is only reached at temperatures of 100--150\,K as the molecule co-desorbs with water from dust grain surfaces. Therefore, we may be underestimating this conversion factor at lower temperatures. However, this does not explain the discrepancy of the profiles to the west because temperatures lower than 100\,K are reached in only a few cases and the discrepancy is only seen at temperatures $>$100\,K.
CO abundances may be further enhanced at higher temperatures when more complex species are destroyed and leave behind CO as one of the final products.

Probably even higher uncertainties exist for the conversion from dust to H\2 due to the unknown dust properties. In Fig.\,\ref{fig:NH2}b we show the dust emissivity spectral index $\beta$, which is directly linked to the dust properties and varies significantly between south and west direction but also along one direction. It is rather unlikely that the dust properties vary by this much in one source. Whether $\beta$ is a constant, however, as is assumed in Fig.\,\ref{fig:NH2}c is not certain, either. Whether these uncertainties can explain the differences seen in the H\2 column density profiles to the west or whether environmental differences (e.g., outflow, accretion through filaments) may play a role remains unclear to now.

\subsection{Segregation of O- and N-bearing COMs}\label{Dss:OandN}

The morphology of COM emission shown in Fig.\,\ref{fig:COM-Maps} suggests a separation of COMs into two maybe even three groups. Emissions of the O-bearing COMs \met, \et, \mf, \dme, and, to some extent, \ad are extended and structured following the (dust) continuum emission, while those of the N-bearing COMs \etc and \vc are more compact and less structured. Morphologies of \fmm and \mic are in between as these are compact, however, with some structure. 

Many observations revealed segregation of O- and N-bearing COMs in other star-forming regions with a variety of explanations.
\citet{Bisschop07} found a segregation of COMs in terms of rotational temperature on the basis of unresolved single-dish observations of seven high-mass young stellar objects. They classified the COMs as either \textit{hot} or \textit{cold} species.
In their classification, \ad (amongst others) is a \textit{cold} species that traces maximum temperatures of $<$100\,K, while \met, \et, \mf, \fmm, \etc, and HNCO are \textit{hot} species because they trace temperatures $>$100\,K. Although primarily, this is not a segregation of O- and N-bearing COMs, all species of the latter type are classified as \textit{hot} species by \citet{Bisschop07}. 
This separation was similarly observed in Orion\,KL \citep[][]{Crockett15} and, with a slightly deviating classification, in the hot corino around the low-mass protostar IRAS\,16293--2422B \citep[][]{Jorgensen18}, where the latter study did only investigate O-bearing COMs. In these last two studies, differences between COMs were also observed in the morphology of the COM emission.
This classification into \textit{hot} and \textit{cold} species was proposed to primarily arise from differences in the binding energy of the molecules. Cold species have low binding energies. Therefore, they can already thermally desorb at lower temperatures and their emissions are more widespread. Hot species desorb at higher temperatures, due to higher binding energies, and their emitting area is more compact around the protostar. Although this theory is supported by results of chemical models \citep[][]{Garrod13} it is not really confirmed observationally because the true desorption temperatures of COMs in interstellar environments are not known.

Deviations from this classification have been observed, for example, in IRAS\,16293--2422A \citep[][]{Manigand20}, where \met and \mf, although being hot species, trace a more extended region. Moreover, most recent observations towards CygX-N30 do not reveal a significant difference in temperature between various COMs in general \citep{vanderWalt21}.
Such a classification that depends on binding energies is not straightforward as their values are uncertain and vary depending on the dust composition. We will discuss the role of binding energies in more detail in Sect.\,\ref{Dss:Ebin}.

A temperature-dependent segregation of O- and N-bearing COMs can result from carbon-grain sublimation, which was proposed to occur inside the so-called `soot' line at $T\gtrsim300$\,K \citep[][]{vantHoff20}.
This mechanism would lead to an increase in the abundance of nitriles and hydrocarbons that add to the abundances that are already present due to thermal desorption at lower temperatures. Accordingly, this leads to the observation of N-bearing species at higher excitation temperatures with enhanced column densities and with a more compact morphology than O-bearing species. Therefore, when only regarding the morphology, where the emission of N-bearing COMs is more compactly distributed around the protostar, this mechanism may be able to explain the differences between the COMs observed in Sgr\,B2\,(N1).

A segregation of O- and N-bearing COMs was also observed in the hot core of the massive protostar G328.2551-0.5321 \citep{Csengeri19}. 
The observed morphologies of the two families of COMs are completely different. O-bearing COMs show two intensity peaks on either side of the protostar while the emission of N-bearing COMs is more compact and spherically distributed around it. The authors concluded that abundances of the O-bearing COMs are enhanced due to a non-thermal desorption process. They interpreted the two observed intensity peaks in the morphology as tracing the location of material falling onto a disk, thereby, inducing accretion shocks, which enhances the abundance of O-bearing COMs. In contrast, the morphology of N-bearing COMs follows the expectations for thermal desorption from dust grains. The morphology of O-bearing COMs in Sgr\,B2\,(N1) does not present such a behaviour as there is no indication of two distinct intensity peaks  on opposite sides of the protostar and at opposite velocity offsets from the source velocity that would imply the presence of a disk (cf. Fig.\,\ref{fig:COM-Maps} and the peak velocity map in Fig.\,\ref{fig:LVINE_O}a). 

Results from most recent chemical models by G22 suggest a separation between O- and N-bearing COMs that is caused by their different main formation pathways on dust grains at low temperatures and in the gas phase, respectively. More details are discussed in Sect.\,\ref{Dss:Xprofiles} when these modelled abundance profiles are compared to our observed ones.
Based on this comparison and the discussion on the temperature profiles below, we try to identify which of the aforementioned mechanisms may cause the segregation of O- and N-bearing COMs seen around Sgr\,B2\,(N1). 

\subsection{Temperature profiles}\label{Dss:Tprofiles}
The variation of the slopes of the rotation temperature profiles derived from various COMs is generally small. They span ranges from $-0.6$ to $-0.8$ for O-bearing COMs and from $-0.4$ to $-0.6$ for N-bearing COMs (except \fmm). In Sect.\,\ref{ss:profiles} we have already shown that our derived values lie in between the theoretical slopes expected for central heating through an envelope containing optically thin or thick dust, for which a range of dust emissivity indices beta of 0.5--2 was assumed.
Assuming $\beta=1.3$, which we used to compute H\2 column densities in Sect.\,\ref{ss:NH2}, the expected power-law index would be $-0.38$ or $-0.93$ for optically thin and thick dust, respectively. Although the values slightly deviate from the observed power-law indices, our results suggest that the COMs in Sgr\,B2\,(N1) along the two analysed directions trace the temperature structure of the collapsing envelope that is heated by the central (proto)star(s). 

Temperature profiles have been derived for 22 massive cores as part of the \textit{CORE} project by \citet{Gieser21} based on observations of H\2CO and CH\3CN. The slopes span a range from $-$0.1 to $-$0.6 with an average value of $-(0.4\pm0.1)$ excluding three outliers with slopes of $-$1 to $-$1.5. With the \textit{CORE}-extension sample \citep[][]{Gieser22}, temperature profiles of another ten cores were shown and slopes of $-$0.1 to $-$1 reported, where a fit was only applied for those cores whose profiles are resolved and that did not show flat profiles.

An observed temperature profile with a slope of $-0.77$ was reported for the hot core G31.41+0.31 \citep{Beltran18} and supported by models \citep{Osorio09}. Models by \citet{Nomura04} suggest steeper profiles at distances considered here for higher H\2 column densities, which correspond to higher dust optical depths. The modelled temperature profile for Sgr\,B2\,(N2) shows a steepening from far distances to closer distances to the protostar, spanning a range of slopes for the optically thin regime to $-0.8$ for an optically thicker regime \citep[][]{Bonfand19}. 

Based on this discussion, it seems that for the scales traced with the ReMoCA survey, the desorption process of COMs in Sgr\,B2\,(N1) is thermally driven, that is, there is no clear sign in the temperature profiles that would indicate shock chemistry induced by the outflow or accretion shocks, at least not along the two analysed directions. A more detailed discussion on the differentiation between thermal and non-thermal desorption is done in Sect.\,\ref{Dss:desorb} also considering the abundance profiles. 
In addition to this result, we concluded in Sect.\,\ref{ss:profiles} that the power-law indices of the temperature profiles are more consistent with expected values for an optically thick dust continuum. 

\subsection{Abundance profiles}\label{Dss:Xprofiles}

We focus the discussion on the abundance profiles on a comparison with most recent astrochemical models performed by G22. The chemistry of the G22 models includes new methods for simulating non-diffusive reaction mechanisms, but the general physical and chemical setup is based on earlier works \citep[][]{Garrod08,Garrod13,Garrod17}. This setup includes two stages: a cold collapse phase (stage 1) and a subsequent warm-up phase (stage 2). In stage 1 the density increases from $3\times10^3$ to $2\times10^8$\,cm$^{-3}$ and the dust temperature decreases from $\sim$15\,K to 8\,K while the gas temperature is kept constant at a value of 10\,K on a timescale of $\lesssim$10$^6$\,yr. In stage 2 the final density of stage 1 is adopted and kept constant. Gas and dust temperatures increase simultaneously as it is assumed that they are well coupled during this stage. Three different timescales of this temperature rise are considered separately: $10^6$\,yr (slow warm-up), $2\times10^5$\,yr (medium warm-up), and $4\times10^4$\,yr (fast warm-up), which corresponds to the time it takes to reach 200\,K, respectively. The models then continue to a maximum temperature of 400\,K. In both stages the chemistry takes into account gas-phase, grain/ice-surface, and bulk ice chemistry, which are all coupled by various thermal and non-thermal processes. Amongst these is the interaction with cosmic rays (CRs) under all conditions, where the ionisation rate is kept constant at the standard value of $\zeta=1.3\times10^{-17}$\,s$^{-1}$.
The most recent model used in G22, however, does include a number of changes compared to the earlier models. We refer the reader to the article for a detailed description of new adjustments and their influence on the COM abundances. Some of the new implementations will, however, be mentioned in the following as they play a crucial role in explaining our observations.

\begin{figure}
    \centering
    \includegraphics[width=0.5\textwidth]{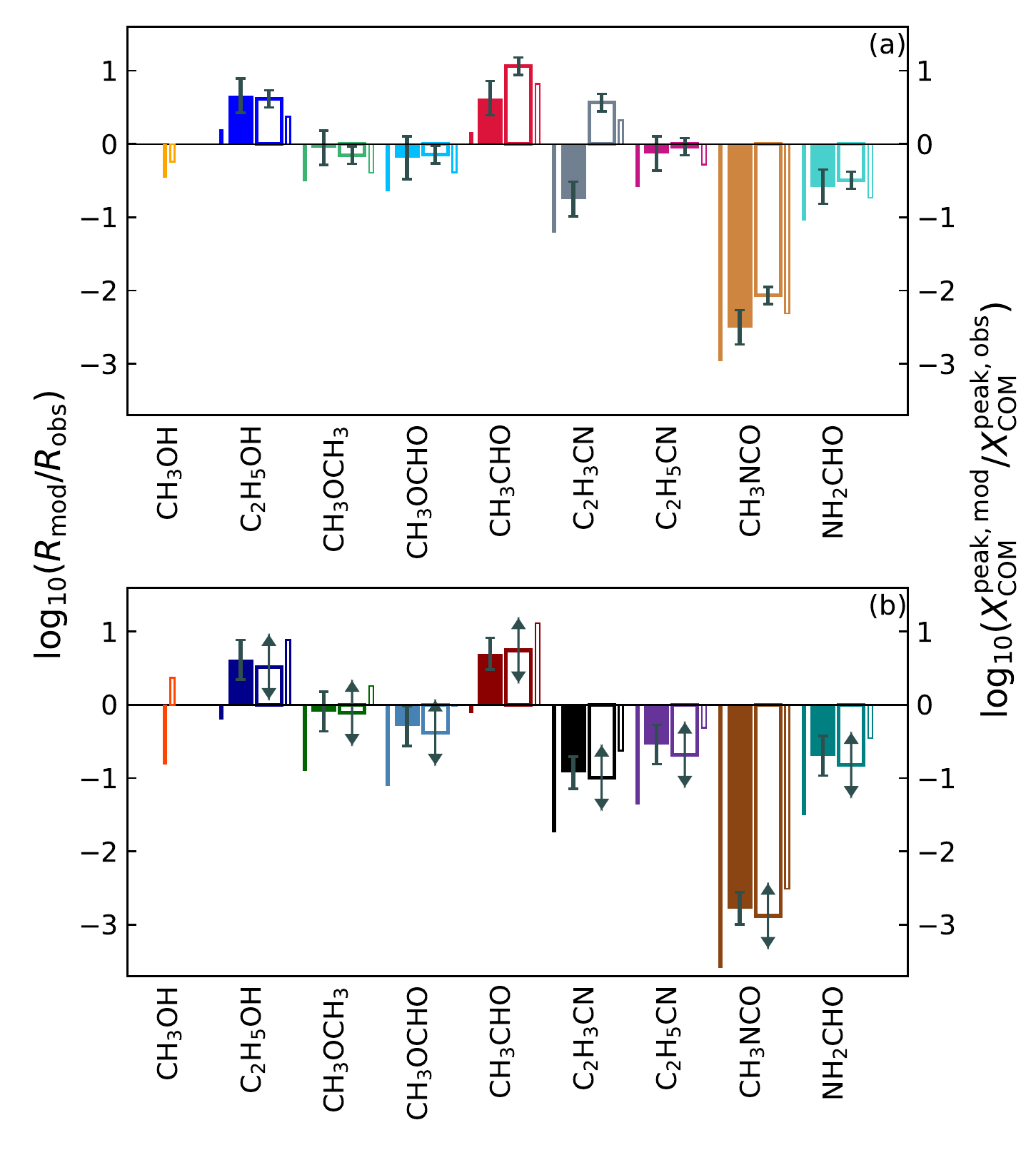}
    \caption{Comparison between observed peak abundances with respect to H\2 and the modelled ones to the a) south and b) west, where H\2 column densities were derived from dust (thick filled bars) or C$^{18}$O emission (thick unfilled bars). The peaks are taken from the profiles shown in Fig.\,\ref{fig:TvsX} and are normalised by the value for \met to obtain $R_{\rm obs}$. Note that the position of the peak can differ between the COM and methanol in all four cases. The peaks with respect to \met in the models, $R_{\rm mod}$, are taken from Table\,17 in \citet{Garrod22}, where we use the values of the slow warm-up model. Thin bars show the ratio of modelled abundance peaks $X^{\rm peak}_{\rm mod}$ and observed ones $X^{\rm peak}_{\rm obs}$. Observed abundances are with respect to H\2 derived from dust (filled bars) or C$^{18}$O 1--0 emission (blank bars). Modelled peak abundances are taken from Table\,16 in \citet{Garrod22} and multiplied by a factor 2 to roughly convert to abundances with respect to H\2 (from total H).  All values are summarised in Table\,\ref{tab:Xpeak}. Arrows pointing downwards or upwards indicate upper or lower limits, respectively.}
    \label{fig:Rch3oh}
\end{figure}

In general, our results are overall best described by the model of the slow warm-up phase, which was previously also found to best explain the observed abundances in Sgr\,B2\,(N2) (G22). 
To compare the modelled results to our observations and, based on that, make implications on the main formation and destruction pathways of COMs, we create a diagram similar to Fig.\,15 of G22, where they show the ratio of peak COM abundances relative to the peak abundance of \met. The result is shown with thick bars in Fig.\,\ref{fig:Rch3oh}, where on the left y axis we plot
\begin{align}
    \frac{R_{\rm mod}}{R_{\rm obs}} = \frac{X_{\rm COM}^{\rm peak,mod}}{X_{{\rm CH}_3{\rm OH}}^{\rm peak,mod}} \frac{X_{{\rm CH}_3{\rm OH}}^{\rm peak,obs}}{X_{\rm COM}^{\rm peak,obs}} 
\end{align}
where $X_{\rm COM}^{\rm peak}$ is the peak COM abundance and $X_{{\rm CH}_3{\rm OH}}^{\rm peak}$ is the peak abundance of \met.
The observed peak abundances with respect to H\2 are extracted from Fig.\,\ref{fig:TvsX}. The modelled peak abundance ratios are taken from Table\,17 in G22.
All values are listed together with the corresponding peak-abundance temperature in Table\,\ref{tab:Xpeak}.  In addition, we show a comparison of modelled and observed peak abundances $\nicefrac{X_{\rm COM}^{\rm peak,mod}}{X_{\rm COM}^{\rm peak,obs}}$ without normalising to \met (thin bars, right y axis). The modelled peak abundances are taken from Table\,16 in G22 and multiplied by a factor 2 to convert to abundances with respect to H\2 (from total H).

Besides peak abundances, the behaviours of the profile are compared to the model predictions. In particular, we use Figs.\,12,  20, and 21 from G22 that present results of the cold and subsequent slow warm-up phase. All three show the time evolution of the COM abundances, that is also linked to the temperature increase due to the evolution of the protostar. Figure\,12 of G22 presents the abundance profile in the solid and gas phase while Figs.\,20--21 of G22 (see also their Figs.\,13--14 to find responsible chemical reactions) show the rate at which a COM is dominantly formed or destroyed at each point of the evolution.

\begin{table*}[]
\caption{Observed (to the south and west) and modelled peak COM abundances and corresponding temperatures.} 
\centering
\small
\begin{tabular}{lclclclcllc}
\hline\hline \\[-0.3cm] 
& \multicolumn{4}{c}{South} & \multicolumn{4}{c}{West} & & \\ \cmidrule(lr){2-5}\cmidrule(lr){6-9} 
 & \multicolumn{2}{c}{$N({\rm H}_2)$ from dust} & \multicolumn{2}{c}{$N({\rm H}_2)$ from C$^{18}$O} & \multicolumn{2}{c}{$N({\rm H}_2)$ from dust} & \multicolumn{2}{c}{$N({\rm H}_2)$ from C$^{18}$O} & & \\ \cmidrule(lr){2-3}\cmidrule(lr){4-5}\cmidrule(lr){6-7}\cmidrule(lr){8-9}
 Molecule & $X_{\rm peak}$ & $T_{\rm rot}$ (K) & $X_{\rm peak}$ & $T_{\rm rot}$ (K) & $X_{\rm peak}$ & $T_{\rm rot}$ (K) & $X_{\rm peak}$ & $T_{\rm rot}$ (K) & $X_{\rm peak}^{\rm mod}$ & $T_{\rm rot}^{\rm mod}$ (K) \\ \hline \\[-0.3cm] 
CH$_3$OH & 4.8$\pm$2.3($-$5) & 104$\pm$1 & 2.9$\pm$0.6($-$5) & 122$\pm$1 & 1.1$\pm$0.5($-$4) & 131$\pm$2 & $\ge$5.7($-$6) &$\ge$149 &1.7($-$5) & 166 \\[0.05cm] 
 C$_2$H$_5$OH & 8.3$\pm$2.1($-$7) & 130$\pm$1 & 5.6$\pm$1.1($-$7) & 119$\pm$2 & 2.1$\pm$0.9($-$6) & 147$\pm$3 & $\ge$1.4($-$7) &$\ge$169 &1.3($-$6) & 166 \\[0.05cm] 
 CH$_3$OCH$_3$ & 1.4$\pm$0.4($-$6) & 131$\pm$1 & 1.1$\pm$0.2($-$6) & 110$\pm$1 & 3.5$\pm$1.5($-$6) & 109$\pm$2 & $\ge$1.9($-$7) &$\ge$128 &4.4($-$7) & 200 \\[0.05cm] 
 CH$_3$OCHO & 2.2$\pm$1.1($-$6) & 111$\pm$2 & 1.2$\pm$0.2($-$6) & 137$\pm$2 & 6.3$\pm$2.8($-$6) & 144$\pm$3 & $\ge$4.1($-$7) &$\ge$162 &5.0($-$7) & 225 \\[0.05cm] 
 CH$_3$CHO & 3.4$\pm$0.9($-$7) & 277$\pm$8 & 7.5$\pm$1.4($-$8) & 168$\pm$7 & 6.5$\pm$1.6($-$7) & 293$\pm$16 & $\ge$3.1($-$8) &$\ge$159 &5.0($-$7) & 257 \\[0.05cm] 
 C$_2$H$_3$CN & 6.5$\pm$1.6($-$7) & 299$\pm$15 & 1.9$\pm$0.4($-$8) & 182$\pm$6 & 2.2$\pm$0.5($-$6) & 182$\pm$6 & $\ge$1.4($-$7) &$\ge$165 &4.0($-$8) & 166 \\[0.05cm] 
 C$_2$H$_5$CN & 1.2$\pm$0.3($-$6) & 169$\pm$1 & 5.7$\pm$1.1($-$7) & 169$\pm$1 & 6.8$\pm$3.0($-$6) & 151$\pm$1 & $\ge$4.9($-$7) &$\ge$167 &3.0($-$7) & 166 \\[0.05cm] 
 CH$_3$NCO & 6.7$\pm$1.7($-$8) & 139$\pm$7 & 1.5$\pm$0.3($-$8) & 106$\pm$7 & 2.9$\pm$0.7($-$7) & 123$\pm$18 & $\ge$1.9($-$8) &$\ge$85 &7.2($-$11) & 343 \\[0.05cm] 
 NH$_2$CHO & 1.2$\pm$0.3($-$6) & 122$\pm$3 & 5.8$\pm$1.1($-$7) & 122$\pm$3 & 3.4$\pm$1.5($-$6) & 136$\pm$3 & $\ge$2.4($-$7) &$\ge$138 &1.1($-$7) & 166 \\[0.05cm] 
 \hline\hline
\end{tabular}
\tablefoot{Values of observed peak abundance $X_{\rm peak}$ are taken from Fig.\,\ref{fig:TvsX} and represent COM column densities with respect to H\2, where we consider H\2 column densities derived from either dust (2nd and 6th columns) or C$^{18}$O 1--0 emission (4th and 8th columns). The modelled peak abundances $X_{\rm peak}^{\rm mod}$ present COM column densities with respect to total H taken from Table\,16 from \citet{Garrod22} multiplied by a factor 2 to roughly convert to abundances with respect to H\2. The values in parentheses show the decimal power, where $x\pm y(z) = (x\pm y)\times 10^z$.}
\label{tab:Xpeak}
\end{table*}

\subsubsection{Methanol \met}
Methanol is the most abundant COM in the observation as well as the model with peak abundances that differ by factors of a few only (cf. Table\,\ref{tab:Xpeak} and thin bars in Fig.\,\ref{fig:Rch3oh}).
According to the model, \met is predominately formed in the bulk ice before the onset of the warm-up as seen in Fig.\,20 of G22. 
After co-desorption with water starting at $\sim$120\,K and finishing at 166\,K, the molecule is destroyed in the gas phase, mainly through proton transfer from H\3O$^+$. This leads to the drop of gas-phase abundance seen towards higher temperatures right after the peak at 166\,K in the model. 
Except for the value of peak abundance, that is underestimated by the model (except for \wco), however only by factors of a few, and temperature of this peak, that is higher in the model by some tens of Kelvin (see Table\,\ref{tab:Xpeak} and Fig.\,\ref{fig:Rch3oh}), the model is in good agreement with the observations (cf. Sect.\,\ref{sss:met}). Therefore, the \met abundance in Sgr\,B2\,(N1) is most likely enhanced in the gas phase due to thermal desorption at $\sim$100\,K supported by the increase that seems fairly sudden (see also Sect.\,\ref{Dss:desorb}).

The similar behaviour of abundances of the main and $^{13}$C isotopologues 
implies that similar formation and destruction mechanisms take place for both
species. This also indicates that our modelling of the
\met emission accounted well for potential optical depth effects. Considering only the abundance plateau, that is, distances $\lesssim$\,2\arcsec to Sgr\,B2\,(N1), we find an observed median $^{12}$C/$^{13}$C isotopic ratio of of 22 and 24 to the south and west, respectively, which is consistent with the value of 25 obtained for methanol and ethanol in Sgr\,B2\,(N2) with the EMoCA survey \citep[][]{Mueller16} and agrees with expected values for the Galactic centre region.

\subsubsection{Ethanol \et} 
The bulk of \et in the model is produced in the cold phase on dust grains and is effectively destroyed after desorption into the gas phase mainly via proton addition by protonated molecules like H\3O$^{+}$, H$\overset{+}{_3}$, or HCO$^{+}$, which is followed by electron recombination (cf. Fig.\,20 of G22). Similar to methanol, some fraction of the protonated molecule does not recombine with an electron, instead, it transfers the proton to NH\3 and goes back to unprotonated ethanol.
The observed slight increase of the abundance of ethanol relative to methanol beyond their abundance peak (cf. Sect.\,\ref{sss:et}) suggests that ethanol is less effectively destroyed than methanol after desorption in the gas phase, which is not accounted for by the current model. 

Observed and modelled values of peak abundance differ by a factor 2.5 at most, except for \wco (factor 10). The observed abundance ratios with respect to methanol shown in Fig.\,\ref{fig:Rch3oh} are overproduced by the model by factors 3--5 in all cases. 
The chemical network of G22 may indeed lack several grain-surface reactions that affect the rate of ethanol production; in particular, reactions between atomic H and the set of C\2H$_X$ species generally result in H addition, while H abstraction by atomic H to produce H\2 and a less hydrogen-rich hydrocarbon is not included. Jin et al. (in prep.) have made such adjustments to the network, which could result in a moderate reduction in ethanol production efficiency on grain surfaces at low temperatures, via, e.g., \\ C\2H\5 + OH $\rightarrow$ \et.
 
\subsubsection{Dimethyl ether \dme}
In the model, the bulk of \dme is produced on dust grains in the cold collapse phase as well as on the surface during the warm-up phase and co-desorbs with water at $T>120$\,K.  
In the gas phase at temperatures $\gtrsim$\,120\,K, \dme is additionally produced via proton transfer from CH\3OCH$\overset{+}{_4}$ to NH\3. This is an efficient formation pathway because the protonated dimethyl ether is the product of the reaction between two abundant reactants: methanol and protonated methanol. 
Eventually, this leads to the abundance peak of \dme at slightly higher temperatures of 200\,K and a flat profile as temperature increases (cf. Fig.\,12 of G22). With this, the modelled profile differs considerably from the observations 
(cf. Sect.\,\ref{sss:dme}). 

Observed and modelled peak abundances differ by factors smaller than 2 for \wco and up to $\sim$8 for \wdu, with \sdu and \sco in between (cf. Table\,\ref{tab:Xpeak}). A similar behaviour of peak abundances between observations and model is seen for methanol, therefore, the observed ratio of both COMs is remarkably comparable to the modelled one in all cases, with differences less than a factor 2 (cf. Fig.\,\ref{fig:Rch3oh}). 
In summary, while the model succeeds in reproducing the observed peak abundance of dimethyl ether with 
respect to methanol, it seems that the destruction of this molecule in the gas phase is more efficient in Sgr\,B2\,(N1) than in the model beyond the position of peak abundance at higher temperatures. Whether that indicates a destruction channel missing in the chemical network of the model or whether this is a result of any geometrical or physical conditions not caught by the model remains uncertain.
It is likely that the regions closer to the centre of Sgr\,B2\,(N1) have, in addition to higher temperatures, also higher densities, in which case the destruction may be accelerated. Such an effect is not considered by the model as the density is fixed during the warm-up. However, this would reduce the discrepancy between the observations and the model only if the destruction of dimethyl ether would be more accelerated than that of methanol.

\subsubsection{Methyl formate \mf}

Similar to the aforementioned O-bearing COMs, methyl formate is predominantly produced on dust grains in the cold phase as well as during the warm-up in the model and desorbs into the gas phase as soon as uncovered from water at $\lesssim$166\,K (cf. Fig.\,20 of G22). Moreover, \mf is efficiently produced in the gas phase at temperatures of water desorption via oxygenation of CH\3OCH\2, which was primarily sourced from \dme. This leads to an abundance peak at a higher temperature than for the other COMs. At $T>200$\,K, Fig.\,20 of G22 suggests some destruction of the molecule, which is only minor and does not lead to a drop in the gas-phase abundance profile. 
Similar to \dme, the observed values of peak abundance agree well with the model for \wco and are higher for \sco, \sdu, and  \wdu by factors 2--4 up to an order of magnitude (cf. Table\,\ref{tab:Xpeak}). However, when normalised to methanol the observations and the model agree again more or less within the error bars (cf. Fig.\,\ref{fig:Rch3oh}).
The observed decrease in abundance towards higher temperatures (cf. Sect.\,\ref{sss:mf}), which is similar to the other O-bearing COMs, is not reproduced by the model and may demand a more efficient destruction rate of the molecule. 
Similar to \dme, higher densities closer to Sgr\,B2\,(N1), not taken into account by the model, may accelerate the destruction process.

\subsubsection{Acetaldehyde \ad}

According to the model (cf. Fig.\,20 of G22), only a minor fraction of \ad is produced on dust grains in the cold phase and during the warm-up phase, and co-desorbs into the gas phase with water. A significant amount is formed in the gas phase at temperatures $\gtrsim$150\,K via the oxygenation of C\2H\5. This leads to a continuously increasing gas-phase abundance profile in the model, which agrees with the observed results for \ad (cf. Sect.\,\ref{sss:ad}). 
Peak abundances to the south agree well with the model predictions while those to the west are overestimated by the model by factors 5 and 10 (cf. Table\,\ref{tab:Xpeak}). When normalised to \met, the model overestimates the abundance ratio by factors 5--10 for all directions (cf. Fig.\,\ref{fig:Rch3oh}). 
In the model the acetaldehyde peak abundance and, in turn, the ratio to methanol are sensitive to the warm-up timescale, where the ratio is higher by a factor 6.4 for the slow than for the fast warm-up (cf. Tables\,16 and 18 in G22).  Therefore, the overestimation of the observed ratio by the model may be an indication that the warm-up in Sgr\,B2\,(N1) proceeds faster than predicted for the slow warm-up model to which we are comparing here. However, for all other COMs, except \et, the difference between model and observations is slightly larger when assuming the fast warm-up.

\subsubsection{Vinyl cyanide \vc}\label{Dss:vc}

As can be seen in Fig.\,21 of G22, there are two main formation phases of \vc in the model. Some formation happens on the dust grains before the warm-up phase, however, the bulk of the molecule is produced in the gas phase at temperatures higher than 160\,K predominantly through the reaction of C\2H\4 and CN. This leads to a continuous increase of abundances in the model and could explain the increase seen in the observations (cf. Sect.\,\ref{sss:vc}).
However, the values of observed peak abundance are underestimated by the model by a factor 2 for \sco up to more than an order of magnitude for \wdu with \sco and \wco in between (cf. Table\,\ref{tab:Xpeak}). This discrepancy is also seen in the abundance ratios with respect to methanol, that is it is underestimated by factors 6--10 in the model for all but \sco, for which the ratio is overestimated by a factor 4 (cf. Fig.\,\ref{fig:Rch3oh}). 
G22 mention that the model tends to underestimate abundances of N-bearing COMs including \vc, which they attribute to a possible overproduction of HC\3N in the model. They propose that this could be counteracted if HC\3N was hydrogenated on grain surfaces at early stages to form \vc and  \etc, eventually.

\subsubsection{Ethyl cyanide \etc}

\etc is exclusively produced on dust grains in the model, partially during the warm-up phase but mainly in the cold phase (cf. Fig.\,21 of G22). As soon as the molecule arrives in the gas phase after co-desorption with water, it is efficiently destroyed via proton transfer from H\3O$^+$ leading to a drop in abundance at high temperatures. However, some of the protonated ethyl cyanide goes back to the unprotonated form by transferring the extra proton to NH\3.
This behaviour at high temperatures differs significantly from the observations (cf. Sect.\,\ref{sss:etc}).  

The observed steeper increase of abundance around 100\,K may point to thermal desorption, however, peak abundances are only reached at much higher temperatures suggesting an increase of abundance in the gas phase.
Although there currently is no efficient gas-phase formation route for \etc included in the model, G22 discussed that the reaction of CH$\overset{+}{_3}$ and CH\3CN again leading to C\2H\5CNH$^{+}$ \citep[see also][]{McEwan89} could be efficient in increasing the ethyl cyanide abundance at $T>100$\,K, when the protonated species transfers the extra proton to NH\3.

The observed values of peak abundance are larger than in the model by factors of a few for all cases except for \wdu, for which abundances differ by more than an order of magnitude (cf. Table\,\ref{tab:Xpeak}). To the south, the ratio with respect to methanol does, however, agree well with what is expected from the model, while to the west, an underestimation by a factor of a few is still evident (cf. Fig.\,\ref{fig:Rch3oh}).
Including the above-mentioned additional gas-phase reaction to the model would probably be able to at least partially counteract the heavy decrease in gas-phase abundance seen in the model and may even somewhat increase the peak abundance to better match the observations. Additionally, 
there may be another formation channel on dust grains via multiple hydrogenation of HC\3N as described for \vc.

Vinyl and ethyl cyanide are often mentioned together as they share a chemical link. However, there is no such strong link in the new model of G22 compared to former studies \citep[][]{Garrod17}. Previously, a large fraction of the \vc gas-phase abundance was produced from C\2H\5CNH$^{+}$, which, in turn, was sourced from ethyl cyanide, leading to decreasing abundances of \etc with increasing temperature in the gas phase. In the new model, the gas-phase reaction mentioned in Sect.\,\ref{Dss:vc} dominates the formation of \vc and weakens the gas-phase relationship of the two COMs (G22). 

\subsubsection{Methyl isocyanate \mic}
According to the model, \mic is exclusively produced on dust grains mainly during the warm-up phase, through the reaction CH\3\,+\,OCN$\,\rightarrow$\,\mic \citep[][]{Belloche17}. Subsequently, a fraction is already destroyed in the grain mantle, where it is hydrogenated to eventually form CH\3NHCHO, a molecule that was recently identified in the interstellar medium \citep[][]{Belloche17,Belloche19}. The other fraction desorbs to the gas phase, where it reacts with H\3O$^{+}$ resulting in further decrease in abundance (cf. Fig.\,\ref{fig:roc_mic}). 
Therefore, we would expect similar behaviour of the abundance profile as for example for \et, for which abundances peak right after the molecule desorbs and decrease significantly towards higher temperatures. If that were the case, it would disagree with the observed abundance profile of \mic (cf. Sect.\,\ref{sss:mic}). 

Moreover, the model underestimates the observed peak abundances by 2--3 orders of magnitude (cf. Fig.\,\ref{fig:Rch3oh}), which is also the case for the abundance ratios with respect to methanol.
G22 argue that this may partly be explained by uncertainties in the activation barriers of reactions occurring on the grain surface. However, the observed trend of the abundance profile may additionally suggest that \mic is also produced in the gas phase. \citet{Halfen15} proposed that the reaction of HNCO with CH\3 would be able to efficiently produce \mic in the gas phase, which is supported by later studies targeting low-mass protostars \citep[][]{Martin-Domenech17,Quenard18}. 
However, as \citet{Halfen15} noted, the CH\3\,+\,HNCO reaction that they proposed could be endothermic, and this seems to be confirmed by the calculations of \citet{Majumdar18}. The latter authors proposed alternative formation mechanisms for the grain-surface production of CH$_3$NCO. These included a barrierless reaction between atomic N and the CH$_3$CO radical, and the reaction of H with HCN that is trapped in a van der Waals complex with surface CO. In the latter case, in spite of a barrier to hydrogenation of HCN, the resultant product, H$_2$CN, would have sufficient energy to overcome the barrier to its own reaction with CO to produce CH$_2$NCO. This radical would then be hydrogenated to methyl isocyanate. The protostellar chemistry model presented by \citeauthor{Majumdar18} indicates that their proposed mechanisms can achieve gas-phase abundances on the order of $10^{-7}$ with respect to H$_2$, following desorption of the ice mantles; these abundances are in line with some of our observed values. It is unclear how the non-diffusive chemistry included in the G22 model might influence the efficiency of mechanisms proposed by \citet{Majumdar18}. Another alternative mechanism for grain-surface production of CH$_3$NCO, proposed by \citet[][submitted]{Vavra22}, would be a reaction between HNCO and the diradical methylene (CH$_2$), which is formed during the course of CH$_4$ production on dust-grain surfaces, and may also be a photoproduct of methane in the bulk ice.
Whatever the mechanism, the lack of an efficient gas-phase process to form CH$_3$NCO would indicate that surface/ice chemistry is at work.

\subsubsection{Formamide \fmm}
According to the model, \fmm is dominantly produced on dust grains in the cold phase and during the warm-up phase (cf. Fig. 21 of G22). 
These production mechanisms continue as \fmm starts to co-desorb with water at $T\gtrsim120$\,K and moreover, its abundance is enhanced by hydrogenation of NH\2CO when this radical is uncovered as water desorbs. 
At these temperatures also gas-phase reactions start to produce some \fmm. However, gas-phase production through the reaction of NH\2 and H\2CO \citep[][]{Barone15} is responsible only for around 9\% of total production in the G22 model. Despite the many formation channels of this COM, the model predicts that it is efficiently destroyed via protonation followed by recombination, which is in disagreement with observed abundance profile (cf. Sect.\,\ref{sss:fmm}). 

Also for this N(+O)-bearing COM, the observed peak abundance is higher than the modelled one by factors of a few to the west and by more than an order of magnitude to the south (cf. Table\,\ref{tab:Xpeak}). Ratios of abundances with respect to methanol are underestimated by the model by factors 3--7. 
So far it is uncertain what causes the difference between model and observations.  Based on the observed abundance profiles and in comparison with other COMs, we may expect both, thermal desorption at $\sim$100\,K, which, however, seems to happen at significantly shorter projected distances than for other COMs, and possibly formation in the gas phase at higher temperatures as also predicted by the model. As there is no indication for a significant drop in the observed abundances with increasing temperature, there may exist an additional supply of \fmm or/and the molecule may be less efficiently destroyed than in the model. 

\subsection{Resolved desorption of COMs in Sgr\,B2\,(N1)}\label{Dss:desorb}

From Sect.\,\ref{Dss:Tprofiles}, we concluded that the temperature profiles derived from the COMs reveal the gradient that is expected from heating of an envelope with optically thick dust by a protostar. This is true for all COMs be they predicted to mainly form in the gas phase like \ad or in the solid phase (and then desorb) like \met. However, the high angular resolution of the ReMoCA survey allowed us to resolve the abundance profiles of the COMs and reveal their differences. This allows us to draw conclusions on the prevalent desorption process.

\subsubsection{Thermal co-desorption with water}\label{Dsss:coH2O}

In the case of thermal desorption, we expect a sudden rise in column density and abundance of a COM at the location of desorption, which can either happen at the co-desorption temperature of water and COMs, or at a specific desorption temperature that depends on the COM's binding energy to the grain surface or to itself.  
Perhaps the greatest change implemented to the current model used in G22 compared to the earlier studies \citep[][]{Garrod13} is that diffusion in the bulk ice as well as from the bulk ice to the surface layers regardless of the temperature is prohibited for all species except H and H\2.  This entails that COMs formed in the bulk ice cannot diffuse to the surface from where they could thermally desorb depending on their binding energy. 
In the new non-diffusive model of G22 the COMs are trapped in the water ice and the bulk of the COMs can only desorb when water does, which happens at 120--170\,K in the model. 
Based on Fig.\,\ref{fig:TvsX}, COMs that present such an
abundance profile with a steep increase at a given temperature in Sgr\,B2\,(N1) and thereby, are in agreement with the model are \met, \et, \fmm, and \etc, and tentatively, \dme and \mf. However, instead of a temperature $\gtrsim$120\,K as predicted by the model, abundances of these species already increase at temperatures $\sim$100\,K.
Provided that this behaviour can be associated with their thermal co-desorption with water, then this process seems to start at these low temperatures of about 100\,K and to already terminate at $\sim$100-150\,K, corresponding to the peak-abundance temperature of the above-mentioned species (see Fig.\,\ref{fig:TvsX}). 
Because the binding energy of water is uncertain as it depends on the precise ice structure, the value in the model could well be lower, leading to a lower desorption temperature of water which would be in better agreement with the observations. Jin et al. (in prep.), who use the model of G22, have in fact recently revised the value downwards and achieved a better agreement between their model and observations of Orion\,KL.

Though all O-bearing COMs except \ad present an abundance increase at $\sim$100\,K, the column densities of dimethyl ether and methyl formate do not increase as steeply as those of methanol and ethanol (cf. Fig.\,\ref{fig:Ncolprofiles}). 
In the model, gas-phase formation of the two starts as soon as water starts desorbing, which may explain the observed behaviour. 
Besides, \dme and \mf have lower binding energies than \met and \et amongst other COMs and water (cf. Table\,\ref{tab:Ebin}), which led to thermal desorption at lower temperatures in earlier models \citep[e.g.,][]{Garrod13}. However, if thermal desorption took place depending on binding energies, we would still expect a steep increase of column density and abundance only at lower temperatures for \dme and \mf, which is not evident in the column density and abundance profiles. 
However, the observed profiles could also be the result of a two-step thermal desorption process that we will elaborate on further in Sects.\,\ref{dsss:2step} and \ref{dsss:gas}.

The abundance profile of ethyl cyanide increases more steeply around 100\,K than at higher temperatures, however, not as steeply as for methanol. 
This may indicate that only some fraction is released from dust grains at around 100\,K, which would be in agreement with desorption either along with water or depending on the binding energy, given its similar values of binding energy as methanol and ethanol (cf. Table\,\ref{tab:Ebin}). 
Formamide abundances increase steeply, however, only  at shorter distances to Sgr\,B2\,(N1) than methanol, that is, potentially, at higher temperatures.

A steep increase is not observed for \ad and \vc, which is consistent with the expectation from the model that they are mainly formed in the gas phase. Although a fraction is produced on dust grains and released during water desorption in the model of G22, the amount may not be significant enough to be detected in the observed abundance profiles. The same would apply if desorption depended on the binding energy.
Temperatures derived from \mic have higher uncertainties, nonetheless, the abundance profiles of this COM are similar to those of \ad suggesting a predominant gas-phase formation. Moreover, although the binding energy of \mic could be as high as 6500\,K, thermal desorption at higher temperatures seems unlikely given the only mild increase of abundance with increasing temperatures.

In summary, the behaviour of the observed column density and abundance profiles for COMs that are expected to (partially) form on dust grain surfaces and to desorb subsequently, suggests that co-desorption with water is likely to take place.
Together with the overall good agreement of the model by G22 with our observations, precisely because the co-desorption with water was implemented amongst other revisions, this process of thermal co-desorption seems to be the prevalent one in Sgr\,B2\,(N1).

\begin{figure}
    \centering
    \includegraphics[width=0.5\textwidth]{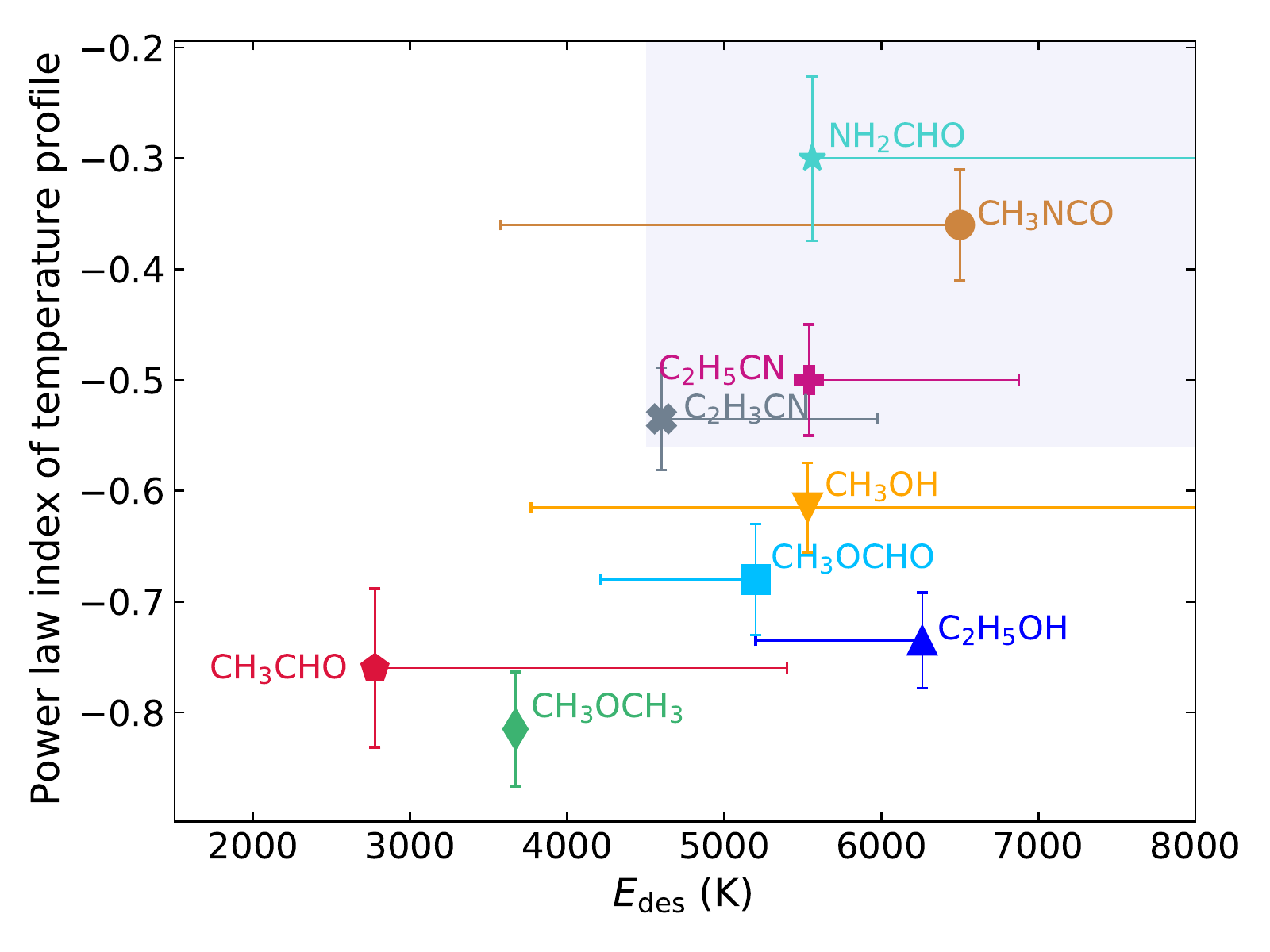}
    \caption{Binding (or desorption) energies versus the power-law indices of the COM temperature profiles shown in Fig.\,\ref{fig:Tprofiles} averaged over the south and west directions. The binding energies are taken from the third column of Table\,\ref{tab:Ebin}, with uncertainties indicating the range of values listed in that table. The lilac rectangle highlights the location of the N-bearing COMs.} 
    \label{fig:EbinSlopes}
\end{figure}

Therefore, we also exclude that the observed segregation of COMs is based on their binding energies as it was proposed for the classification into \textit{hot} and \textit{cold} species as introduced in Sect.\,\ref{Dss:OandN}. 
We observe a small difference between O- and N-bearing COMs in the power law indices of the temperature profiles, where the latter have shallower slopes than the former and therefore, do not trace as high temperatures on average. Naively, for this situation, we would infer that when observed along the same line of sight, N-bearing (or N+O-bearing) COMs show lower temperatures because they are destroyed in the hotter regions that are traced by the O-bearing COMs. 
This is rather the opposite of what is predicted by the chemical models of G22. However, we cannot exclude this kind of behaviour given that the model does not take into account density changes and neither the model nor the observations consider the three-dimensional (3-D) structure of the hot core. 
There also exists a rough dependence of the power-law indices on the binding energy of the COMs as shown in Fig.\,\ref{fig:EbinSlopes}, where N-bearing (or N+O-bearing) COMs have higher binding energies on average and thus, can stick to the grain surface until higher temperatures are reached. Why this would lead to N-bearing (or N+O-bearing) COMs tracing lower temperatures is unclear to us.

\subsubsection{Transition from non-thermal to thermal desorption?}\label{Dsss:ntdesorp}
Assuming that the increase in abundance at $\sim$100\,K represents the thermal co-desorption of COMs and water into the gas phase, non-zero abundance values at lower temperature, such as the value of 10$^{-8}$ at 80\,K for \met or at $\sim$55\,K for $^{13}$\met in Fig.\,\ref{fig:TvsX}f, the value $\sim$10$^{-9}$ at 55\,K for \etc in panel (i) of the same figure, and low-temperature values around 10$^{-9}$ for \fmm in panel (j), would be the consequence of 
another desorption process.
One possibility would be non-thermal desorption, which would be able to enhance COM abundances at temperatures lower than $\sim$100\,K.
The cold-phase chemistry as described in detail by \citet{Jin20} is included in the model of G22, and this includes both chemical desorption and UV-driven desorption of surface molecules. The latter mechanism tends to show only limited ability to desorb molecules at visual extinction values greater than around unity, under the assumption of the standard interstellar radiation field and CR ionisation rate. Chemical desorption \citep[i.e. desorption induced by the release of chemical energy upon formation of a molecule; see][]{Garrod07}, on the other hand, is able to drive substantial COM desorption at low temperatures. However, strong desorption occurs during the periods of strongest ice production; at later times in the models, prior to thermal desorption, abundances of COMs relative to H in the gas phase remain below a value of 10$^{-11}$, which is at least two orders of magnitude lower than the observed values that we suspect may indicate non-thermal desorption of COMs.

However, with its location in the Galactic centre, Sgr\,B2\,(N) resides in a violent environment with enhanced turbulence and enhanced CR fluxes that influence the star-formation process. 
The enhanced CR flux in Sgr\,B2\,(N) \citep[][]{Bonfand19} can work in both directions. CRs and the secondary UV field that they provoke can support formation and non-thermal desorption of some COMs while they destroy others when upon collision with the dust grains. 
Non-thermal desorption due to a shock passing and turbulence is most probably the reason for the complex chemistry observed in G+0.693$-$0.027, a source not far from Sgr\,B2\,(N) without any signposts of active star formation \citep[e.g.,][]{Zeng18,Zeng20} although the most recent study of this object propose that it may be on the verge of star formation \citep[][]{Colzi22}. Shocks are proposed to be induced through a cloud-cloud collision. Multiple studies suggest that these interactions between clouds not only happen in this source but in Sgr\,B2 in general \citep{Hasegawa94,Sato00,AA20,Zeng20} and that these may even have triggered the star formation process. Therefore, besides enhanced CR flux, enhanced turbulence could increase the abundance of COMs via non-thermal desorption in the colder gas phase of Sgr\,B2\,(N1). 
Specifically for this hot core, shocks may also arise from accretion because along the directions analysed here, filaments have been located through which accretion towards Sgr\,B2\,(N1) may happen \citep{Schwoerer19}. However, because we do not see any evidence for accretion shocks in the morphology \citep[as e.g.,][did in G328.2551--0.5321]{Csengeri19}, we excluded this possibility. Moreover, although we avoided the outflow axes for the analysis in this work, an influence of the outflow of Sgr\,B2\,(N1) can probably not be excluded a priori, especially at shortest distances to the hot core's centre. In this case, however, we would expect a simultaneous increase of temperature in the shocked gas. Because the temperature profiles do not show any evidence for that, we expect the influence of the outflow to also be minor if existent at all along the directions analysed in this work. 

\subsubsection{Thermal desorption in two steps?}\label{dsss:2step}
Thermal desorption may also take place in two steps. Figure\,8 in G22 shows that, although water is still abundant, CO is the main constituent of the outermost ice layers at the end of the cold collapse phase. Therefore, these outermost ice layers can possibly desorb at much lower temperatures (20--30\,K) given the low binding energy of CO. If some COMs are also abundant in these layers, their binding energies will be reduced compared to the value they have for binding to pure water ice. 
As a result these COMs may be able to desorb according to their reduced binding energy at temperatures $<$100\,K. 
If, after desorption of these CO-rich ice layers, the remaining ice were still mostly CO-related COMs, then the ice would be less polar and binding would presumably be weaker, still allowing for desorption at lower temperatures. Based on Fig.\,9 of G22, COMs that seem to be present in these outer ice layers are \met, \et, \dme, \mf, \ad but also \etc, which is not CO-related. Hence, thermal desorption due to reduced binding energies may not only apply to O-bearing COMs.
This process may be able to explain the non-zero abundances at low temperatures for \met and \etc as well as the shallower increase for \mf and \dme. The slightly different behavior shown by methyl formate and dimethyl ether may be due to their somewhat lower binding energies (compared with the other molecules mentioned above), even on water ice. Strong trapping of COMs by H$_2$O in this picture would then manifest once the upper layers, which are rich in CO, had desorbed along with some COMs. Some COMs would still be present in the more water-rich ice below, to be released at higher temperatures when water desorbs. 

\subsubsection{Gas-phase production}\label{dsss:gas}
Gas-phase formation may also present an option to enhance COM abundances at low temperatures. Gas-phase reactions for \dme and \mf may be able to explain their overall shallower increase in abundance. However, it is questionable whether gas-phase formation, e.g., for \fmm, can explain the non-zero abundance values at low temperatures because, although included in the model, the abundances prior to thermal desorption remain lower than the observed values at low temperatures. This may indicate that gas-phase formation is simply not efficient enough to reproduce the observed abundances or/and that the reactants are not sufficiently abundant, which is also plausible given the low temperatures. Moreover, for methanol, there is currently no efficient gas-phase formation route that would explain the COM's non-zero abundance at low temperatures. Eventually, we would exclude gas-phase reactions as the main reason for the non-zero abundances at low temperatures.

\subsection{Implications for COM segregation}
\begin{figure*}
    \centering
    \includegraphics[width=0.45\textwidth]{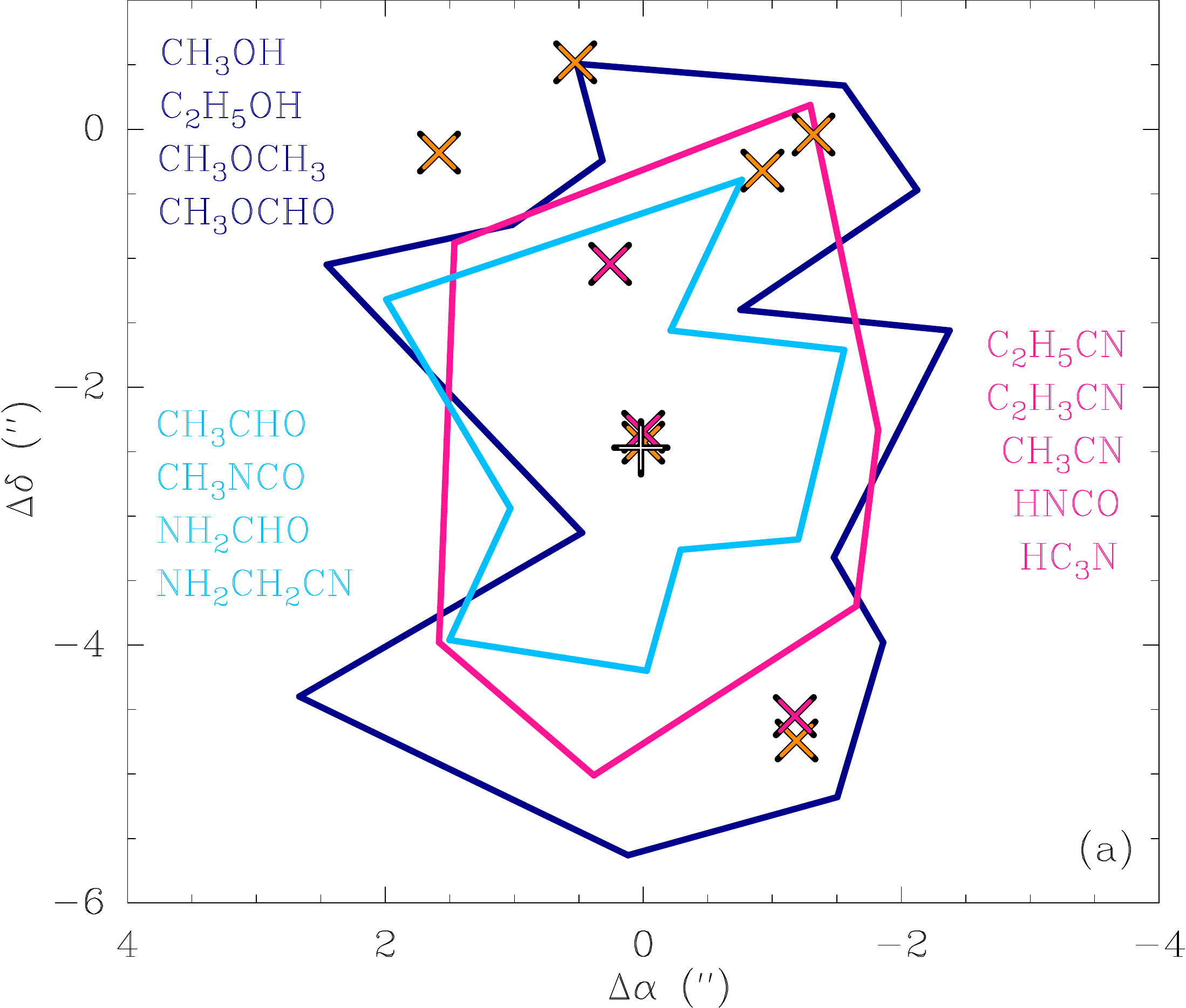}
    \includegraphics[width=0.53\textwidth]{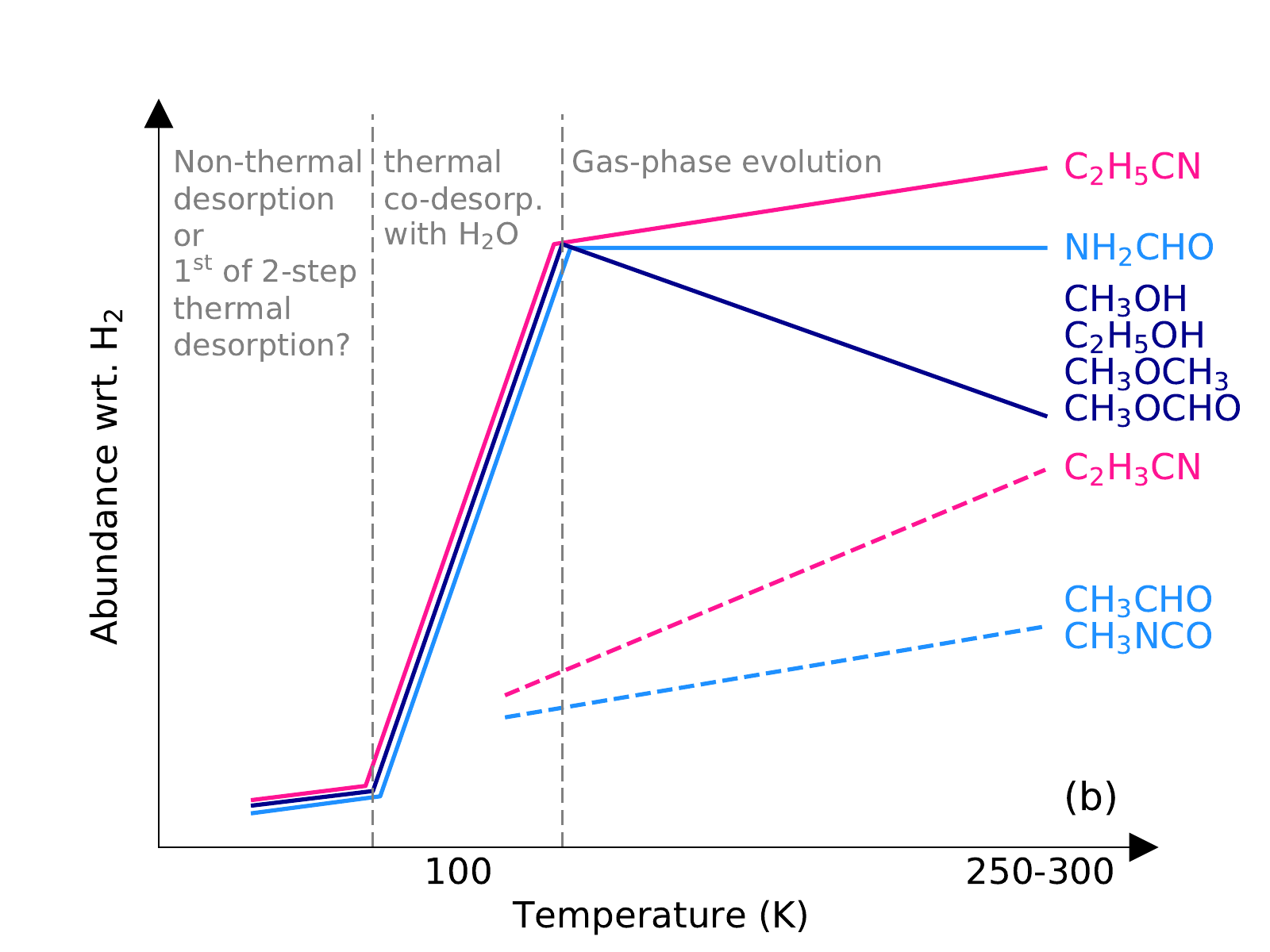}
    \caption{\textbf{(a)} Overview sketch of the three different types of morphology observed in the integrated intensity maps of molecular emission built with the LVINE method as shown in Fig.\,\ref{fig:COM-Maps}. The molecules, COMs and simpler ones, that show the respective morphology are listed in the colour of the contour. The markers are the same as in Fig.\,\ref{fig:continuum-maps}. \textbf{(b)} Overview sketch of the different types of abundance profile as a function of temperature (taken from Fig.\,\ref{fig:TvsX}) and their interpretation in terms of desorption process and gas-phase evolution (see Sect.\,\ref{Dss:Xprofiles}). The colours are the same as in (a) and indicate the observed emission morphology for the respective COM. Solid and dashed lines of one colour indicate a similar morphology but different types of abundance profile. }
    \label{fig:sketches}
\end{figure*}
Taking into account all observational uncertainties and considering the simplicity of the model of G22 in terms of dynamics and geometry, it is striking how well the model can explain the majority of observed results. Moreover, for some of the differences there may already exist potential explanations that remain to be proven, however. There are plausible reactions that, when added to the chemical network, may already be able to explain differences seen for COMs like \mic and \etc. 
Furthermore, although our results are best reproduced under the assumption of a slow warm-up, this may not be the case for Sgr\,B2\,(N1) in reality as perhaps indicated by results from \ad. According to G22, similar results are possibly obtained for a faster warm-up  with enhanced CR flux because the main difference between all three warm-up phases is the exposure time to CRs and their induced secondary UV field and because CRs present one way of destruction of some COMs on the grain surface as well as in the gas phase. \citet{Bonfand19} found that a cosmic ray ionisation rate 50 times that of the standard value best described the COM abundances in the hot cores Sgr\,B2\,(N2--N5) derived on the basis of the lower-angular-resolution survey EMoCA. However, although similar results may be expected for a slow warm-up and a faster warm-up with enhanced CR flux, we can only know for sure if the model was run with these conditions.
As already mentioned above, geometry and changes of the density in the warm-up phase are not treated by the current model. On the other hand, our observational results do only consider projected distances and not the underlying 3-D structure (e.g., filaments and the outflow). 

In Fig.\,\ref{fig:sketches}a we summarise the three different morphologies of COM emission regions that we have identified based on the LVINE maps shown in Fig.\,\ref{fig:COM-Maps}: a) extended and structured  (dark blue, O-bearing COMs), b) more compact and uniform (pink, N-bearing COMs, HC\3N, and HNCO), and c) compact and structured (light blue, \ad, \mic, \fmm, \aan). We have also observed differences in the behaviours of the abundance profiles of the various COMs (see Sect.\,\ref{Rss:Xprofiles}). Figure\,\ref{fig:sketches}b shows an overview of the trends seen in the abundance profiles as a function of temperature (taken from Fig.\,\ref{fig:TvsX}). 
The comparison with model predictions of G22, done for each COM separately in Sect.\,\ref{Dss:Xprofiles}, indicates that the observed differences between COM abundance profiles trace back to the COMs' desorption processes at temperatures around 100\,K (thermal co-desorption with H\2O) and at lower temperatures (non-thermal or thermal desorption) and their individual formation and destruction pathways at higher temperatures. 

Although the differences seen in the COM abundance profiles do not simply match those seen in the morphology of COM emission, the above reasoning may also be able to explain the morphology to some extent. The extended and structured morphology of the emission regions for the four O-bearing COMs that are mainly formed on dust grains closely follows the dust continuum emission suggesting a close relation of the dust and these COMs. Emission from the N-bearing (or N+O-bearing) COMs and \ad is more compactly distributed around the protostar, where higher temperatures, which are required for the gas-phase reactions to become efficient or to take place at all, are expected and observed. Further differences amongst the latter group may be attributed to a mix of solid- and gas-phase formation with different efficiencies. 
However, whether the N-bearing (or N+O-bearing) molecules that are not analysed in detail follow this classification cannot be concluded from their morphology alone. For example, CH\3CN and HNCO show a similar morphology as ethyl or vinyl cyanide (cf. Fig.\,\ref{fig:COM-Maps}) and according to the chemical model, CH\3CN is indeed efficiently produced in the gas phase while HNCO seems to rather be formed on dust grains and destroyed when arriving in the gas phase (see Fig.\,21 of G22).

Based on our observational results, we do not identify an influence on the COM temperature and abundance profiles due to carbon-grain sublimation (see Sect.\,\ref{Dss:OandN}). Although the more compact emission region of N-bearing COMs could have suggested it, the only slightly larger abundance values at highest measured temperatures do not point to a significant enhancement of N-bearing COMs as expected from this process. However, the location of the soot line at $\sim$300\,K is only a rough estimate \citep[][]{vantHoff20} and could well be at higher temperatures, that is, closer to the protostar such that we cannot observe it due to insufficient angular resolution and the high continuum optical depth.

\subsection{The role of binding energies}\label{Dss:Ebin}
\begin{figure*}
    \centering
    \includegraphics[width=\textwidth]{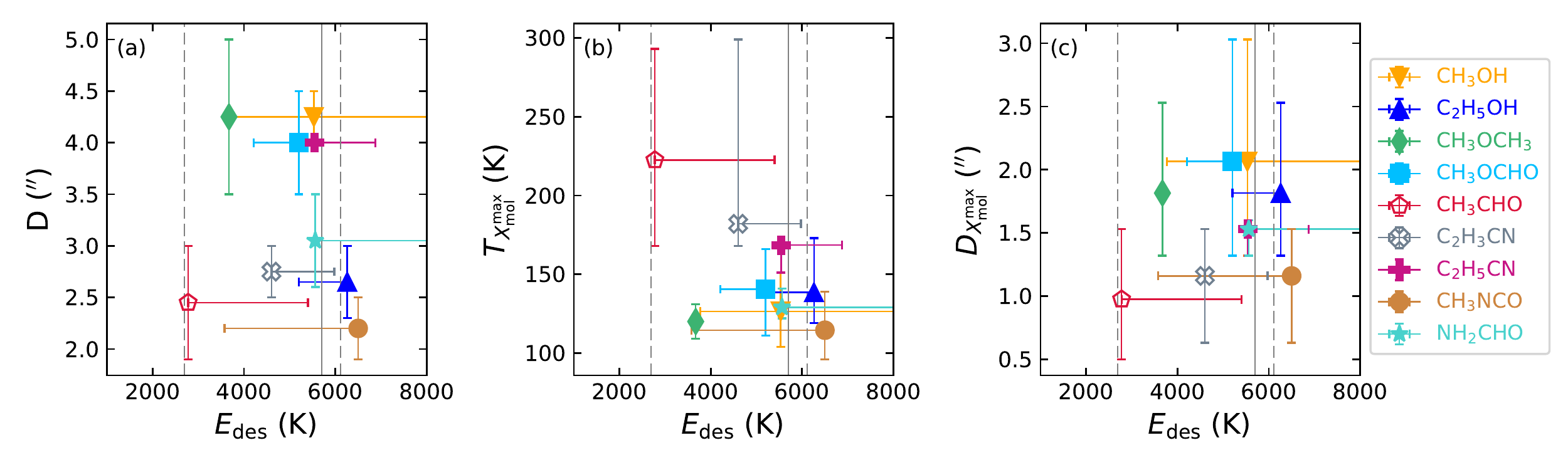}
    \caption{\textbf{(a)} Binding (or desorption) energy versus maximum distance to which the respective COM is still detected. The binding energies are taken from the third column of Table\,\ref{tab:Ebin}, with uncertainties indicating the range of values listed in that table, and distances are taken from Fig.\,\ref{fig:Ncolprofiles}. They represent the mean value from both direction for each COM, respectively. The error bars show the spread in the values. \textbf{(b)} Binding energy versus rotation temperature at which the COM abundance relative to H\2 peak. The values are taken from Fig.\,\ref{fig:TvsX}. They represent the median value from both direction and both methods of H\2 column density derivation for each COM, respectively. The error bars show the spread in the values. \textbf{(c)} Same as (b), but here, the distance at which the COM peaks is shown, which is taken from Figs.\,\ref{fig:DvsX} and \ref{fig:DvsX_co}.
    In all panels the solid grey line indicates a binding energy of water of 5700\,K while the dashed lines show the possible range of values based on other studies (see Table\,\ref{tab:Ebin}). Unfilled symbols indicate species that are mainly formed in the gas phase according to the model by \citet{Garrod22}.}
    \label{fig:Ebin}
\end{figure*}

Although, based on our observational results, thermal desorption of the bulk of a COM seems to happen alongside water and does not depend on binding energies as concluded from Sect.\,\ref{Dss:desorb}, the latter may still have an influence on the observed profiles.  
First, the desorption temperature of water and all other molecules depends on their binding energies. Table\,\ref{tab:Ebin} lists values of binding energies either measured in laboratory experiments \citep{Wakelam17,Das18,Ferrero20} or assumed by chemical models \citep[][]{Garrod13, Garrod22}. For those species, for which multiple values are available, it is evident that these can vary greatly. The binding energy depends on a number of parameters, amongst these are the dust surface composition, that is the material the molecule is bound to, and the surface structure, that is, depending on the binding site a COM may stick more strongly to it. The latter can lead to a range of binding energies for a given COM and dust composition as shown by \citet{Ferrero20}.
In Sect.\,\ref{Dsss:coH2O}, we already mentioned that a lower binding energy in the model is able to better explain not only our results, but also observations in Orion\,KL (Jin et al, in prep). 

\begin{table}[]
    \caption{Binding energies of water and complex organic molecules investigated in this work from various studies.}
    \hspace{-0.1cm}
    \begin{tabular}{lccccc}
     \hline\hline\\[-0.3cm]
     \small{Molecule} & \multicolumn{5}{c}{$E_\mathrm{bin}$ (K)} \\	\cmidrule{2-6}
      & G13 & B19/G22 & W17 & D18 & F20\\\hline\\[-0.3cm]
     \small{H\2O} & 5700 & * & 5600 & 2690 & 3600--6110 \\
     \small{\met} & 5530 & * & 5000 & 4370 & 3770--8620 \\
     \small{\et} & 6260 & * & 5200 & -- & -- \\
     \small{\dme} & 3675 & * & -- & -- & -- \\
     \small{\mf} & 4210 & 5200 & -- & -- & -- \\
     \small{\ad} & 2775 & * & 5400 & 3850 & -- \\
     \small{\mic} & 3575 & 6500 & 4700 & -- & -- \\
     \small{\etc} & 6875 & 5540 & -- & -- & -- \\
     \small{\vc} & 5975 & 4600 & -- & -- & -- \\
     \small{\fmm} & 5560 & * & 6300 & -- & 5790--11000 \\
     \hline\hline
    \end{tabular}
    \tablefoot{G13: \citet{Garrod13}. B19: \citet{Belloche19}; only for \mic. G22: \citet{Garrod22}. W17: \citet{Wakelam17}. D18: \citet{Das18}. F20: \citet{Ferrero20}. \\ (*) Same value as in G13.}
    \label{tab:Ebin}
\end{table}

We investigate possible correlations of binding energy with various parameters in Fig.\,\ref{fig:Ebin}, where binding energies are taken from the third column in Table\,\ref{tab:Ebin}.
In Fig.\,\ref{fig:Ebin}a, we look at the correlation of binding energy with the maximum spatial extent of the COM emission, where we use the maximum projected distance to which COM column densities could be determined from Figs.\,\ref{fig:Ncolprofiles} and \ref{fig:Dmax-COM}. 
We show the average of the values obtained to the west and south and error bars represent the range of possible values.
Although we show all analysed COMs, a relation may not be expected for those that are mainly produced in the gas phase, which are \ad and \vc (unfilled symbols), which are, therefore, not considered in the following discussion, and possibly also \etc, \mic, and \fmm. There is no obvious trend visible; however, \et and \mic, which show the smallest spatial extents, have a higher binding energy than water, which could mean that they remain on the grain surface though uncovered from water because even higher temperatures are required for their desorption. 
All COMs with a smaller binding energy than water co-desorb with it in the model as soon as they are uncovered which may explain the small variance amongst them in Fig.\,\ref{fig:Ebin}a. However, we cannot exclude a contribution to the abundance by non-thermal desorption processes (Sect.\,\ref{Dsss:ntdesorp}) or thermal desorption due to reduced binding energies (Sect.\,\ref{dsss:2step}) at these large distances from Sgr\,B2\,(N1), that is low temperatures, which could introduce a bias to this correlation.

Based on the assumption that \et and \mic have a higher binding energy than water and that
they, therefore, stick longer to the grain surface, we may expect a correlation of the binding energy with the temperature of peak abundance, which is shown in Fig.\,\ref{fig:Ebin}b. 
Here, we use the median value obtained from \sdu, \wdu, \sco, ad \wco extracted from Fig.\,\ref{fig:TvsX}.  
The error bars indicate the range that is spanned by the four values. 
No trend is visible, also because the spread in observed temperature values for each COM is large and all COMs could as well peak at the same temperature.
Moreover, for most of these species, except methanol and ethanol, we may expect further enhancement of abundances through gas-phase reactions, either based on the predictions of the model by G22 or based on the observed column density and abundance profiles or both. 
Additionally, we show the projected distance at which the COM abundances peak in Fig.\,\ref{fig:Ebin}c, where values are taken from Figs.\,\ref{fig:DvsX} and \ref{fig:DvsX_co}. In this case no correlation is evident at all also because the spread in this parameter for most COMs is large.

Based on the discussion above and in Sect.\,\ref{Dss:desorb}, binding energies may not have a great impact on the thermal desorption that is associated with these COMs' peak abundances in the gas phase and they are not the driving source of segregation between COMs in Sgr\,B2\,(N1). 
Nonetheless, they may have a minor influence on the power-law indices of the temperature profiles, the temperature of peak abundance, and on the maximum spatial extent of COMs at least for those mainly formed on dust grains and despite further enhancement of their abundance in the gas phase and (non-)thermal desorption processes. However, as seen in Fig.\,\ref{fig:Ebin}, not only temperatures and distances show a large spread for each COM, also the values of binding energy do, which makes the discussion let alone final conclusion on the role of binding energies difficult.

\section{Conclusion}\label{s:conclusion}

Thanks to the high-angular resolution of the ReMoCA spectral line survey we were able to resolve the hot core region of the main hub Sgr\,B2\,(N1) located in the star-forming protocluster Sgr\,B2\,(N). Our goal was to shed light on the desorption process that releases COMs from dust grain surfaces, where most of them are believed to form, into the gas phase. For each of a selected number of COMs, we derived rotation temperature, column density, and abundance profiles going south and west from the continuum peak position of Sgr\,B2\,(N1). The two directions were chosen after an inspection of the morphology of the continuum and COM emission and to avoid the outflow driven by Sgr\,B2\,(N1). Abundances have been derived with respect to methanol as well as H\2, where column densities of the latter were derived from the dust continuum emission as well as from C$^{18}$O\,1--0 emission. Finally, the observational results were  compared to state-of-art astrochemical models performed by \citet{Garrod22}.
The main results of this article are the following:
\begin{enumerate}
    \item Based on the morphology of their emissions, the COMs can be classified into two groups. The O-bearing COMs \met, \et, \dme, and \mf show structured emission that closely follows that of the dust continuum emission. In contrast, N-bearing organic molecules such as \etc, \vc, CH\3CN, and HC\3N have less structured and more compact morphologies. COMs like \ad, \fmm, and \mic reveal morphologies that are somewhere in between.
    \item The temperature profiles derived from all COMs have power-law indices that vary from $-$0.4 to $-$0.8, which is in agreement with other observations as well as models and which is consistent with a temperature gradient due to protostellar heating of an 
    envelope with optically thick dust. 
    \item The southbound H\2 column density profiles derived from dust and C$^{18}$O emission show similar behaviours with values that agree within the error bars, while the westbound profile from C$^{18}$O yields higher values by factors of a few than the one from dust, which is surprising and still demands an explanation.
    \item COM column density and abundance profiles support the segregation of COMs seen in the emission morphology. The comparison with astrochemical models indicates that the gas-phase abundance for the majority of O-bearing COMs is dominantly set by their solid-phase abundance as they are mainly formed on dust grain surfaces during the cold collapse phase. 
    \ad and \vc rather trace the warmer gas-phase chemistry as this is where they are mainly formed. \fmm seems to be produced in both phases and while, up to now, \mic and \etc are exclusively formed on dust grains in the model, observational results suggest (partial) gas-phase formation for these COMs. There exist reactions, that are not yet included in the model but could be promising candidates for the efficient production of these COMs.
    \item Most abundance profiles show a steep increase at $\sim$100\,K, and then either a plateau, smooth increase, or drop at higher temperatures. The steep increase can probably be associated with the COMs' thermal desorption from dust grains and given that it occurs at roughly the same temperature for the respective 
    COMs, we suggest that it represents thermal co-desorption of these COMs with 
    water, rather than thermal desorption depending on binding energies. Co-desorption of COMs and water is in agreement with predictions by the models of \citet{Garrod22}, in which this outcome is a consequence of the deactivation of the ability of COMs to diffuse within the bulk ice mantles. No convincing correlation with binding energies has been found. 
    \item Non-zero abundance values for some COMs at temperatures below $\sim$100\,K, that is before the steep increase associated with co-desorption with water, suggest another desorption process at work at these low temperatures. One explanation would be a partial thermal desorption of molecules from the outer, water-poor (and initially CO-rich) layers of the ice mantles, due to weaker binding caused by the lower water content in those layers. CO would desorb at low temperatures (20--30\,K), leaving behind COMs that would be bound less strongly to each other than they would be in a water-rich ice, allowing them to desorb at temperatures $<$100\,K. Most of the COMs in the ice would remain trapped in the water-rich layers beneath. Another explanation would be non-thermal desorption, which could include chemical desorption or photo-desorption. In either case, our observations have resolved and revealed for the first time the transition between two regimes of desorption of COMs in a hot core.
\end{enumerate}

Given that our observational results only consider the projected distance from Sgr\,B2\,(N1) and that the model  does not take into account any geometrical changes, it is striking how well the two agree. The comparison to the model were focussed on the results for the slow warm-up phase as they best described the observations, however, we may expect similar results from a faster warm-up with enhanced cosmic-ray flux. The latter is confirmed to exist in Sgr\,B2\,(N1) and may be able to still alter the modelled results if taken into account. 

\begin{acknowledgements}
This paper makes use of the following ALMA data: ADS/JAO. ALMA\#2016.1.00074.S. ALMA is a partnership of ESO (representing its member states), NSF (USA), and NINS (Japan), together with NRC (Canada), NSC and ASIAA (Taiwan), and KASI (Republic of Korea), in cooperation with the Republic of Chile. The Joint ALMA Observatory is operated by ESO, AUI/NRAO, and NAOJ. The interferometric data are available in the ALMA archive at https://almascience.eso.org/aq/. Part of this work has been
carried out within the Collaborative Research Centre 956, sub-project B3, funded by the Deutsche Forschungsgemeinschaft (DFG) – project ID 184018867. RTG acknowledges funding through the Astronomy \& Astrophysics program of the National Science Foundation (grant No. AST 19-06489). 
The authors would like to thank the developers of the many Python libraries, made available as open-source software, in particular this research has made use of NumPy \citep[][]{numpy}, matplotlib \citep[][]{matplotlib}, and SciPy \citep[][]{scipy}.
\end{acknowledgements}

\bibliographystyle{aa} 
\bibliography{refs.bib} 

\begin{thebibliography}{134}
\expandafter\ifx\csname natexlab\endcsname\relax\def\natexlab#1{#1}\fi

\bibitem[{{Ag{\'u}ndez} {et~al.}(2021){Ag{\'u}ndez}, {Marcelino}, {Tercero},
  {Cabezas}, {de Vicente}, \& {Cernicharo}}]{Agundez21}
{Ag{\'u}ndez}, M., {Marcelino}, N., {Tercero}, B., {et~al.} 2021, \aap, 649, L4

\bibitem[{{ALMA Partnership} {et~al.}(2016){ALMA Partnership}, {Asayama},
  {Biggs}, {de Gregorio}, {Dent}, {Di Francesco}, E., {Hales}, {Humphries},
  {Kameno}, {Müller}, {Vila Vilaro}, {Villard}, \& {Stoehr}}]{ALMAc4}
{ALMA Partnership}, {Asayama}, S., {Biggs}, A., {et~al.} 2016, ALMA Cycle 4
  Technical Handbook

\bibitem[{{Anderson} {et~al.}(1990{\natexlab{a}}){Anderson}, {De Lucia}, \&
  {Herbst}}]{CH3OH_rot_1990}
{Anderson}, T., {De Lucia}, F., \& {Herbst}, E. 1990{\natexlab{a}}, \apjs, 72,
  797

\bibitem[{{Anderson} {et~al.}(1987){Anderson}, {Herbst}, \& {De
  Lucia}}]{13CH3OH_rot_1987}
{Anderson}, T., {Herbst}, E., \& {De Lucia}, F.~C. 1987, \apjs, 64, 703

\bibitem[{{Anderson} {et~al.}(1990{\natexlab{b}}){Anderson}, {Herbst}, \& {De
  Lucia}}]{13CH3OH_rot_1990}
{Anderson}, T., {Herbst}, E., \& {De Lucia}, F.~C. 1990{\natexlab{b}}, \apjs,
  74, 647

\bibitem[{{Armijos-Abenda{\~n}o} {et~al.}(2020){Armijos-Abenda{\~n}o},
  {Banda-Barrag{\'a}n}, {Mart{\'\i}n-Pintado}, {D{\'e}nes}, {Federrath}, \&
  {Requena-Torres}}]{AA20}
{Armijos-Abenda{\~n}o}, J., {Banda-Barrag{\'a}n}, W.~E., {Mart{\'\i}n-Pintado},
  J., {et~al.} 2020, \mnras, 499, 4918

\bibitem[{{Bacmann} {et~al.}(2012){Bacmann}, {Taquet}, {Faure}, {Kahane}, \&
  {Ceccarelli}}]{Bacmann12}
{Bacmann}, A., {Taquet}, V., {Faure}, A., {Kahane}, C., \& {Ceccarelli}, C.
  2012, \aap, 541, L12

\bibitem[{{Barone} {et~al.}(2015){Barone}, {Latouche}, {Skouteris}, {Vazart},
  {Balucani}, {Ceccarelli}, \& {Lefloch}}]{Barone15}
{Barone}, V., {Latouche}, C., {Skouteris}, D., {et~al.} 2015, \mnras, 453, L31

\bibitem[{{Baskakov} {et~al.}(1996){Baskakov}, {Dyubko}, {Ilyushin},
  {Efimenko}, {Efremov}, {Podnos}, \& {Alekseev}}]{VyCN_rot_1996}
{Baskakov}, O.~I., {Dyubko}, S.~F., {Ilyushin}, V.~V., {et~al.} 1996, J. Mol.
  Spectrosc., 179, 94

\bibitem[{{Belloche} {et~al.}(2014){Belloche}, {Garrod}, {M{\"u}ller}, \&
  {Menten}}]{Belloche14}
{Belloche}, A., {Garrod}, R.~T., {M{\"u}ller}, H. S.~P., \& {Menten}, K.~M.
  2014, Science, 345, 1584

\bibitem[{{Belloche} {et~al.}(2019){Belloche}, {Garrod}, {M{\"u}ller},
  {Menten}, {Medvedev}, {Thomas}, \& {Kisiel}}]{Belloche19}
{Belloche}, A., {Garrod}, R.~T., {M{\"u}ller}, H.~S.~P., {et~al.} 2019, \aap,
  628, A10

\bibitem[{{Belloche} {et~al.}(2017){Belloche}, {Meshcheryakov}, {Garrod},
  {Ilyushin}, {Alekseev}, {Motiyenko}, {Margul{\`e}s}, {M{\"u}ller}, \&
  {Menten}}]{Belloche17}
{Belloche}, A., {Meshcheryakov}, A.~A., {Garrod}, R.~T., {et~al.} 2017, \aap,
  601, A49

\bibitem[{{Belloche} {et~al.}(2016){Belloche}, {M{\"u}ller}, {Garrod}, \&
  {Menten}}]{Belloche16}
{Belloche}, A., {M{\"u}ller}, H.~S.~P., {Garrod}, R.~T., \& {Menten}, K.~M.
  2016, \aap, 587, A91

\bibitem[{{Belloche} {et~al.}(2013){Belloche}, {M{\"u}ller}, {Menten},
  {Schilke}, \& {Comito}}]{Belloche13}
{Belloche}, A., {M{\"u}ller}, H.~S.~P., {Menten}, K.~M., {Schilke}, P., \&
  {Comito}, C. 2013, \aap, 559, A47

\bibitem[{{Beltr{\'a}n} {et~al.}(2018){Beltr{\'a}n}, {Cesaroni}, {Rivilla},
  {S{\'a}nchez-Monge}, {Moscadelli}, {Ahmadi}, {Allen}, {Beuther}, {Etoka},
  {Galli}, {Galv{\'a}n-Madrid}, {Goddi}, {Johnston}, {Klaassen},
  {K{\"o}lligan}, {Kuiper}, {Kumar}, {Maud}, {Mottram}, {Peters}, {Schilke},
  {Testi}, {van der Tak}, \& {Walmsley}}]{Beltran18}
{Beltr{\'a}n}, M.~T., {Cesaroni}, R., {Rivilla}, V.~M., {et~al.} 2018, \aap,
  615, A141

\bibitem[{{Bisschop} {et~al.}(2007){Bisschop}, {J{\o}rgensen}, {van Dishoeck},
  \& {de Wachter}}]{Bisschop07}
{Bisschop}, S.~E., {J{\o}rgensen}, J.~K., {van Dishoeck}, E.~F., \& {de
  Wachter}, E.~B.~M. 2007, \aap, 465, 913

\bibitem[{{Blake} {et~al.}(1984){Blake}, {Sutton}, {Masson}, {Phillips},
  {Herbst}, {Plummer}, \& {De Lucia}}]{13CH3OH_OMC-1_1984}
{Blake}, G.~A., {Sutton}, E.~C., {Masson}, C.~R., {et~al.} 1984, \apj, 286, 586

\bibitem[{{Bonfand} {et~al.}(2019){Bonfand}, {Belloche}, {Garrod}, {Menten},
  {Willis}, {St{\'e}phan}, \& {M{\"u}ller}}]{Bonfand19}
{Bonfand}, M., {Belloche}, A., {Garrod}, R.~T., {et~al.} 2019, \aap, 628, A27

\bibitem[{{Bonfand} {et~al.}(2017){Bonfand}, {Belloche}, {Menten}, {Garrod}, \&
  {M{\"u}ller}}]{Bonfand17}
{Bonfand}, M., {Belloche}, A., {Menten}, K.~M., {Garrod}, R.~T., \&
  {M{\"u}ller}, H.~S.~P. 2017, \aap, 604, A60

\bibitem[{{Boogert} {et~al.}(2015){Boogert}, {Gerakines}, \&
  {Whittet}}]{Boogert15}
{Boogert}, A.~C.~A., {Gerakines}, P.~A., \& {Whittet}, D. C.~B. 2015, \araa,
  53, 541

\bibitem[{{Brauer} {et~al.}(2009){Brauer}, {Pearson}, {Drouin}, \&
  {Yu}}]{EtCN_rot_2009}
{Brauer}, C.~S., {Pearson}, J.~C., {Drouin}, B.~J., \& {Yu}, S. 2009, \apjs,
  184, 133

\bibitem[{{Calcutt} {et~al.}(2018){Calcutt}, {J{\o}rgensen}, {M{\"u}ller},
  {Kristensen}, {Coutens}, {Bourke}, {Garrod}, {Persson}, {van der Wiel}, {van
  Dishoeck}, \& {Wampfler}}]{Calcutt18}
{Calcutt}, H., {J{\o}rgensen}, J.~K., {M{\"u}ller}, H.~S.~P., {et~al.} 2018,
  \aap, 616, A90

\bibitem[{{Cazzoli} \& {Kisiel}(1988)}]{VyCN_rot_1988}
{Cazzoli}, G. \& {Kisiel}, Z. 1988, J. Mol. Spectrosc., 130, 303

\bibitem[{{Cernicharo} {et~al.}(2016){Cernicharo}, {Kisiel}, {Tercero},
  {Kolesnikov{\'a}}, {Medvedev}, {L{\'o}pez}, {Fortman}, {Winnewisser}, {de
  Lucia}, {Alonso}, \& {Guillemin}}]{MeNCO_rot_2016}
{Cernicharo}, J., {Kisiel}, Z., {Tercero}, B., {et~al.} 2016, \aap, 587, L4

\bibitem[{{Codella} {et~al.}(2017){Codella}, {Ceccarelli}, {Caselli},
  {Balucani}, {Barone}, {Fontani}, {Lefloch}, {Podio}, {Viti}, {Feng},
  {Bachiller}, {Bianchi}, {Dulieu}, {Jim{\'e}nez-Serra}, {Holdship}, {Neri},
  {Pineda}, {Pon}, {Sims}, {Spezzano}, {Vasyunin}, {Alves}, {Bizzocchi},
  {Bottinelli}, {Caux}, {Chac{\'o}n-Tanarro}, {Choudhury}, {Coutens}, {Favre},
  {Hily-Blant}, {Kahane}, {Jaber Al-Edhari}, {Laas}, {L{\'o}pez-Sepulcre},
  {Ospina}, {Oya}, {Punanova}, {Puzzarini}, {Quenard}, {Rimola}, {Sakai},
  {Skouteris}, {Taquet}, {Testi}, {Theul{\'e}}, {Ugliengo}, {Vastel}, {Vazart},
  {Wiesenfeld}, \& {Yamamoto}}]{Codella17}
{Codella}, C., {Ceccarelli}, C., {Caselli}, P., {et~al.} 2017, \aap, 605, L3

\bibitem[{{Codella} {et~al.}(2015){Codella}, {Fontani}, {Ceccarelli}, {Podio},
  {Viti}, {Bachiller}, {Benedettini}, \& {Lefloch}}]{Codella15}
{Codella}, C., {Fontani}, F., {Ceccarelli}, C., {et~al.} 2015, \mnras, 449, L11

\bibitem[{{Collings} {et~al.}(2004){Collings}, {Anderson}, {Chen}, {Dever},
  {Viti}, {Williams}, \& {McCoustra}}]{Collings04}
{Collings}, M.~P., {Anderson}, M.~A., {Chen}, R., {et~al.} 2004, \mnras, 354,
  1133

\bibitem[{{Colzi} {et~al.}(2022){Colzi}, {Mart{\'\i}n-Pintado}, {Rivilla},
  {Jim{\'e}nez-Serra}, {Zeng}, {Rodr{\'\i}guez-Almeida}, {Rico-Villas},
  {Mart{\'\i}n}, \& {Requena-Torres}}]{Colzi22}
{Colzi}, L., {Mart{\'\i}n-Pintado}, J., {Rivilla}, V.~M., {et~al.} 2022, \apjl,
  926, L22

\bibitem[{{Coudert} {et~al.}(2002){Coudert}, {{\c{C}}ar{\c{c}}abal},
  {Chevalier}, {Broquier}, {Hepp}, \& {Herman}}]{DME_IR_2002}
{Coudert}, L.~H., {{\c{C}}ar{\c{c}}abal}, P., {Chevalier}, M., {et~al.} 2002,
  J. Mol. Spectrosc., 212, 203

\bibitem[{{Crockett} {et~al.}(2015){Crockett}, {Bergin}, {Neill}, {Favre},
  {Blake}, {Herbst}, {Anderson}, \& {Hassel}}]{Crockett15}
{Crockett}, N.~R., {Bergin}, E.~A., {Neill}, J.~L., {et~al.} 2015, \apj, 806,
  239

\bibitem[{{Csengeri} {et~al.}(2019){Csengeri}, {Belloche}, {Bontemps},
  {Wyrowski}, {Menten}, \& {Bouscasse}}]{Csengeri19}
{Csengeri}, T., {Belloche}, A., {Bontemps}, S., {et~al.} 2019, \aap, 632, A57

\bibitem[{{Daly} {et~al.}(2013){Daly}, {Berm{\'u}dez}, {L{\'o}pez}, {Tercero},
  {Pearson}, {Marcelino}, {Alonso}, \& {Cernicharo}}]{EtCN_rot_2013}
{Daly}, A.~M., {Berm{\'u}dez}, C., {L{\'o}pez}, A., {et~al.} 2013, \apj, 768,
  81

\bibitem[{{Das} {et~al.}(2018){Das}, {Sil}, {Gorai}, {Chakrabarti}, \&
  {Loison}}]{Das18}
{Das}, A., {Sil}, M., {Gorai}, P., {Chakrabarti}, S.~K., \& {Loison}, J.~C.
  2018, \apjs, 237, 9

\bibitem[{{De Pree} {et~al.}(2015){De Pree}, {Peters}, {Mac Low}, {Wilner},
  {Goss}, {Galv{\'a}n-Madrid}, {Keto}, {Klessen}, \& {Monsrud}}]{DePree15}
{De Pree}, C.~G., {Peters}, T., {Mac Low}, M.~M., {et~al.} 2015, \apj, 815, 123

\bibitem[{{Durig} {et~al.}(2011){Durig}, {Deeb}, {Darkhalil}, {Klaassen},
  {Gounev}, \& {Ganguly}}]{EtOH_IR_etc_2011}
{Durig}, J.~R., {Deeb}, H., {Darkhalil}, I.~D., {et~al.} 2011, J. Mol. Struct.,
  985, 202

\bibitem[{{Endres} {et~al.}(2009){Endres}, {Drouin}, {Pearson}, {M{\"u}ller},
  {Lewen}, {Schlemmer}, \& {Giesen}}]{DME_rot_2009}
{Endres}, C.~P., {Drouin}, B.~J., {Pearson}, J.~C., {et~al.} 2009, \aap, 504,
  635

\bibitem[{{Endres} {et~al.}(2016){Endres}, {Schlemmer}, {Schilke}, {Stutzki},
  \& {M{\"u}ller}}]{CDMS}
{Endres}, C.~P., {Schlemmer}, S., {Schilke}, P., {Stutzki}, J., \&
  {M{\"u}ller}, H. S.~P. 2016, Journal of Molecular Spectroscopy, 327, 95

\bibitem[{{Fedoseev} {et~al.}(2015){Fedoseev}, {Cuppen}, {Ioppolo}, {Lamberts},
  \& {Linnartz}}]{Fedoseev15}
{Fedoseev}, G., {Cuppen}, H.~M., {Ioppolo}, S., {Lamberts}, T., \& {Linnartz},
  H. 2015, \mnras, 448, 1288

\bibitem[{{Fern{\'a}ndez} {et~al.}(2019){Fern{\'a}ndez}, {Tejeda}, {Carvajal},
  \& {Senent}}]{DME_FIR_2019}
{Fern{\'a}ndez}, J.~M., {Tejeda}, G., {Carvajal}, M., \& {Senent}, M.~L. 2019,
  \apjs, 241, 13

\bibitem[{{Ferrero} {et~al.}(2020){Ferrero}, {Zamirri}, {Ceccarelli}, {Witzel},
  {Rimola}, \& {Ugliengo}}]{Ferrero20}
{Ferrero}, S., {Zamirri}, L., {Ceccarelli}, C., {et~al.} 2020, \apj, 904, 11

\bibitem[{{Fukuyama} {et~al.}(1996){Fukuyama}, {Odashima}, {Takagi}, \&
  {Tsunekawa}}]{EtCN_rot_1996}
{Fukuyama}, Y., {Odashima}, H., {Takagi}, K., \& {Tsunekawa}, S. 1996, \apjs,
  104, 329

\bibitem[{{Fukuyama} {et~al.}(1999){Fukuyama}, {Omori}, {Odashima}, {Takagi},
  \& {Tsunekawa}}]{EtCN_rot_1999}
{Fukuyama}, Y., {Omori}, K., {Odashima}, H., {Takagi}, K., \& {Tsunekawa}, S.
  1999, J. Mol. Spectrosc., 193, 72

\bibitem[{{Garrod}(2008)}]{Garrod08}
{Garrod}, R.~T. 2008, \aap, 491, 239

\bibitem[{{Garrod}(2013)}]{Garrod13}
{Garrod}, R.~T. 2013, \apj, 765, 60

\bibitem[{{Garrod} {et~al.}(2017){Garrod}, {Belloche}, {M{\"u}ller}, \&
  {Menten}}]{Garrod17}
{Garrod}, R.~T., {Belloche}, A., {M{\"u}ller}, H.~S.~P., \& {Menten}, K.~M.
  2017, \aap, 601, A48

\bibitem[{{Garrod} \& {Herbst}(2006)}]{Garrod06}
{Garrod}, R.~T. \& {Herbst}, E. 2006, \aap, 457, 927

\bibitem[{Garrod {et~al.}(2022)Garrod, Jin, Matis, Jones, Willis, \&
  Herbst}]{Garrod22}
Garrod, R.~T., Jin, M., Matis, K.~A., {et~al.} 2022, The Astrophysical Journal
  Supplement Series, 259, 1

\bibitem[{{Garrod} {et~al.}(2007){Garrod}, {Wakelam}, \& {Herbst}}]{Garrod07}
{Garrod}, R.~T., {Wakelam}, V., \& {Herbst}, E. 2007, \aap, 467, 1103

\bibitem[{{Gaume} {et~al.}(1995){Gaume}, {Claussen}, {de Pree}, {Goss}, \&
  {Mehringer}}]{Gaume95}
{Gaume}, R.~A., {Claussen}, M.~J., {de Pree}, C.~G., {Goss}, W.~M., \&
  {Mehringer}, D.~M. 1995, \apj, 449, 663

\bibitem[{{Gerry} \& {Winnewisser}(1973)}]{VyCN_rot_1973}
{Gerry}, M.~C.~L. \& {Winnewisser}, G. 1973, J. Mol. Spectrosc., 48, 1

\bibitem[{{Gieser} {et~al.}(2021){Gieser}, {Beuther}, {Semenov}, {Ahmadi},
  {Suri}, {M{\"o}ller}, {Beltr{\'a}n}, {Klaassen}, {Zhang}, {Urquhart},
  {Henning}, {Feng}, {Galv{\'a}n-Madrid}, {de Souza Magalh{\~a}es},
  {Moscadelli}, {Longmore}, {Leurini}, {Kuiper}, {Peters}, {Menten},
  {Csengeri}, {Fuller}, {Wyrowski}, {Lumsden}, {S{\'a}nchez-Monge}, {Maud},
  {Linz}, {Palau}, {Schilke}, {Pety}, {Pudritz}, {Winters}, \&
  {Pi{\'e}tu}}]{Gieser21}
{Gieser}, C., {Beuther}, H., {Semenov}, D., {et~al.} 2021, \aap, 648, A66

\bibitem[{{Gieser} {et~al.}(2022){Gieser}, {Beuther}, {Semenov}, {Suri},
  {Soler}, {Linz}, {Syed}, {Henning}, {Feng}, {M{\"o}ller}, {Palau}, {Winters},
  {Beltr{\'a}n}, {Kuiper}, {Moscadelli}, {Klaassen}, {Urquhart}, {Peters},
  {Longmore}, {S{\'a}nchez-Monge}, {Galv{\'a}n-Madrid}, {Pudritz}, \&
  {Johnston}}]{Gieser22}
{Gieser}, C., {Beuther}, H., {Semenov}, D., {et~al.} 2022, \aap, 657, A3

\bibitem[{{Goldsmith} \& {Langer}(1999)}]{Goldsmith99}
{Goldsmith}, P.~F. \& {Langer}, W.~D. 1999, \apj, 517, 209

\bibitem[{{Halfen} {et~al.}(2015){Halfen}, {Ilyushin}, \& {Ziurys}}]{Halfen15}
{Halfen}, D.~T., {Ilyushin}, V.~V., \& {Ziurys}, L.~M. 2015, \apjl, 812, L5

\bibitem[{{Haque} {et~al.}(1974){Haque}, {Lees}, {Saint Clair}, {Beers}, \&
  {Johnson}}]{13CH3OH_rot_1974}
{Haque}, S.~S., {Lees}, R.~M., {Saint Clair}, J.~M., {Beers}, Y., \& {Johnson},
  D.~R. 1974, \apjl, 187, L15

\bibitem[{{Harris} {et~al.}(2020){Harris}, {Millman}, {van der Walt},
  {Gommers}, {Virtanen}, {Cournapeau}, {Wieser}, {Taylor}, {Berg}, {Smith},
  {Kern}, {Picus}, {Hoyer}, {van Kerkwijk}, {Brett}, {Haldane}, {del R{\'\i}o},
  {Wiebe}, {Peterson}, {G{\'e}rard-Marchant}, {Sheppard}, {Reddy}, {Weckesser},
  {Abbasi}, {Gohlke}, \& {Oliphant}}]{numpy}
{Harris}, C.~R., {Millman}, K.~J., {van der Walt}, S.~J., {et~al.} 2020, \nat,
  585, 357

\bibitem[{{Hasegawa} {et~al.}(1994){Hasegawa}, {Sato}, {Whiteoak}, \&
  {Miyawaki}}]{Hasegawa94}
{Hasegawa}, T., {Sato}, F., {Whiteoak}, J.~B., \& {Miyawaki}, R. 1994, \apjl,
  429, L77

\bibitem[{{Heise} {et~al.}(1981){Heise}, {Winther}, \& {Lutz}}]{EtCN_IR_1981}
{Heise}, H.~M., {Winther}, F., \& {Lutz}, H. 1981, J. Mol. Spectrosc., 90, 531

\bibitem[{{Henkel} {et~al.}(1994){Henkel}, {Wilson}, {Langer}, {Chin}, \&
  {Mauersberger}}]{Henkel94}
{Henkel}, C., {Wilson}, T.~L., {Langer}, N., {Chin}, Y.~N., \& {Mauersberger},
  R. 1994, {Interstellar CNO Isotope Ratios}, ed. T.~L. {Wilson} \& K.~J.
  {Johnston}, Vol. 439, 72--88

\bibitem[{{Herbst} {et~al.}(1984){Herbst}, {Messer}, {De Lucia}, \&
  {Helminger}}]{CH3OH_rot_Herbst_1984}
{Herbst}, E., {Messer}, J.~K., {De Lucia}, F.~C., \& {Helminger}, P. 1984, J.
  Mol. Spectrosc., 108, 42

\bibitem[{{Herbst} \& {van Dishoeck}(2009)}]{Herbst09}
{Herbst}, E. \& {van Dishoeck}, E.~F. 2009, \araa, 47, 427

\bibitem[{{Higuchi} {et~al.}(2015){Higuchi}, {Hasegawa}, {Saigo}, {Sanhueza},
  \& {Chibueze}}]{Higuchi15}
{Higuchi}, A.~E., {Hasegawa}, T., {Saigo}, K., {Sanhueza}, P., \& {Chibueze},
  J.~O. 2015, \apj, 815, 106

\bibitem[{{Hunter}(2007)}]{matplotlib}
{Hunter}, J.~D. 2007, Computing in Science and Engineering, 9, 90

\bibitem[{{Ilyushin} {et~al.}(2009){Ilyushin}, {Kryvda}, \&
  {Alekseev}}]{MeFo_rot_2009}
{Ilyushin}, V., {Kryvda}, A., \& {Alekseev}, E. 2009, J. Mol. Spectrosc., 255,
  32

\bibitem[{{Jim{\'e}nez-Serra} {et~al.}(2016){Jim{\'e}nez-Serra}, {Vasyunin},
  {Caselli}, {Marcelino}, {Billot}, {Viti}, {Testi}, {Vastel}, {Lefloch}, \&
  {Bachiller}}]{Jimenez-Serra16}
{Jim{\'e}nez-Serra}, I., {Vasyunin}, A.~I., {Caselli}, P., {et~al.} 2016,
  \apjl, 830, L6

\bibitem[{{Jim{\'e}nez-Serra} {et~al.}(2021){Jim{\'e}nez-Serra}, {Vasyunin},
  {Spezzano}, {Caselli}, {Cosentino}, \& {Viti}}]{Jimenez-Serra21}
{Jim{\'e}nez-Serra}, I., {Vasyunin}, A.~I., {Spezzano}, S., {et~al.} 2021,
  \apj, 917, 44

\bibitem[{{Jin} \& {Garrod}(2020)}]{Jin20}
{Jin}, M. \& {Garrod}, R.~T. 2020, \apjs, 249, 26

\bibitem[{{J{\o}rgensen} {et~al.}(2020){J{\o}rgensen}, {Belloche}, \&
  {Garrod}}]{Jorgensen20}
{J{\o}rgensen}, J.~K., {Belloche}, A., \& {Garrod}, R.~T. 2020, \araa, 58, 727

\bibitem[{{J{\o}rgensen} {et~al.}(2018){J{\o}rgensen}, {M{\"u}ller}, {Calcutt},
  {Coutens}, {Drozdovskaya}, {{\"O}berg}, {Persson}, {Taquet}, {van Dishoeck},
  \& {Wampfler}}]{Jorgensen18}
{J{\o}rgensen}, J.~K., {M{\"u}ller}, H.~S.~P., {Calcutt}, H., {et~al.} 2018,
  \aap, 620, A170

\bibitem[{{Karakawa} {et~al.}(2001){Karakawa}, {Oka}, {Odashima}, {Takagi}, \&
  {Tsunekawa}}]{MeFo_rot_2001}
{Karakawa}, Y., {Oka}, K., {Odashima}, H., {Takagi}, K., \& {Tsunekawa}, S.
  2001, J. Mol. Spectrosc., 210, 196

\bibitem[{{Kenyon} {et~al.}(1993){Kenyon}, {Calvet}, \& {Hartmann}}]{Kenyon93}
{Kenyon}, S.~J., {Calvet}, N., \& {Hartmann}, L. 1993, \apj, 414, 676

\bibitem[{{Khlifi} {et~al.}(1999){Khlifi}, {Nollet}, {Paillous}, {Bruston},
  {Raulin}, {B{\'e}nilan}, \& {Khanna}}]{VyCN_IR_1999}
{Khlifi}, M., {Nollet}, M., {Paillous}, P., {et~al.} 1999, J. Mol. Spectrosc.,
  194, 206

\bibitem[{{Kisiel} {et~al.}(2015){Kisiel}, {Martin-Drumel}, \&
  {Pirali}}]{VyCN_FIR_2015}
{Kisiel}, Z., {Martin-Drumel}, M.-A., \& {Pirali}, O. 2015, J. Mol. Spectrosc.,
  315, 83

\bibitem[{{Kobayashi} {et~al.}(2020){Kobayashi}, {Sakai}, {Fujitake},
  {Tokaryk}, {Billinghurst}, \& {Ohashi}}]{MeFo_rot_FIR_2020}
{Kobayashi}, K., {Sakai}, Y., {Fujitake}, M., {et~al.} 2020, Can. J. Phys., 98,
  551

\bibitem[{{Kryvda} {et~al.}(2009){Kryvda}, {Gerasimov}, {Dyubko}, {Alekseev},
  \& {Motiyenko}}]{FA_rot_2009}
{Kryvda}, A.~V., {Gerasimov}, V.~G., {Dyubko}, S.~F., {Alekseev}, E.~A., \&
  {Motiyenko}, R.~A. 2009, J. Mol. Spectrosc., 254, 28

\bibitem[{{Kuriyama} {et~al.}(1986){Kuriyama}, {Takagi}, {Takeo}, \&
  {Matsumura}}]{13CH3OH_rot_1986}
{Kuriyama}, H., {Takagi}, K., {Takeo}, H., \& {Matsumura}, C. 1986, \apj, 311,
  1073

\bibitem[{{Kutzer} {et~al.}(2016){Kutzer}, {Weismann}, {Wa{\ss}muth}, {Pirali},
  {Roy}, {Yamada}, \& {Giesen}}]{DME_FIR_2016}
{Kutzer}, P., {Weismann}, D., {Wa{\ss}muth}, B., {et~al.} 2016, J. Mol.
  Spectrosc., 329, 28

\bibitem[{{Kwon} {et~al.}(2009){Kwon}, {Looney}, {Mundy}, {Chiang}, \&
  {Kemball}}]{Kwon09}
{Kwon}, W., {Looney}, L.~W., {Mundy}, L.~G., {Chiang}, H.-F., \& {Kemball},
  A.~J. 2009, \apj, 696, 841

\bibitem[{{Lees} \& {Baker}(1968)}]{3isos_rot_1968}
{Lees}, R.~M. \& {Baker}, J.~G. 1968, \jcp, 48, 5299

\bibitem[{{Li} {et~al.}(2017){Li}, {Liu}, {Hasegawa}, \& {Hirano}}]{Li17}
{Li}, J. I.-H., {Liu}, H.~B., {Hasegawa}, Y., \& {Hirano}, N. 2017, \apj, 840,
  72

\bibitem[{{Lovas} {et~al.}(1979){Lovas}, {Lutz}, \&
  {Dreizler}}]{DME-compilation_1979}
{Lovas}, F.~J., {Lutz}, H., \& {Dreizler}, H. 1979, J. Phys. Chem. Ref. Data,
  8, 1051

\bibitem[{{Majumdar} {et~al.}(2018){Majumdar}, {Loison}, {Ruaud}, {Gratier},
  {Wakelam}, \& {Coutens}}]{Majumdar18}
{Majumdar}, L., {Loison}, J.~C., {Ruaud}, M., {et~al.} 2018, \mnras, 473, L59

\bibitem[{{Mangum} \& {Shirley}(2015)}]{Mangum15}
{Mangum}, J.~G. \& {Shirley}, Y.~L. 2015, \pasp, 127, 266

\bibitem[{{Manigand} {et~al.}(2020){Manigand}, {J{\o}rgensen}, {Calcutt},
  {M{\"u}ller}, {Ligterink}, {Coutens}, {Drozdovskaya}, {van Dishoeck}, \&
  {Wampfler}}]{Manigand20}
{Manigand}, S., {J{\o}rgensen}, J.~K., {Calcutt}, H., {et~al.} 2020, \aap, 635,
  A48

\bibitem[{{Maret} {et~al.}(2011){Maret}, {Hily-Blant}, {Pety}, {Bardeau}, \&
  {Reynier}}]{Maret11}
{Maret}, S., {Hily-Blant}, P., {Pety}, J., {Bardeau}, S., \& {Reynier}, E.
  2011, \aap, 526, A47

\bibitem[{{Mart{\'\i}n-Dom{\'e}nech} {et~al.}(2014){Mart{\'\i}n-Dom{\'e}nech},
  {Mu{\~n}oz Caro}, {Bueno}, \& {Goesmann}}]{Martin-Domenech14}
{Mart{\'\i}n-Dom{\'e}nech}, R., {Mu{\~n}oz Caro}, G.~M., {Bueno}, J., \&
  {Goesmann}, F. 2014, \aap, 564, A8

\bibitem[{{Mart{\'\i}n-Dom{\'e}nech} {et~al.}(2017){Mart{\'\i}n-Dom{\'e}nech},
  {Rivilla}, {Jim{\'e}nez-Serra}, {Qu{\'e}nard}, {Testi}, \&
  {Mart{\'\i}n-Pintado}}]{Martin-Domenech17}
{Mart{\'\i}n-Dom{\'e}nech}, R., {Rivilla}, V.~M., {Jim{\'e}nez-Serra}, I.,
  {et~al.} 2017, \mnras, 469, 2230

\bibitem[{{McEwan} {et~al.}(1989){McEwan}, {Denison}, {Huntress}, {Anicich},
  {Snodgrass}, \& {Bowers}}]{McEwan89}
{McEwan}, M.\, J., {Denison}, A.\, B., {Huntress}, W.\, T., {et~al.} 1989, The
  Journal of Physical Chemistry A, 93, 4064

\bibitem[{{McGuire}(2022)}]{McGuire22}
{McGuire}, B.~A. 2022, {2021 Census of Interstellar, Circumstellar,
  Extragalactic, Protoplanetary Disk, and Exoplanetary Molecules}

\bibitem[{{McNaughton} {et~al.}(1999){McNaughton}, {Evans}, {Lane}, \&
  {Nielsen}}]{FA_IR_1999}
{McNaughton}, D., {Evans}, C.~J., {Lane}, S., \& {Nielsen}, C.~J. 1999, J. Mol.
  Spectrosc., 193, 104

\bibitem[{{Melosso} {et~al.}(2020){Melosso}, {Belloche}, {Martin-Drumel},
  {Pirali}, {Tamassia}, {Bizzocchi}, {Garrod}, {M{\"u}ller}, {Menten}, {Dore},
  \& {Puzzarini}}]{Melosso20}
{Melosso}, M., {Belloche}, A., {Martin-Drumel}, M.~A., {et~al.} 2020, \aap,
  641, A160

\bibitem[{{Motiyenko} {et~al.}(2020){Motiyenko}, {Belloche}, {Garrod},
  {Margul{\`e}s}, {M{\"u}ller}, {Menten}, \& {Guillemin}}]{Motiyenko20}
{Motiyenko}, R.~A., {Belloche}, A., {Garrod}, R.~T., {et~al.} 2020, \aap, 642,
  A29

\bibitem[{{Motiyenko} {et~al.}(2012){Motiyenko}, {Tercero}, {Cernicharo}, \&
  {Margul{\`e}s}}]{FA_rot_2012}
{Motiyenko}, R.~A., {Tercero}, B., {Cernicharo}, J., \& {Margul{\`e}s}, L.
  2012, \aap, 548, A71

\bibitem[{{M{\"u}ller} {et~al.}(2008){M{\"u}ller}, {Belloche}, {Menten},
  {Comito}, \& {Schilke}}]{VyCN_rot_2008}
{M{\"u}ller}, H. S.~P., {Belloche}, A., {Menten}, K.~M., {Comito}, C., \&
  {Schilke}, P. 2008, J. Mol. Spectrosc., 251, 319

\bibitem[{{M{\"u}ller} {et~al.}(2016){M{\"u}ller}, {Belloche}, {Xu}, {Lees},
  {Garrod}, {Walters}, {van Wijngaarden}, {Lewen}, {Schlemmer}, \&
  {Menten}}]{Mueller16}
{M{\"u}ller}, H. S.~P., {Belloche}, A., {Xu}, L.-H., {et~al.} 2016, \aap, 587,
  A92

\bibitem[{{M{\"u}ller} {et~al.}(2004){M{\"u}ller}, {Menten}, \&
  {M{\"a}der}}]{CH3OH_MMM_2004}
{M{\"u}ller}, H.~S.~P., {Menten}, K.~M., \& {M{\"a}der}, H. 2004, \aap, 428,
  1019

\bibitem[{{Neustock} {et~al.}(1990){Neustock}, {Guarnieri}, {Demaison}, \&
  {Wlodarczak}}]{DME_rot_1990}
{Neustock}, W., {Guarnieri}, A., {Demaison}, J., \& {Wlodarczak}, G. 1990, Z.
  Naturforsch. A, 45, 702

\bibitem[{{Nomura} \& {Millar}(2004)}]{Nomura04}
{Nomura}, H. \& {Millar}, T.~J. 2004, \aap, 414, 409

\bibitem[{{Osorio} {et~al.}(2009){Osorio}, {Anglada}, {Lizano}, \&
  {D'Alessio}}]{Osorio09}
{Osorio}, M., {Anglada}, G., {Lizano}, S., \& {D'Alessio}, P. 2009, \apj, 694,
  29

\bibitem[{{Ossenkopf} \& {Henning}(1994)}]{Ossenkopf94}
{Ossenkopf}, V. \& {Henning}, T. 1994, \aap, 291, 943

\bibitem[{{Paulive} {et~al.}(2021){Paulive}, {Shingledecker}, \&
  {Herbst}}]{Paulive21}
{Paulive}, A., {Shingledecker}, C.~N., \& {Herbst}, E. 2021, \mnras, 500, 3414

\bibitem[{{Pearson} {et~al.}(2008){Pearson}, {Brauer}, \&
  {Drouin}}]{EtOH_rot_2008}
{Pearson}, J.~C., {Brauer}, C.~S., \& {Drouin}, B.~J. 2008, J. Mol. Spectrosc.,
  251, 394

\bibitem[{{Pearson} {et~al.}(2010){Pearson}, {M{\"u}ller}, {Pickett}, {Cohen},
  \& {Drouin}}]{JPL}
{Pearson}, J.~C., {M{\"u}ller}, H.~S.~P., {Pickett}, H.~M., {Cohen}, E.~A., \&
  {Drouin}, B.~J. 2010, \jqsrt, 111, 1614

\bibitem[{{Pearson} {et~al.}(1996){Pearson}, {Sastry}, {Herbst}, \& {De
  Lucia}}]{g-EtOH_rot_1996}
{Pearson}, J.~C., {Sastry}, K.~V.~L.~N., {Herbst}, E., \& {De Lucia}, F.~C.
  1996, J. Mol. Spectrosc., 175, 246

\bibitem[{{Pearson} {et~al.}(1997){Pearson}, {Sastry}, {Herbst}, \& {De
  Lucia}}]{g-EtOH_rot_1997}
{Pearson}, J.~C., {Sastry}, K.~V.~L.~N., {Herbst}, E., \& {De Lucia}, F.~C.
  1997, \apj, 480, 420

\bibitem[{{Pearson} {et~al.}(1995){Pearson}, {Sastry}, {Winnewisser}, {Herbst},
  \& {De Lucia}}]{a-EtOH_rot_1995}
{Pearson}, J.~C., {Sastry}, K.~V.~L.~N., {Winnewisser}, M., {Herbst}, E., \&
  {De Lucia}, F.~C. 1995, J. Phys. Chem. Ref. Data, 24, 1

\bibitem[{{Pickett} {et~al.}(1981){Pickett}, {Cohen}, {Brinza}, \&
  {Schaefer}}]{CH3OH_rot_1981}
{Pickett}, H.~M., {Cohen}, E.~A., {Brinza}, D.~E., \& {Schaefer}, M.~M. 1981,
  J. Mol. Spectrosc., 89, 542

\bibitem[{{Qu{\'e}nard} {et~al.}(2018){Qu{\'e}nard}, {Jim{\'e}nez-Serra},
  {Viti}, {Holdship}, \& {Coutens}}]{Quenard18}
{Qu{\'e}nard}, D., {Jim{\'e}nez-Serra}, I., {Viti}, S., {Holdship}, J., \&
  {Coutens}, A. 2018, \mnras, 474, 2796

\bibitem[{{Reid} {et~al.}(2019){Reid}, {Menten}, {Brunthaler}, {Zheng}, {Dame},
  {Xu}, {Li}, {Sakai}, {Wu}, {Immer}, {Zhang}, {Sanna}, {Moscadelli}, {Rygl},
  {Bartkiewicz}, {Hu}, {Quiroga-Nu{\~n}ez}, \& {van Langevelde}}]{Reid19}
{Reid}, M.~J., {Menten}, K.~M., {Brunthaler}, A., {et~al.} 2019, \apj, 885, 131

\bibitem[{{Requena-Torres} {et~al.}(2006){Requena-Torres},
  {Mart{\'\i}n-Pintado}, {Rodr{\'\i}guez-Franco}, {Mart{\'\i}n},
  {Rodr{\'\i}guez-Fern{\'a}ndez}, \& {de Vicente}}]{Requena-Torres06}
{Requena-Torres}, M.~A., {Mart{\'\i}n-Pintado}, J., {Rodr{\'\i}guez-Franco},
  A., {et~al.} 2006, \aap, 455, 971

\bibitem[{{Rodr{\'\i}guez-Fern{\'a}ndez}
  {et~al.}(2001){Rodr{\'\i}guez-Fern{\'a}ndez}, {Mart{\'\i}n-Pintado},
  {Fuente}, {de Vicente}, {Wilson}, \&
  {H{\"u}ttemeister}}]{Rodriguez-Fernandez01}
{Rodr{\'\i}guez-Fern{\'a}ndez}, N.~J., {Mart{\'\i}n-Pintado}, J., {Fuente}, A.,
  {et~al.} 2001, \aap, 365, 174

\bibitem[{{Rolffs} {et~al.}(2011){Rolffs}, {Schilke}, {Wyrowski}, {Dullemond},
  {Menten}, {Thorwirth}, \& {Belloche}}]{Rolffs11a}
{Rolffs}, R., {Schilke}, P., {Wyrowski}, F., {et~al.} 2011, \aap, 529, A76

\bibitem[{{Ruaud} {et~al.}(2015){Ruaud}, {Loison}, {Hickson}, {Gratier},
  {Hersant}, \& {Wakelam}}]{Ruaud15}
{Ruaud}, M., {Loison}, J.~C., {Hickson}, K.~M., {et~al.} 2015, \mnras, 447,
  4004

\bibitem[{{S{\'a}nchez-Monge} {et~al.}(2017){S{\'a}nchez-Monge}, {Schilke},
  {Schmiedeke}, {Ginsburg}, {Cesaroni}, {Lis}, {Qin}, {M{\"u}ller}, {Bergin},
  {Comito}, \& {M{\"o}ller}}]{Sanchez-Monge17}
{S{\'a}nchez-Monge}, {\'A}., {Schilke}, P., {Schmiedeke}, A., {et~al.} 2017,
  \aap, 604, A6

\bibitem[{{Sastry} {et~al.}(1984){Sastry}, {Lees}, \& {De
  Lucia}}]{CH3OH_rot_Sastry_1984}
{Sastry}, K.~V.~L.~N., {Lees}, R.~M., \& {De Lucia}, F.~C. 1984, J. Mol.
  Spectrosc., 103, 486

\bibitem[{{Sato} {et~al.}(2000){Sato}, {Hasegawa}, {Whiteoak}, \&
  {Miyawaki}}]{Sato00}
{Sato}, F., {Hasegawa}, T., {Whiteoak}, J.~B., \& {Miyawaki}, R. 2000, \apj,
  535, 857

\bibitem[{{Schw{\"o}rer} {et~al.}(2019){Schw{\"o}rer}, {S{\'a}nchez-Monge},
  {Schilke}, {M{\"o}ller}, {Ginsburg}, {Meng}, {Schmiedeke}, {M{\"u}ller},
  {Lis}, \& {Qin}}]{Schwoerer19}
{Schw{\"o}rer}, A., {S{\'a}nchez-Monge}, {\'A}., {Schilke}, P., {et~al.} 2019,
  \aap, 628, A6

\bibitem[{{Scibelli} \& {Shirley}(2020)}]{Scibelli20}
{Scibelli}, S. \& {Shirley}, Y. 2020, \apj, 891, 73

\bibitem[{{Shimanouchi}(1972)}]{Shimanouchi72}
{Shimanouchi}, T. 1972, Nat. Stand. Ref. Data Ser., Nat. Bur. Stand. (U.S.),
  39, 1

\bibitem[{{Shingledecker} {et~al.}(2018){Shingledecker}, {Tennis}, {Le Gal}, \&
  {Herbst}}]{Shingledecker18}
{Shingledecker}, C.~N., {Tennis}, J., {Le Gal}, R., \& {Herbst}, E. 2018, \apj,
  861, 20

\bibitem[{{Shu}(1977)}]{Shu77}
{Shu}, F.~H. 1977, \apj, 214, 488

\bibitem[{{Smirnov} {et~al.}(2014){Smirnov}, {Alekseev}, {Ilyushin},
  {Margul{\'e}s}, {Motiyenko}, \& {Drouin}}]{CH3CHO_rot_2014}
{Smirnov}, I.~A., {Alekseev}, E.~A., {Ilyushin}, V.~V., {et~al.} 2014, J. Mol.
  Spectrosc., 295, 44

\bibitem[{{Thiel} {et~al.}(2017){Thiel}, {Belloche}, {Menten}, {Garrod}, \&
  {M{\"u}ller}}]{Thiel17}
{Thiel}, V., {Belloche}, A., {Menten}, K.~M., {Garrod}, R.~T., \& {M{\"u}ller},
  H.~S.~P. 2017, \aap, 605, L6

\bibitem[{{Tudorie} {et~al.}(2012){Tudorie}, {Ilyushin}, {Vander Auwera},
  {Pirali}, {Roy}, \& {Huet}}]{MeFo_rot_FIR_2012}
{Tudorie}, M., {Ilyushin}, V., {Vander Auwera}, J., {et~al.} 2012, \jcp, 137,
  064304

\bibitem[{{van der Walt} {et~al.}(2021){van der Walt}, {Kristensen},
  {J{\o}rgensen}, {Calcutt}, {Manigand}, {el Akel}, {Garrod}, \&
  {Qiu}}]{vanderWalt21}
{van der Walt}, S.~J., {Kristensen}, L.~E., {J{\o}rgensen}, J.~K., {et~al.}
  2021, \aap, 655, A86

\bibitem[{{van 't Hoff} {et~al.}(2020){van 't Hoff}, {Bergin}, {J{\o}rgensen},
  \& {Blake}}]{vantHoff20}
{van 't Hoff}, M. L.~R., {Bergin}, E.~A., {J{\o}rgensen}, J.~K., \& {Blake},
  G.~A. 2020, \apjl, 897, L38

\bibitem[{{V\'avra} {et~al.}(2022){V\'avra}, {Kolesnikov\'a}, \&
  {Belloche}}]{Vavra22}
{V\'avra}, K., {Kolesnikov\'a}, L., \& {Belloche}, A. 2022, \aap, submitted

\bibitem[{Virtanen {et~al.}(2020)Virtanen, Gommers, Oliphant, Haberland, Reddy,
  Cournapeau, Burovski, Peterson, Weckesser, Bright, {van der Walt}, Brett,
  Wilson, Millman, Mayorov, Nelson, Jones, Kern, Larson, Carey, Polat, Feng,
  Moore, {VanderPlas}, Laxalde, Perktold, Cimrman, Henriksen, Quintero, Harris,
  Archibald, Ribeiro, Pedregosa, {van Mulbregt}, \& {SciPy 1.0
  Contributors}}]{scipy}
Virtanen, P., Gommers, R., Oliphant, T.~E., {et~al.} 2020, Nature Methods, 17,
  261

\bibitem[{{Viti} {et~al.}(2004){Viti}, {Collings}, {Dever}, {McCoustra}, \&
  {Williams}}]{Viti04}
{Viti}, S., {Collings}, M.~P., {Dever}, J.~W., {McCoustra}, M. R.~S., \&
  {Williams}, D.~A. 2004, \mnras, 354, 1141

\bibitem[{{Wakelam} {et~al.}(2017){Wakelam}, {Loison}, {Mereau}, \&
  {Ruaud}}]{Wakelam17}
{Wakelam}, V., {Loison}, J.~C., {Mereau}, R., \& {Ruaud}, M. 2017, Molecular
  Astrophysics, 6, 22

\bibitem[{{Xu} {et~al.}(2008){Xu}, {Fisher}, {Lees}, {Shi}, {Hougen},
  {Pearson}, {Drouin}, {Blake}, \& {Braakman}}]{CH3OH_rot_FIR_2008}
{Xu}, L.-H., {Fisher}, J., {Lees}, R.~M., {et~al.} 2008, J. Mol. Spectrosc.,
  251, 305

\bibitem[{{Xu} \& {Lovas}(1997)}]{12_13CH3OH-compilation_1997}
{Xu}, L.-H. \& {Lovas}, F.~J. 1997, J. Phys. Chem. Ref. Data, 26, 17

\bibitem[{{Zeng} {et~al.}(2018){Zeng}, {Jim{\'e}nez-Serra}, {Rivilla},
  {Mart{\'\i}n}, {Mart{\'\i}n-Pintado}, {Requena-Torres},
  {Armijos-Abenda{\~n}o}, {Riquelme}, \& {Aladro}}]{Zeng18}
{Zeng}, S., {Jim{\'e}nez-Serra}, I., {Rivilla}, V.~M., {et~al.} 2018, \mnras,
  478, 2962

\bibitem[{{Zeng} {et~al.}(2020){Zeng}, {Zhang}, {Jim{\'e}nez-Serra}, {Tercero},
  {Lu}, {Mart{\'\i}n-Pintado}, {de Vicente}, {Rivilla}, \& {Li}}]{Zeng20}
{Zeng}, S., {Zhang}, Q., {Jim{\'e}nez-Serra}, I., {et~al.} 2020, \mnras, 497,
  4896

\end{thebibliography}

\begin{appendix}

\section{Spectroscopic calculations} \label{app:specs}

The CH$_3$OH data were taken from the CDMS (tag 032504). Version 3 is a recalculation of the original 
work by \citet{CH3OH_rot_FIR_2008}. The entry includes transitions up to $\varv_{\rm t} = 2$, and we 
used data from these three states. Transitions in the range of our survey were contributed by 
\citet{3isos_rot_1968}, \citet{CH3OH_rot_1981}, \citet{CH3OH_rot_Sastry_1984}, \citet{CH3OH_rot_Herbst_1984}, \citet{CH3OH_rot_1990}, \citet{CH3OH_MMM_2004}, 
and \citet{CH3OH_rot_FIR_2008}. This entry resolves intensity issues that prevailed in 
earlier entries of the CDMS and JPL catalogues. The partition function takes into account 
energies up to $\varv_{\rm t} = 3$; it should be converged up to $\sim$200~K and should be 
quite good at 300~K.

We used the $^{13}$CH$_3$OH data from the CDMS (tag 033502). Version 2 is a recalculation of 
\citet{12_13CH3OH-compilation_1997} with contributions from 
\citet{13CH3OH_rot_1974}, \citet{13CH3OH_OMC-1_1984}, \citet{13CH3OH_rot_1986}, \cite{13CH3OH_rot_1987}, and \citet{13CH3OH_rot_1990}. 
The intensities were calculated employing scaled partition function values from the main isotopic species. 
The errors introduced by the scaling are small compared to available alternatives. 

The ethanol calculation is from the CDMS catalogue (tag 046524, version 1). It is based on the extensive work 
by \citet{EtOH_rot_2008} with additional data from \citet{a-EtOH_rot_1995,g-EtOH_rot_1996,g-EtOH_rot_1997}. 
The main difference with respect to the JPL catalogue entry are in part very different intensities as 
shown by \citet{Mueller16}. The calculated frequencies are at least nearly identical to those in the JPL 
catalogue, but no experimental frequencies were merged in the CDMS entry.  Vibrational corrections are 
based on the fundamental frequencies in \citet{EtOH_IR_etc_2011}. We accounted for contributions by 
overtone and combination states by applying the harmonic oscillator approximation. The errors by this 
approximation are usually small, especially because the exact energies of low-lying vibrations are 
frequently not known or if so mostly only for the fundamental vibrations.

We employed the CDMS entry for dimethyl ether (tag 046514, version 1). It is based on the analysis of 
\citet{DME_rot_2009} with additional rest frequencies in the range of our survey from the compilation of 
\citet{DME-compilation_1979} and from \citet{DME_rot_1990}. We also considered data in the two lowest 
excited states, $\varv_{11} = 1$ and $\varv_{15} = 1$. These data were provided by C.~P. Endres 
and involve unpublished data. The partition function in the CDMS takes 
into account vibrational contributions up to 500~cm$^{-1}$. This is sufficient at 100~K and reasonable 
around 150~K. In order to account for contributions of higher vibrations, we evaluated the ground state 
contributions first and corrected them subsequently in the harmonic approximation. 
Most of the fundamental vibrations were taken from \citet{DME_IR_2002}. We took information on the 
torsional fundamentals from \citet{DME_FIR_2019} and on the COC bending mode from \citet{DME_FIR_2016}. 

We use the JPL entry (tag 060003, version 2) for methyl formate, based on \citet{MeFo_rot_2009} with 
additional data from \citet{MeFo_rot_2001}. 
The partition function of methyl formate considered $\varv_{\rm t} = 1$. 
Its contribution was first subtracted and subsequently vibrational corrections added 
in the harmonic approximation. The torsional band origin was determined in 
\citet{MeFo_rot_FIR_2012}, additional low-lying vibrations  were mentioned in 
\citet{MeFo_rot_FIR_2020}, higher lying vibrations were summarised by \citet{Shimanouchi72}.

The acetaldehyde data were communicated by R.~A. Motiyenko. They are based on \citet{CH3CHO_rot_2014} 
with additional data extending beyond $\varv_{\rm t} = 2$.  Vibrational contributions were evaluated in the harmonic approximation employing the compilation of \citet{Shimanouchi72}.

We used the CDMS entries (tags 057505 and 057506, version 1 in each case) to identify methyl isocyanate. 
The entries were generated from the compilation of assignments in \citet{MeNCO_rot_2016}. 
The rotational part of the partition function was determined employing the rigid rotor approximation; 
this was the best to do given the highly perturbed rotational spectrum. Vibrational corrections were 
applied in \citet{MeNCO_rot_2016} for states up to 580~K. The errors are small up to $\sim$150~K, 
but reach the 10\% level at 200~K. The vibrational spectrum is unfortunately not known well enough 
to employ further corrections. 

We identified ethyl cyanide employing the CDMS entry (tag 055502, version 2). It is based on 
\citet{EtCN_rot_2009} with frequencies in the range of our survey from \citet{EtCN_rot_1996}. 
We also used excited state data $\varv_{20} = 1(A)$ (tag 055513, version 1) and $\varv_{12} = 1(A)$
(tag 055514, version 1) based on \citet{EtCN_rot_2013} with previously published $\varv_{20} = 1$ 
data from \citet{EtCN_rot_1999}. The vibrational correction factors are available through 
the CDMS documentation; they were evaluated using \citet{EtCN_IR_1981}.

The vinyl cyanide data are from the CDMS (tag 053515, version 1). It is based on \citet{VyCN_rot_2008} 
with data in the range of our survey mainly from \citet{VyCN_rot_1973}, \citet{VyCN_rot_1988}, and \citet{VyCN_rot_1996}. Data on the two lowest excited vibrational states, $\varv_{11} = 1$ and 
$\varv_{15} = 1$, are from unpublished data by one of us (HSPM) and takes also into account published 
data from \citet{VyCN_rot_1988}. We applied vibrational corrections using low-lying fundamentals of 
\citet{VyCN_FIR_2015} and higher lying fundamentals of \citet{VyCN_IR_1999}. The entry needs 
to be updated, see \citet{VyCN_FIR_2015} and references therein with data also on several excited 
vibrational states, but the current data are sufficient for the purpose of our study. 

We use the CDMS entries for formamide in its ground vibrational state (tag 045512, version 2) 
and in $\varv_{12} = 1$ (tag 045516, version 1). They are based on \citet{FA_rot_2012} and 
contain data in the range of our survey mostly from \citet{FA_rot_2009}. We consulted 
\citet{FA_IR_1999} for vibrational correction. We summed up first the contributions of the 
anharmonic NH$_2$ out of plane vibration and its overtones and applied further vibrational corrections 
in the harmonic approximation.

\section{LVINE maps}\label{app:LVINE}
\begin{figure}[h]
    \includegraphics[width=0.5\textwidth]{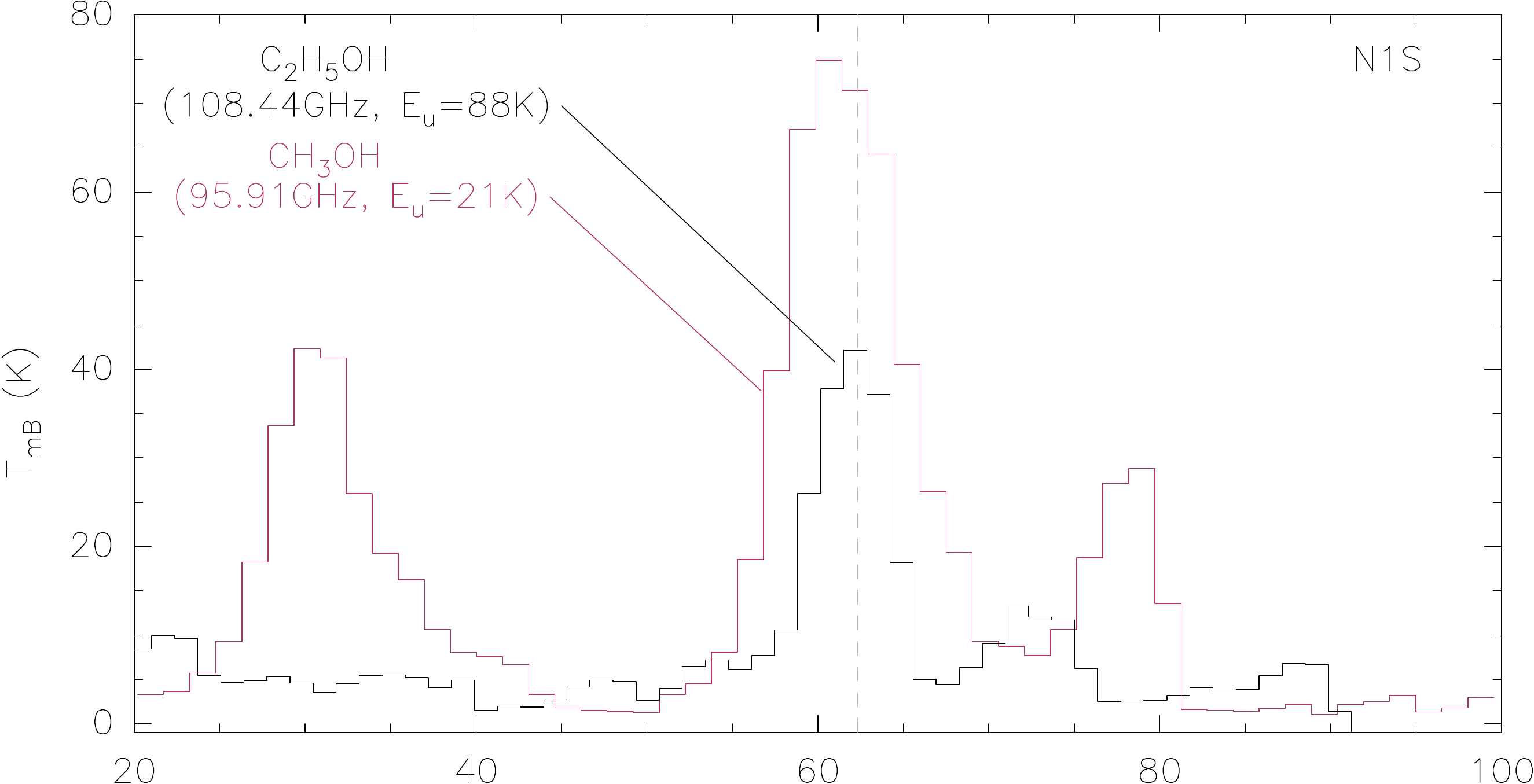}\vspace{0.2cm}
    \includegraphics[width=0.5\textwidth]{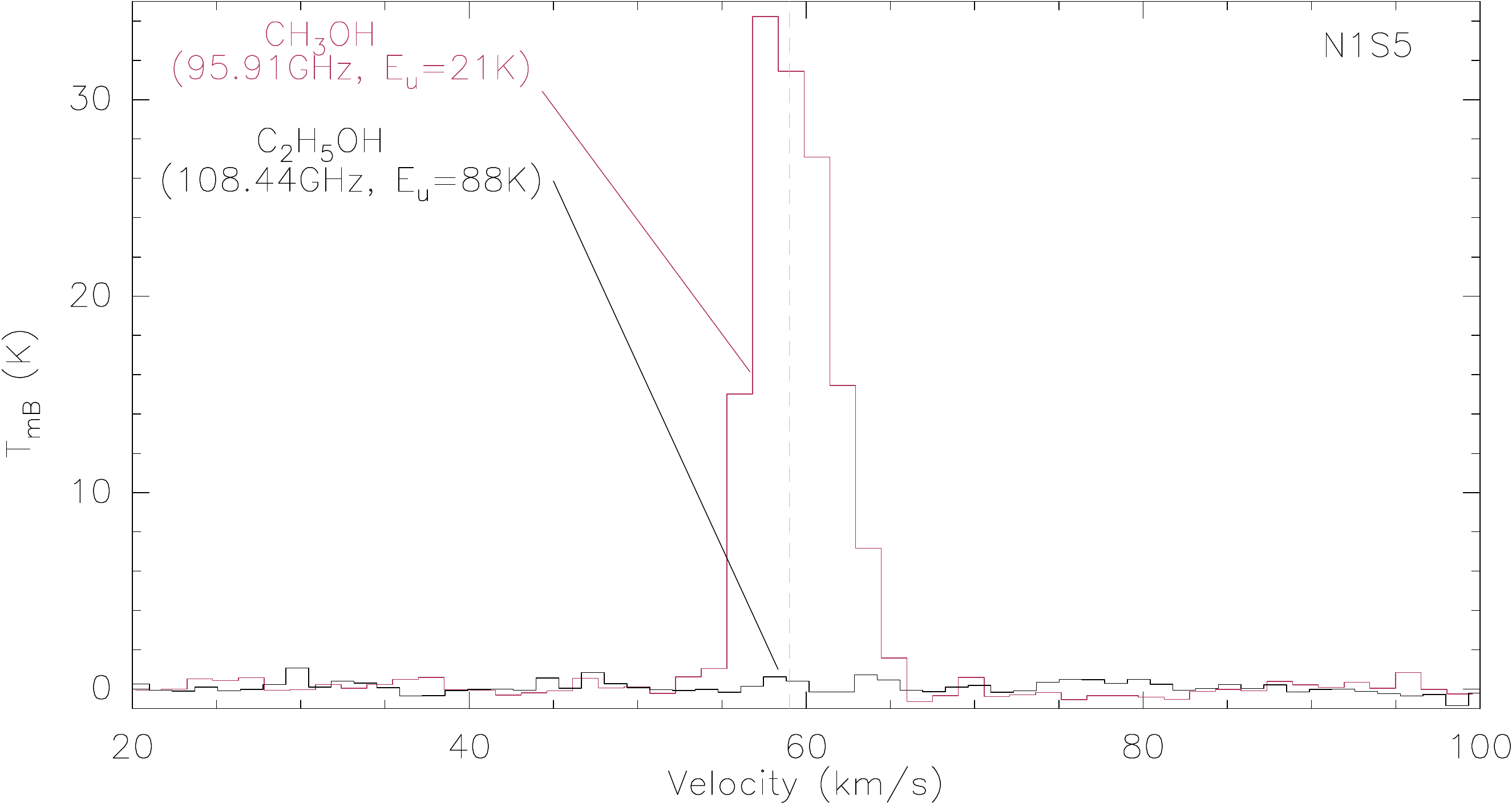}
    \caption{Spectra at positions N1S and N1S5 of the transitions of \et (black) and \met (maroon) that are used to create maps of peak velocity and line width, which are shown in Fig.\,\ref{fig:LVINE_O}. The rest frequency and upper level energy of the transition are indicated with the respective molecule and in the respective colour in the upper left corner. The grey dashed line indicates the peak velocities (62.3\,\kms at N1S and 59\,\kms at N1S5).}
    \label{fig:LVINE_O_spectra}
\end{figure}
\begin{figure*}[h]
    \includegraphics[width=0.35\textwidth]{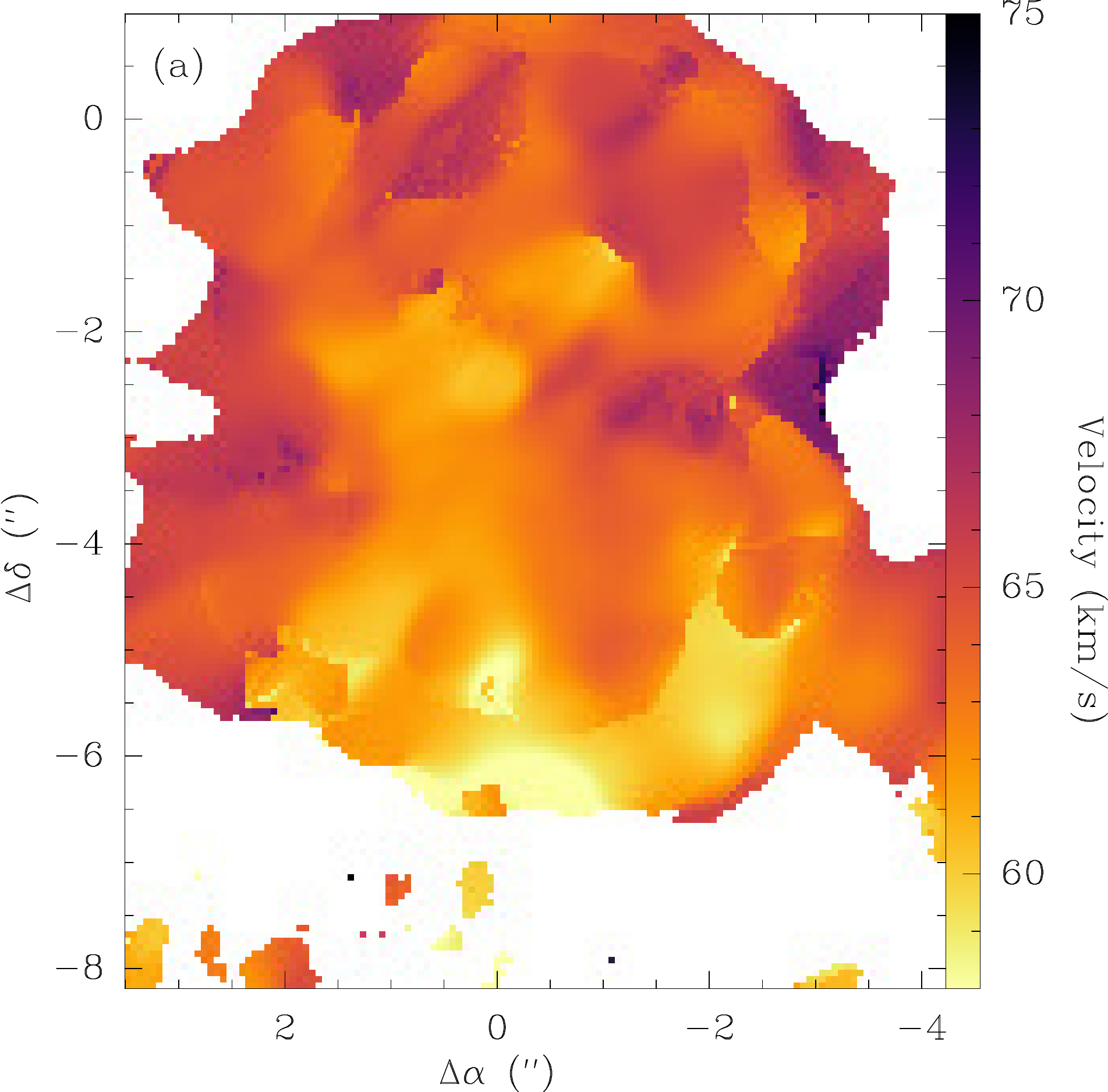}\hspace{0.1cm}
    \includegraphics[width=0.35\textwidth]{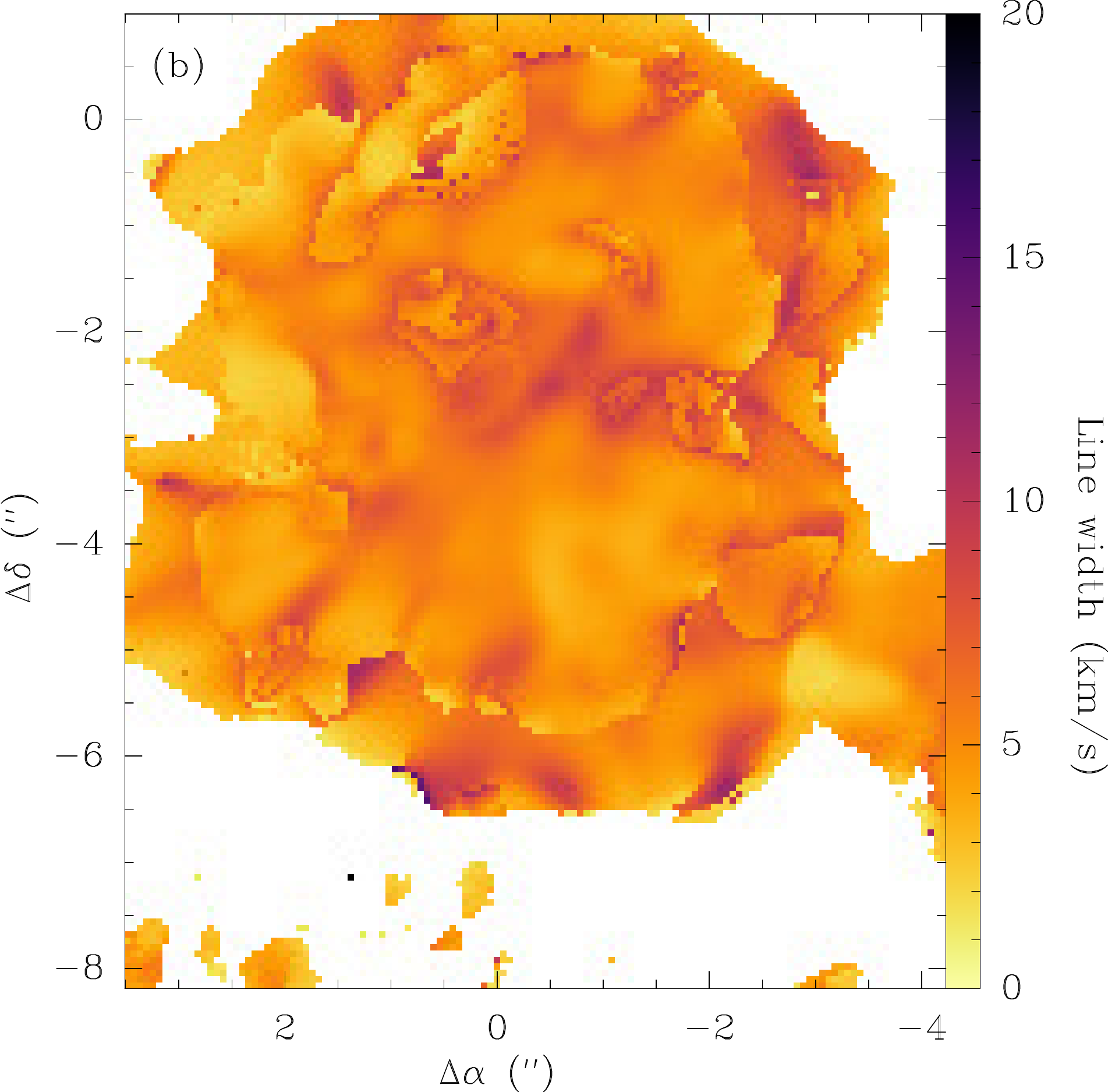}\hspace{0.1cm}
    \includegraphics[width=0.3\textwidth]{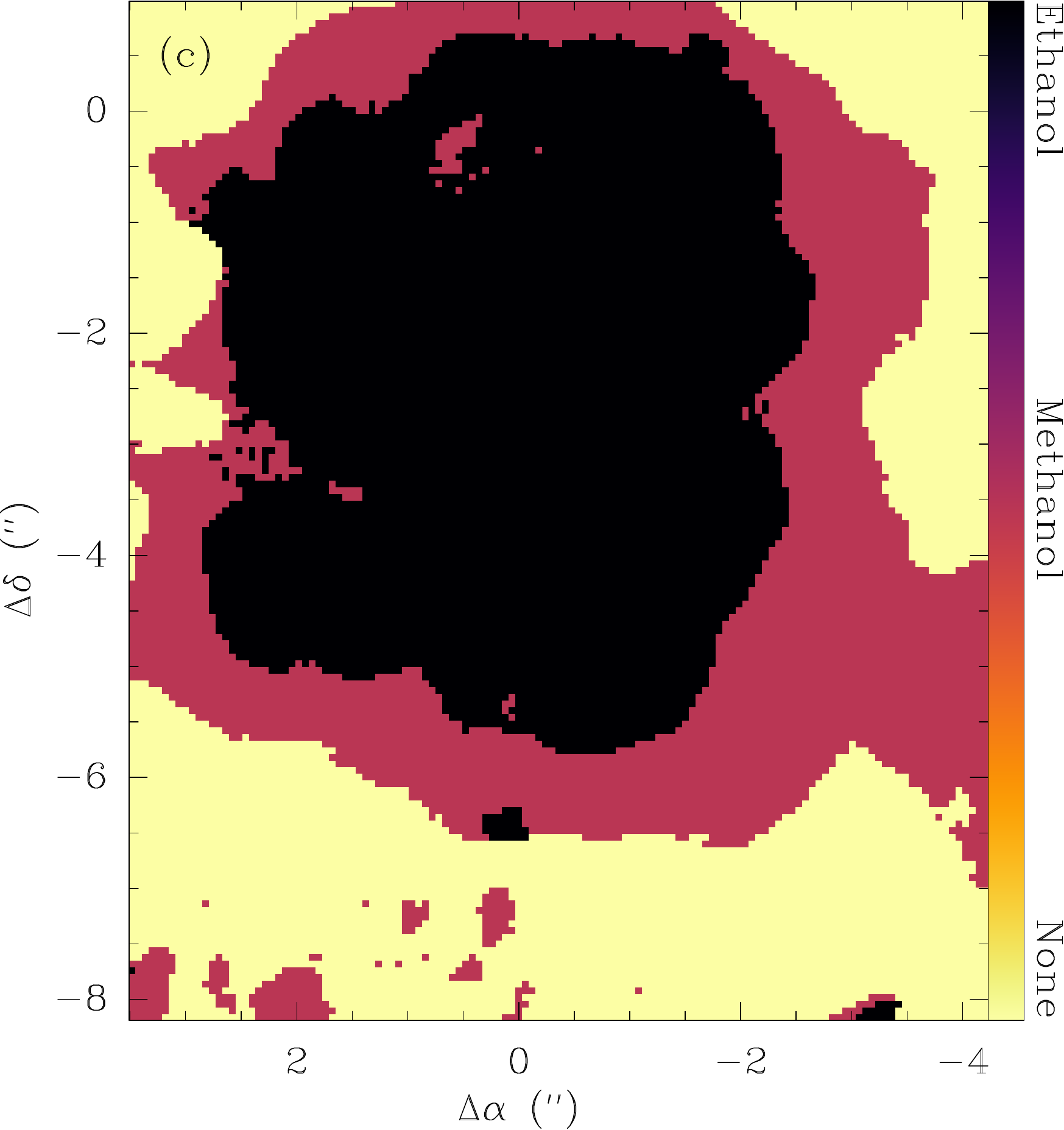}
    \caption{\textbf{(a)} Peak velocity map of Sgr\,B2\,(N1) created using the 108.44\,GHz ($E_u=88$\,K) transition of \et (black region in c) and the 95.91\,GHz ($E_u=21$\,K) transition of \met (maroon region in c). \textbf{(b)} Line width map created using the same transitions as in (a). \textbf{(c)} Map showing at which positions in the map either \et (black) or \met (maroon) is used. The transitions from ethanol to methanol happens when the fitted peak intensity of the ethanol line is less than 5$\sigma$, where $\sigma$ is the median rms noise level listed in Table\,2 in \citet{Belloche19}. At positions where neither \et nor \met are detected above this threshold (yellow region) the peak velocity and line width are fixed to values of 64\,\kms and 4\,\kms, respectively, to produce the LVINE maps shown in Fig.\,\ref{fig:COM-Maps} and are blanked in a) and b).}
    \label{fig:LVINE_O}
\end{figure*}

Figures\,\ref{fig:1mm+et} and \ref{fig:COM-Maps} show Line-width and Velocity-corrected INtegrated Emission (LVINE) maps. 
The LVINE method is an extension of the VINE method introduced by \citet{Calcutt18}. The authors determined the peak velocity of a bright methanol line at every position in their map in order to integrate intensities of other molecules/transitions over a fixed interval that then is varied around the derived peak velocity.
We extend this method by also varying the interval in which intensities are integrated based on the line width at each position.

In order to create maps of peak velocity and line width we look for a strong line that is neither contaminated nor optically thick. Because it is difficult to find a spectral line that combines all three criteria, we use two lines: an optically thin yet not too strong ethanol transition at 108.44\,GHz ($E_u=88$\,K) for the region closer to Sgr~B2~(N1) and a strong methanol transition at 95.91\,GHz ($E_u=21$\,K), at far distances from Sgr~B2~(N1). The methanol line is optically thick at positions close to Sgr~B2~(N1) ($\tau\sim3$ at N1S), however, at positions used here, its opacity is moderate ($\tau\sim0.4$ at N1S5). The two transitions are shown in Fig.\,\ref{fig:LVINE_O_spectra} at a position close to Sgr~B2~(N1) (N1S), where \et is used, and a position farther away (N1S5), where \met is used.
At each position in the map, that is, for each pixel, a 1D Gaussian profile is fitted from which the peak velocity and line width for the respective map is extracted. If the fitted peak intensity of the ethanol transition at one position drops below a threshold of 5$\sigma$, where $\sigma$ is the median rms noise level listed in Table\,2 in \citet{Belloche19}, then the methanol transition is used instead. 
The final maps of peak velocity and line width as well as a map that indicates where ethanol, methanol, or neither of the two (None) are used are shown in Fig.\,\ref{fig:LVINE_O}\,a--c, respectively. When neither the ethanol nor the methanol line have an intensity above the threshold, a fixed value of 64\,\kms is allocated to the peak velocity and a value of 4\,\kms to the line width for this pixel.

The Gaussian fitting happened unsupervised for most parts. However, in some regions an adjustment of input parameters and/or the refinement of the window, in which the fitting is done, is necessary due to, for example, additional velocity components that lead to contamination and possibly bias the resulting velocity and line width when taken into account during the fitting process. The decision on the \textit{true} velocity components is made by inspecting neighbouring positions. The decision on when to intervene is also made in the same way. 
Although we were able to avoid fatal failures of the fitting procedure in that way, some uncertainties remain hidden. 
However, with our velocity map we trace the major velocity gradients seen in Sgr\,B2\,(N1), that is, the lower velocities ($<$\,62\,\kms) seen to the south and the sudden rise in velocity to $\sim$70\,\kms to the west. 
Other less pronounced velocity changes, which are associated with filaments \citep{Schwoerer19}, are evident as well. The line widths for the considered velocity components can be as narrow as 2--3\,\kms and remain generally below a value of 10\,\kms.

\section{Continuum masking}\label{app:Cmask}

\begin{figure*}[h]
    \centering
    \includegraphics[width=0.31\textwidth]{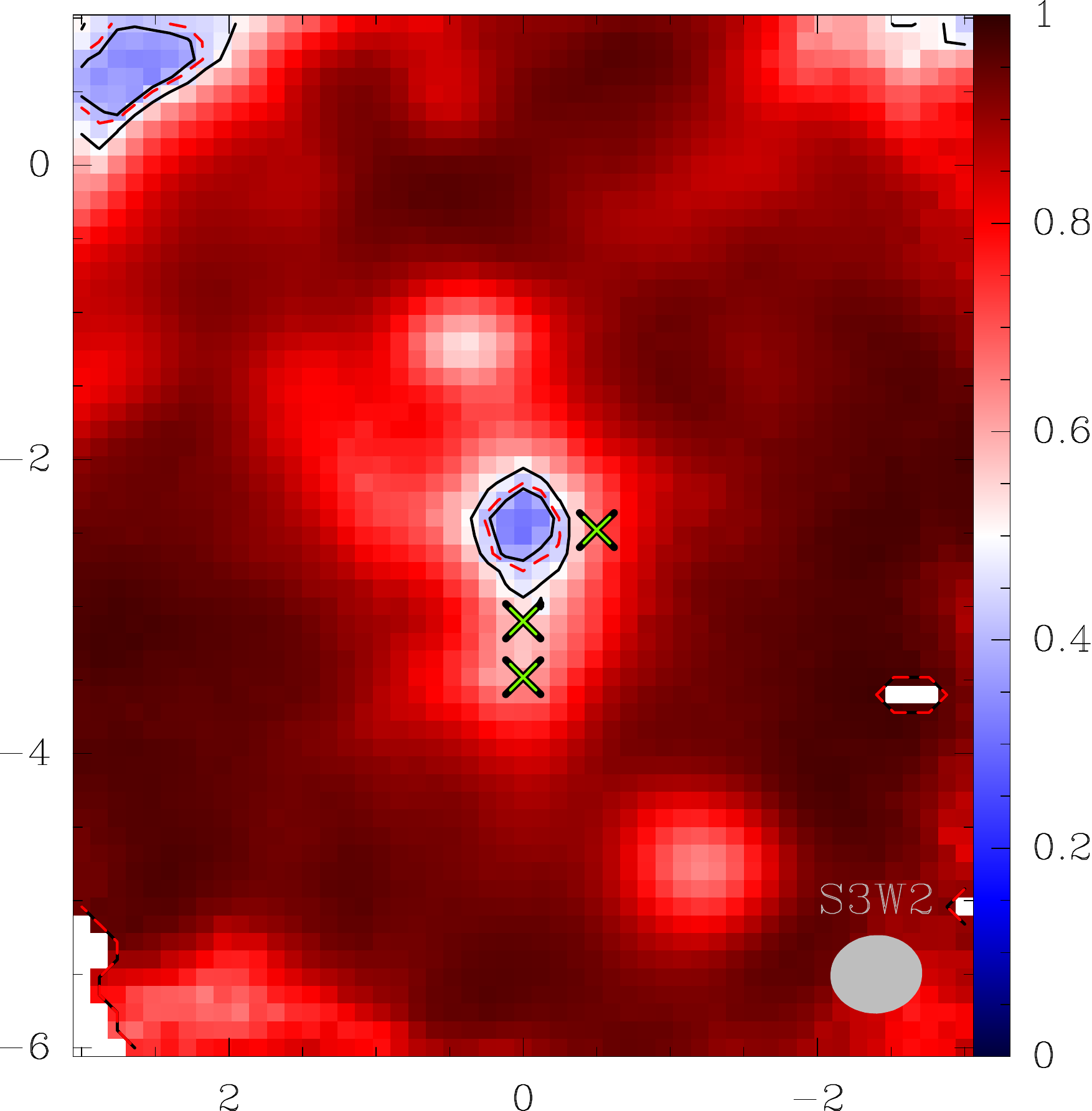}\hspace{0.1cm}
    \includegraphics[width=0.31\textwidth]{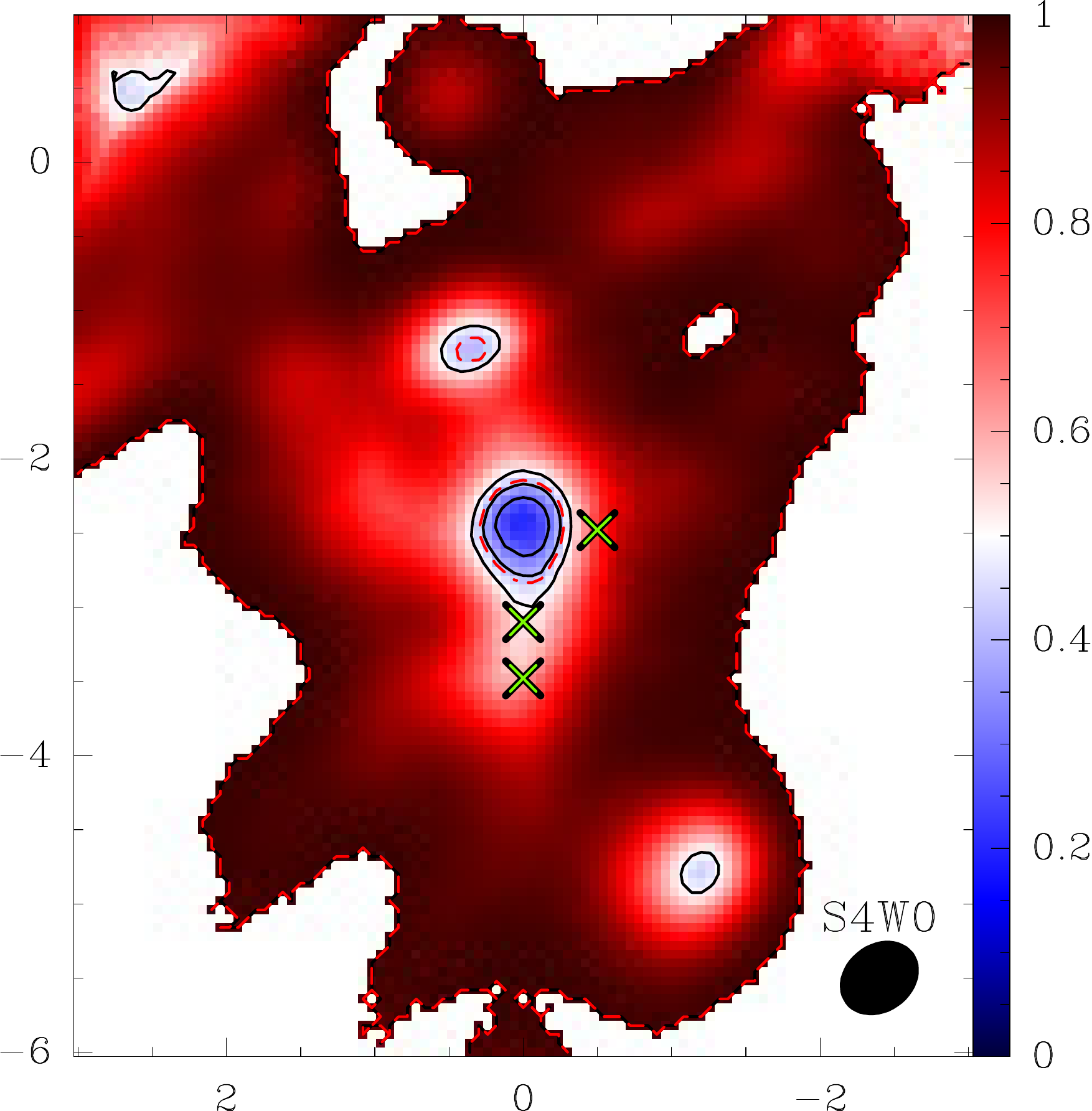}\hspace{0.1cm}
    \includegraphics[width=0.31\textwidth]{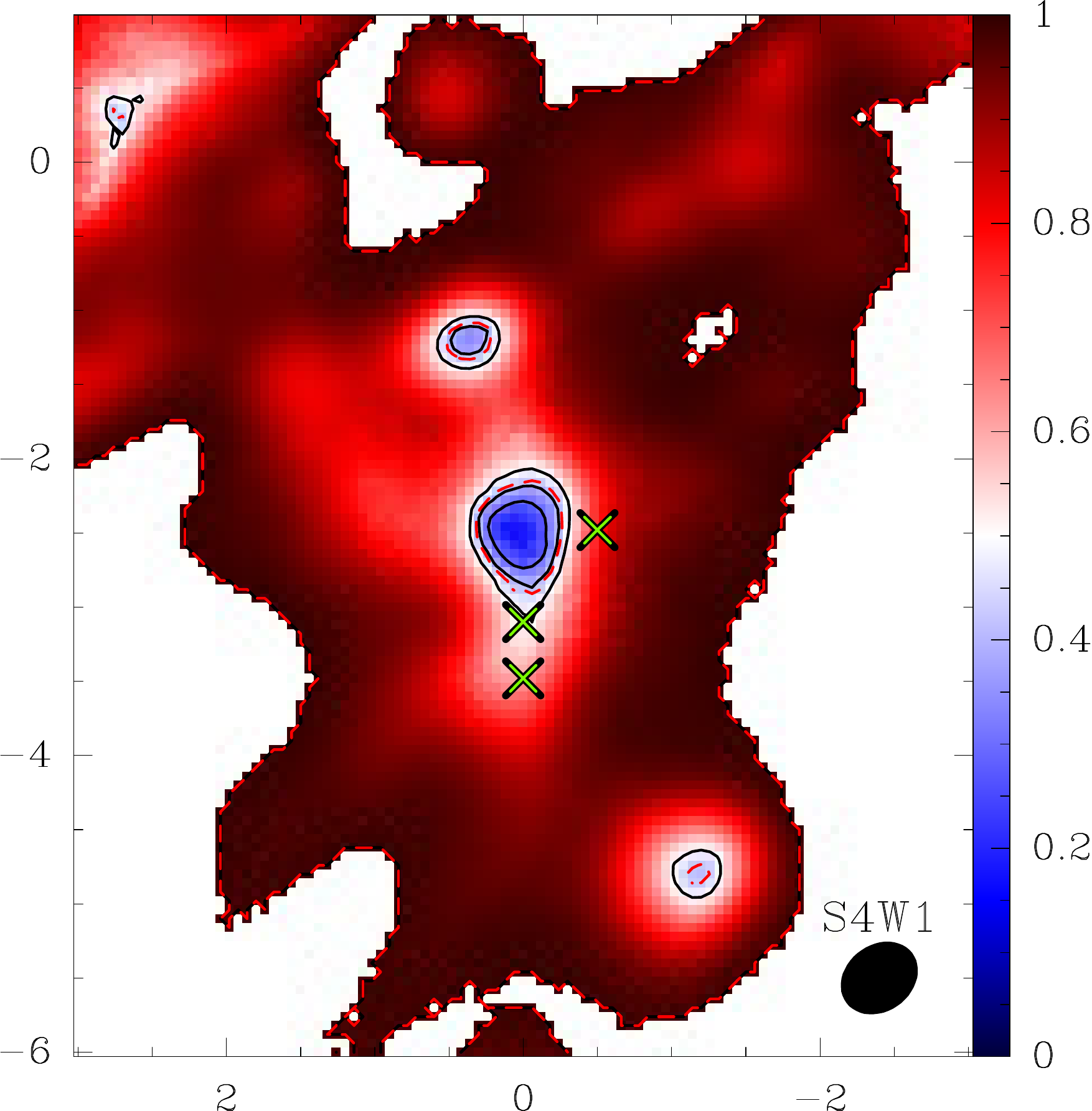}\\
    \includegraphics[width=0.31\textwidth]{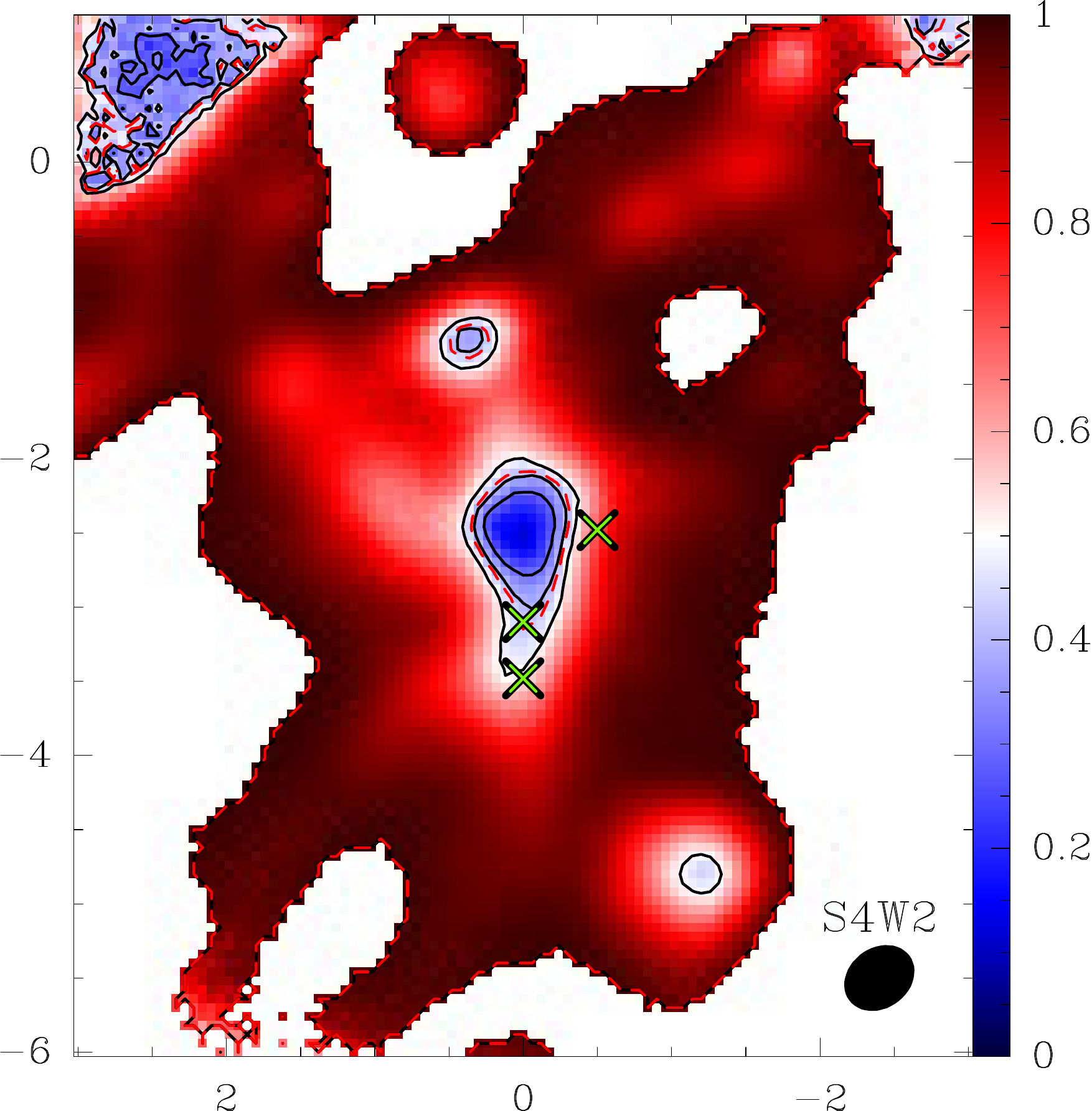}\hspace{0.1cm}
    \includegraphics[width=0.31\textwidth]{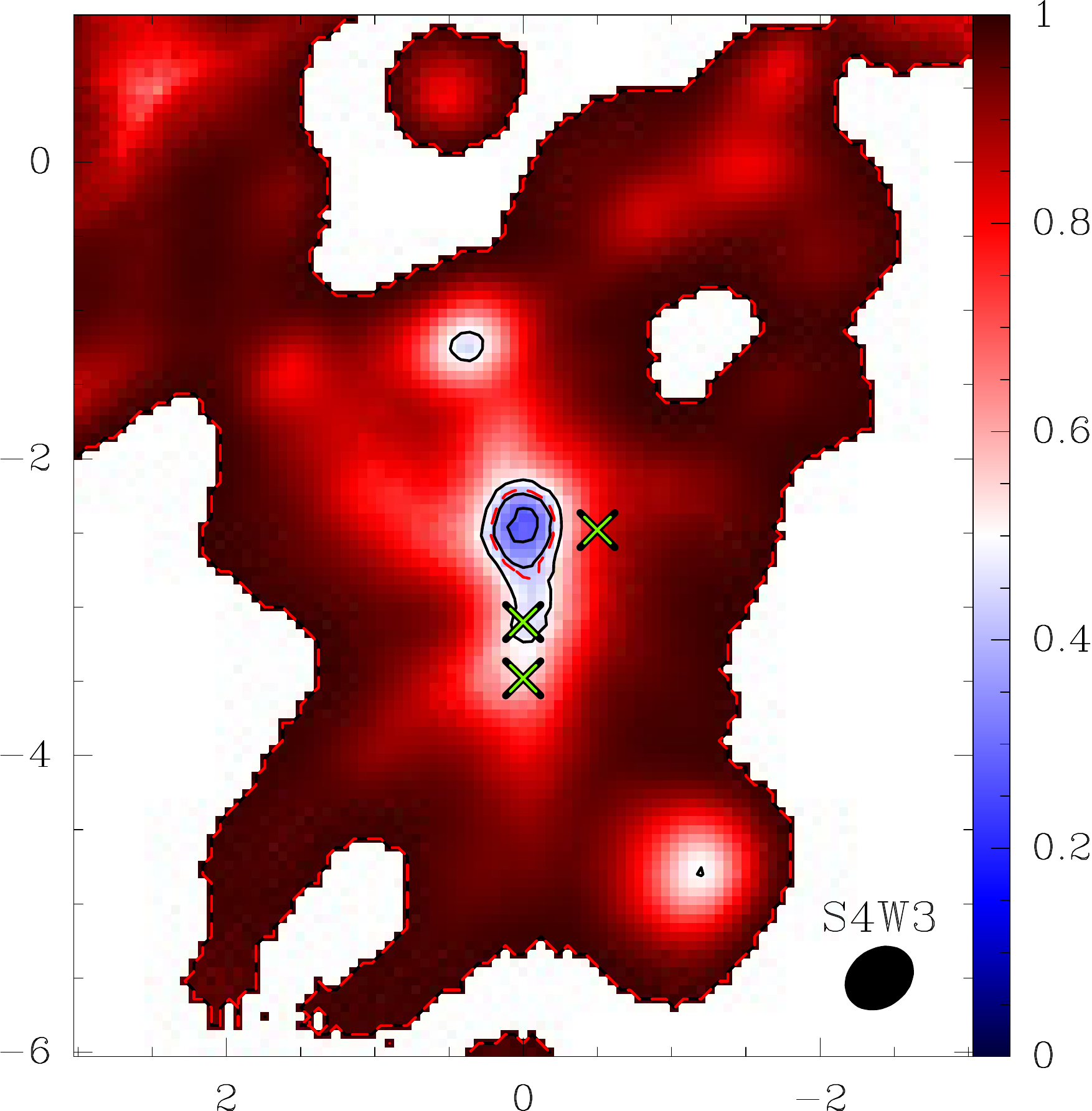}\hspace{0.1cm}
    \includegraphics[width=0.31\textwidth]{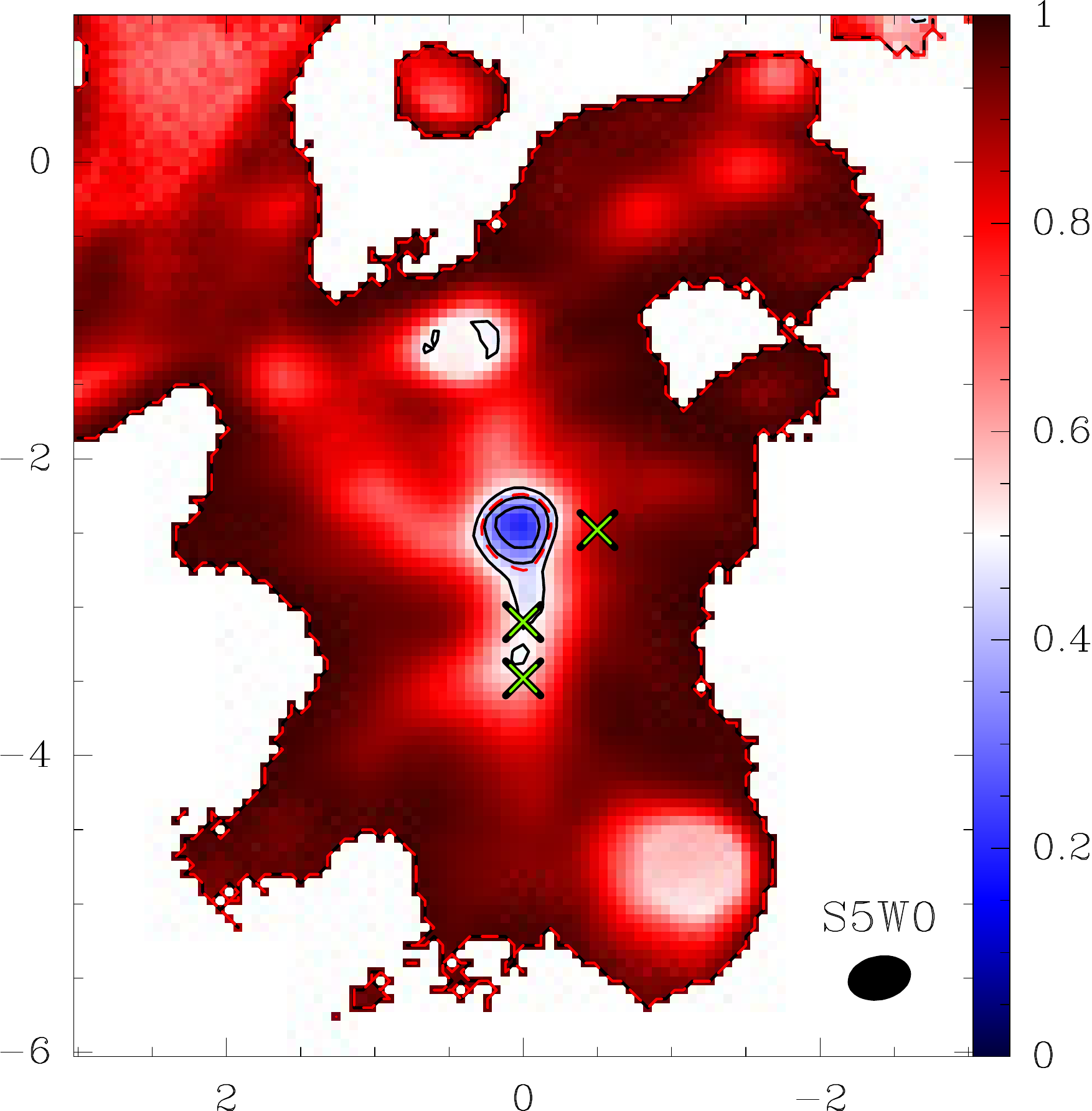}\\ 
    \includegraphics[width=0.33\textwidth]{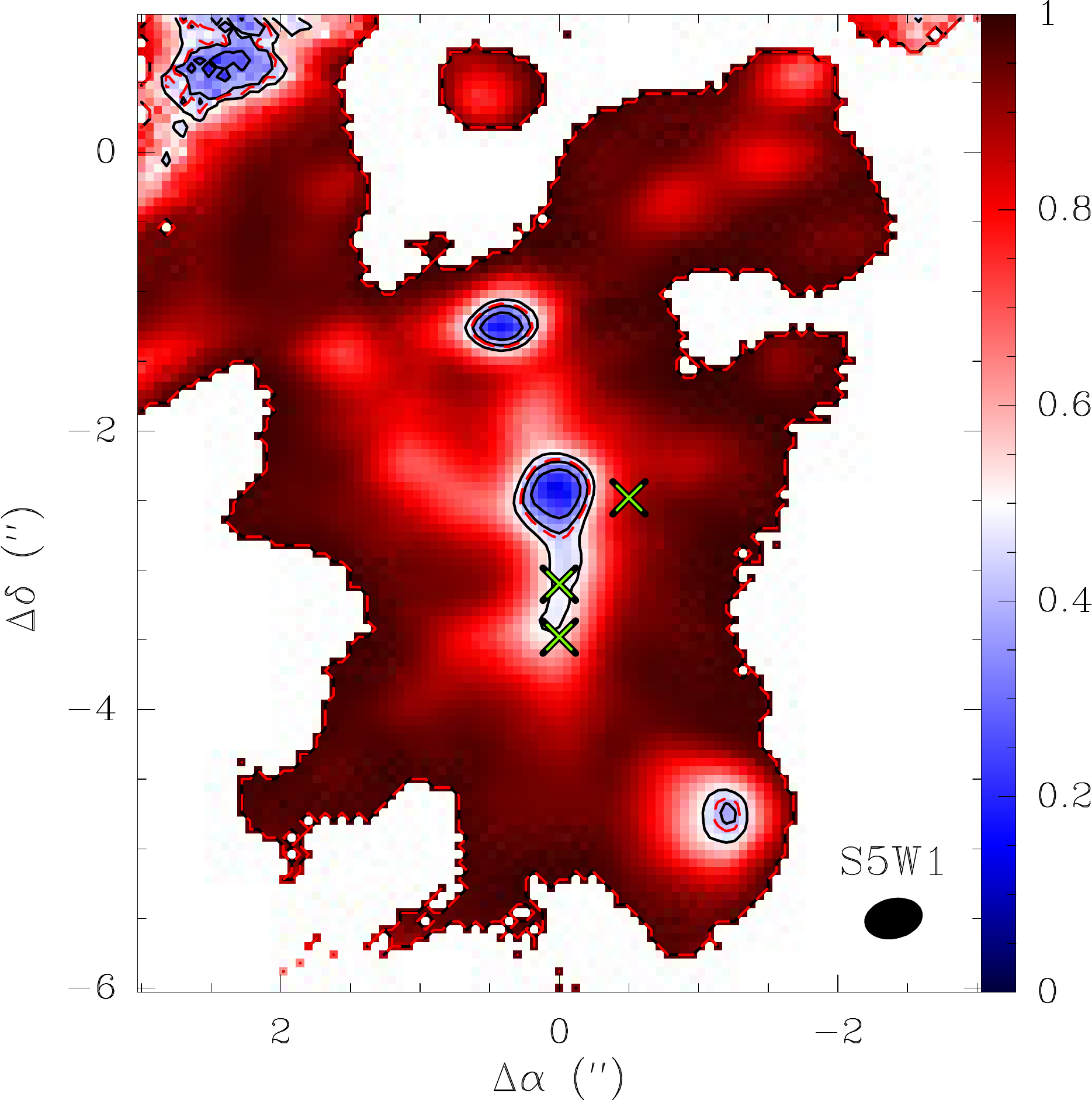}\hspace{0.1cm}
    \includegraphics[width=0.31\textwidth]{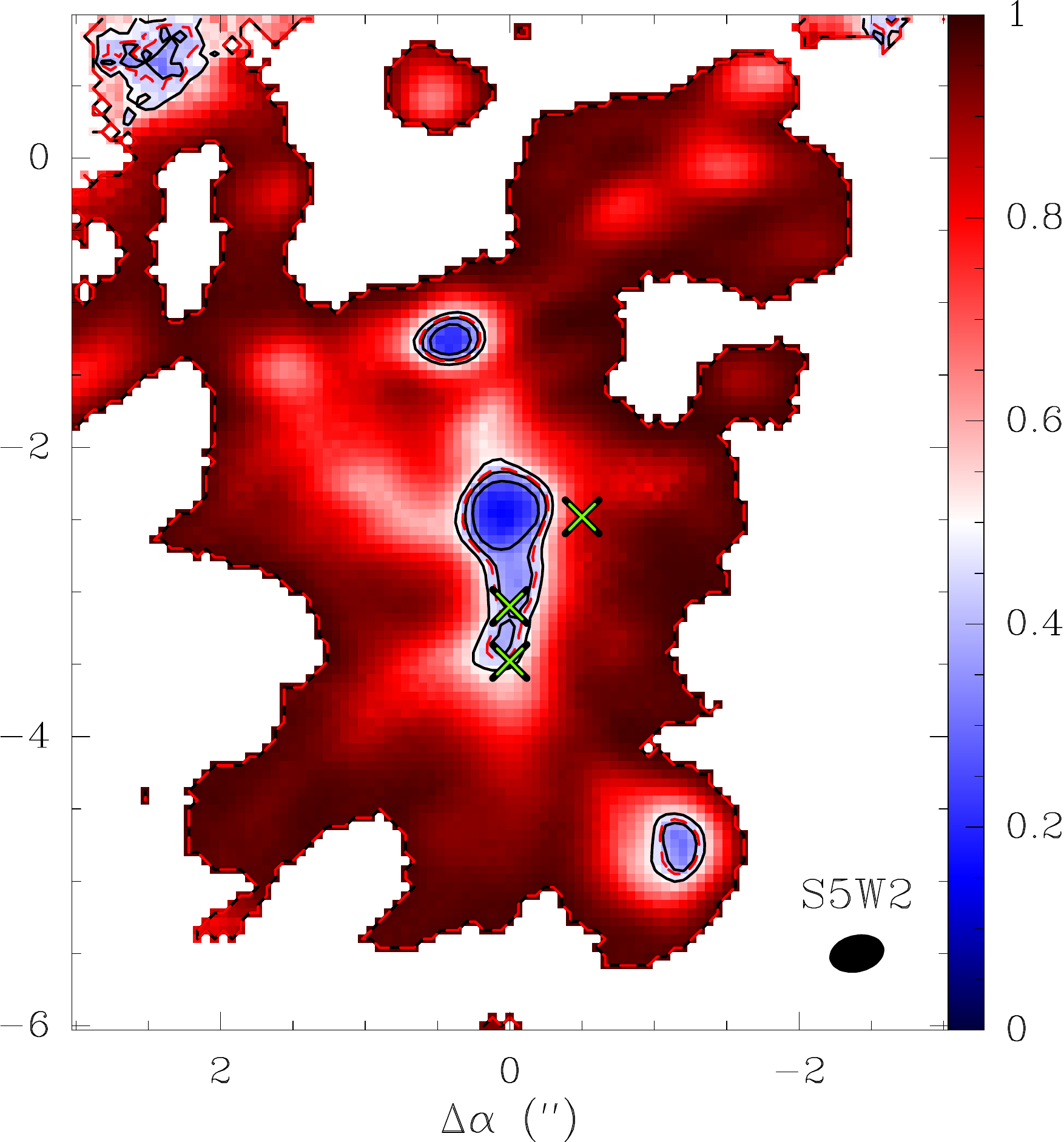}\hspace{0.1cm}
    \includegraphics[width=0.33\textwidth]{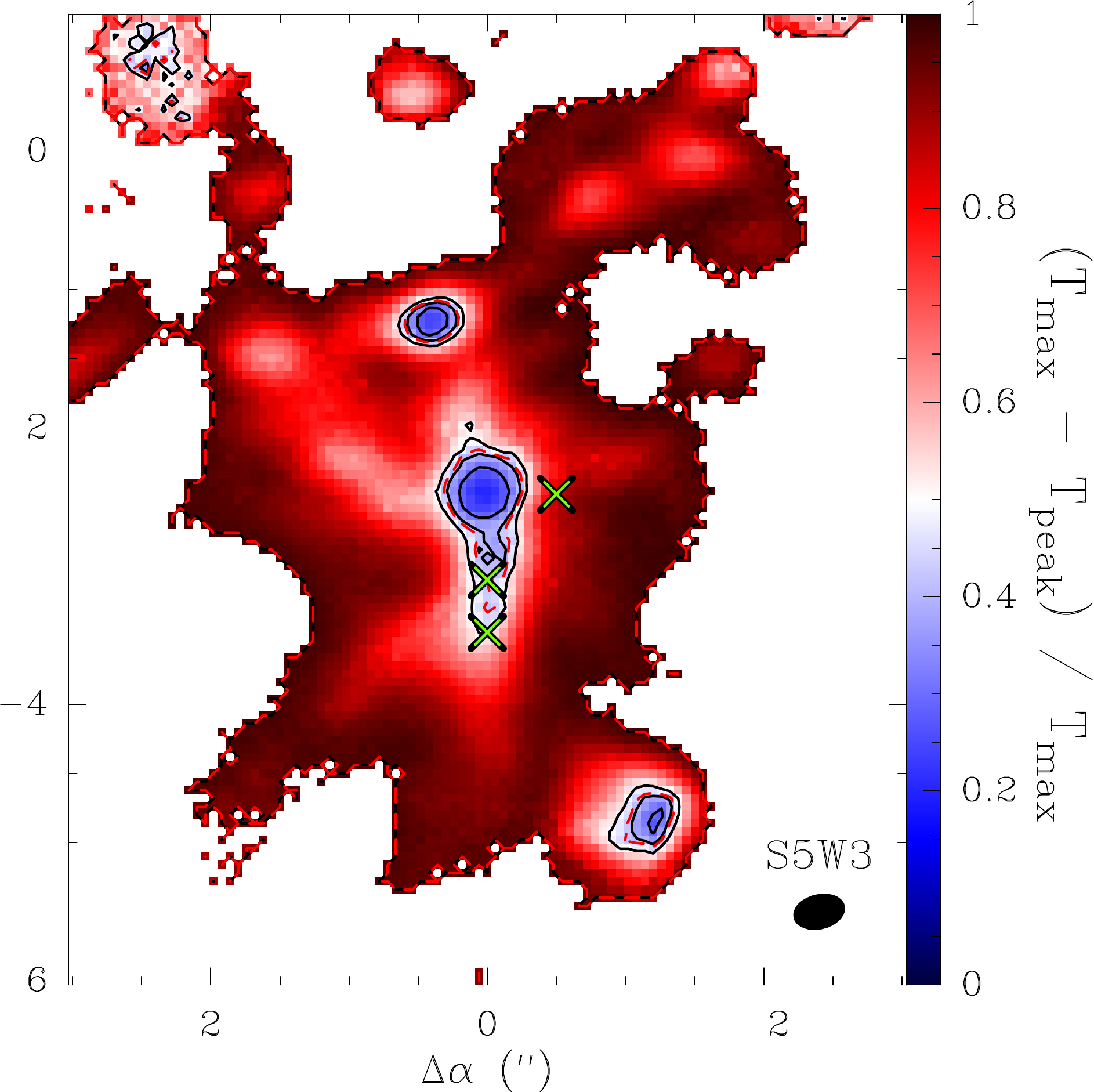}
    \caption{Visualisation of the determination of the continuum emission mask needed because of high continuum optical depth towards Sgr\,B2\,(N1). From the channel-intensity distribution of the continuum-included spectra, the intensity at the peak $T_{\rm peak}$ and the maximum intensity $T_{\rm max}$ are determined. The maps in colour-scale and the contours show the difference $T_{\rm max}-T_{\rm peak}$ normalised by $T_{\max}$. This ratio is only computed when $T_{\rm peak}>3\sigma$, where $\sigma$ is the average noise level in the continuum emission map. The contour steps (black) are 0.3, 0.4, and 0.5. The red dashed contour at a value of 0.43 indicates the boundary of the adopted mask. Green crosses show positions N1S-1 and N1S to the south and N1W-1 to the west. The observational setup and spectral window are shown in the lower right corner together with the respective beam sizes. Only those spectral windows are shown from which molecular transitions are used to create integrated intensity maps (cf. Figs.\,\ref{fig:1mm+et} and \ref{fig:COM-Maps}).}
    \label{fig:opacityC}
\end{figure*}

Although for most parts the continuum emission is optically thin (cf. Fig.\,\ref{fig:spectral-index}), we identified an increase in optical depth of the continuum at distances $\lesssim$0.5\arcsec to Sgr\,B2\,(N1). When the continuum becomes optically thick it obscures the line emission from COMs. Because the Weeds model cannot treat this effect properly an analysis of COM emission in this region is impossible to do. Therefore, we exclude this closest region around Sgr~B2~(N1) by creating masks. The definition of the mask size is challenging as the continuum emission depends on angular resolution, hence on the observational setup and frequency. 

To create the masks we compute the channel-intensity distribution for the continuum-included spectra of each spectral window and determine the intensity at the peak of this distribution $T_{\rm peak}$ and the maximum intensity $T_{\rm max}$. We plot the difference $T_{\rm max}-T_{\rm peak}$ normalised by $T_{\rm max}$ in Fig.\,\ref{fig:opacityC}. A position in the map is blanked when $T_{\rm peak}<3\sigma$, where $\sigma$ is the average noise level in the continuum emission map.
That way we can identify at which distance from Sgr~B2~(N1) the emission lines get attenuated due to the increasing continuum optical thickness. As expected the observational setups of higher angular resolution and higher frequencies are more affected. We decided to use the contour at a value of 0.43 (see red dashed contour in Fig.\,\ref{fig:opacityC}) to define the mask for each setup. This decision is made based on these maps and also on the comparison to the spectra. 
It allows us to keep position N1S in our analysis for all observational setups and N1S-1 for all setups except spectral windows 2 and 3 of setup 5. For these two spectral windows, the Weeds model systematically overestimates intensities at N1S-1 although the model seems to fit otherwise. At this close distance to Sgr~B2~(N1) the increasing continuum optical thickness can be expected to attenuate the observed intensity of spectral lines untreated by Weeds. Other reasons may be presented by an overestimation of the source size or filtering of emission by the interferometre.

The approach applied here is not the most accurate because, for example, it does not take into account absorption lines or extremely strong lines from very abundant molecules (e.g., $^{13}$CO). However, it is sufficient for our purposes. 
Accordingly, spectral windows 2 and 3 of setup 5 are excluded from the analysis of position N1S-1 and integrated intensities within the mask are blanked in all respective maps.
Figure\,\ref{fig:opacityC} further suggests to mask emission originating from the H{\small II} regions K3 and K1 for some spectral windows. However, inspecting spectra at the continuum peak positions of the two shows that the continuum is still sufficiently optically thin and the spectral lines are simply weak. Therefore, we do not apply any mask to the H{\small II} regions.

\section{Additional figures}\label{app:extrafigs}
\begin{figure*}[h]
    \centering
    \includegraphics[width=\textwidth]{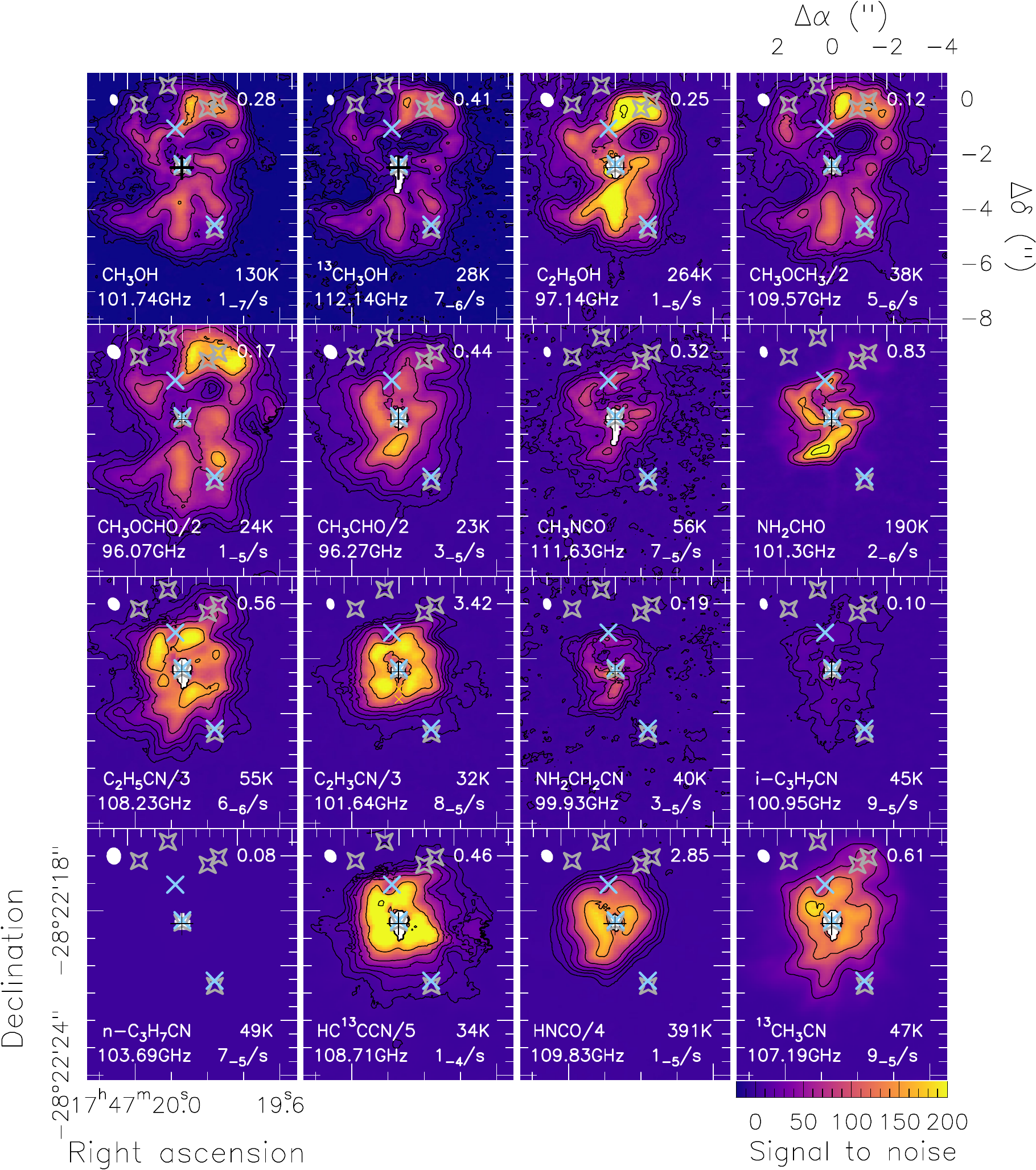}
    \caption{Same as Fig.\,\ref{fig:COM-Maps} except that instead of integrated intensities the signal-to-noise ratio is shown. The noise is computed by $\sqrt{N}\times\Delta\varv\times\sigma$, where $N$ is the number of channels over which integration was performed, $\Delta\varv$ is the channel separation in \kms, and $\sigma$ is the median rms noise level as shown in Table\,2 in \citet{Belloche19}. Values of \dme, \mf, and \ad are scaled down by a factor of 2, those of \etc and \vc by a factor 3, \cyano by a factor 5, HNCO by a factor 4. The contour steps are 3, 9, 18, 36, 72, 144, 288, and 576.}
    \label{fig:snr}
\end{figure*}

\begin{figure}[]
    \hspace{-0.2cm}
    \includegraphics[width=.5\textwidth]{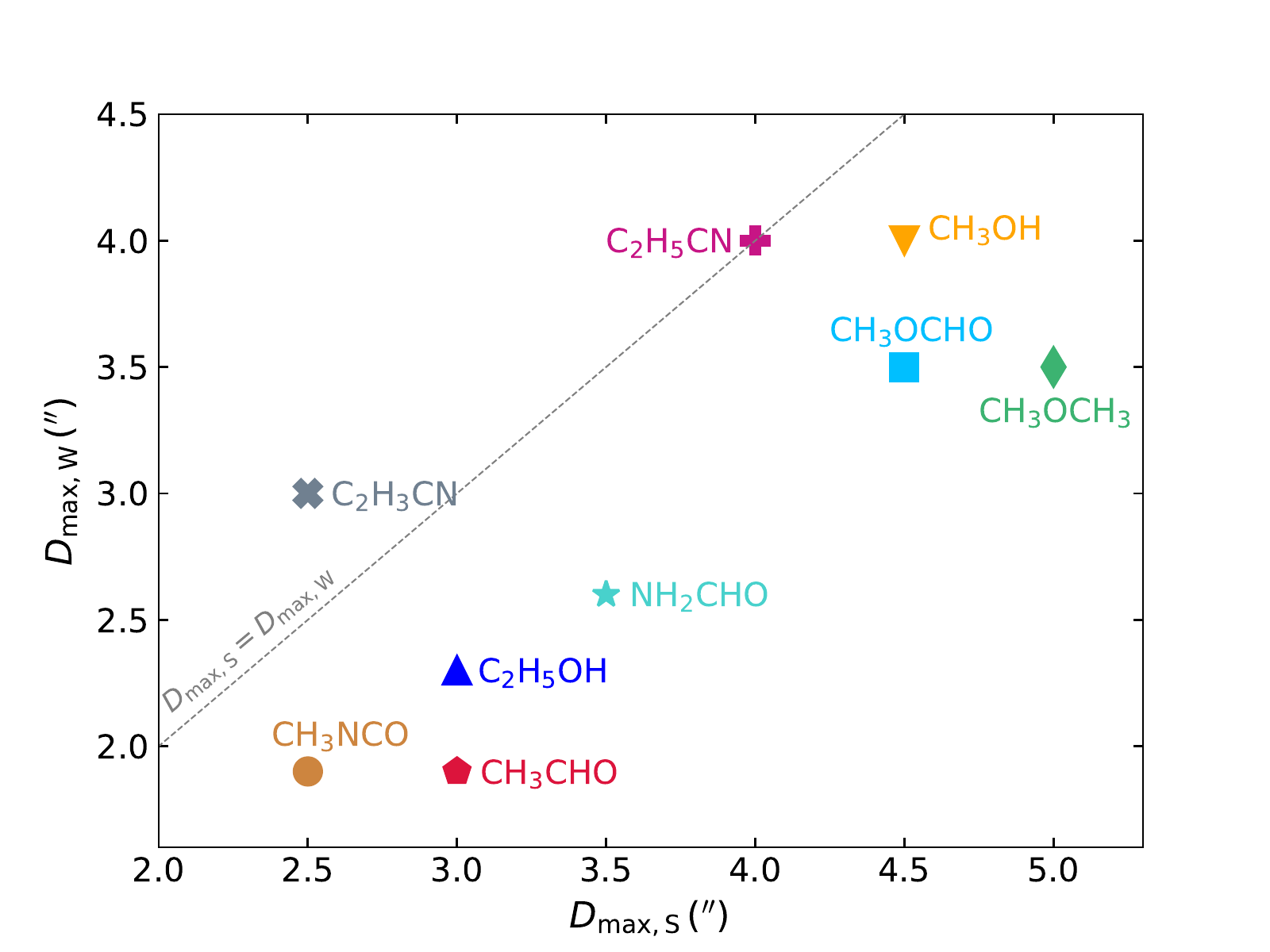}
    \caption{Maximum distances to the south $D_{\rm max,S}$ and west $D_{\rm max,W}$ from Sgr\,B2\,(N1) to which a COM is still detected and beyond which only upper limits for column density can be derived. These distances correspond to vertical dashed lines in Fig.\,\ref{fig:Ncolprofiles}.}
    \label{fig:Dmax-COM}
\end{figure}

\begin{figure}
    \includegraphics[width=0.5\textwidth]{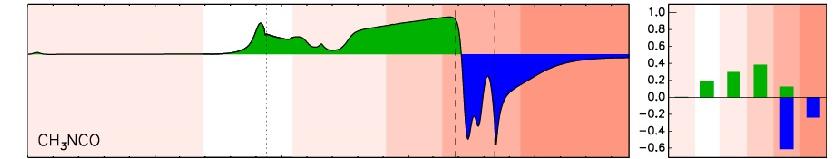}\\[-0.1cm]
    \includegraphics[width=0.5\textwidth]{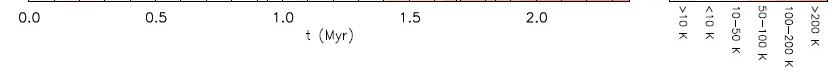}
    \caption{Same as in Fig.\,13 in \citet{Garrod22}. It shows the net rate of change in \mic abundances, summed over all chemical phases, during the cold collapse phase and the subsequent \textit{slow} warm-up.}
    \label{fig:roc_mic}
\end{figure}

Because the noise level within each LVINE map shown in Fig.\,\ref{fig:COM-Maps} is not uniform due to the pixel-dependent integration limits, that is, the number of channels over which the integration is performed is different in each pixel, we show in Fig.\,\ref{fig:snr} signal-to-noise maps complementary to the LVINE maps.
Figure\,\ref{fig:Dmax-COM} highlights the maximum distances to the south and west to which emission of a certain COM is still detectable.
Figure\,\ref{fig:roc_mic} shows the same as in Fig.\,13 in \citet{Garrod22} for \mic, that is, it illustrates the net rate of change in abundance of the COM starting from the cold phase and assuming a subsequent \textit{slow} warm-up. 

\section{Additional tables}\label{app:tables}
Table\,\ref{tab:positions} lists the positions selected for analysis, which are also indictad in Fig.\,\ref{fig:1mm+et}. Tables\,\ref{tab:n1s-1}--\ref{tab:n1w8} provide the parameters used for the LTE modelling with Weeds of each COM at each position, respectively. Moreover, the results obtained from the population diagrams are shown. Table\,\ref{tab:H2} lists the H\2 column densities derived from either dust continuum emission or C$^{18}$O at each position, respectively, along with the dust continuum flux, continuum optical depth, and the model parameters of C$^{18}$O. 
\begin{table}[]
    \caption{Selected positions for analysis, also shown in Fig.\,\ref{fig:1mm+et} in terms of equatorial offsets from the continuum peak position of Sgr\,B2\,(N1).}
    \centering
    \begin{tabular}{lcc}
       \hline\hline\\[-0.2cm]
         Position &  $\Delta\alpha$ & $\Delta\beta$ \\
         &  ($^{\prime\prime}$) &  ($^{\prime\prime}$) \\[0.1cm]\hline\\[-0.3cm]
         N1S-1 & 0.0 & $-$3.1 \\
         N1S & 0.0 & $-$3.5 \\
         N1S1 & 0.0 & $-$4.0 \\
         N1S2 & 0.0 & $-$4.5 \\
         N1S3 & 0.0 & $-$5.0 \\
         N1S4 & 0.0 & $-$5.5 \\
         N1S5 & 0.0 & $-$6.0 \\
         N1S6 & 0.0 & $-$6.2 \\
         N1S7 & 0.0 & $-$6.5 \\
         N1S8 & 0.0 & $-$7.0 \\
         N1S9 & 0.0 & $-$7.5 \\
         N1S10 & 0.0 & $-$8.0 \\
         N1W-1 & $-$0.5 & $-$2.5 \\
         N1W & $-$1.0 & $-$2.5 \\
         N1W1 & $-$1.3 & $-$2.5 \\
         N1W2 & $-$1.6 & $-$2.5 \\
         N1W3 & $-$1.9 & $-$2.5 \\
         N1W4 & $-$2.3 & $-$2.5 \\
         N1W5 & $-$2.6 & $-$2.5 \\
         N1W6 & $-$3.0 & $-$2.5 \\
         N1W7 & $-$3.5 & $-$2.5 \\
         N1W8 & $-$4.0 & $-$2.5 \\
         \hline\hline
    \end{tabular}
    \label{tab:positions}
\end{table}

\begin{table*}[h!]
\caption{Weeds parameters used for COMs at N1S-1.} 
\centering
\begin{tabular}{rrrrrrrrrrr}
\hline\hline \\[-0.3cm] 
Molecule\tablefootmark{a} & Size\tablefootmark{b} & $T_{\rm rot,W}$\tablefootmark{c} & $T_{\rm rot,obs}$\tablefootmark{d} & $N_{\rm W}\tablefootmark{e}$ & $N_{\rm obs}\tablefootmark{f}$ & $C_{\rm vib}$\tablefootmark{g} & $\Delta\varv$\tablefootmark{h} & $\varv_{\rm off}$\tablefootmark{i} \\ 
 & $(^{\prime\prime})$ & (K) & (K) & (cm$^{-2}$) & (cm$^{-2}$) &  & (km\,s$^{-1}$) & (km\,s$^{-1}$) \\\hline \\[-0.3cm] 
CH$_3$OH, $v=0$ & $2.0$ & $250$ & $262\pm 10$ & $2.5(19)$ & $(1.5\pm 0.2)(19)$ & $1.00$ & $6.0$ & $0.0$ \\ 
 $v=1$ & $2.0$ & $250$ & $262\pm 10$ & $2.5(19)$ & $(1.5\pm 0.2)(19)$ & $1.00$ & $6.0$ & $0.0$ \\ 
 $v=2$ & $2.0$ & $250$ & $262\pm 10$ & $2.5(19)$ & $(1.5\pm 0.2)(19)$ & $1.00$ & $6.0$ & $0.0$ \\ 
 $^{13}$CH$_3$OH, $v=0$ & $2.0$ & $250$ & $262\pm 27$ & $1.0(18)$ & $(8.8\pm 2.1)(17)$ & $1.00$ & $6.0$ & $0.5$ \\ 
 \hline \\[-0.35cm] 
C$_2$H$_5$OH, $v=0$ & $2.0$ & $300$ & $301\pm 15$ & $2.5(18)$ & $(2.0\pm 0.2)(18)$ & $2.88$ & $6.3$ & $0.3$ \\ 
 \hline \\[-0.35cm] 
CH$_3$OCH$_3$, $v=0$ & $2.0$ & $254$ & $254\pm 17$ & $2.3(18)$ & $(1.4\pm 0.3)(18)$ & $1.16$ & $4.5$ & $1.5$ \\ 
 \hline \\[-0.35cm] 
CH$_3$OCHO, $v=0$ & $2.0$ & $320$ & $317\pm 35$ & $3.1(18)$ & $(2.2\pm 0.4)(18)$ & $2.62$ & $5.0$ & $0.5$ \\ 
 $v=1$ & $2.0$ & $320$ & $317\pm 35$ & $3.1(18)$ & $(2.2\pm 0.4)(18)$ & $2.62$ & $5.0$ & $0.5$ \\ 
 \hline \\[-0.35cm] 
CH$_3$CHO, $v=0$ & $2.0$ & $280$ & $277\pm 8$ & $1.6(18)$ & $(5.5\pm 0.4)(17)$ & $1.13$ & $6.0$ & $0.0$ \\ 
 $v=1$ & $2.0$ & $280$ & $277\pm 8$ & $1.6(18)$ & $(5.5\pm 0.4)(17)$ & $1.13$ & $6.0$ & $0.0$ \\ 
 \hline \\[-0.35cm] 
CH$_3$NCO, $v=0$ & $2.0$ & $150$ & $139\pm 7$ & $3.1(17)$ & $(1.2\pm 0.2)(17)$ & $1.00$ & $5.5$ & $-0.2$ \\ 
 \hline \\[-0.35cm] 
C$_2$H$_5$CN, $v=0$ & $2.0$ & $230$ & $231\pm 5$ & $5.1(18)$ & $(3.1\pm 0.2)(18)$ & $2.21$ & $6.8$ & $-0.7$ \\ 
 $v_{12}=1$ & $2.0$ & $230$ & $231\pm 5$ & $5.1(18)$ & $(3.1\pm 0.2)(18)$ & $2.21$ & $6.8$ & $-0.7$ \\ 
 $v_{13}=2$ & $2.0$ & $230$ & $231\pm 5$ & $5.1(18)$ & $(3.1\pm 0.2)(18)$ & $2.21$ & $6.8$ & $-0.7$ \\ 
 $v_{20}=1$ & $2.0$ & $230$ & $231\pm 5$ & $5.1(18)$ & $(3.1\pm 0.2)(18)$ & $2.21$ & $6.8$ & $-0.7$ \\ 
 $v_{21}=2$ & $2.0$ & $230$ & $231\pm 5$ & $5.1(18)$ & $(3.1\pm 0.2)(18)$ & $2.21$ & $6.8$ & $-0.7$ \\ 
 \hline \\[-0.35cm] 
C$_2$H$_3$CN, $v=0$ & $2.0$ & $300$ & $299\pm 15$ & $3.0(18)$ & $(2.6\pm 0.3)(18)$ & $1.20$ & $6.0$ & $-0.5$ \\ 
 $v_{11}=1$ & $2.0$ & $300$ & $299\pm 15$ & $3.0(18)$ & $(2.6\pm 0.3)(18)$ & $1.20$ & $6.0$ & $-0.5$ \\ 
 $v_{15}=1$ & $2.0$ & $300$ & $299\pm 15$ & $3.0(18)$ & $(2.6\pm 0.3)(18)$ & $1.20$ & $6.0$ & $-0.5$ \\ 
 \hline \\[-0.35cm] 
NH$_2$CHO, $v=0$ & $2.0$ & $170$ & $166\pm 5$ & $4.1(18)$ & $(3.2\pm 0.3)(18)$ & $1.10$ & $5.3$ & $0.1$ \\ 
 $v_{12}=1$ & $2.0$ & $170$ & $166\pm 5$ & $4.1(18)$ & $(3.2\pm 0.3)(18)$ & $1.10$ & $5.3$ & $0.1$ \\ 
 \hline\hline
\end{tabular}
\tablefoot{\tablefoottext{a}{COMs and vibrational states used to derive population diagrams. }\tablefoottext{b}{Size of the emitting region.}\tablefoottext{c}{Rotation temperature used for the Weeds model.}\tablefoottext{d}{Rotation temperature derived from the population diagram.}\tablefoottext{e}{Column density used for the Weeds model.}\tablefoottext{f}{Column density derived from the population diagram.}\tablefoottext{g}{Vibrational state correction applied when partition function does not account for higher-excited vibrational states, where $C_{\rm vib}=C_{\rm vib}(T_{\rm rot,obs})$.}\tablefoottext{h}{$FWHM$ of the transitions.}\tablefoottext{i}{Offset from the source systemic velocity, which was set to 62\kms.}\\ Values in parentheses show the decimal power, where $x(z) = x\times 10^z$ or $(x\pm y)(z) = (x\pm y)\times 10^z$. \\  Upper limits on $N_W$ indicate that a population diagram could not be derived, either because too many transitions are contaminated or the molecule is not detected. The temperatures used for the Weeds model at these positions are determined by extrapolating the temperature profile of the respective molecule that was derived in Sect.\,\ref{ss:profiles}.  }\label{tab:n1s-1}
\end{table*}

\begin{table*}[h!]
\caption{Same as Table\,\ref{tab:n1s-1}, but for position N1S.} 
\centering
\begin{tabular}{rrrrrrrrrrr}
\hline\hline \\[-0.3cm] 
Molecule & Size & $T_{\rm rot,W}$ & $T_{\rm rot,obs}$ & $N_{\rm W}$ & $N_{\rm obs}$ & $C_{\rm vib}$ & $\Delta\varv$ & $\varv_{\rm off}$ \\ 
 & $(^{\prime\prime})$ & (K) & (K) & (cm$^{-2}$) & (cm$^{-2}$) &  & (km\,s$^{-1}$) & (km\,s$^{-1}$) \\\hline \\[-0.3cm] 
CH$_3$OH, $v=0$ & $2.0$ & $220$ & $228\pm 5$ & $3.0(19)$ & $(2.2\pm 0.2)(19)$ & $1.00$ & $5.0$ & $0.5$ \\ 
 $v=1$ & $2.0$ & $220$ & $228\pm 5$ & $3.0(19)$ & $(2.2\pm 0.2)(19)$ & $1.00$ & $5.0$ & $0.5$ \\ 
 $v=2$ & $2.0$ & $220$ & $228\pm 5$ & $3.0(19)$ & $(2.2\pm 0.2)(19)$ & $1.00$ & $5.0$ & $0.5$ \\ 
 $^{13}$CH$_3$OH, $v=0$ & $2.0$ & $215$ & $206\pm 10$ & $1.2(18)$ & $(1.1\pm 0.1)(18)$ & $1.00$ & $4.5$ & $0.5$ \\ 
 \hline \\[-0.35cm] 
C$_2$H$_5$OH, $v=0$ & $2.0$ & $215$ & $220\pm 7$ & $1.8(18)$ & $(1.6\pm 0.1)(18)$ & $1.85$ & $5.4$ & $0.3$ \\ 
 \hline \\[-0.35cm] 
CH$_3$OCH$_3$, $v=0$ & $2.0$ & $250$ & $225\pm 9$ & $1.3(18)$ & $(8.8\pm 1.0)(17)$ & $1.09$ & $4.5$ & $0.5$ \\ 
 $v_{11}=1$ & $2.0$ & $250$ & $225\pm 9$ & $1.3(18)$ & $(8.8\pm 1.0)(17)$ & $1.09$ & $4.5$ & $0.5$ \\ 
 \hline \\[-0.35cm] 
CH$_3$OCHO, $v=0$ & $2.0$ & $280$ & $294\pm 24$ & $3.0(18)$ & $(2.8\pm 0.4)(18)$ & $2.32$ & $4.0$ & $0.5$ \\ 
 $v=1$ & $2.0$ & $280$ & $294\pm 24$ & $3.0(18)$ & $(2.8\pm 0.4)(18)$ & $2.32$ & $4.0$ & $0.5$ \\ 
 \hline \\[-0.35cm] 
CH$_3$CHO, $v=0$ & $2.0$ & $200$ & $214\pm 6$ & $9.4(17)$ & $(4.1\pm 0.3)(17)$ & $1.05$ & $5.0$ & $0.0$ \\ 
 $v=1$ & $2.0$ & $200$ & $214\pm 6$ & $9.4(17)$ & $(4.1\pm 0.3)(17)$ & $1.05$ & $5.0$ & $0.0$ \\ 
 \hline \\[-0.35cm] 
CH$_3$NCO, $v=0$ & $2.0$ & $150$ & $154\pm 25$ & $2.4(17)$ & $(1.5\pm 0.6)(17)$ & $1.00$ & $5.0$ & $-0.2$ \\ 
 \hline \\[-0.35cm] 
C$_2$H$_5$CN, $v=0$ & $2.0$ & $190$ & $186\pm 4$ & $3.7(18)$ & $(3.0\pm 0.2)(18)$ & $1.69$ & $6.0$ & $-0.1$ \\ 
 $v_{13}=2$ & $2.0$ & $190$ & $186\pm 4$ & $3.7(18)$ & $(3.0\pm 0.2)(18)$ & $1.69$ & $6.0$ & $-0.1$ \\ 
 $v_{20}=1$ & $2.0$ & $190$ & $186\pm 4$ & $3.7(18)$ & $(3.0\pm 0.2)(18)$ & $1.69$ & $6.0$ & $-0.1$ \\ 
 $v_{21}=2$ & $2.0$ & $190$ & $186\pm 4$ & $3.7(18)$ & $(3.0\pm 0.2)(18)$ & $1.69$ & $6.0$ & $-0.1$ \\ 
 \hline \\[-0.35cm] 
C$_2$H$_3$CN, $v=0$ & $2.0$ & $220$ & $217\pm 7$ & $9.4(17)$ & $(6.5\pm 0.6)(17)$ & $1.05$ & $6.0$ & $-0.5$ \\ 
 $v_{11}=1$ & $2.0$ & $220$ & $217\pm 7$ & $9.4(17)$ & $(6.5\pm 0.6)(17)$ & $1.05$ & $6.0$ & $-0.5$ \\ 
 $v_{15}=1$ & $2.0$ & $220$ & $217\pm 7$ & $9.4(17)$ & $(6.5\pm 0.6)(17)$ & $1.05$ & $6.0$ & $-0.5$ \\ 
 \hline \\[-0.35cm] 
NH$_2$CHO, $v=0$ & $2.0$ & $160$ & $157\pm 4$ & $4.0(18)$ & $(3.4\pm 0.2)(18)$ & $1.08$ & $5.7$ & $0.1$ \\ 
 $v_{12}=1$ & $2.0$ & $160$ & $157\pm 4$ & $4.0(18)$ & $(3.4\pm 0.2)(18)$ & $1.08$ & $5.7$ & $0.1$ \\ 
 \hline\hline
\end{tabular}
\label{tab:n1s}
\end{table*}

\begin{table*}[h!]
\caption{Same as Table\,\ref{tab:n1s-1}, but for position N1S1.} 
\centering
\begin{tabular}{rrrrrrrrrrr}
\hline\hline \\[-0.3cm] 
Molecule & Size & $T_{\rm rot,W}$ & $T_{\rm rot,obs}$ & $N_{\rm W}$ & $N_{\rm obs}$ & $C_{\rm vib}$ & $\Delta\varv$ & $\varv_{\rm off}$ \\ 
 & $(^{\prime\prime})$ & (K) & (K) & (cm$^{-2}$) & (cm$^{-2}$) &  & (km\,s$^{-1}$) & (km\,s$^{-1}$) \\\hline \\[-0.3cm] 
CH$_3$OH, $v=0$ & $2.0$ & $173$ & $168\pm 3$ & $2.2(19)$ & $(2.0\pm 0.2)(19)$ & $1.00$ & $4.2$ & $0.0$ \\ 
 $v=1$ & $2.0$ & $173$ & $168\pm 3$ & $2.2(19)$ & $(2.0\pm 0.2)(19)$ & $1.00$ & $4.2$ & $0.0$ \\ 
 $v=2$ & $2.0$ & $173$ & $168\pm 3$ & $2.2(19)$ & $(2.0\pm 0.2)(19)$ & $1.00$ & $4.2$ & $0.0$ \\ 
 $^{13}$CH$_3$OH, $v=0$ & $2.0$ & $170$ & $169\pm 3$ & $8.0(17)$ & $(8.2\pm 0.5)(17)$ & $1.00$ & $4.2$ & $0.3$ \\ 
 \hline \\[-0.35cm] 
C$_2$H$_5$OH, $v=0$ & $2.0$ & $170$ & $173\pm 2$ & $1.1(18)$ & $(1.15\pm 0.05)(18)$ & $1.46$ & $4.5$ & $0.0$ \\ 
 \hline \\[-0.35cm] 
CH$_3$OCH$_3$, $v=0$ & $2.0$ & $160$ & $154\pm 2$ & $1.0(18)$ & $(1.0\pm 0.1)(18)$ & $1.01$ & $3.5$ & $-0.2$ \\ 
 $v_{11}=1$ & $2.0$ & $160$ & $154\pm 2$ & $1.0(18)$ & $(1.0\pm 0.1)(18)$ & $1.01$ & $3.5$ & $-0.2$ \\ 
 $v_{15}=1$ & $2.0$ & $160$ & $154\pm 2$ & $1.0(18)$ & $(1.0\pm 0.1)(18)$ & $1.01$ & $3.5$ & $-0.2$ \\ 
 \hline \\[-0.35cm] 
CH$_3$OCHO, $v=0$ & $2.0$ & $200$ & $206\pm 9$ & $1.5(18)$ & $(1.5\pm 0.1)(18)$ & $1.50$ & $4.5$ & $-0.3$ \\ 
 $v=1$ & $2.0$ & $200$ & $206\pm 9$ & $1.5(18)$ & $(1.5\pm 0.1)(18)$ & $1.50$ & $4.5$ & $-0.3$ \\ 
 \hline \\[-0.35cm] 
CH$_3$CHO, $v=0$ & $2.0$ & $150$ & $168\pm 7$ & $2.0(17)$ & $(1.0\pm 0.1)(17)$ & $1.02$ & $4.5$ & $0.0$ \\ 
 $v=1$ & $2.0$ & $150$ & $168\pm 7$ & $2.0(17)$ & $(1.0\pm 0.1)(17)$ & $1.02$ & $4.5$ & $0.0$ \\ 
 $v=2$ & $2.0$ & $150$ & $168\pm 7$ & $2.0(17)$ & $(1.0\pm 0.1)(17)$ & $1.02$ & $4.5$ & $0.0$ \\ 
 \hline \\[-0.35cm] 
CH$_3$NCO, $v=0$ & $2.0$ & $110$ & $106\pm 7$ & $4.0(16)$ & $(4.0\pm 0.7)(16)$ & $1.00$ & $4.5$ & $0.3$ \\ 
 \hline \\[-0.35cm] 
C$_2$H$_5$CN, $v=0$ & $2.0$ & $170$ & $169\pm 1$ & $1.5(18)$ & $(1.53\pm 0.04)(18)$ & $1.53$ & $5.0$ & $0.7$ \\ 
 $v_{12}=1$ & $2.0$ & $170$ & $169\pm 1$ & $1.5(18)$ & $(1.53\pm 0.04)(18)$ & $1.53$ & $5.0$ & $0.7$ \\ 
 $v_{13}=2$ & $2.0$ & $170$ & $169\pm 1$ & $1.5(18)$ & $(1.53\pm 0.04)(18)$ & $1.53$ & $5.0$ & $0.7$ \\ 
 $v_{20}=1$ & $2.0$ & $170$ & $169\pm 1$ & $1.5(18)$ & $(1.53\pm 0.04)(18)$ & $1.53$ & $5.0$ & $0.7$ \\ 
 $v_{21}=2$ & $2.0$ & $170$ & $169\pm 1$ & $1.5(18)$ & $(1.53\pm 0.04)(18)$ & $1.53$ & $5.0$ & $0.7$ \\ 
 \hline \\[-0.35cm] 
C$_2$H$_3$CN, $v=0$ & $2.0$ & $175$ & $182\pm 6$ & $5.1(16)$ & $(5.5\pm 3.6)(16)$ & $1.02$ & $4.5$ & $0.5$ \\ 
 $v_{11}=1$ & $2.0$ & $175$ & $182\pm 6$ & $5.1(16)$ & $(5.5\pm 3.6)(16)$ & $1.02$ & $4.5$ & $0.5$ \\ 
 \hline \\[-0.35cm] 
NH$_2$CHO, $v=0$ & $2.0$ & $110$ & $122\pm 3$ & $1.6(18)$ & $(1.2\pm 0.1)(18)$ & $1.04$ & $5.3$ & $0.1$ \\ 
 $v_{12}=1$ & $2.0$ & $110$ & $122\pm 3$ & $1.6(18)$ & $(1.2\pm 0.1)(18)$ & $1.04$ & $5.3$ & $0.1$ \\ 
 \hline\hline
\end{tabular}
\label{tab:n1s1}
\end{table*}

\begin{table*}[h!]
\caption{Same as Table\,\ref{tab:n1s-1}, but for position N1S2.} 
\centering
\begin{tabular}{rrrrrrrrrrr}
\hline\hline \\[-0.3cm] 
Molecule & Size & $T_{\rm rot,W}$ & $T_{\rm rot,obs}$ & $N_{\rm W}$ & $N_{\rm obs}$ & $C_{\rm vib}$ & $\Delta\varv$ & $\varv_{\rm off}$ \\ 
 & $(^{\prime\prime})$ & (K) & (K) & (cm$^{-2}$) & (cm$^{-2}$) &  & (km\,s$^{-1}$) & (km\,s$^{-1}$) \\\hline \\[-0.3cm] 
CH$_3$OH, $v=0$ & $2.0$ & $140$ & $140\pm 1$ & $1.5(19)$ & $(1.6\pm 0.1)(19)$ & $1.00$ & $4.0$ & $-0.5$ \\ 
 $v=1$ & $2.0$ & $140$ & $140\pm 1$ & $1.5(19)$ & $(1.6\pm 0.1)(19)$ & $1.00$ & $4.0$ & $-0.5$ \\ 
 $v=2$ & $2.0$ & $140$ & $140\pm 1$ & $1.5(19)$ & $(1.6\pm 0.1)(19)$ & $1.00$ & $4.0$ & $-0.5$ \\ 
 $^{13}$CH$_3$OH, $v=0$ & $2.0$ & $140$ & $137\pm 3$ & $5.5(17)$ & $(5.5\pm 0.4)(17)$ & $1.00$ & $4.5$ & $0.0$ \\ 
 \hline \\[-0.35cm] 
C$_2$H$_5$OH, $v=0$ & $2.0$ & $130$ & $130\pm 1$ & $5.7(17)$ & $(6.5\pm 0.2)(17)$ & $1.21$ & $4.5$ & $-0.8$ \\ 
 \hline \\[-0.35cm] 
CH$_3$OCH$_3$, $v=0$ & $2.0$ & $130$ & $131\pm 1$ & $9.5(17)$ & $(9.4\pm 0.4)(17)$ & $1.01$ & $4.0$ & $-0.3$ \\ 
 $v_{11}=1$ & $2.0$ & $130$ & $131\pm 1$ & $9.5(17)$ & $(9.4\pm 0.4)(17)$ & $1.01$ & $4.0$ & $-0.3$ \\ 
 \hline \\[-0.35cm] 
CH$_3$OCHO, $v=0$ & $2.0$ & $170$ & $167\pm 5$ & $8.9(17)$ & $(9.4\pm 0.6)(17)$ & $1.27$ & $4.5$ & $-0.8$ \\ 
 $v=1$ & $2.0$ & $170$ & $167\pm 5$ & $8.9(17)$ & $(9.4\pm 0.6)(17)$ & $1.27$ & $4.5$ & $-0.8$ \\ 
 \hline \\[-0.35cm] 
CH$_3$CHO, $v=0$ & $2.0$ & $120$ & $118\pm 4$ & $5.0(16)$ & $(2.5\pm 0.2)(16)$ & $1.00$ & $4.5$ & $-0.4$ \\ 
 $v=1$ & $2.0$ & $120$ & $118\pm 4$ & $5.0(16)$ & $(2.5\pm 0.2)(16)$ & $1.00$ & $4.5$ & $-0.4$ \\ 
 \hline \\[-0.35cm] 
CH$_3$NCO, $v=0$ & $2.0$ & $110$ & $107\pm 12$ & $1.5(16)$ & $(1.4\pm 0.4)(16)$ & $1.00$ & $4.5$ & $0.3$ \\ 
 \hline \\[-0.35cm] 
C$_2$H$_5$CN, $v=0$ & $2.0$ & $140$ & $136\pm 1$ & $5.2(17)$ & $(5.2\pm 0.2)(17)$ & $1.29$ & $5.5$ & $1.0$ \\ 
 $v_{13}=2$ & $2.0$ & $140$ & $136\pm 1$ & $5.2(17)$ & $(5.2\pm 0.2)(17)$ & $1.29$ & $5.5$ & $1.0$ \\ 
 $v_{20}=1$ & $2.0$ & $140$ & $136\pm 1$ & $5.2(17)$ & $(5.2\pm 0.2)(17)$ & $1.29$ & $5.5$ & $1.0$ \\ 
 $v_{21}=2$ & $2.0$ & $140$ & $136\pm 1$ & $5.2(17)$ & $(5.2\pm 0.2)(17)$ & $1.29$ & $5.5$ & $1.0$ \\ 
 \hline \\[-0.35cm] 
C$_2$H$_3$CN, $v=0$ & $2.0$ & $140$ & $129\pm 12$ & $5.0(15)$ & $(4.7\pm 0.9)(15)$ & $1.00$ & $4.0$ & $0.4$ \\ 
 \hline \\[-0.35cm] 
NH$_2$CHO, $v=0$ & $2.0$ & $140$ & $147\pm 10$ & $5.3(16)$ & $(4.4\pm 0.9)(16)$ & $1.07$ & $3.5$ & $0.0$ \\ 
 $v_{12}=1$ & $2.0$ & $140$ & $147\pm 10$ & $5.3(16)$ & $(4.4\pm 0.9)(16)$ & $1.07$ & $3.5$ & $0.0$ \\ 
 \hline\hline
\end{tabular}
\label{tab:n1s2}
\end{table*}

\begin{table*}[h!]
\caption{Same as Table\,\ref{tab:n1s-1}, but for position N1S3.} 
\centering
\begin{tabular}{rrrrrrrrrrr}
\hline\hline \\[-0.3cm] 
Molecule & Size & $T_{\rm rot,W}$ & $T_{\rm rot,obs}$ & $N_{\rm W}$ & $N_{\rm obs}$ & $C_{\rm vib}$ & $\Delta\varv$ & $\varv_{\rm off}$ \\ 
 & $(^{\prime\prime})$ & (K) & (K) & (cm$^{-2}$) & (cm$^{-2}$) &  & (km\,s$^{-1}$) & (km\,s$^{-1}$) \\\hline \\[-0.3cm] 
CH$_3$OH, $v=0$ & $2.0$ & $120$ & $122\pm 1$ & $1.5(19)$ & $(1.4\pm 0.1)(19)$ & $1.00$ & $4.0$ & $-1.6$ \\ 
 $v=1$ & $2.0$ & $120$ & $122\pm 1$ & $1.5(19)$ & $(1.4\pm 0.1)(19)$ & $1.00$ & $4.0$ & $-1.6$ \\ 
 $v=2$ & $2.0$ & $120$ & $122\pm 1$ & $1.5(19)$ & $(1.4\pm 0.1)(19)$ & $1.00$ & $4.0$ & $-1.6$ \\ 
 $^{13}$CH$_3$OH, $v=0$ & $2.0$ & $120$ & $121\pm 2$ & $3.0(17)$ & $(3.5\pm 0.2)(17)$ & $1.00$ & $4.5$ & $-1.0$ \\ 
 \hline \\[-0.35cm] 
C$_2$H$_5$OH, $v=0$ & $2.0$ & $115$ & $119\pm 2$ & $2.9(17)$ & $(3.0\pm 0.1)(17)$ & $1.16$ & $5.5$ & $-1.8$ \\ 
 \hline \\[-0.35cm] 
CH$_3$OCH$_3$, $v=0$ & $2.0$ & $110$ & $110\pm 1$ & $5.5(17)$ & $(5.9\pm 0.2)(17)$ & $1.00$ & $3.8$ & $-1.4$ \\ 
 $v_{11}=1$ & $2.0$ & $110$ & $110\pm 1$ & $5.5(17)$ & $(5.9\pm 0.2)(17)$ & $1.00$ & $3.8$ & $-1.4$ \\ 
 \hline \\[-0.35cm] 
CH$_3$OCHO, $v=0$ & $2.0$ & $140$ & $137\pm 2$ & $6.3(17)$ & $(6.5\pm 0.3)(17)$ & $1.14$ & $4.5$ & $-2.0$ \\ 
 $v=1$ & $2.0$ & $140$ & $137\pm 2$ & $6.3(17)$ & $(6.5\pm 0.3)(17)$ & $1.14$ & $4.5$ & $-2.0$ \\ 
 \hline \\[-0.35cm] 
CH$_3$CHO, $v=0$ & $2.0$ & $115$ & $108\pm 5$ & $2.0(16)$ & $(9.3\pm 1.0)(15)$ & $1.00$ & $4.0$ & $-1.7$ \\ 
 $v=1$ & $2.0$ & $115$ & $108\pm 5$ & $2.0(16)$ & $(9.3\pm 1.0)(15)$ & $1.00$ & $4.0$ & $-1.7$ \\ 
 \hline \\[-0.35cm] 
CH$_3$NCO, $v=0$ & $2.0$ & $90$ & $90\pm 8$ & $2.7(15)$ & $(3.1\pm 0.8)(15)$ & $1.00$ & $3.5$ & $0.0$ \\ 
 \hline \\[-0.35cm] 
C$_2$H$_5$CN, $v=0$ & $2.0$ & $115$ & $119\pm 2$ & $1.4(17)$ & $(1.3\pm 0.1)(17)$ & $1.20$ & $5.5$ & $-1.0$ \\ 
 $v_{20}=1$ & $2.0$ & $115$ & $119\pm 2$ & $1.4(17)$ & $(1.3\pm 0.1)(17)$ & $1.20$ & $5.5$ & $-1.0$ \\ 
 \hline \\[-0.35cm] 
C$_2$H$_3$CN, $v=0$ & $2.0$ & $40$ & $35\pm 2$ & $6.0(14)$ & $(6.1\pm 1.0)(14)$ & $1.00$ & $4.0$ & $0.8$ \\ 
 \hline \\[-0.35cm] 
NH$_2$CHO, $v=0$ & $2.0$ & $112$ & $112\pm 16$ & $1.0(15)$ & $(1.0\pm 0.05)(15)$ & $1.03$ & $3.5$ & $-1.0$ \\ 
 \hline\hline
\end{tabular}
\label{tab:n1s3}
\end{table*}

\begin{table*}[h!]
\caption{Same as Table\,\ref{tab:n1s-1}, but for position N1S4.} 
\centering
\begin{tabular}{rrrrrrrrrrr}
\hline\hline \\[-0.3cm] 
Molecule & Size & $T_{\rm rot,W}$ & $T_{\rm rot,obs}$ & $N_{\rm W}$ & $N_{\rm obs}$ & $C_{\rm vib}$ & $\Delta\varv$ & $\varv_{\rm off}$ \\ 
 & $(^{\prime\prime})$ & (K) & (K) & (cm$^{-2}$) & (cm$^{-2}$) &  & (km\,s$^{-1}$) & (km\,s$^{-1}$) \\\hline \\[-0.3cm] 
CH$_3$OH, $v=0$ & $2.0$ & $100$ & $104\pm 1$ & $6.0(18)$ & $(5.5\pm 0.4)(18)$ & $1.00$ & $4.0$ & $-2.0$ \\ 
 $v=1$ & $2.0$ & $100$ & $104\pm 1$ & $6.0(18)$ & $(5.5\pm 0.4)(18)$ & $1.00$ & $4.0$ & $-2.0$ \\ 
 $v=2$ & $2.0$ & $100$ & $104\pm 1$ & $6.0(18)$ & $(5.5\pm 0.4)(18)$ & $1.00$ & $4.0$ & $-2.0$ \\ 
 $^{13}$CH$_3$OH, $v=0$ & $2.0$ & $100$ & $102\pm 2$ & $1.2(17)$ & $(1.4\pm 0.1)(17)$ & $1.00$ & $4.2$ & $-1.5$ \\ 
 \hline \\[-0.35cm] 
C$_2$H$_5$OH, $v=0$ & $2.0$ & $100$ & $103\pm 2$ & $2.0(16)$ & $(2.3\pm 0.2)(16)$ & $1.10$ & $4.0$ & $-3.0$ \\ 
 \hline \\[-0.35cm] 
CH$_3$OCH$_3$, $v=0$ & $2.0$ & $100$ & $98\pm 4$ & $1.6(17)$ & $(1.6\pm 0.2)(17)$ & $1.00$ & $3.5$ & $-1.0$ \\ 
 \hline \\[-0.35cm] 
CH$_3$OCHO, $v=0$ & $2.0$ & $115$ & $111\pm 2$ & $2.8(17)$ & $(3.0\pm 0.1)(17)$ & $1.07$ & $3.5$ & $-1.8$ \\ 
 $v=1$ & $2.0$ & $115$ & $111\pm 2$ & $2.8(17)$ & $(3.0\pm 0.1)(17)$ & $1.07$ & $3.5$ & $-1.8$ \\ 
 \hline \\[-0.35cm] 
CH$_3$CHO, $v=0$ & $2.0$ & $80$ & $112\pm 56$ & $1.5(15)$ & $(1.51\pm 1.47)(15)$ & $1.00$ & $4.0$ & $-2.5$ \\ 
 \hline \\[-0.35cm] 
CH$_3$NCO, $v=0$ & $2.0$ & $89$ & $89\pm 6$ & $\leq8.0(14)$ &  & $1.00$ & $4.1$ & $-2.0$ \\ 
 \hline \\[-0.35cm] 
C$_2$H$_5$CN, $v=0$ & $2.0$ & $100$ & $100\pm 5$ & $2.1(16)$ & $(2.2\pm 0.2)(16)$ & $1.11$ & $5.0$ & $-2.0$ \\ 
 \hline \\[-0.35cm] 
C$_2$H$_3$CN, $v=0$ & $2.0$ & $117$ & $117\pm 12$ & $\leq3.0(14)$ &  & $1.00$ & $4.1$ & $-2.0$ \\ 
 \hline \\[-0.35cm] 
NH$_2$CHO, $v=0$ & $2.0$ & $106$ & $106\pm 17$ & $8.2(14)$ & $(8.2\pm 1.2)(14)$ & $1.02$ & $4.5$ & $-1.5$ \\ 
 \hline\hline
\end{tabular}
\label{tab:n1s4}
\end{table*}

\begin{table*}[h!]
\caption{Same as Table\,\ref{tab:n1s-1}, but for position N1S5.} 
\centering
\begin{tabular}{rrrrrrrrrrr}
\hline\hline \\[-0.3cm] 
Molecule & Size & $T_{\rm rot,W}$ & $T_{\rm rot,obs}$ & $N_{\rm W}$ & $N_{\rm obs}$ & $C_{\rm vib}$ & $\Delta\varv$ & $\varv_{\rm off}$ \\ 
 & $(^{\prime\prime})$ & (K) & (K) & (cm$^{-2}$) & (cm$^{-2}$) &  & (km\,s$^{-1}$) & (km\,s$^{-1}$) \\\hline \\[-0.3cm] 
CH$_3$OH, $v=0$ & $2.0$ & $110$ & $105\pm 4$ & $4.5(17)$ & $(3.7\pm 0.5)(17)$ & $1.00$ & $4.3$ & $-3.0$ \\ 
 $v=1$ & $2.0$ & $110$ & $105\pm 4$ & $4.5(17)$ & $(3.7\pm 0.5)(17)$ & $1.00$ & $4.3$ & $-3.0$ \\ 
 $^{13}$CH$_3$OH, $v=0$ & $2.0$ & $90$ & $94\pm 8$ & $2.2(16)$ & $(2.7\pm 0.5)(16)$ & $1.00$ & $4.0$ & $-2.5$ \\ 
 \hline \\[-0.35cm] 
C$_2$H$_5$OH, $v=0$ & $2.0$ & $89$ & $89\pm 9$ & $\leq6.4(15)$ &  & $1.06$ & $4.1$ & $-2.6$ \\ 
 \hline \\[-0.35cm] 
CH$_3$OCH$_3$, $v=0$ & $2.0$ & $85$ & $84\pm 5$ & $3.0(16)$ & $(2.5\pm 0.3)(16)$ & $1.00$ & $3.5$ & $-2.5$ \\ 
 \hline \\[-0.35cm] 
CH$_3$OCHO, $v=0$ & $2.0$ & $100$ & $95\pm 1$ & $1.1(17)$ & $(1.14\pm 0.04)(17)$ & $1.03$ & $3.5$ & $-2.3$ \\ 
 $v=1$ & $2.0$ & $100$ & $95\pm 1$ & $1.1(17)$ & $(1.14\pm 0.04)(17)$ & $1.03$ & $3.5$ & $-2.3$ \\ 
 \hline \\[-0.35cm] 
CH$_3$CHO, $v=0$ & $2.0$ & $86$ & $86\pm 7$ & $\leq8.0(14)$ &  & $1.00$ & $3.9$ & $-2.6$ \\ 
 \hline \\[-0.35cm] 
CH$_3$NCO, $v=0$ & $2.0$ & $85$ & $85\pm 7$ & $\leq6.2(14)$ &  & $1.00$ & $3.9$ & $-2.6$ \\ 
 \hline \\[-0.35cm] 
C$_2$H$_5$CN, $v=0$ & $2.0$ & $98$ & $88\pm 12$ & $2.1(15)$ & $(2.0\pm 0.4)(15)$ & $1.07$ & $4.0$ & $-2.6$ \\ 
 \hline \\[-0.35cm] 
C$_2$H$_3$CN, $v=0$ & $2.0$ & $107$ & $107\pm 12$ & $\leq3.0(14)$ &  & $1.00$ & $3.9$ & $-2.6$ \\ 
 \hline \\[-0.35cm] 
NH$_2$CHO, $v=0$ & $2.0$ & $102$ & $102\pm 19$ & $5.1(14)$ & $(5.1\pm 0.4)(14)$ & $1.02$ & $4.0$ & $-2.0$ \\ 
 \hline\hline
\end{tabular}
\label{tab:n1s5}
\end{table*}

\begin{table*}[h!]
\caption{Same as Table\,\ref{tab:n1s-1}, but for position N1S6.} 
\centering
\begin{tabular}{rrrrrrrrrrr}
\hline\hline \\[-0.3cm] 
Molecule & Size & $T_{\rm rot,W}$ & $T_{\rm rot,obs}$ & $N_{\rm W}$ & $N_{\rm obs}$ & $C_{\rm vib}$ & $\Delta\varv$ & $\varv_{\rm off}$ \\ 
 & $(^{\prime\prime})$ & (K) & (K) & (cm$^{-2}$) & (cm$^{-2}$) &  & (km\,s$^{-1}$) & (km\,s$^{-1}$) \\\hline \\[-0.3cm] 
CH$_3$OH, $v=0$ & $2.0$ & $100$ & $104\pm 4$ & $2.0(17)$ & $(1.8\pm 0.2)(17)$ & $1.00$ & $3.0$ & $-2.8$ \\ 
 $v=1$ & $2.0$ & $100$ & $104\pm 4$ & $2.0(17)$ & $(1.8\pm 0.2)(17)$ & $1.00$ & $3.0$ & $-2.8$ \\ 
 $^{13}$CH$_3$OH, $v=0$ & $2.0$ & $70$ & $54\pm 6$ & $8.0(15)$ & $(5.8\pm 1.2)(15)$ & $1.00$ & $4.0$ & $-2.5$ \\ 
 \hline \\[-0.35cm] 
C$_2$H$_5$OH, $v=0$ & $2.0$ & $85$ & $85\pm 9$ & $\leq6.3(15)$ &  & $1.05$ & $3.9$ & $-2.8$ \\ 
 \hline \\[-0.35cm] 
CH$_3$OCH$_3$, $v=0$ & $2.0$ & $85$ & $86\pm 6$ & $2.5(16)$ & $(2.7\pm 0.4)(16)$ & $1.00$ & $3.5$ & $-2.5$ \\ 
 \hline \\[-0.35cm] 
CH$_3$OCHO, $v=0$ & $2.0$ & $100$ & $92\pm 2$ & $9.3(16)$ & $(1.0\pm 0.5)(17)$ & $1.03$ & $3.0$ & $-2.5$ \\ 
 $v=1$ & $2.0$ & $100$ & $92\pm 2$ & $9.3(16)$ & $(1.0\pm 0.5)(17)$ & $1.03$ & $3.0$ & $-2.5$ \\ 
 \hline \\[-0.35cm] 
CH$_3$CHO, $v=0$ & $2.0$ & $83$ & $83\pm 7$ & $\leq7.0(14)$ &  & $1.00$ & $3.3$ & $-2.8$ \\ 
 \hline \\[-0.35cm] 
CH$_3$NCO, $v=0$ & $2.0$ & $84$ & $84\pm 7$ & $\leq5.0(14)$ &  & $1.00$ & $3.3$ & $-2.8$ \\ 
 \hline \\[-0.35cm] 
C$_2$H$_5$CN, $v=0$ & $2.0$ & $90$ & $83\pm 23$ & $8.5(14)$ & $(8.8\pm 3.8)(14)$ & $1.06$ & $3.0$ & $-3.5$ \\ 
 \hline \\[-0.35cm] 
C$_2$H$_3$CN, $v=0$ & $2.0$ & $104$ & $104\pm 13$ & $\leq2.7(14)$ &  & $1.00$ & $3.3$ & $-2.8$ \\ 
 \hline \\[-0.35cm] 
NH$_2$CHO, $v=0$ & $2.0$ & $100$ & $100\pm 19$ & $\leq2.5(14)$ &  & $1.02$ & $3.3$ & $-2.8$ \\ 
 \hline\hline
\end{tabular}
\label{tab:n1s6}
\end{table*}

\begin{table*}[h!]
\caption{Same as Table\,\ref{tab:n1s-1}, but for position N1S7.} 
\centering
\begin{tabular}{rrrrrrrrrrr}
\hline\hline \\[-0.3cm] 
Molecule & Size & $T_{\rm rot,W}$ & $T_{\rm rot,obs}$ & $N_{\rm W}$ & $N_{\rm obs}$ & $C_{\rm vib}$ & $\Delta\varv$ & $\varv_{\rm off}$ \\ 
 & $(^{\prime\prime})$ & (K) & (K) & (cm$^{-2}$) & (cm$^{-2}$) &  & (km\,s$^{-1}$) & (km\,s$^{-1}$) \\\hline \\[-0.3cm] 
CH$_3$OH, $v=0$ & $2.0$ & $100$ & $103\pm 8$ & $8.0(16)$ & $(6.4\pm 1.6)(16)$ & $1.00$ & $2.5$ & $-3.2$ \\ 
 $v=1$ & $2.0$ & $100$ & $103\pm 8$ & $8.0(16)$ & $(6.4\pm 1.6)(16)$ & $1.00$ & $2.5$ & $-3.2$ \\ 
 $^{13}$CH$_3$OH, $v=0$ & $2.0$ & $87$ & $87\pm 5$ & $\leq2.0(15)$ &  & $1.00$ & $2.3$ & $-3.1$ \\ 
 \hline \\[-0.35cm] 
C$_2$H$_5$OH, $v=0$ & $2.0$ & $81$ & $81\pm 9$ & $\leq4.2(15)$ &  & $1.04$ & $2.3$ & $-3.1$ \\ 
 \hline \\[-0.35cm] 
CH$_3$OCH$_3$, $v=0$ & $2.0$ & $90$ & $90\pm 8$ & $2.3(16)$ & $(2.7\pm 0.5)(16)$ & $1.00$ & $2.3$ & $-2.5$ \\ 
 \hline \\[-0.35cm] 
CH$_3$OCHO, $v=0$ & $2.0$ & $100$ & $102\pm 3$ & $7.3(16)$ & $(8.4\pm 0.7)(16)$ & $1.05$ & $2.5$ & $-2.8$ \\ 
 $v=1$ & $2.0$ & $100$ & $102\pm 3$ & $7.3(16)$ & $(8.4\pm 0.7)(16)$ & $1.05$ & $2.5$ & $-2.8$ \\ 
 \hline \\[-0.35cm] 
CH$_3$CHO, $v=0$ & $2.0$ & $79$ & $79\pm 7$ & $\leq4.5(14)$ &  & $1.00$ & $2.3$ & $-3.1$ \\ 
 \hline \\[-0.35cm] 
CH$_3$NCO, $v=0$ & $2.0$ & $82$ & $82\pm 7$ & $\leq3.5(14)$ &  & $1.00$ & $2.3$ & $-3.1$ \\ 
 \hline \\[-0.35cm] 
C$_2$H$_5$CN, $v=0$ & $2.0$ & $60$ & $57\pm 12$ & $2.8(14)$ & $(2.6\pm 0.9)(14)$ & $1.01$ & $2.0$ & $-4.0$ \\ 
 \hline \\[-0.35cm] 
C$_2$H$_3$CN, $v=0$ & $2.0$ & $99$ & $99\pm 13$ & $\leq2.0(14)$ &  & $1.00$ & $2.3$ & $-3.1$ \\ 
 \hline \\[-0.35cm] 
NH$_2$CHO, $v=0$ & $2.0$ & $98$ & $98\pm 20$ & $\leq2.0(14)$ &  & $1.01$ & $2.3$ & $-3.1$ \\ 
 \hline\hline
\end{tabular}
\label{tab:n1s7}
\end{table*}

\begin{table*}[h!]
\caption{Same as Table\,\ref{tab:n1s-1}, but for position N1S8.} 
\centering
\begin{tabular}{rrrrrrrrrrr}
\hline\hline \\[-0.3cm] 
Molecule & Size & $T_{\rm rot,W}$ & $T_{\rm rot,obs}$ & $N_{\rm W}$ & $N_{\rm obs}$ & $C_{\rm vib}$ & $\Delta\varv$ & $\varv_{\rm off}$ \\ 
 & $(^{\prime\prime})$ & (K) & (K) & (cm$^{-2}$) & (cm$^{-2}$) &  & (km\,s$^{-1}$) & (km\,s$^{-1}$) \\\hline \\[-0.3cm] 
CH$_3$OH, $v=0$ & $2.0$ & $81$ & $81\pm 6$ & $8.0(15)$ & $(6.2\pm 1.1)(15)$ & $1.00$ & $2.5$ & $-3.5$ \\ 
 $^{13}$CH$_3$OH, $v=0$ & $2.0$ & $80$ & $80\pm 5$ & $\leq2.0(15)$ &  & $1.00$ & $2.3$ & $-3.3$ \\ 
 \hline \\[-0.35cm] 
C$_2$H$_5$OH, $v=0$ & $2.0$ & $74$ & $74\pm 9$ & $\leq4.1(15)$ &  & $1.03$ & $2.3$ & $-3.3$ \\ 
 \hline \\[-0.35cm] 
CH$_3$OCH$_3$, $v=0$ & $2.0$ & $90$ & $98\pm 21$ & $1.0(16)$ & $(1.1\pm 0.5)(16)$ & $1.00$ & $2.0$ & $-3.5$ \\ 
 \hline \\[-0.35cm] 
CH$_3$OCHO, $v=0$ & $2.0$ & $40$ & $36\pm 8$ & $3.0(15)$ & $(2.8\pm 1.1)(15)$ & $1.00$ & $2.5$ & $-3.0$ \\ 
 \hline \\[-0.35cm] 
CH$_3$CHO, $v=0$ & $2.0$ & $72$ & $72\pm 7$ & $\leq3.7(14)$ &  & $1.00$ & $2.3$ & $-3.3$ \\ 
 \hline \\[-0.35cm] 
CH$_3$NCO, $v=0$ & $2.0$ & $79$ & $79\pm 7$ & $\leq3.3(14)$ &  & $1.00$ & $2.3$ & $-3.3$ \\ 
 \hline \\[-0.35cm] 
C$_2$H$_5$CN, $v=0$ & $2.0$ & $92$ & $92\pm 12$ & $\leq2.2(14)$ &  & $1.08$ & $2.3$ & $-3.3$ \\ 
 \hline \\[-0.35cm] 
C$_2$H$_3$CN, $v=0$ & $2.0$ & $92$ & $92\pm 13$ & $\leq1.5(14)$ &  & $1.00$ & $2.3$ & $-3.3$ \\ 
 \hline \\[-0.35cm] 
NH$_2$CHO, $v=0$ & $2.0$ & $94$ & $94\pm 20$ & $\leq2.0(14)$ &  & $1.01$ & $2.3$ & $-3.3$ \\ 
 \hline\hline
\end{tabular}
\label{tab:n1s8}
\end{table*}

\begin{table*}[h!]
\caption{Same as Table\,\ref{tab:n1s-1}, but for position N1S9.} 
\centering
\begin{tabular}{rrrrrrrrrrr}
\hline\hline \\[-0.3cm] 
Molecule & Size & $T_{\rm rot,W}$ & $T_{\rm rot,obs}$ & $N_{\rm W}$ & $N_{\rm obs}$ & $C_{\rm vib}$ & $\Delta\varv$ & $\varv_{\rm off}$ \\ 
 & $(^{\prime\prime})$ & (K) & (K) & (cm$^{-2}$) & (cm$^{-2}$) &  & (km\,s$^{-1}$) & (km\,s$^{-1}$) \\\hline \\[-0.3cm] 
CH$_3$OH, $v=0$ & $2.0$ & $76$ & $76\pm 6$ & $\leq3.3(15)$ &  & $1.00$ & $2.8$ & $-3.5$ \\ 
 $^{13}$CH$_3$OH, $v=0$ & $2.0$ & $75$ & $75\pm 5$ & $\leq1.5(15)$ &  & $1.00$ & $2.0$ & $-3.5$ \\ 
 \hline \\[-0.35cm] 
CH$_3$OCH$_3$, $v=0$ & $2.0$ & $69$ & $69\pm 6$ & $3.5(15)$ & $(3.6\pm 0.6)(15)$ & $1.00$ & $2.0$ & $-3.5$ \\ 
 \hline \\[-0.35cm] 
CH$_3$OCHO, $v=0$ & $2.0$ & $75$ & $75\pm 11$ & $\leq2.0(15)$ &  & $1.01$ & $2.0$ & $-3.5$ \\ 
 \hline \\[-0.35cm] 
C$_2$H$_5$CN, $v=0$ & $2.0$ & $88$ & $88\pm 12$ & $\leq1.6(14)$ &  & $1.07$ & $2.0$ & $-3.5$ \\ 
 \hline\hline
\end{tabular}
\label{tab:n1s9}
\end{table*}

\begin{table*}[h!]
\caption{Same as Table\,\ref{tab:n1s-1}, but for position N1S10.} 
\centering
\begin{tabular}{rrrrrrrrrrr}
\hline\hline \\[-0.3cm] 
Molecule & Size & $T_{\rm rot,W}$ & $T_{\rm rot,obs}$ & $N_{\rm W}$ & $N_{\rm obs}$ & $C_{\rm vib}$ & $\Delta\varv$ & $\varv_{\rm off}$ \\ 
 & $(^{\prime\prime})$ & (K) & (K) & (cm$^{-2}$) & (cm$^{-2}$) &  & (km\,s$^{-1}$) & (km\,s$^{-1}$) \\\hline \\[-0.3cm] 
CH$_3$OH, $v=0$ & $2.0$ & $71$ & $71\pm 6$ & $\leq2.5(15)$ &  & $1.00$ & $2.3$ & $-3.5$ \\ 
 $^{13}$CH$_3$OH, $v=0$ & $2.0$ & $70$ & $70\pm 5$ & $\leq1.5(15)$ &  & $1.00$ & $2.0$ & $-3.5$ \\ 
 \hline \\[-0.35cm] 
CH$_3$OCH$_3$, $v=0$ & $2.0$ & $65$ & $65\pm 6$ & $\leq3.0(15)$ &  & $1.00$ & $2.0$ & $-3.5$ \\ 
 \hline \\[-0.35cm] 
CH$_3$OCHO, $v=0$ & $2.0$ & $70$ & $70\pm 11$ & $\leq1.5(15)$ &  & $1.01$ & $2.0$ & $-3.5$ \\ 
 \hline \\[-0.35cm] 
C$_2$H$_5$CN, $v=0$ & $2.0$ & $83$ & $83\pm 12$ & $\leq1.1(14)$ &  & $1.06$ & $2.0$ & $-3.5$ \\ 
 \hline\hline
\end{tabular}
\label{tab:n1s10}
\end{table*}

\begin{table*}[h!]
\caption{Same as Table\,\ref{tab:n1s-1}, but for position N1W-1.} 
\centering
\begin{tabular}{rrrrrrrrrrr}
\hline\hline \\[-0.3cm] 
Molecule & Size & $T_{\rm rot,W}$ & $T_{\rm rot,obs}$ & $N_{\rm W}$ & $N_{\rm obs}$ & $C_{\rm vib}$ & $\Delta\varv$ & $\varv_{\rm off}$ \\ 
 & $(^{\prime\prime})$ & (K) & (K) & (cm$^{-2}$) & (cm$^{-2}$) &  & (km\,s$^{-1}$) & (km\,s$^{-1}$) \\\hline \\[-0.3cm] 
CH$_3$OH, $v=0$ & $2.0$ & $240$ & $231\pm 6$ & $2.3(19)$ & $(1.9\pm 0.2)(19)$ & $1.00$ & $6.0$ & $3.2$ \\ 
 $v=1$ & $2.0$ & $240$ & $231\pm 6$ & $2.3(19)$ & $(1.9\pm 0.2)(19)$ & $1.00$ & $6.0$ & $3.2$ \\ 
 $v=2$ & $2.0$ & $240$ & $231\pm 6$ & $2.3(19)$ & $(1.9\pm 0.2)(19)$ & $1.00$ & $6.0$ & $3.2$ \\ 
 $^{13}$CH$_3$OH, $v=0$ & $2.0$ & $270$ & $268\pm 19$ & $9.0(17)$ & $(7.2\pm 1.2)(17)$ & $1.00$ & $6.5$ & $2.2$ \\ 
 \hline \\[-0.35cm] 
C$_2$H$_5$OH, $v=0$ & $2.0$ & $320$ & $319\pm 16$ & $1.4(18)$ & $(1.5\pm 0.1)(18)$ & $3.20$ & $5.0$ & $2.6$ \\ 
 \hline \\[-0.35cm] 
CH$_3$OCH$_3$, $v=0$ & $2.0$ & $315$ & $315\pm 20$ & $\leq2.7(17)$ &  & $1.35$ & $5.5$ & $3.5$ \\ 
 \hline \\[-0.35cm] 
CH$_3$OCHO, $v=0$ & $2.0$ & $220$ & $250\pm 31$ & $1.3(18)$ & $(1.1\pm 0.3)(18)$ & $1.85$ & $5.0$ & $3.0$ \\ 
 $v=1$ & $2.0$ & $220$ & $250\pm 31$ & $1.3(18)$ & $(1.1\pm 0.3)(18)$ & $1.85$ & $5.0$ & $3.0$ \\ 
 \hline \\[-0.35cm] 
CH$_3$CHO, $v=0$ & $2.0$ & $300$ & $293\pm 16$ & $6.9(17)$ & $(3.1\pm 0.4)(17)$ & $1.15$ & $5.0$ & $2.5$ \\ 
 $v=1$ & $2.0$ & $300$ & $293\pm 16$ & $6.9(17)$ & $(3.1\pm 0.4)(17)$ & $1.15$ & $5.0$ & $2.5$ \\ 
 $v=2$ & $2.0$ & $300$ & $293\pm 16$ & $6.9(17)$ & $(3.1\pm 0.4)(17)$ & $1.15$ & $5.0$ & $2.5$ \\ 
 \hline \\[-0.35cm] 
CH$_3$NCO, $v=0$ & $2.0$ & $170$ & $141\pm 18$ & $3.0(17)$ & $(1.7\pm 0.5)(17)$ & $1.00$ & $7.4$ & $4.0$ \\ 
 \hline \\[-0.35cm] 
C$_2$H$_5$CN, $v=0$ & $2.0$ & $230$ & $237\pm 5$ & $2.5(18)$ & $(2.1\pm 0.1)(18)$ & $2.30$ & $7.2$ & $2.8$ \\ 
 $v_{12}=1$ & $2.0$ & $230$ & $237\pm 5$ & $2.5(18)$ & $(2.1\pm 0.1)(18)$ & $2.30$ & $7.2$ & $2.8$ \\ 
 $v_{13}=2$ & $2.0$ & $230$ & $237\pm 5$ & $2.5(18)$ & $(2.1\pm 0.1)(18)$ & $2.30$ & $7.2$ & $2.8$ \\ 
 $v_{20}=1$ & $2.0$ & $230$ & $237\pm 5$ & $2.5(18)$ & $(2.1\pm 0.1)(18)$ & $2.30$ & $7.2$ & $2.8$ \\ 
 $v_{21}=2$ & $2.0$ & $230$ & $237\pm 5$ & $2.5(18)$ & $(2.1\pm 0.1)(18)$ & $2.30$ & $7.2$ & $2.8$ \\ 
 \hline \\[-0.35cm] 
C$_2$H$_3$CN, $v=0$ & $2.0$ & $250$ & $263\pm 6$ & $1.7(18)$ & $(1.8\pm 0.1)(18)$ & $1.11$ & $6.5$ & $3.5$ \\ 
 $v_{11}=1$ & $2.0$ & $250$ & $263\pm 6$ & $1.7(18)$ & $(1.8\pm 0.1)(18)$ & $1.11$ & $6.5$ & $3.5$ \\ 
 $v_{15}=1$ & $2.0$ & $250$ & $263\pm 6$ & $1.7(18)$ & $(1.8\pm 0.1)(18)$ & $1.11$ & $6.5$ & $3.5$ \\ 
 \hline \\[-0.35cm] 
NH$_2$CHO, $v=0$ & $2.0$ & $180$ & $188\pm 5$ & $2.3(18)$ & $(1.8\pm 0.1)(18)$ & $1.14$ & $5.7$ & $2.3$ \\ 
 $v_{12}=1$ & $2.0$ & $170$ & $188\pm 5$ & $2.3(18)$ & $(1.8\pm 0.1)(18)$ & $1.14$ & $5.7$ & $2.3$ \\ 
 \hline\hline
\end{tabular}
\label{tab:n1w-1}
\end{table*}

\begin{table*}[h!]
\caption{Same as Table\,\ref{tab:n1s-1}, but for position N1W.} 
\centering
\begin{tabular}{rrrrrrrrrrr}
\hline\hline \\[-0.3cm] 
Molecule & Size & $T_{\rm rot,W}$ & $T_{\rm rot,obs}$ & $N_{\rm W}$ & $N_{\rm obs}$ & $C_{\rm vib}$ & $\Delta\varv$ & $\varv_{\rm off}$ \\ 
 & $(^{\prime\prime})$ & (K) & (K) & (cm$^{-2}$) & (cm$^{-2}$) &  & (km\,s$^{-1}$) & (km\,s$^{-1}$) \\\hline \\[-0.3cm] 
CH$_3$OH, $v=0$ & $2.0$ & $180$ & $183\pm 5$ & $1.5(19)$ & $(1.4\pm 0.2)(19)$ & $1.00$ & $6.0$ & $3.0$ \\ 
 $v=1$ & $2.0$ & $180$ & $183\pm 5$ & $1.5(19)$ & $(1.4\pm 0.2)(19)$ & $1.00$ & $6.0$ & $3.0$ \\ 
 $v=2$ & $2.0$ & $180$ & $183\pm 5$ & $1.5(19)$ & $(1.4\pm 0.2)(19)$ & $1.00$ & $6.0$ & $3.0$ \\ 
 $^{13}$CH$_3$OH, $v=0$ & $2.0$ & $190$ & $193\pm 7$ & $7.0(17)$ & $(6.8\pm 0.7)(17)$ & $1.00$ & $6.0$ & $2.6$ \\ 
 \hline \\[-0.35cm] 
C$_2$H$_5$OH, $v=0$ & $2.0$ & $230$ & $218\pm 5$ & $6.0(17)$ & $(6.4\pm 0.4)(17)$ & $1.83$ & $5.0$ & $2.6$ \\ 
 \hline \\[-0.35cm] 
CH$_3$OCH$_3$, $v=0$ & $2.0$ & $175$ & $173\pm 8$ & $2.1(17)$ & $(1.8\pm 0.2)(17)$ & $1.03$ & $4.5$ & $3.0$ \\ 
 \hline \\[-0.35cm] 
CH$_3$OCHO, $v=0$ & $2.0$ & $180$ & $180\pm 12$ & $9.4(17)$ & $(8.3\pm 1.1)(17)$ & $1.34$ & $5.0$ & $3.5$ \\ 
 $v=1$ & $2.0$ & $180$ & $180\pm 12$ & $9.4(17)$ & $(8.3\pm 1.1)(17)$ & $1.34$ & $5.0$ & $3.5$ \\ 
 \hline \\[-0.35cm] 
CH$_3$CHO, $v=0$ & $2.0$ & $200$ & $197\pm 9$ & $2.0(17)$ & $(1.0\pm 0.1)(17)$ & $1.03$ & $5.0$ & $2.0$ \\ 
 $v=1$ & $2.0$ & $200$ & $197\pm 9$ & $2.0(17)$ & $(1.0\pm 0.1)(17)$ & $1.03$ & $5.0$ & $2.0$ \\ 
 \hline \\[-0.35cm] 
CH$_3$NCO, $v=0$ & $2.0$ & $120$ & $123\pm 18$ & $1.2(17)$ & $(1.1\pm 0.4)(17)$ & $1.00$ & $7.0$ & $3.0$ \\ 
 \hline \\[-0.35cm] 
C$_2$H$_5$CN, $v=0$ & $2.0$ & $200$ & $196\pm 2$ & $1.3(18)$ & $(1.38\pm 0.04)(18)$ & $1.79$ & $7.2$ & $1.7$ \\ 
 $v_{12}=1$ & $2.0$ & $200$ & $196\pm 2$ & $1.3(18)$ & $(1.38\pm 0.04)(18)$ & $1.79$ & $7.2$ & $1.7$ \\ 
 $v_{13}=2$ & $2.0$ & $200$ & $196\pm 2$ & $1.3(18)$ & $(1.38\pm 0.04)(18)$ & $1.79$ & $7.2$ & $1.7$ \\ 
 $v_{20}=1$ & $2.0$ & $200$ & $196\pm 2$ & $1.3(18)$ & $(1.38\pm 0.04)(18)$ & $1.79$ & $7.2$ & $1.7$ \\ 
 $v_{21}=2$ & $2.0$ & $200$ & $196\pm 2$ & $1.3(18)$ & $(1.38\pm 0.04)(18)$ & $1.79$ & $7.2$ & $1.7$ \\ 
 \hline \\[-0.35cm] 
C$_2$H$_3$CN, $v=0$ & $2.0$ & $170$ & $182\pm 6$ & $9.2(17)$ & $(8.2\pm 1.0)(17)$ & $1.02$ & $6.0$ & $2.0$ \\ 
 $v_{11}=1$ & $2.0$ & $170$ & $182\pm 6$ & $9.2(17)$ & $(8.2\pm 1.0)(17)$ & $1.02$ & $6.0$ & $2.0$ \\ 
 $v_{15}=1$ & $2.0$ & $170$ & $182\pm 6$ & $9.2(17)$ & $(8.2\pm 1.0)(17)$ & $1.02$ & $6.0$ & $2.0$ \\ 
 \hline \\[-0.35cm] 
NH$_2$CHO, $v=0$ & $2.0$ & $160$ & $165\pm 3$ & $1.4(18)$ & $(1.4\pm 0.1)(18)$ & $1.10$ & $5.7$ & $2.3$ \\ 
 $v_{12}=1$ & $2.0$ & $160$ & $165\pm 3$ & $1.4(18)$ & $(1.4\pm 0.1)(18)$ & $1.10$ & $5.7$ & $2.3$ \\ 
 \hline\hline
\end{tabular}
\label{tab:n1w}
\end{table*}

\begin{table*}[h!]
\caption{Same as Table\,\ref{tab:n1s-1}, but for position N1W1.} 
\centering
\begin{tabular}{rrrrrrrrrrr}
\hline\hline \\[-0.3cm] 
Molecule & Size & $T_{\rm rot,W}$ & $T_{\rm rot,obs}$ & $N_{\rm W}$ & $N_{\rm obs}$ & $C_{\rm vib}$ & $\Delta\varv$ & $\varv_{\rm off}$ \\ 
 & $(^{\prime\prime})$ & (K) & (K) & (cm$^{-2}$) & (cm$^{-2}$) &  & (km\,s$^{-1}$) & (km\,s$^{-1}$) \\\hline \\[-0.3cm] 
CH$_3$OH, $v=0$ & $2.0$ & $155$ & $151\pm 2$ & $1.5(19)$ & $(1.4\pm 0.1)(19)$ & $1.00$ & $5.5$ & $3.0$ \\ 
 $v=1$ & $2.0$ & $155$ & $151\pm 2$ & $1.5(19)$ & $(1.4\pm 0.1)(19)$ & $1.00$ & $5.5$ & $3.0$ \\ 
 $v=2$ & $2.0$ & $155$ & $151\pm 2$ & $1.5(19)$ & $(1.4\pm 0.1)(19)$ & $1.00$ & $5.5$ & $3.0$ \\ 
 $^{13}$CH$_3$OH, $v=0$ & $2.0$ & $160$ & $155\pm 4$ & $5.5(17)$ & $(5.2\pm 0.4)(17)$ & $1.00$ & $6.0$ & $3.0$ \\ 
 \hline \\[-0.35cm] 
C$_2$H$_5$OH, $v=0$ & $2.0$ & $170$ & $173\pm 4$ & $3.7(17)$ & $(4.2\pm 0.3)(17)$ & $1.46$ & $5.0$ & $3.0$ \\ 
 \hline \\[-0.35cm] 
CH$_3$OCH$_3$, $v=0$ & $2.0$ & $135$ & $130\pm 2$ & $5.0(17)$ & $(5.2\pm 0.3)(17)$ & $1.00$ & $5.0$ & $3.5$ \\ 
 $v_{11}=1$ & $2.0$ & $135$ & $130\pm 2$ & $5.0(17)$ & $(5.2\pm 0.3)(17)$ & $1.00$ & $5.0$ & $3.5$ \\ 
 \hline \\[-0.35cm] 
CH$_3$OCHO, $v=0$ & $2.0$ & $180$ & $166\pm 4$ & $1.1(18)$ & $(1.1\pm 0.1)(18)$ & $1.27$ & $5.0$ & $3.5$ \\ 
 $v=1$ & $2.0$ & $180$ & $166\pm 4$ & $1.1(18)$ & $(1.1\pm 0.1)(18)$ & $1.27$ & $5.0$ & $3.5$ \\ 
 \hline \\[-0.35cm] 
CH$_3$CHO, $v=0$ & $2.0$ & $170$ & $168\pm 9$ & $8.1(16)$ & $(4.0\pm 0.5)(16)$ & $1.02$ & $4.5$ & $2.0$ \\ 
 $v=1$ & $2.0$ & $170$ & $168\pm 9$ & $8.1(16)$ & $(4.0\pm 0.5)(16)$ & $1.02$ & $4.5$ & $2.0$ \\ 
 \hline \\[-0.35cm] 
CH$_3$NCO, $v=0$ & $2.0$ & $100$ & $96\pm 11$ & $5.0(16)$ & $(4.5\pm 1.3)(16)$ & $1.00$ & $6.4$ & $2.5$ \\ 
 \hline \\[-0.35cm] 
C$_2$H$_5$CN, $v=0$ & $2.0$ & $170$ & $168\pm 1$ & $1.3(18)$ & $(1.34\pm 0.04)(18)$ & $1.52$ & $8.0$ & $2.0$ \\ 
 $v_{12}=1$ & $2.0$ & $170$ & $168\pm 1$ & $1.3(18)$ & $(1.34\pm 0.04)(18)$ & $1.52$ & $8.0$ & $2.0$ \\ 
 $v_{13}=2$ & $2.0$ & $170$ & $168\pm 1$ & $1.3(18)$ & $(1.34\pm 0.04)(18)$ & $1.52$ & $8.0$ & $2.0$ \\ 
 $v_{20}=1$ & $2.0$ & $170$ & $168\pm 1$ & $1.3(18)$ & $(1.34\pm 0.04)(18)$ & $1.52$ & $8.0$ & $2.0$ \\ 
 $v_{21}=2$ & $2.0$ & $170$ & $168\pm 1$ & $1.3(18)$ & $(1.34\pm 0.04)(18)$ & $1.52$ & $8.0$ & $2.0$ \\ 
 \hline \\[-0.35cm] 
C$_2$H$_3$CN, $v=0$ & $2.0$ & $160$ & $168\pm 3$ & $3.5(17)$ & $(3.3\pm 0.2)(17)$ & $1.01$ & $6.5$ & $2.6$ \\ 
 $v_{11}=1$ & $2.0$ & $160$ & $168\pm 3$ & $3.5(17)$ & $(3.3\pm 0.2)(17)$ & $1.01$ & $6.5$ & $2.6$ \\ 
 $v_{15}=1$ & $2.0$ & $160$ & $168\pm 3$ & $3.5(17)$ & $(3.3\pm 0.2)(17)$ & $1.01$ & $6.5$ & $2.6$ \\ 
 \hline \\[-0.35cm] 
NH$_2$CHO, $v=0$ & $2.0$ & $130$ & $141\pm 3$ & $6.4(17)$ & $(6.4\pm 0.4)(17)$ & $1.06$ & $6.0$ & $2.3$ \\ 
 $v_{12}=1$ & $2.0$ & $130$ & $141\pm 3$ & $6.4(17)$ & $(6.4\pm 0.4)(17)$ & $1.06$ & $6.0$ & $2.3$ \\ 
 \hline\hline
\end{tabular}
\label{tab:n1w1}
\end{table*}

\begin{table*}[h!]
\caption{Same as Table\,\ref{tab:n1s-1}, but for position N1W2.} 
\centering
\begin{tabular}{rrrrrrrrrrr}
\hline\hline \\[-0.3cm] 
Molecule & Size & $T_{\rm rot,W}$ & $T_{\rm rot,obs}$ & $N_{\rm W}$ & $N_{\rm obs}$ & $C_{\rm vib}$ & $\Delta\varv$ & $\varv_{\rm off}$ \\ 
 & $(^{\prime\prime})$ & (K) & (K) & (cm$^{-2}$) & (cm$^{-2}$) &  & (km\,s$^{-1}$) & (km\,s$^{-1}$) \\\hline \\[-0.3cm] 
CH$_3$OH, $v=0$ & $2.0$ & $135$ & $131\pm 2$ & $1.0(19)$ & $(8.4\pm 0.8)(18)$ & $1.00$ & $5.5$ & $3.0$ \\ 
 $v=1$ & $2.0$ & $135$ & $131\pm 2$ & $1.0(19)$ & $(8.4\pm 0.8)(18)$ & $1.00$ & $5.5$ & $3.0$ \\ 
 $v=2$ & $2.0$ & $135$ & $131\pm 2$ & $1.0(19)$ & $(8.4\pm 0.8)(18)$ & $1.00$ & $5.5$ & $3.0$ \\ 
 $^{13}$CH$_3$OH, $v=0$ & $2.0$ & $140$ & $135\pm 4$ & $3.3(17)$ & $(3.5\pm 0.3)(17)$ & $1.00$ & $5.0$ & $3.0$ \\ 
 \hline \\[-0.35cm] 
C$_2$H$_5$OH, $v=0$ & $2.0$ & $150$ & $147\pm 3$ & $1.9(17)$ & $(2.2\pm 0.2)(17)$ & $1.30$ & $5.0$ & $3.5$ \\ 
 \hline \\[-0.35cm] 
CH$_3$OCH$_3$, $v=0$ & $2.0$ & $110$ & $109\pm 2$ & $3.2(17)$ & $(3.5\pm 0.2)(17)$ & $1.00$ & $4.0$ & $3.5$ \\ 
 $v_{11}=1$ & $2.0$ & $110$ & $109\pm 2$ & $3.2(17)$ & $(3.5\pm 0.2)(17)$ & $1.00$ & $4.0$ & $3.5$ \\ 
 \hline \\[-0.35cm] 
CH$_3$OCHO, $v=0$ & $2.0$ & $150$ & $144\pm 3$ & $5.9(17)$ & $(6.2\pm 0.3)(17)$ & $1.17$ & $4.0$ & $3.5$ \\ 
 $v=1$ & $2.0$ & $150$ & $144\pm 3$ & $5.9(17)$ & $(6.2\pm 0.3)(17)$ & $1.17$ & $4.0$ & $3.5$ \\ 
 \hline \\[-0.35cm] 
CH$_3$CHO, $v=0$ & $2.0$ & $125$ & $115\pm 4$ & $2.0(16)$ & $(1.0\pm 0.1)(16)$ & $1.00$ & $4.5$ & $2.0$ \\ 
 $v=1$ & $2.0$ & $125$ & $115\pm 4$ & $2.0(16)$ & $(1.0\pm 0.1)(16)$ & $1.00$ & $4.5$ & $2.0$ \\ 
 \hline \\[-0.35cm] 
CH$_3$NCO, $v=0$ & $2.0$ & $90$ & $86\pm 8$ & $1.3(16)$ & $(1.1\pm 0.3)(16)$ & $1.00$ & $6.4$ & $2.5$ \\ 
 \hline \\[-0.35cm] 
C$_2$H$_5$CN, $v=0$ & $2.0$ & $150$ & $151\pm 1$ & $6.3(17)$ & $(6.3\pm 0.2)(17)$ & $1.39$ & $7.3$ & $2.0$ \\ 
 $v_{12}=1$ & $2.0$ & $150$ & $151\pm 1$ & $6.3(17)$ & $(6.3\pm 0.2)(17)$ & $1.39$ & $7.3$ & $2.0$ \\ 
 $v_{13}=2$ & $2.0$ & $150$ & $151\pm 1$ & $6.3(17)$ & $(6.3\pm 0.2)(17)$ & $1.39$ & $7.3$ & $2.0$ \\ 
 $v_{20}=1$ & $2.0$ & $150$ & $151\pm 1$ & $6.3(17)$ & $(6.3\pm 0.2)(17)$ & $1.39$ & $7.3$ & $1.2$ \\ 
 $v_{21}=2$ & $2.0$ & $150$ & $151\pm 1$ & $6.3(17)$ & $(6.3\pm 0.2)(17)$ & $1.39$ & $7.3$ & $2.0$ \\ 
 \hline \\[-0.35cm] 
C$_2$H$_3$CN, $v=0$ & $2.0$ & $150$ & $153\pm 5$ & $2.5(16)$ & $(2.7\pm 0.2)(16)$ & $1.01$ & $4.5$ & $2.0$ \\ 
 $v_{11}=1$ & $2.0$ & $150$ & $153\pm 5$ & $2.5(16)$ & $(2.7\pm 0.2)(16)$ & $1.01$ & $4.5$ & $2.0$ \\ 
 \hline \\[-0.35cm] 
NH$_2$CHO, $v=0$ & $2.0$ & $130$ & $136\pm 3$ & $3.2(17)$ & $(3.2\pm 0.2)(17)$ & $1.05$ & $5.5$ & $2.0$ \\ 
 $v_{12}=1$ & $2.0$ & $120$ & $136\pm 3$ & $3.7(17)$ & $(3.2\pm 0.2)(17)$ & $1.05$ & $6.0$ & $2.0$ \\ 
 \hline\hline
\end{tabular}
\label{tab:n1w2}
\end{table*}

\begin{table*}[h!]
\caption{Same as Table\,\ref{tab:n1s-1}, but for position N1W3.} 
\centering
\begin{tabular}{rrrrrrrrrrr}
\hline\hline \\[-0.3cm] 
Molecule & Size & $T_{\rm rot,W}$ & $T_{\rm rot,obs}$ & $N_{\rm W}$ & $N_{\rm obs}$ & $C_{\rm vib}$ & $\Delta\varv$ & $\varv_{\rm off}$ \\ 
 & $(^{\prime\prime})$ & (K) & (K) & (cm$^{-2}$) & (cm$^{-2}$) &  & (km\,s$^{-1}$) & (km\,s$^{-1}$) \\\hline \\[-0.3cm] 
CH$_3$OH, $v=0$ & $2.0$ & $110$ & $107\pm 4$ & $1.5(18)$ & $(1.3\pm 0.1)(18)$ & $1.00$ & $5.0$ & $3.0$ \\ 
 $^{13}$CH$_3$OH, $v=0$ & $2.0$ & $120$ & $122\pm 4$ & $1.5(17)$ & $(1.6\pm 0.2)(17)$ & $1.00$ & $5.0$ & $3.0$ \\ 
 \hline \\[-0.35cm] 
C$_2$H$_5$OH, $v=0$ & $2.0$ & $130$ & $133\pm 4$ & $4.9(16)$ & $(5.1\pm 0.5)(16)$ & $1.23$ & $4.5$ & $2.6$ \\ 
 \hline \\[-0.35cm] 
CH$_3$OCH$_3$, $v=0$ & $2.0$ & $90$ & $85\pm 6$ & $4.3(16)$ & $(4.9\pm 0.8)(16)$ & $1.00$ & $3.5$ & $3.0$ \\ 
 \hline \\[-0.35cm] 
CH$_3$OCHO, $v=0$ & $2.0$ & $140$ & $135\pm 8$ & $1.3(17)$ & $(1.3\pm 0.2)(17)$ & $1.14$ & $5.0$ & $3.5$ \\ 
 $v=1$ & $2.0$ & $140$ & $135\pm 8$ & $1.3(17)$ & $(1.3\pm 0.2)(17)$ & $1.14$ & $5.0$ & $3.5$ \\ 
 \hline \\[-0.35cm] 
CH$_3$CHO, $v=0$ & $2.0$ & $100$ & $117\pm 28$ & $3.5(15)$ & $(2.6\pm 1.2)(15)$ & $1.00$ & $4.0$ & $2.5$ \\ 
 \hline \\[-0.35cm] 
CH$_3$NCO, $v=0$ & $2.0$ & $90$ & $89\pm 21$ & $3.0(15)$ & $(2.8\pm 1.6)(15)$ & $1.00$ & $5.5$ & $2.0$ \\ 
 \hline \\[-0.35cm] 
C$_2$H$_5$CN, $v=0$ & $2.0$ & $130$ & $131\pm 2$ & $1.3(17)$ & $(1.3\pm 0.1)(17)$ & $1.26$ & $7.3$ & $2.0$ \\ 
 $v_{20}=1$ & $2.0$ & $130$ & $131\pm 2$ & $1.3(17)$ & $(1.3\pm 0.1)(17)$ & $1.26$ & $7.3$ & $2.0$ \\ 
 \hline \\[-0.35cm] 
C$_2$H$_3$CN, $v=0$ & $2.0$ & $130$ & $138\pm 11$ & $3.0(15)$ & $(2.9\pm 0.5)(15)$ & $1.00$ & $5.0$ & $3.9$ \\ 
 \hline \\[-0.35cm] 
NH$_2$CHO, $v=0$ & $2.0$ & $125$ & $131\pm 4$ & $5.2(16)$ & $(5.4\pm 0.6)(16)$ & $1.05$ & $5.0$ & $1.5$ \\ 
 $v_{12}=1$ & $2.0$ & $125$ & $131\pm 4$ & $5.2(16)$ & $(5.4\pm 0.6)(16)$ & $1.05$ & $5.0$ & $1.5$ \\ 
 \hline\hline
\end{tabular}
\label{tab:n1w3}
\end{table*}

\begin{table*}[h!]
\caption{Same as Table\,\ref{tab:n1s-1}, but for position N1W4.} 
\centering
\begin{tabular}{rrrrrrrrrrr}
\hline\hline \\[-0.3cm] 
Molecule & Size & $T_{\rm rot,W}$ & $T_{\rm rot,obs}$ & $N_{\rm W}$ & $N_{\rm obs}$ & $C_{\rm vib}$ & $\Delta\varv$ & $\varv_{\rm off}$ \\ 
 & $(^{\prime\prime})$ & (K) & (K) & (cm$^{-2}$) & (cm$^{-2}$) &  & (km\,s$^{-1}$) & (km\,s$^{-1}$) \\\hline \\[-0.3cm] 
CH$_3$OH, $v=0$ & $2.0$ & $100$ & $90\pm 5$ & $2.5(17)$ & $(1.7\pm 0.3)(17)$ & $1.00$ & $6.0$ & $2.0$ \\ 
 $^{13}$CH$_3$OH, $v=0$ & $2.0$ & $100$ & $97\pm 6$ & $4.0(16)$ & $(4.2\pm 0.6)(16)$ & $1.00$ & $4.5$ & $2.6$ \\ 
 \hline \\[-0.35cm] 
C$_2$H$_5$OH, $v=0$ & $2.0$ & $115$ & $115\pm 6$ & $9.2(15)$ & $(8.4\pm 1.3)(15)$ & $1.15$ & $5.5$ & $2.0$ \\ 
 \hline \\[-0.35cm] 
CH$_3$OCH$_3$, $v=0$ & $2.0$ & $77$ & $77\pm 6$ & $\leq6.5(15)$ &  & $1.00$ & $4.9$ & $3.4$ \\ 
 \hline \\[-0.35cm] 
CH$_3$OCHO, $v=0$ & $2.0$ & $30$ & $23\pm 2$ & $3.0(15)$ & $(3.0\pm 0.7)(15)$ & $1.00$ & $4.0$ & $3.5$ \\ 
 \hline \\[-0.35cm] 
CH$_3$CHO, $v=0$ & $2.0$ & $93$ & $93\pm 11$ & $\leq1.5(15)$ &  & $1.00$ & $4.9$ & $3.4$ \\ 
 \hline \\[-0.35cm] 
CH$_3$NCO, $v=0$ & $2.0$ & $76$ & $76\pm 6$ & $\leq7.5(14)$ &  & $1.00$ & $4.9$ & $3.4$ \\ 
 \hline \\[-0.35cm] 
C$_2$H$_5$CN, $v=0$ & $2.0$ & $127$ & $127\pm 7$ & $\leq8.1(15)$ &  & $1.24$ & $4.5$ & $2.5$ \\ 
 \hline \\[-0.35cm] 
C$_2$H$_3$CN, $v=0$ & $2.0$ & $128$ & $128\pm 2$ & $\leq1.0(15)$ &  & $1.00$ & $4.9$ & $3.4$ \\ 
 \hline \\[-0.35cm] 
NH$_2$CHO, $v=0$ & $2.0$ & $123$ & $123\pm 6$ & $\leq1.3(15)$ &  & $1.04$ & $5.0$ & $3.0$ \\ 
 \hline\hline
\end{tabular}
\label{tab:n1w4}
\end{table*}

\begin{table*}[h!]
\caption{Same as Table\,\ref{tab:n1s-1}, but for position N1W5.} 
\centering
\begin{tabular}{rrrrrrrrrrr}
\hline\hline \\[-0.3cm] 
Molecule & Size & $T_{\rm rot,W}$ & $T_{\rm rot,obs}$ & $N_{\rm W}$ & $N_{\rm obs}$ & $C_{\rm vib}$ & $\Delta\varv$ & $\varv_{\rm off}$ \\ 
 & $(^{\prime\prime})$ & (K) & (K) & (cm$^{-2}$) & (cm$^{-2}$) &  & (km\,s$^{-1}$) & (km\,s$^{-1}$) \\\hline \\[-0.3cm] 
CH$_3$OH, $v=0$ & $2.0$ & $94$ & $92\pm 2$ & $1.8(17)$ & $(1.5\pm 0.1)(17)$ & $1.00$ & $5.5$ & $7.0$ \\ 
 $^{13}$CH$_3$OH, $v=0$ & $2.0$ & $100$ & $96\pm 18$ & $1.0(16)$ & $(1.1\pm 0.4)(16)$ & $1.00$ & $2.5$ & $7.0$ \\ 
 \hline \\[-0.35cm] 
C$_2$H$_5$OH, $v=0$ & $2.0$ & $105$ & $105\pm 6$ & $\leq6.1(15)$ &  & $1.11$ & $3.7$ & $7.4$ \\ 
 \hline \\[-0.35cm] 
CH$_3$OCH$_3$, $v=0$ & $2.0$ & $69$ & $69\pm 6$ & $\leq5.0(15)$ &  & $1.00$ & $3.7$ & $7.4$ \\ 
 \hline \\[-0.35cm] 
CH$_3$OCHO, $v=0$ & $2.0$ & $50$ & $44\pm 8$ & $5.0(15)$ & $(5.5\pm 2.0)(15)$ & $1.00$ & $2.5$ & $7.0$ \\ 
 \hline \\[-0.35cm] 
CH$_3$CHO, $v=0$ & $2.0$ & $84$ & $84\pm 11$ & $\leq8.0(14)$ &  & $1.00$ & $3.7$ & $7.4$ \\ 
 \hline \\[-0.35cm] 
CH$_3$NCO, $v=0$ & $2.0$ & $72$ & $72\pm 7$ & $\leq5.5(14)$ &  & $1.00$ & $3.7$ & $7.4$ \\ 
 \hline \\[-0.35cm] 
C$_2$H$_5$CN, $v=0$ & $2.0$ & $90$ & $94\pm 9$ & $7.6(15)$ & $(8.2\pm 1.4)(15)$ & $1.09$ & $5.5$ & $8.0$ \\ 
 \hline \\[-0.35cm] 
C$_2$H$_3$CN, $v=0$ & $2.0$ & $120$ & $120\pm 2$ & $\leq1.0(15)$ &  & $1.00$ & $3.7$ & $7.4$ \\ 
 \hline \\[-0.35cm] 
NH$_2$CHO, $v=0$ & $2.0$ & $119$ & $119\pm 6$ & $\leq8.3(14)$ &  & $1.03$ & $4.5$ & $7.4$ \\ 
 \hline\hline
\end{tabular}
\label{tab:n1w5}
\end{table*}

\begin{table*}[h!]
\caption{Same as Table\,\ref{tab:n1s-1}, but for position N1W6.} 
\centering
\begin{tabular}{rrrrrrrrrrr}
\hline\hline \\[-0.3cm] 
Molecule & Size & $T_{\rm rot,W}$ & $T_{\rm rot,obs}$ & $N_{\rm W}$ & $N_{\rm obs}$ & $C_{\rm vib}$ & $\Delta\varv$ & $\varv_{\rm off}$ \\ 
 & $(^{\prime\prime})$ & (K) & (K) & (cm$^{-2}$) & (cm$^{-2}$) &  & (km\,s$^{-1}$) & (km\,s$^{-1}$) \\\hline \\[-0.3cm] 
CH$_3$OH, $v=0$ & $2.0$ & $90$ & $90\pm 3$ & $1.1(18)$ & $(8.0\pm 1.1)(17)$ & $1.00$ & $2.5$ & $7.0$ \\ 
 $^{13}$CH$_3$OH, $v=0$ & $2.0$ & $85$ & $81\pm 8$ & $1.3(16)$ & $(1.5\pm 0.3)(16)$ & $1.00$ & $2.9$ & $7.1$ \\ 
 \hline \\[-0.35cm] 
C$_2$H$_5$OH, $v=0$ & $2.0$ & $95$ & $95\pm 6$ & $\leq4.3(15)$ &  & $1.08$ & $2.6$ & $7.3$ \\ 
 \hline \\[-0.35cm] 
CH$_3$OCH$_3$, $v=0$ & $2.0$ & $90$ & $89\pm 5$ & $4.5(16)$ & $(4.1\pm 0.5)(16)$ & $1.00$ & $2.0$ & $7.0$ \\ 
 \hline \\[-0.35cm] 
CH$_3$OCHO, $v=0$ & $2.0$ & $110$ & $109\pm 4$ & $1.3(17)$ & $(1.3\pm 0.1)(17)$ & $1.06$ & $2.5$ & $7.0$ \\ 
 $v=1$ & $2.0$ & $110$ & $109\pm 4$ & $1.3(17)$ & $(1.3\pm 0.1)(17)$ & $1.06$ & $2.5$ & $7.0$ \\ 
 \hline \\[-0.35cm] 
CH$_3$CHO, $v=0$ & $2.0$ & $75$ & $75\pm 11$ & $\leq5.0(14)$ &  & $1.00$ & $2.6$ & $7.3$ \\ 
 \hline \\[-0.35cm] 
CH$_3$NCO, $v=0$ & $2.0$ & $68$ & $68\pm 7$ & $\leq3.5(14)$ &  & $1.00$ & $2.6$ & $7.3$ \\ 
 \hline \\[-0.35cm] 
C$_2$H$_5$CN, $v=0$ & $2.0$ & $60$ & $68\pm 19$ & $1.3(15)$ & $(1.5\pm 0.7)(15)$ & $1.02$ & $4.3$ & $7.2$ \\ 
 \hline \\[-0.35cm] 
C$_2$H$_3$CN, $v=0$ & $2.0$ & $50$ & $39\pm 4$ & $4.0(14)$ & $(4.6\pm 1.0)(14)$ & $1.00$ & $2.5$ & $8.0$ \\ 
 \hline \\[-0.35cm] 
NH$_2$CHO, $v=0$ & $2.0$ & $114$ & $114\pm 7$ & $\leq2.1(14)$ &  & $1.03$ & $2.6$ & $7.3$ \\ 
 \hline\hline
\end{tabular}
\label{tab:n1w6}
\end{table*}

\begin{table*}[h!]
\caption{Same as Table\,\ref{tab:n1s-1}, but for position N1W7.} 
\centering
\begin{tabular}{rrrrrrrrrrr}
\hline\hline \\[-0.3cm] 
Molecule & Size & $T_{\rm rot,W}$ & $T_{\rm rot,obs}$ & $N_{\rm W}$ & $N_{\rm obs}$ & $C_{\rm vib}$ & $\Delta\varv$ & $\varv_{\rm off}$ \\ 
 & $(^{\prime\prime})$ & (K) & (K) & (cm$^{-2}$) & (cm$^{-2}$) &  & (km\,s$^{-1}$) & (km\,s$^{-1}$) \\\hline \\[-0.3cm] 
CH$_3$OH, $v=0$ & $2.0$ & $90$ & $97\pm 8$ & $3.0(17)$ & $(1.9\pm 0.6)(17)$ & $1.00$ & $2.5$ & $7.0$ \\ 
 $^{13}$CH$_3$OH, $v=0$ & $2.0$ & $85$ & $81\pm 7$ & $1.1(16)$ & $(1.2\pm 0.2)(16)$ & $1.00$ & $2.9$ & $7.1$ \\ 
 \hline \\[-0.35cm] 
C$_2$H$_5$OH, $v=0$ & $2.0$ & $85$ & $85\pm 6$ & $\leq3.7(15)$ &  & $1.05$ & $2.6$ & $7.3$ \\ 
 \hline \\[-0.35cm] 
CH$_3$OCH$_3$, $v=0$ & $2.0$ & $77$ & $76\pm 7$ & $1.5(16)$ & $(1.4\pm 0.3)(16)$ & $1.00$ & $2.0$ & $7.5$ \\ 
 \hline \\[-0.35cm] 
CH$_3$OCHO, $v=0$ & $2.0$ & $105$ & $102\pm 2$ & $6.3(16)$ & $(6.7\pm 0.4)(16)$ & $1.05$ & $2.5$ & $7.0$ \\ 
 $v=1$ & $2.0$ & $105$ & $102\pm 2$ & $6.3(16)$ & $(6.7\pm 0.4)(16)$ & $1.05$ & $2.5$ & $7.0$ \\ 
 \hline \\[-0.35cm] 
CH$_3$CHO, $v=0$ & $2.0$ & $66$ & $66\pm 11$ & $\leq4.0(14)$ &  & $1.00$ & $2.6$ & $7.3$ \\ 
 \hline \\[-0.35cm] 
CH$_3$NCO, $v=0$ & $2.0$ & $64$ & $64\pm 7$ & $\leq3.5(14)$ &  & $1.00$ & $2.6$ & $7.3$ \\ 
 \hline \\[-0.35cm] 
C$_2$H$_5$CN, $v=0$ & $2.0$ & $90$ & $111\pm 42$ & $7.0(14)$ & $(1.0\pm 0.6)(15)$ & $1.16$ & $3.0$ & $7.7$ \\ 
 \hline \\[-0.35cm] 
C$_2$H$_3$CN, $v=0$ & $2.0$ & $104$ & $104\pm 3$ & $\leq5.0(14)$ &  & $1.00$ & $2.6$ & $7.3$ \\ 
 \hline \\[-0.35cm] 
NH$_2$CHO, $v=0$ & $2.0$ & $109$ & $109\pm 7$ & $\leq1.5(14)$ &  & $1.02$ & $2.6$ & $7.3$ \\ 
 \hline\hline
\end{tabular}
\label{tab:n1w7}
\end{table*}

\begin{table*}[h!]
\caption{Same as Table\,\ref{tab:n1s-1}, but for position N1W8.} 
\centering
\begin{tabular}{rrrrrrrrrrr}
\hline\hline \\[-0.3cm] 
Molecule & Size & $T_{\rm rot,W}$ & $T_{\rm rot,obs}$ & $N_{\rm W}$ & $N_{\rm obs}$ & $C_{\rm vib}$ & $\Delta\varv$ & $\varv_{\rm off}$ \\ 
 & $(^{\prime\prime})$ & (K) & (K) & (cm$^{-2}$) & (cm$^{-2}$) &  & (km\,s$^{-1}$) & (km\,s$^{-1}$) \\\hline \\[-0.3cm] 
CH$_3$OH, $v=0$ & $2.0$ & $79$ & $79\pm 8$ & $1.8(16)$ & $(1.4\pm 0.1)(16)$ & $1.00$ & $2.5$ & $7.5$ \\ 
 $^{13}$CH$_3$OH, $v=0$ & $2.0$ & $73$ & $73\pm 5$ & $\leq2.5(15)$ &  & $1.00$ & $2.5$ & $7.9$ \\ 
 \hline \\[-0.35cm] 
C$_2$H$_5$OH, $v=0$ & $2.0$ & $77$ & $77\pm 6$ & $\leq3.6(15)$ &  & $1.03$ & $2.5$ & $7.9$ \\ 
 \hline \\[-0.35cm] 
CH$_3$OCH$_3$, $v=0$ & $2.0$ & $46$ & $46\pm 6$ & $\leq3.3(15)$ &  & $1.00$ & $2.5$ & $7.9$ \\ 
 \hline \\[-0.35cm] 
CH$_3$OCHO, $v=0$ & $2.0$ & $91$ & $91\pm 8$ & $\leq2.7(15)$ &  & $1.03$ & $2.5$ & $7.9$ \\ 
 \hline \\[-0.35cm] 
CH$_3$CHO, $v=0$ & $2.7$ & $59$ & $59\pm 11$ & $\leq4.0(14)$ &  & $1.00$ & $2.5$ & $7.9$ \\ 
 \hline \\[-0.35cm] 
CH$_3$NCO, $v=0$ & $2.0$ & $60$ & $60\pm 7$ & $\leq3.0(14)$ &  & $1.00$ & $2.5$ & $7.9$ \\ 
 \hline \\[-0.35cm] 
C$_2$H$_5$CN, $v=0$ & $2.0$ & $98$ & $98\pm 8$ & $2.8(14)$ & $(2.8\pm 0.5)(14)$ & $1.11$ & $3.0$ & $7.0$ \\ 
 \hline \\[-0.35cm] 
C$_2$H$_3$CN, $v=0$ & $2.0$ & $98$ & $98\pm 3$ & $\leq2.0(14)$ &  & $1.00$ & $2.5$ & $7.9$ \\ 
 \hline \\[-0.35cm] 
NH$_2$CHO, $v=0$ & $2.0$ & $105$ & $105\pm 8$ & $\leq1.0(14)$ &  & $1.02$ & $2.5$ & $7.9$ \\ 
 \hline\hline
\end{tabular}
\label{tab:n1w8}
\end{table*}

\begin{table*}[h!]
\caption{Intensity of the continuum corrected for free-free emission, C$^{18}$O and H\2 column densities for all positions shown in Fig.\,\ref{fig:1mm+et}.} 
\centering
\begin{tabular}{lrrr|rrrrr}
\hline\hline \\[-0.3cm] 
Position & $S_{\rm dust}$\tablefootmark{a} & $\tau_{\rm dust}$\tablefootmark{b} & $N({\rm H}_2)_{\rm dust}$\tablefootmark{c} & $N({{\rm C}^{18}{\rm O}})$\tablefootmark{d} & $N({\rm H}_2)_{{\rm C}^{18}{\rm O}}$\tablefootmark{e} & Size\tablefootmark{f} & $\Delta\varv$\tablefootmark{g} & $\varv_{\rm off}$\tablefootmark{h} \\ [0.1cm]
 & (K) & & ($10^{24}$\,cm$^{-2}$) & ($10^{17}$cm$^{-2}$) & ($10^{24}$cm$^{-2}$) & ($^{\prime\prime}$) & (\kms) & (\kms) \\ \hline\\[-0.3cm]
N1S-1 & 103.6 & 0.40(10) & 12.8(4) & -- & -- & -- & -- & -- \\
N1S & 65.9 & 0.35(9) & 11.4(3) & -- & -- & -- & -- & -- \\
N1S1 & 16.0 & 0.10(2) & 3.3(8) & 10.8 & 2.7(5) & 2.0 & 4.5 & $-0.5$ \\
N1S2 & 6.7 & 0.05(1) & 1.7(4) & 4.8 & 1.2(2) & 2.0 & 4.5 & $-0.5$ \\
N1S3 & 3.3 & 0.030(7) & 1.0(3) & 2.1 & 0.5(1) & 2.0 & 4.0 & $-1.8$\\
N1S4 & 0.9 & 0.010(4) & 0.3(1) & 1.0 & 0.26(5) & 2.0 & 3.5 & $-3.0$\\
N1S5 & 0.87 & 0.010(5) & $\leq$0.3(2) & 1.6 & 0.40(8) & 2.0 & 2.5 & $-2.6$ \\
N1S6 & 2.7 & 0.030(8) & $\leq$1.1(3) & 1.6 & 0.40(8) & 2.0 & 2.5 & $-2.8$ \\
N1S7 & 5.6 & 0.07(2) & $\leq$2.4(6) & 1.6 & 0.41(8) & 2.0 & 2.5 & $-3.1$ \\
N1S8 & -- & -- & -- & 1.6 & 0.39(7) & 2.0 & 2.0 & $-3.3$ \\\hline\\[-0.3cm]
N1W-1 & 27.5 & 0.29(7) & 2.7(7) & $\leq$12.3 & $\leq$3.1(6) & 2.0 & 4.5 & $2.6$ \\
N1W & 6.6 & 0.23(6) & 1.0(3) & $\leq$9.3 & $\leq$2.3(4) & 2.0 & 4.5 & $2.6$ \\
N1W1 & 3.8 & 0.18(4) & 0.7(2) & 8.5 & 2.1(4) & 2.0 & 4.5 & $3.0$\\
N1W2 & 1.0 & 0.07(2) & 0.2(1) & 6.3 & 1.6(3) & 2.0 & 4.5 & $3.0$ \\
N1W3 & $\leq$3.3 & $\leq$0.02 & $\leq$0.8 & 2.1 & 0.5(1) & 2.0 & 4.5 & $3.0$\\
N1W4 & $\leq$3.3 & $\leq$0.03 & $\leq$1.0 & 0.8 & 0.20(4) & 2.0 & 4.5 & $2.0$\\
N1W5 & -- & -- & -- & 0.3 & 0.06(1) & 2.0 & 4.0 & $7.4$ \\
N1W6 & -- & -- & -- & 1.2 & 0.31(6) & 2.0 & 3.0 & $7.0$\\
N1W7 & -- & -- & -- & 3.2 & 0.8(2) & 2.0 & 2.5 & $7.3$ \\
N1W8 & -- & -- & -- & 1.5 & 0.38(7) & 2.0 & 2.0 & $7.9$ \\
 \hline\hline
\end{tabular}
\tablefoot{The values in parentheses show the uncertainties in units of the last digit. \\ \tablefoottext{a}{Measured continuum intensity corrected for free-free emission.}\tablefoottext{b}{Opacity of the dust emission assuming $T_{\rm dust}=T_{\rm rot,obs}$(\et) with $\Delta T_{\rm dust} = 0.2 \times T_{\rm rot,obs}$(\et).}\tablefoottext{c}{H\2 column densities derived from dust emission.}\tablefoottext{d}{Opacity-corrected C$^{18}$O column densities taken from the Weeds model at the respective position with $\Delta N({{\rm C}^{18}{\rm O}}) = 0.15\times N({{\rm C}^{18}{\rm O}})$.}\tablefoottext{e}{H\2 column densities derived from C$^{18}$O\,1--0 emission.}\tablefoottext{f}{Size of the emitting region.}\tablefoottext{g}{$FWHM$ of the transitions.}\tablefoottext{h}{Offset from the source systemic velocity, which was set to 62\kms.}}\label{tab:H2}
\end{table*}

\section{Additional figures: Spectra at N1S}\label{app:spectra}
\begin{figure*}
    \includegraphics[width=\textwidth]{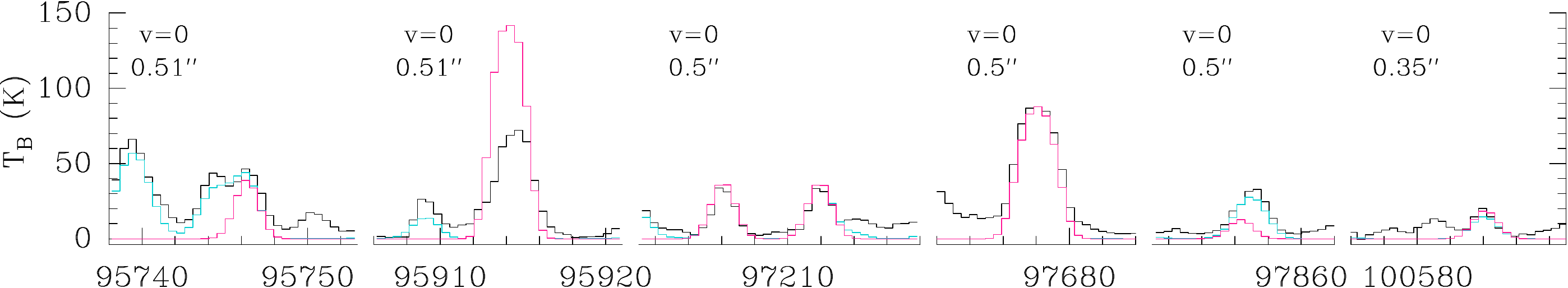}\\[0.2cm]
    \includegraphics[width=\textwidth]{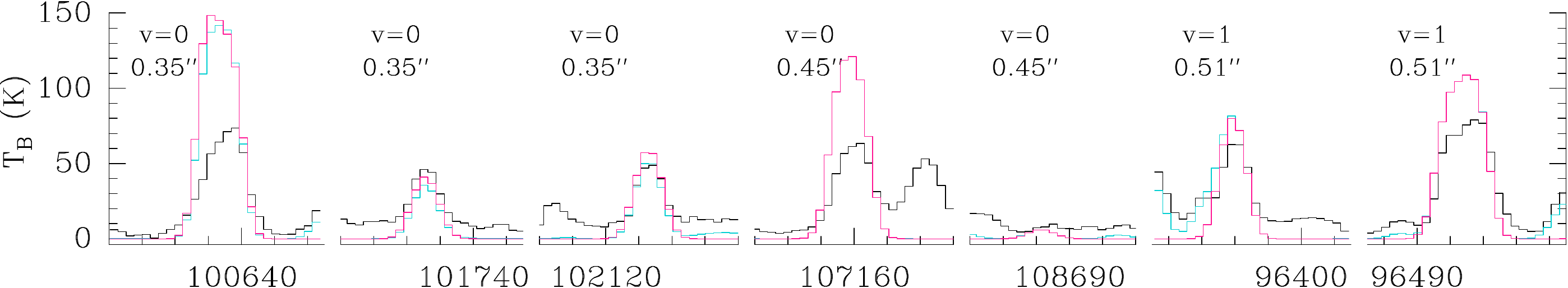}\\[0.2cm]
    \includegraphics[width=\textwidth]{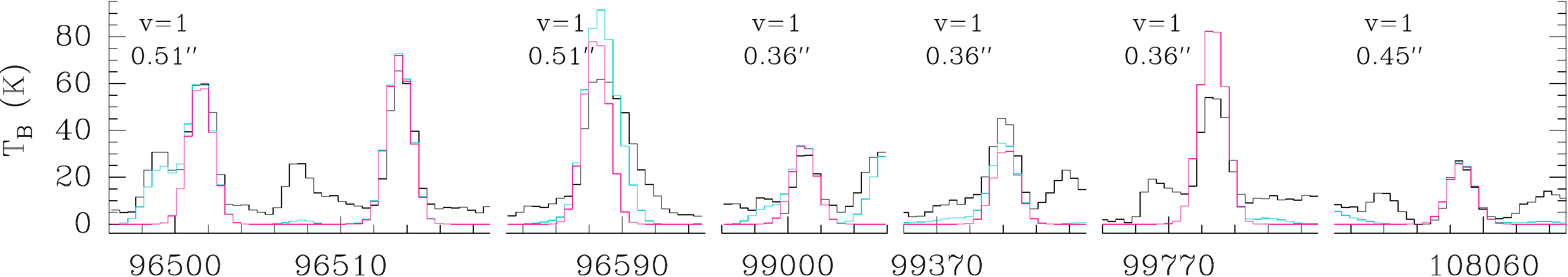}\\[0.2cm]
    \includegraphics[width=0.73\textwidth]{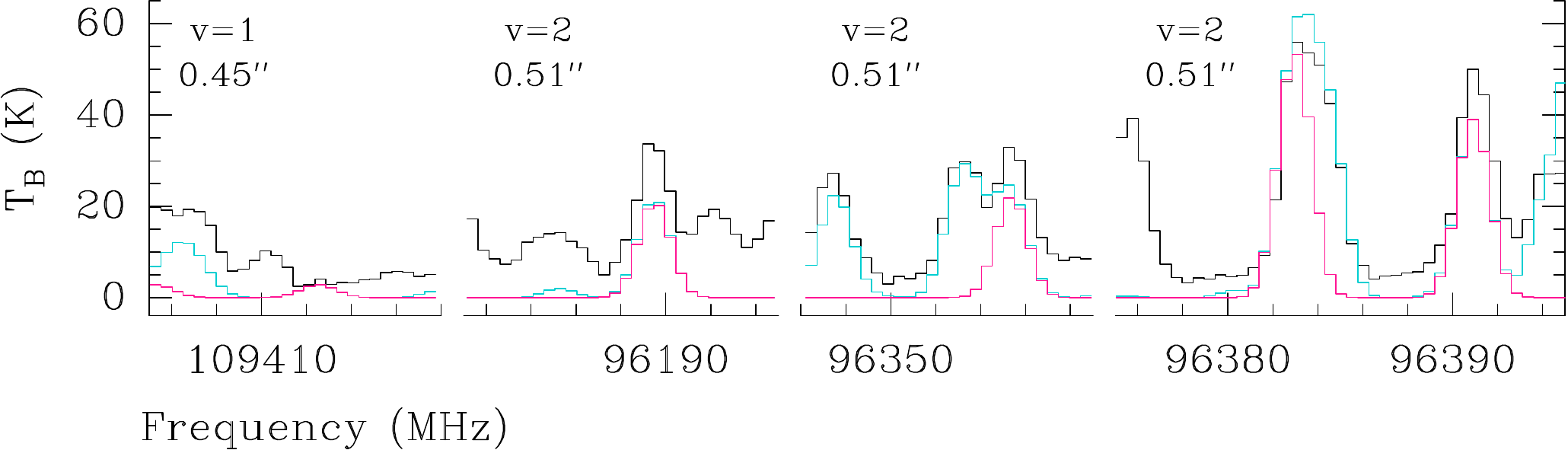}
    \caption{Transitions of \met used to produce the population diagram at position N1S. The observed spectrum is shown in black, the synthetic spectrum of the shown molecule in pink, and the cumulative Weeds model of all COMs, whose spectra have been modelled in this work, in turquoise.  The vibrational state and the beam size of the respective spectral window are shown in the upper left corner. The frequency axis shows steps of 2\,MHz. Transitions marked with pink stars in Figs.\,\ref{fig:spec_13met}--\ref{fig:spec_fmm} are not used for the population diagram.}
    \label{fig:spec_met}
\end{figure*}

\begin{figure*}
    \includegraphics[width=\textwidth]{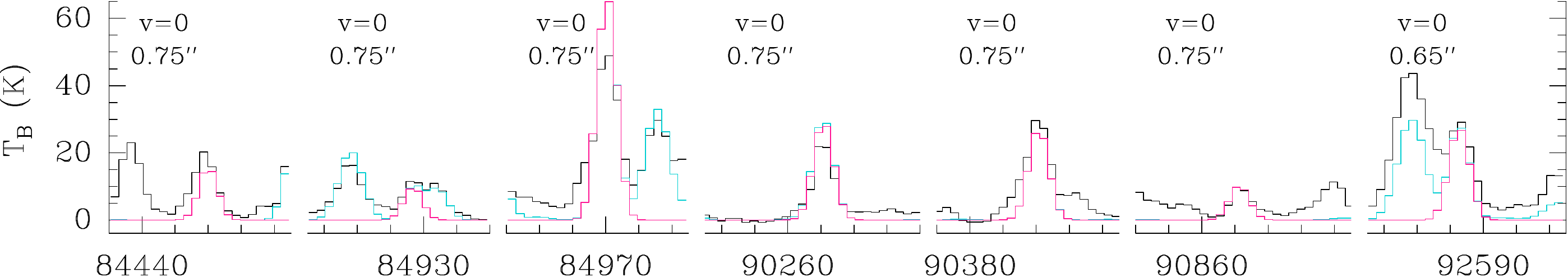}\\[0.2cm]
    \includegraphics[width=0.88\textwidth]{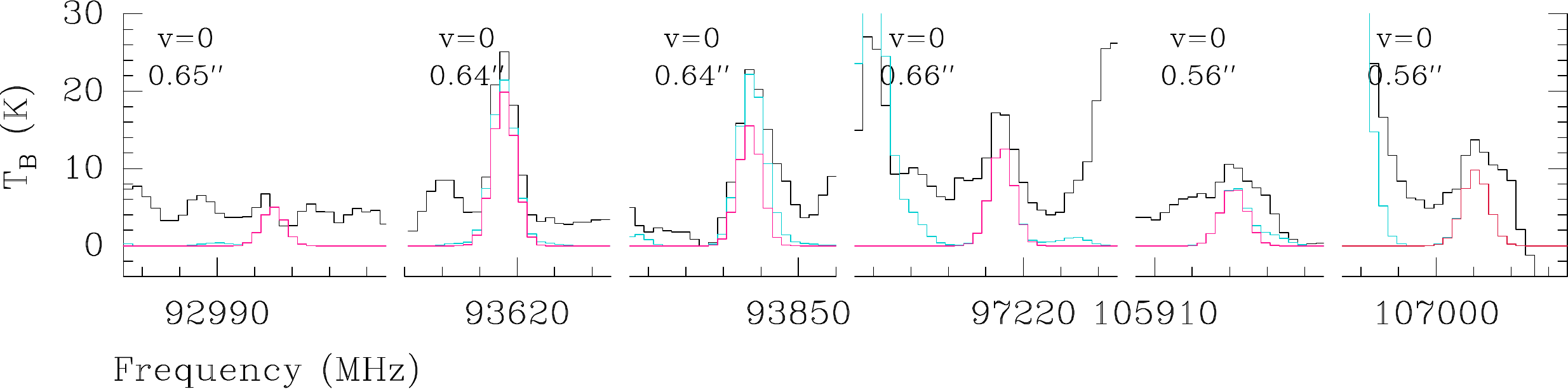}
    \caption{Same as Fig.\,\ref{fig:spec_met}, but for $^{13}$\met.}
    \label{fig:spec_13met}
\end{figure*}

\begin{figure*}
    \includegraphics[width=\textwidth]{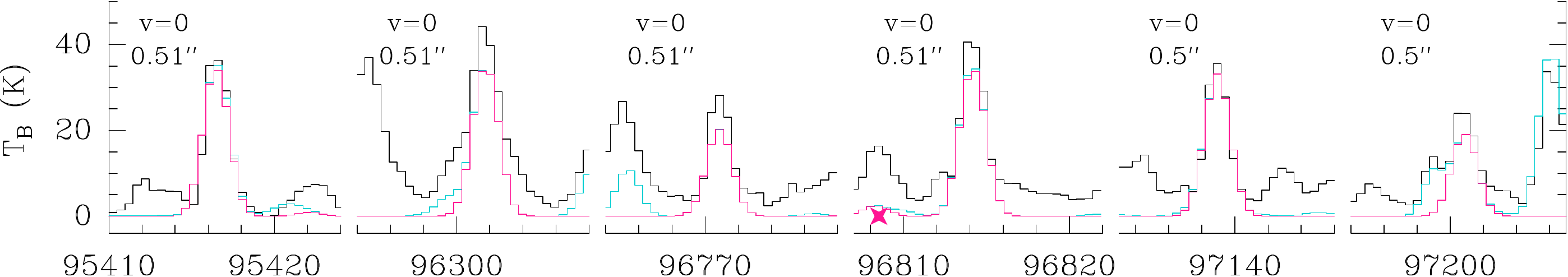}\\[0.2cm]
    \includegraphics[width=\textwidth]{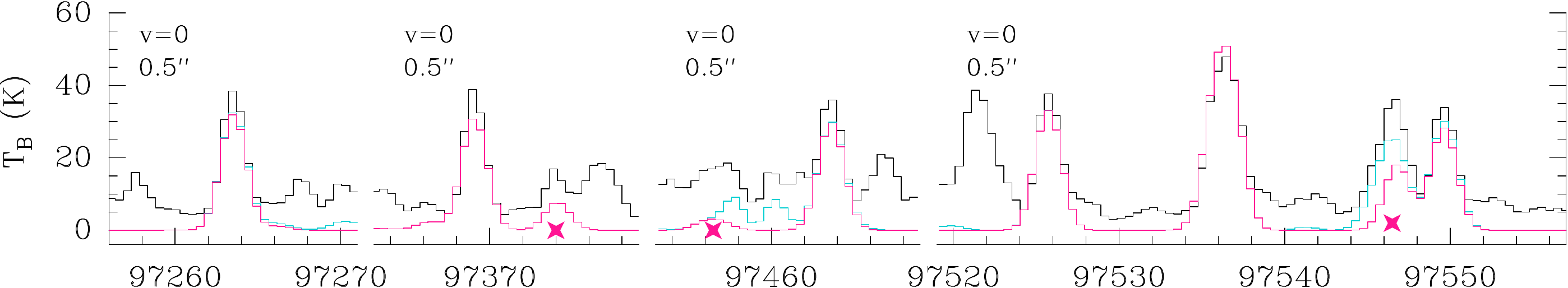}\\[0.2cm]
    \includegraphics[width=\textwidth]{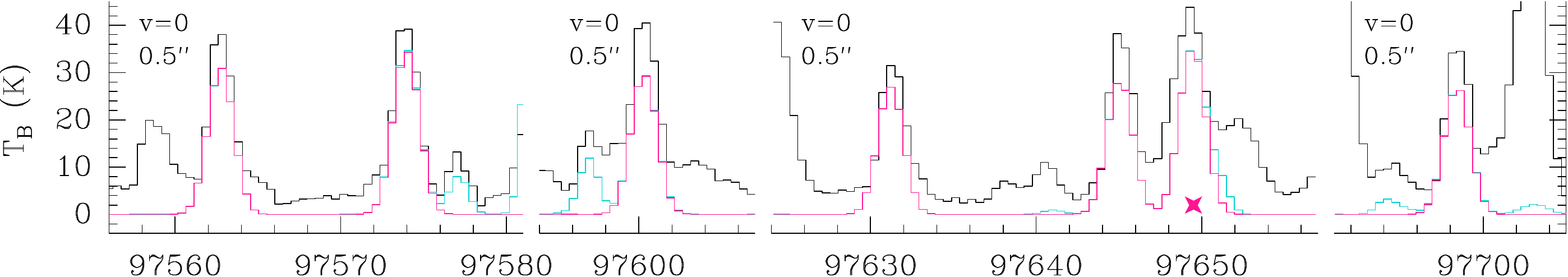}\\[0.2cm]
    \includegraphics[width=\textwidth]{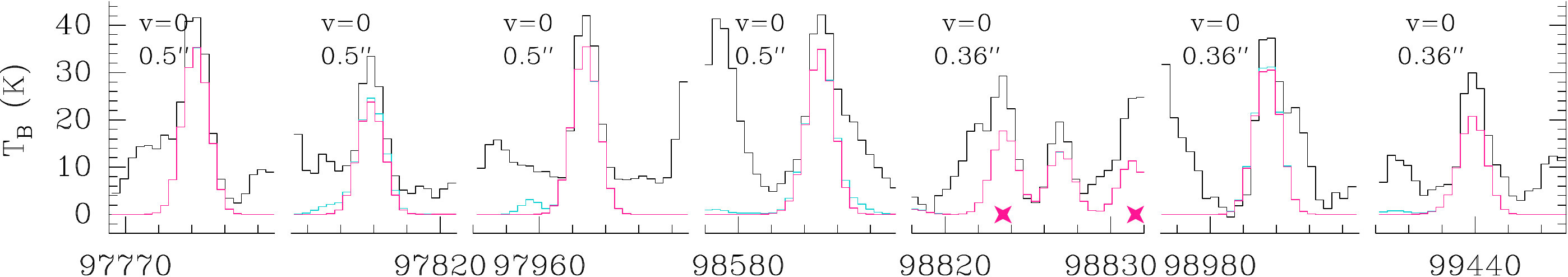}\\[0.2cm]
    \includegraphics[width=\textwidth]{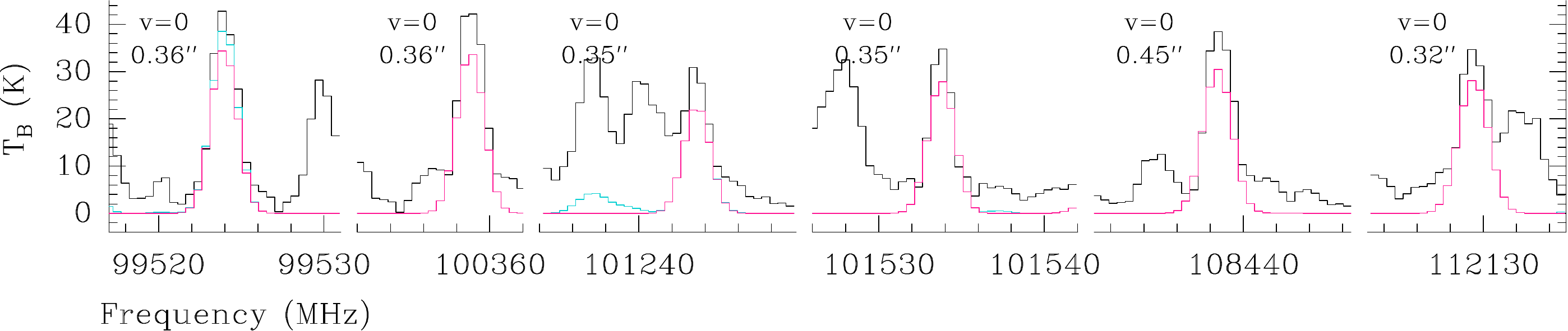}
    \caption{Same as Fig.\,\ref{fig:spec_met}, but for \et.} 
    \label{fig:spec_et}
\end{figure*}

\begin{figure*}
    \includegraphics[width=\textwidth]{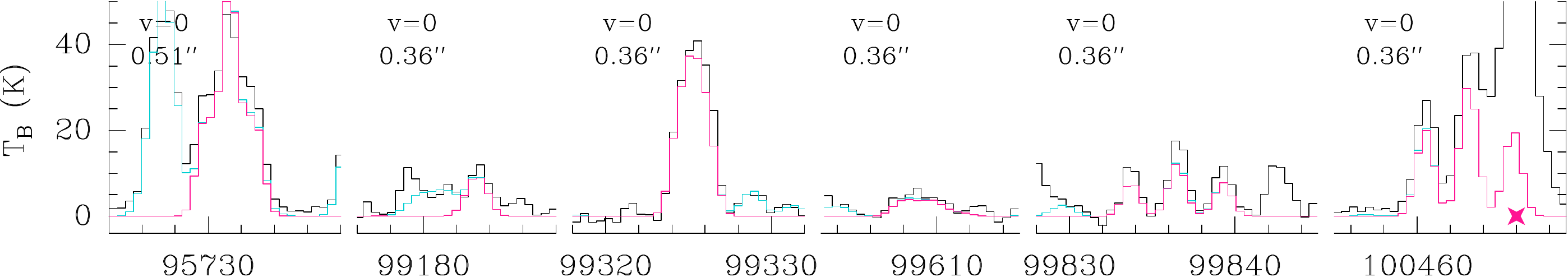}\\[0.2cm]
    \includegraphics[width=\textwidth]{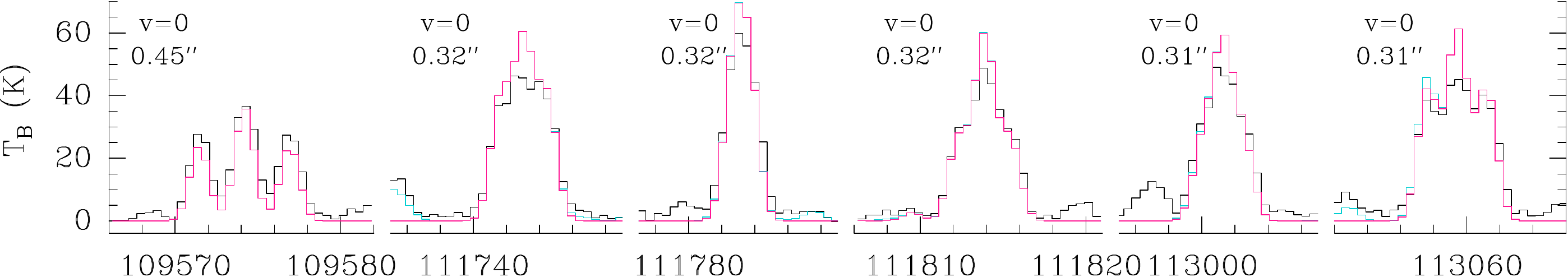}\\[0.2cm]
    \includegraphics[width=\textwidth]{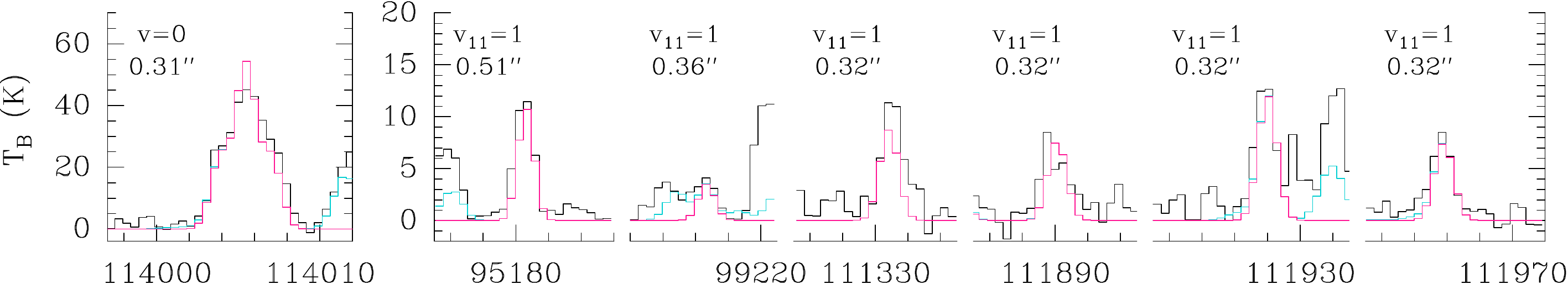}\\[0.2cm]
    \includegraphics[width=.77\textwidth]{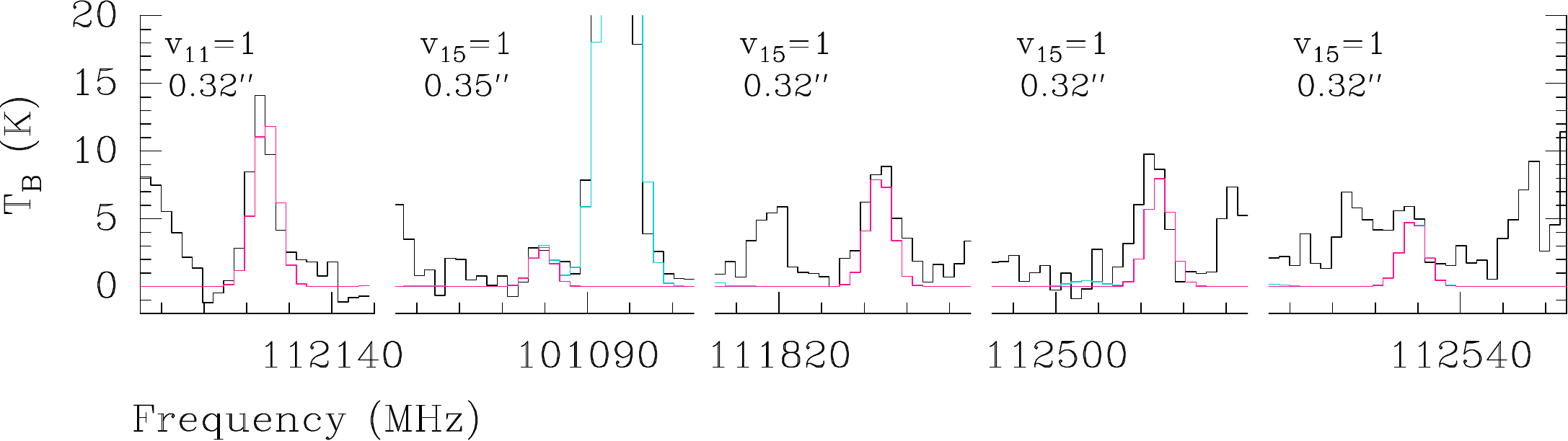}
    \caption{Same as Fig.\,\ref{fig:spec_met}, but for \dme and at position N1S1.}
    \label{fig:spec_dme}
\end{figure*}

\begin{figure*}
    \includegraphics[width=\textwidth]{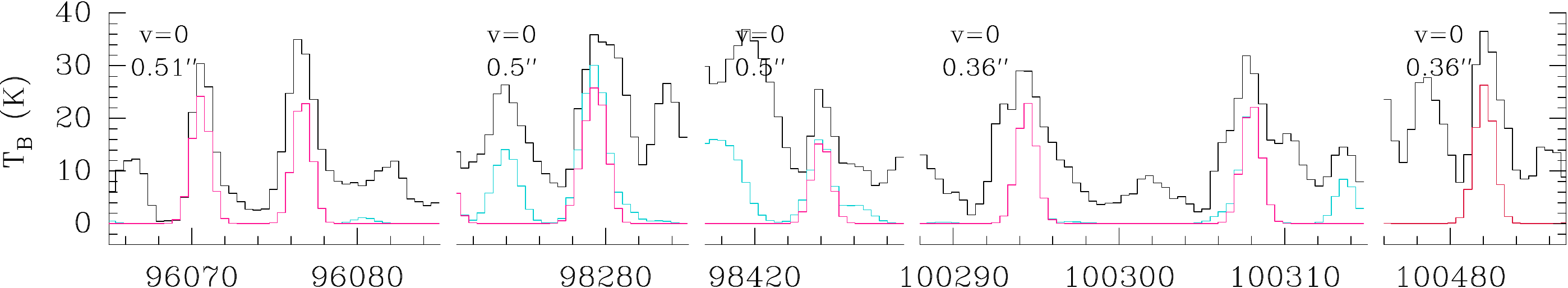}\\[0.2cm]
    \includegraphics[width=\textwidth]{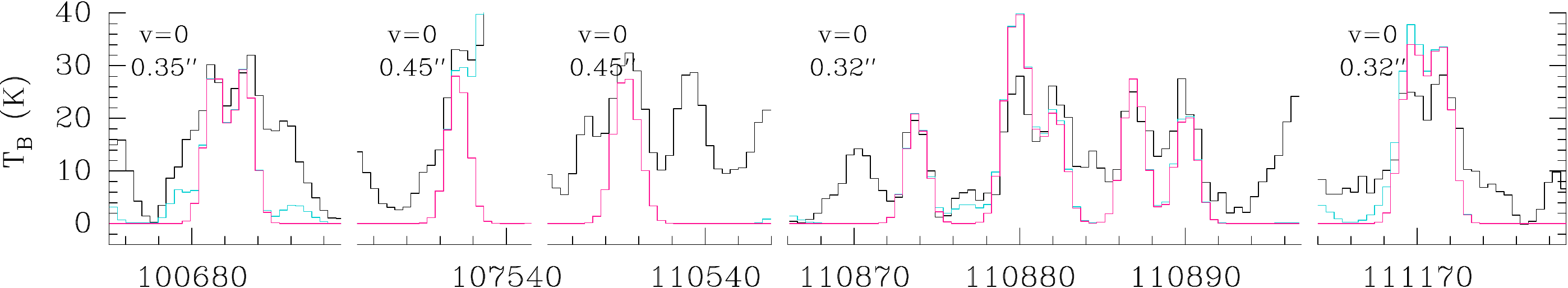}\\[0.2cm]
    \includegraphics[width=\textwidth]{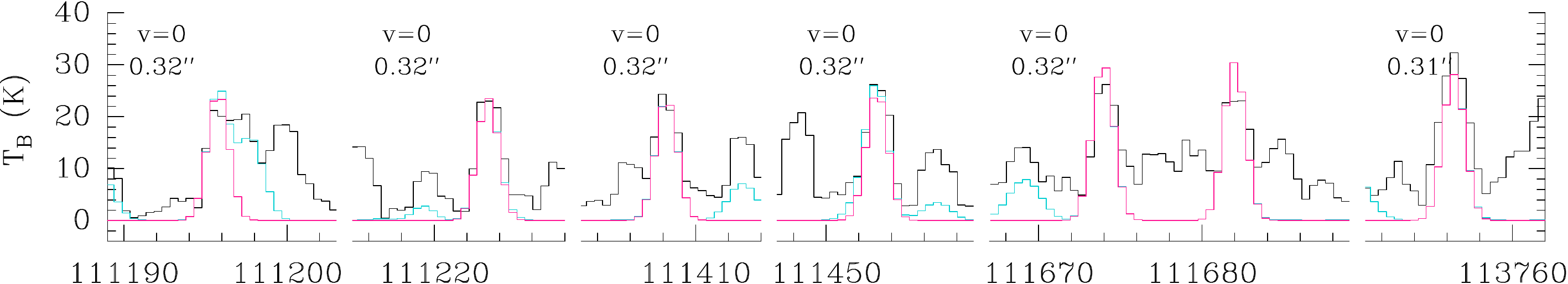}\\[0.2cm]
    \includegraphics[width=\textwidth]{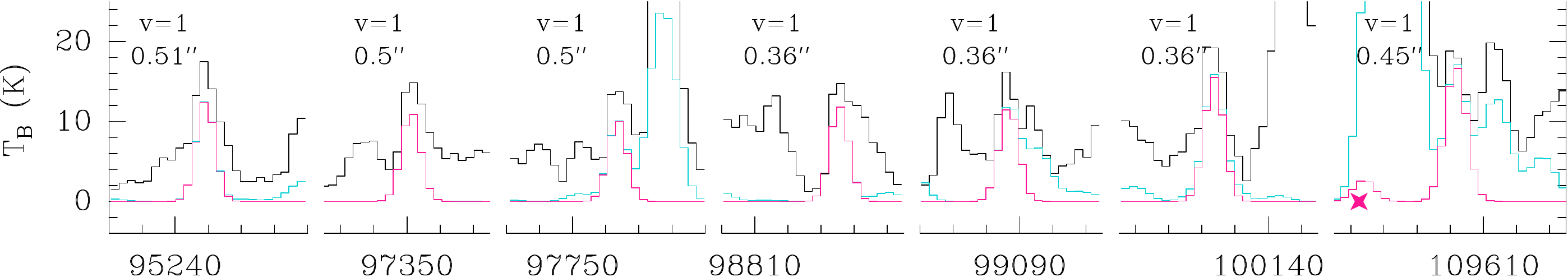}\\[0.2cm]
    \includegraphics[width=0.74\textwidth]{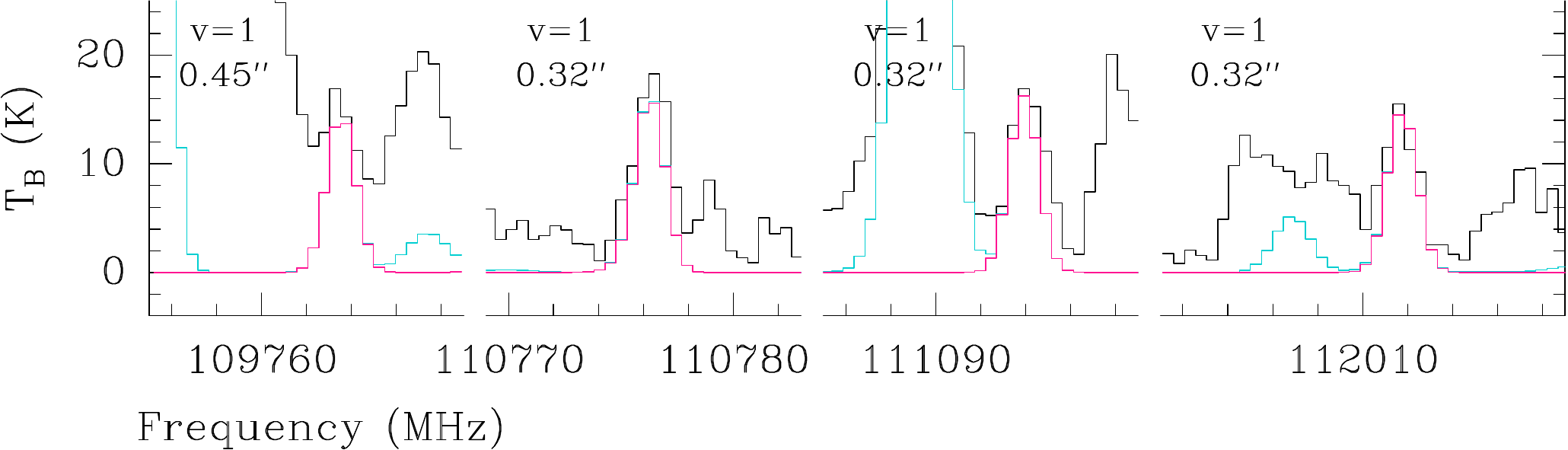}
    \caption{Same as Fig.\,\ref{fig:spec_met}, but for \mf.}
    \label{fig:spec_mf}
\end{figure*}

\begin{figure*}
    \includegraphics[width=\textwidth]{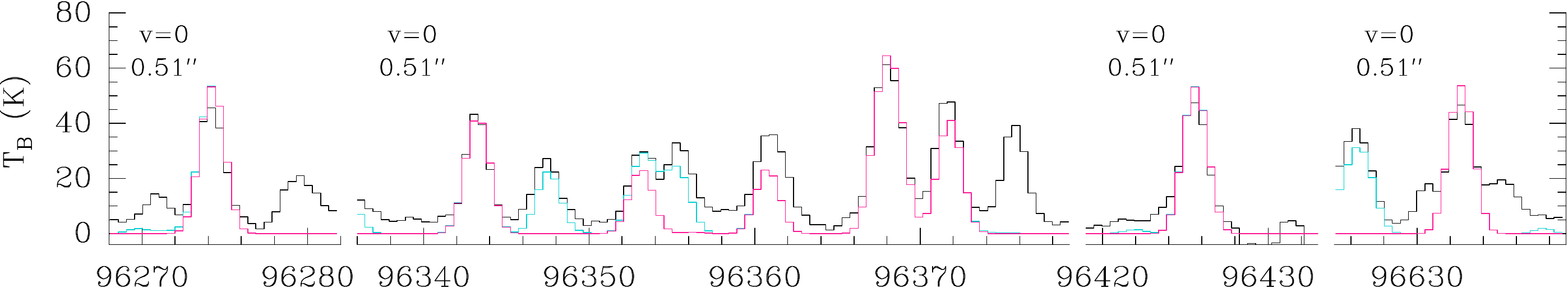}\\[0.2cm]
    \includegraphics[width=\textwidth]{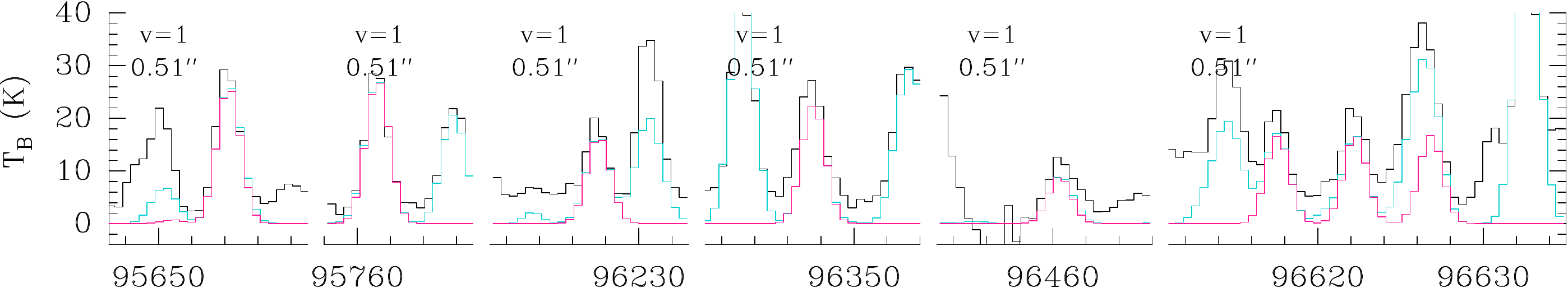}\\[0.2cm]
    \includegraphics[width=0.74\textwidth]{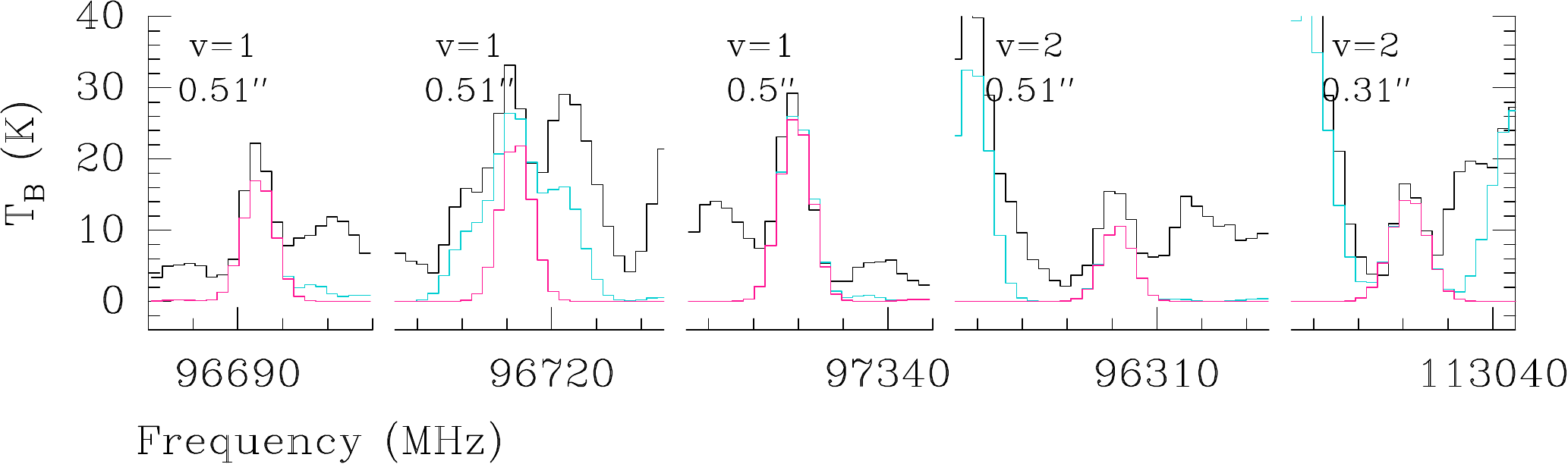}
    \caption{Same as Fig.\,\ref{fig:spec_met}, but for \ad.}
    \label{fig:spec_ad}
\end{figure*}

\begin{figure*}
    \includegraphics[width=\textwidth]{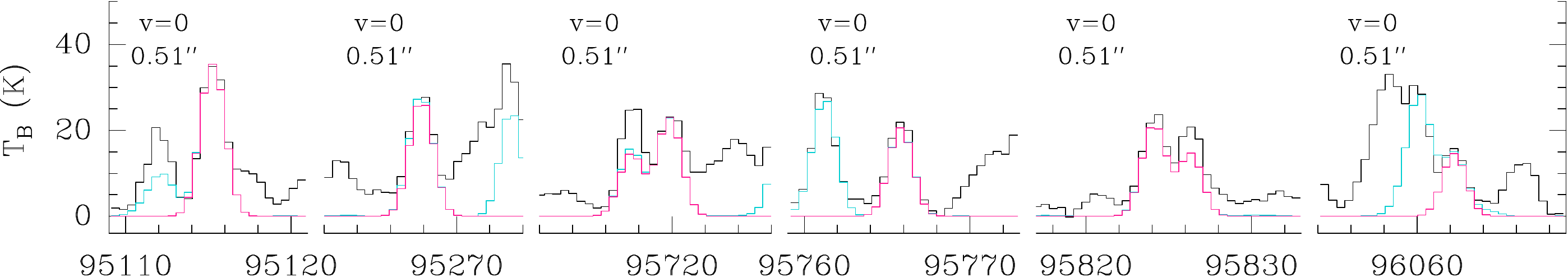}\\[0.2cm]
    \includegraphics[width=\textwidth]{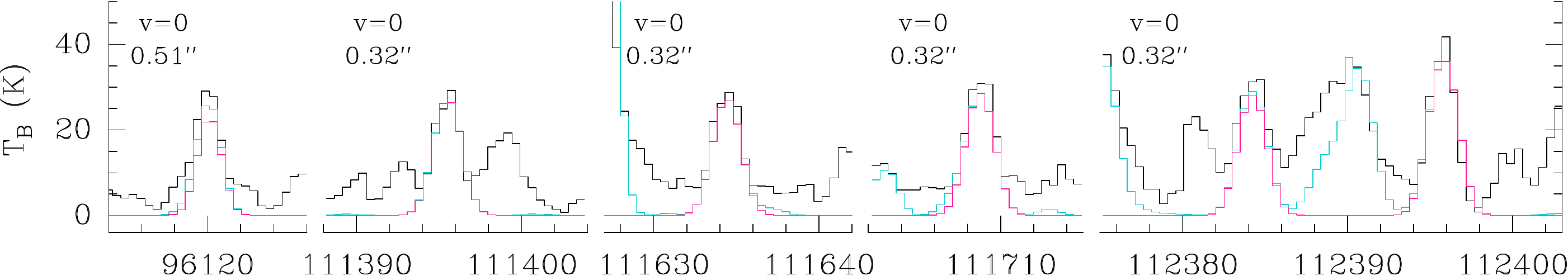}\\[0.2cm]
    \includegraphics[width=0.36\textwidth]{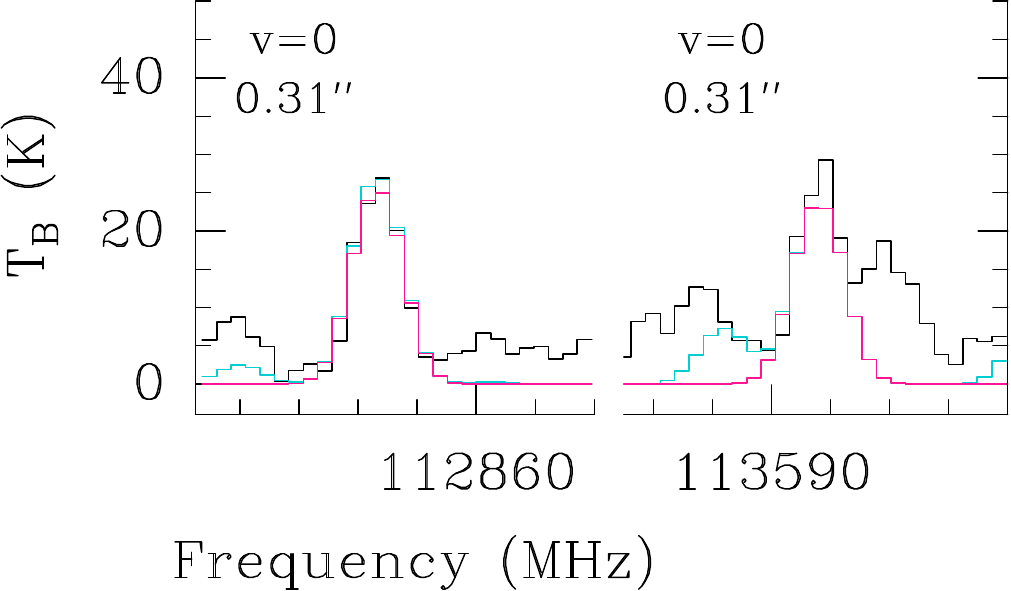}
    \caption{Same as Fig.\,\ref{fig:spec_met}, but for \mic.}
    \label{fig:spec_mic}
\end{figure*}

\begin{figure*}
    \includegraphics[width=\textwidth]{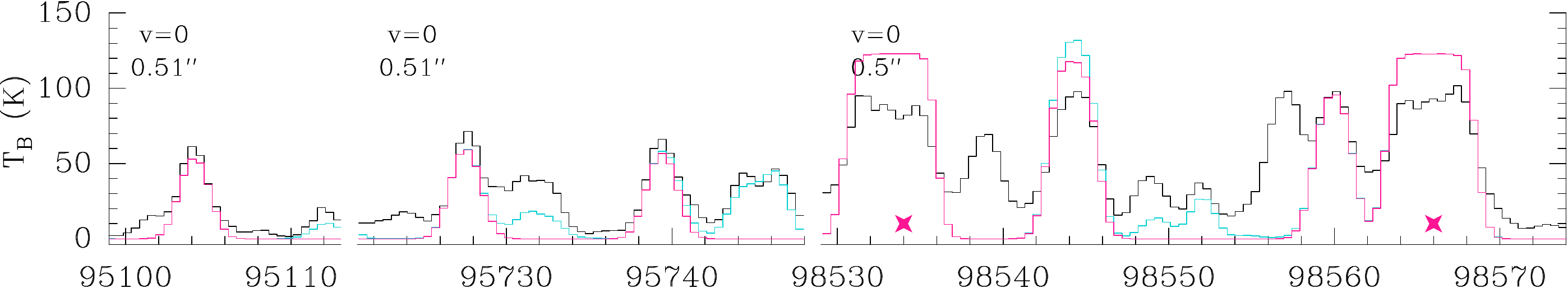}\\[0.2cm]
    \includegraphics[width=\textwidth]{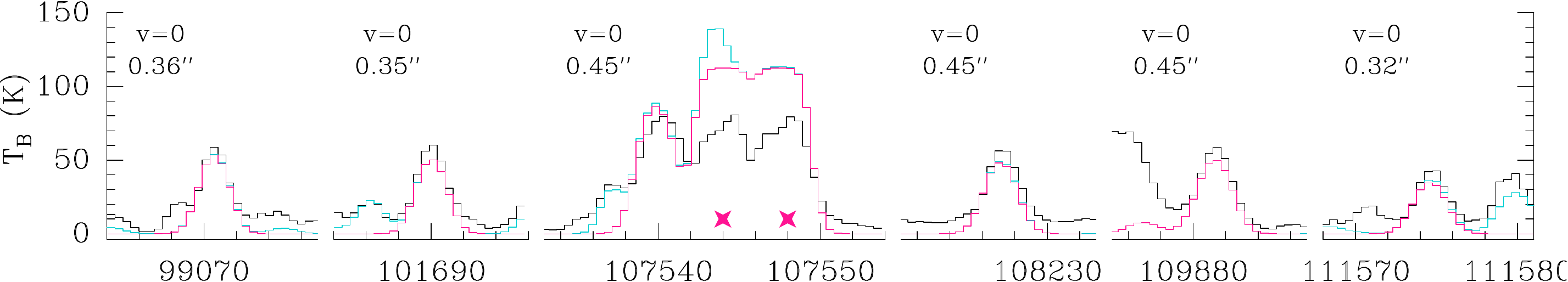}\\[0.2cm]
    \includegraphics[width=\textwidth]{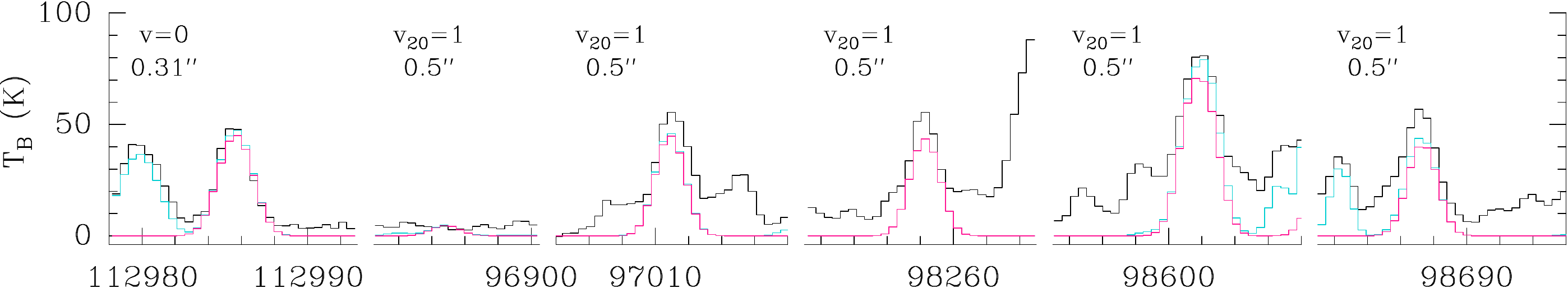}\\[0.2cm]
    \includegraphics[width=\textwidth]{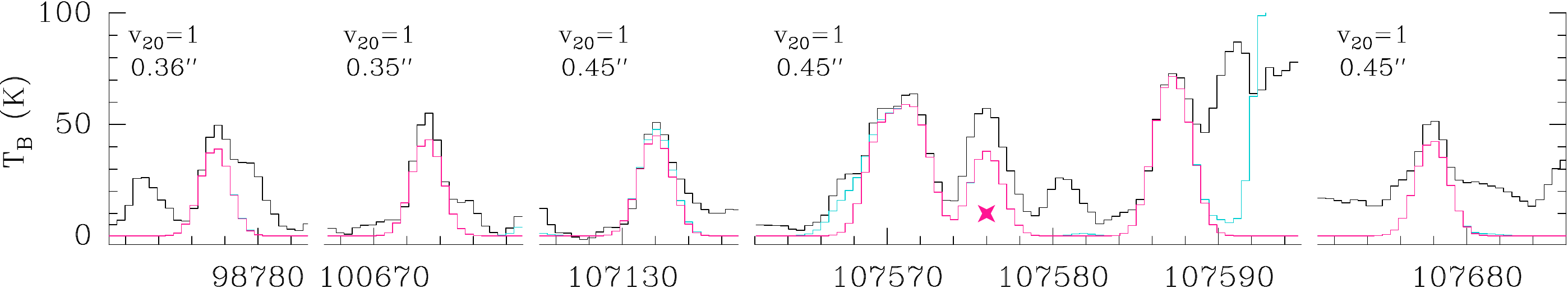}\\[0.2cm]
    \includegraphics[width=\textwidth]{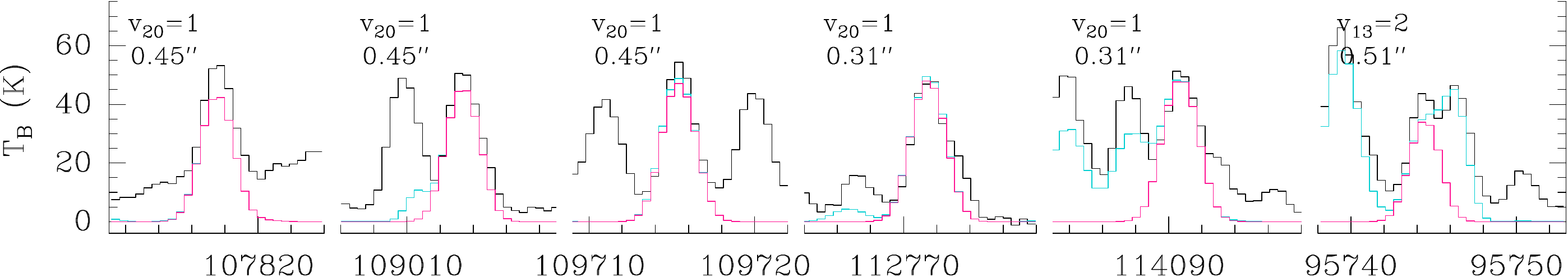}\\[0.2cm]
    \includegraphics[width=0.73\textwidth]{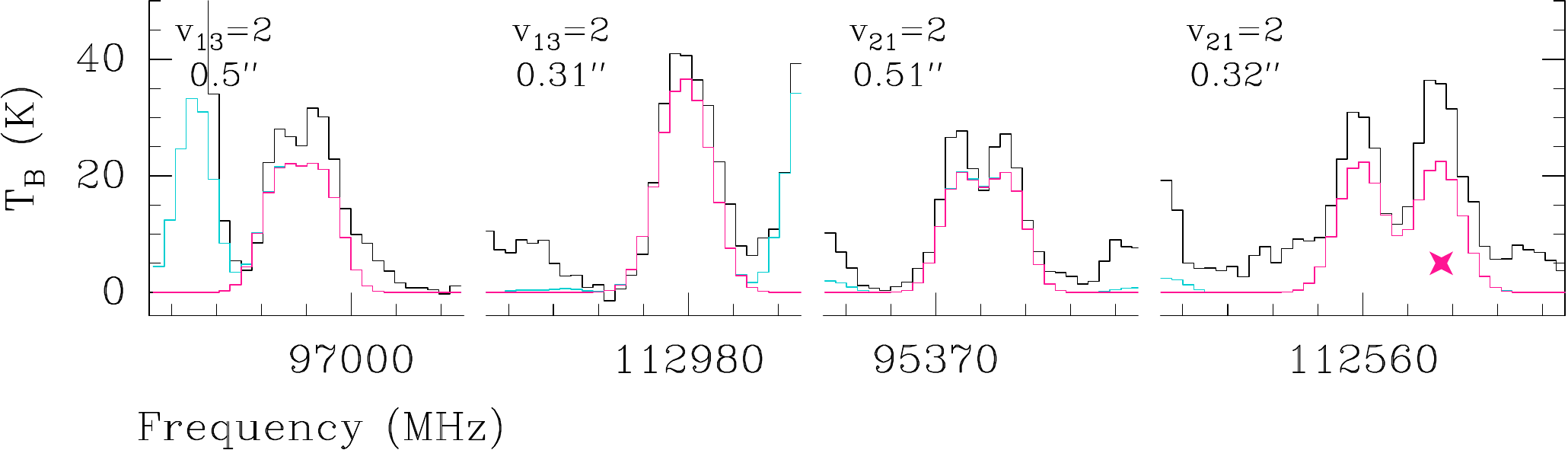}
    \caption{Same as Fig.\,\ref{fig:spec_met}, but for \etc.}
    \label{fig:spec_etc}
\end{figure*}

\begin{figure*}
    \includegraphics[width=\textwidth]{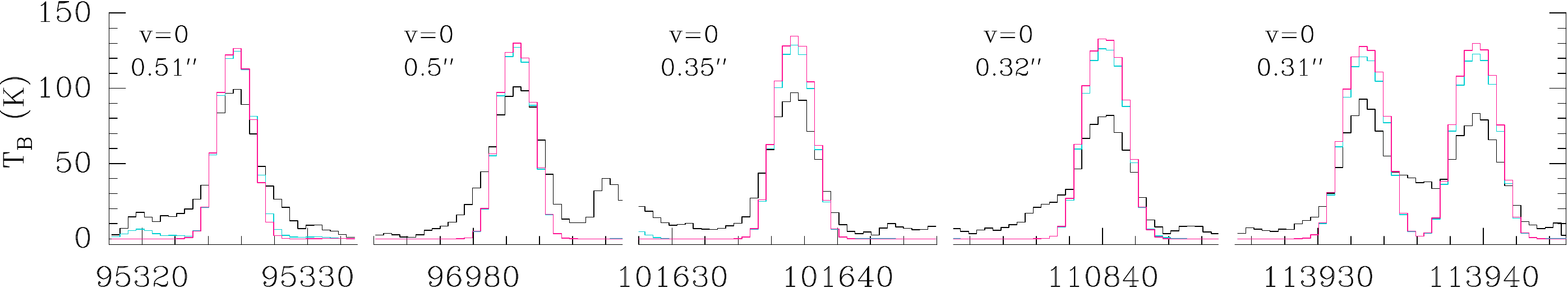}\\[0.2cm]
    \includegraphics[width=\textwidth]{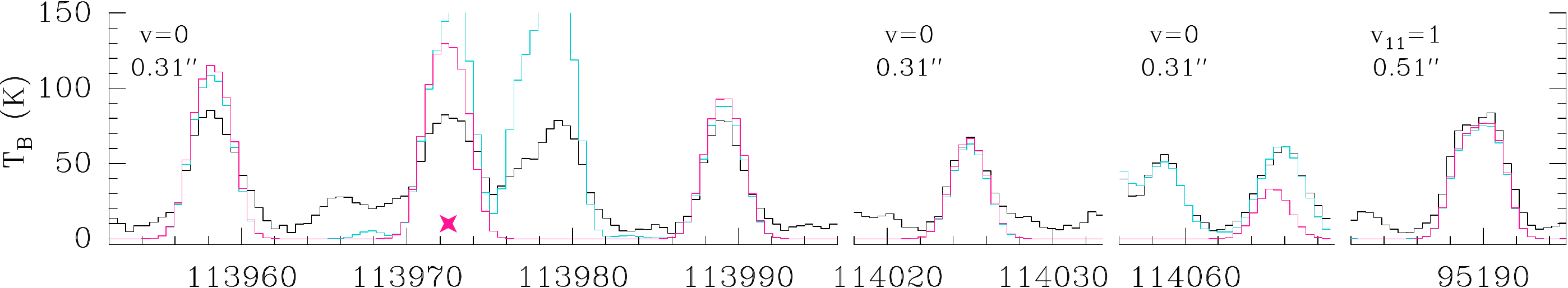}\\[0.2cm]
    \includegraphics[width=\textwidth]{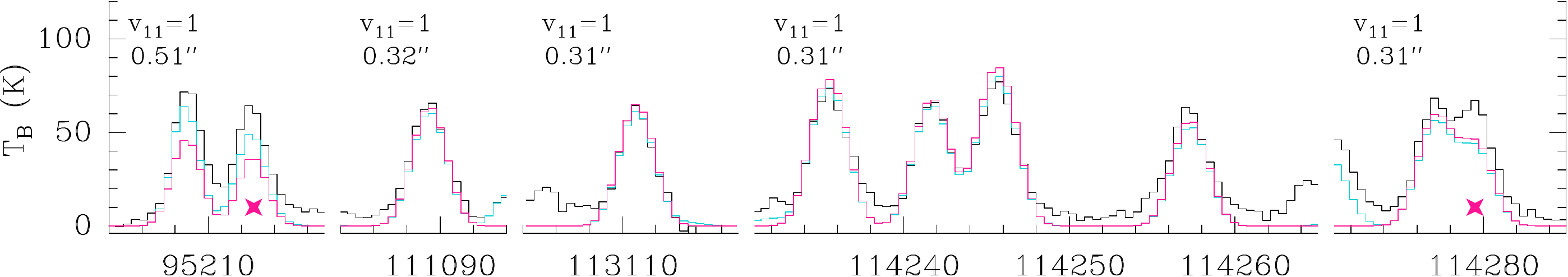}\\[0.2cm]
    \includegraphics[width=.74\textwidth]{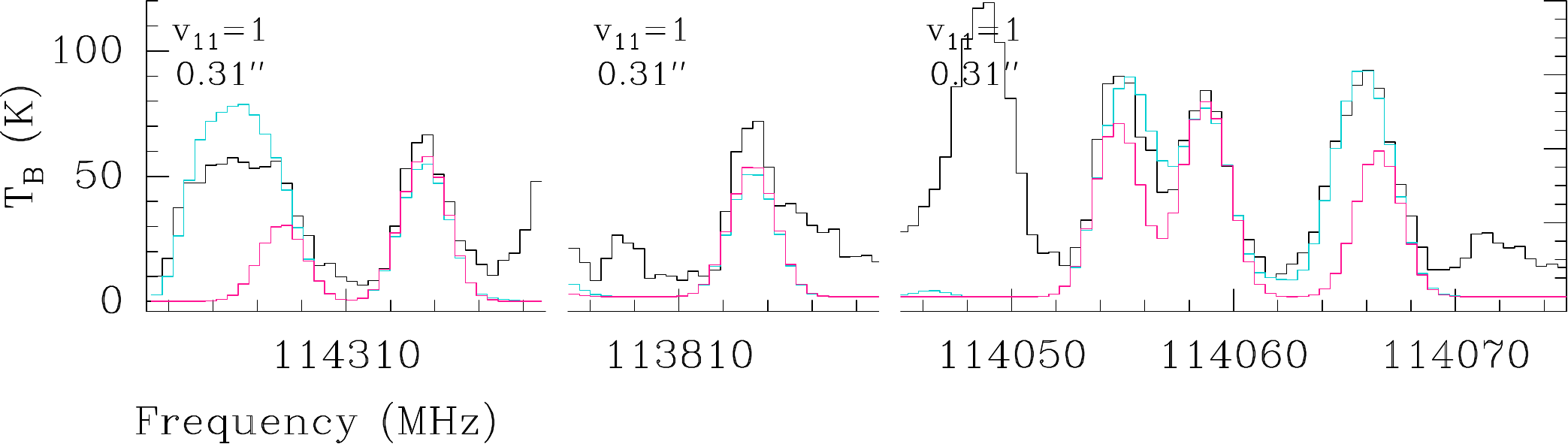}\\
    \caption{Same as Fig.\,\ref{fig:spec_met}, but for \vc.}
    \label{fig:spec_vc}
\end{figure*}

\begin{figure*}
    \includegraphics[width=\textwidth]{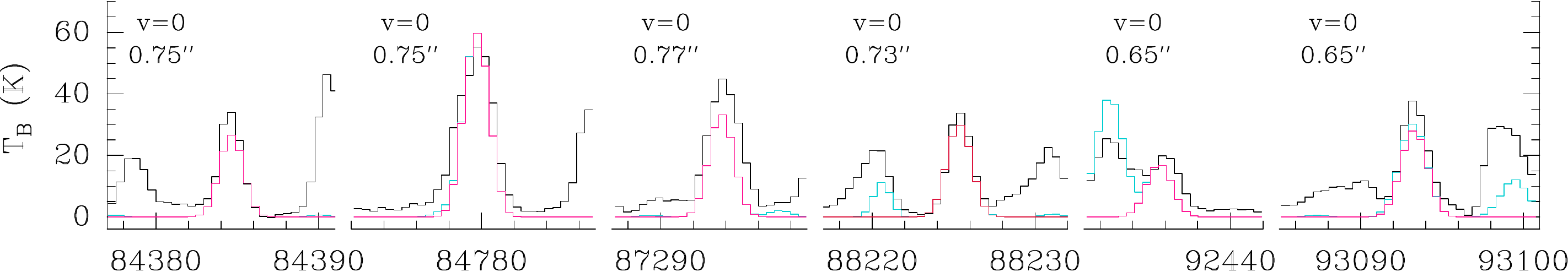}\\[0.2cm]
    \includegraphics[width=\textwidth]{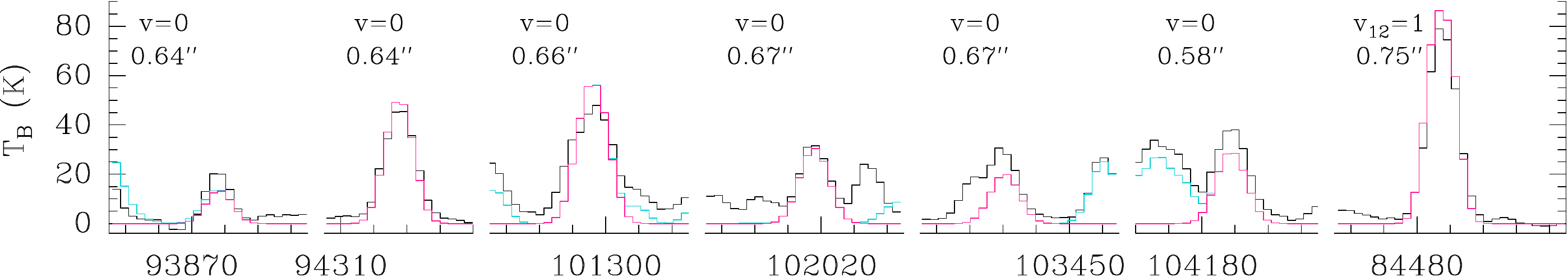}\\[0.2cm]
    \includegraphics[width=\textwidth]{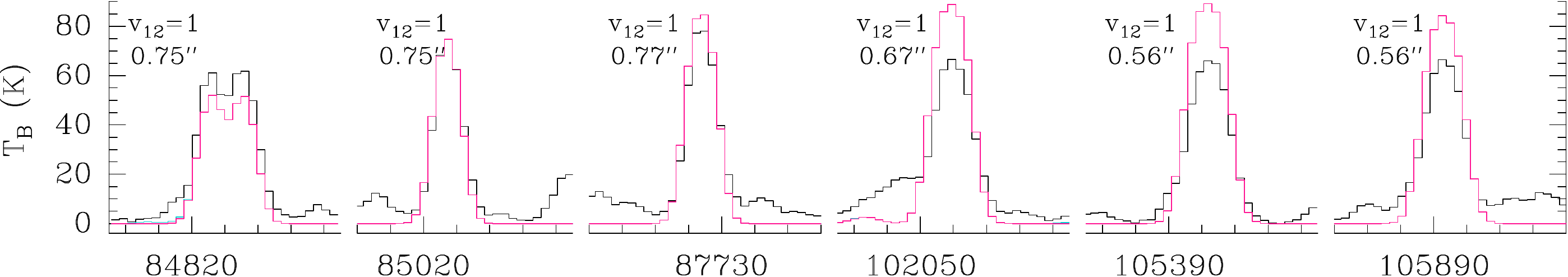}\\[0.2cm]
    \includegraphics[width=.68\textwidth]{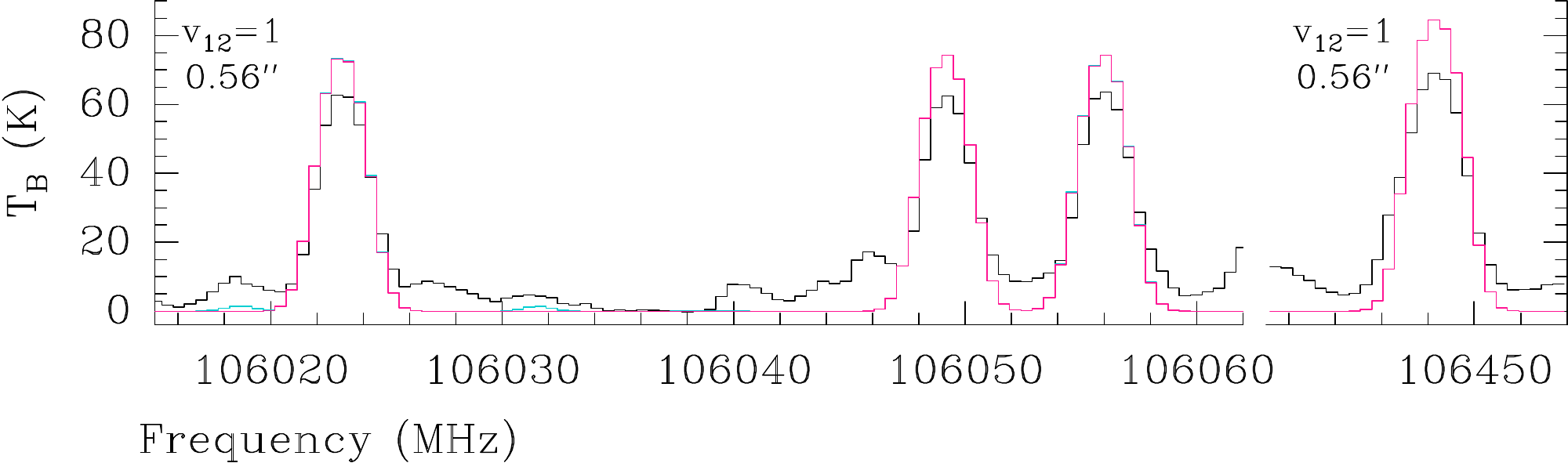}\\
    \caption{Same as Fig.\,\ref{fig:spec_met}, but for \fmm.}
    \label{fig:spec_fmm}
\end{figure*}

Figures\,\ref{fig:spec_met}--\ref{fig:spec_fmm} show observed and synthetic spectra of the analysed COMs at position N1S or, in case of \dme, N1S1. Only transitions that are used to create the respective population diagram are shown.

\section{Population diagrams}\label{app:popdiagrams}
\begin{figure*}[h!]
    \centering
    \includegraphics[width=0.48\textwidth]{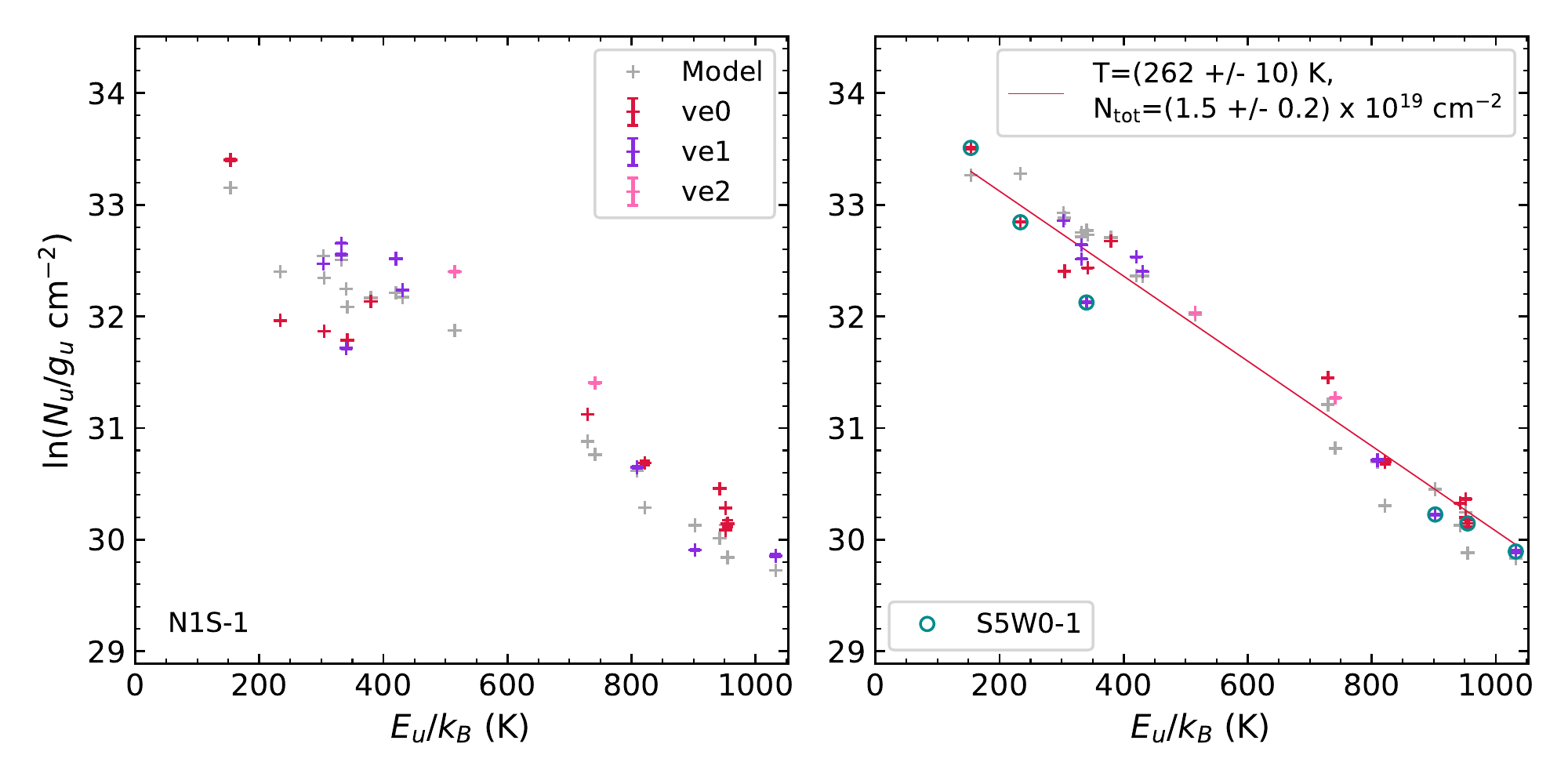}
    \includegraphics[width=0.48\textwidth]{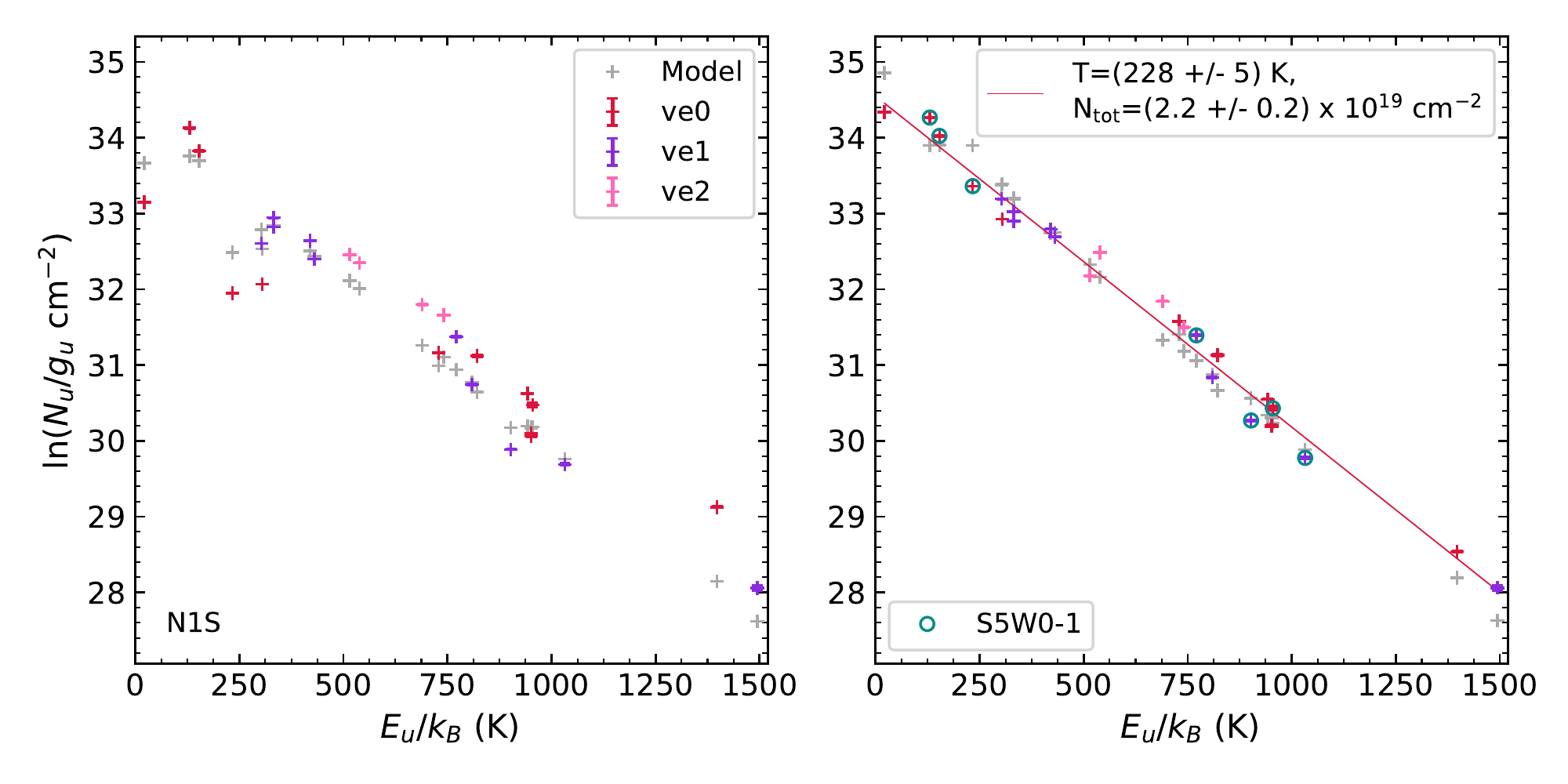}
    \includegraphics[width=0.48\textwidth]{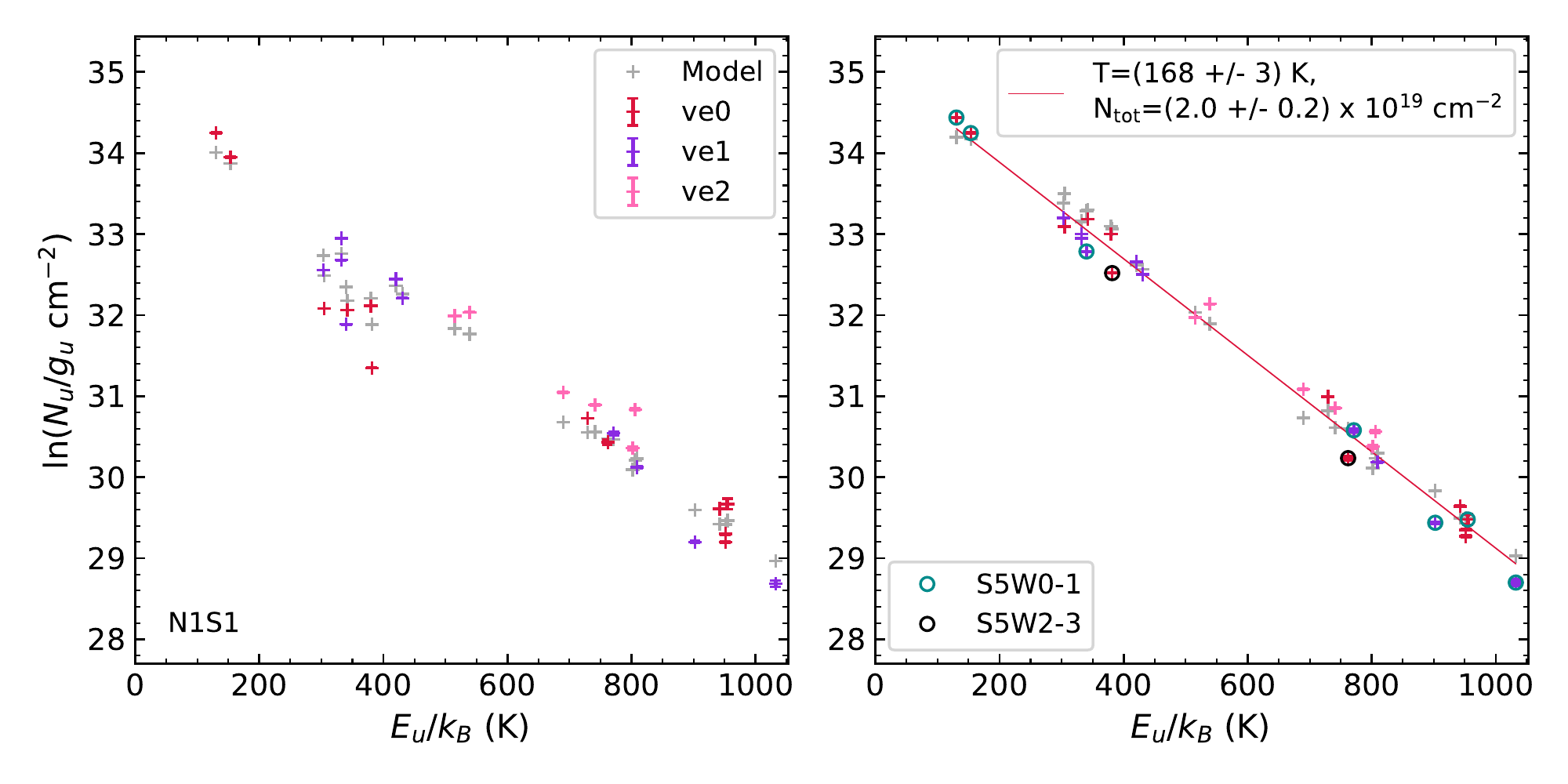}
    \includegraphics[width=0.48\textwidth]{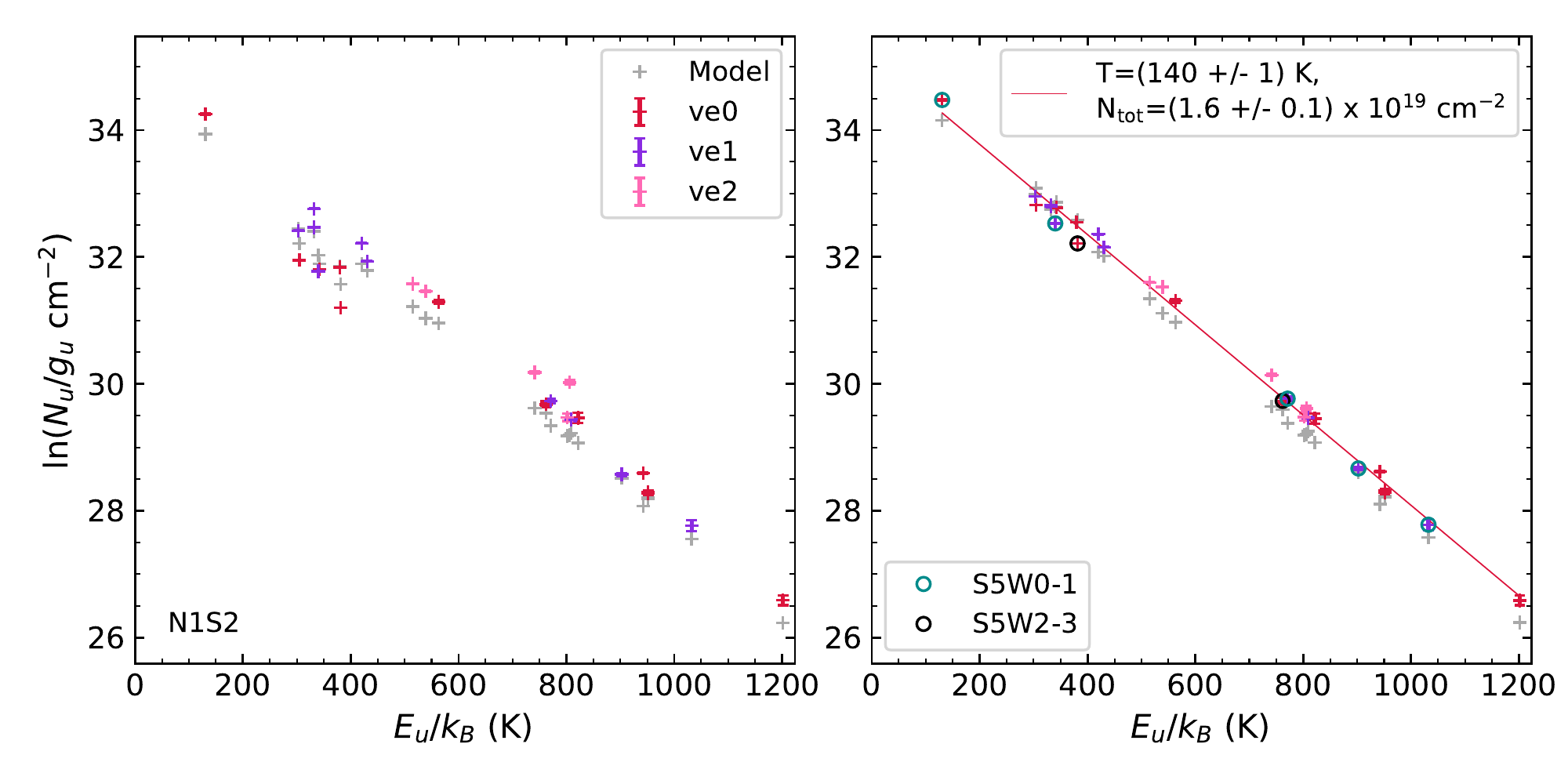}
    \includegraphics[width=0.48\textwidth]{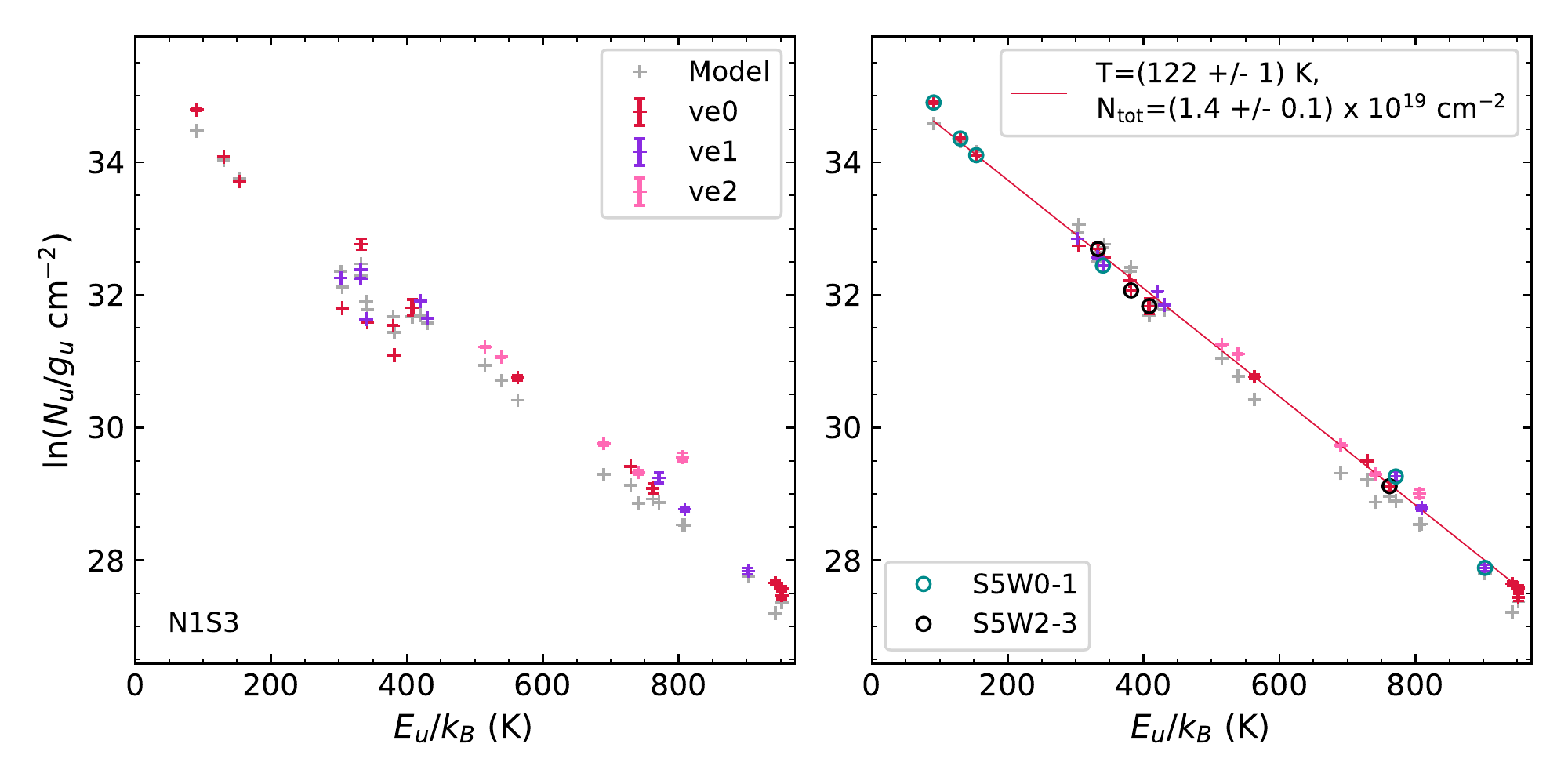}
    \includegraphics[width=0.48\textwidth]{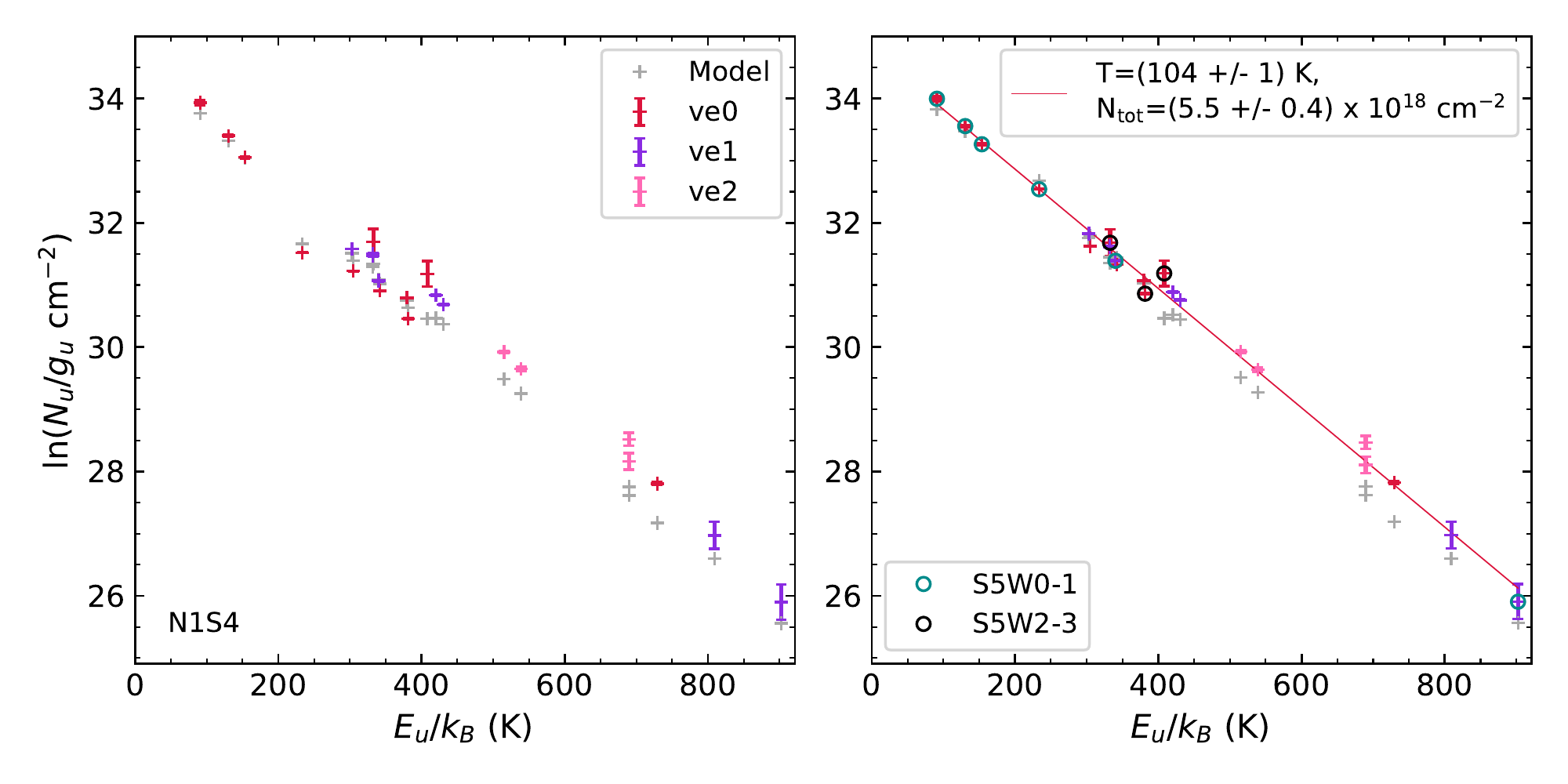}
    \includegraphics[width=0.48\textwidth]{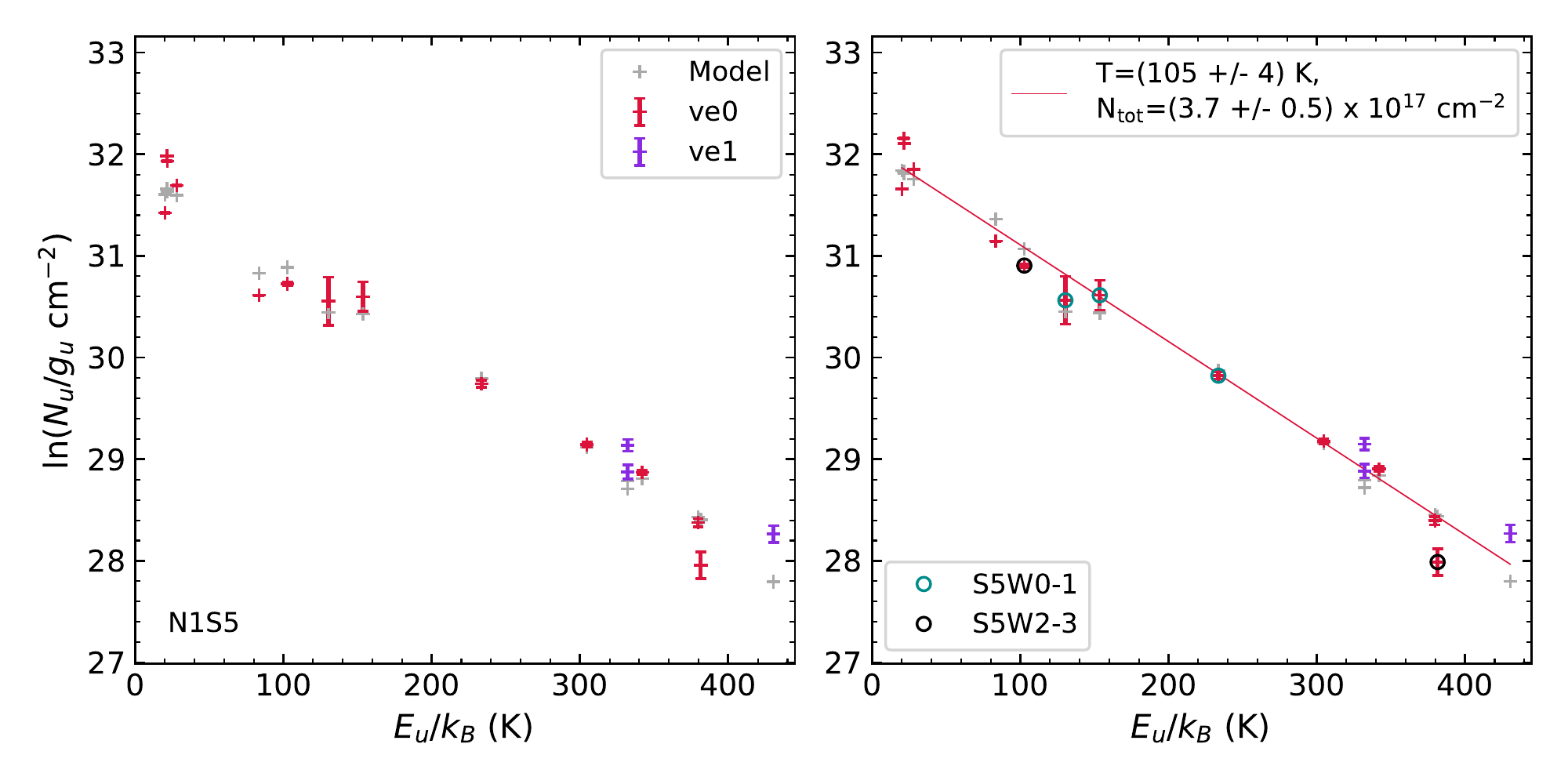}
    \includegraphics[width=0.48\textwidth]{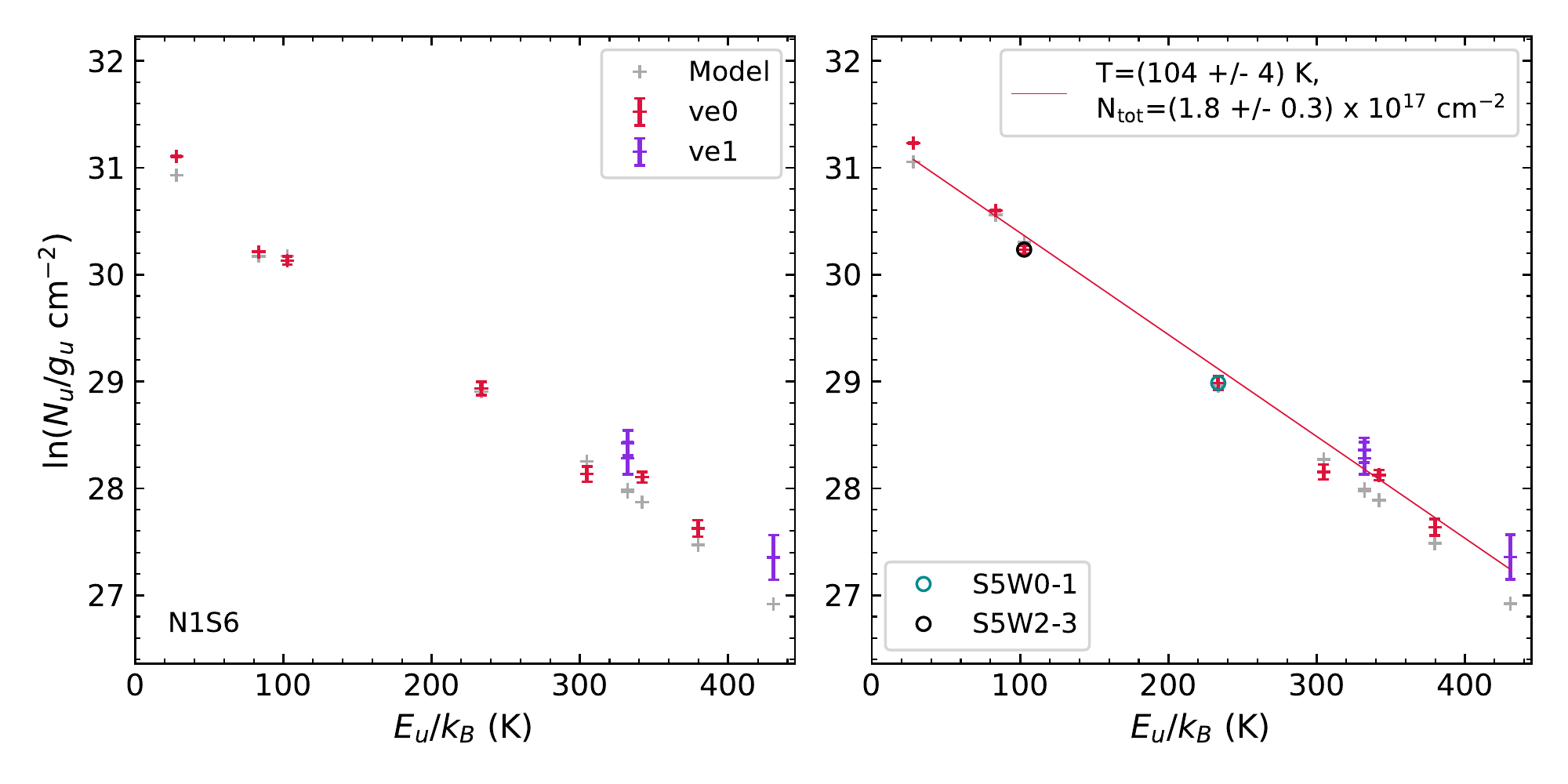}
    \includegraphics[width=0.48\textwidth]{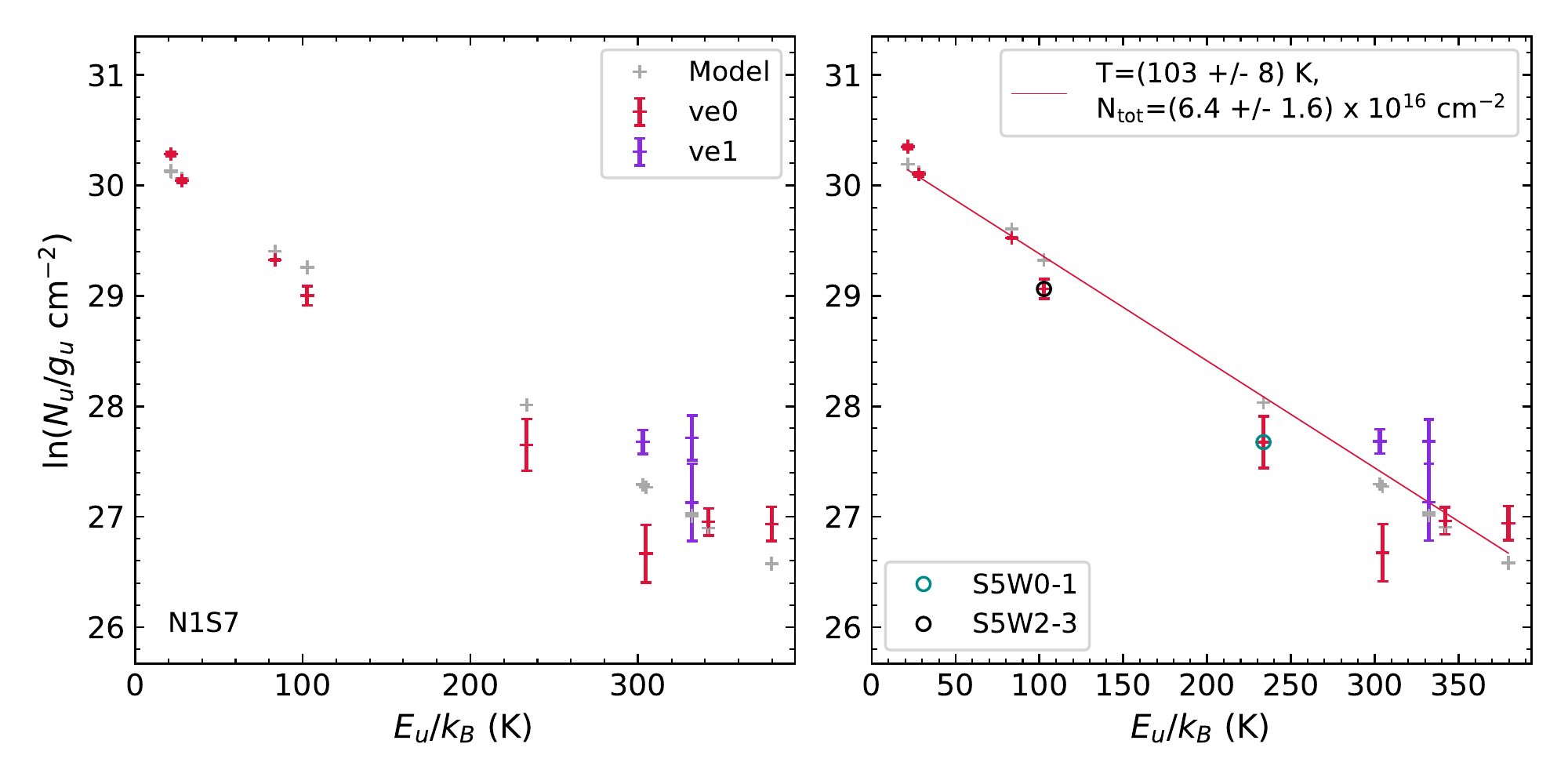}
    \includegraphics[width=0.48\textwidth]{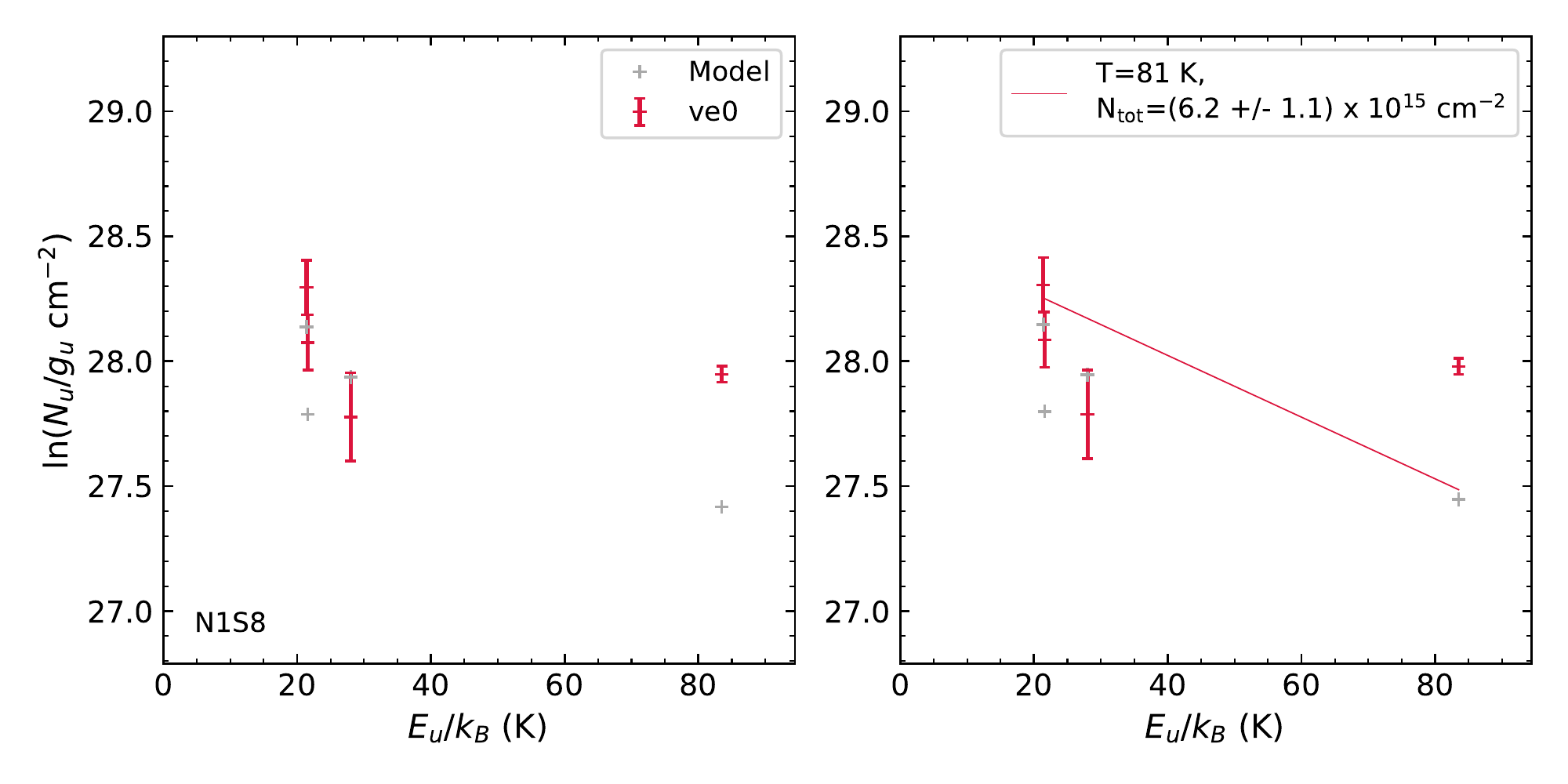}
    \caption{Population diagrams of \met at all positions to the south where the molecule is detected in setups 4--5. Each position takes two plots: the one with the name of the position indicated in the lower left corner and the one to the right. Observed data points are shown in various colours as indicated in the upper right corner of the respective left plot while synthetic data points are shown in grey. No corrections are applied in the respective left panels, while in the right plots corrections for opacity and contamination by other molecules have been considered for both the observed and synthetic populations. The red line is a linear fit to all observed data points (in linear-logarithmic space). The results of the fits are shown in the respective right panels. Teal and black circles show observed data points from spectral windows 0--1 and 2--3 of observational setup 5, respectively, as indicated in the lower left corner in the right panels.}
    \label{fig:PD_met}
\end{figure*}

\begin{figure*}[h]
    \includegraphics[width=0.49\textwidth]{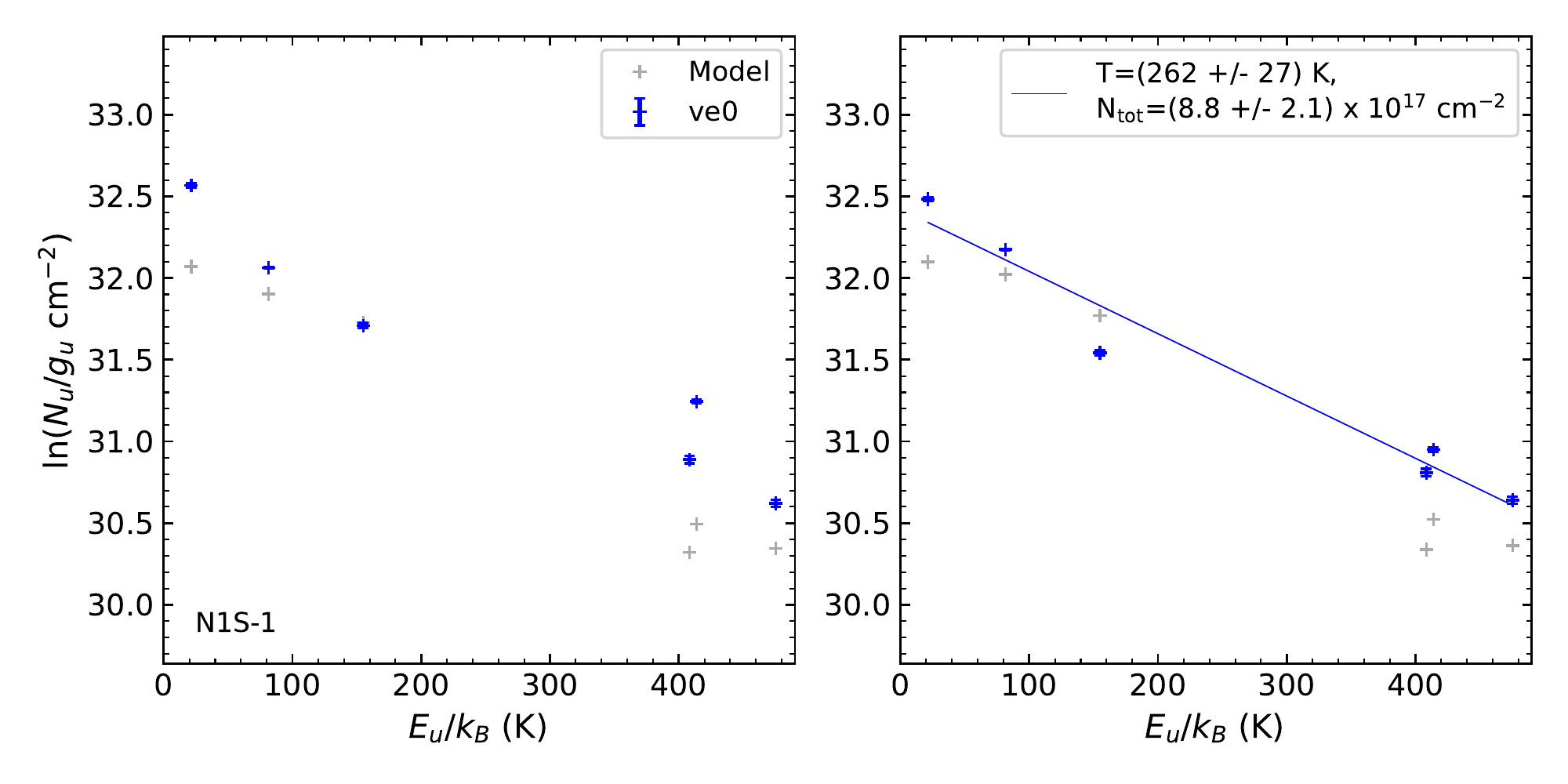}
    \includegraphics[width=0.49\textwidth]{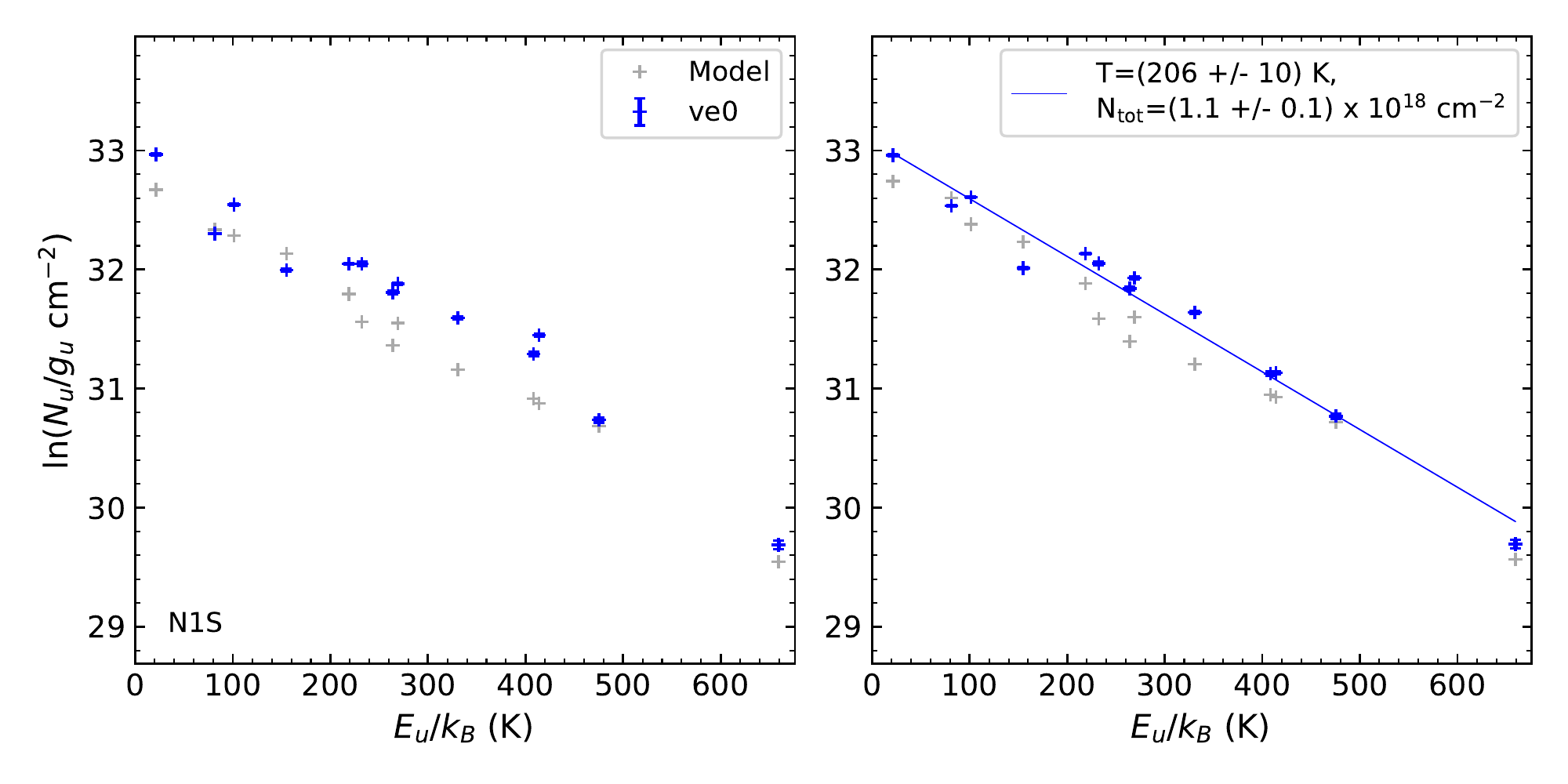}
    \includegraphics[width=0.49\textwidth]{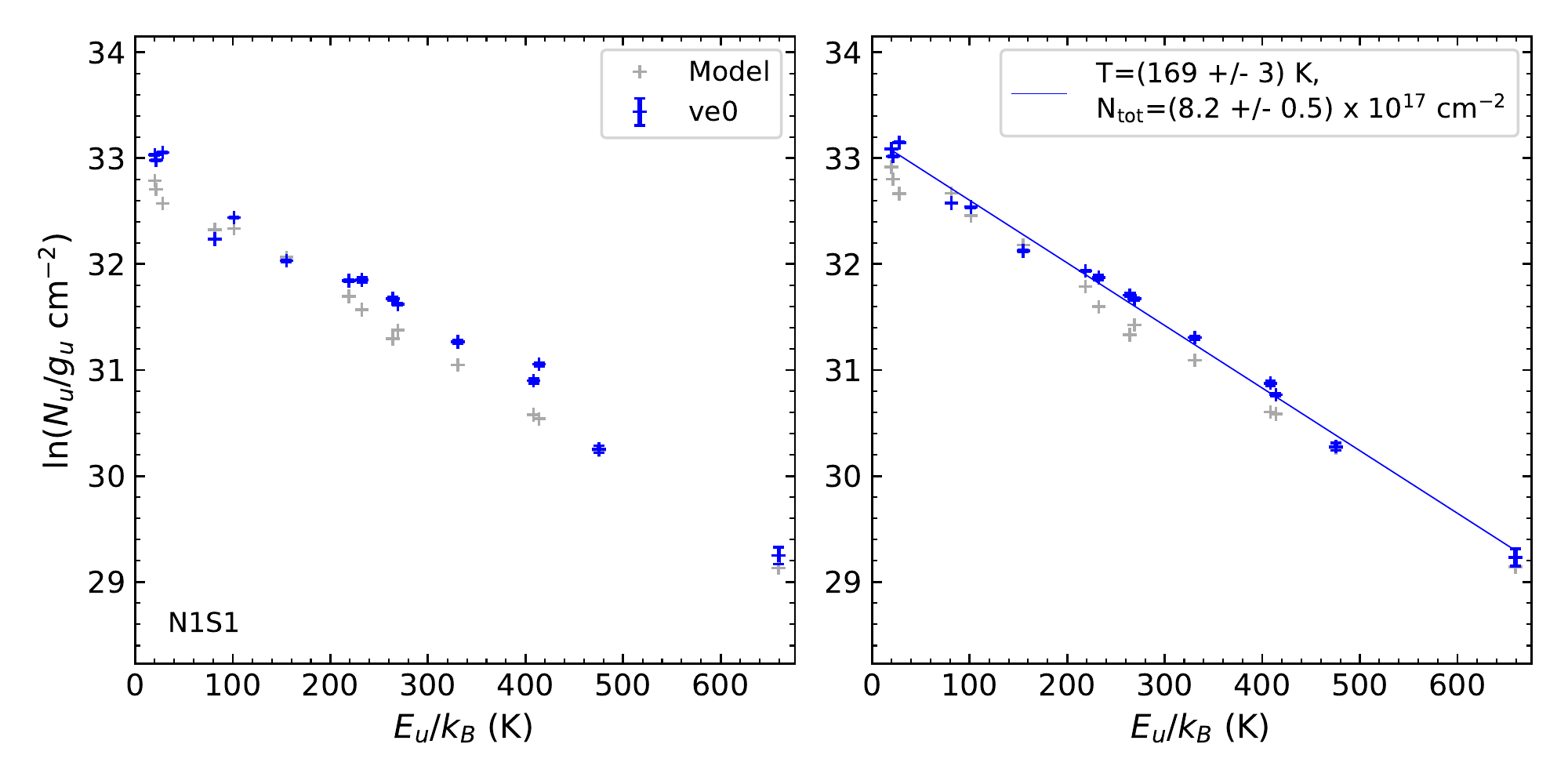}
    \includegraphics[width=0.49\textwidth]{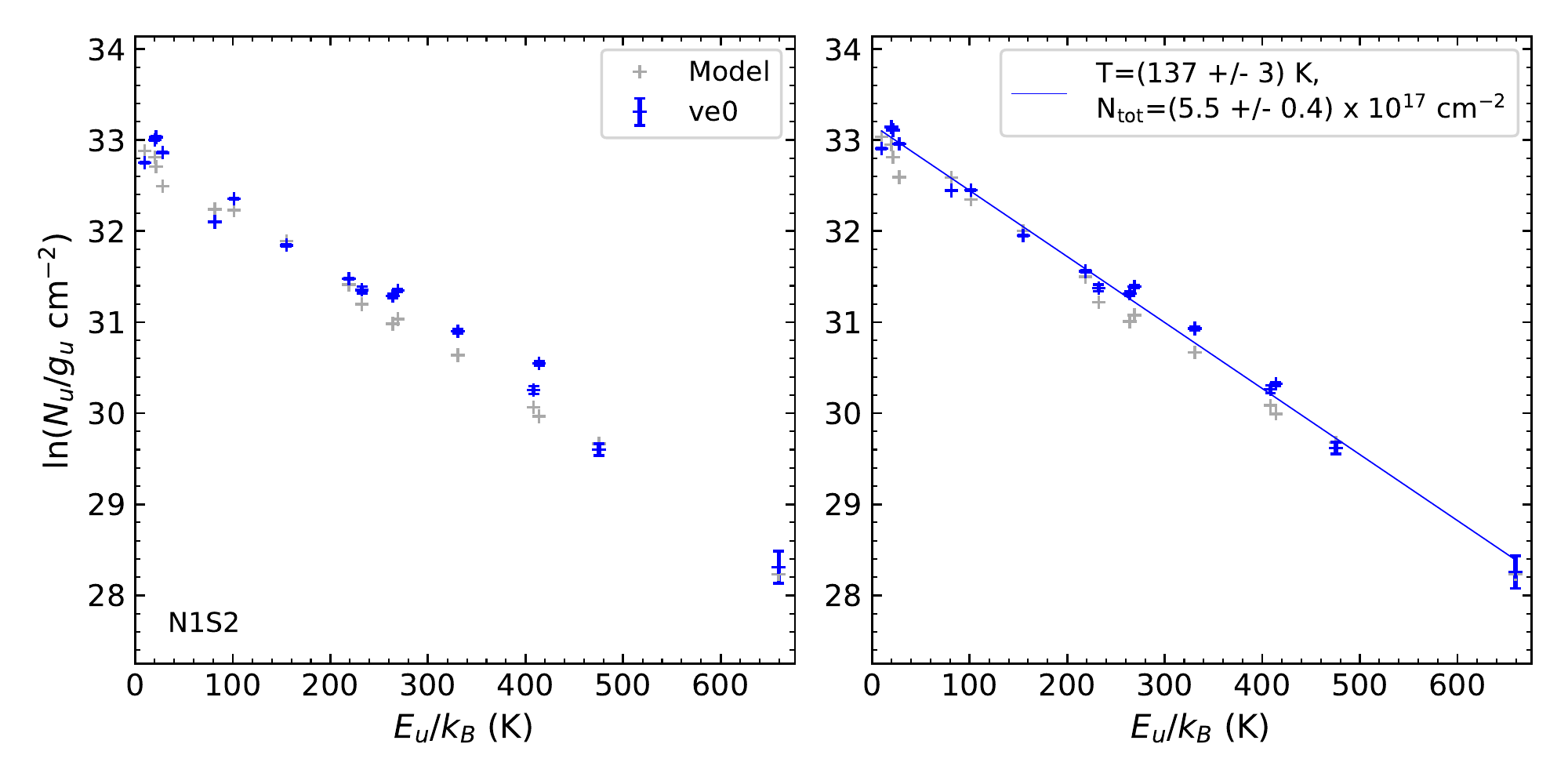}
    \includegraphics[width=0.49\textwidth]{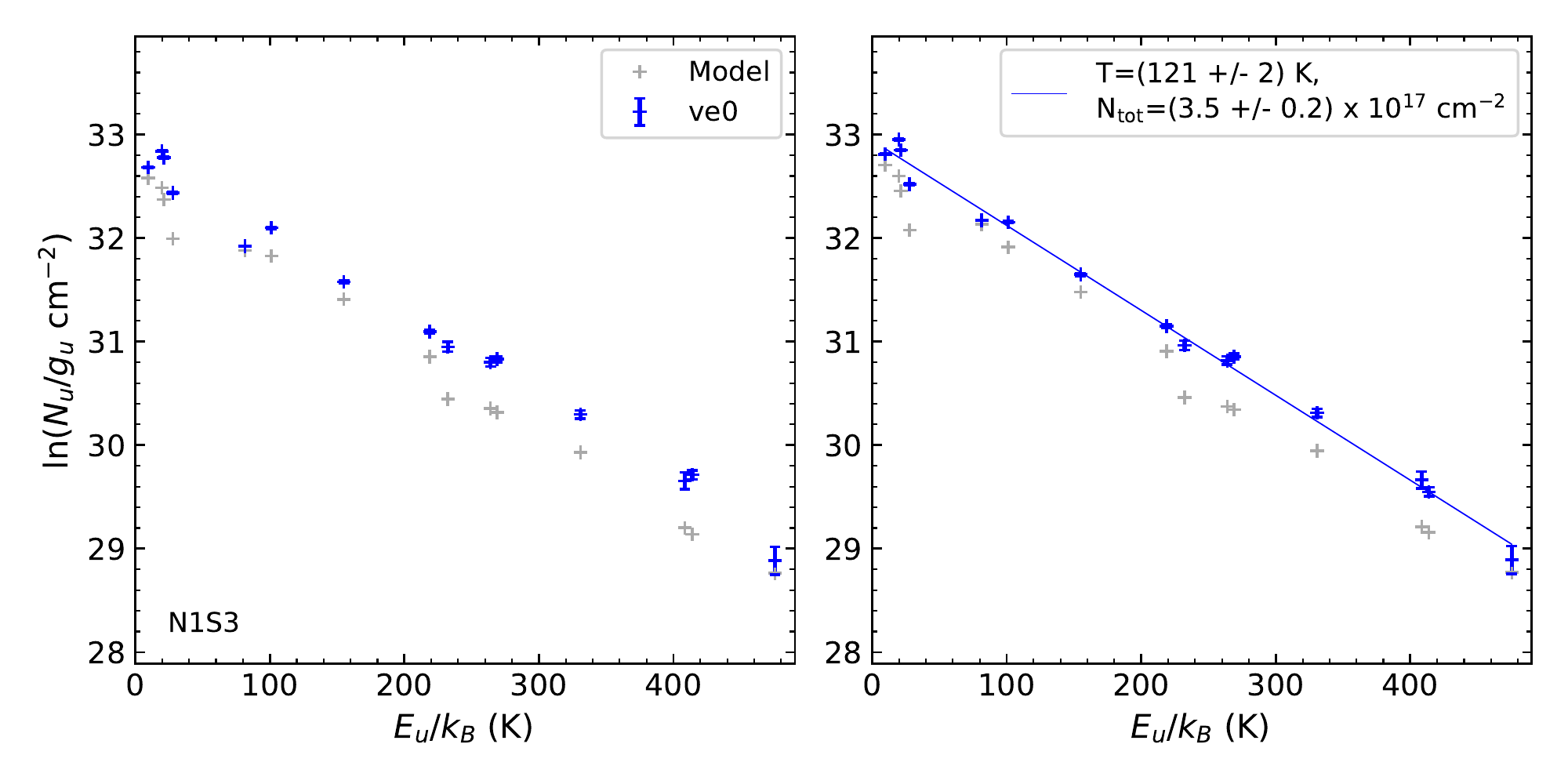}
    \includegraphics[width=0.49\textwidth]{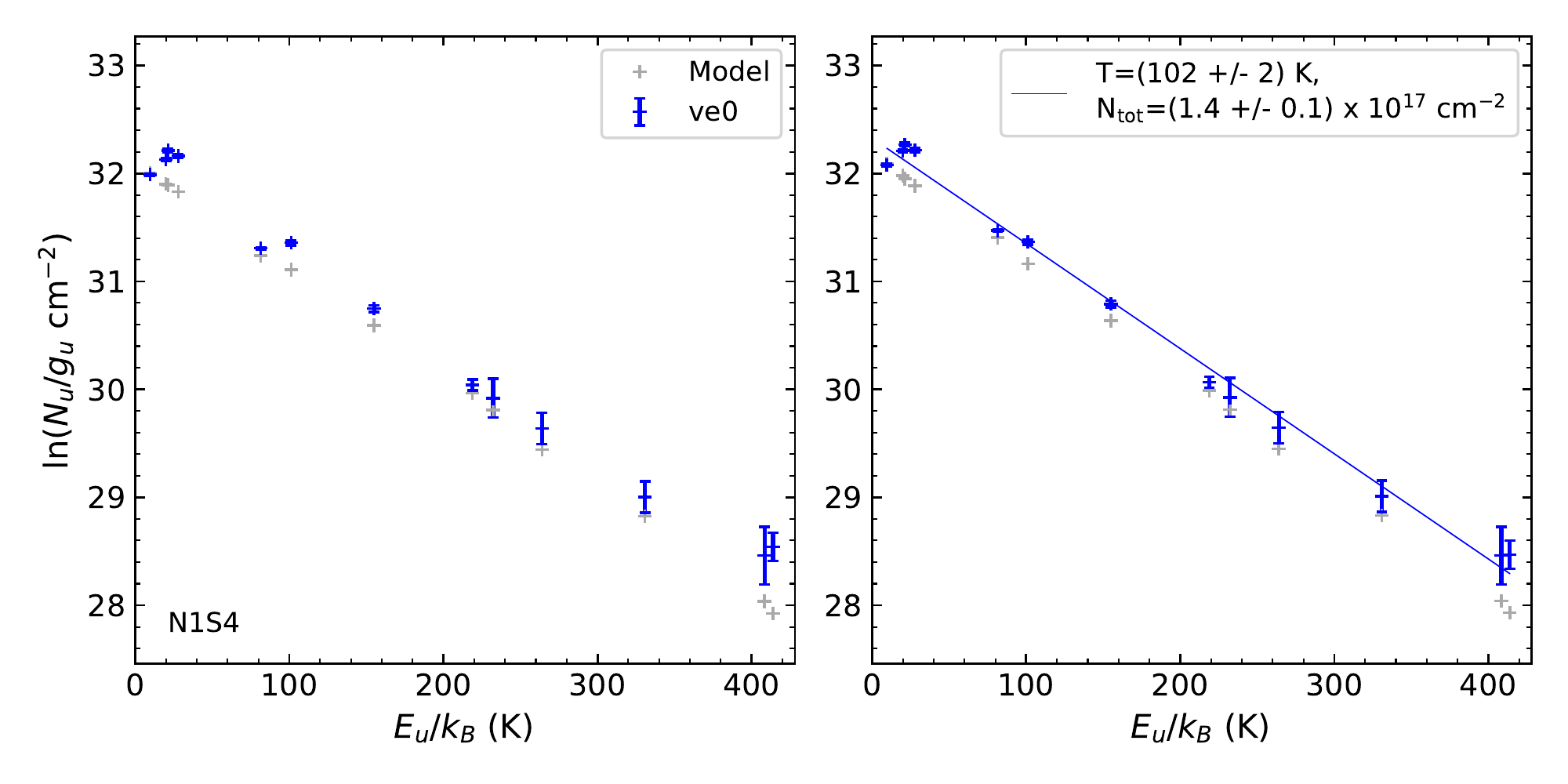}
    \includegraphics[width=0.49\textwidth]{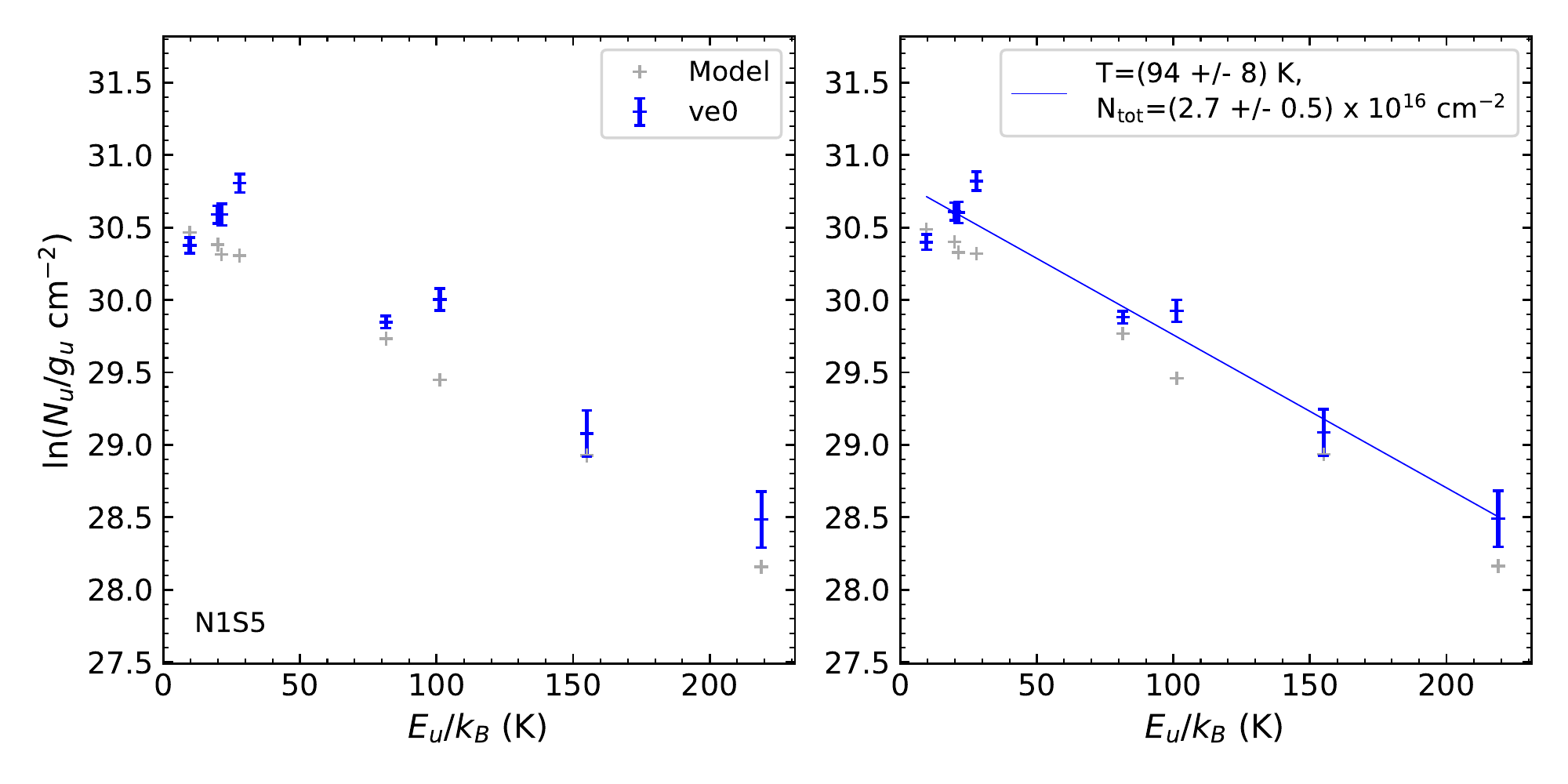}
    \includegraphics[width=0.49\textwidth]{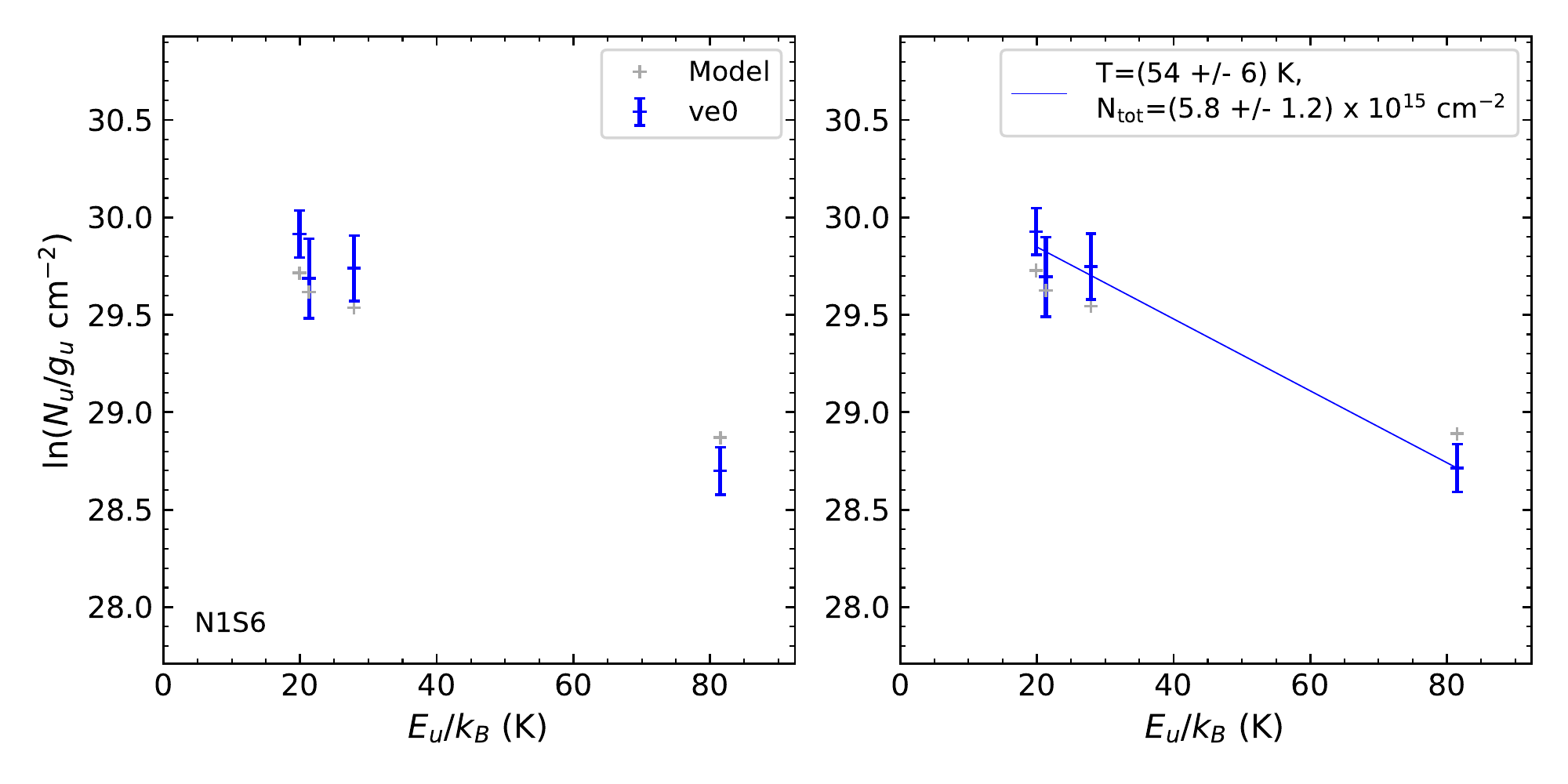}
    \caption{Same as Fig.\,\ref{fig:PD_met}, but for $^{13}$\met, except that setups 1--3 had to be used due to lack of transitions in setups 4--5. The linear fit to observed data points is shown in blue.}
    \label{fig:PD_13met}
\end{figure*}

\begin{figure*}[h]
    \includegraphics[width=0.49\textwidth]{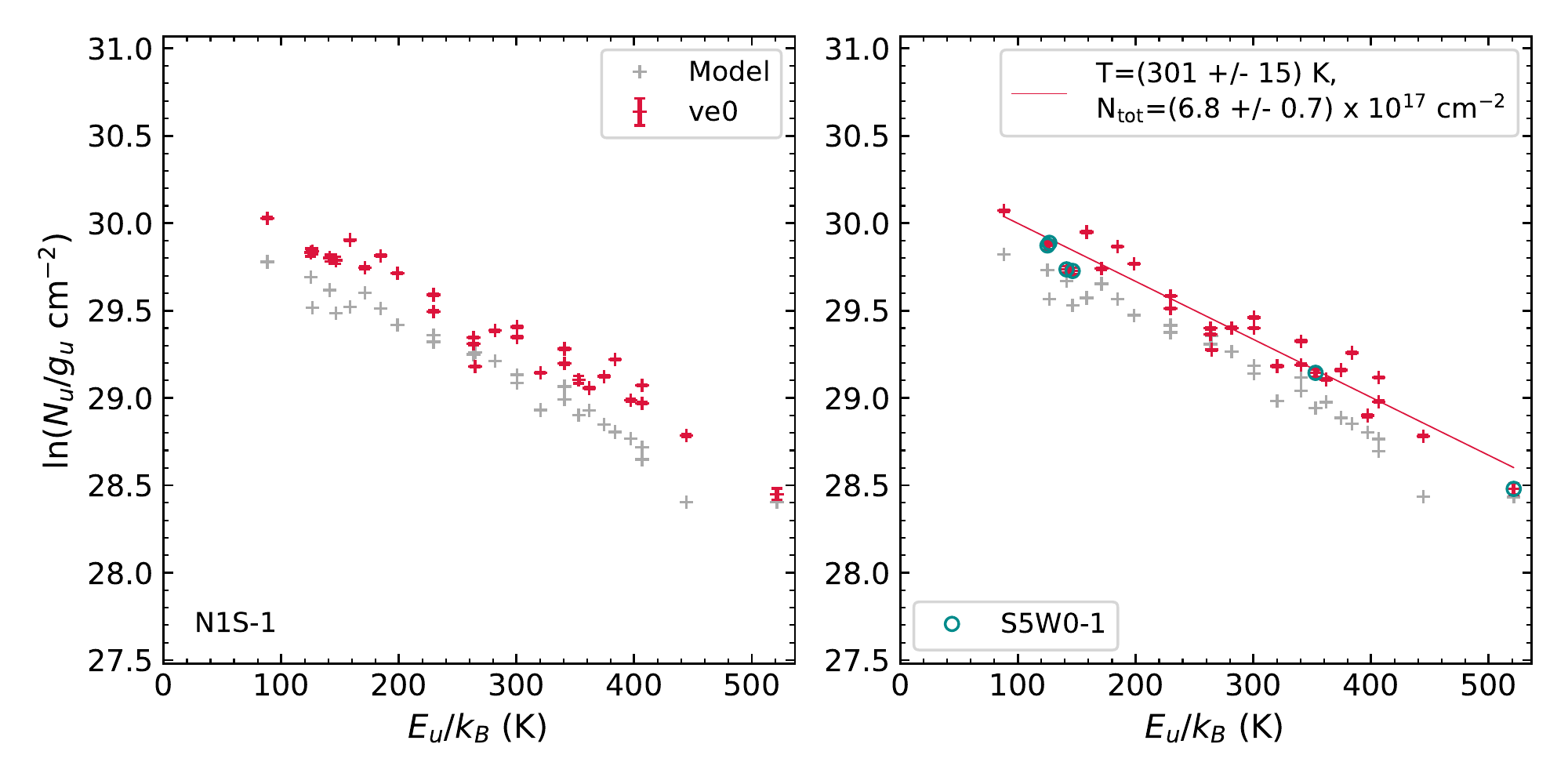}
    \includegraphics[width=0.49\textwidth]{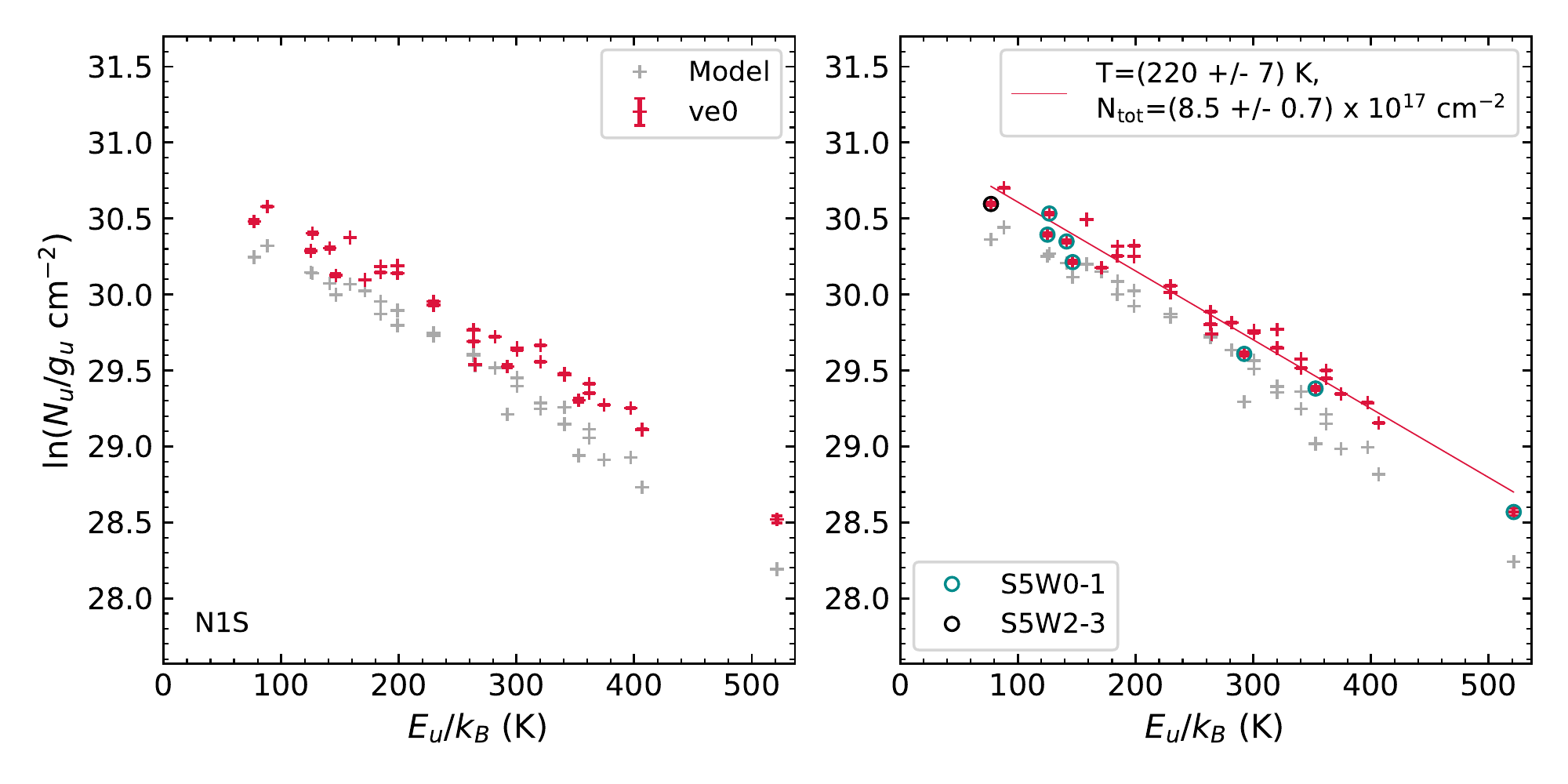}
    \includegraphics[width=0.49\textwidth]{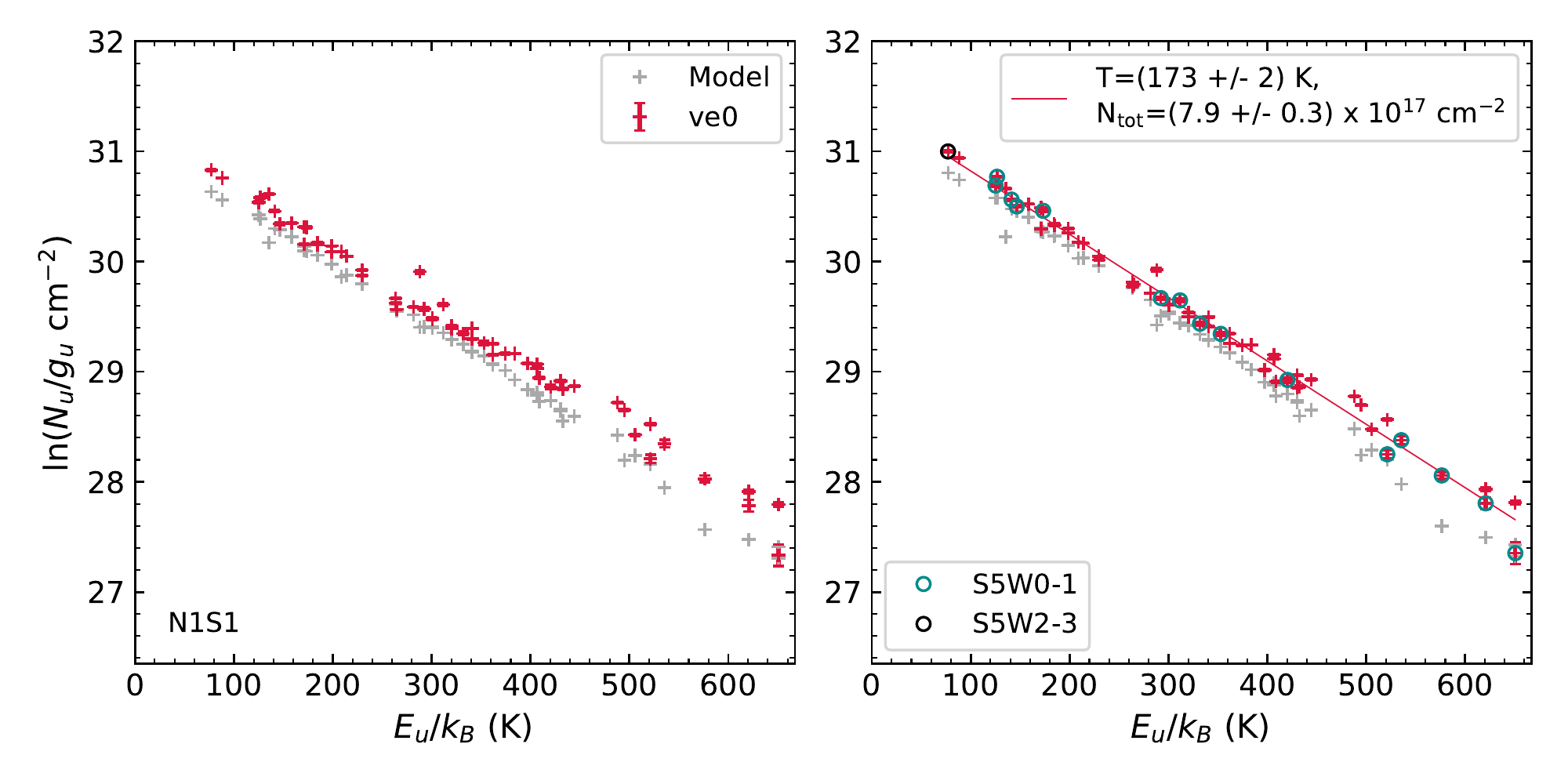}
    \includegraphics[width=0.49\textwidth]{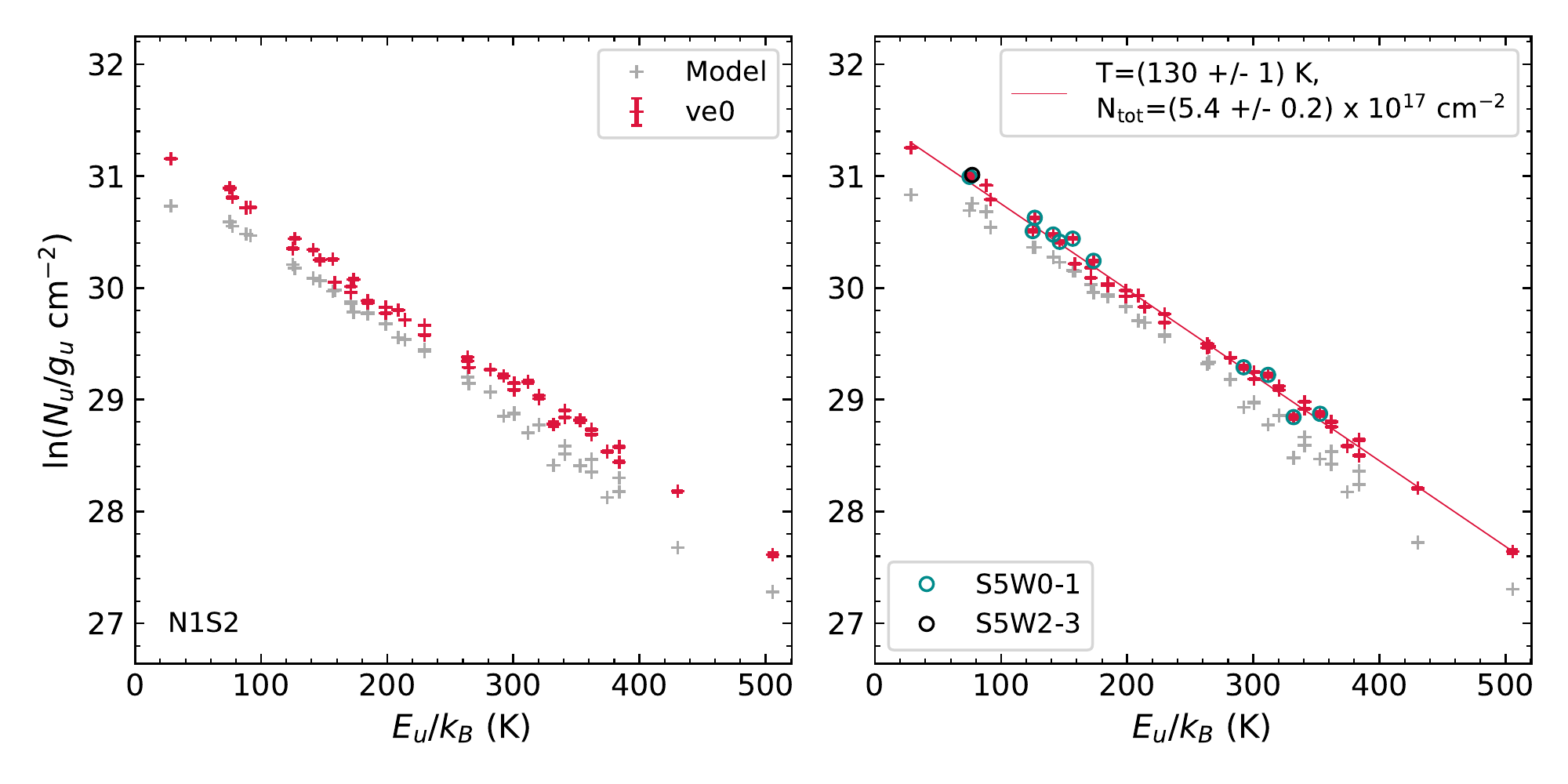}
    \includegraphics[width=0.49\textwidth]{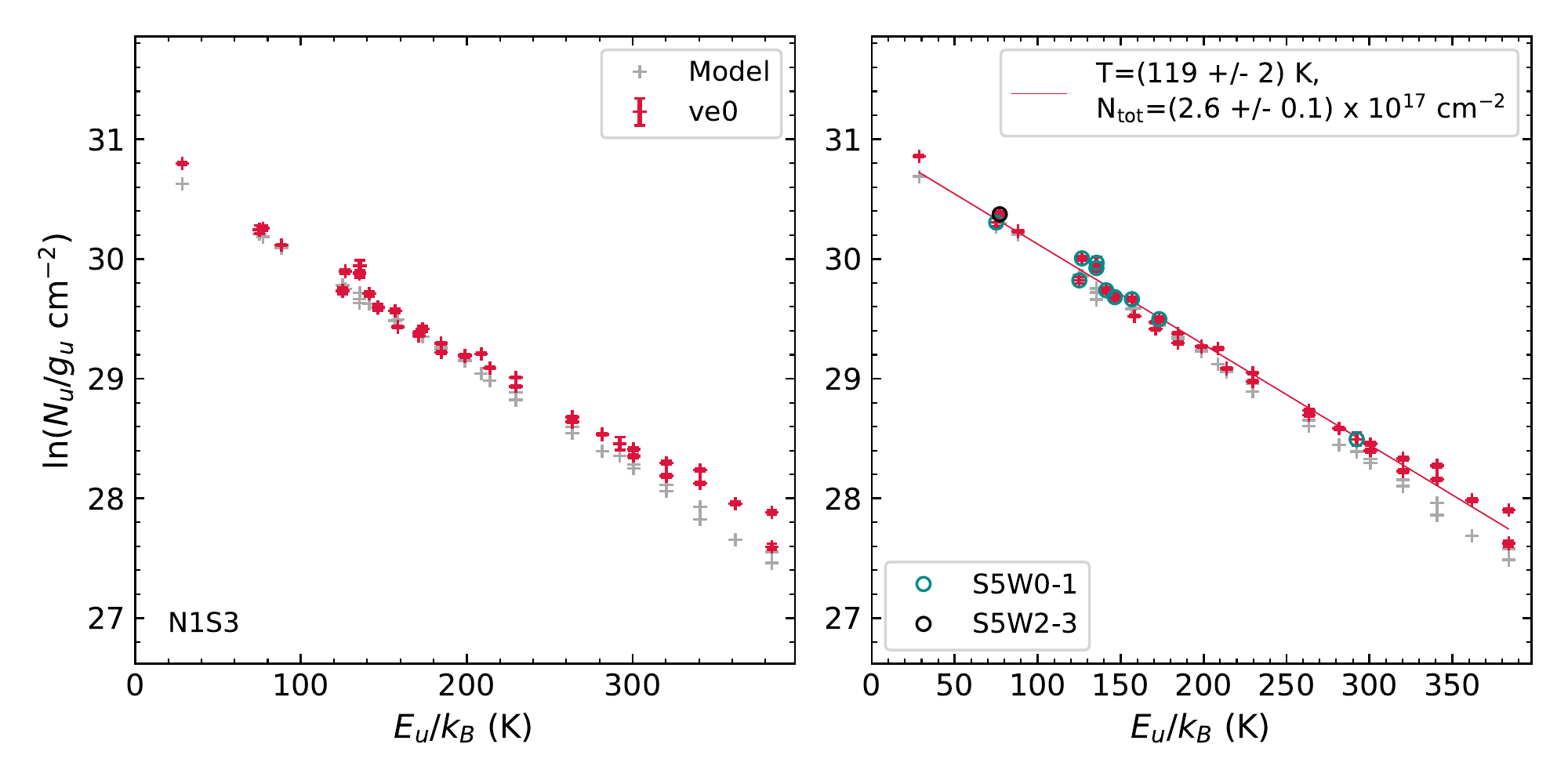}
    \includegraphics[width=0.49\textwidth]{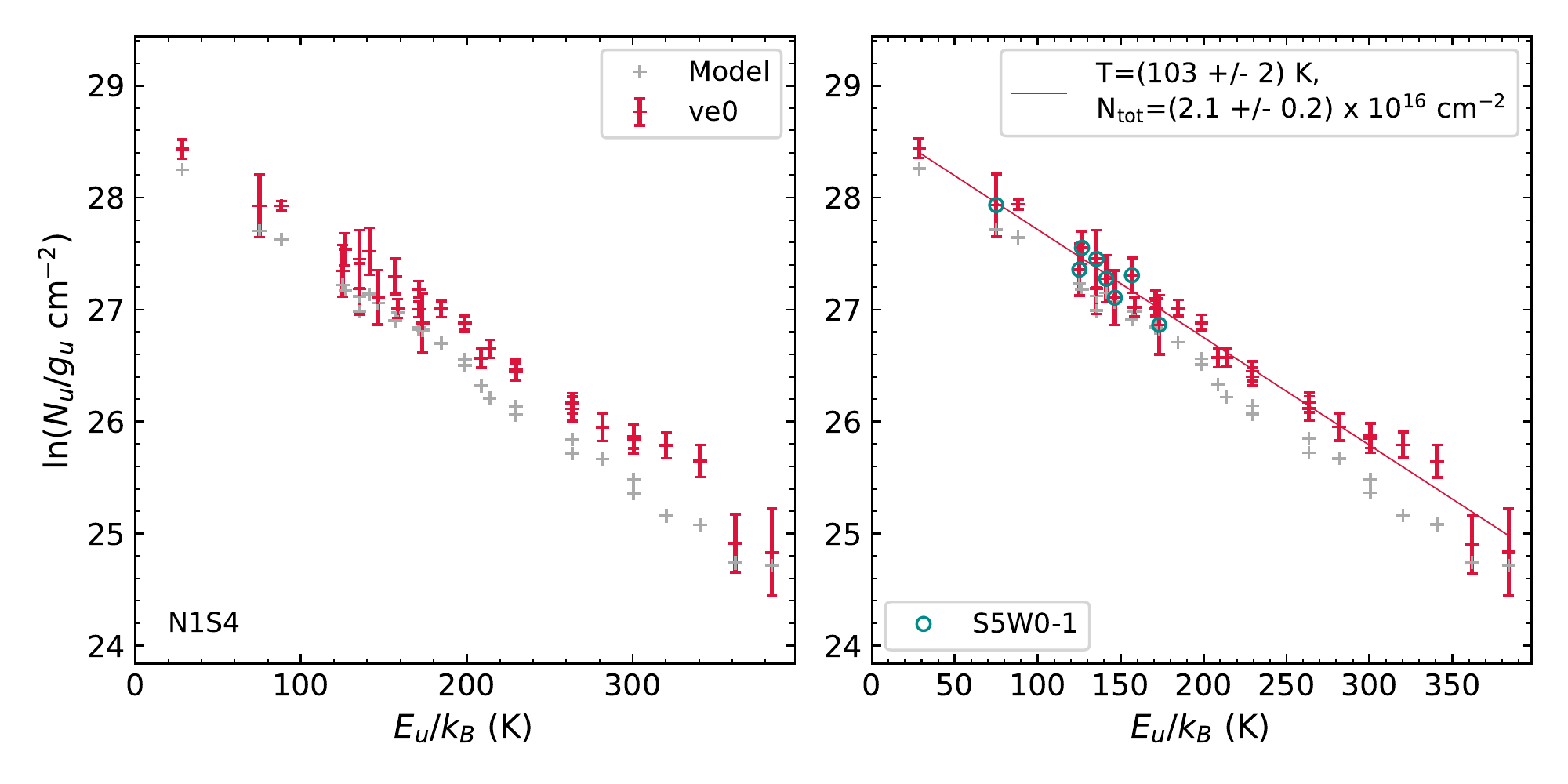}
    \caption{Same as Fig.\,\ref{fig:PD_met}, but for \et.}
    \label{fig:PD_et}
\end{figure*}

\begin{figure*}[h]
    \includegraphics[width=0.49\textwidth]{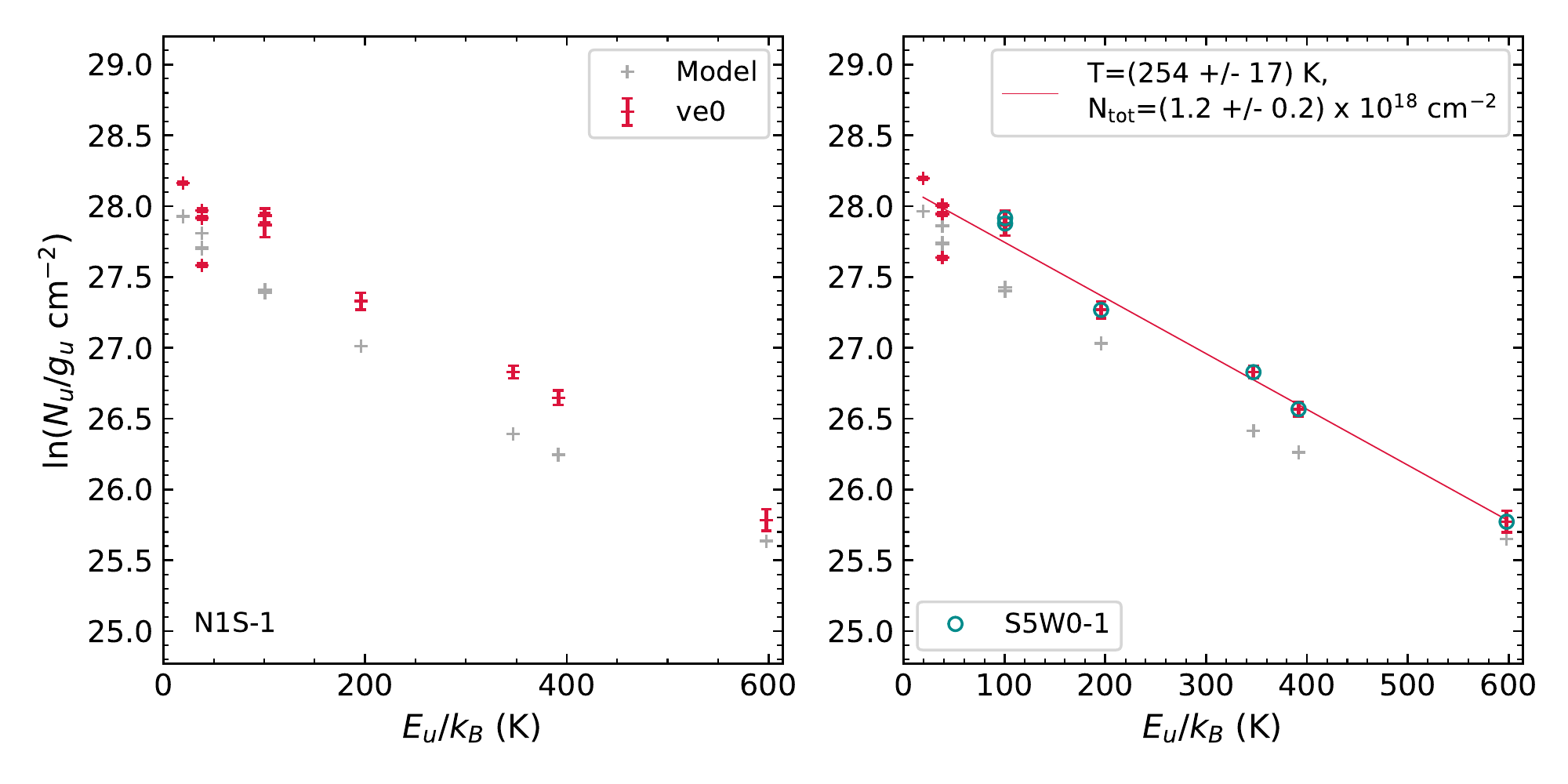}
    \includegraphics[width=0.49\textwidth]{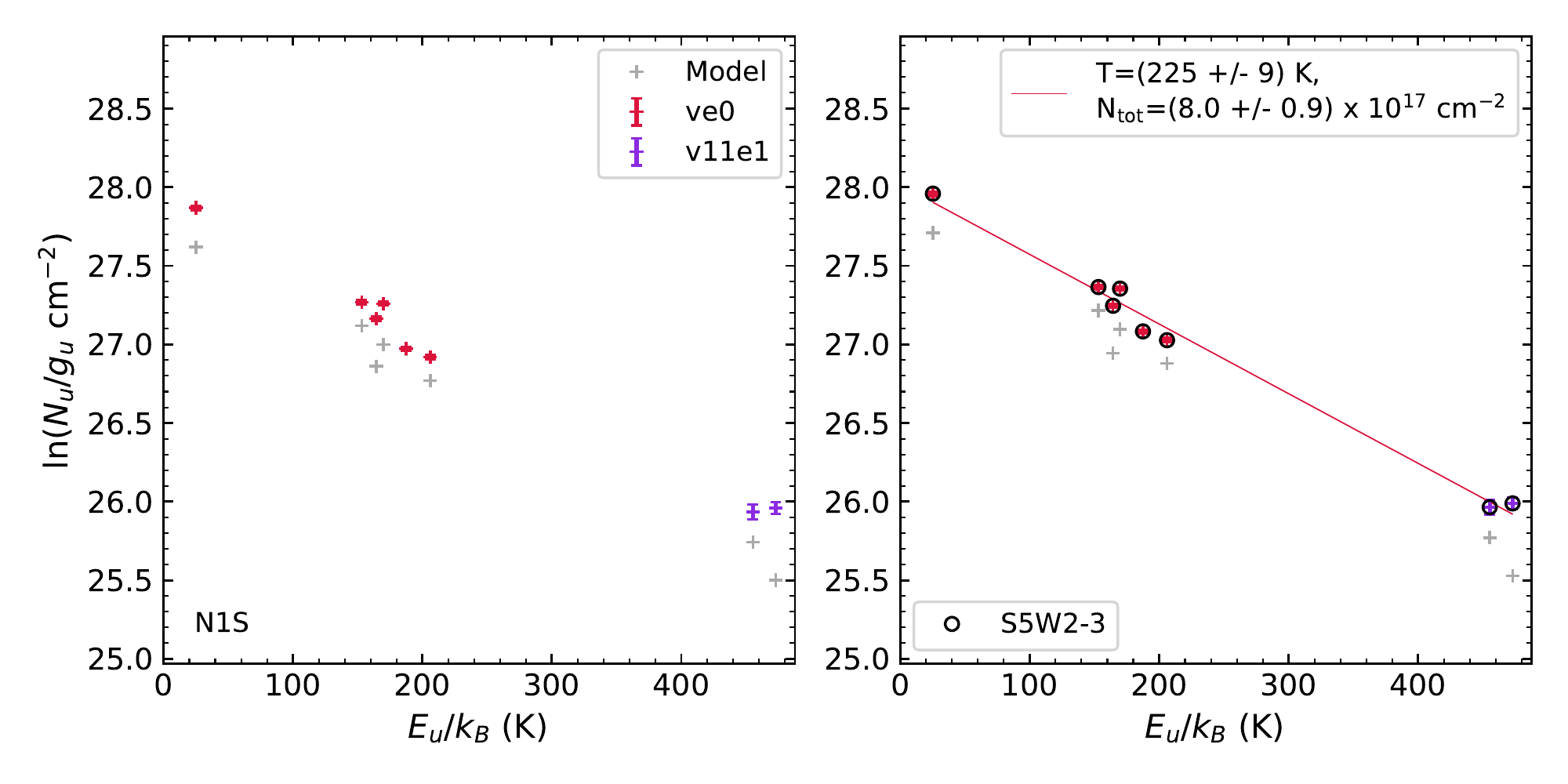}
    \caption{Same as Fig.\,\ref{fig:PD_met}, but for \dme.}
    \label{fig:PD_dme}
\end{figure*}

\begin{figure*}
    \includegraphics[width=0.49\textwidth]{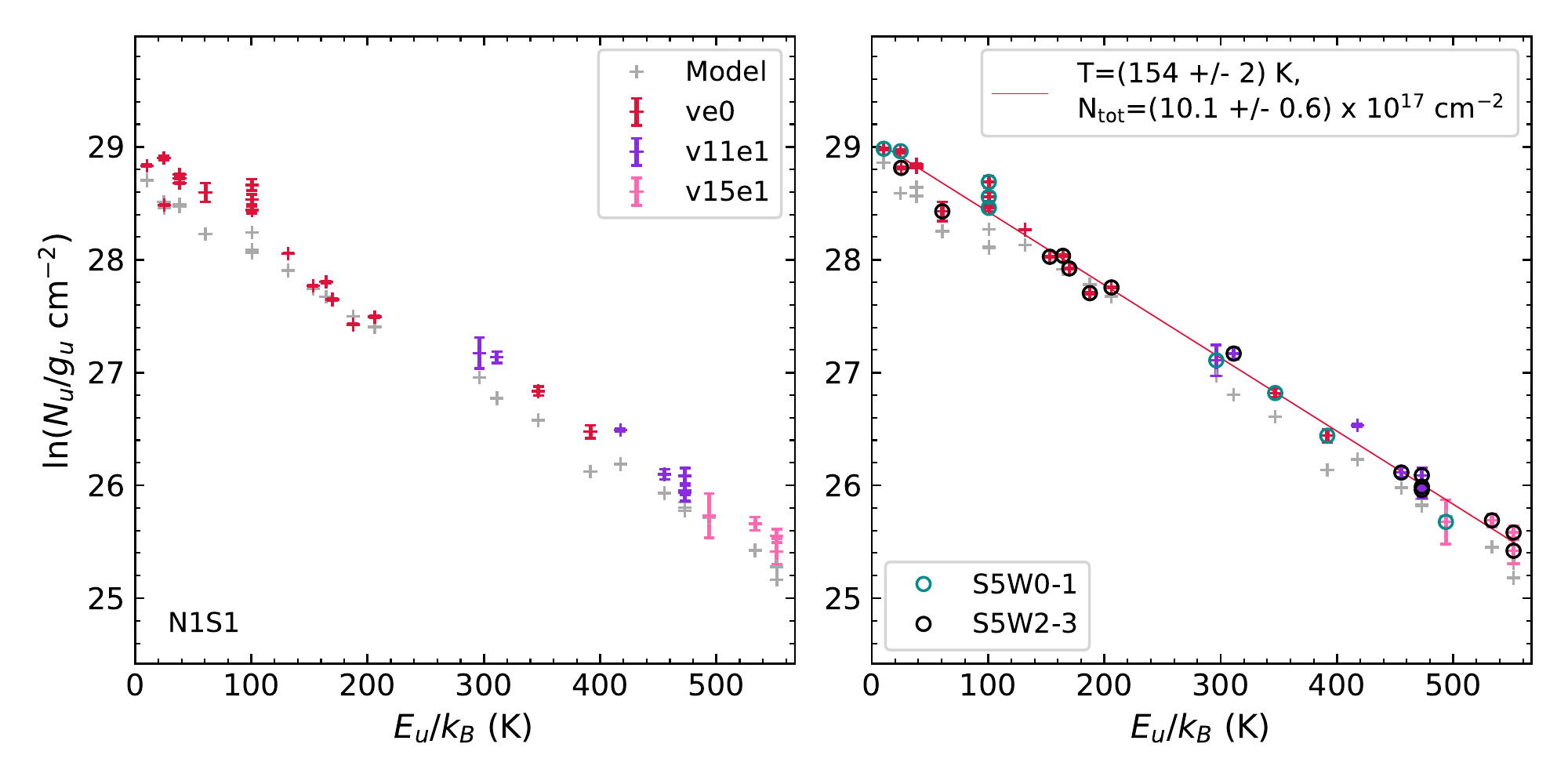}
    \includegraphics[width=0.49\textwidth]{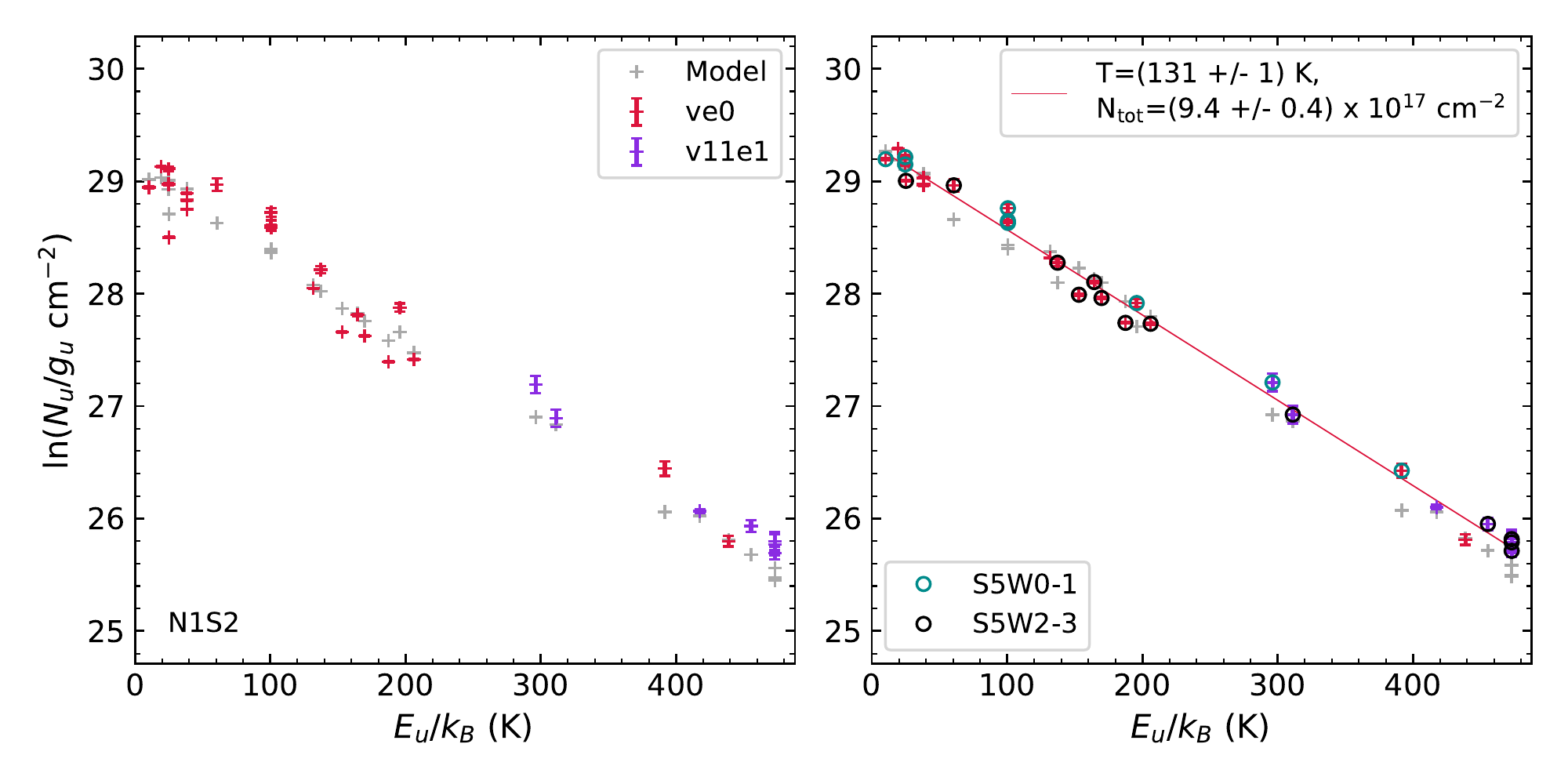}
    \includegraphics[width=0.49\textwidth]{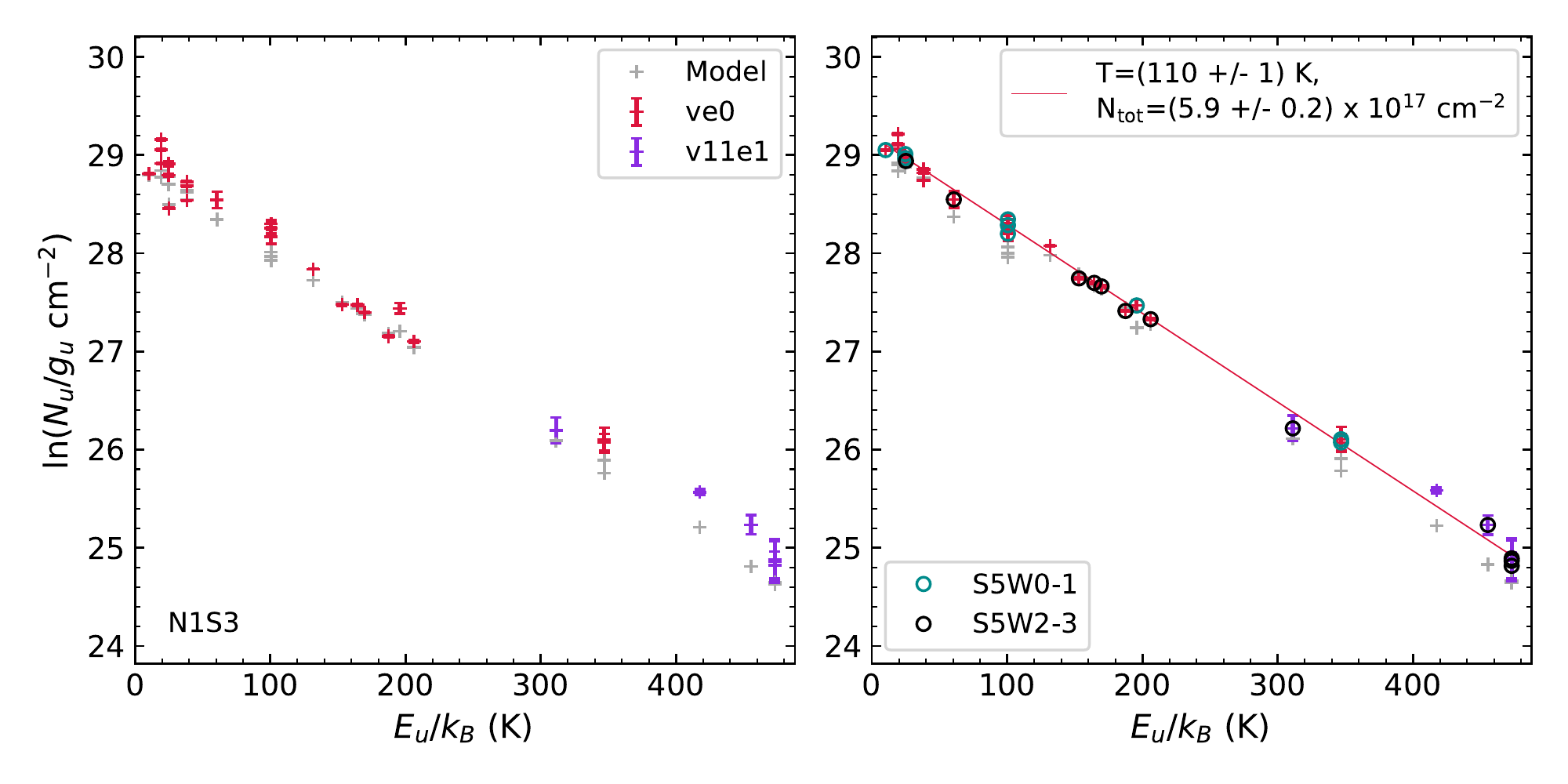}
    \includegraphics[width=0.49\textwidth]{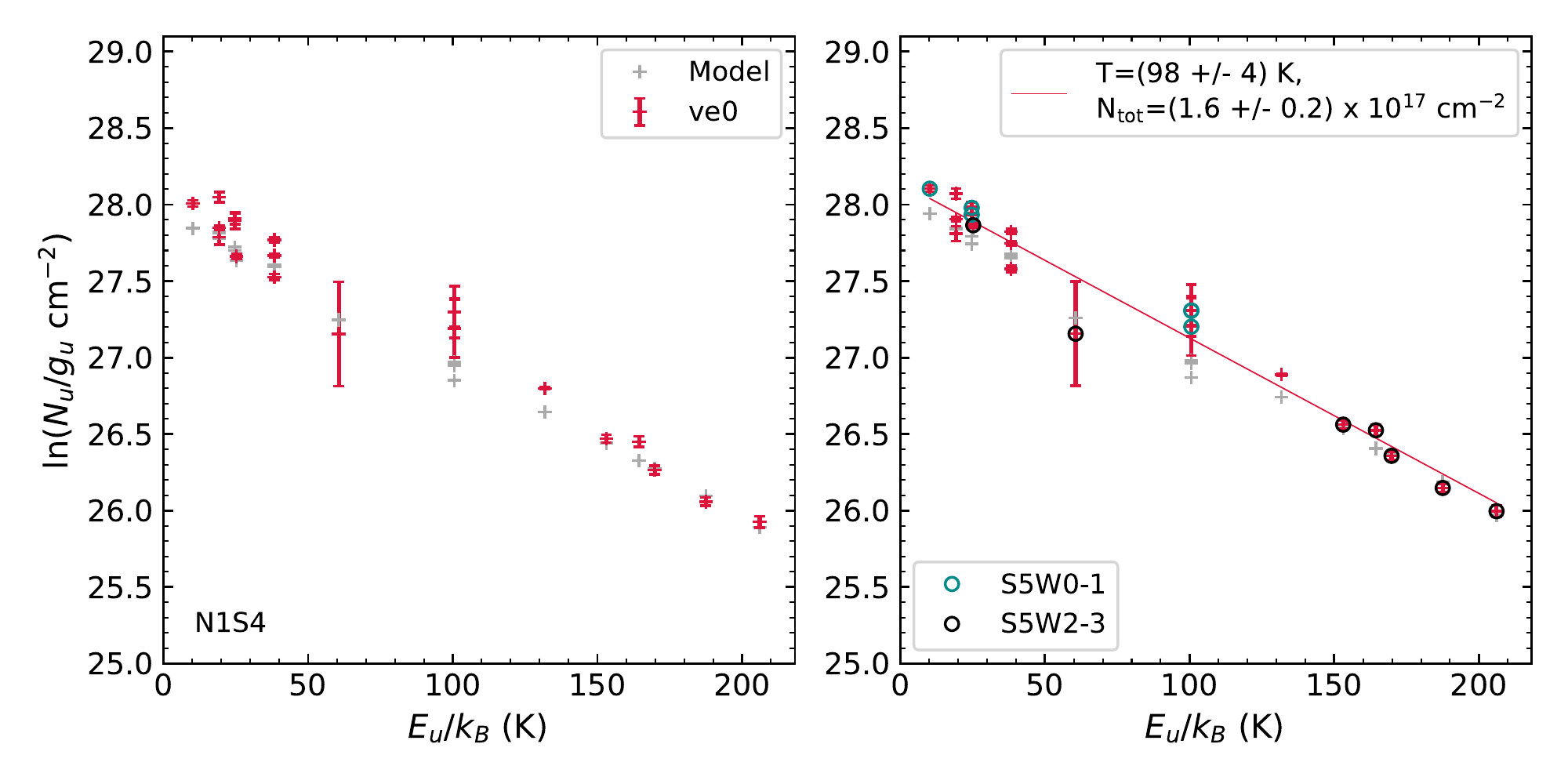}
    \includegraphics[width=0.49\textwidth]{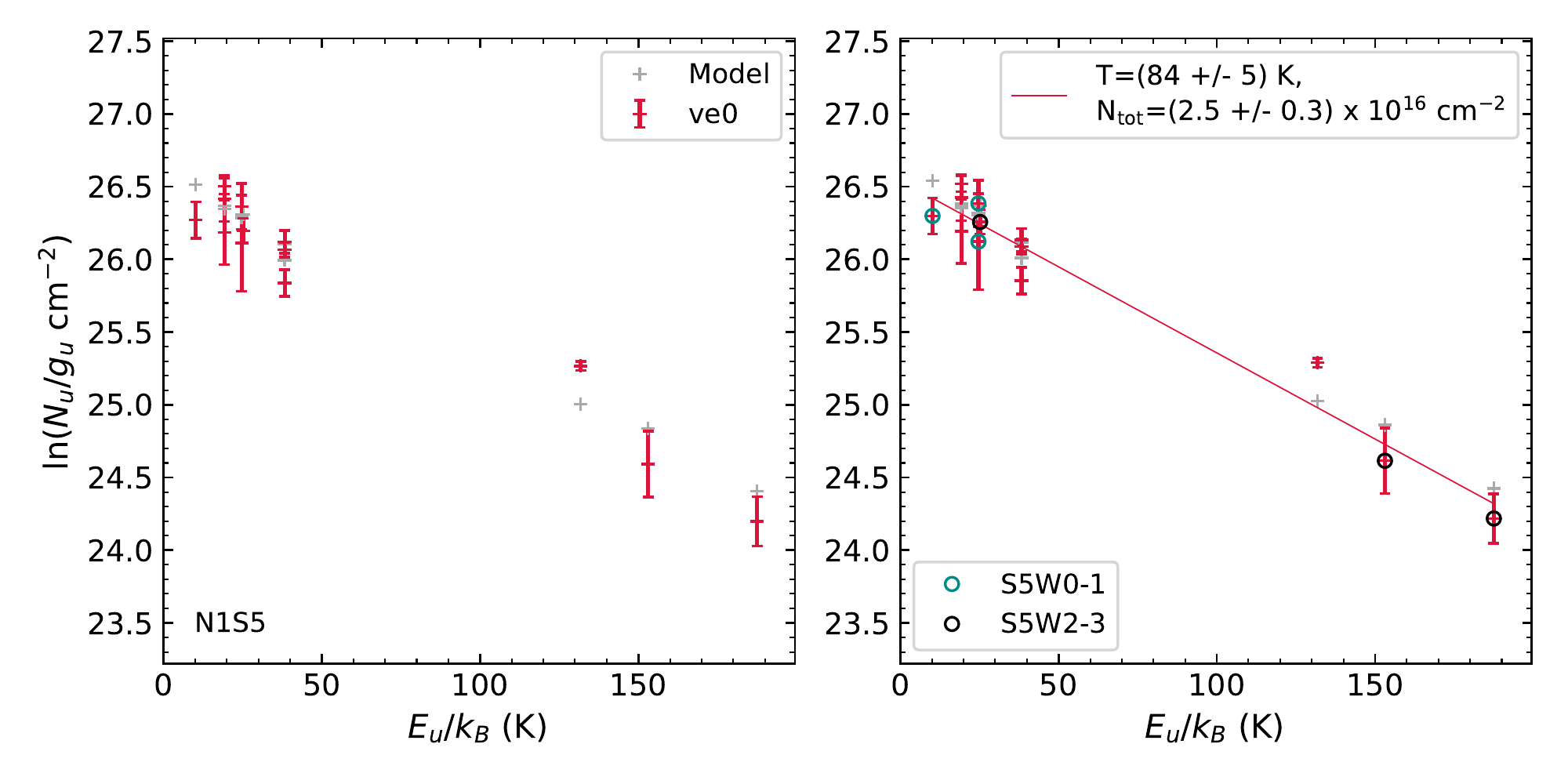}
    \includegraphics[width=0.49\textwidth]{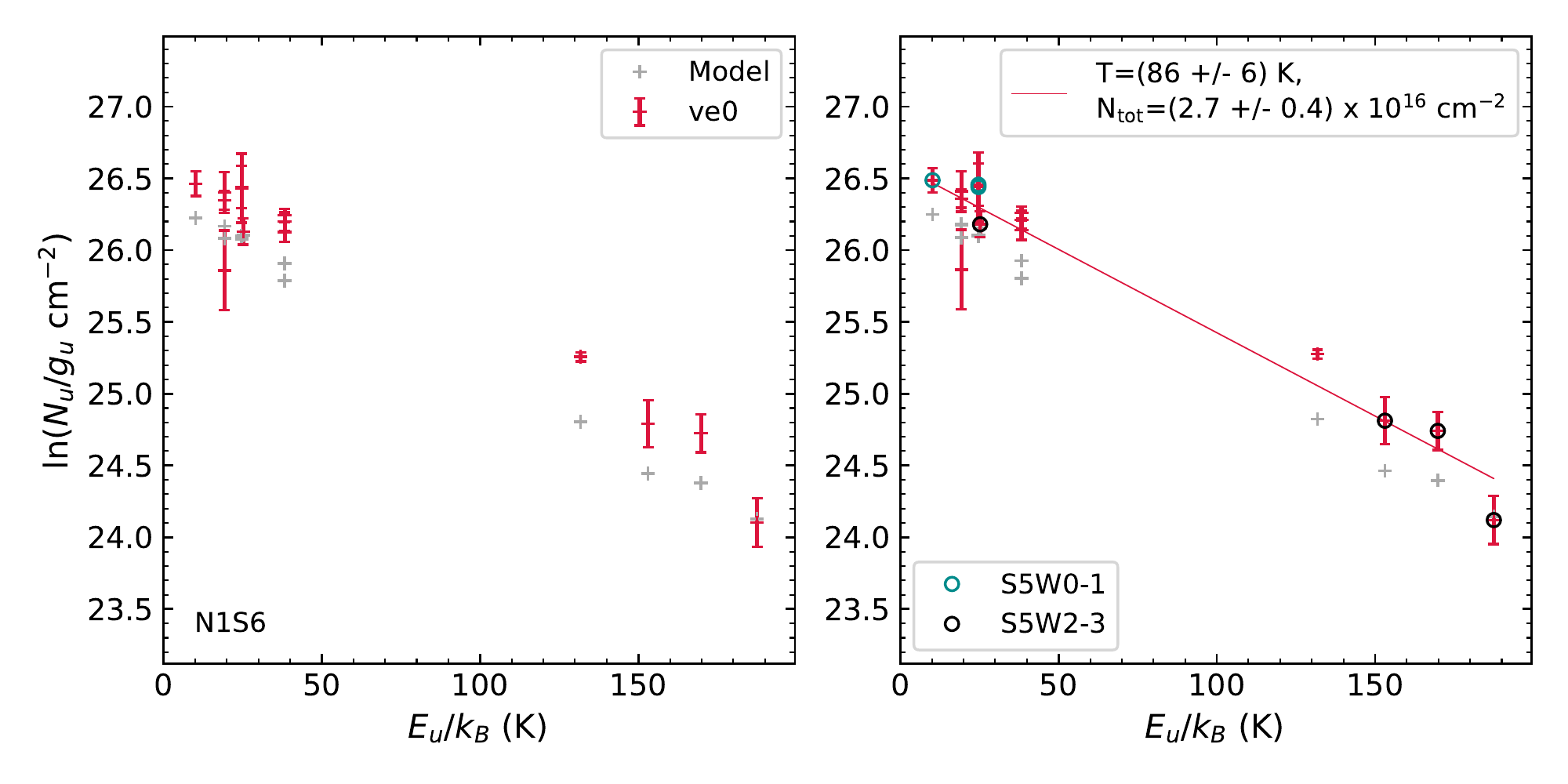}
    \includegraphics[width=0.49\textwidth]{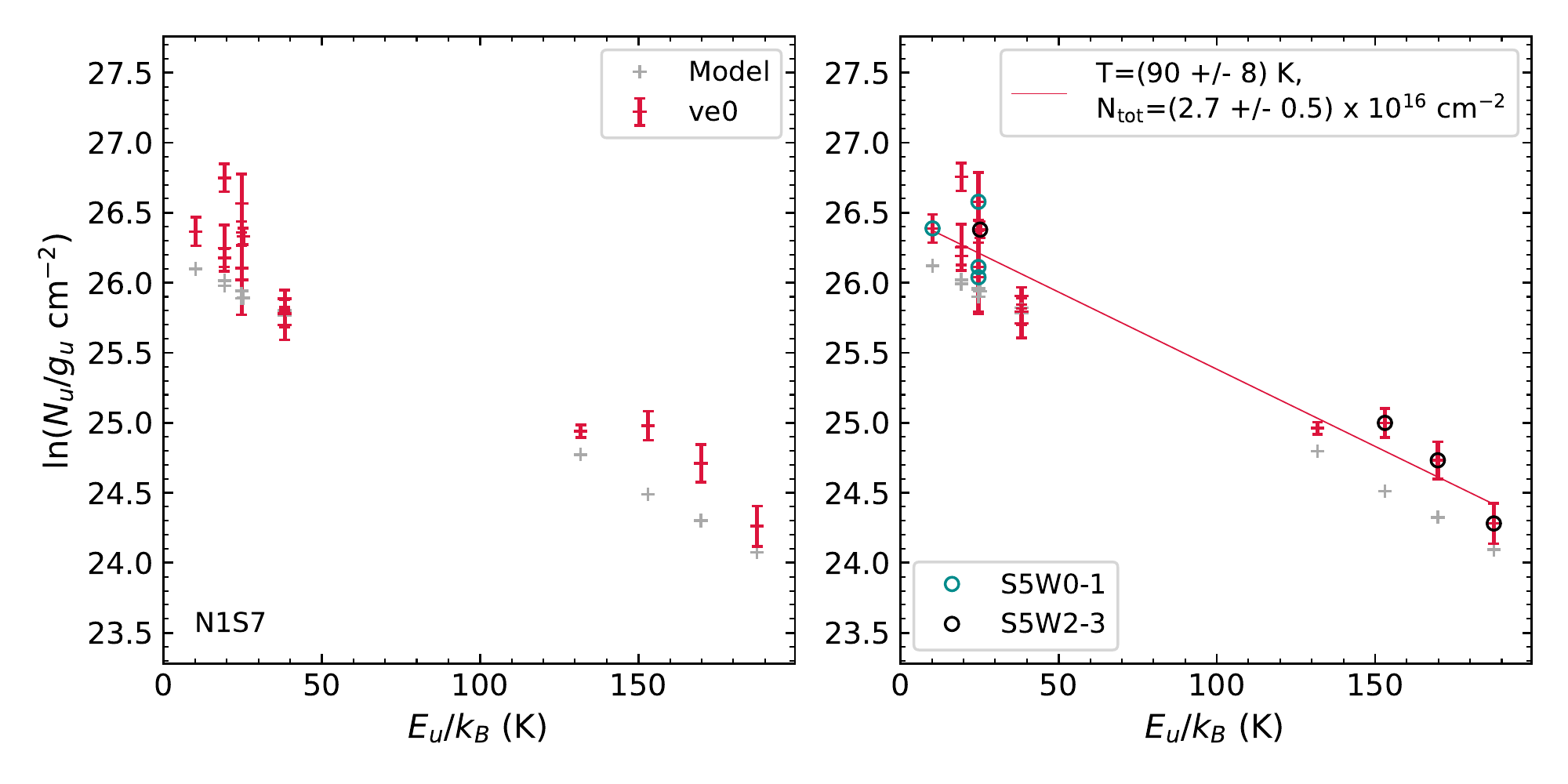}
    \includegraphics[width=0.49\textwidth]{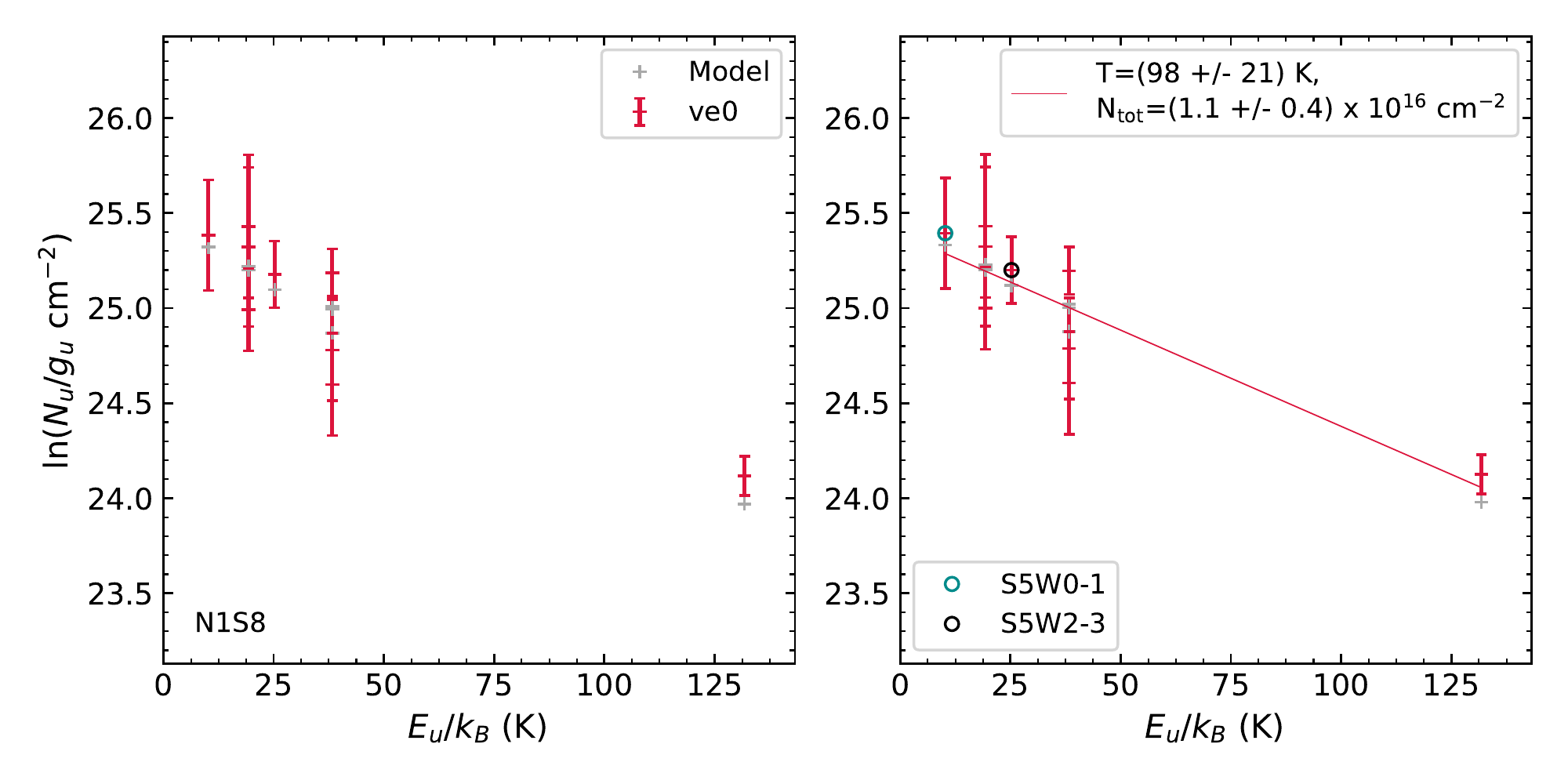}
    \includegraphics[width=0.49\textwidth]{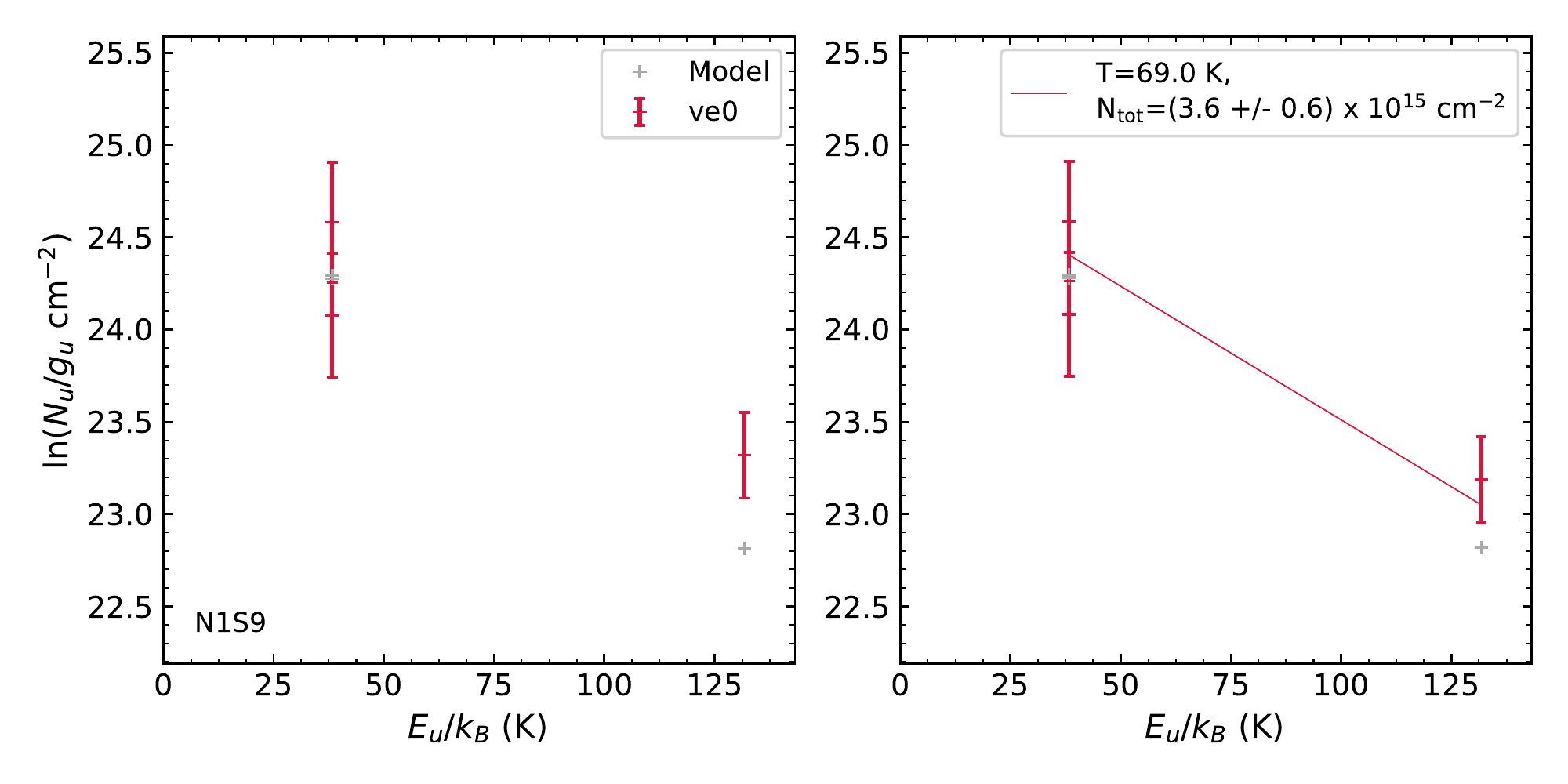}
    \caption{Figure\,\ref{fig:PD_dme} continued.}
    \label{fig:PD_dme2}
\end{figure*}

\begin{figure*}[h]
    \includegraphics[width=0.49\textwidth]{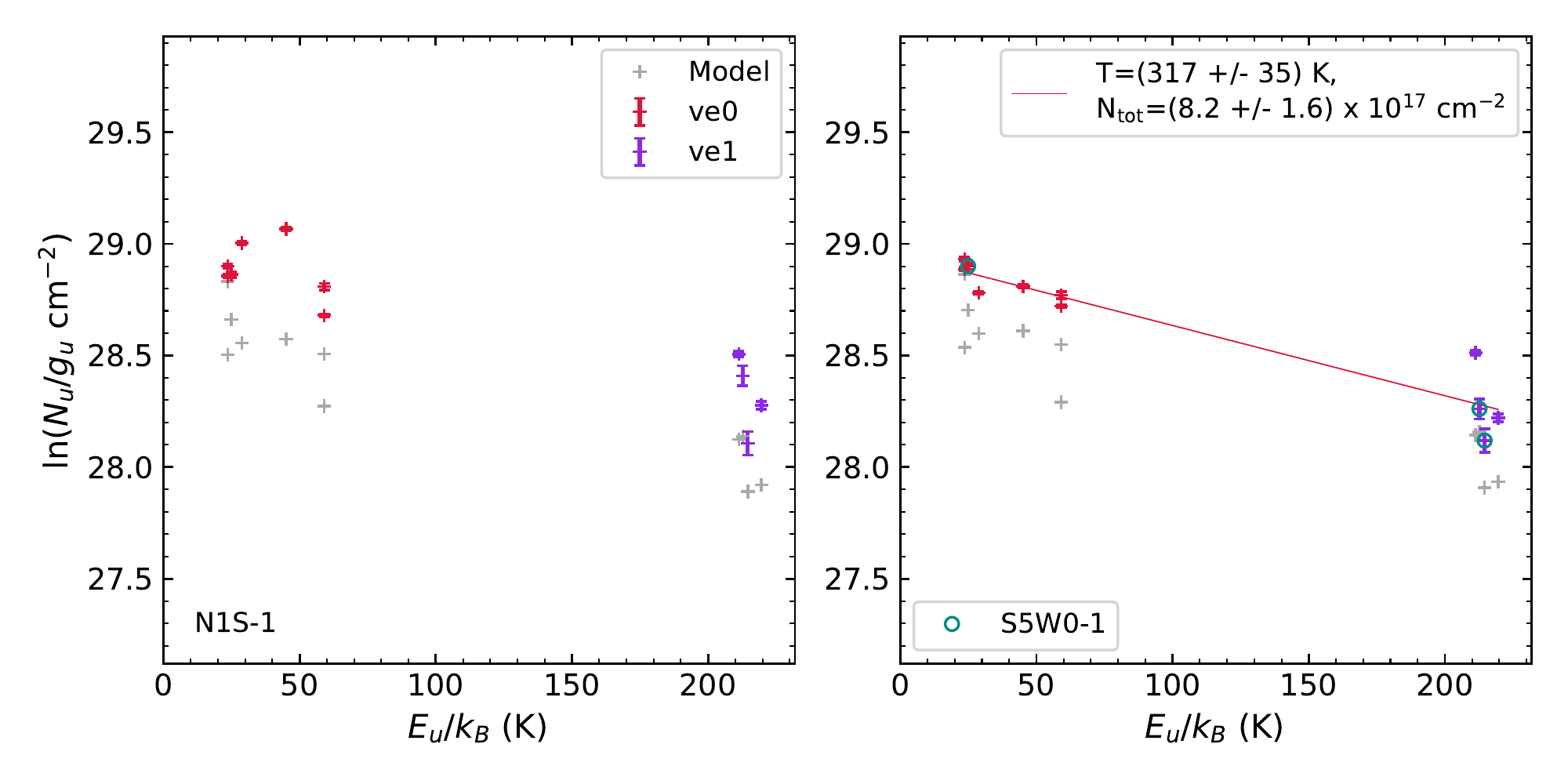}
    \includegraphics[width=0.49\textwidth]{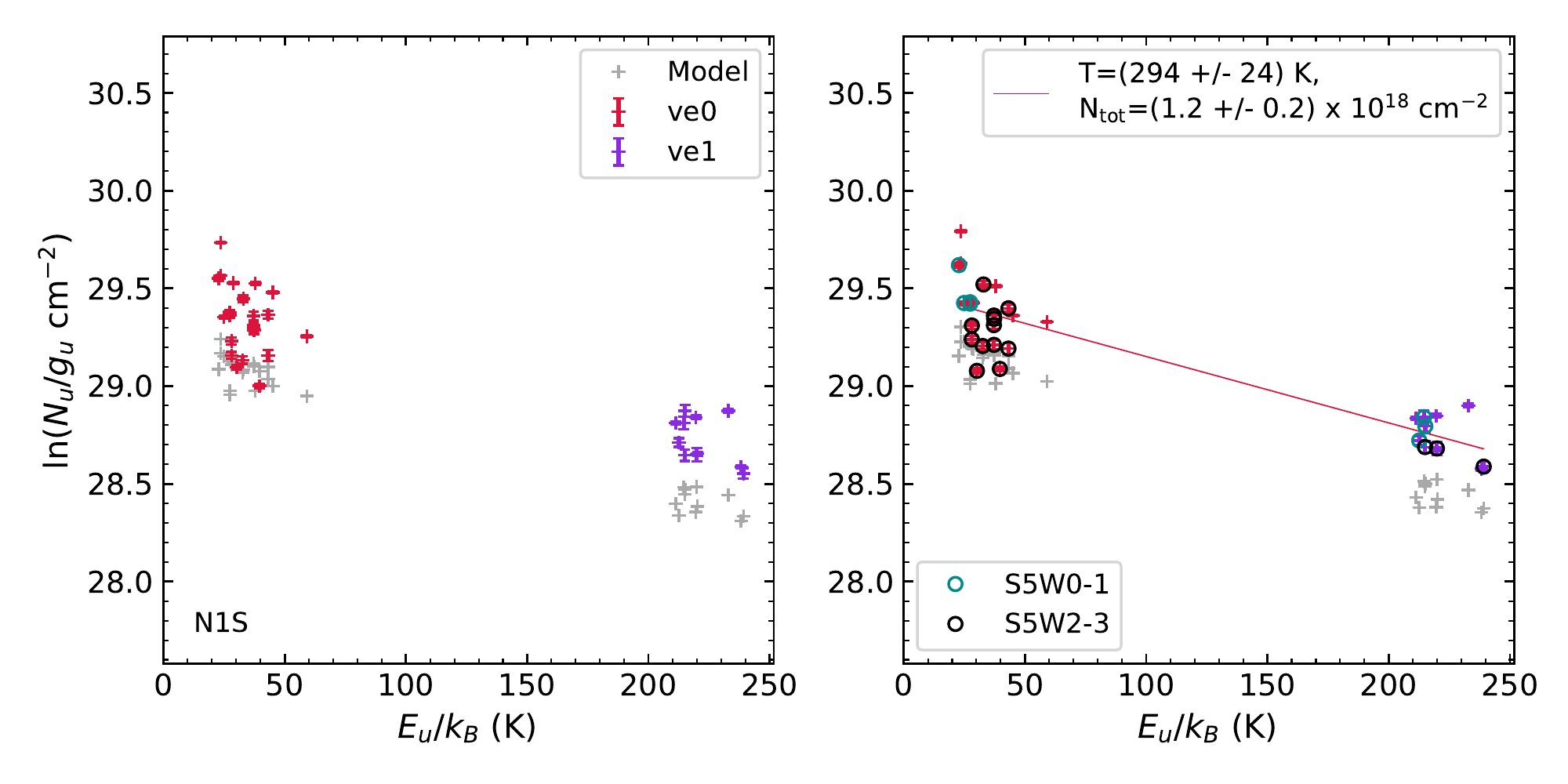}
    \includegraphics[width=0.49\textwidth]{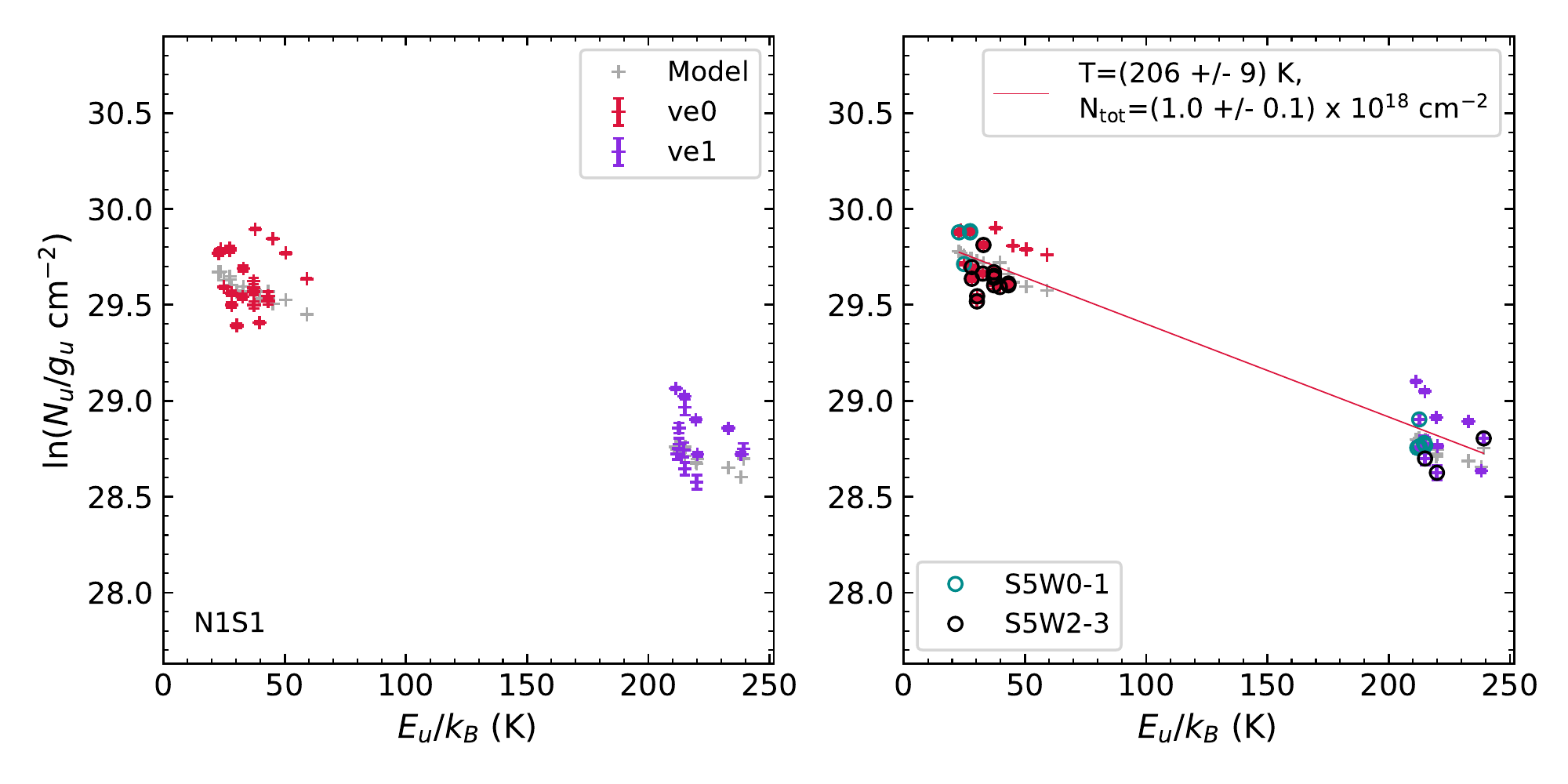}
    \includegraphics[width=0.49\textwidth]{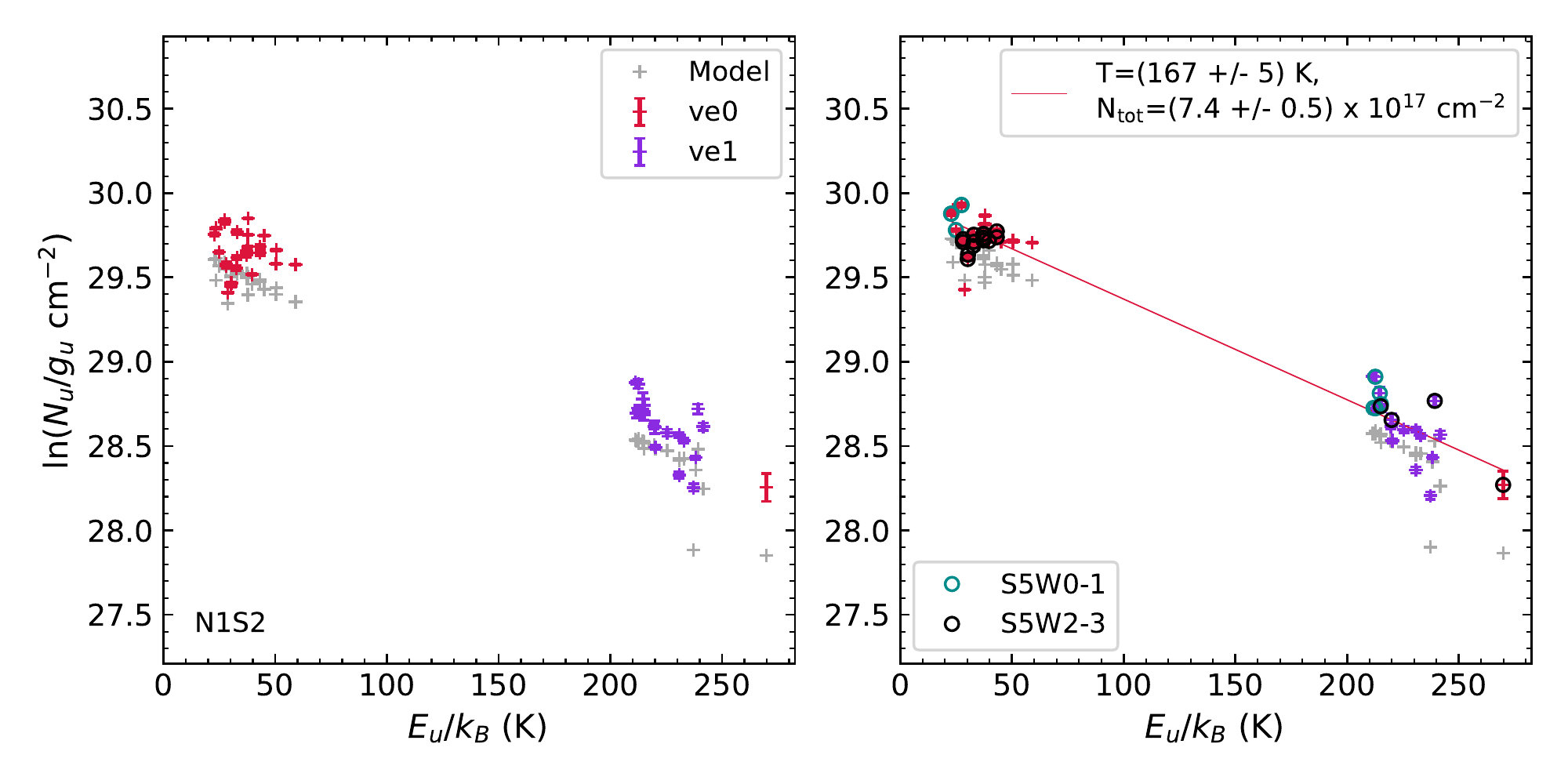}
    \includegraphics[width=0.49\textwidth]{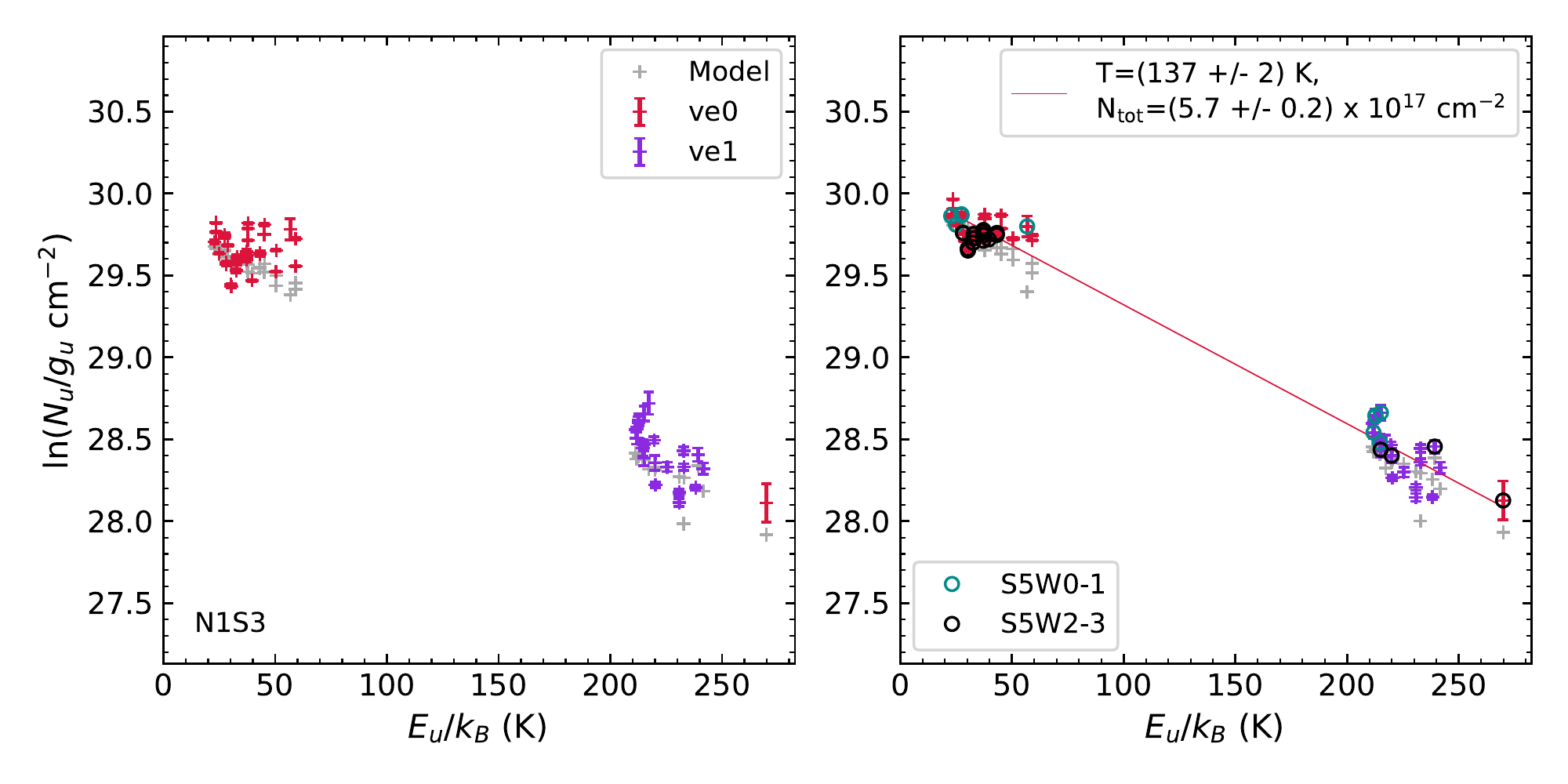}
    \includegraphics[width=0.49\textwidth]{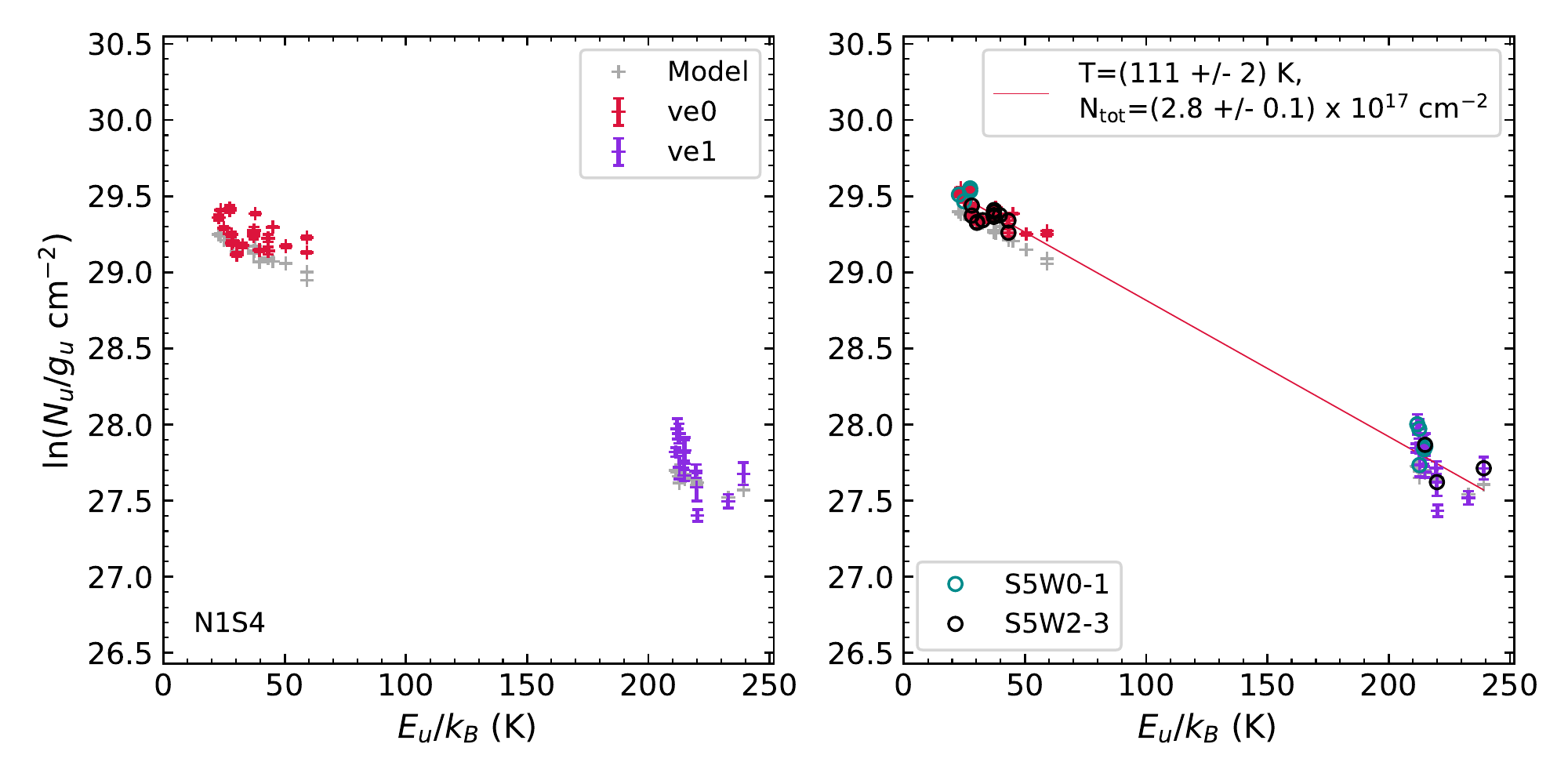}
    \includegraphics[width=0.49\textwidth]{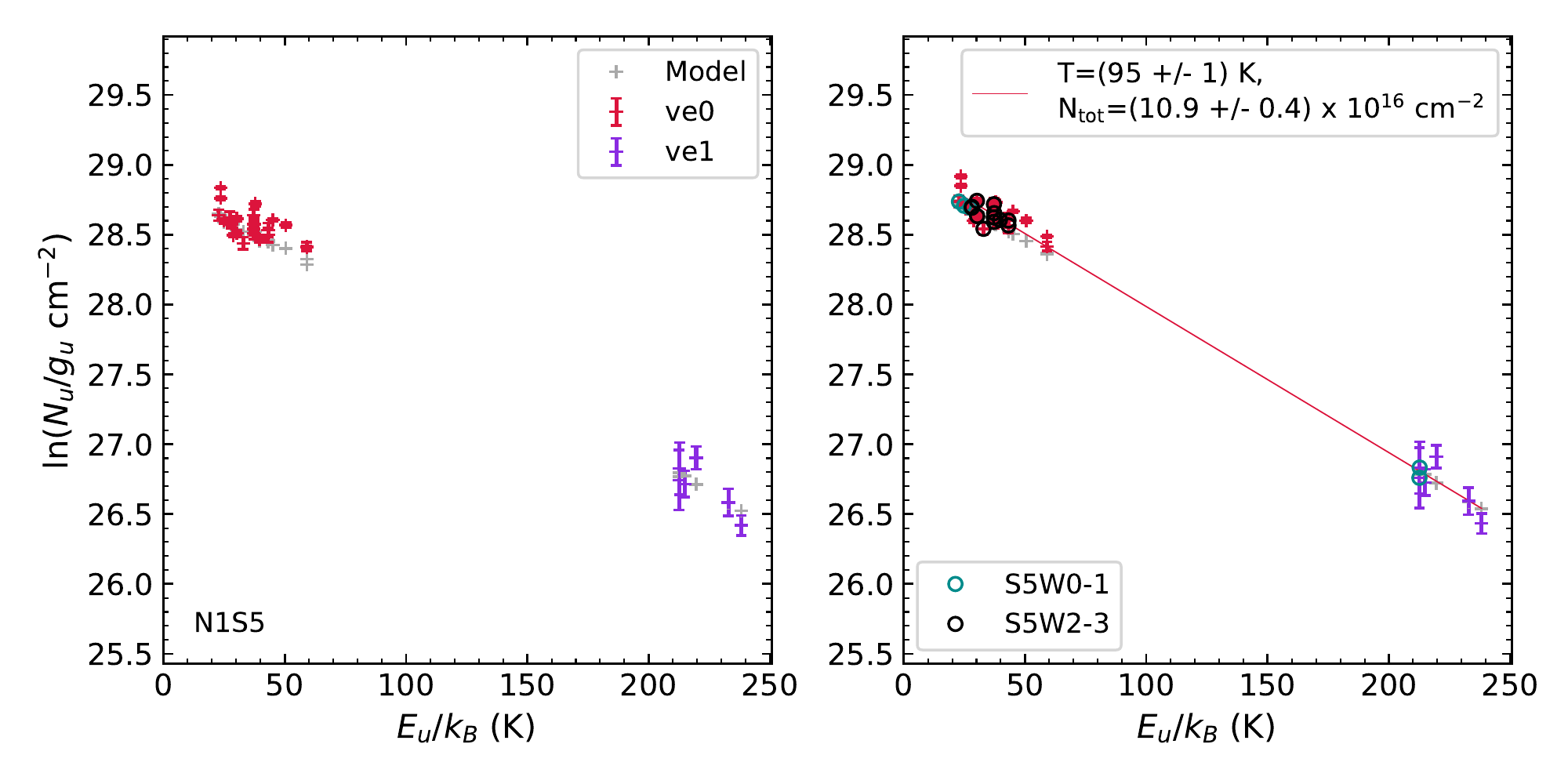}
    \includegraphics[width=0.49\textwidth]{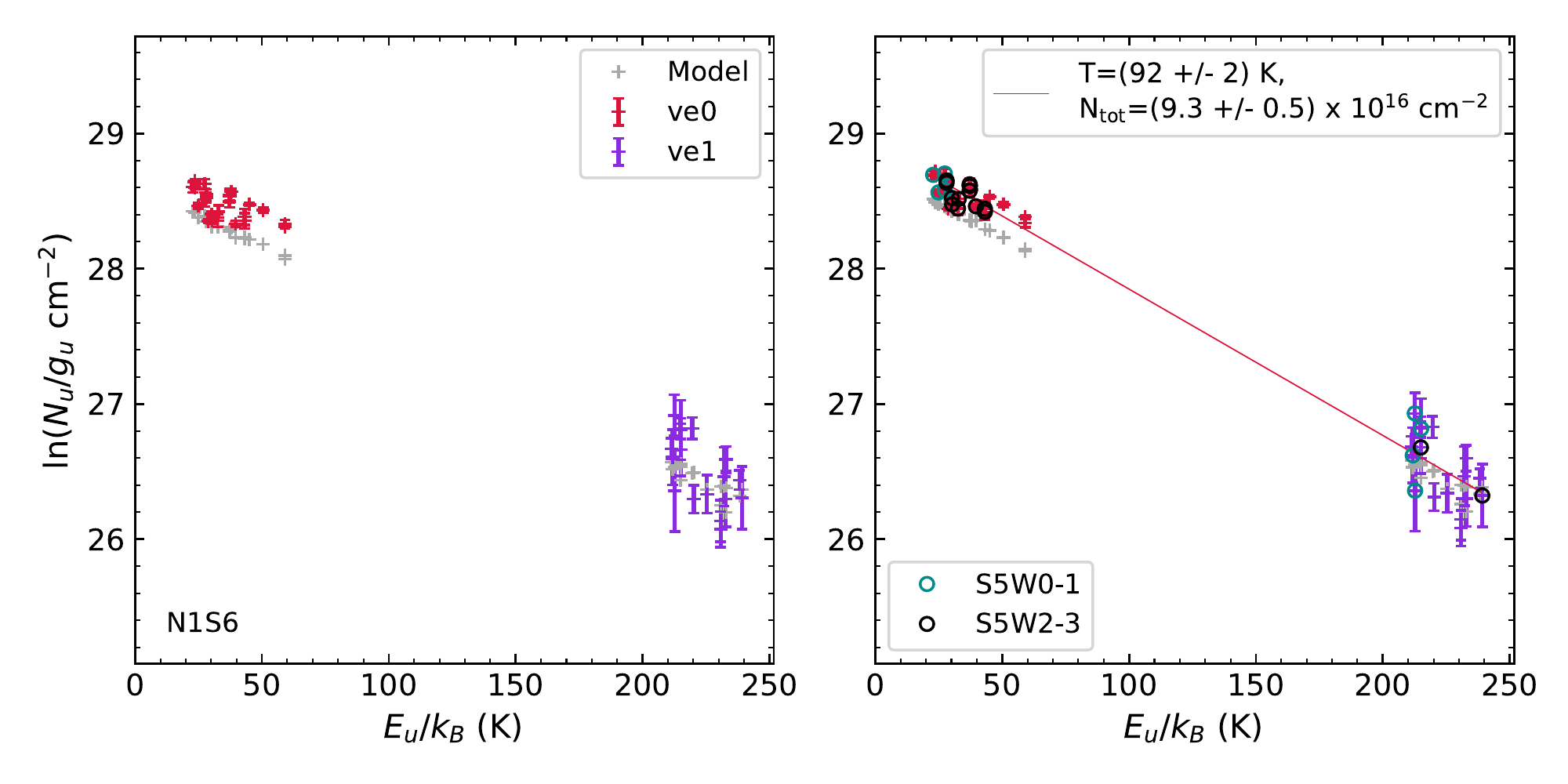}
    \includegraphics[width=0.49\textwidth]{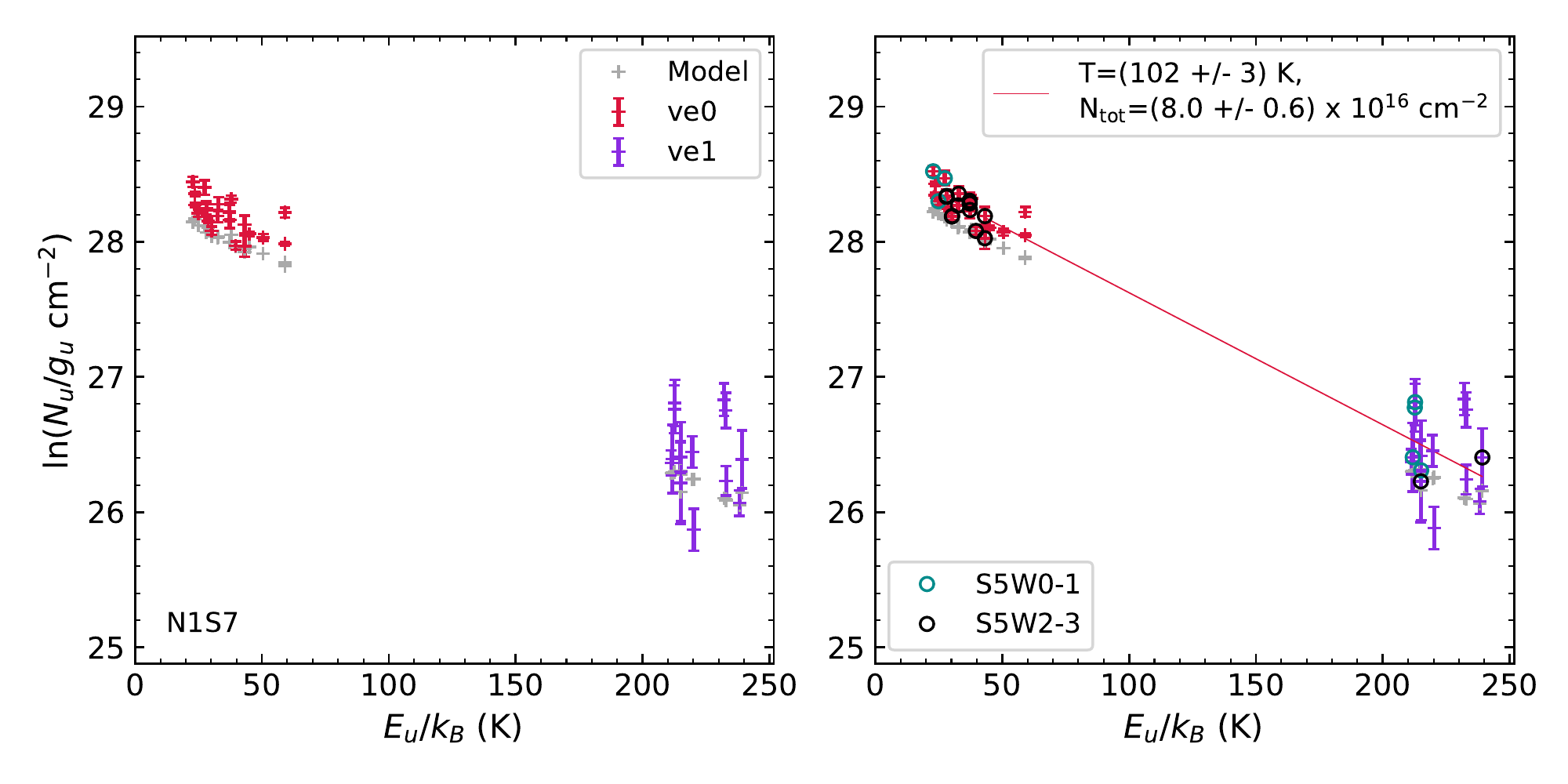}
    \includegraphics[width=0.49\textwidth]{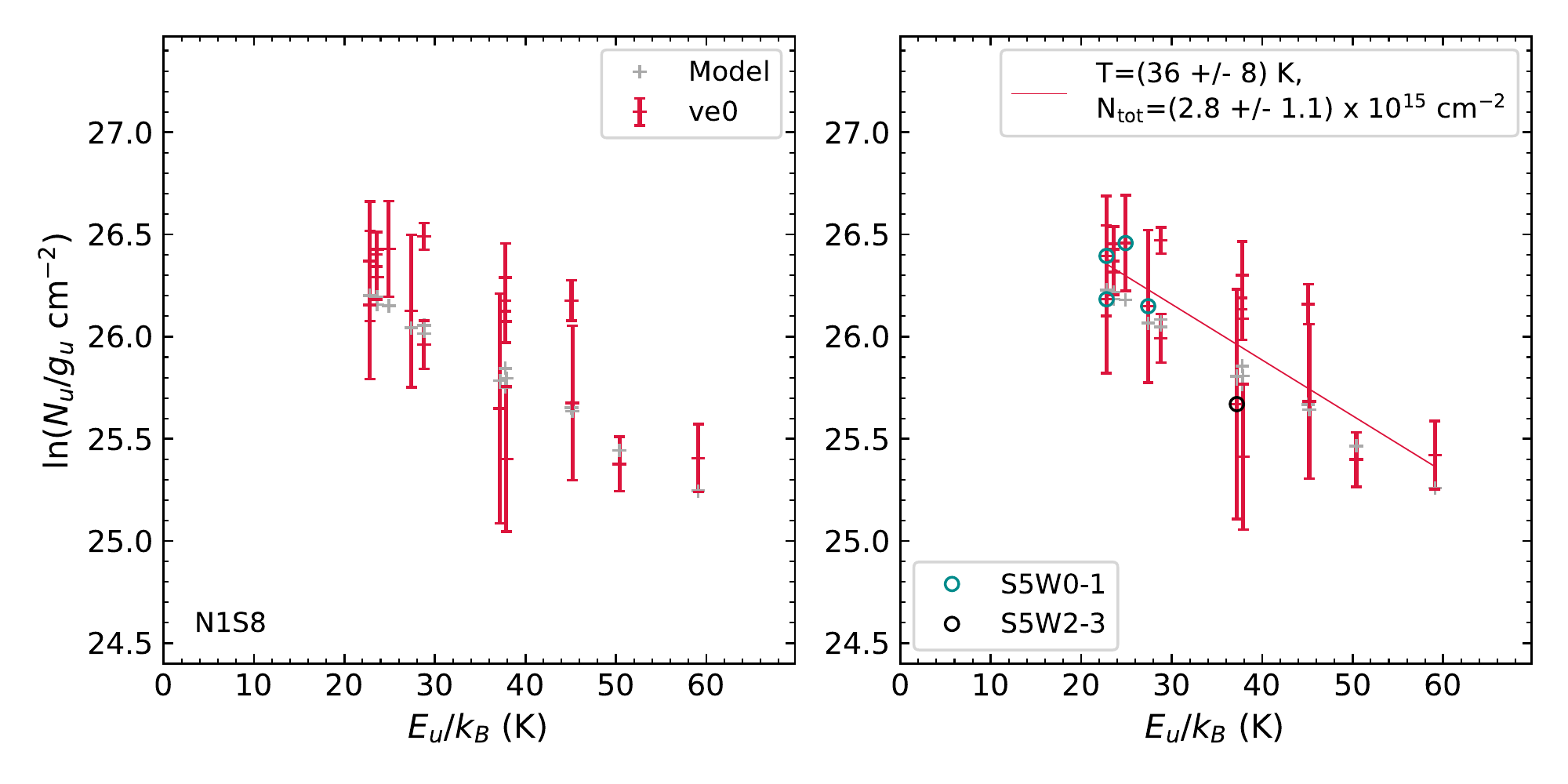}
    \caption{Same as Fig.\,\ref{fig:PD_met}, but for \mf.}
    \label{fig:PD_mf}
\end{figure*}

\begin{figure*}[h]
    \includegraphics[width=0.49\textwidth]{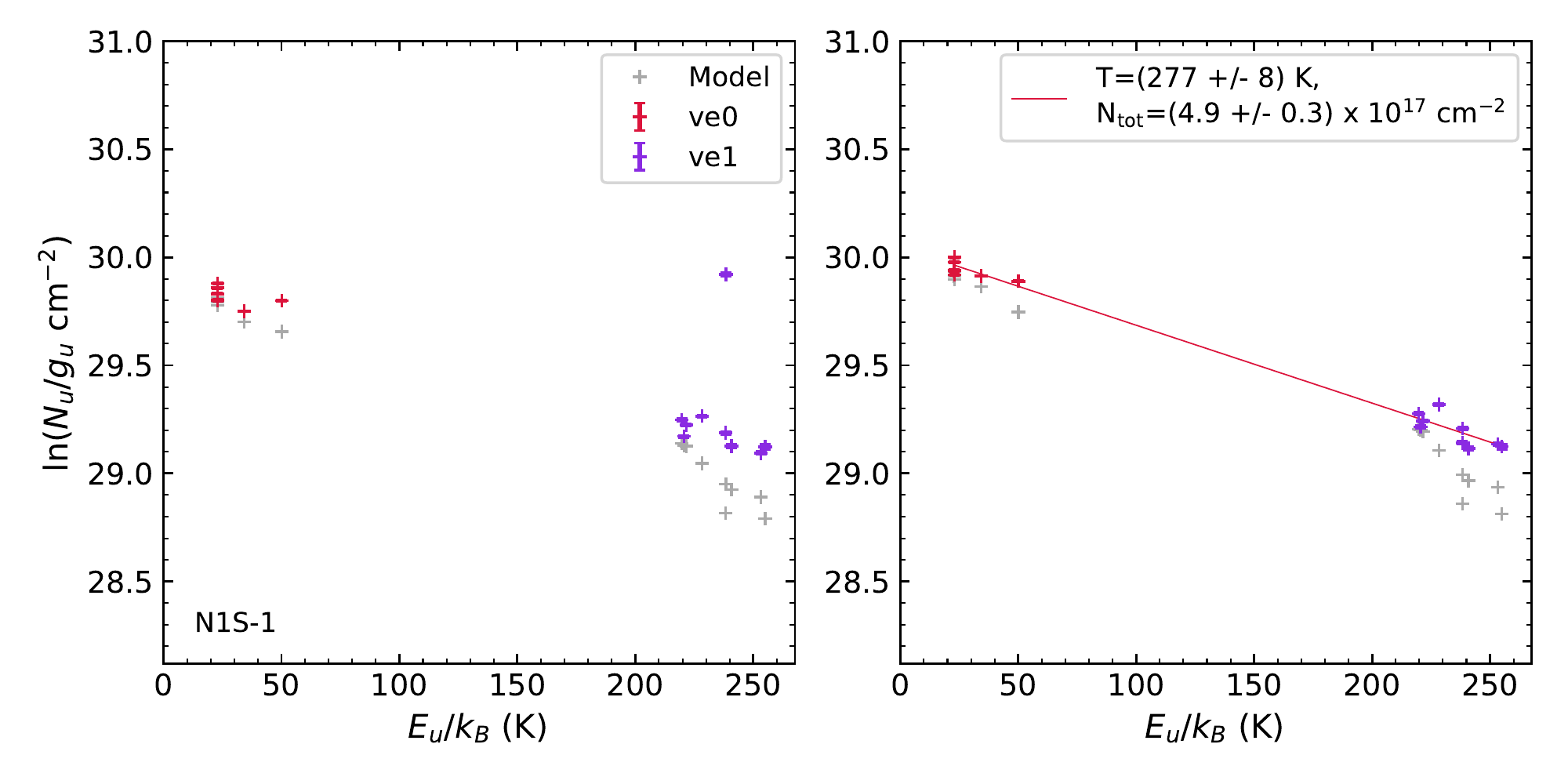}
    \includegraphics[width=0.49\textwidth]{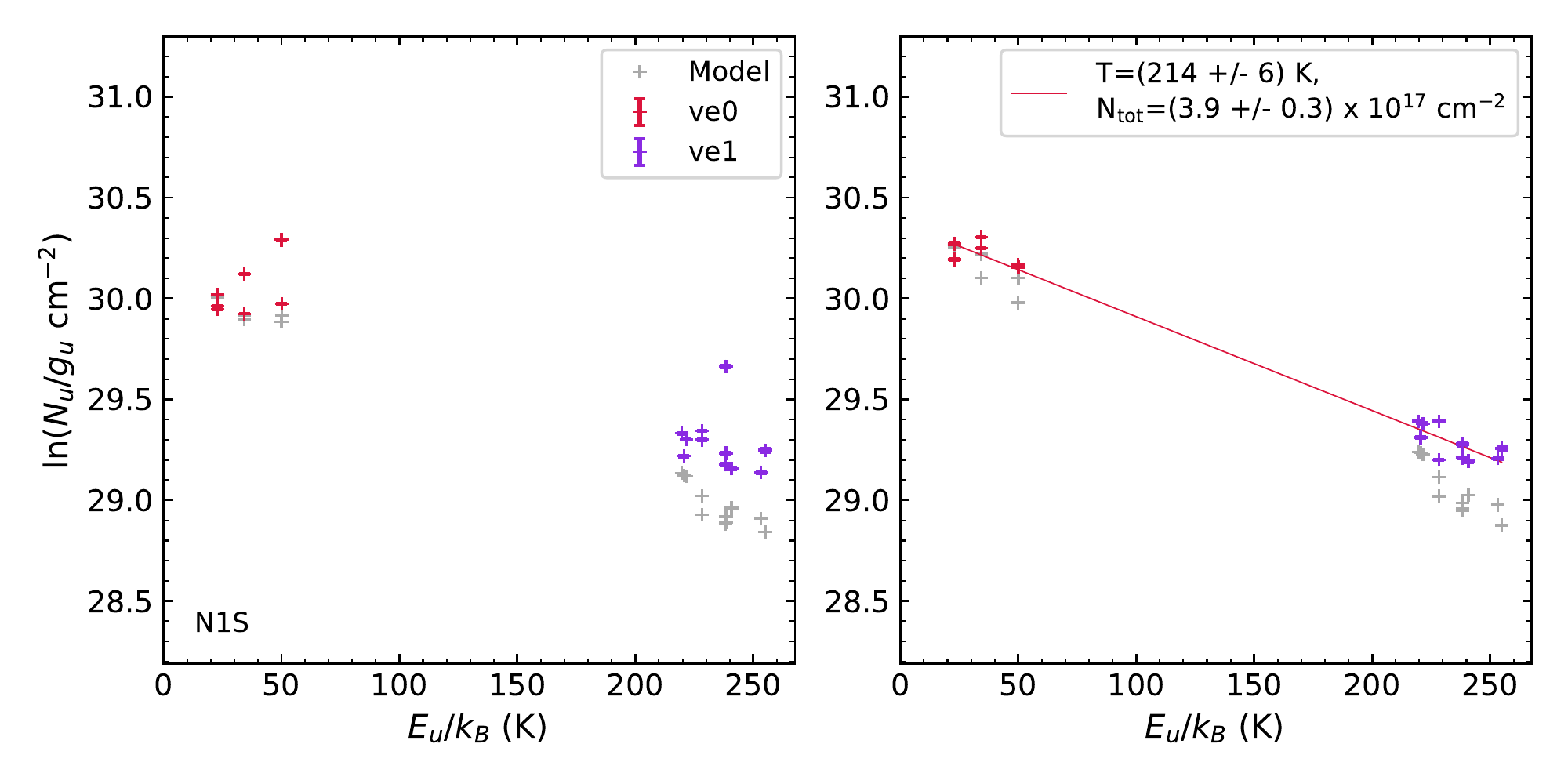}
    \includegraphics[width=0.49\textwidth]{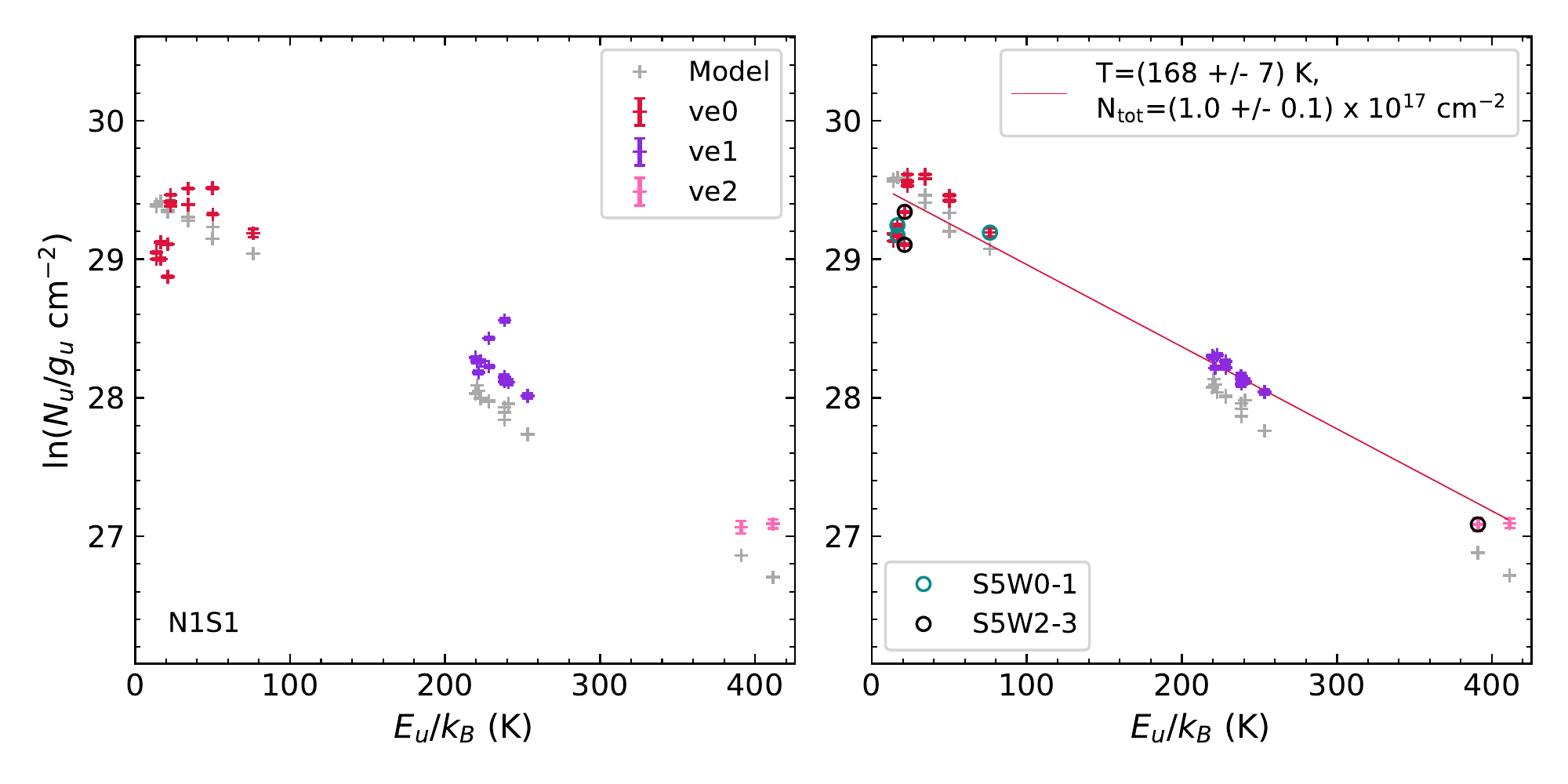}
    \includegraphics[width=0.49\textwidth]{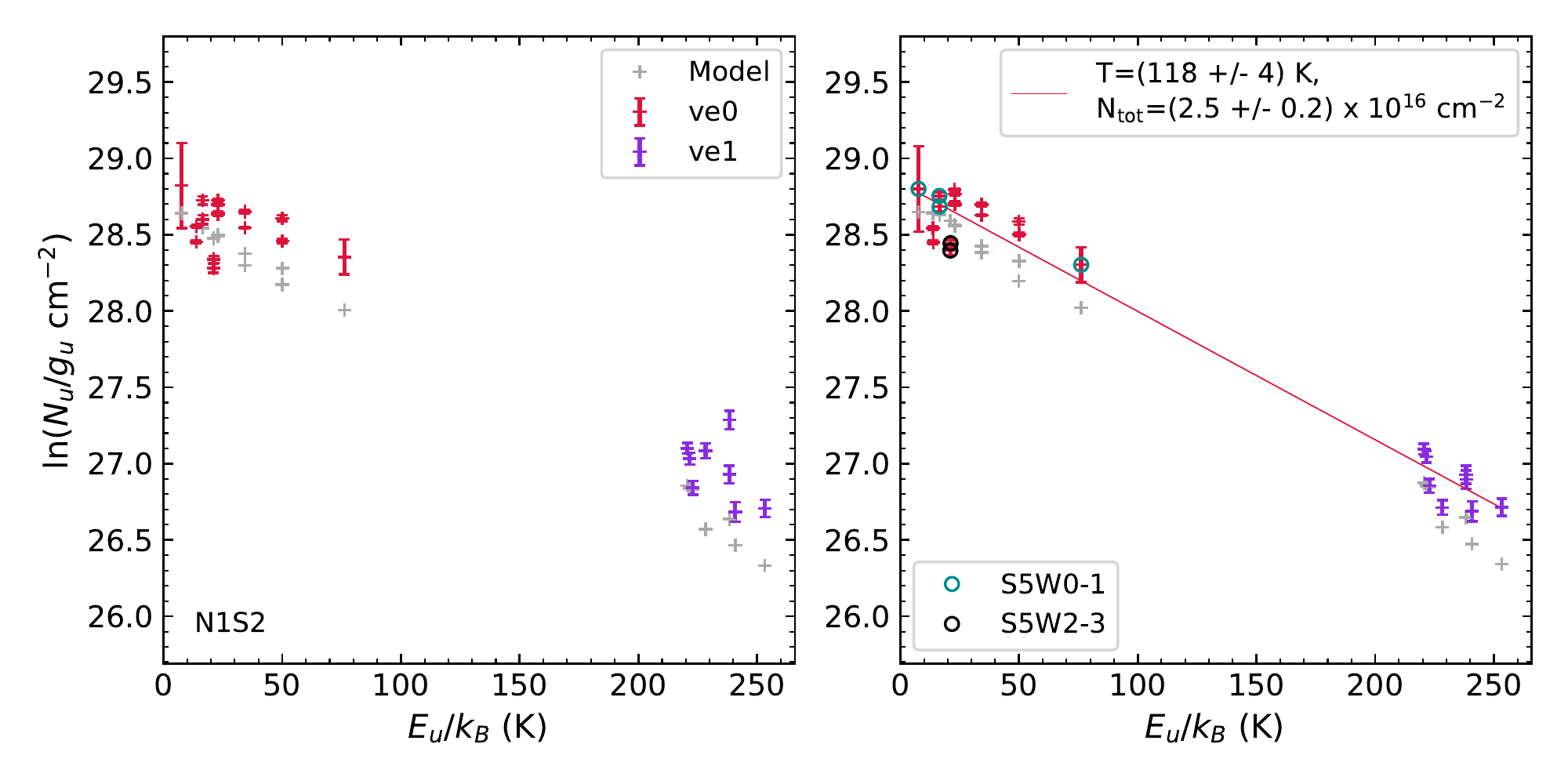}
    \includegraphics[width=0.49\textwidth]{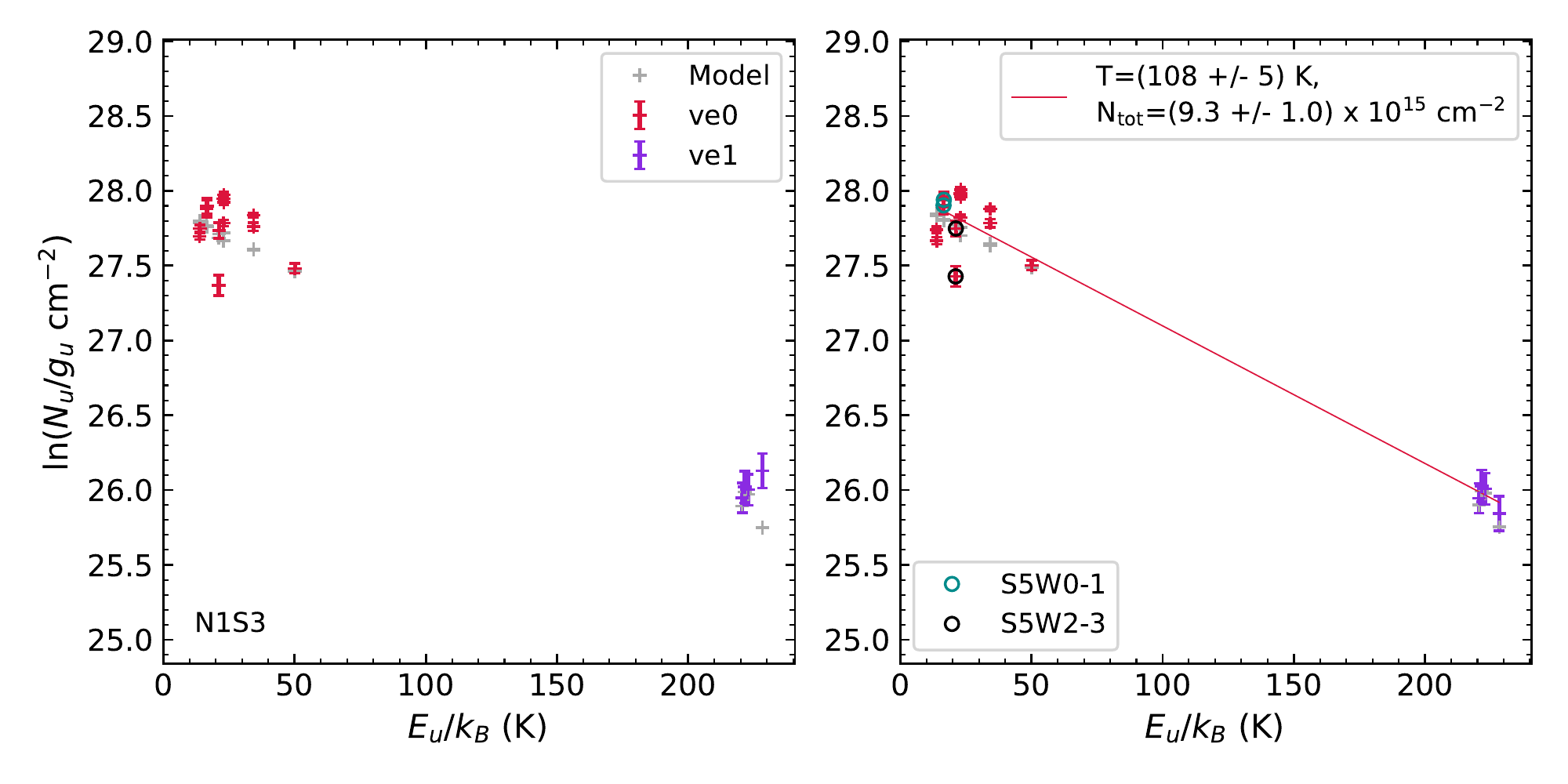}
    \includegraphics[width=0.49\textwidth]{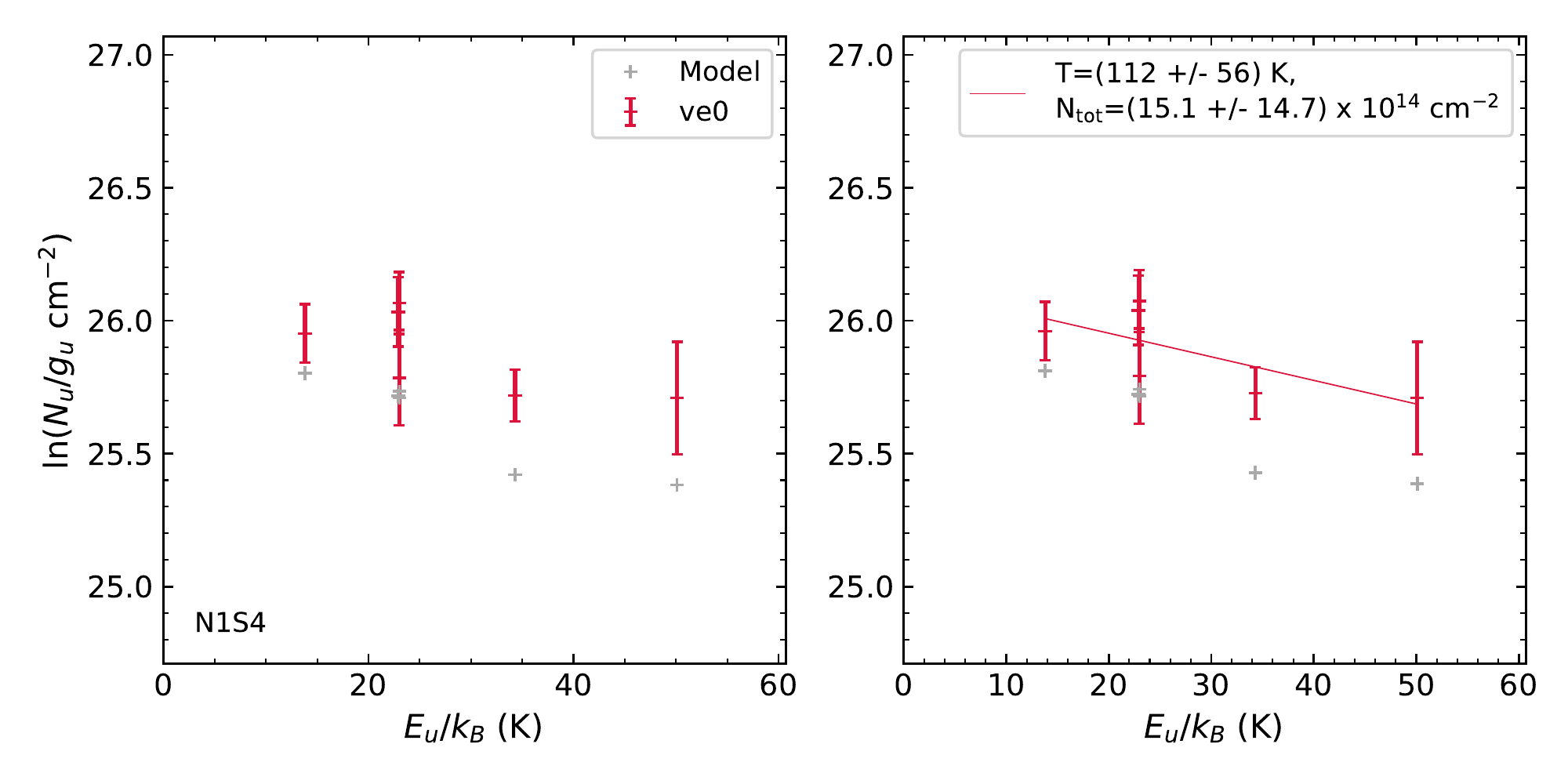}
    \caption{Same as Fig.\,\ref{fig:PD_met}, but for \ad.}
    \label{fig:PD_ad}
\end{figure*}

\begin{figure*}[h]
    \includegraphics[width=0.49\textwidth]{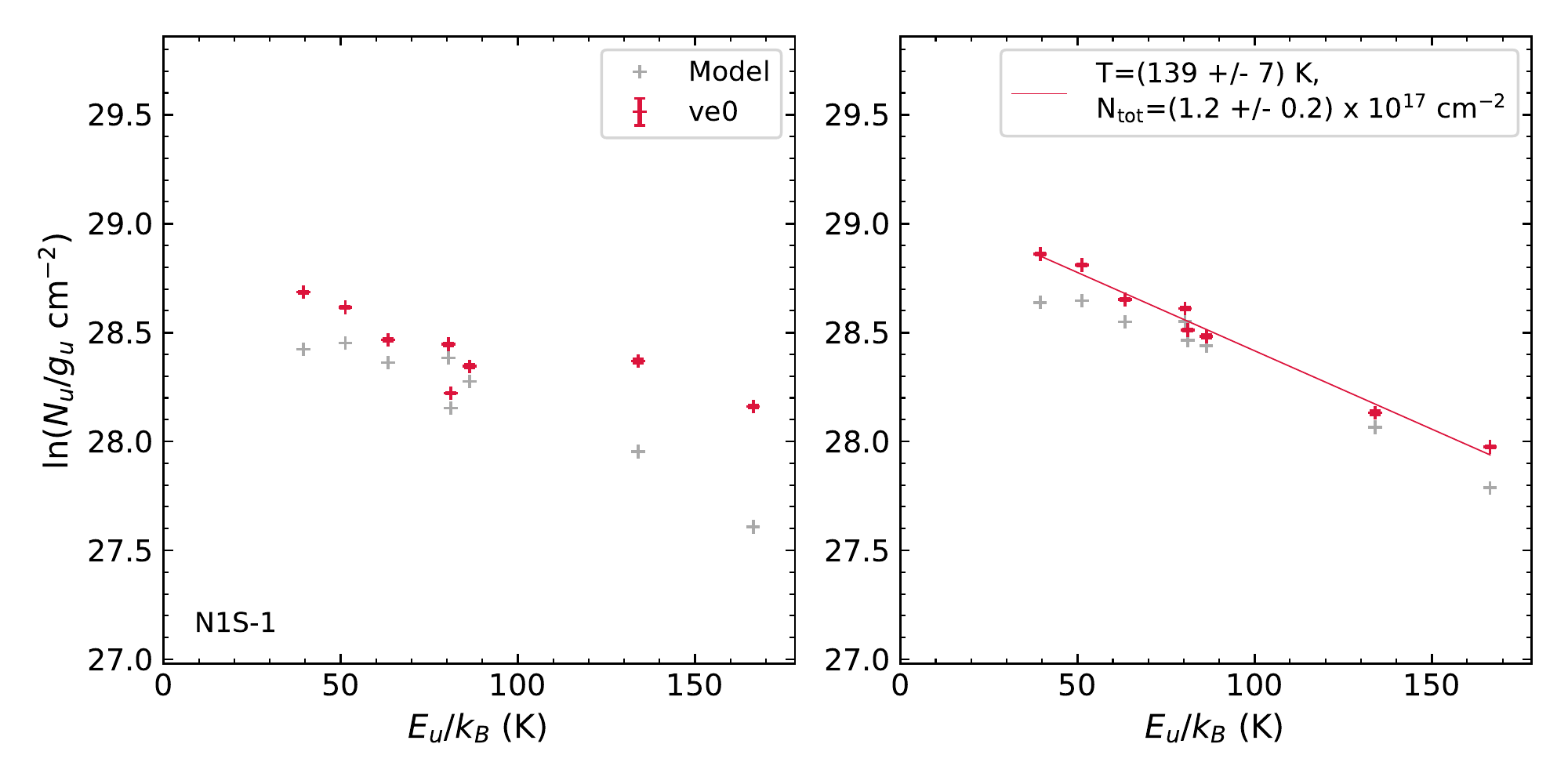}
    \includegraphics[width=0.49\textwidth]{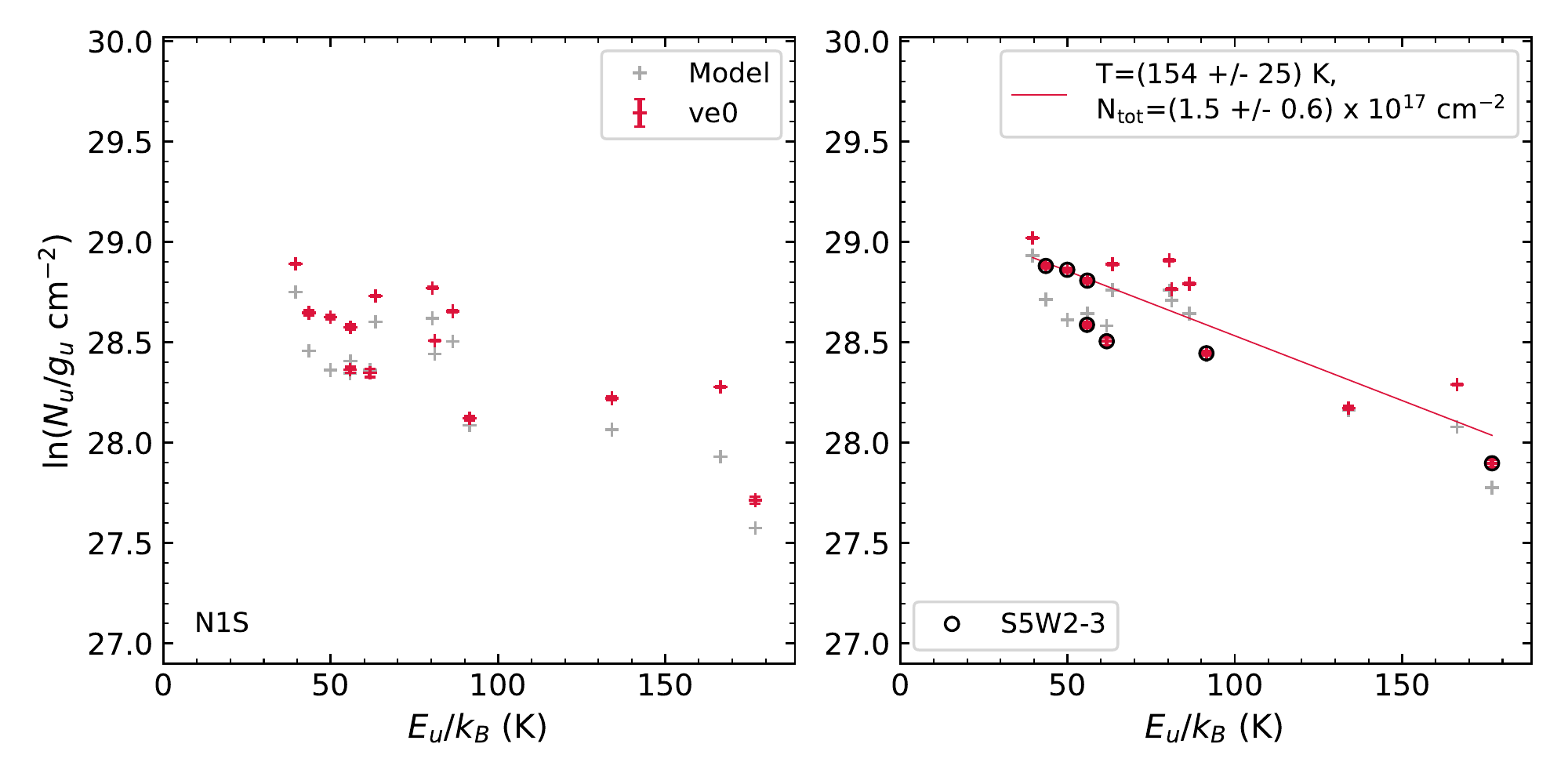}
    \includegraphics[width=0.49\textwidth]{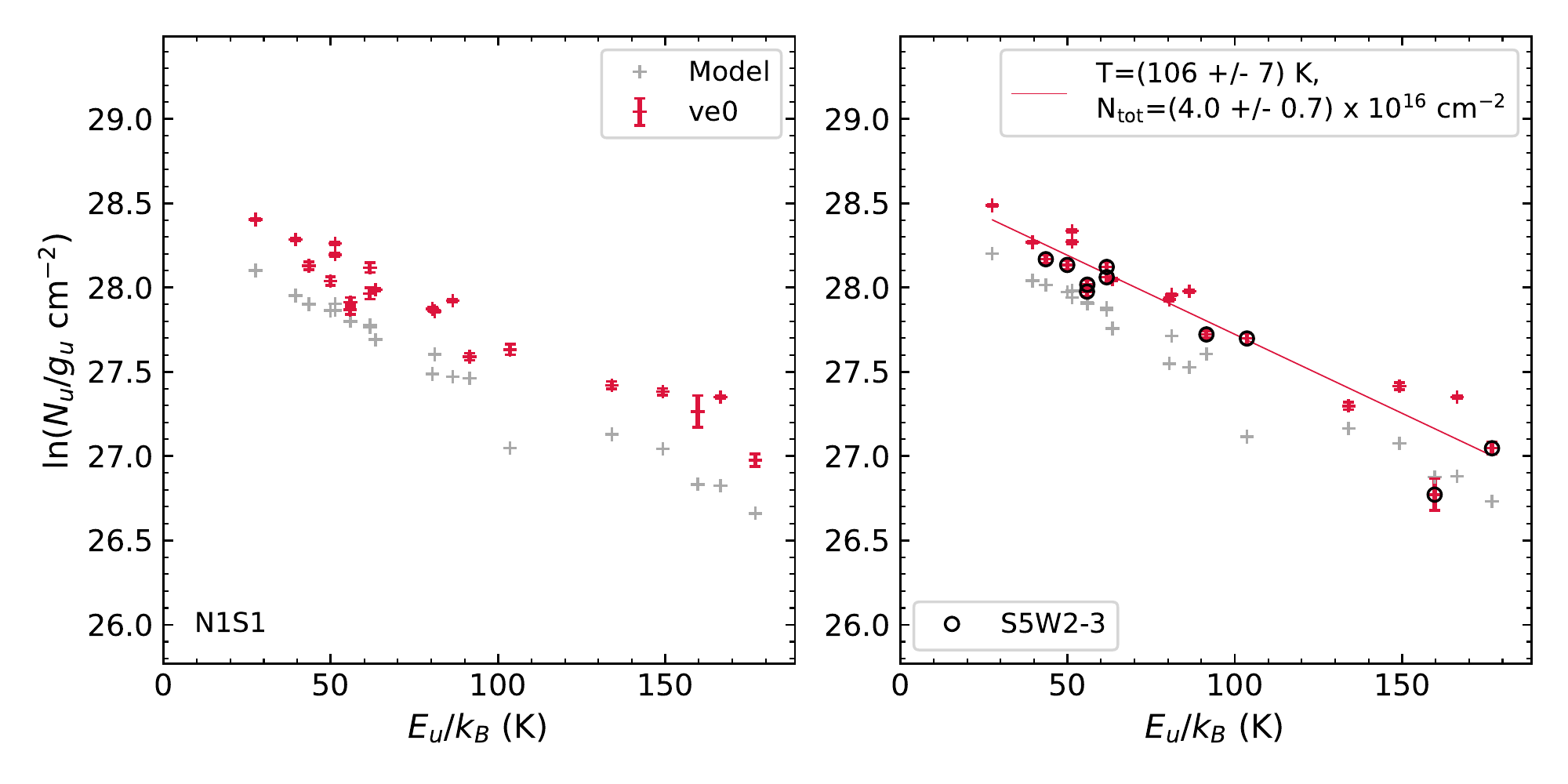}
    \includegraphics[width=0.49\textwidth]{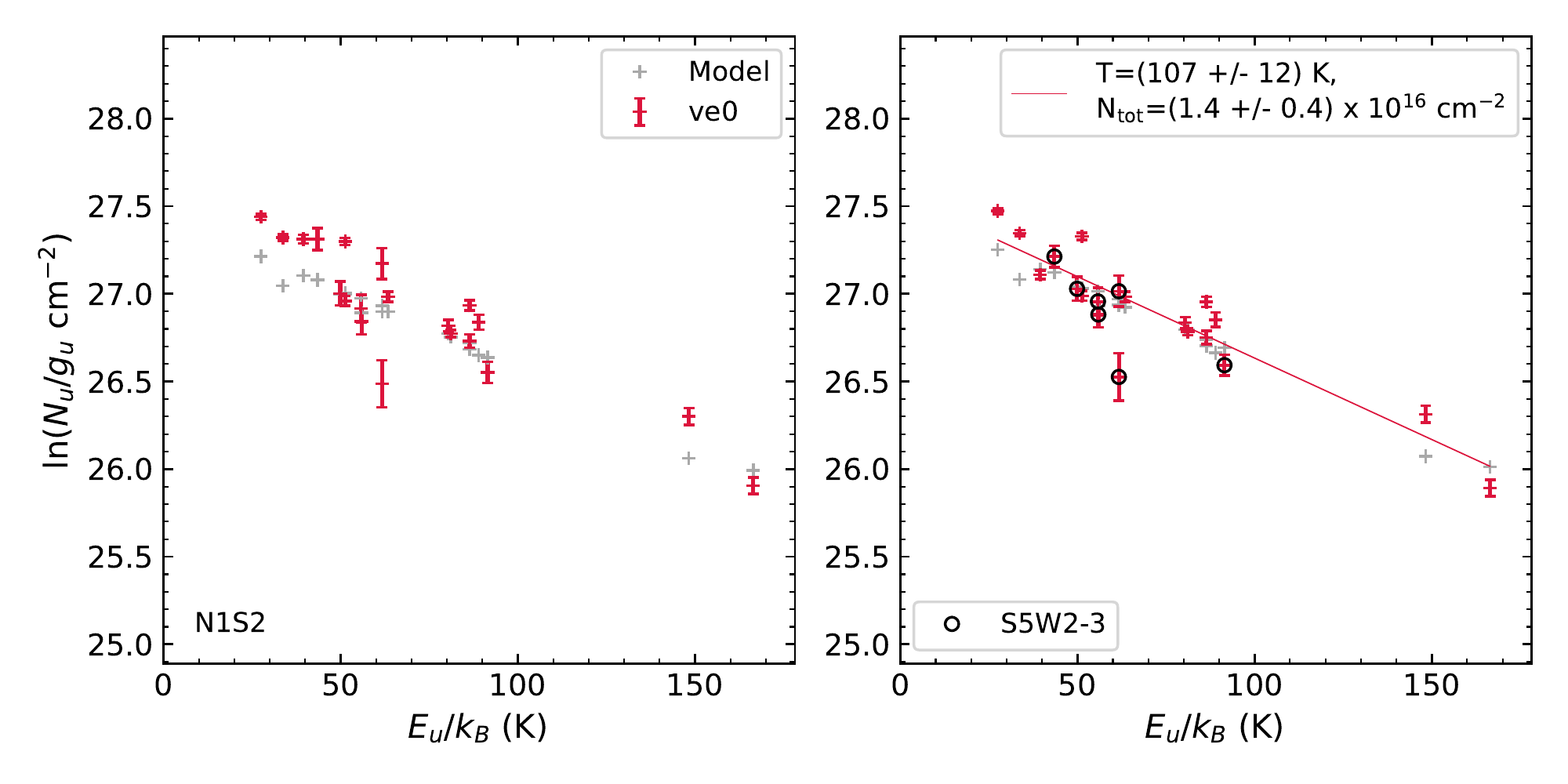}
    \includegraphics[width=0.49\textwidth]{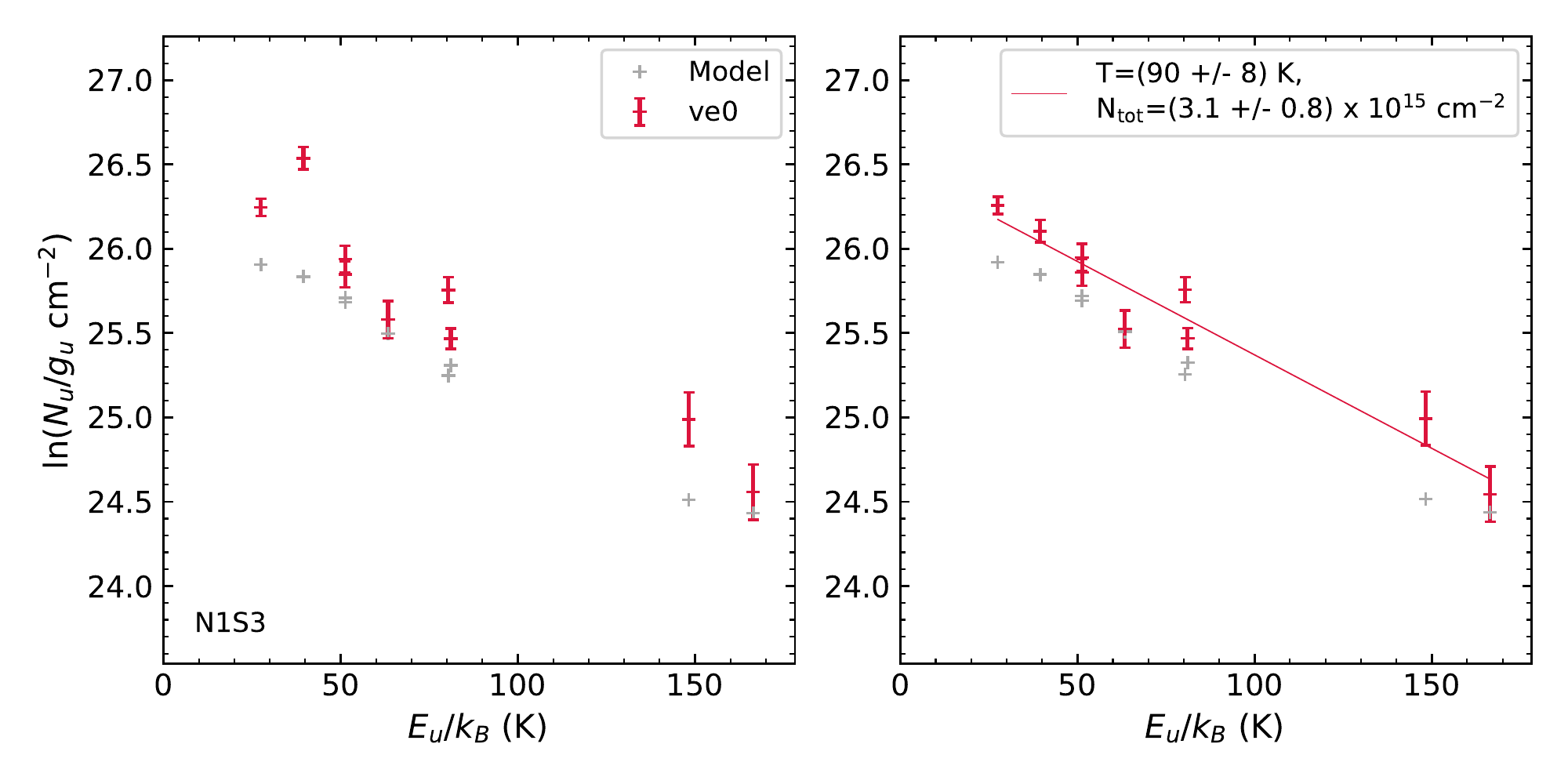}
    \caption{Same as Fig.\,\ref{fig:PD_met}, but for \mic.}
    \label{fig:PD_mic}
\end{figure*}

\begin{figure*}[h]
    \includegraphics[width=0.49\textwidth]{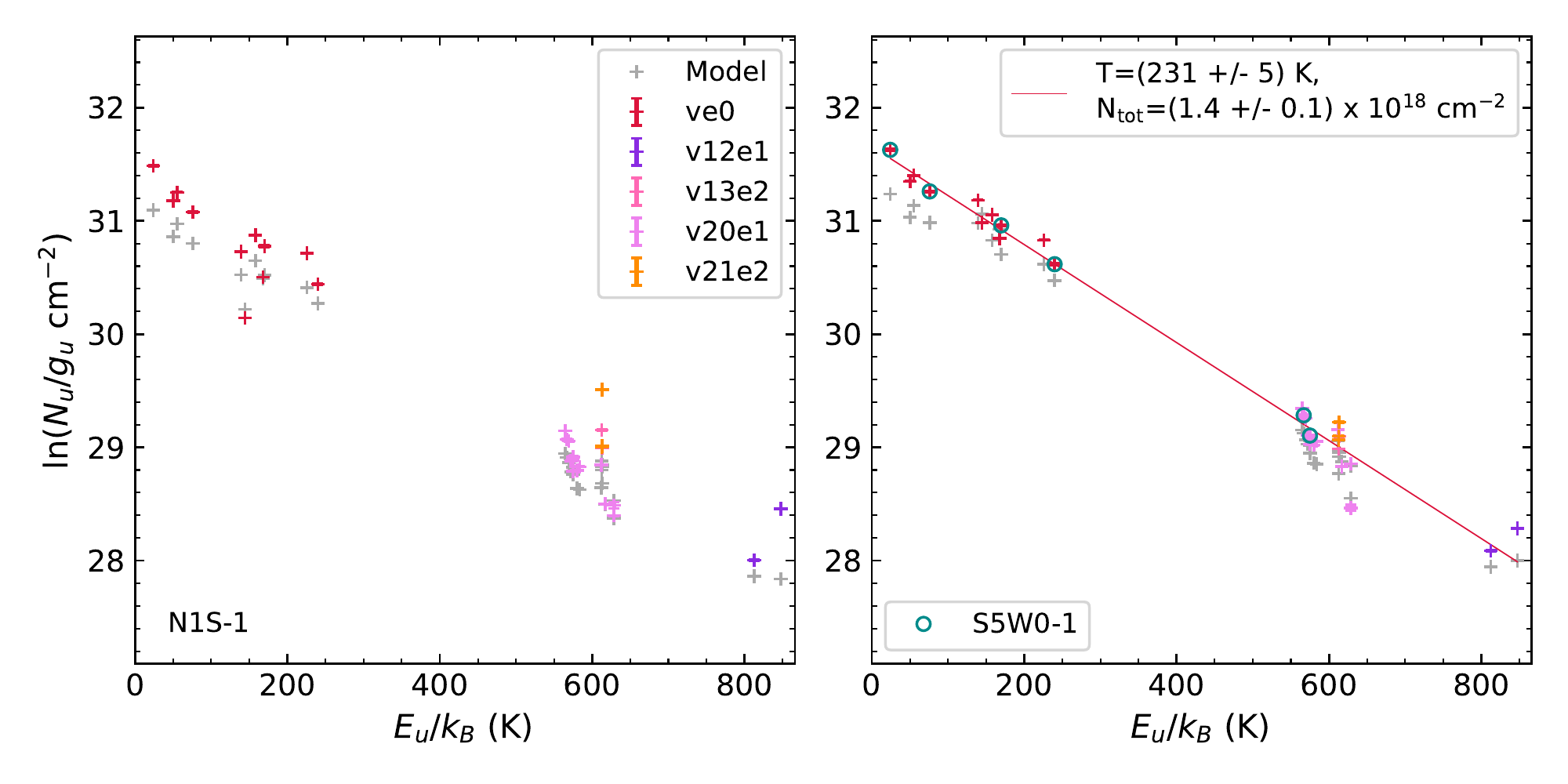}
    \includegraphics[width=0.49\textwidth]{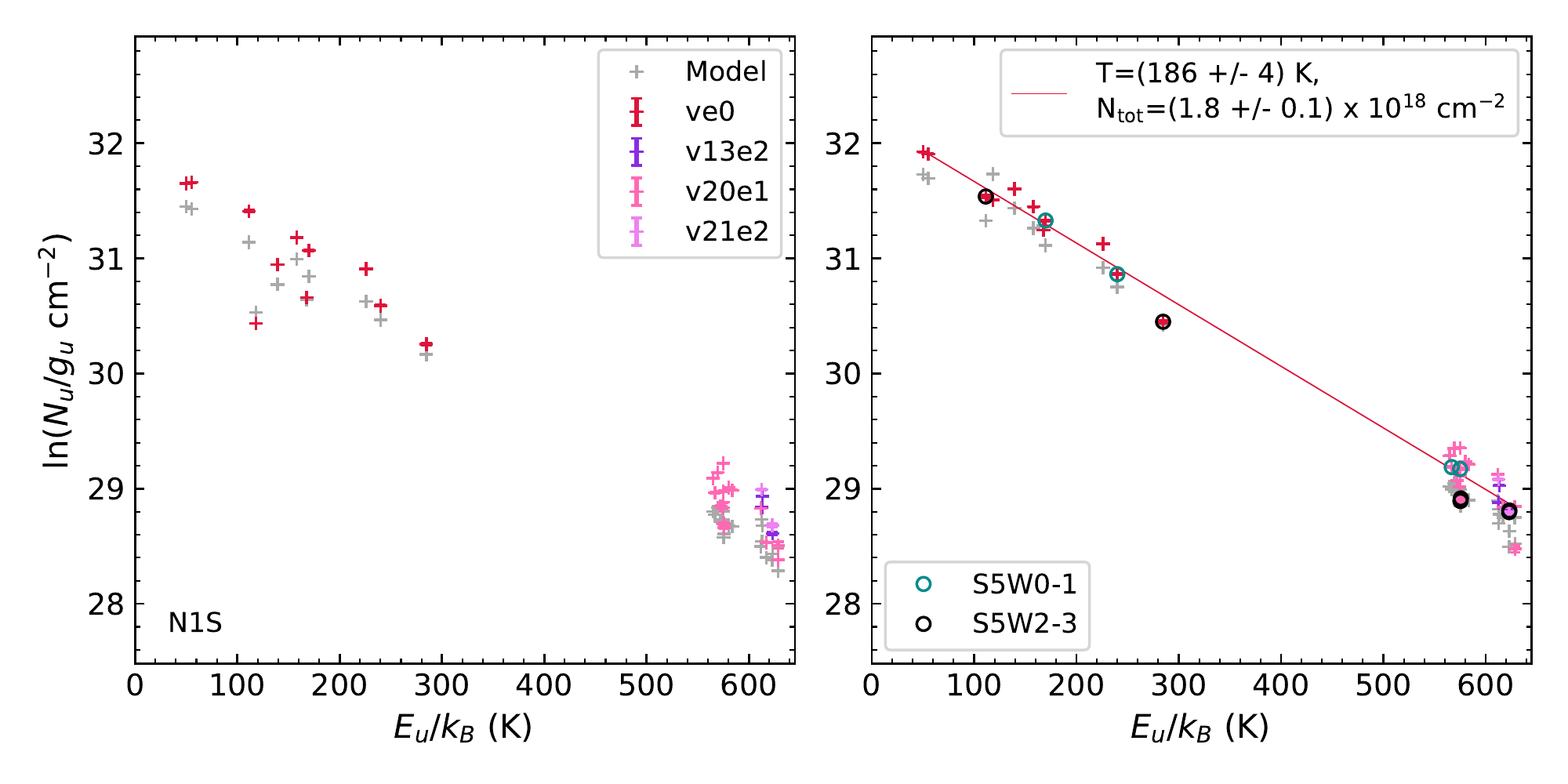}
    \includegraphics[width=0.49\textwidth]{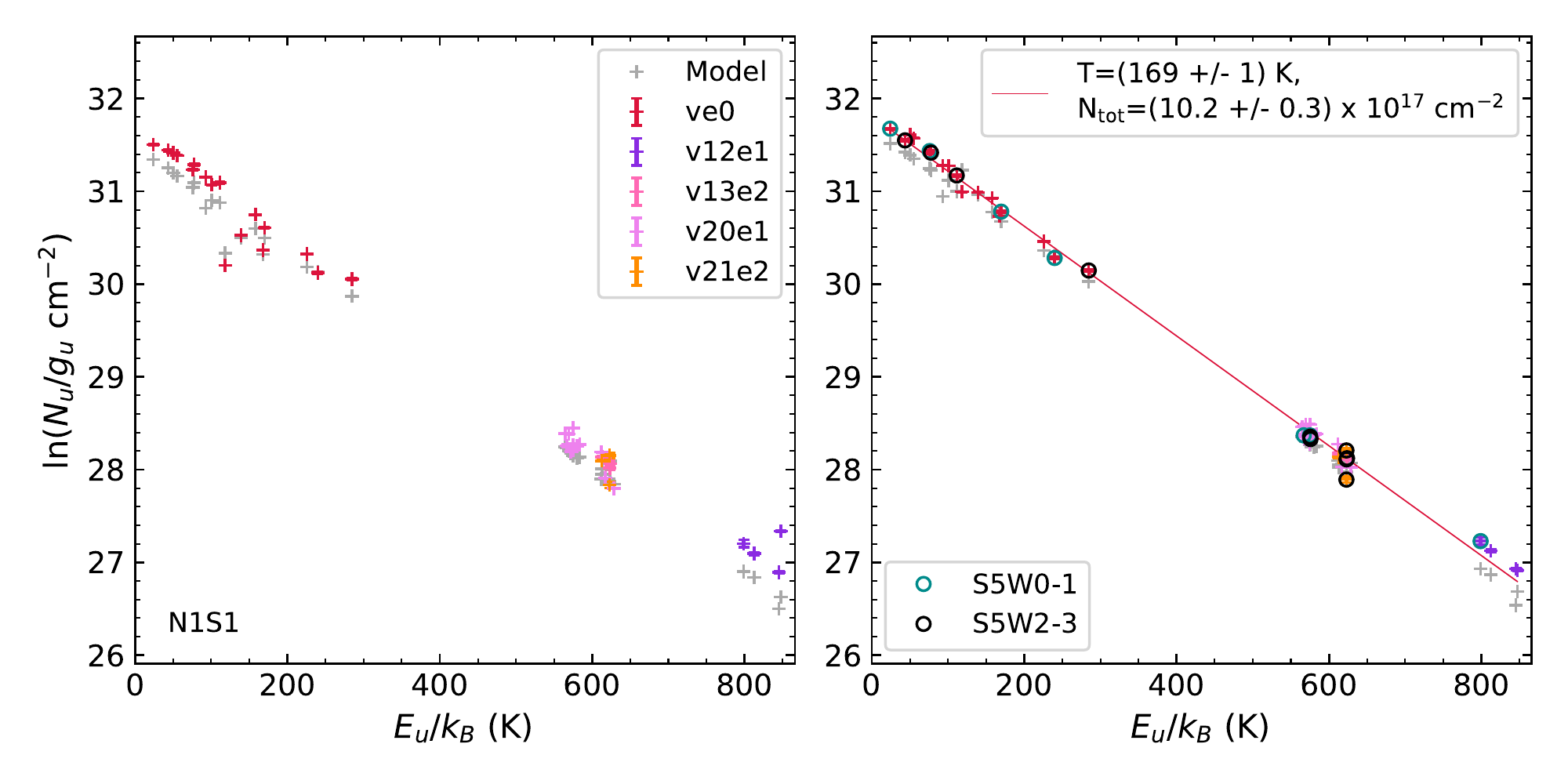}
    \includegraphics[width=0.49\textwidth]{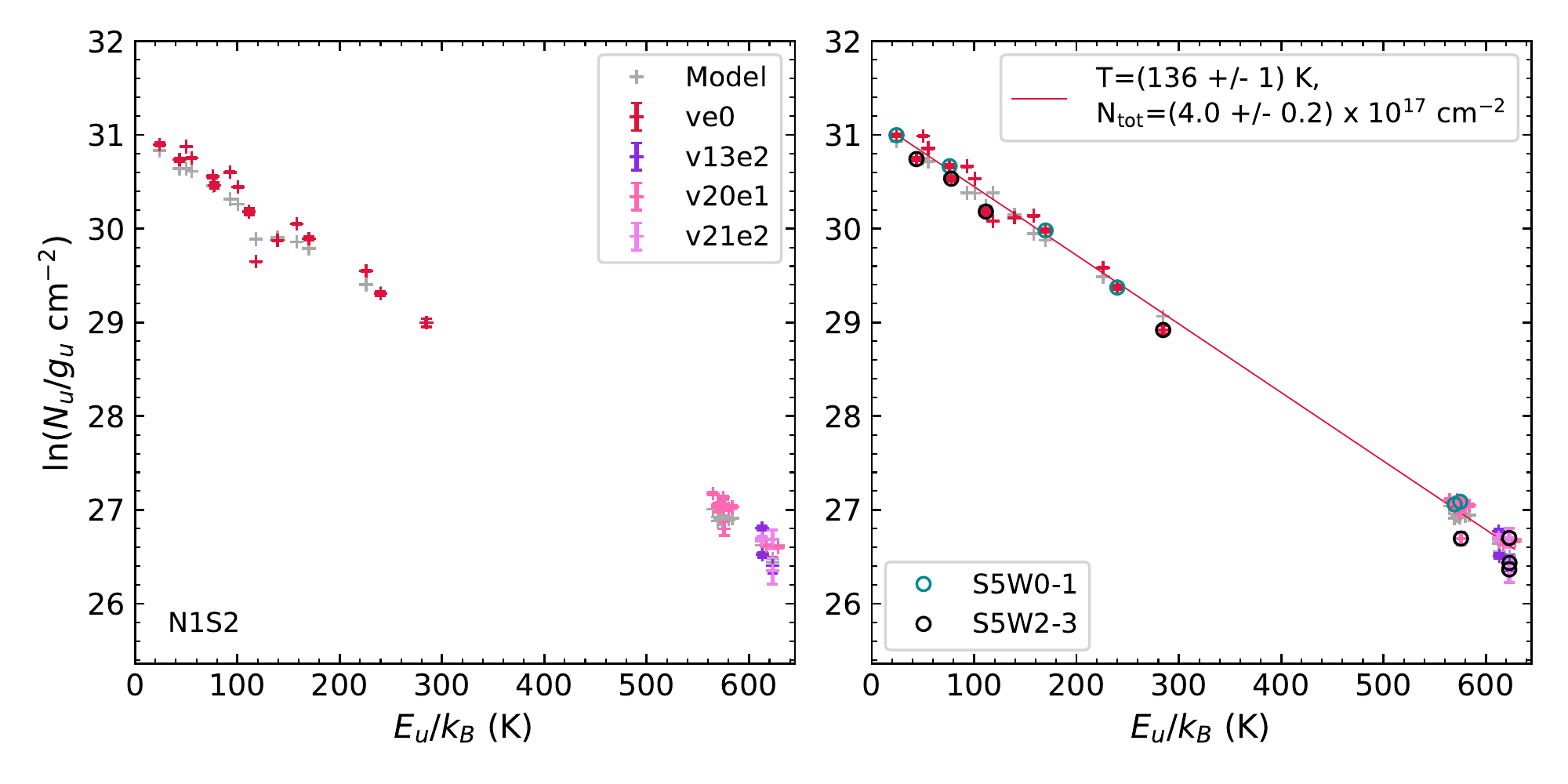}
    \includegraphics[width=0.49\textwidth]{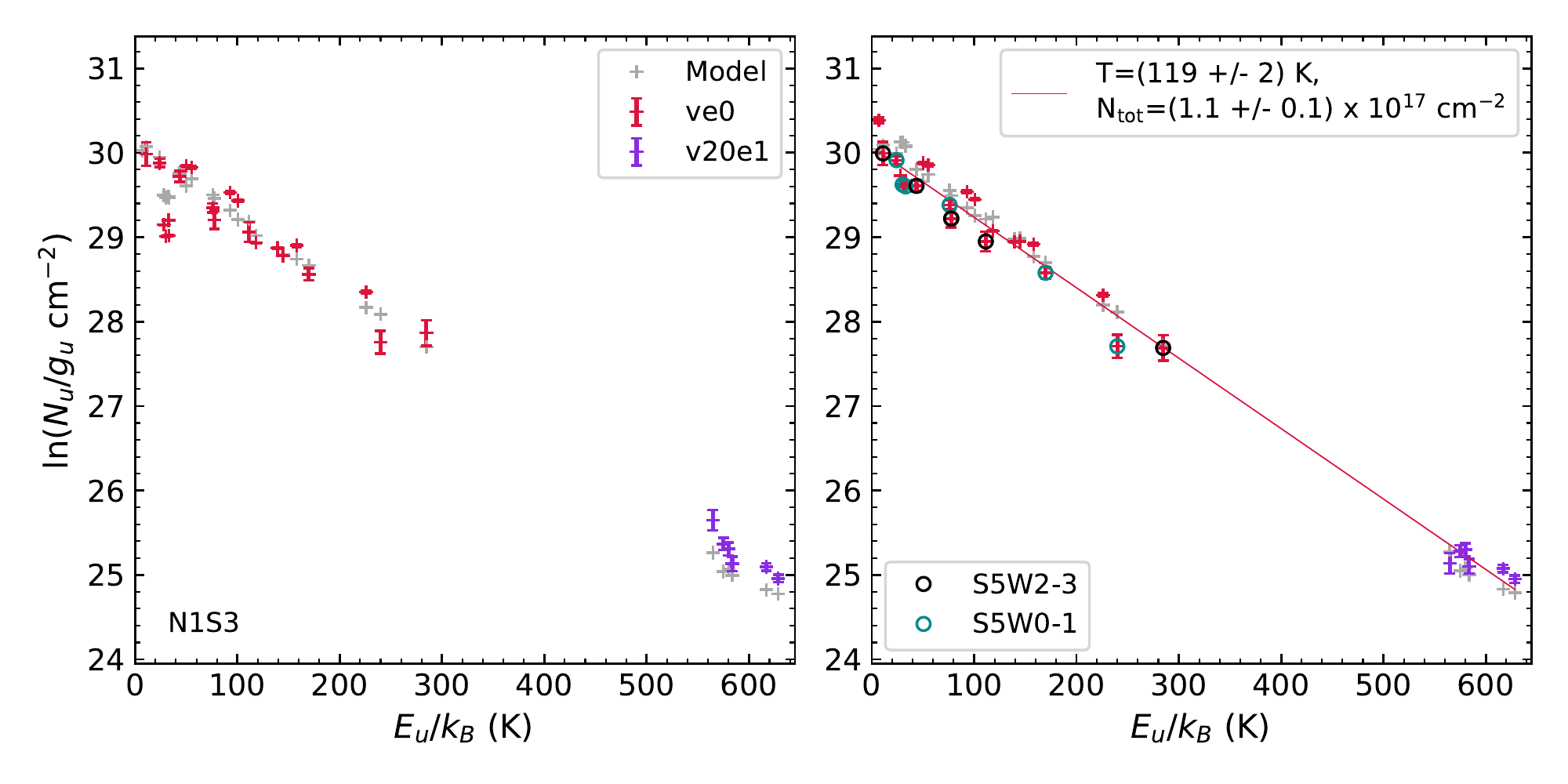}
    \includegraphics[width=0.49\textwidth]{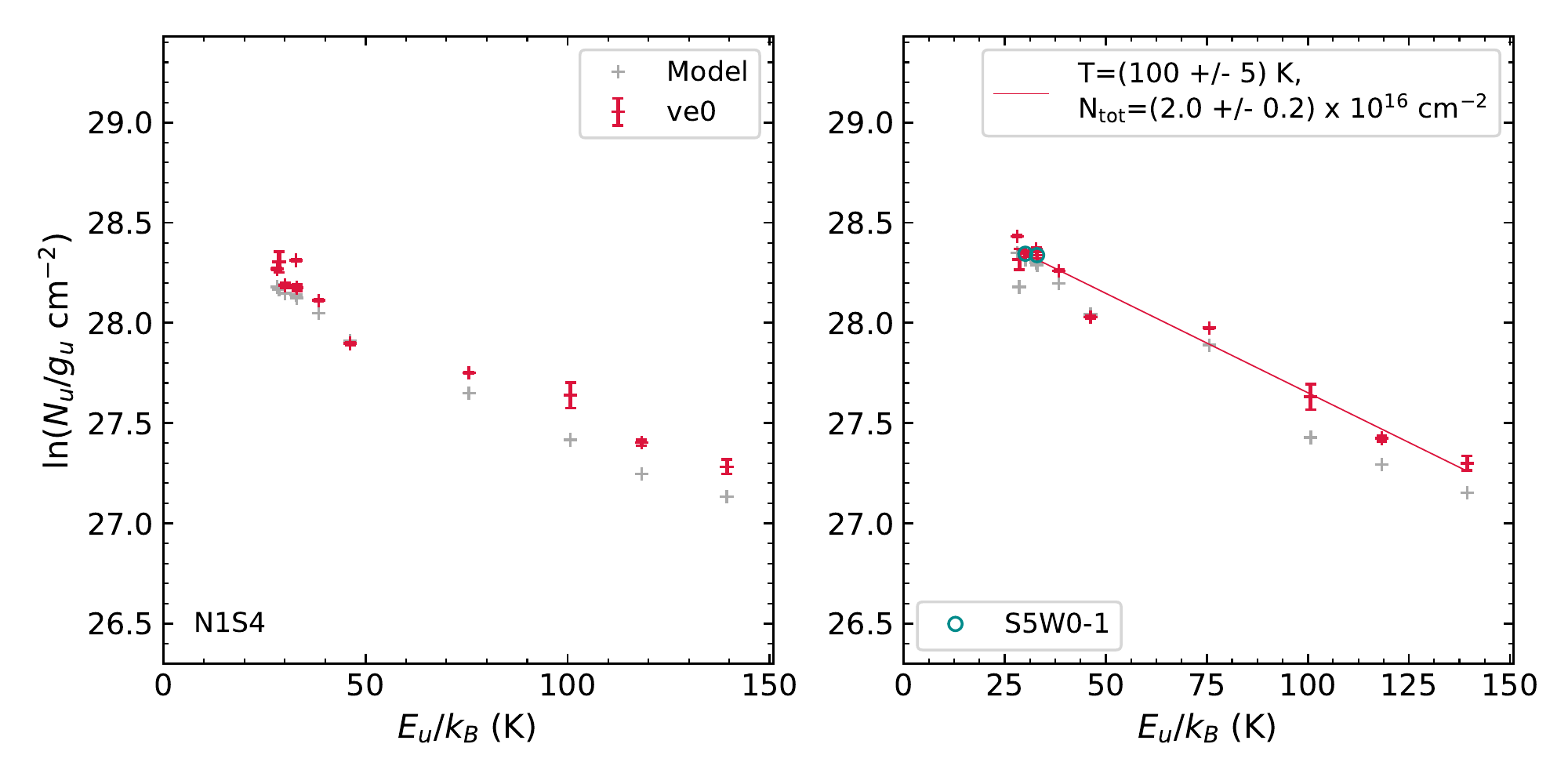}
    \includegraphics[width=0.49\textwidth]{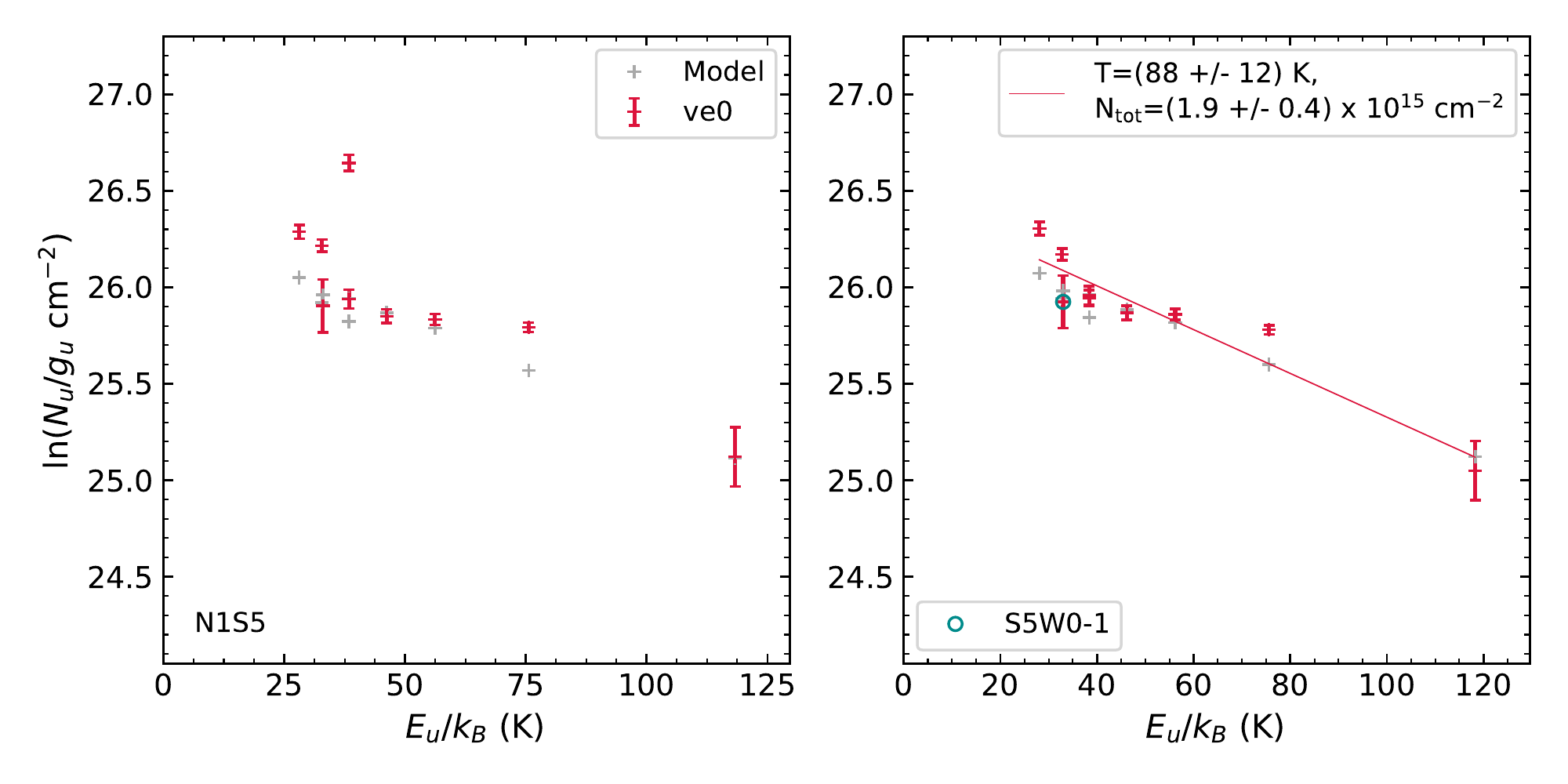}
    \includegraphics[width=0.49\textwidth]{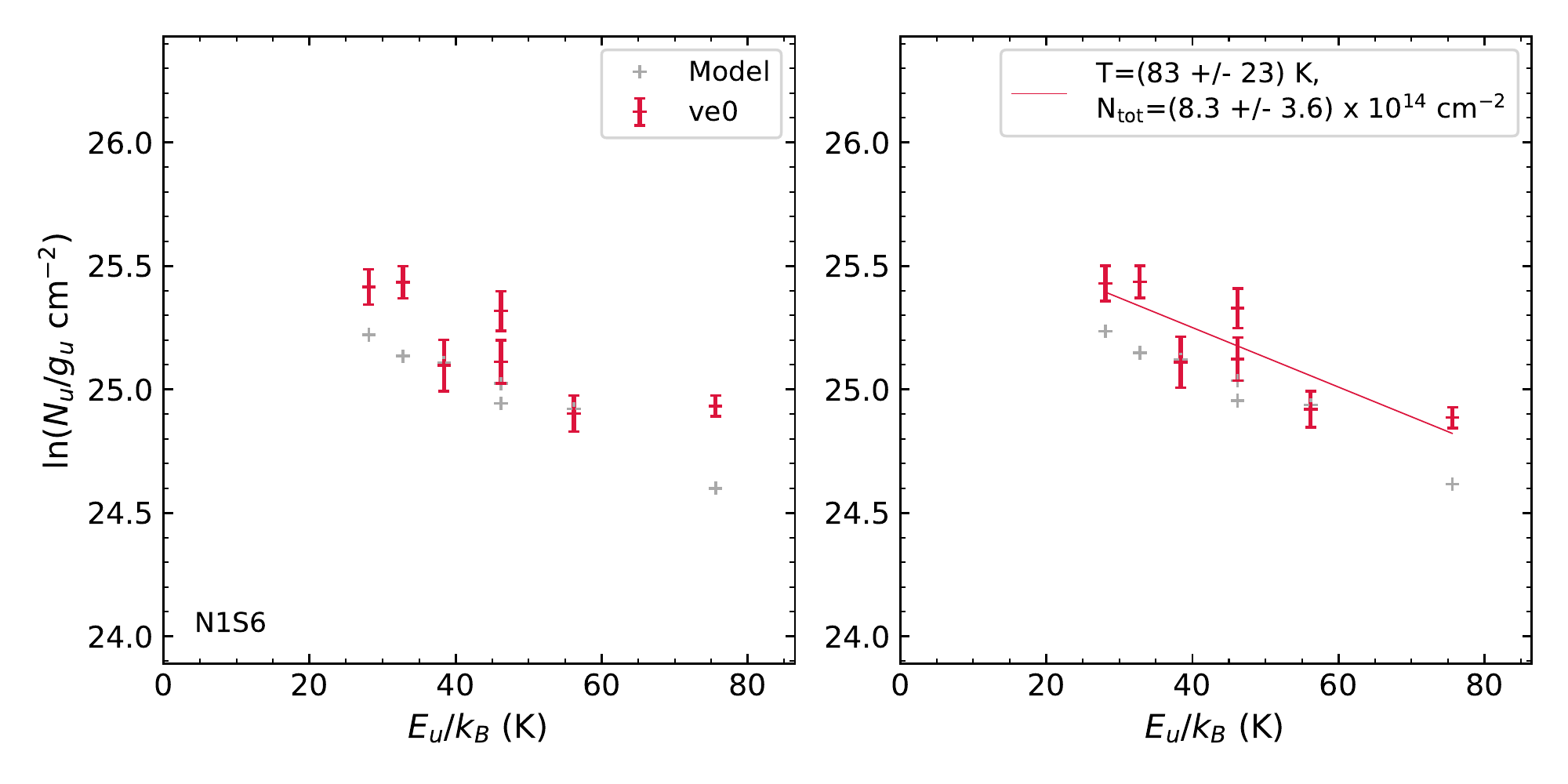}
    \includegraphics[width=0.49\textwidth]{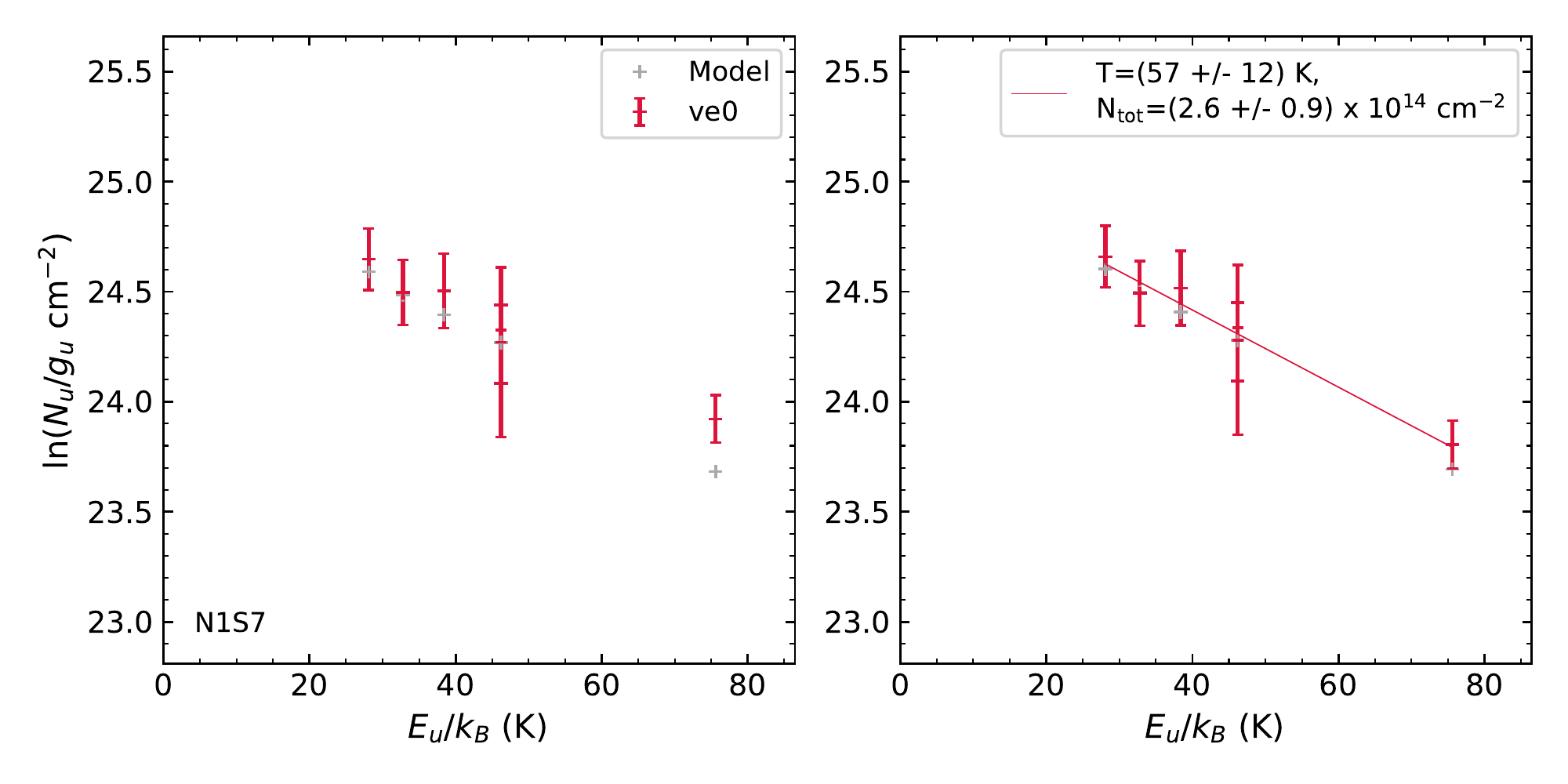}
    \caption{Same as Fig.\,\ref{fig:PD_met}, but for \etc.}
    \label{fig:PD_etc}
\end{figure*}

\begin{figure*}[h]
    \includegraphics[width=0.49\textwidth]{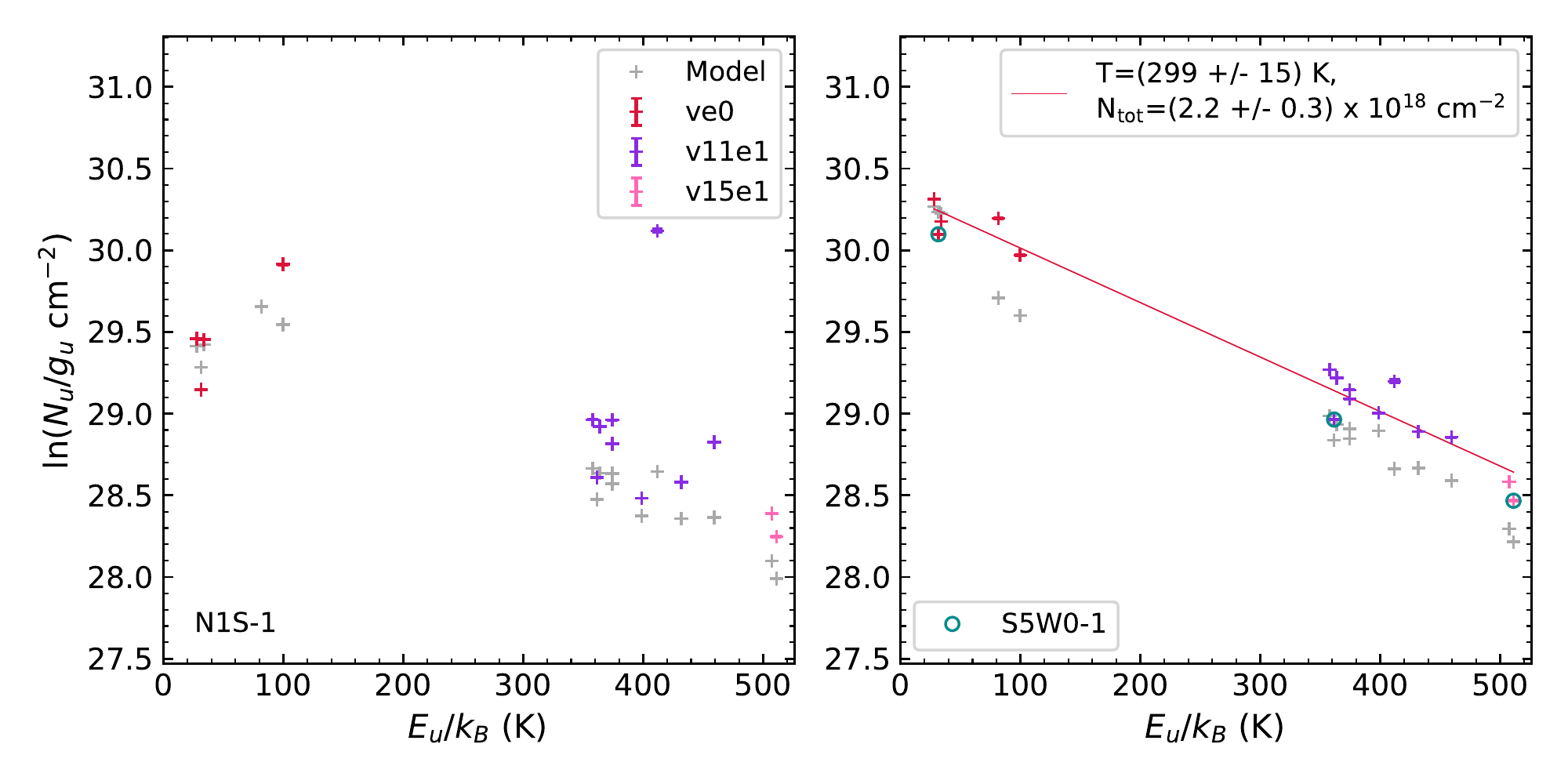}
    \includegraphics[width=0.49\textwidth]{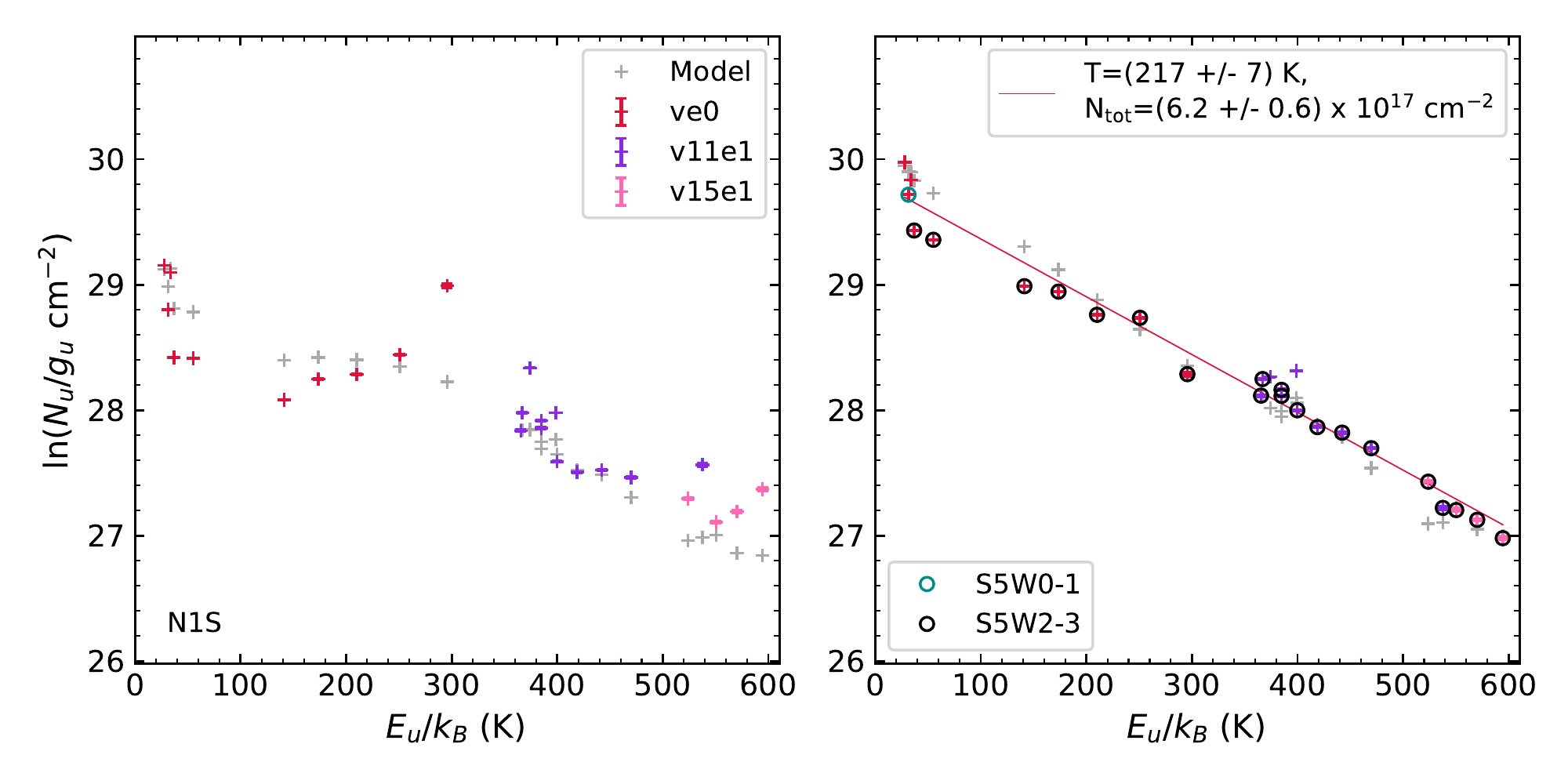}
    \includegraphics[width=0.49\textwidth]{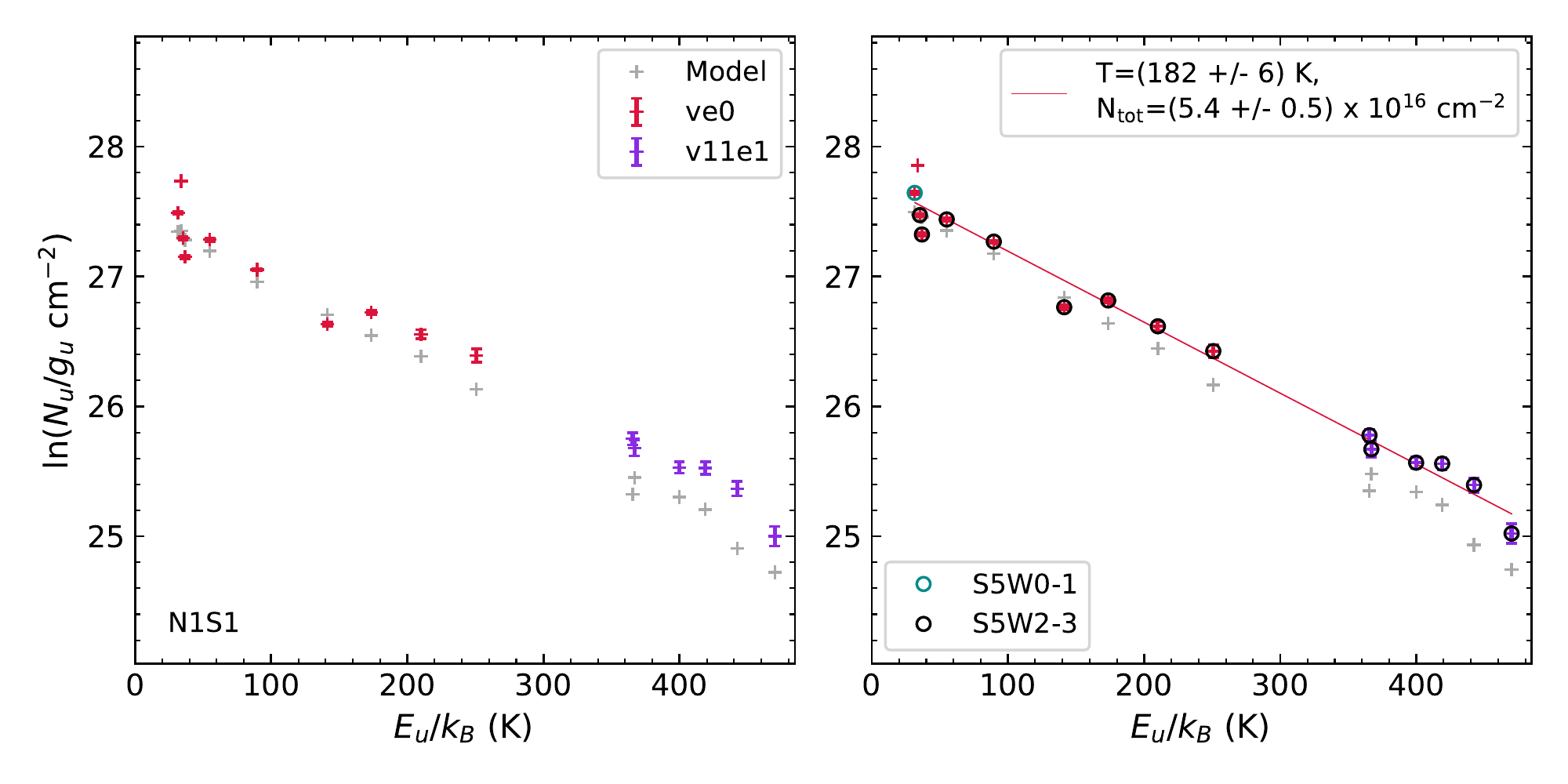}
    \includegraphics[width=0.49\textwidth]{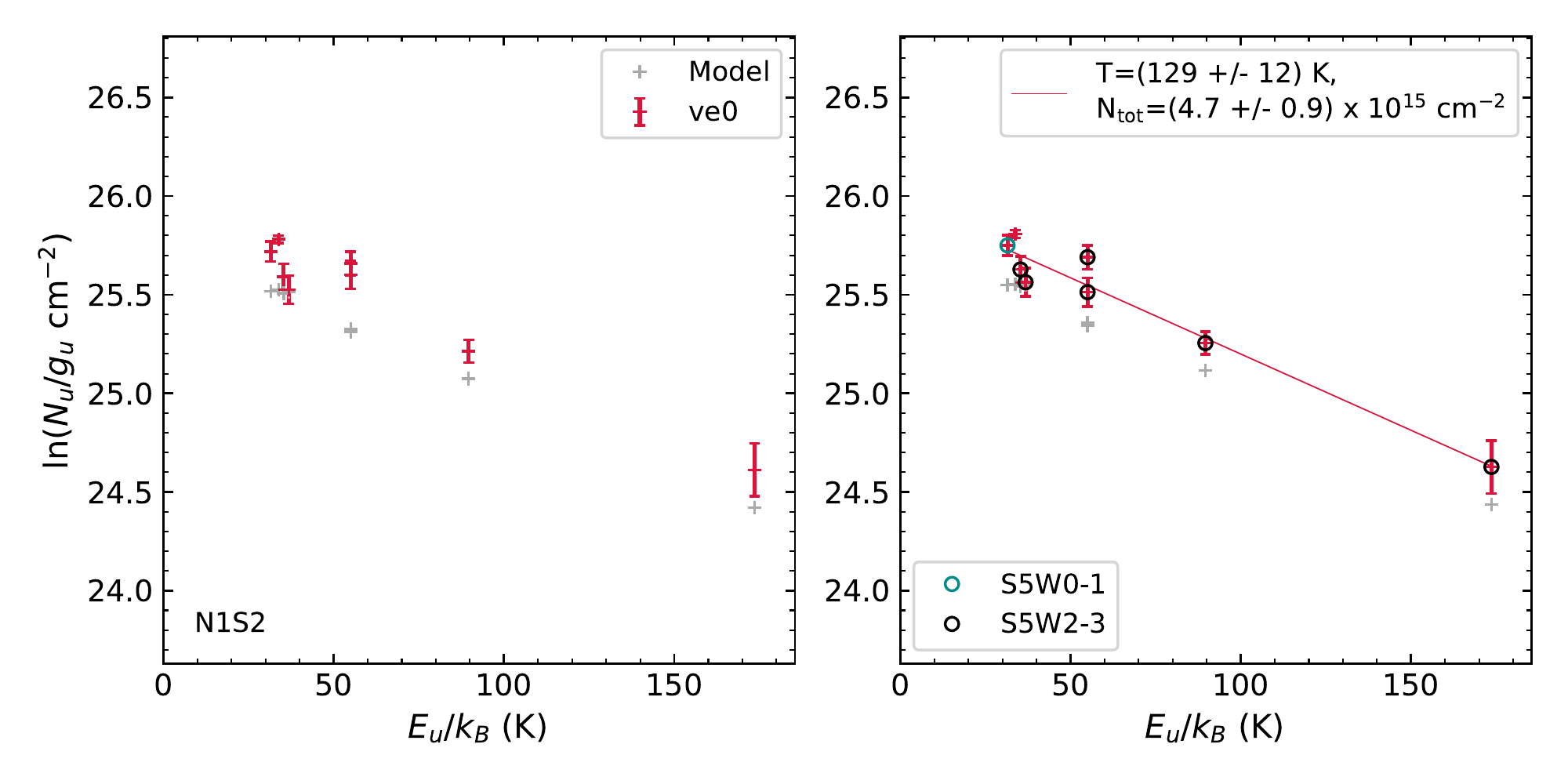}
    \includegraphics[width=0.49\textwidth]{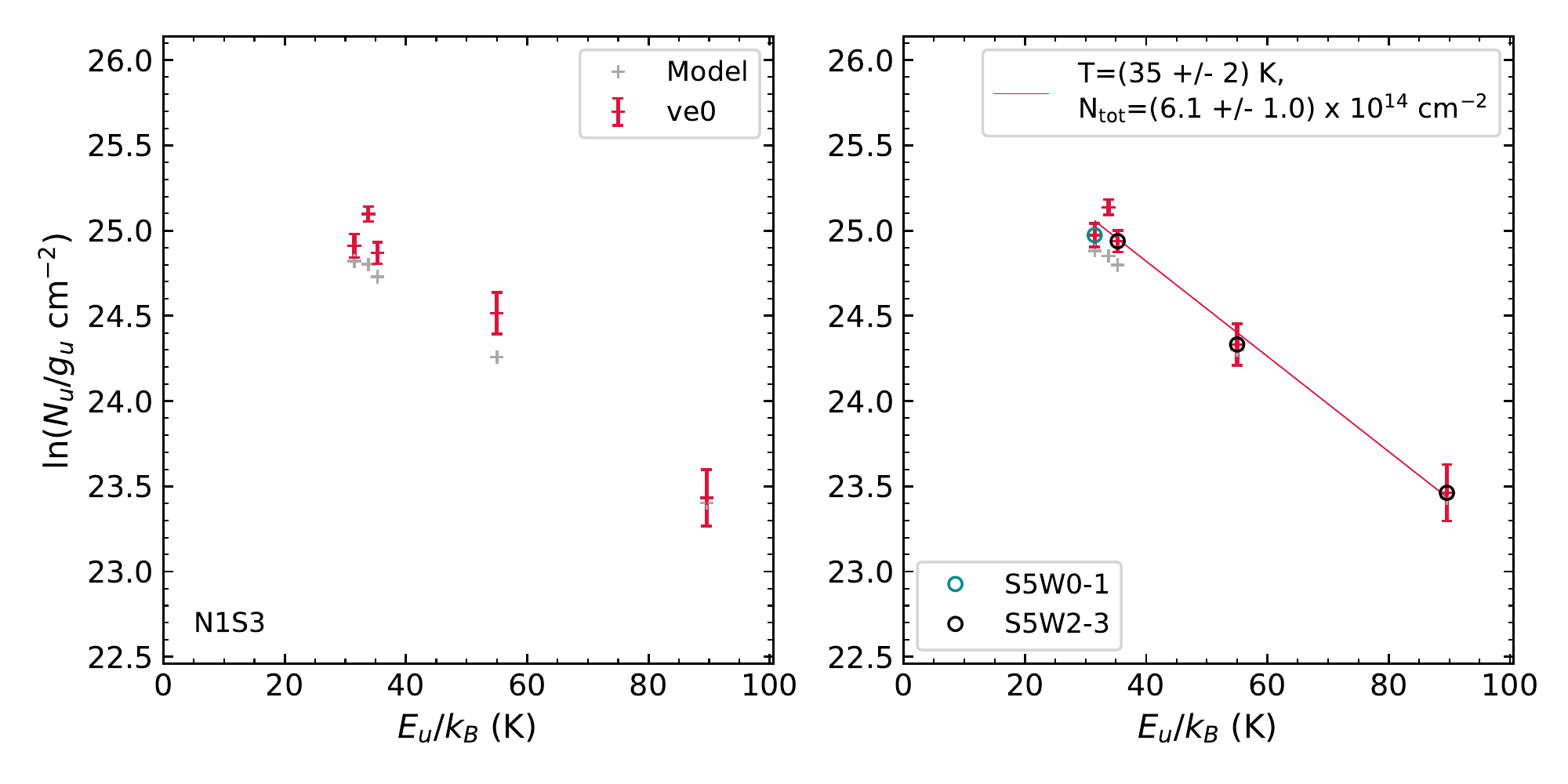}
    \caption{Same as Fig.\,\ref{fig:PD_met}, but for \vc.}
    \label{fig:PD_vc}
\end{figure*}

\begin{figure*}[h]
    \includegraphics[width=0.49\textwidth]{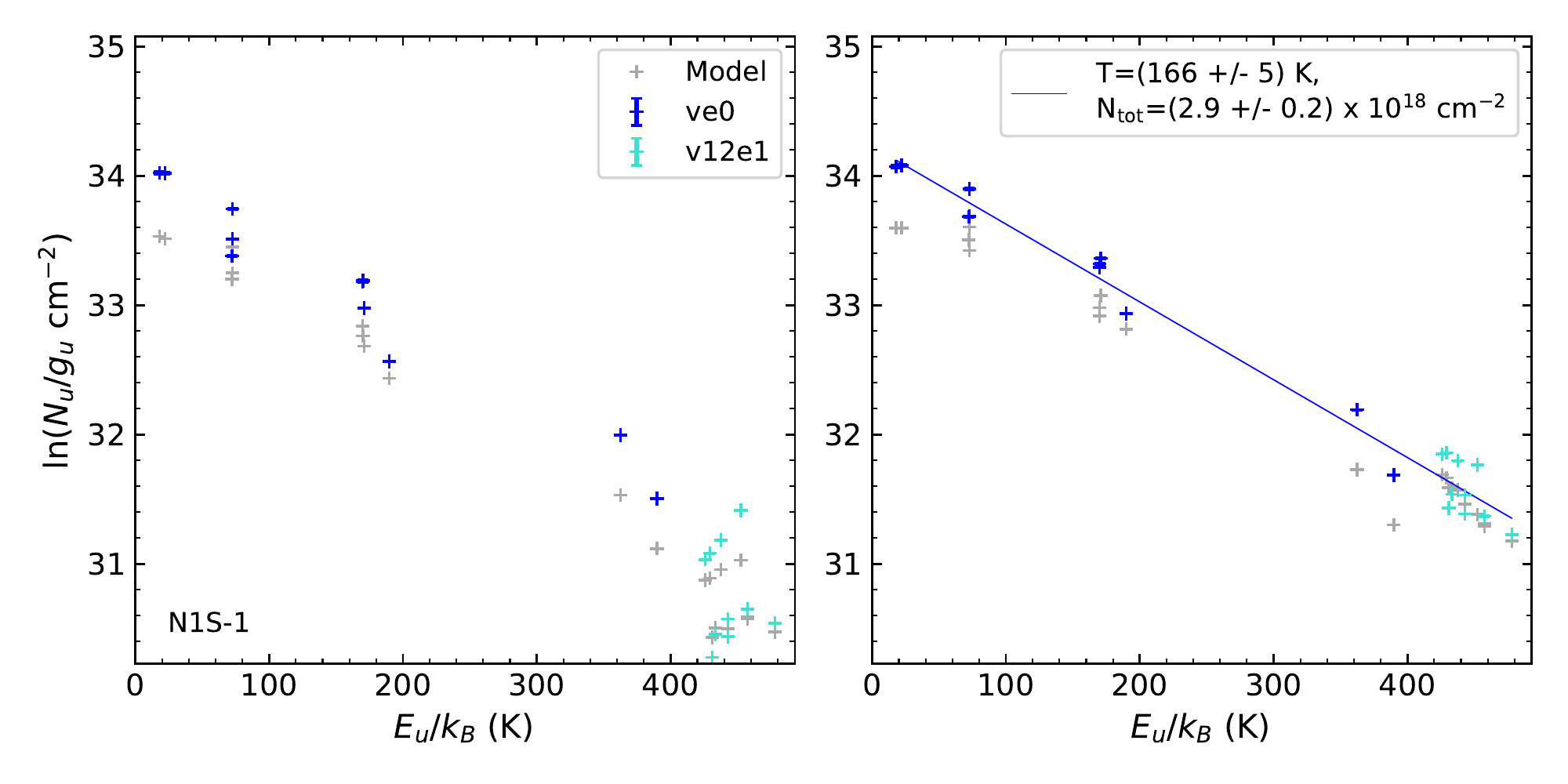}
    \includegraphics[width=0.49\textwidth]{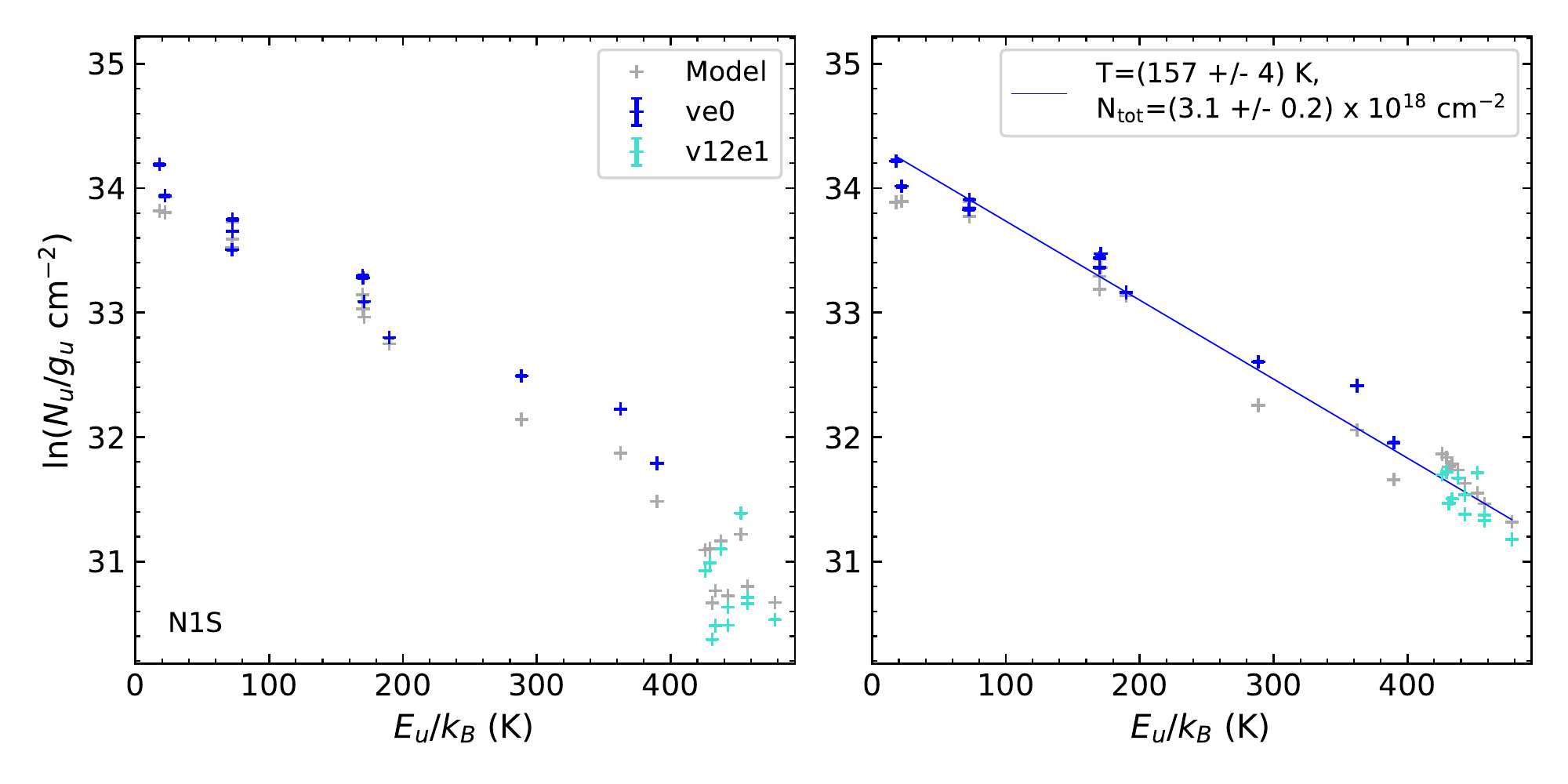}
    \includegraphics[width=0.49\textwidth]{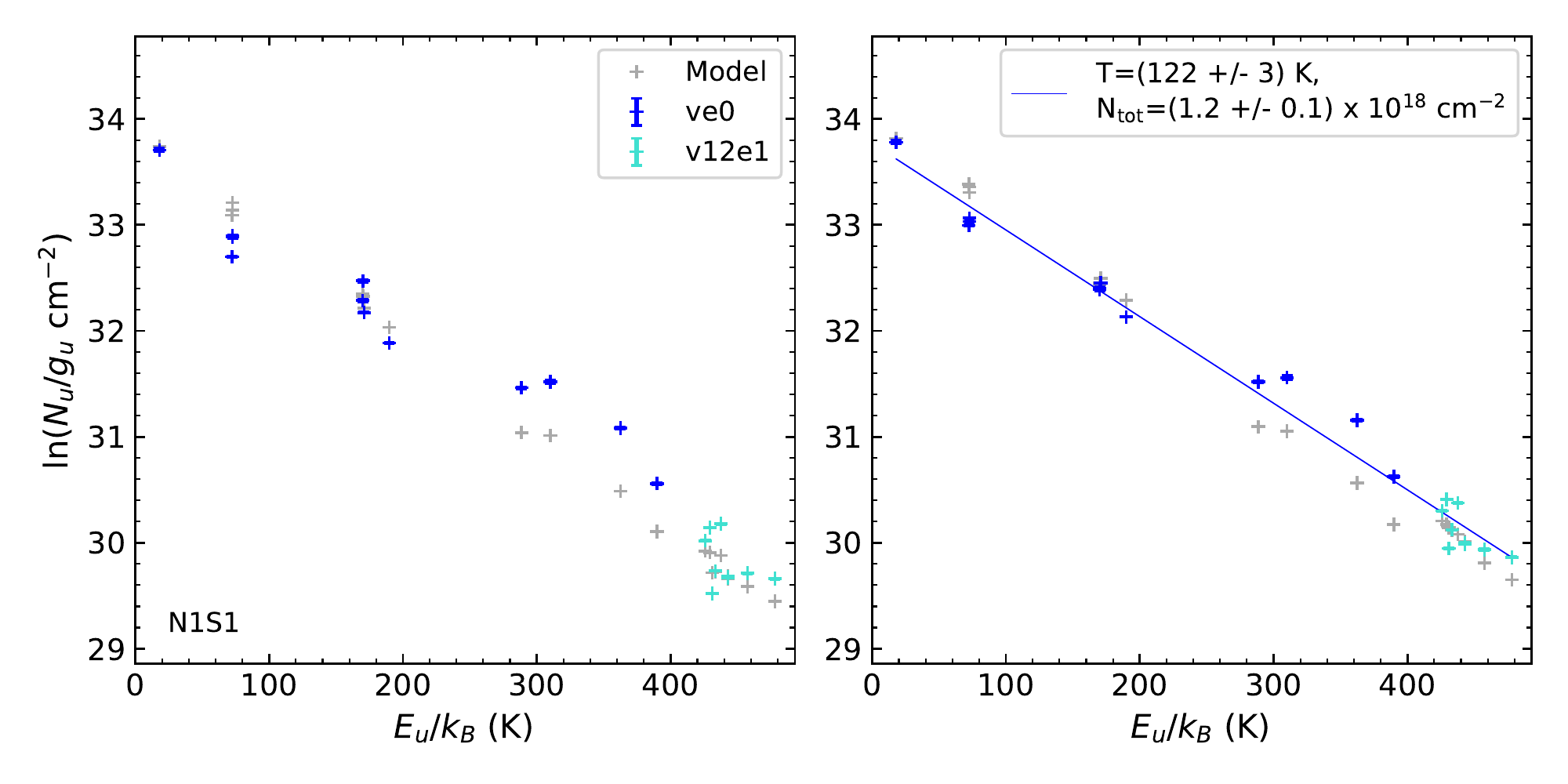}
    \includegraphics[width=0.49\textwidth]{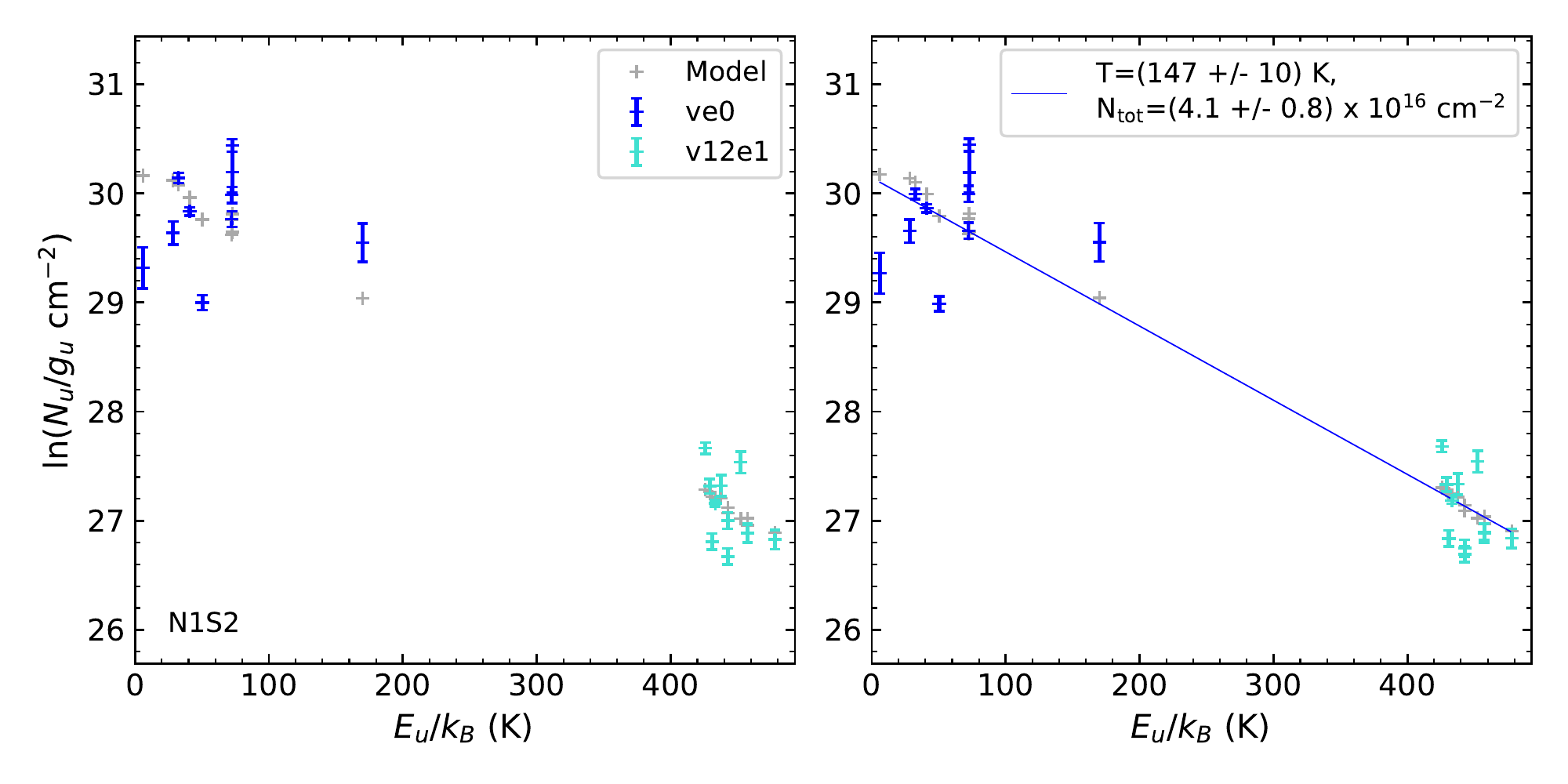}
    \includegraphics[width=0.49\textwidth]{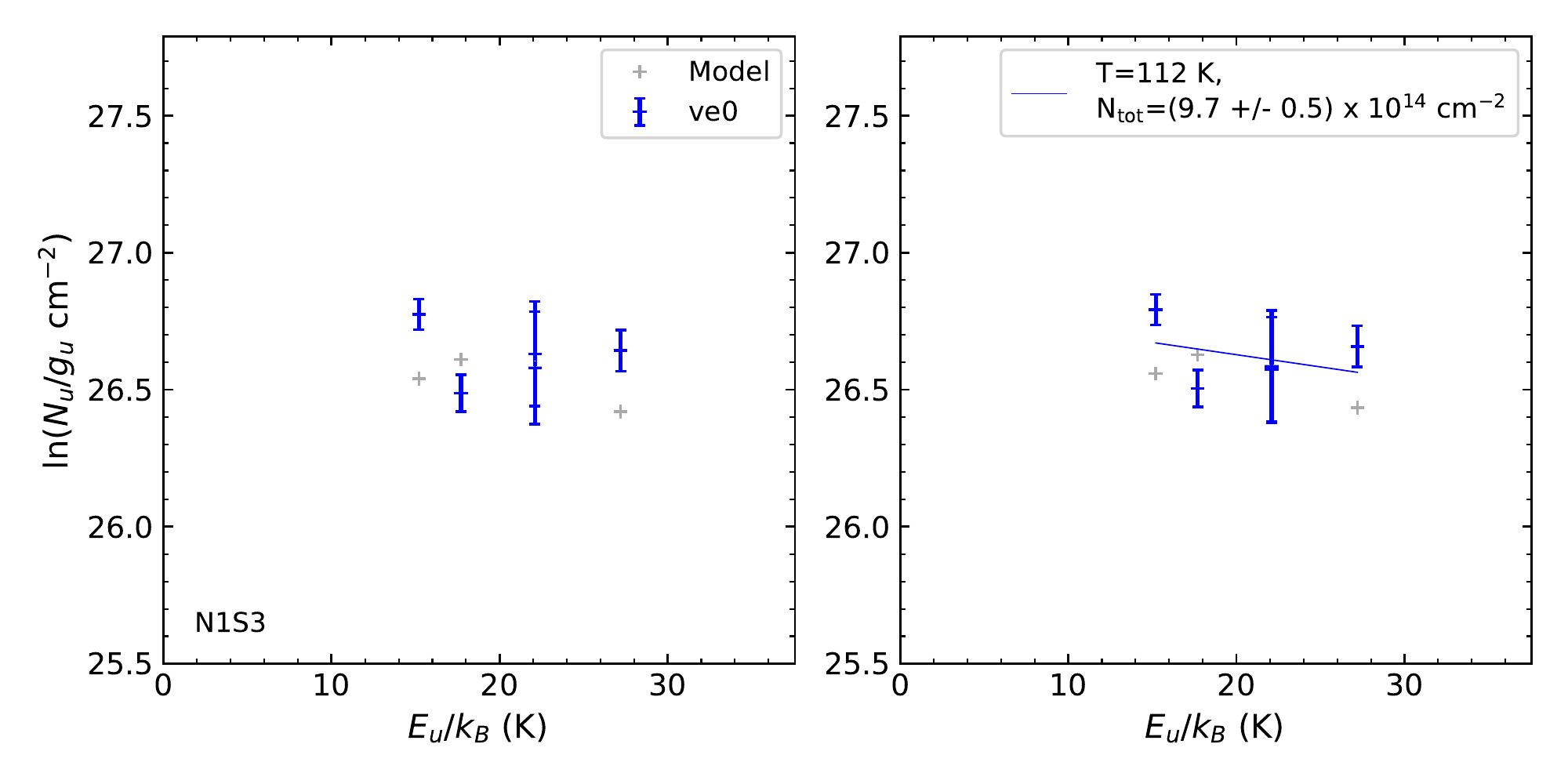}
    \includegraphics[width=0.49\textwidth]{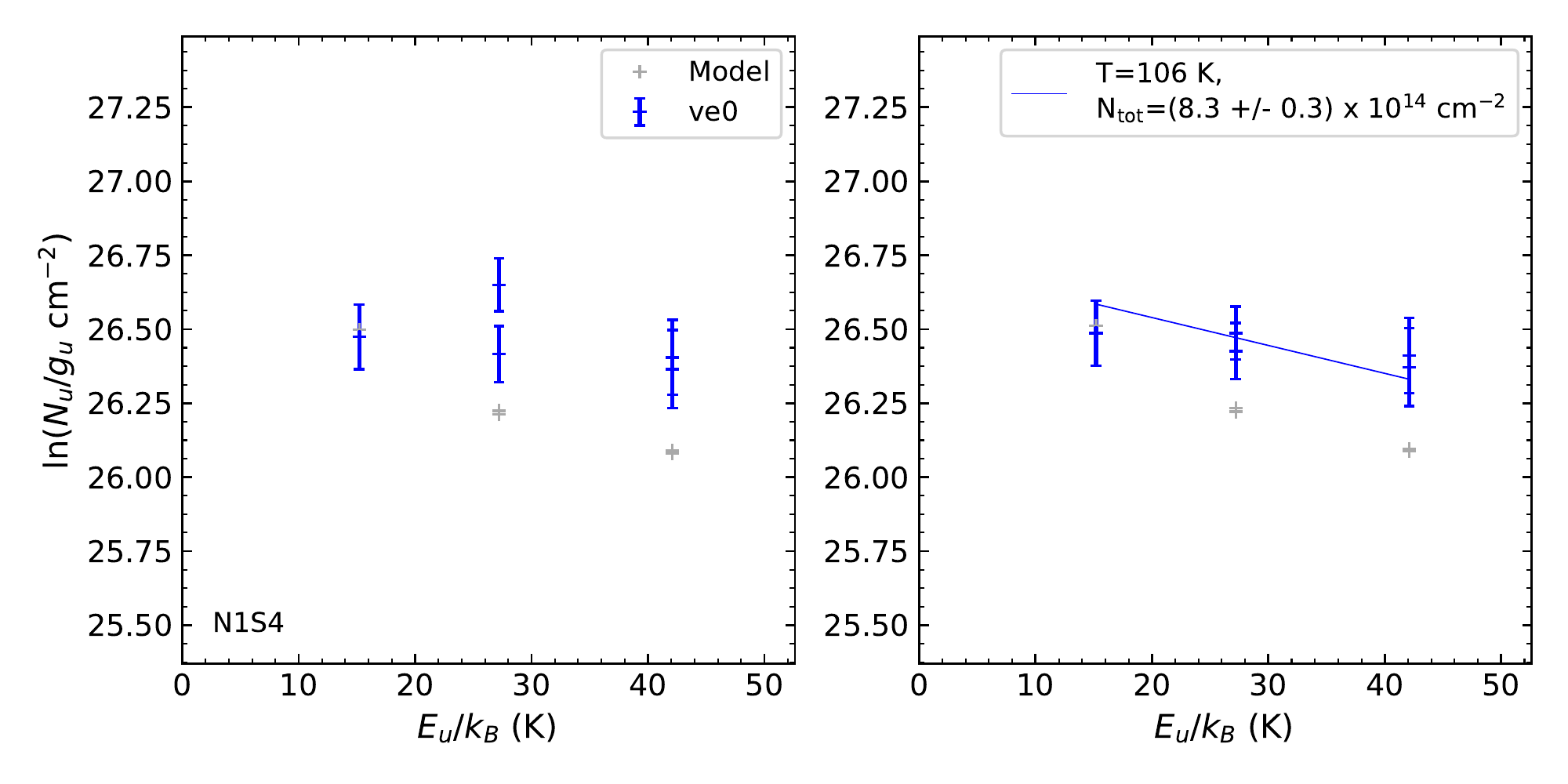}
    \includegraphics[width=0.49\textwidth]{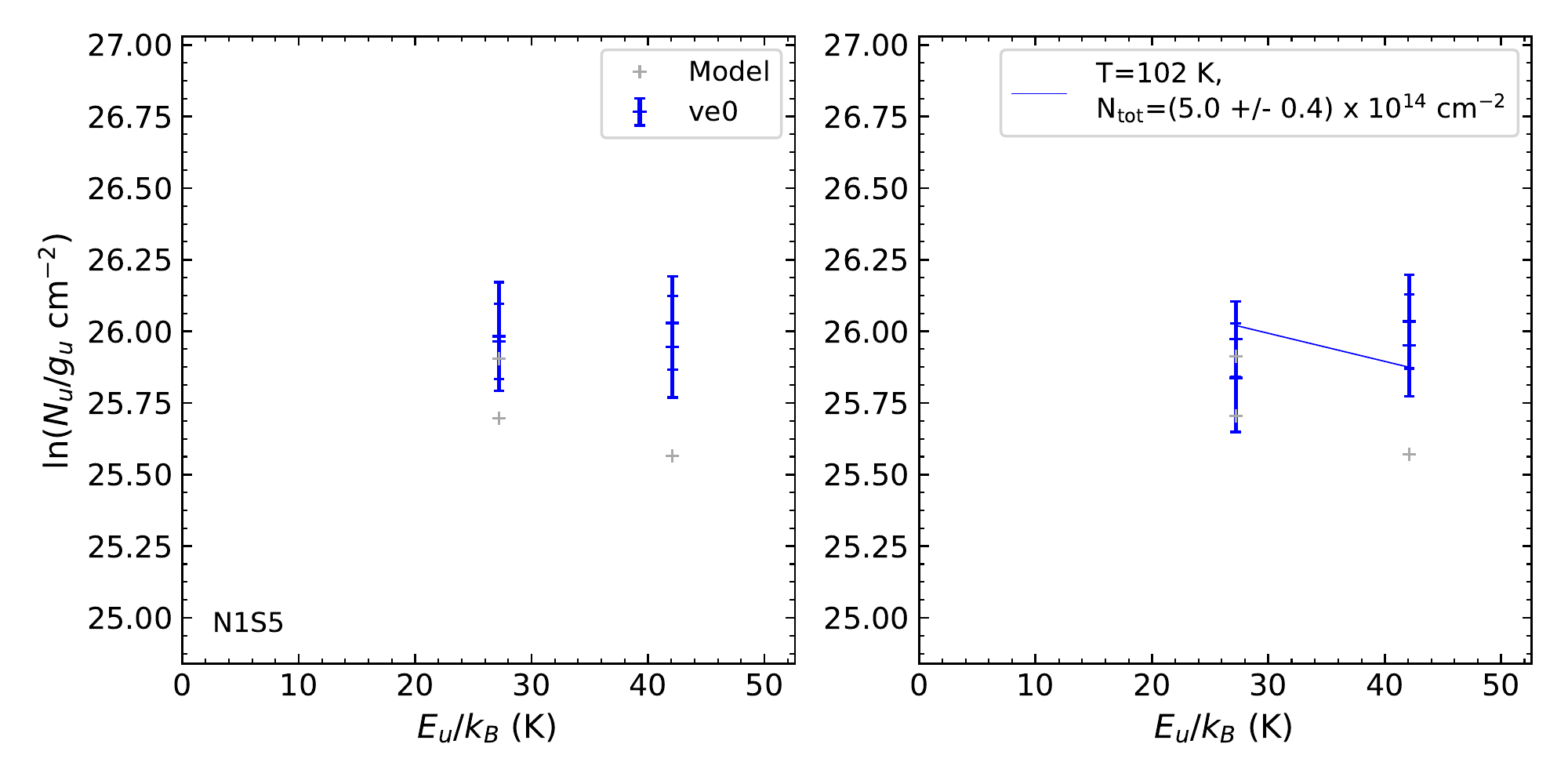}
    \caption{Same as Fig.\,\ref{fig:PD_met}, but for \fmm.}
    \label{fig:PD_fmm}
\end{figure*}

\begin{figure*}[h]
    \includegraphics[width=0.49\textwidth]{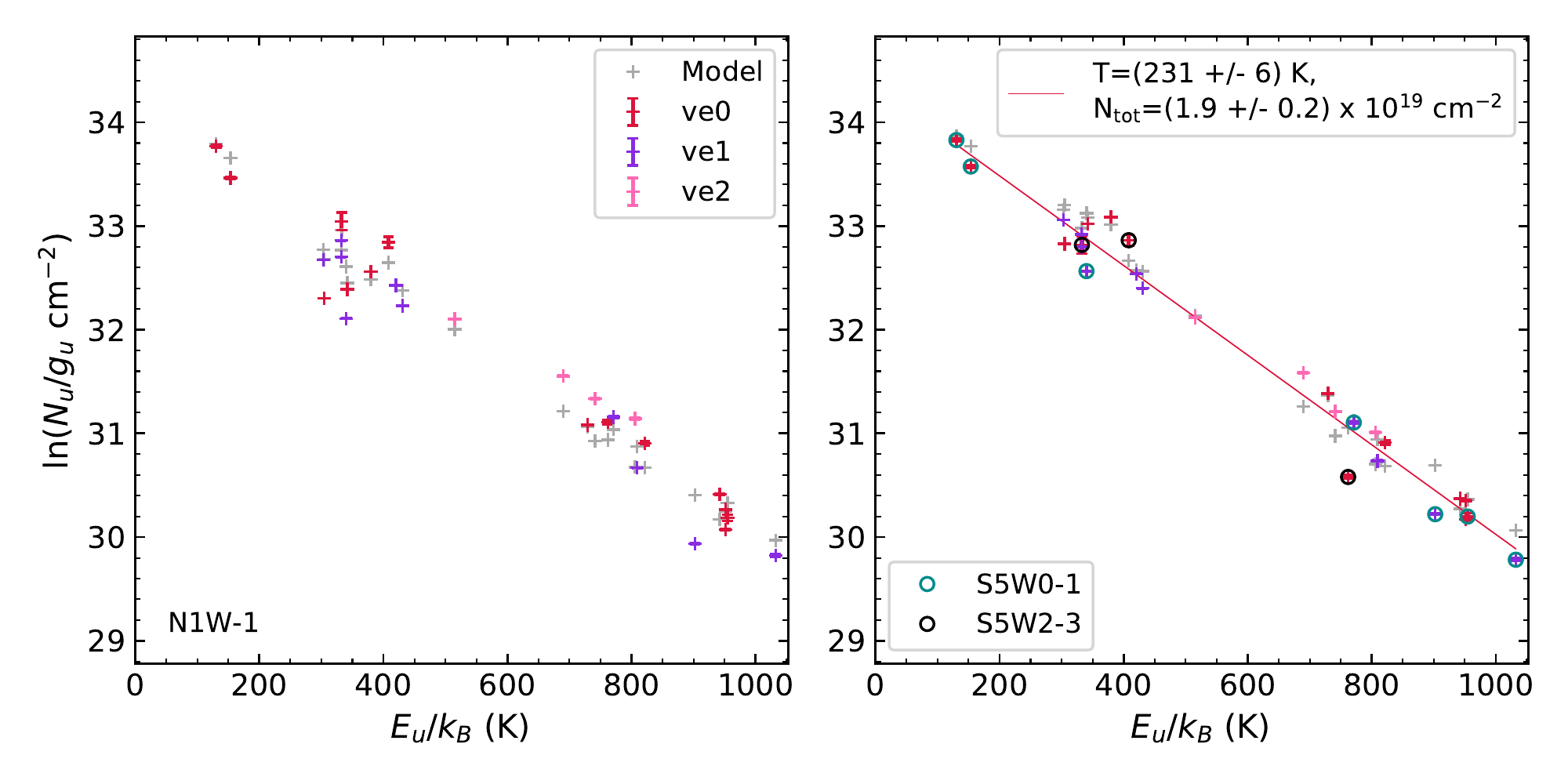}
    \includegraphics[width=0.49\textwidth]{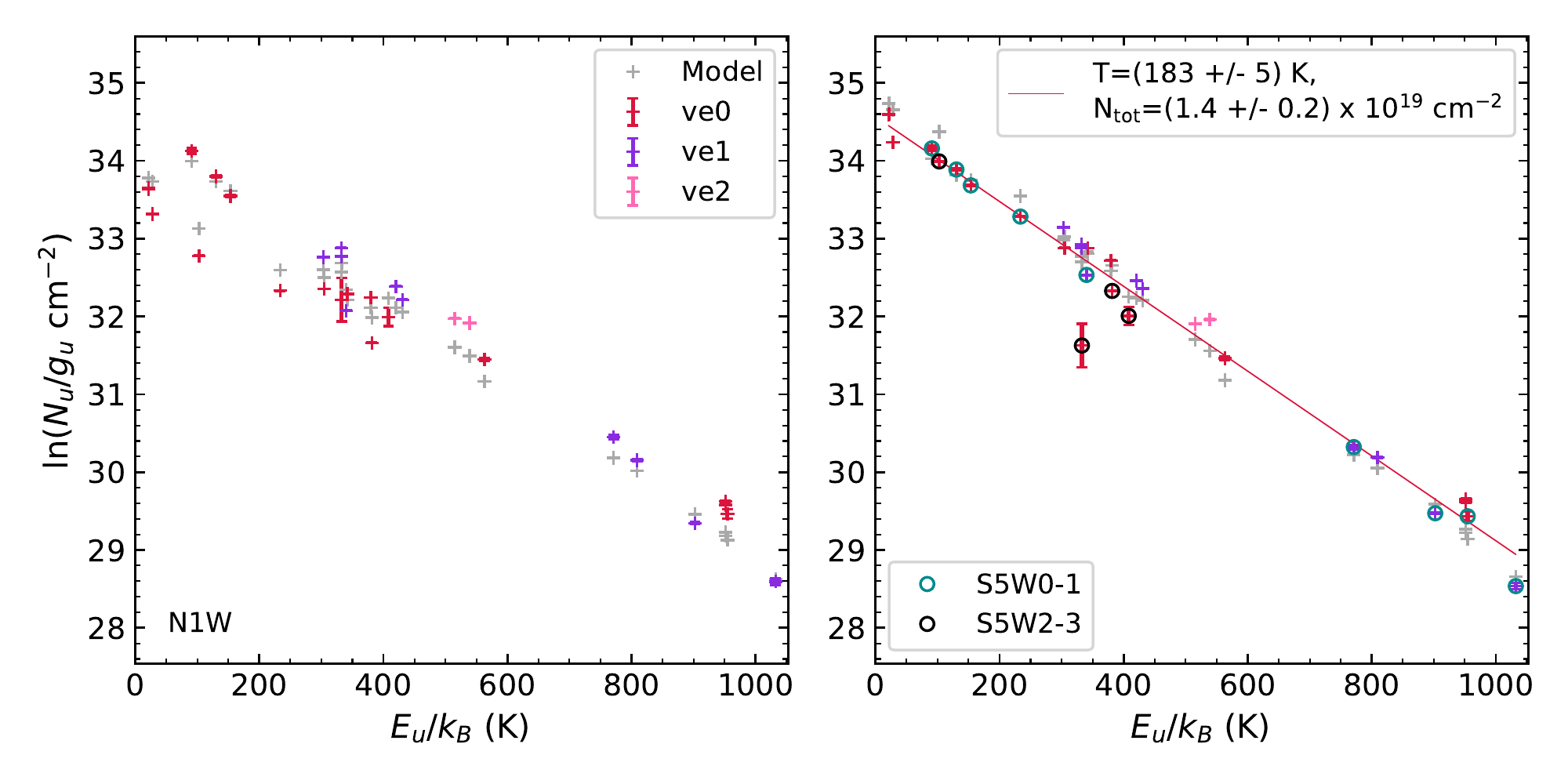}
    \includegraphics[width=0.49\textwidth]{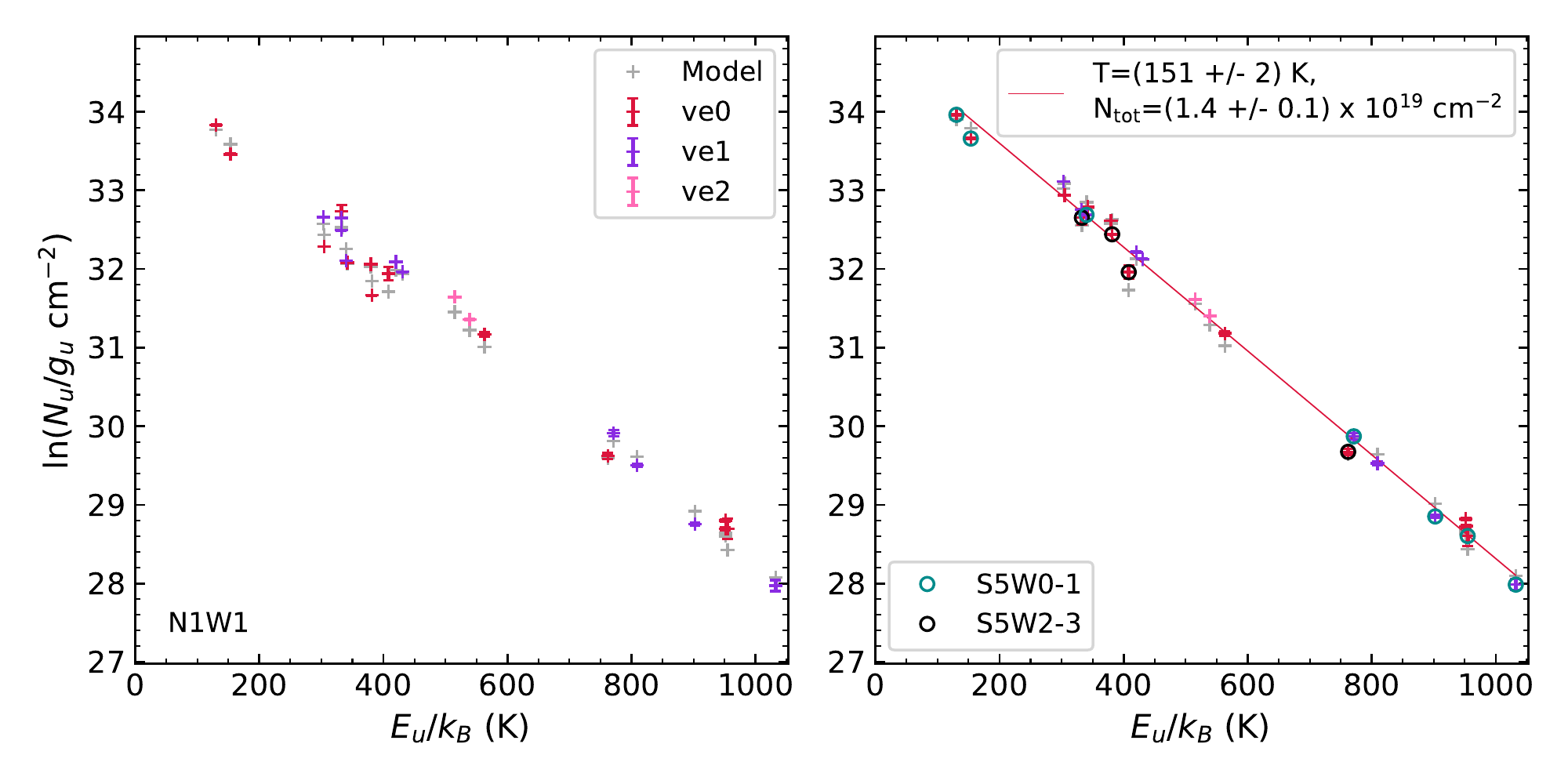}
    \includegraphics[width=0.49\textwidth]{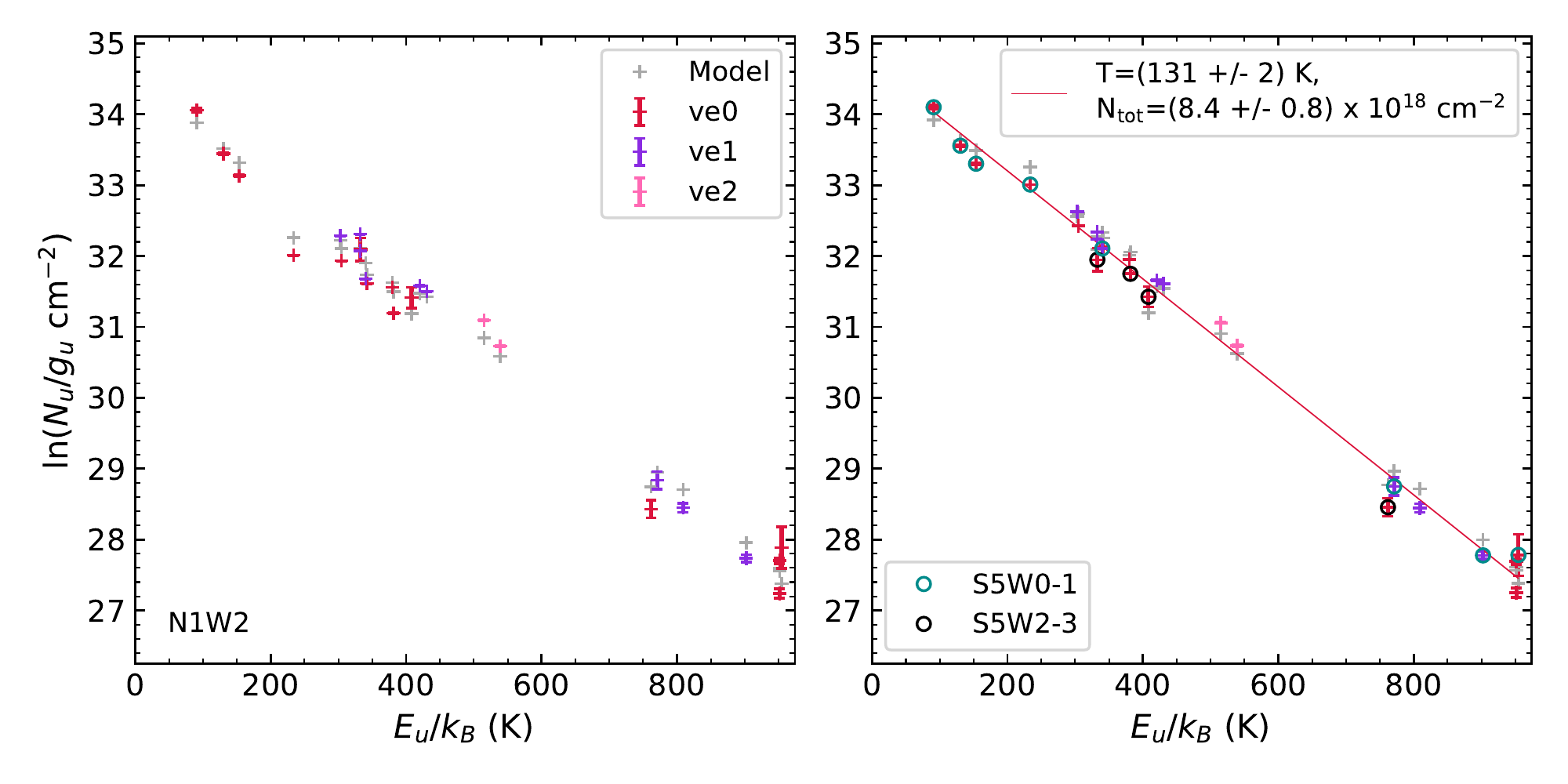}
    \includegraphics[width=0.49\textwidth]{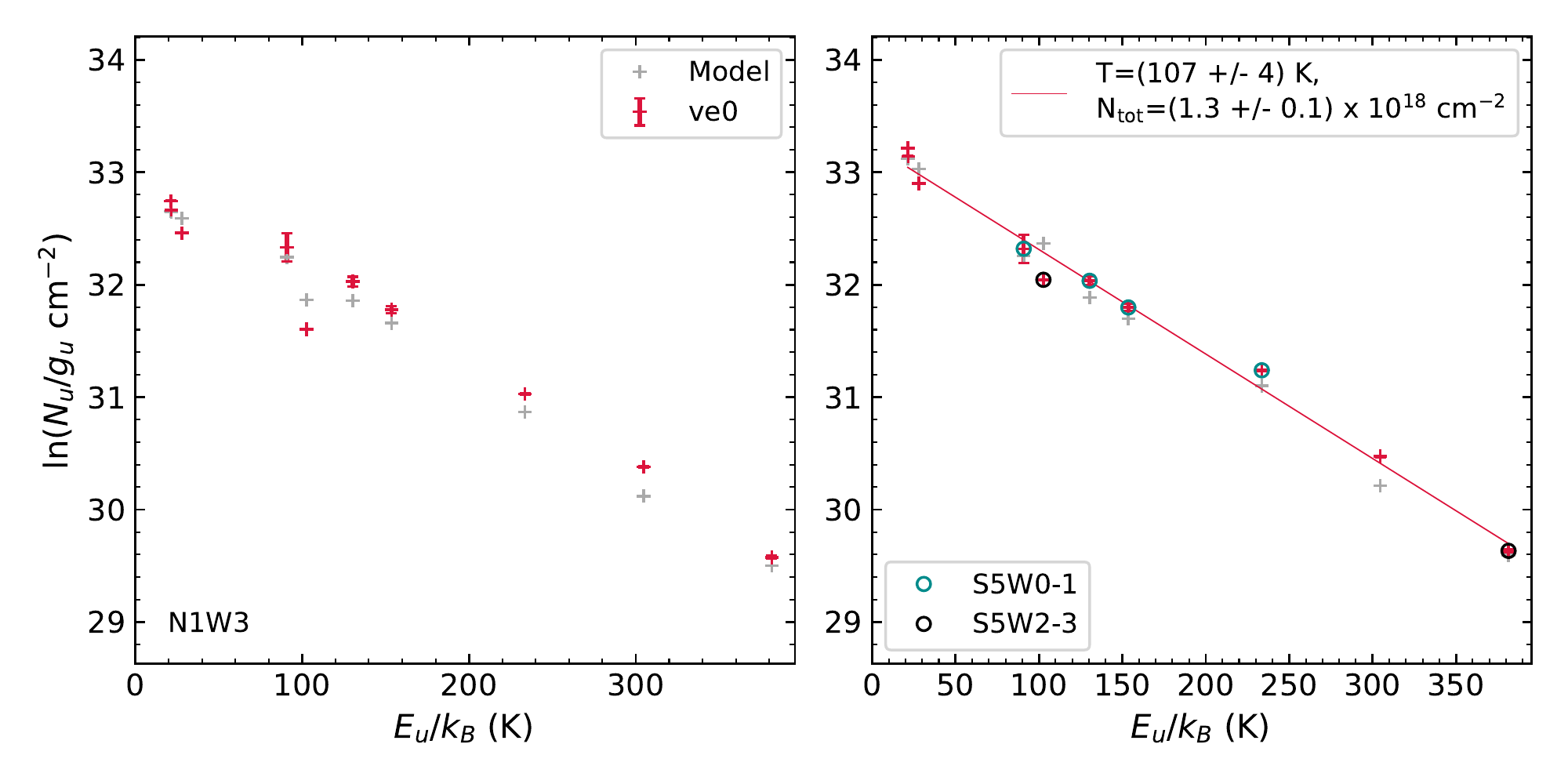}
    \includegraphics[width=0.49\textwidth]{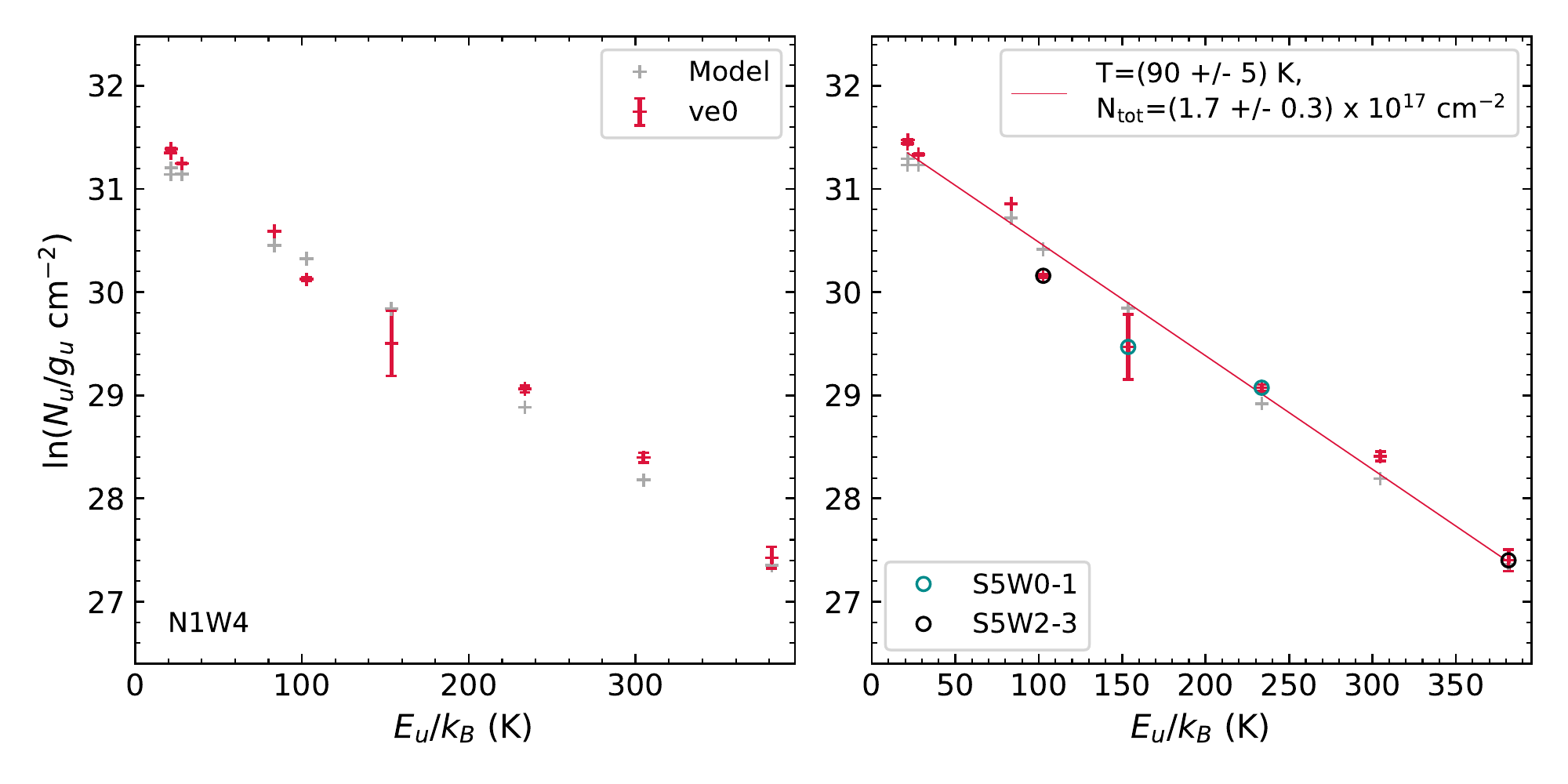}
    \includegraphics[width=0.49\textwidth]{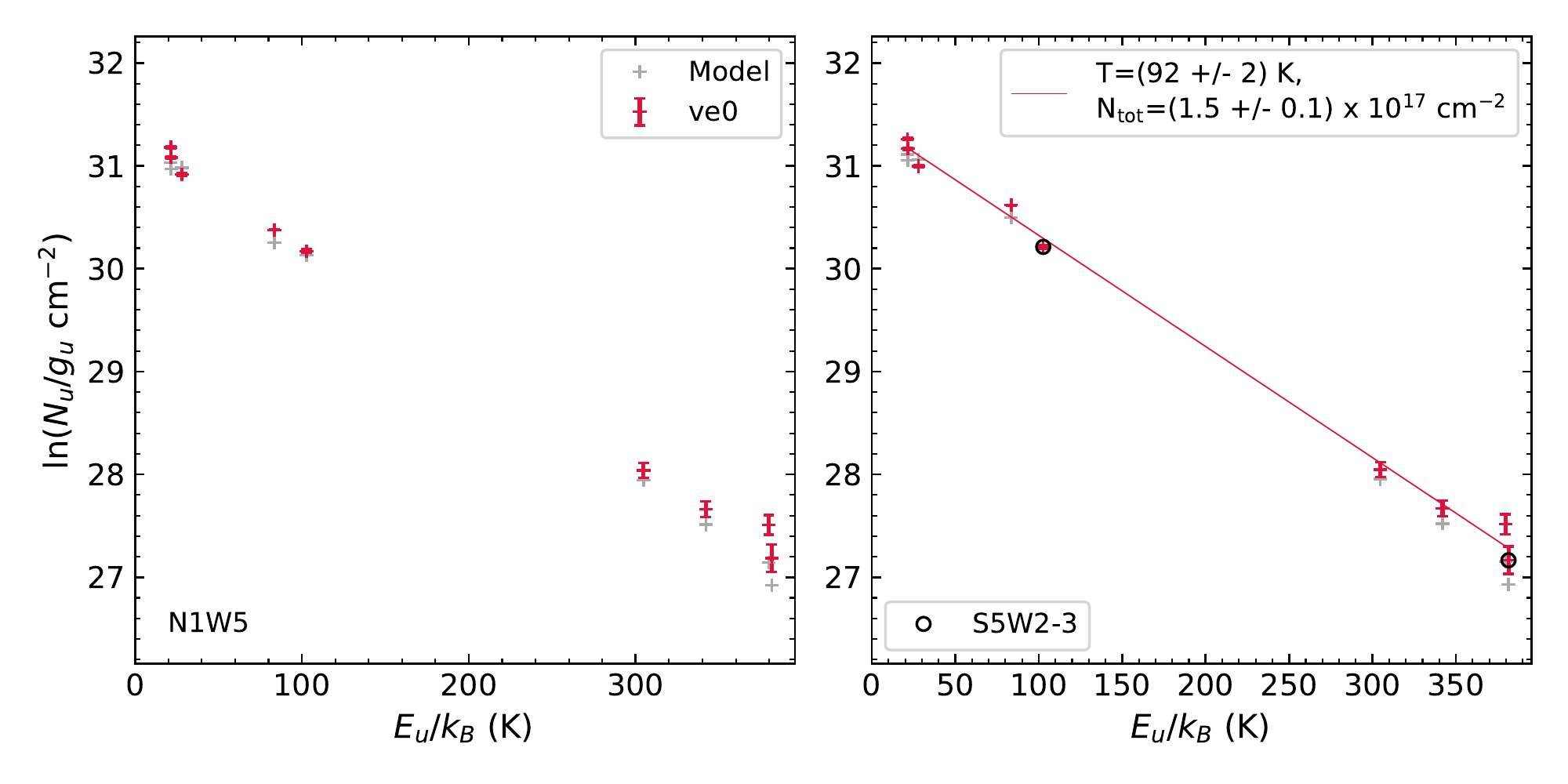}
    \includegraphics[width=0.49\textwidth]{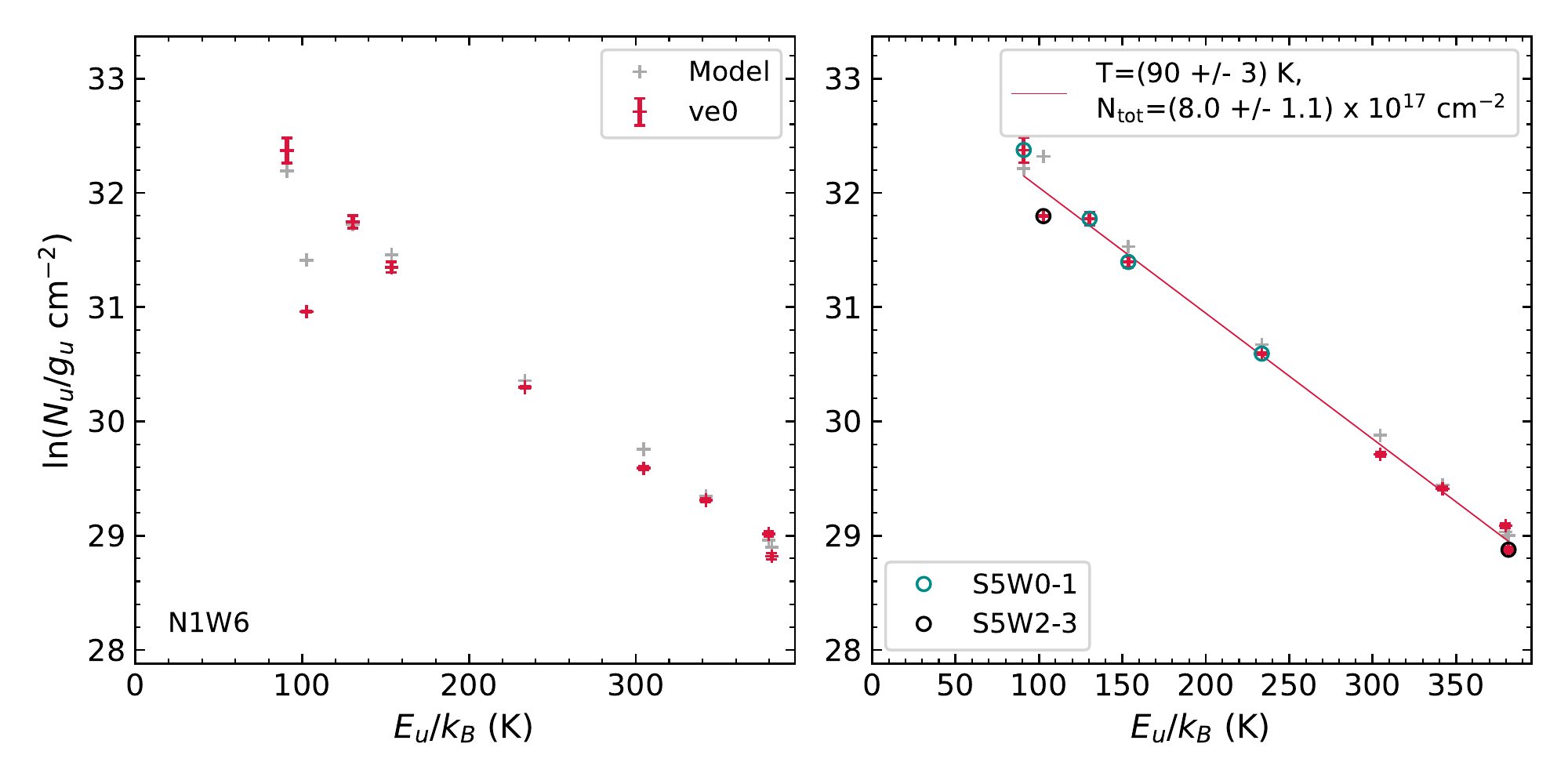}
    \includegraphics[width=0.49\textwidth]{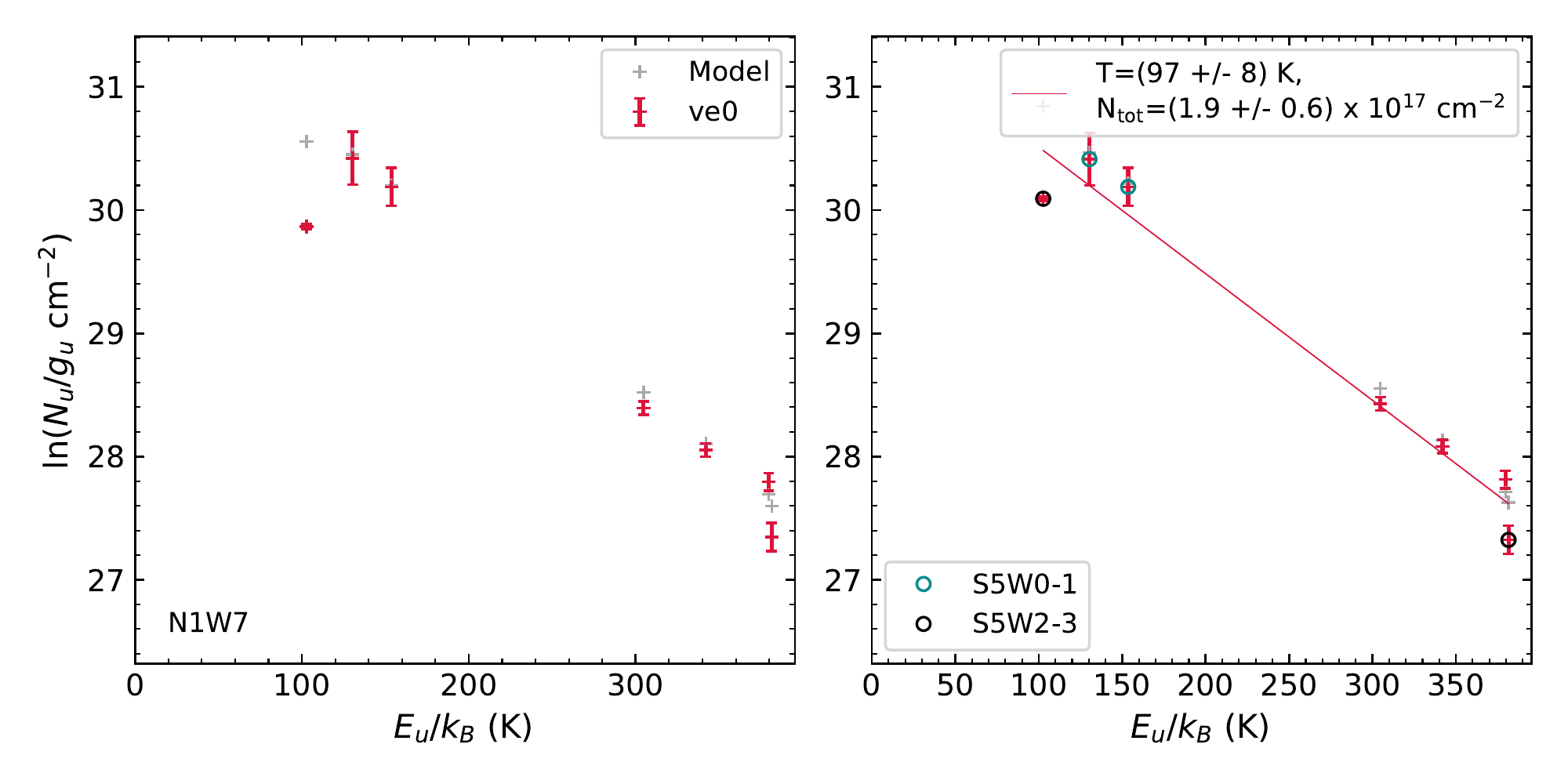}
    \includegraphics[width=0.49\textwidth]{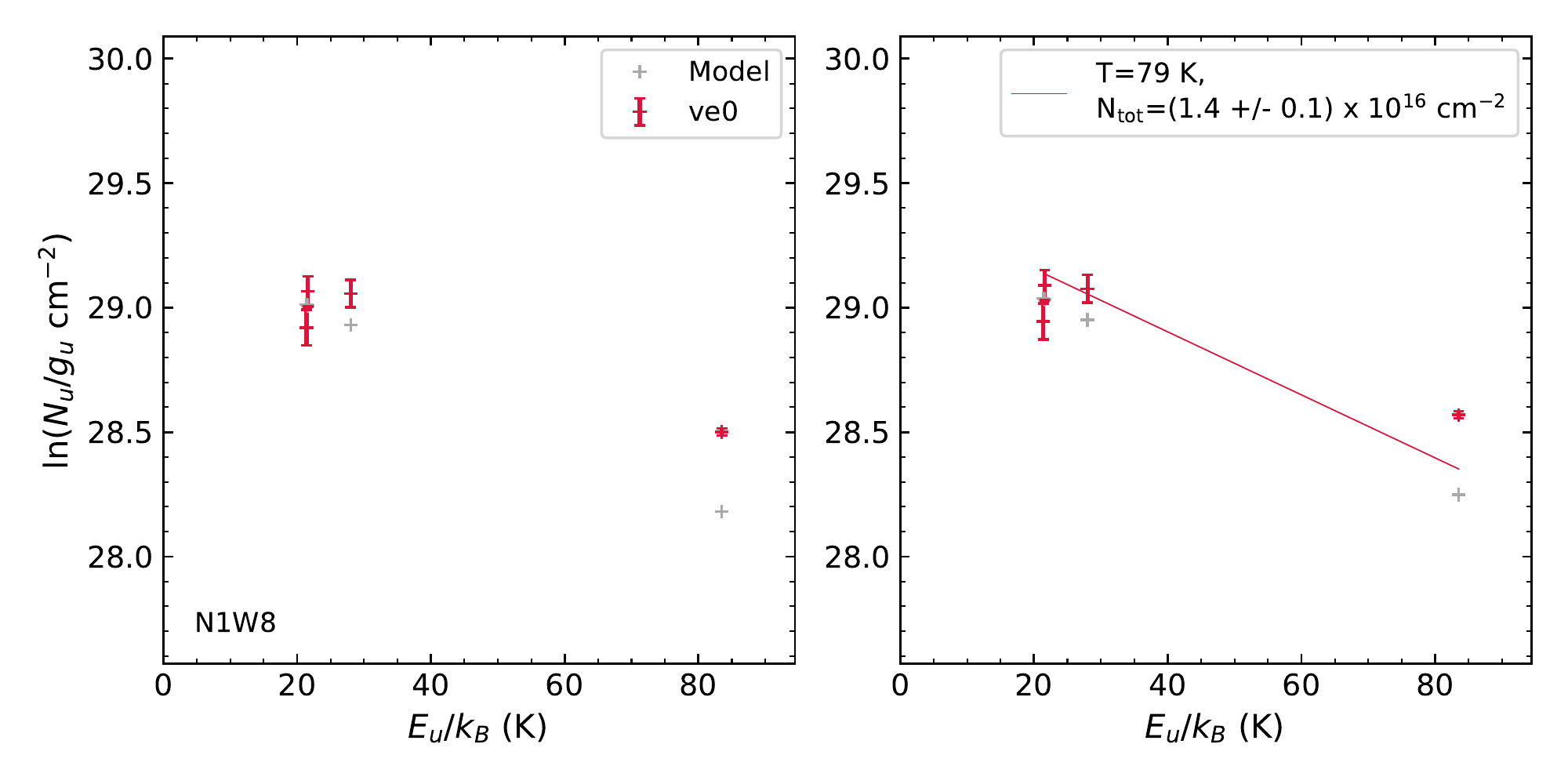}
    \caption{Same as Fig.\,\ref{fig:PD_met}, but for all positions to the west where the molecule is detected.}
    \label{fig:wPD_met}
\end{figure*}

\begin{figure*}[h]
    \includegraphics[width=0.49\textwidth]{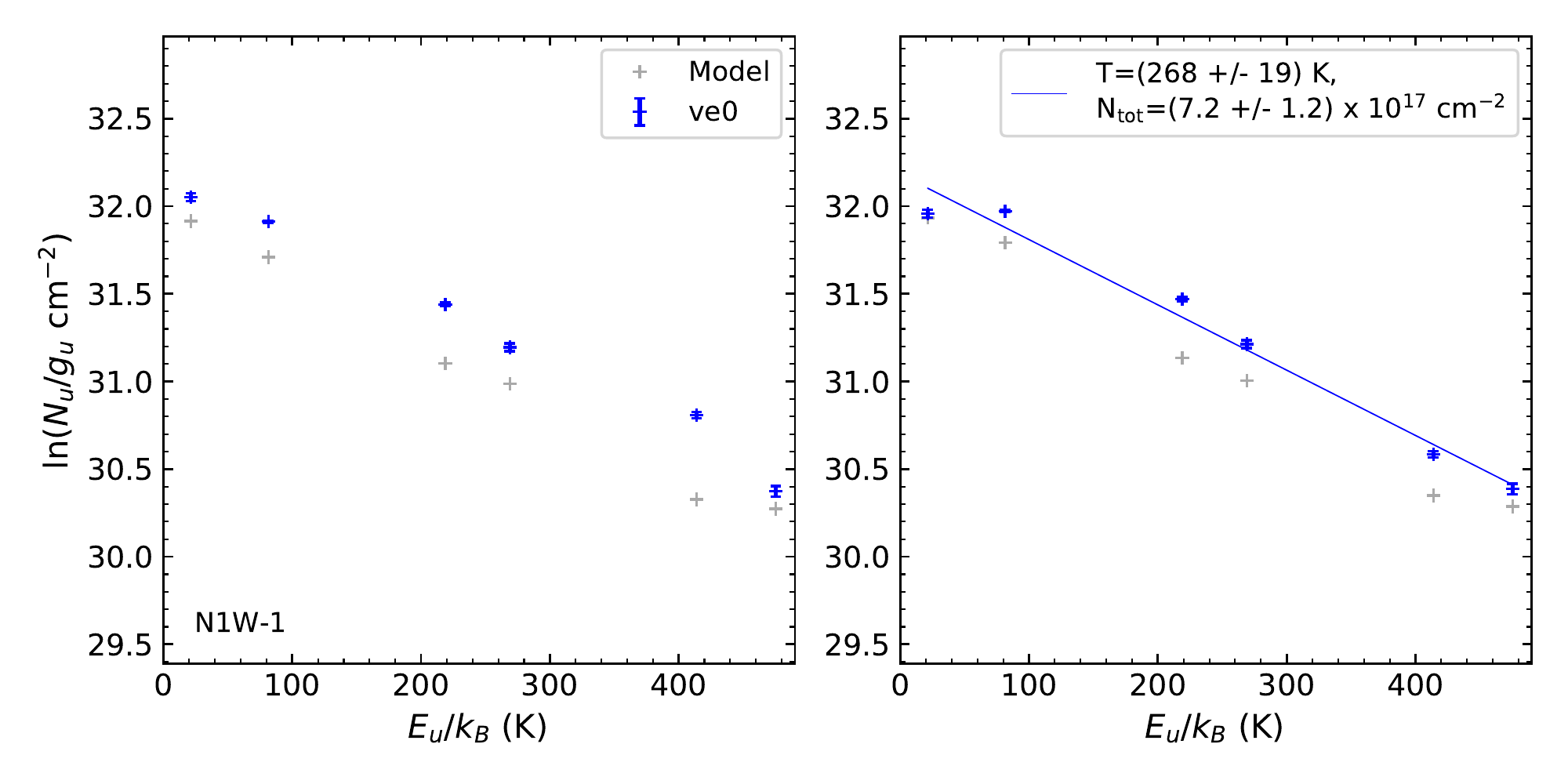}
    \includegraphics[width=0.49\textwidth]{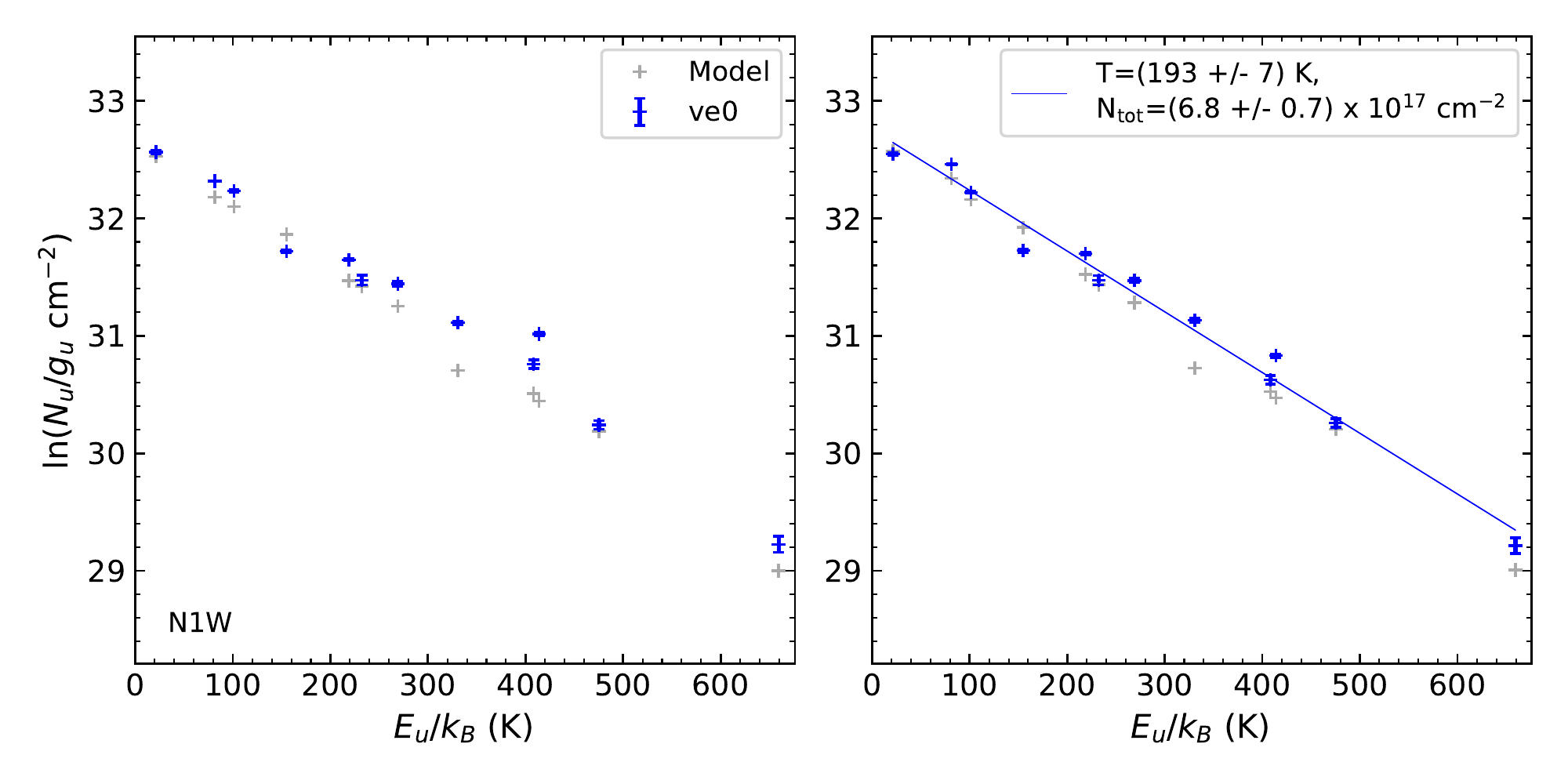}
    \includegraphics[width=0.49\textwidth]{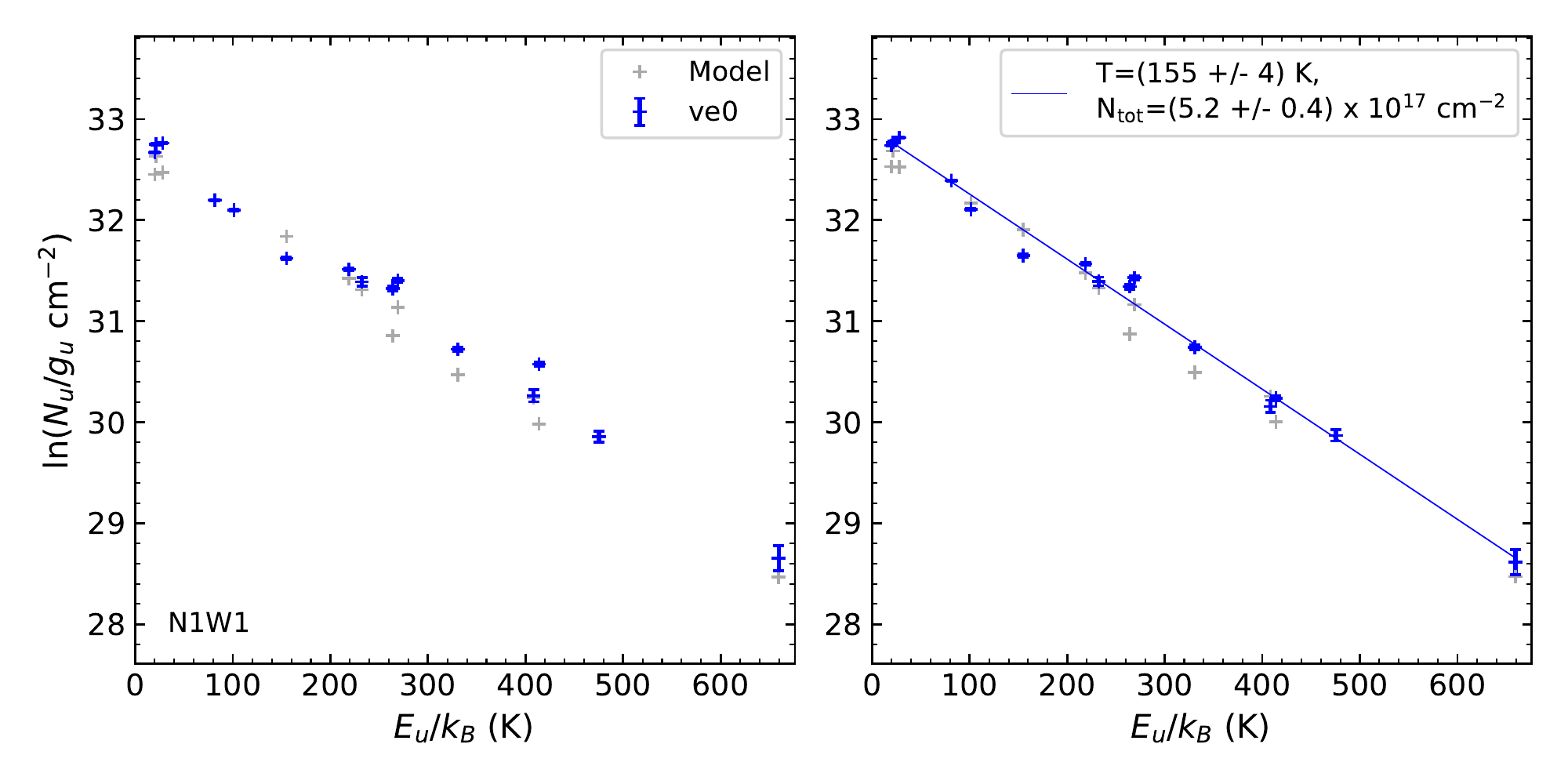}
    \includegraphics[width=0.49\textwidth]{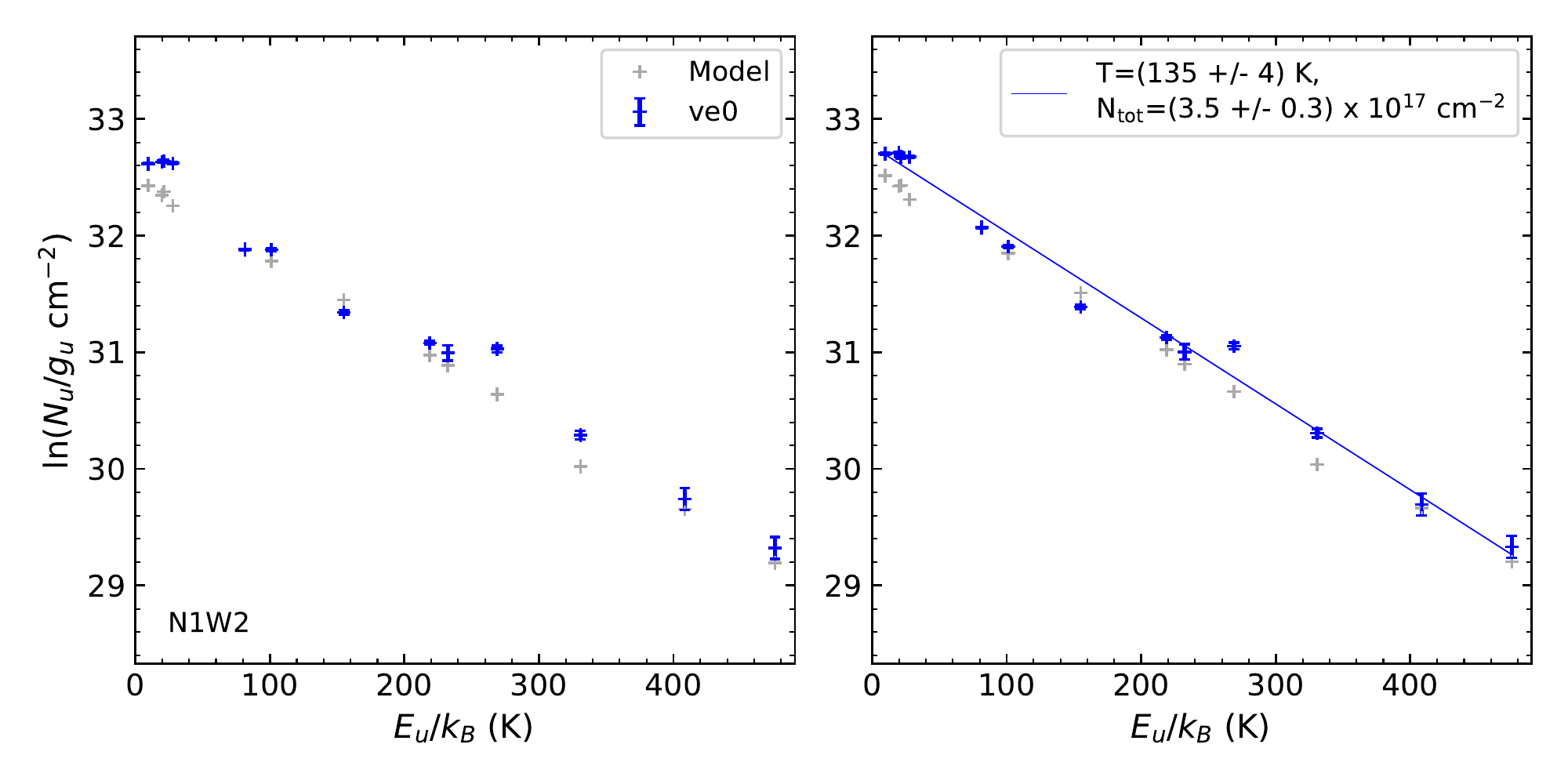}
    \includegraphics[width=0.49\textwidth]{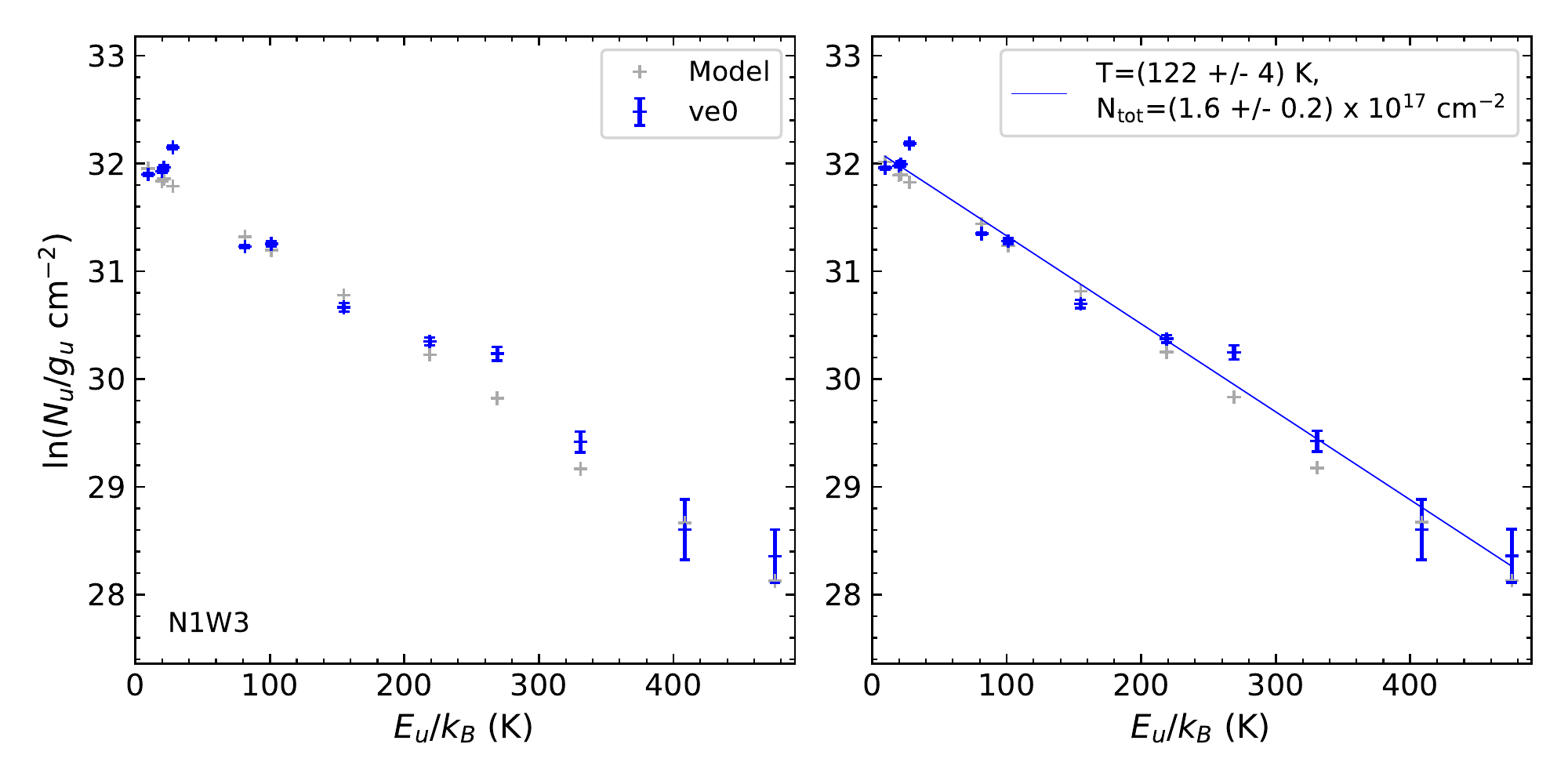}
    \includegraphics[width=0.49\textwidth]{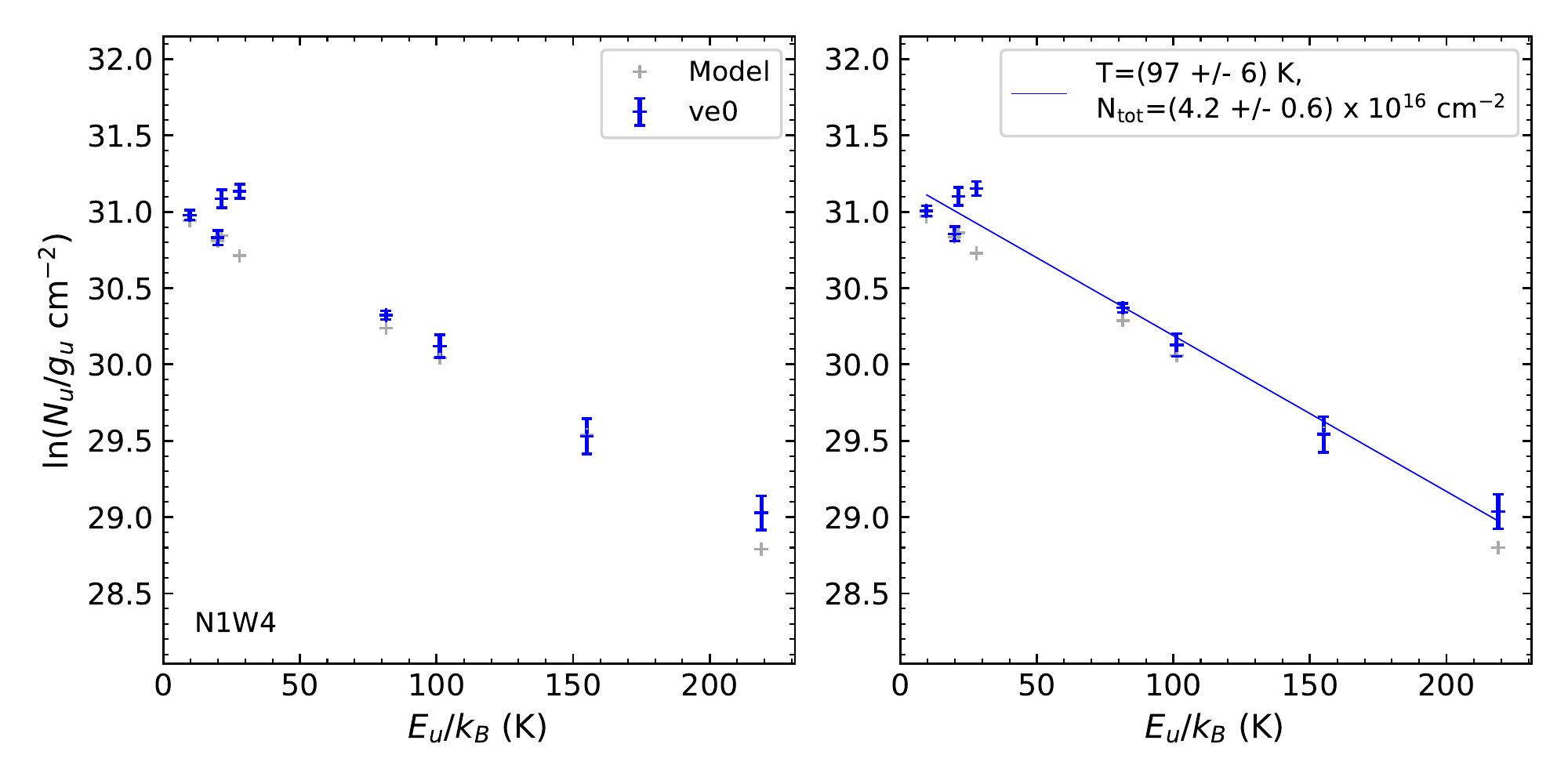}
    \includegraphics[width=0.49\textwidth]{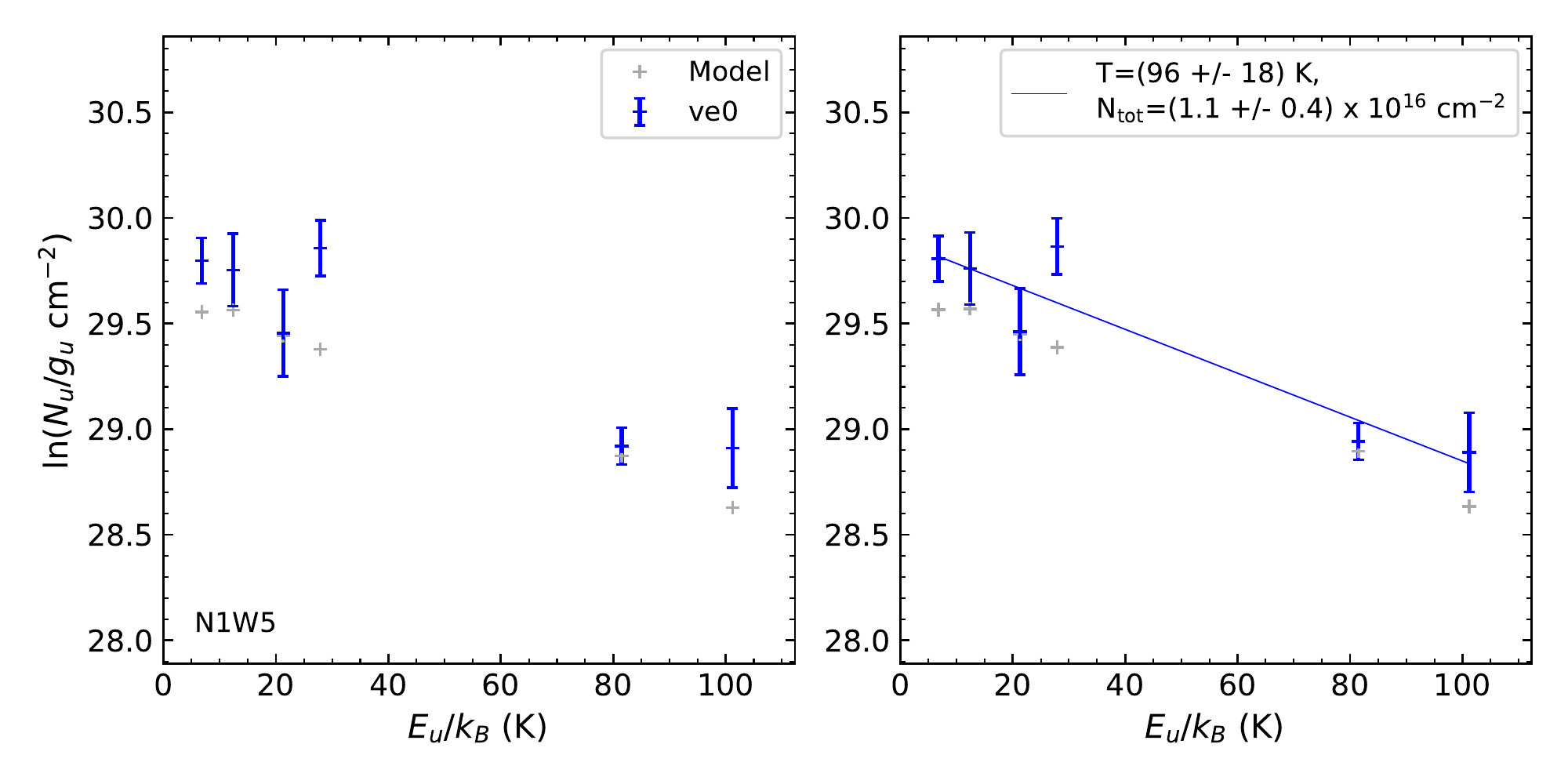}
    \includegraphics[width=0.49\textwidth]{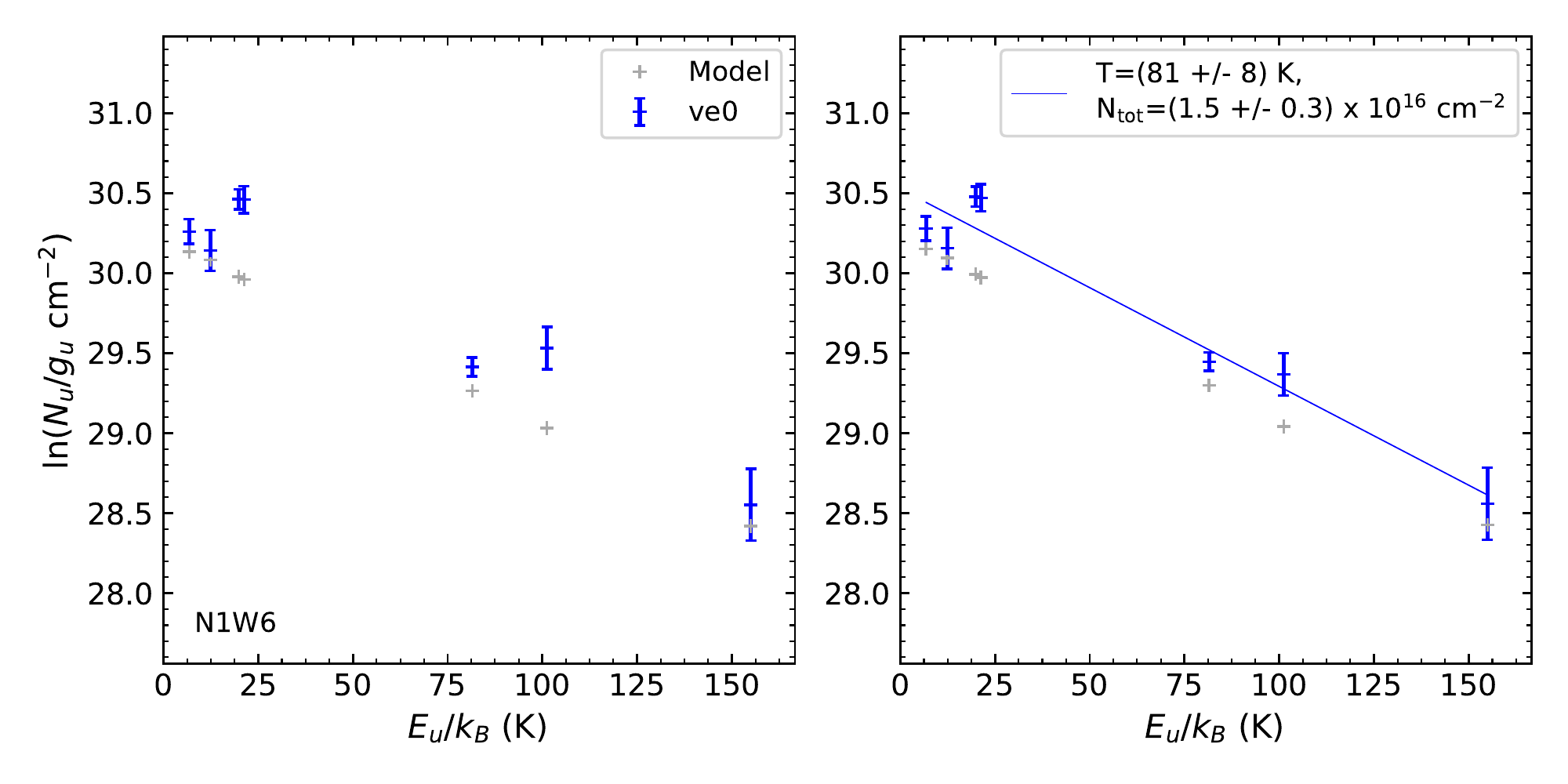}
    \includegraphics[width=0.49\textwidth]{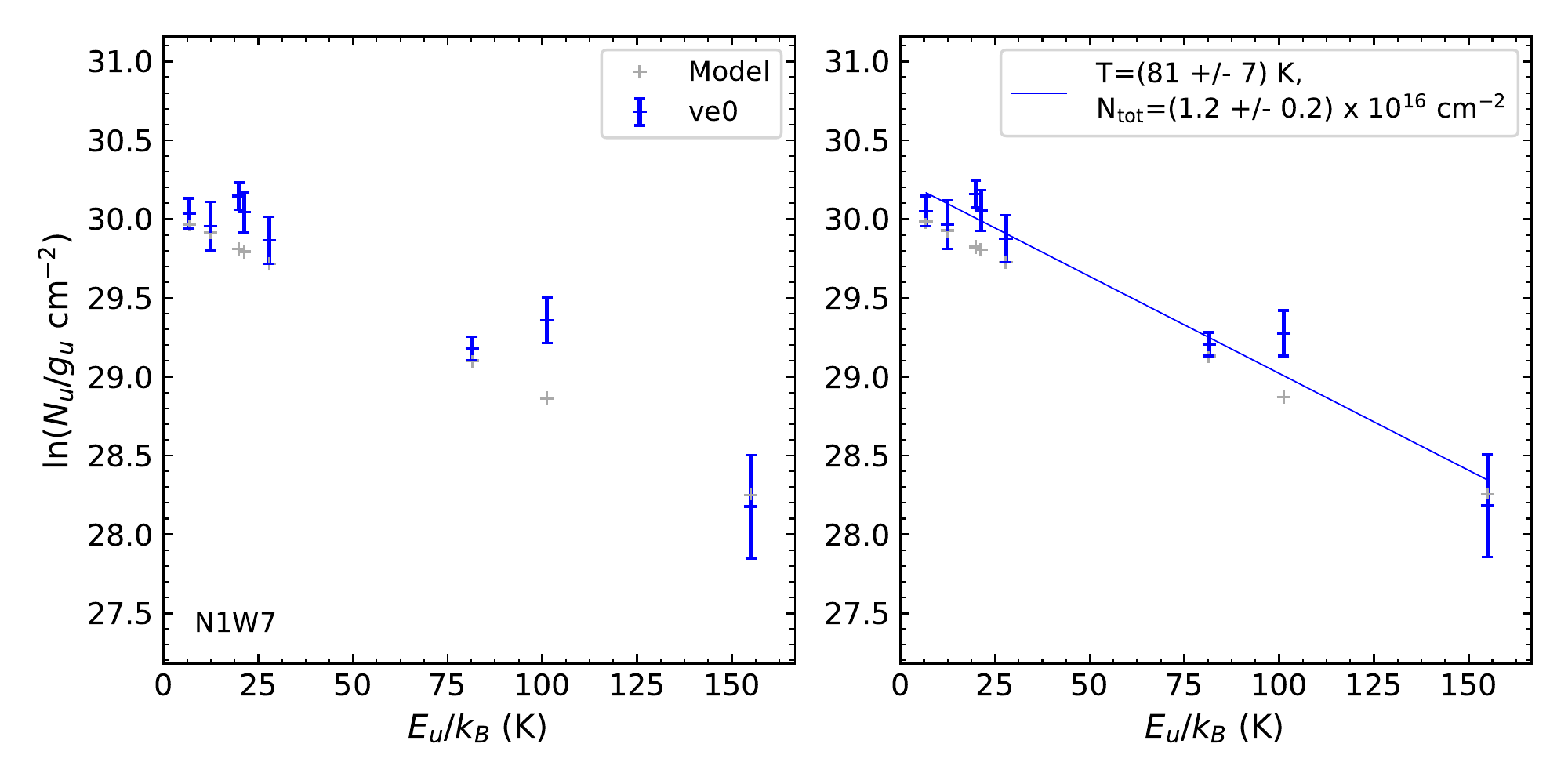}
    \caption{Same as Fig.\,\ref{fig:PD_met}, but for $^{13}$\met and for all positions to the west where the molecule is detected, except that setups 1--3 had to be used due to lack of transitions in setups 4--5. The linear fit to observed data points is shown in blue.}
    \label{fig:wPD_13met}
\end{figure*}

\begin{figure*}[h]
    \includegraphics[width=0.49\textwidth]{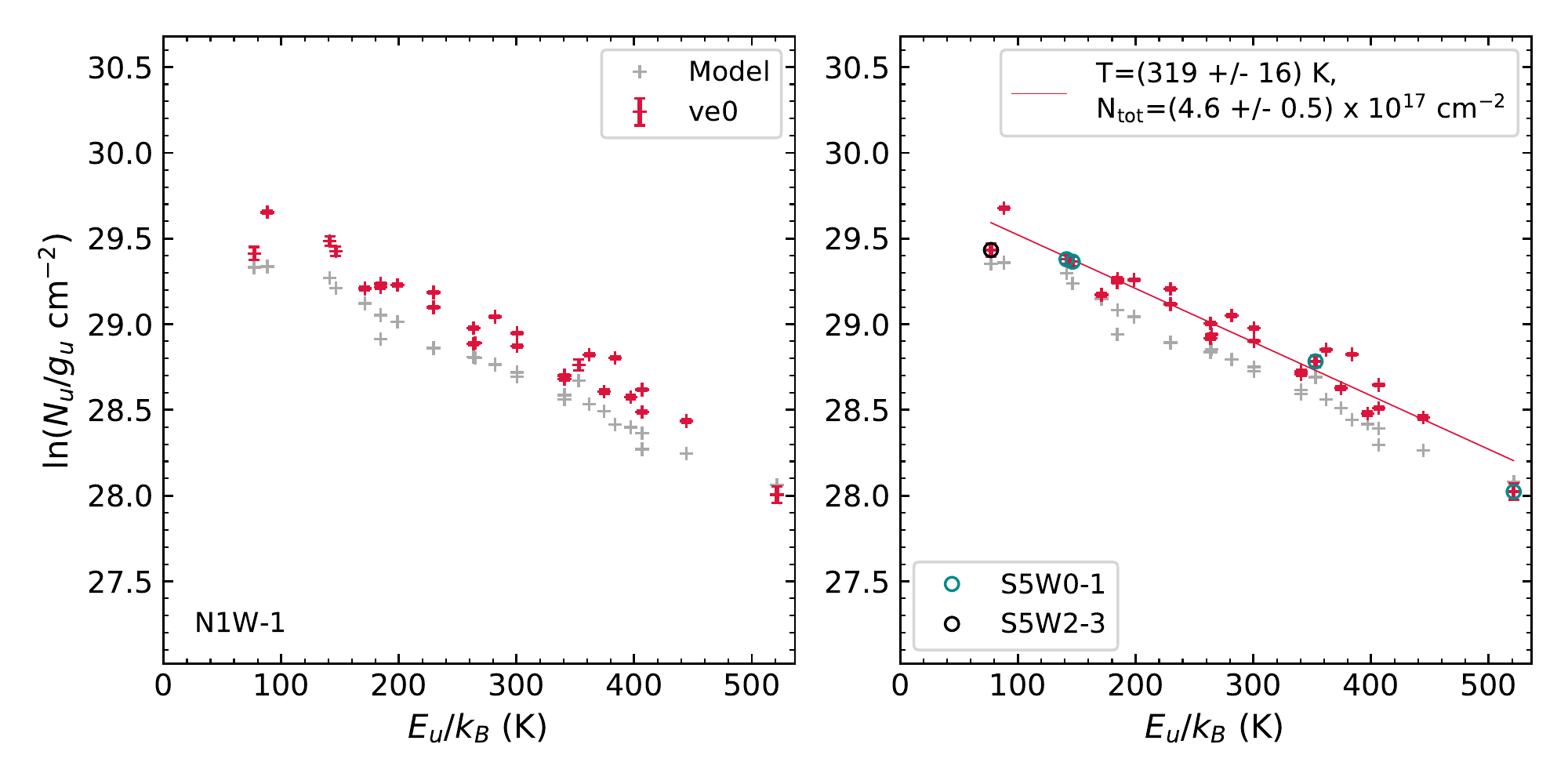}
    \includegraphics[width=0.49\textwidth]{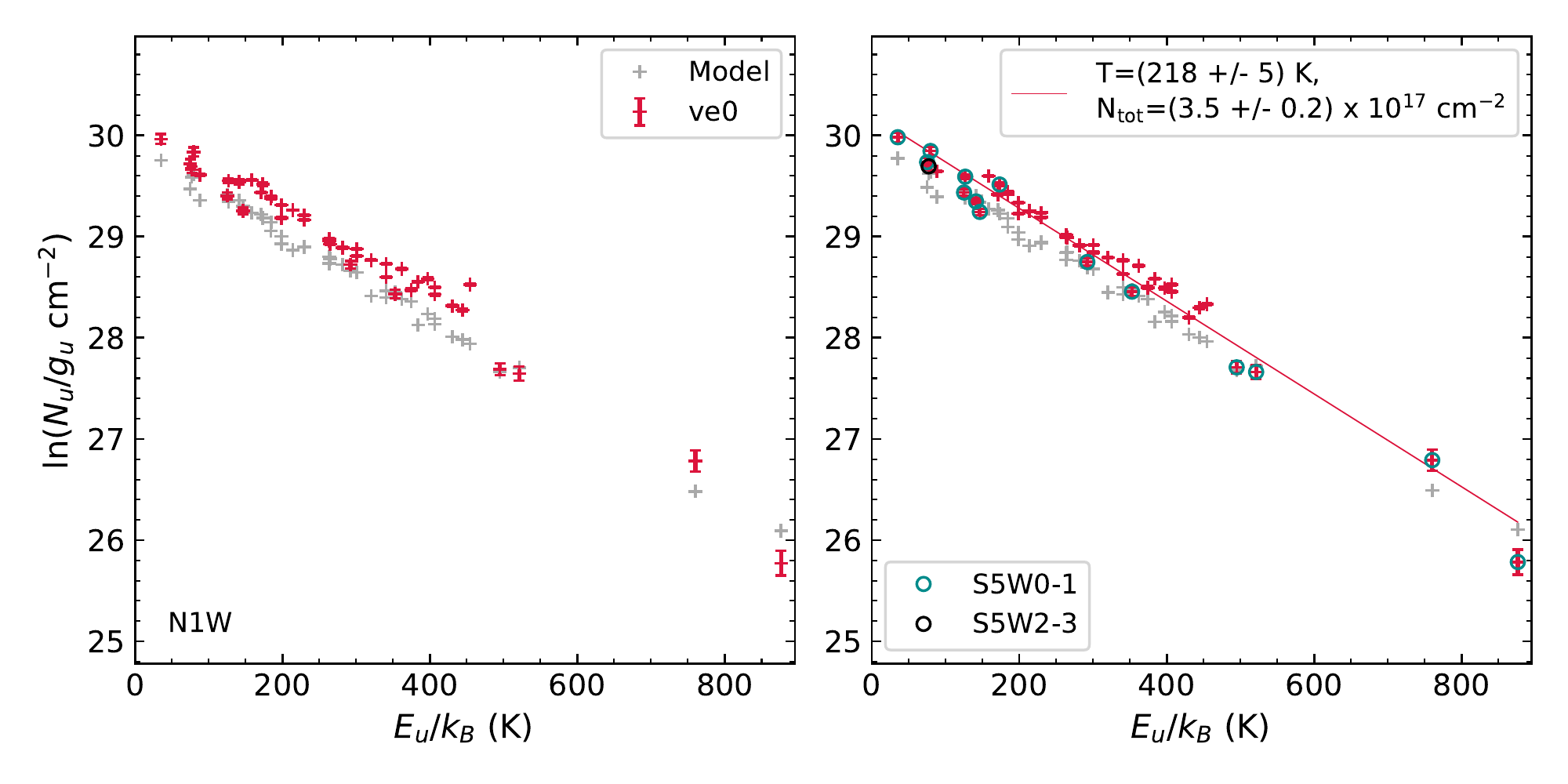}
    \includegraphics[width=0.49\textwidth]{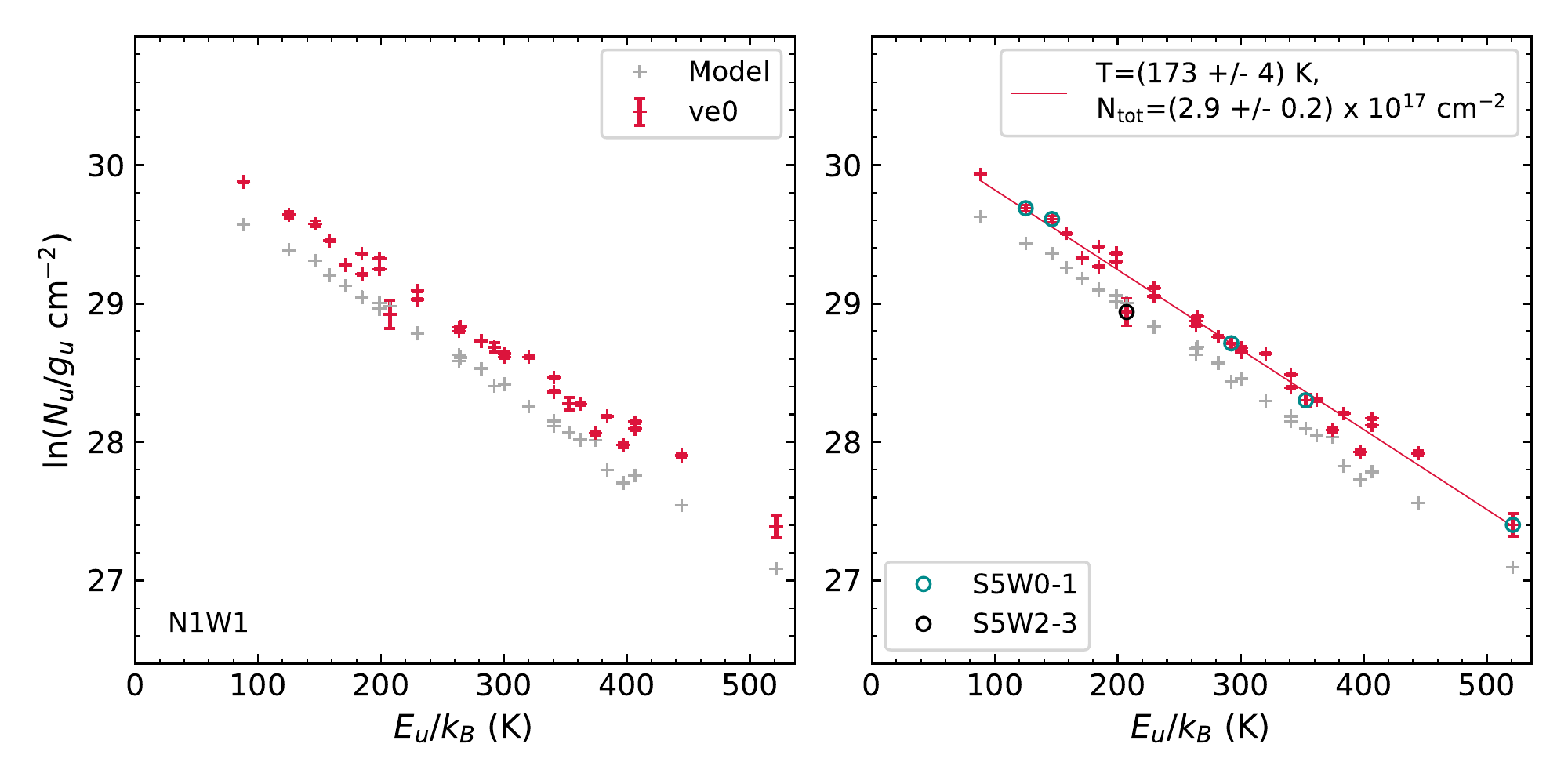}
    \includegraphics[width=0.49\textwidth]{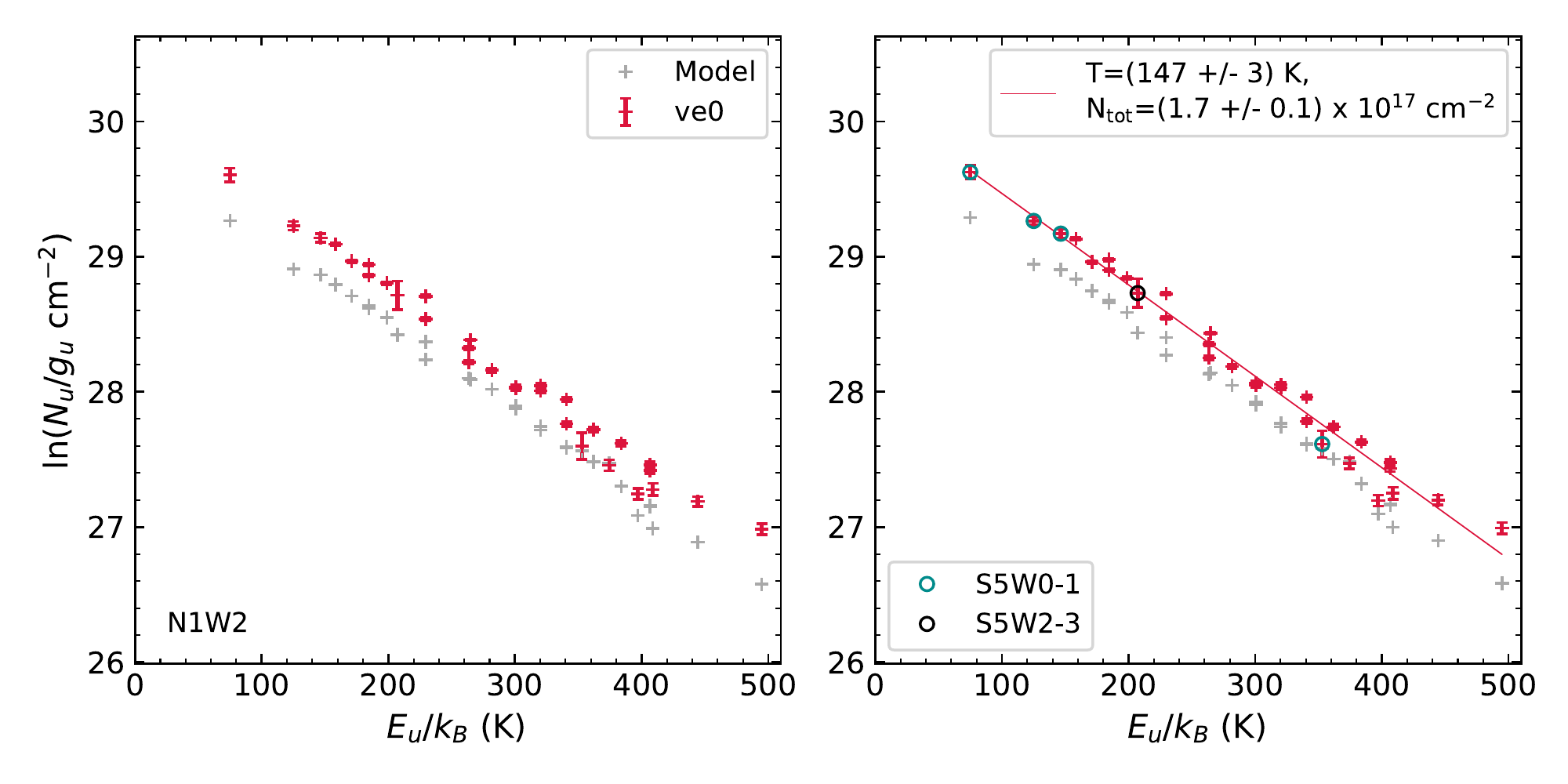}
    \includegraphics[width=0.49\textwidth]{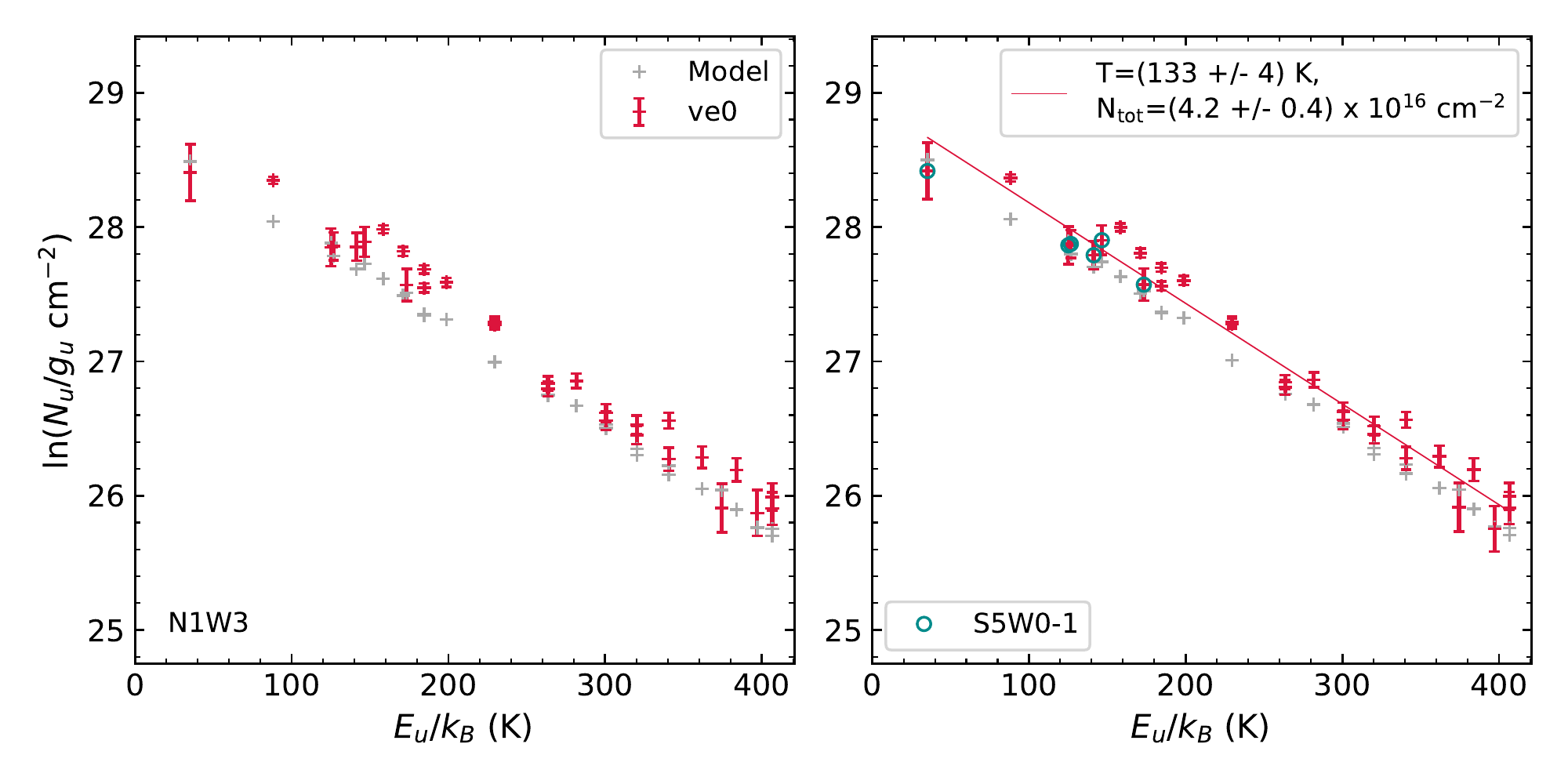}
    \includegraphics[width=0.49\textwidth]{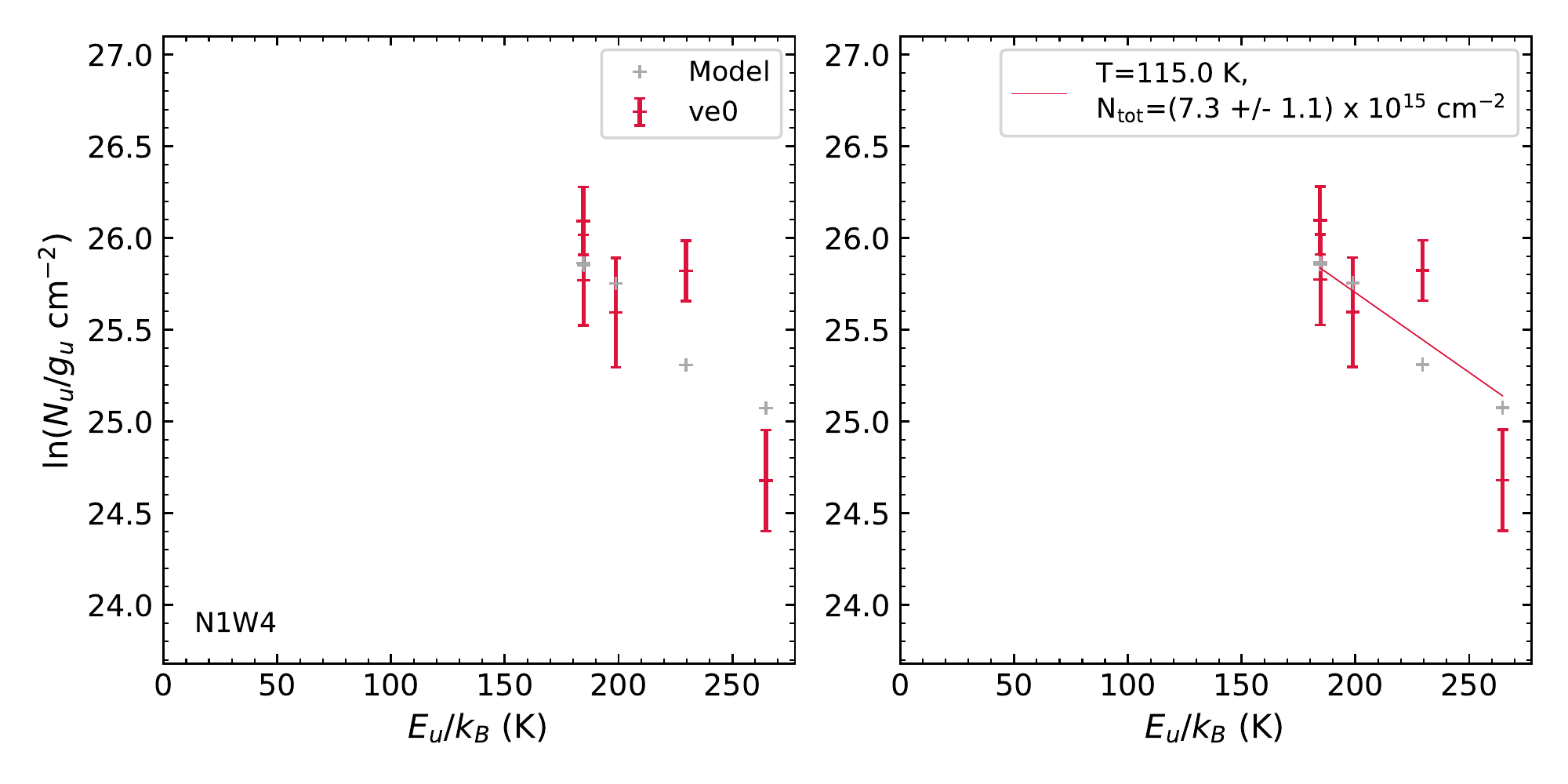}
    \caption{Same as Fig.\,\ref{fig:PD_met}, but for \et and for all positions to the west where the molecule is detected.}
    \label{fig:wPD_et}
\end{figure*}

\begin{figure*}[h]
    \includegraphics[width=0.49\textwidth]{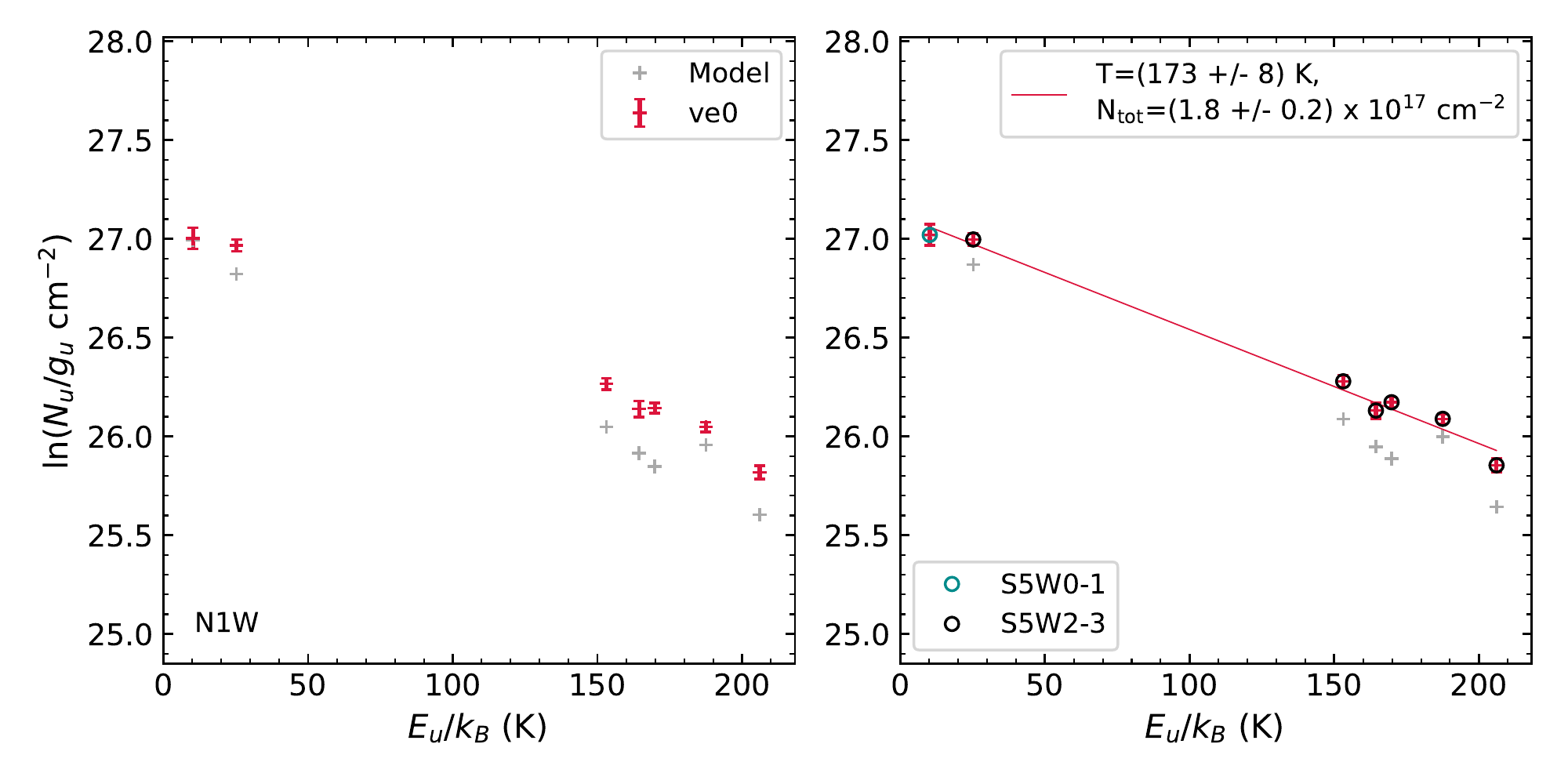}
    \includegraphics[width=0.49\textwidth]{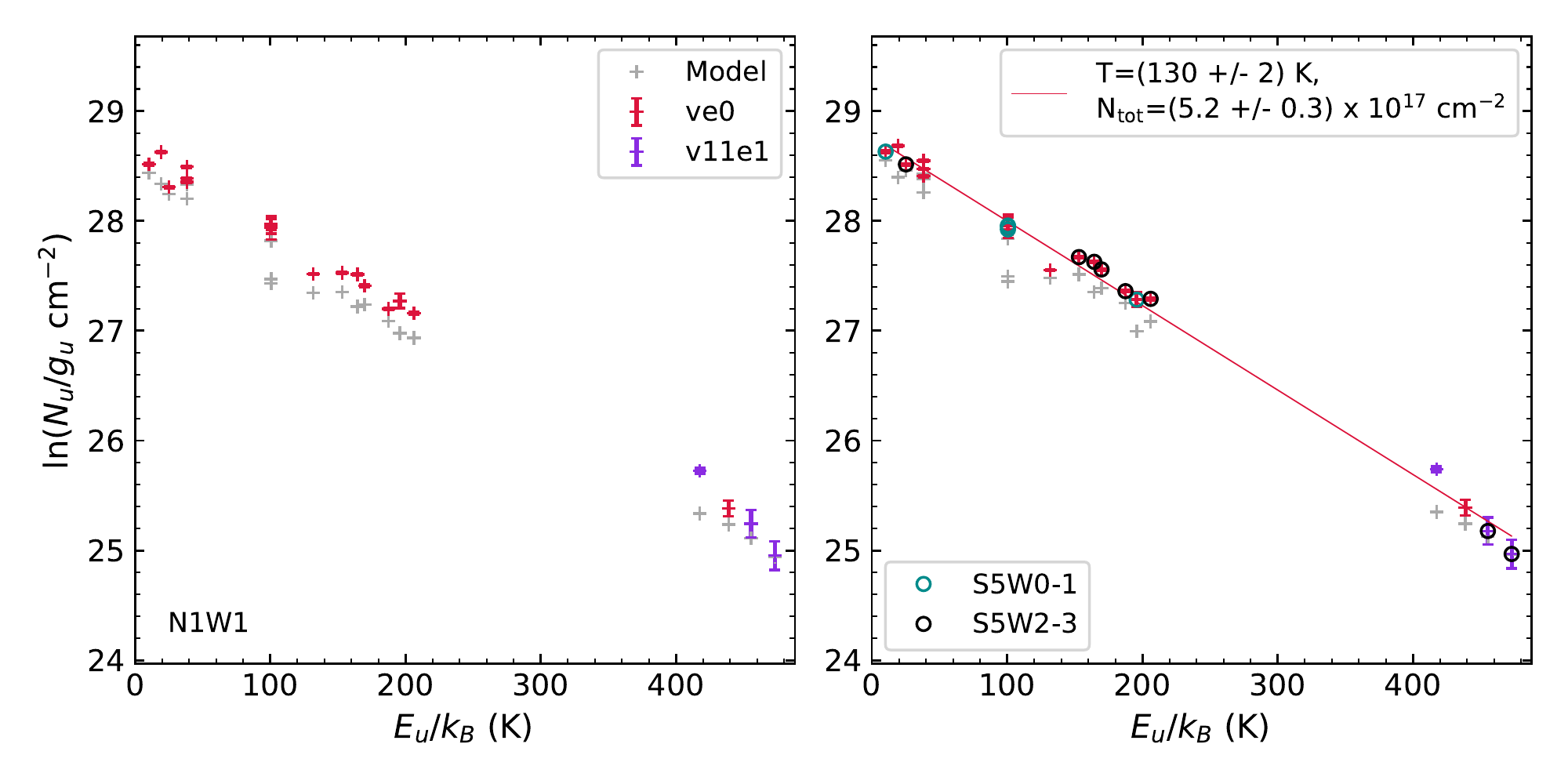}
    \includegraphics[width=0.49\textwidth]{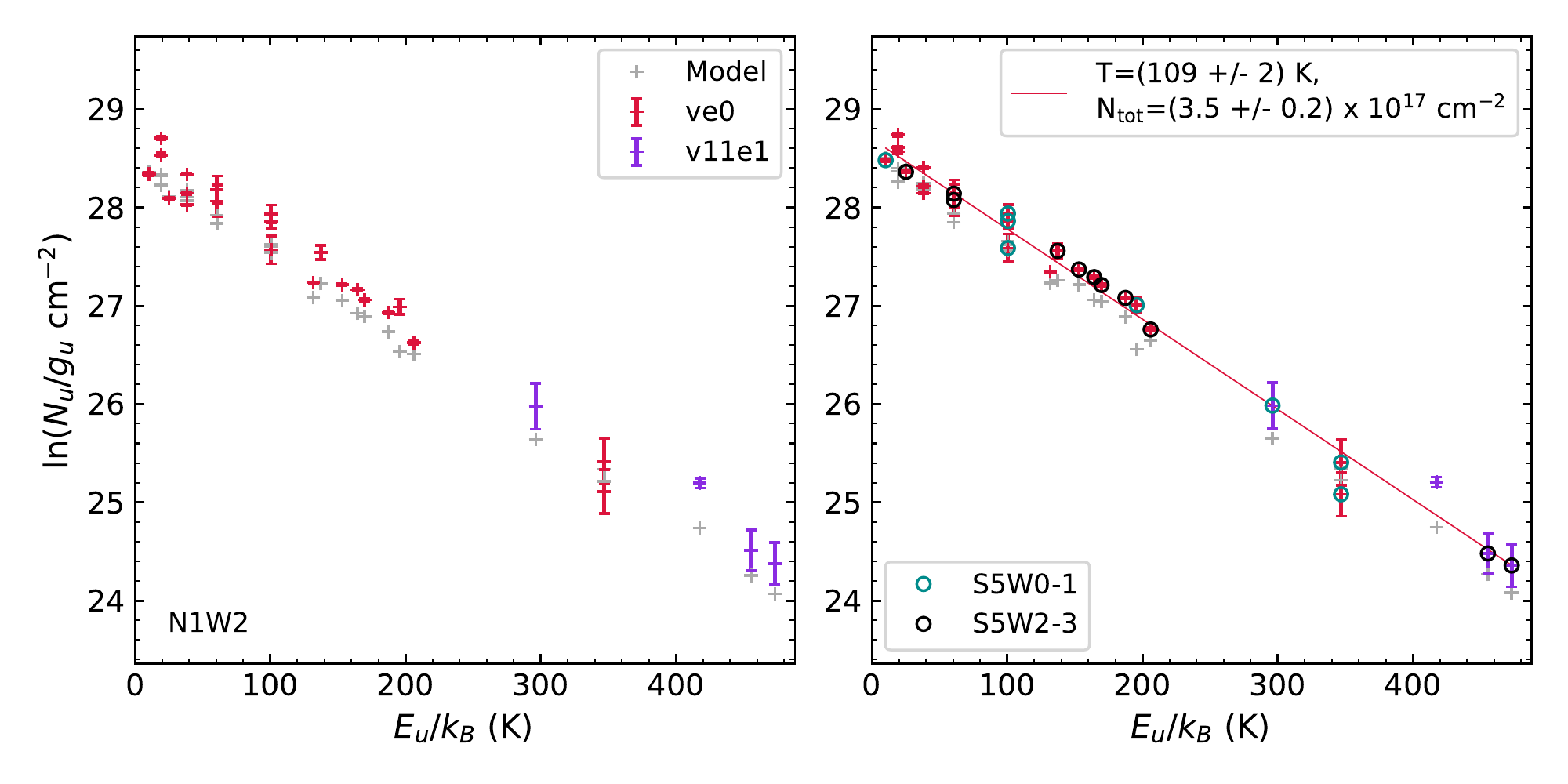}
    \includegraphics[width=0.49\textwidth]{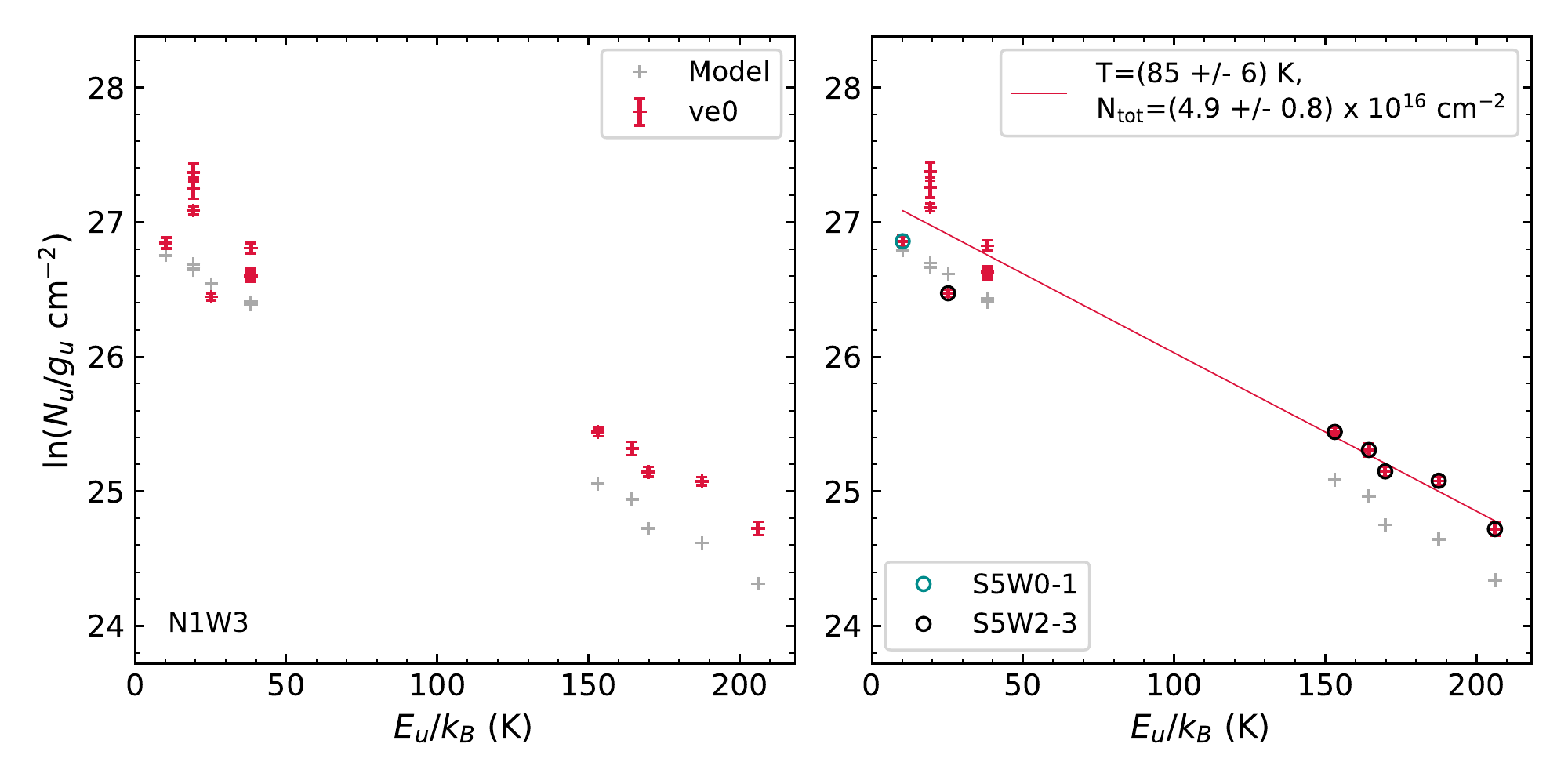}
    \includegraphics[width=0.49\textwidth]{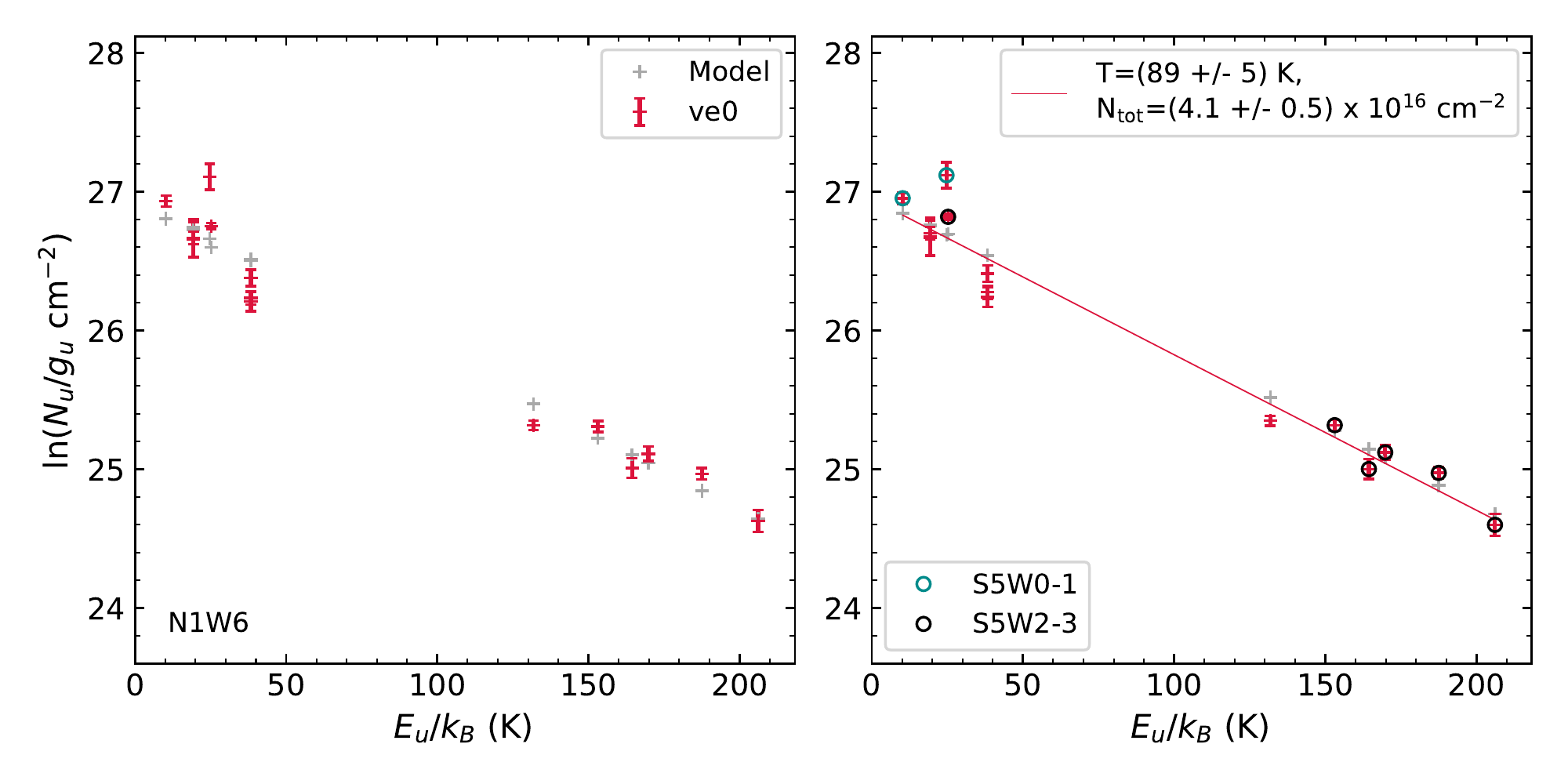}
    \includegraphics[width=0.49\textwidth]{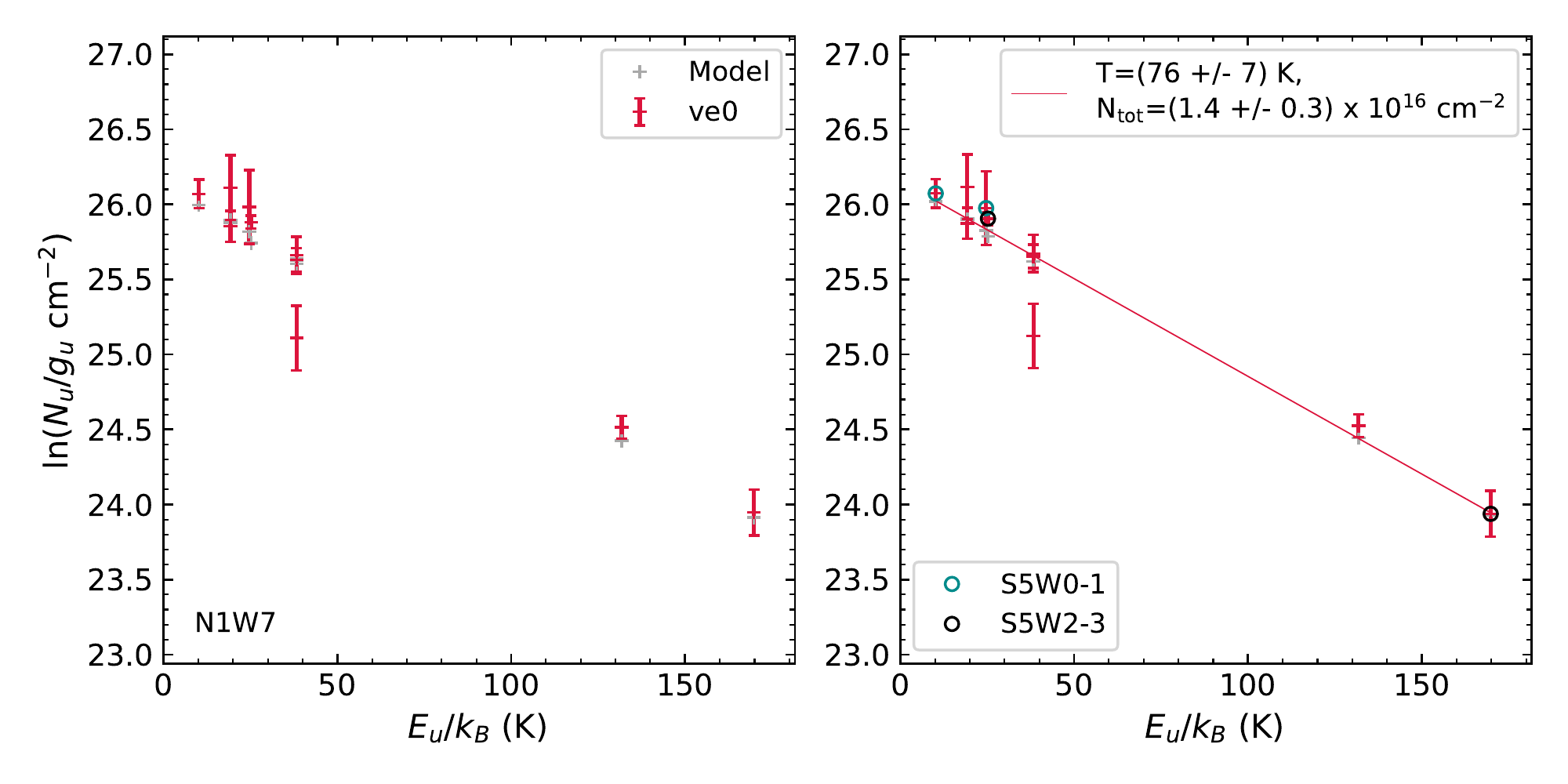}
    \caption{Same as Fig.\,\ref{fig:PD_met}, but for \dme and for all positions to the west where the molecule is detected.}
    \label{fig:wPD_dme}
\end{figure*}

\begin{figure*}[h]
    \includegraphics[width=0.49\textwidth]{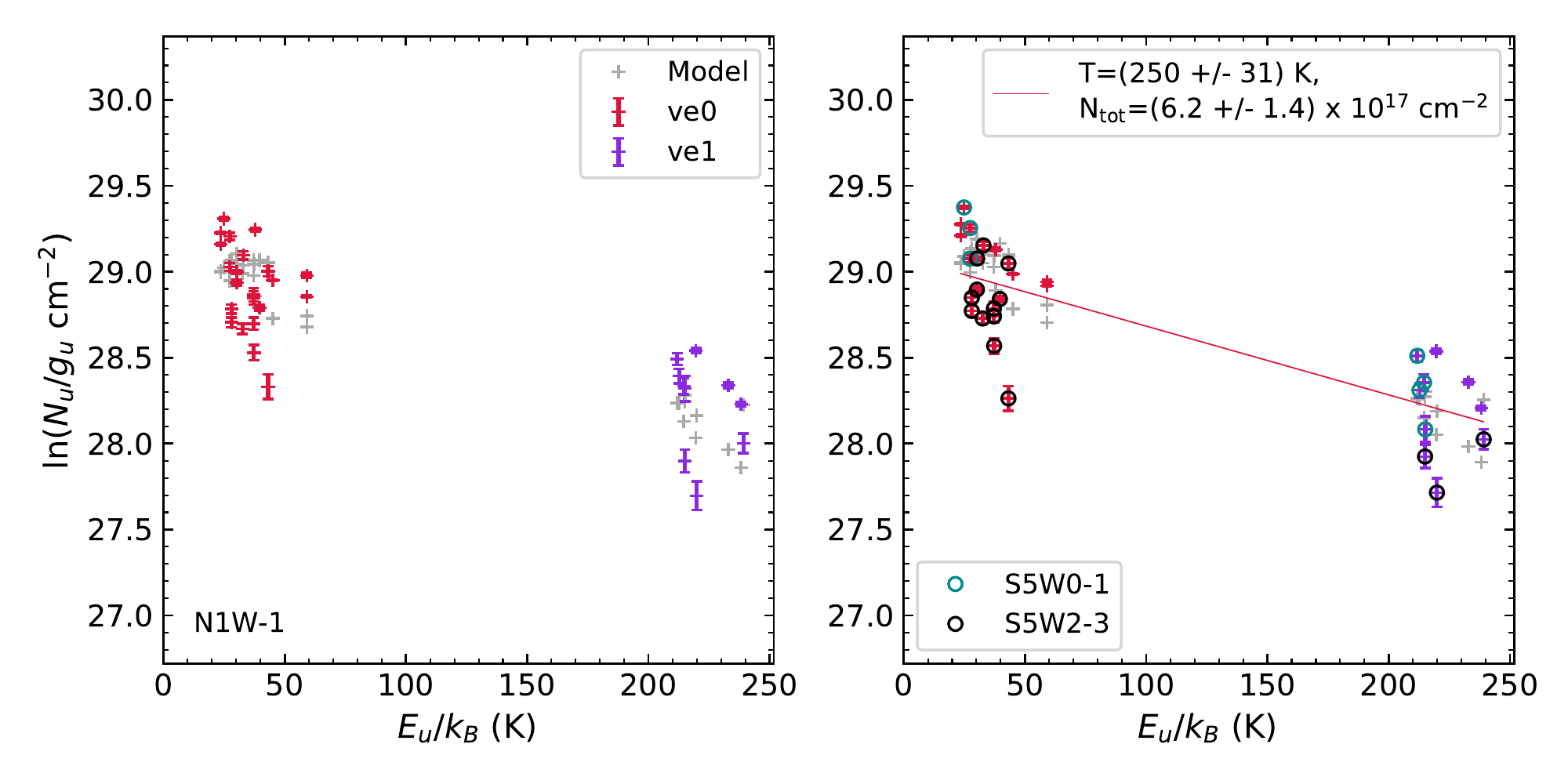}
    \includegraphics[width=0.49\textwidth]{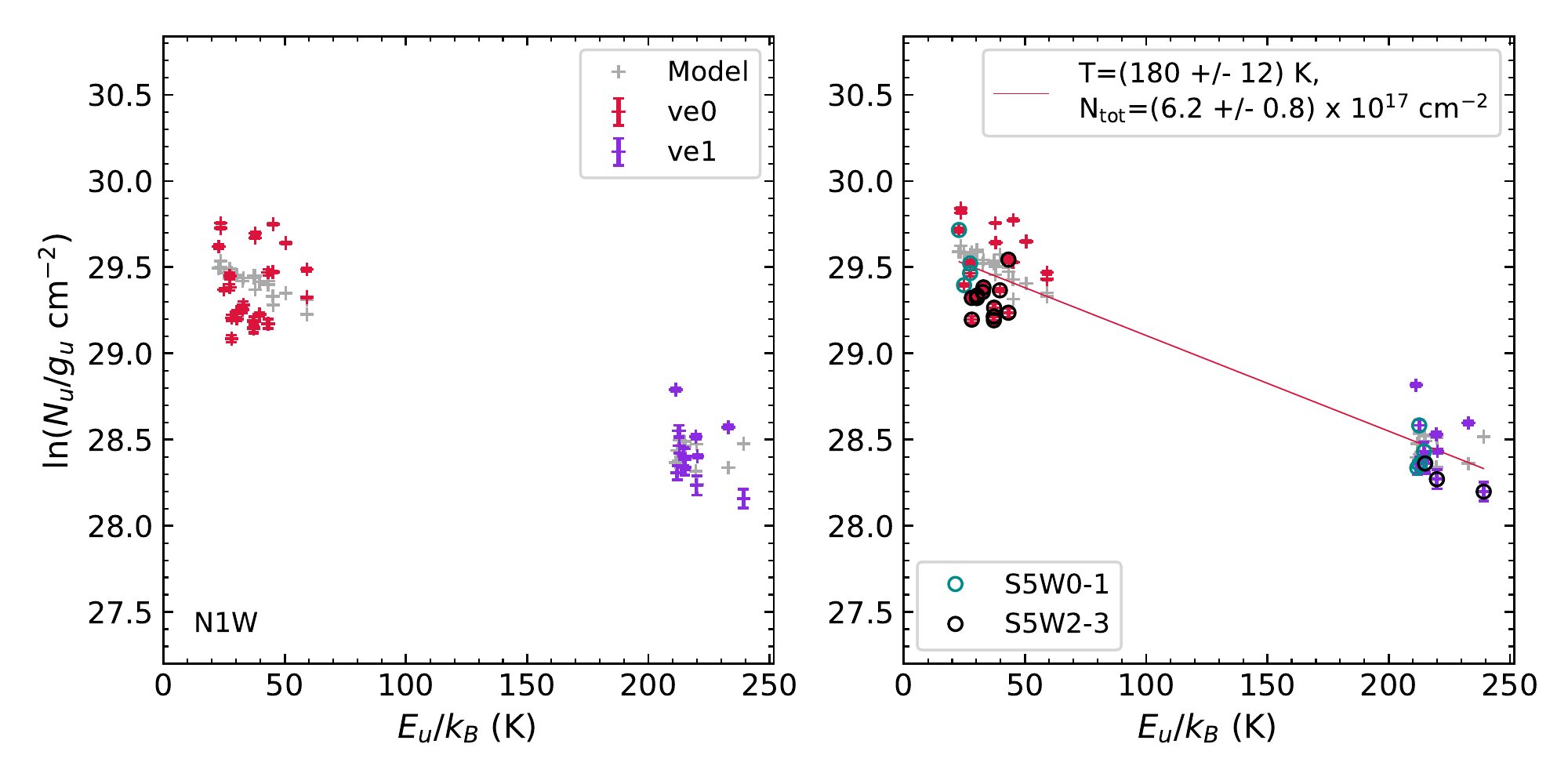}
    \includegraphics[width=0.49\textwidth]{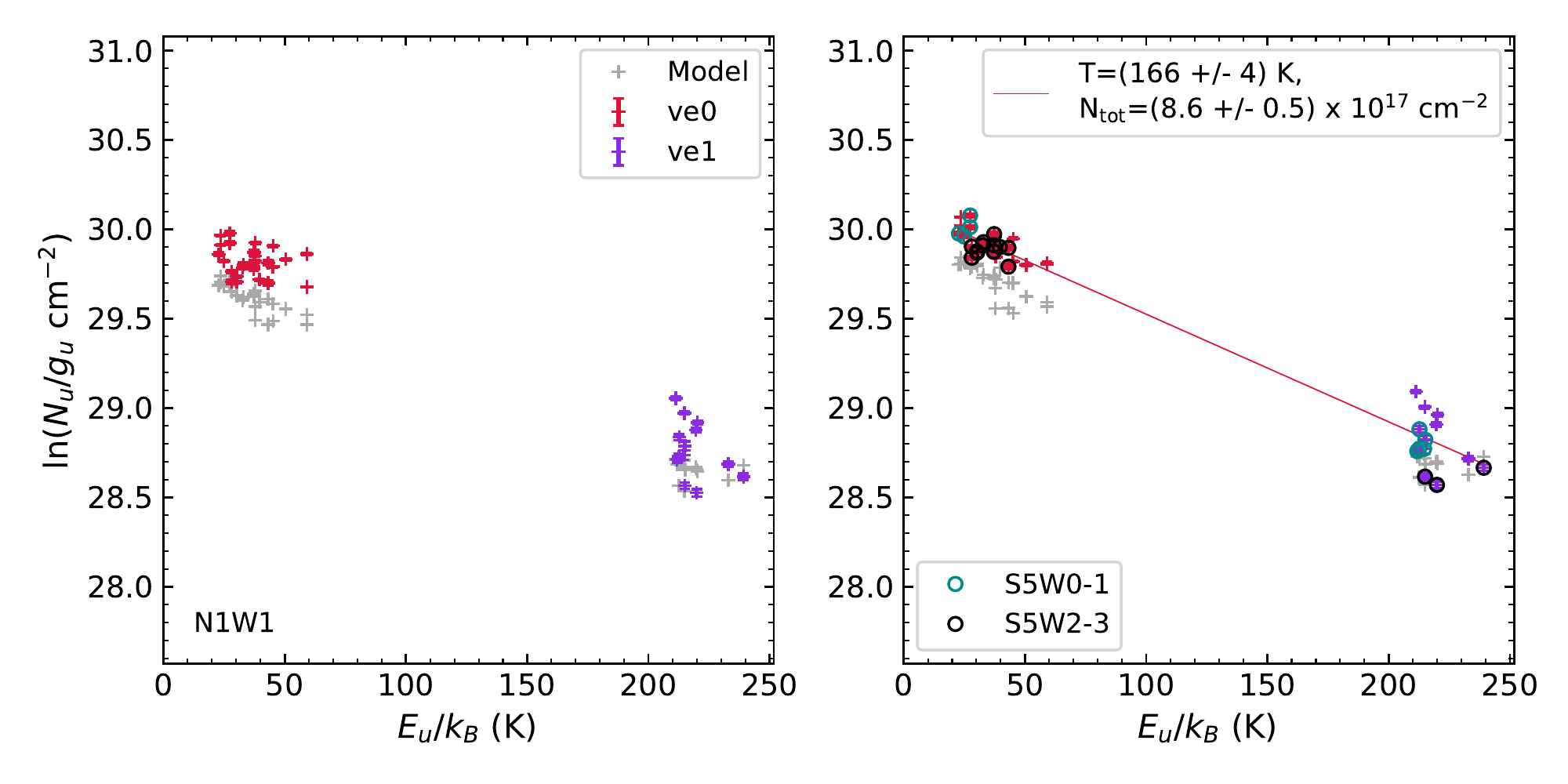}
    \includegraphics[width=0.49\textwidth]{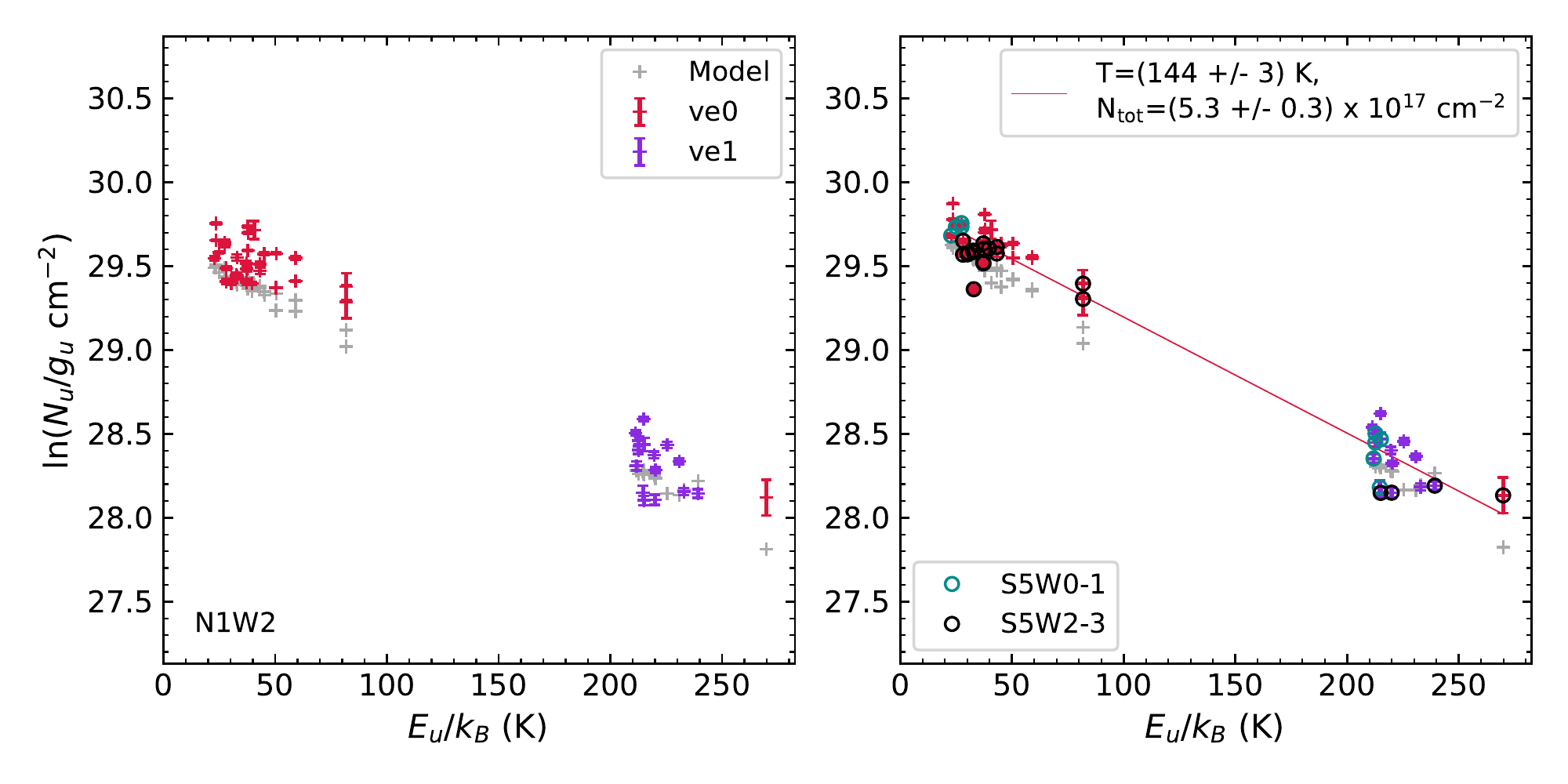}
    \includegraphics[width=0.49\textwidth]{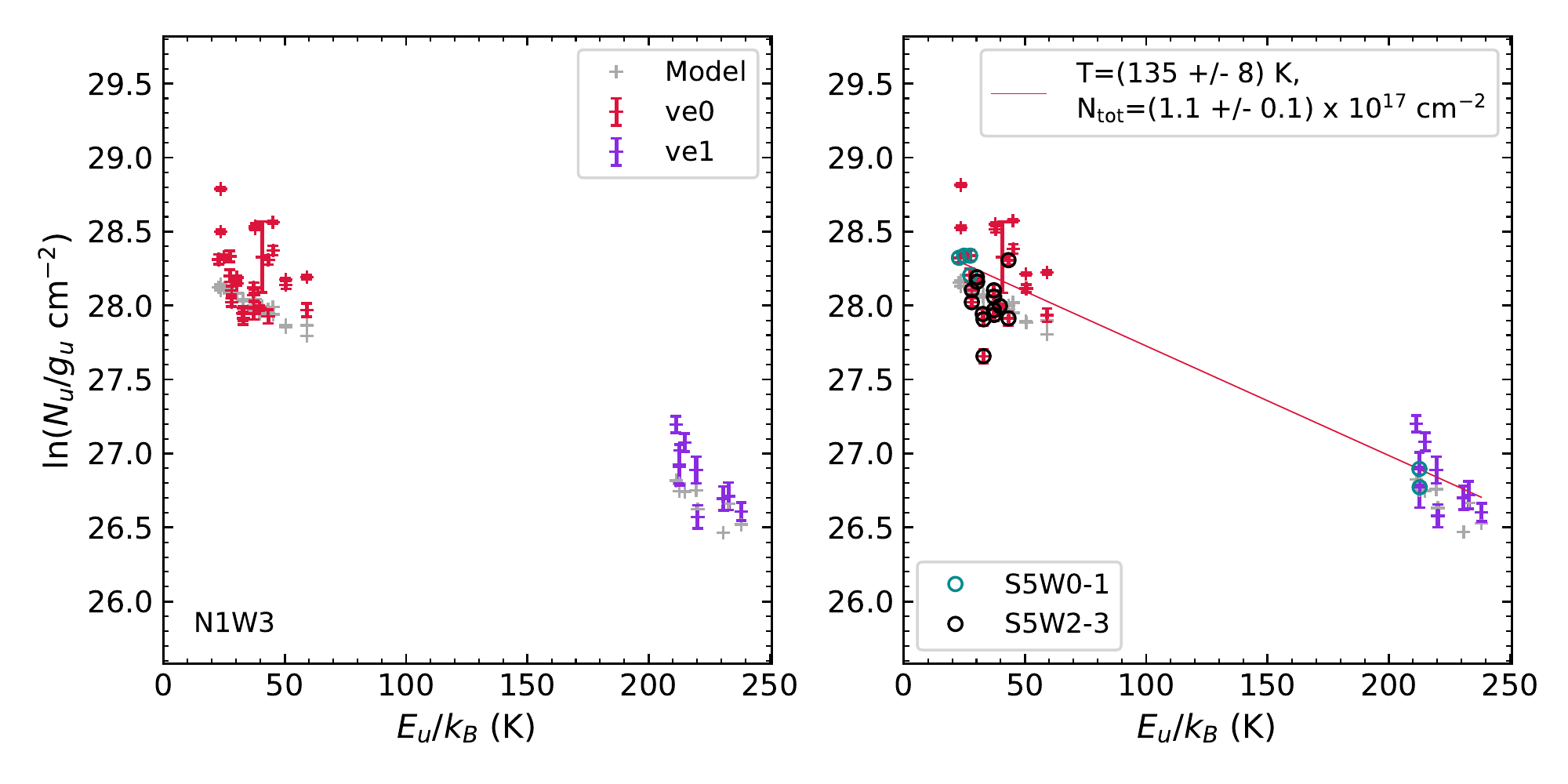}
    \includegraphics[width=0.49\textwidth]{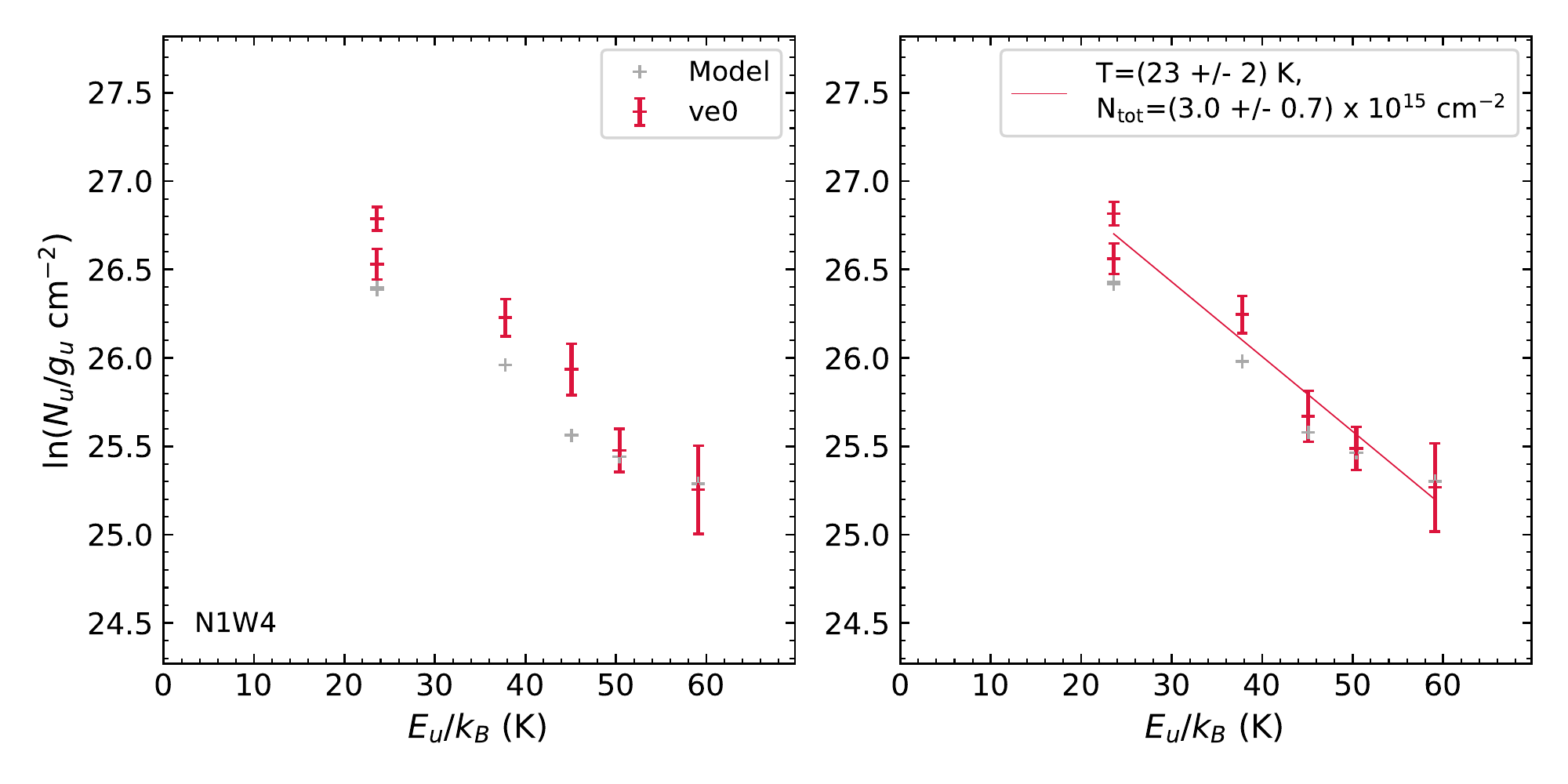}
    \includegraphics[width=0.49\textwidth]{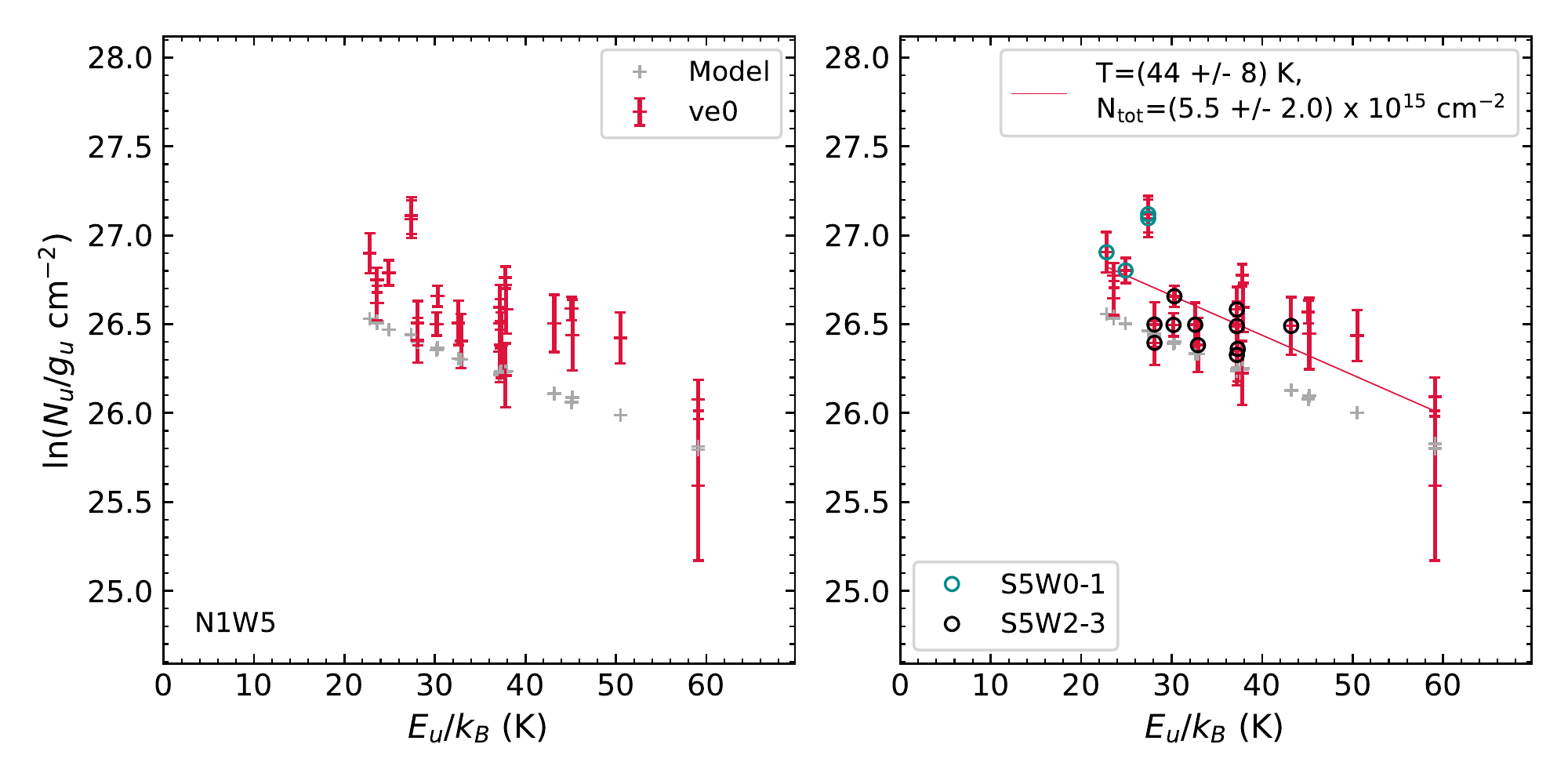}
    \includegraphics[width=0.49\textwidth]{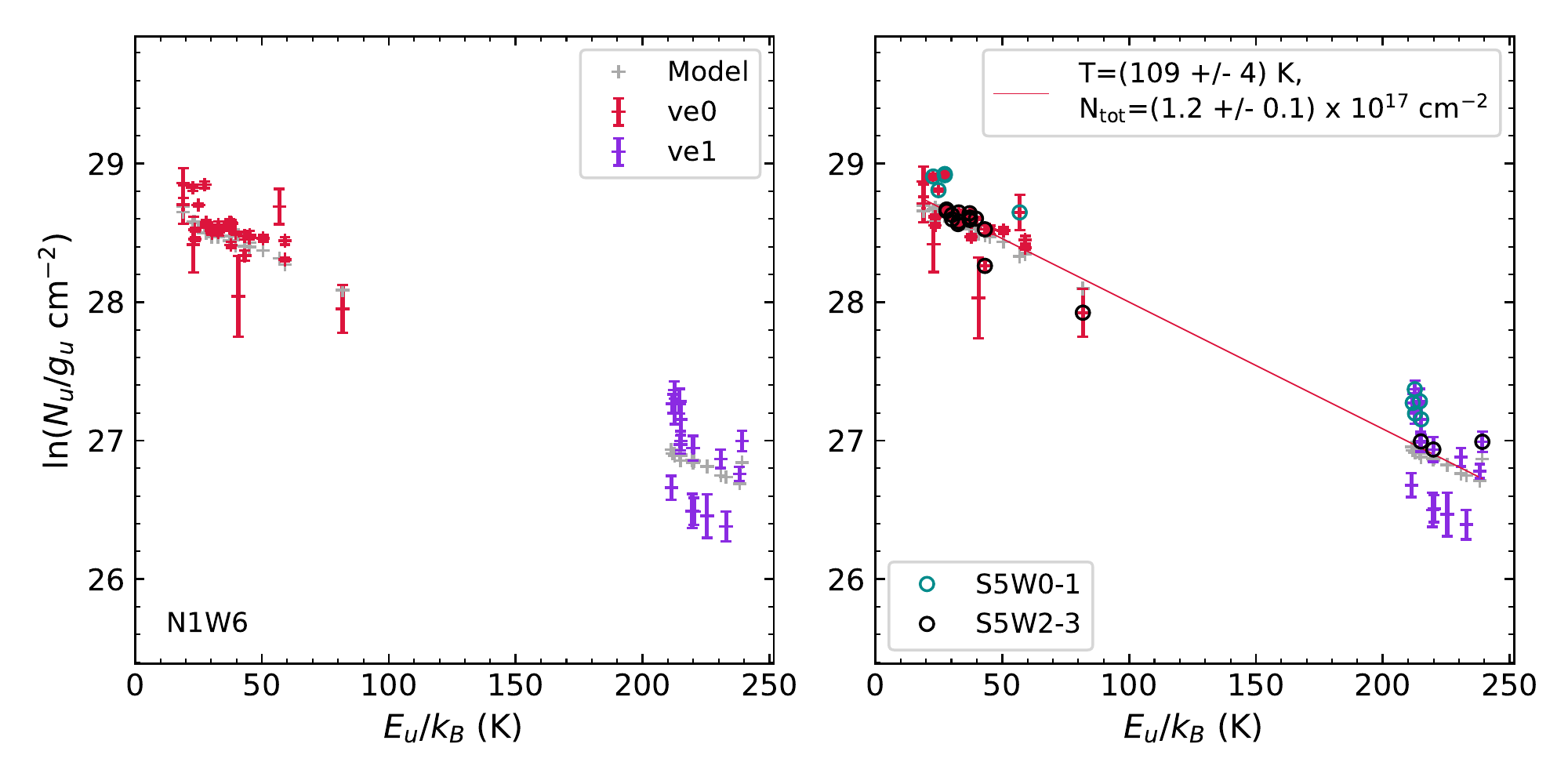}
    \includegraphics[width=0.49\textwidth]{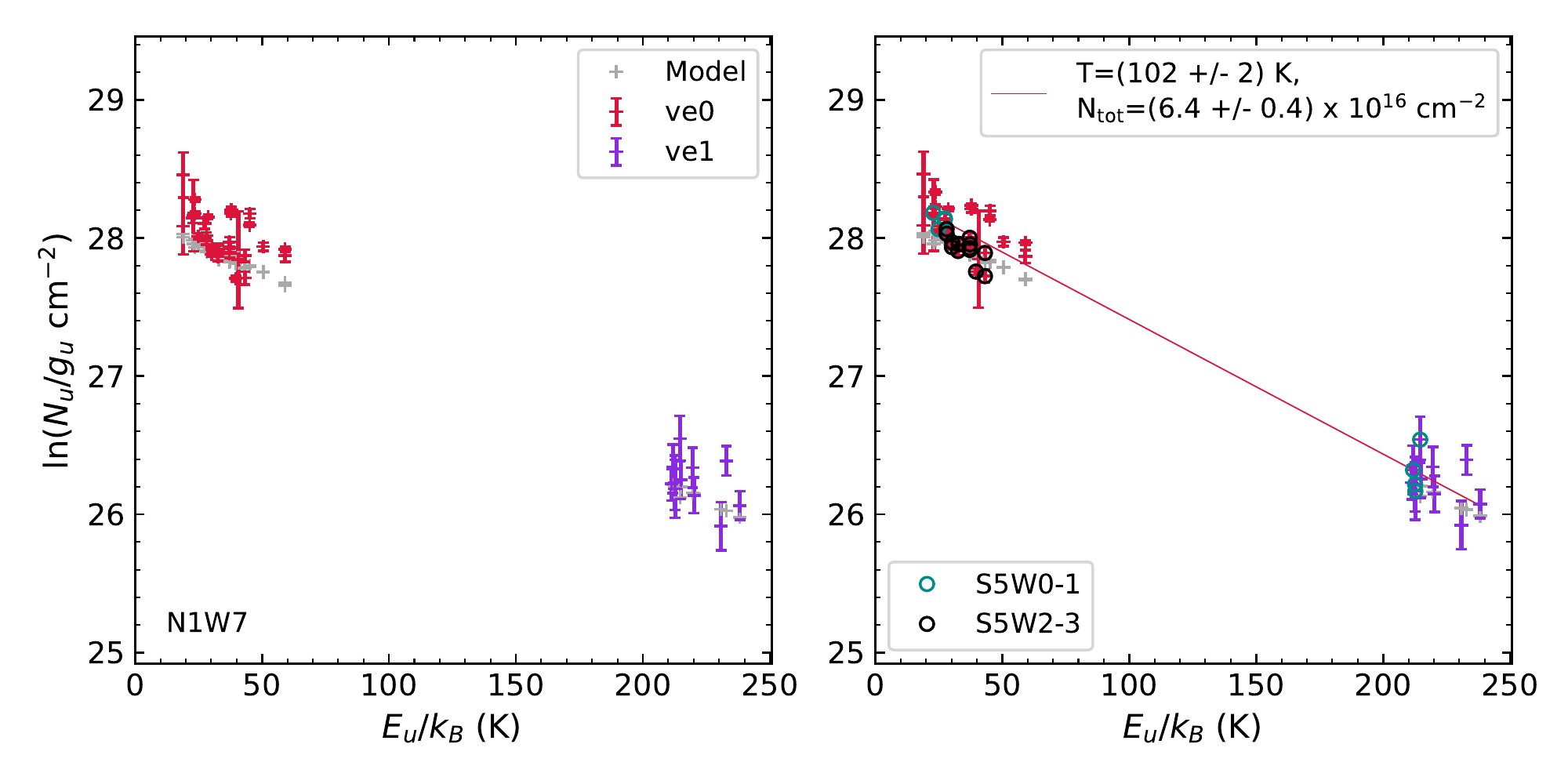}
    \caption{Same as Fig.\,\ref{fig:PD_met}, but for \mf and for all positions to the west where the molecule is detected.}
    \label{fig:wPD_mf}
\end{figure*}

\begin{figure*}[h]
    \includegraphics[width=0.49\textwidth]{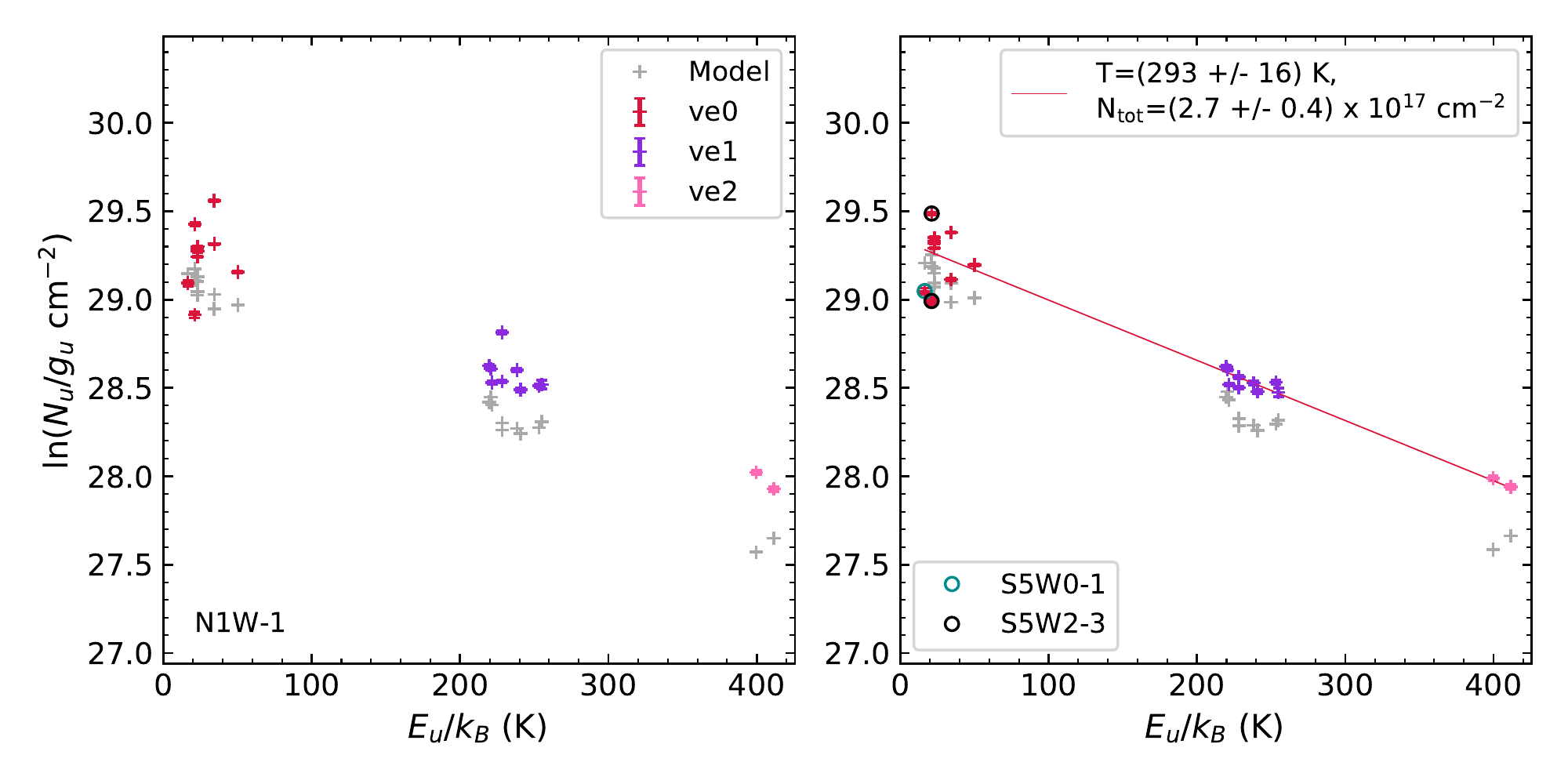}
    \includegraphics[width=0.49\textwidth]{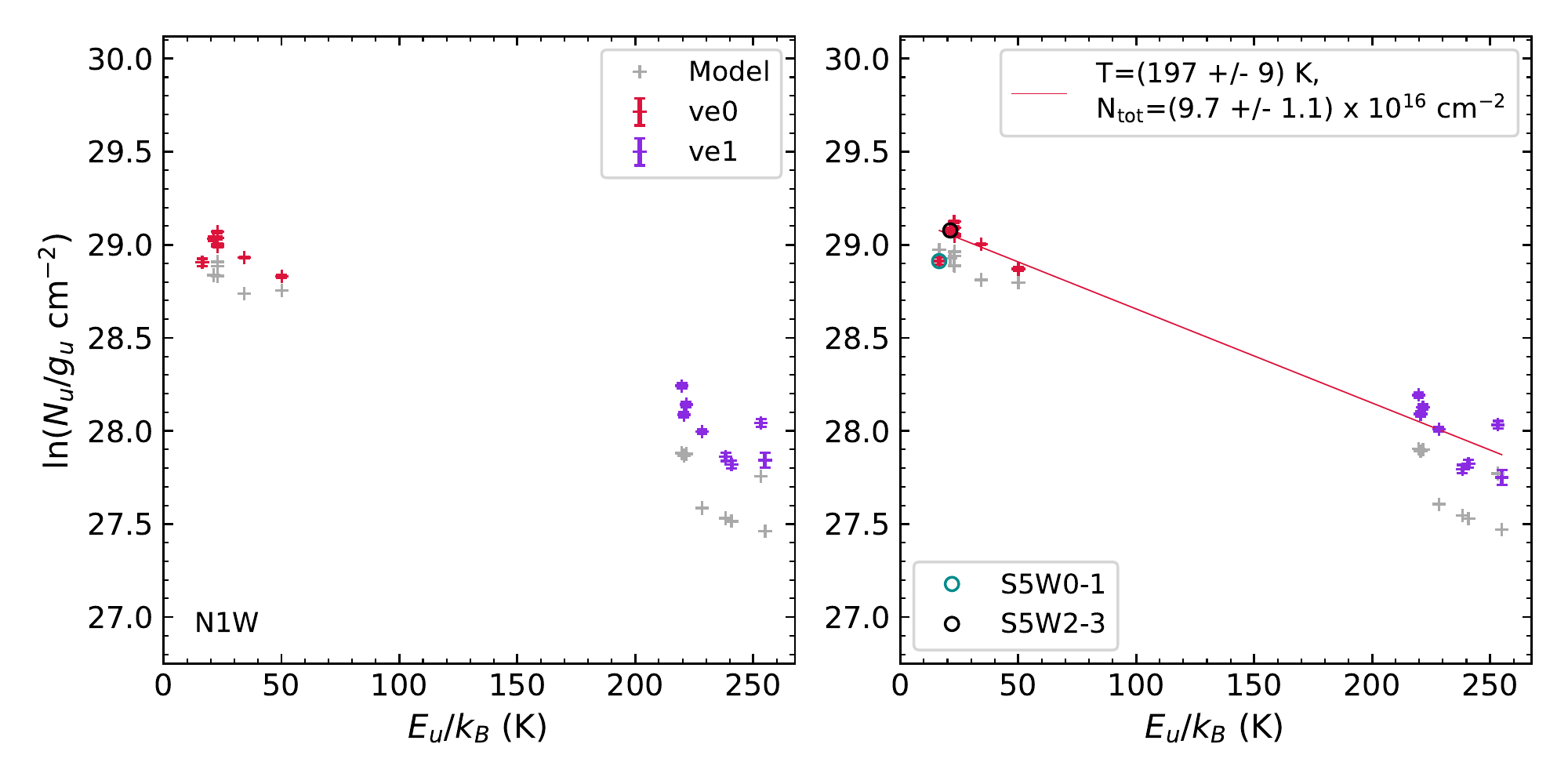}
    \includegraphics[width=0.49\textwidth]{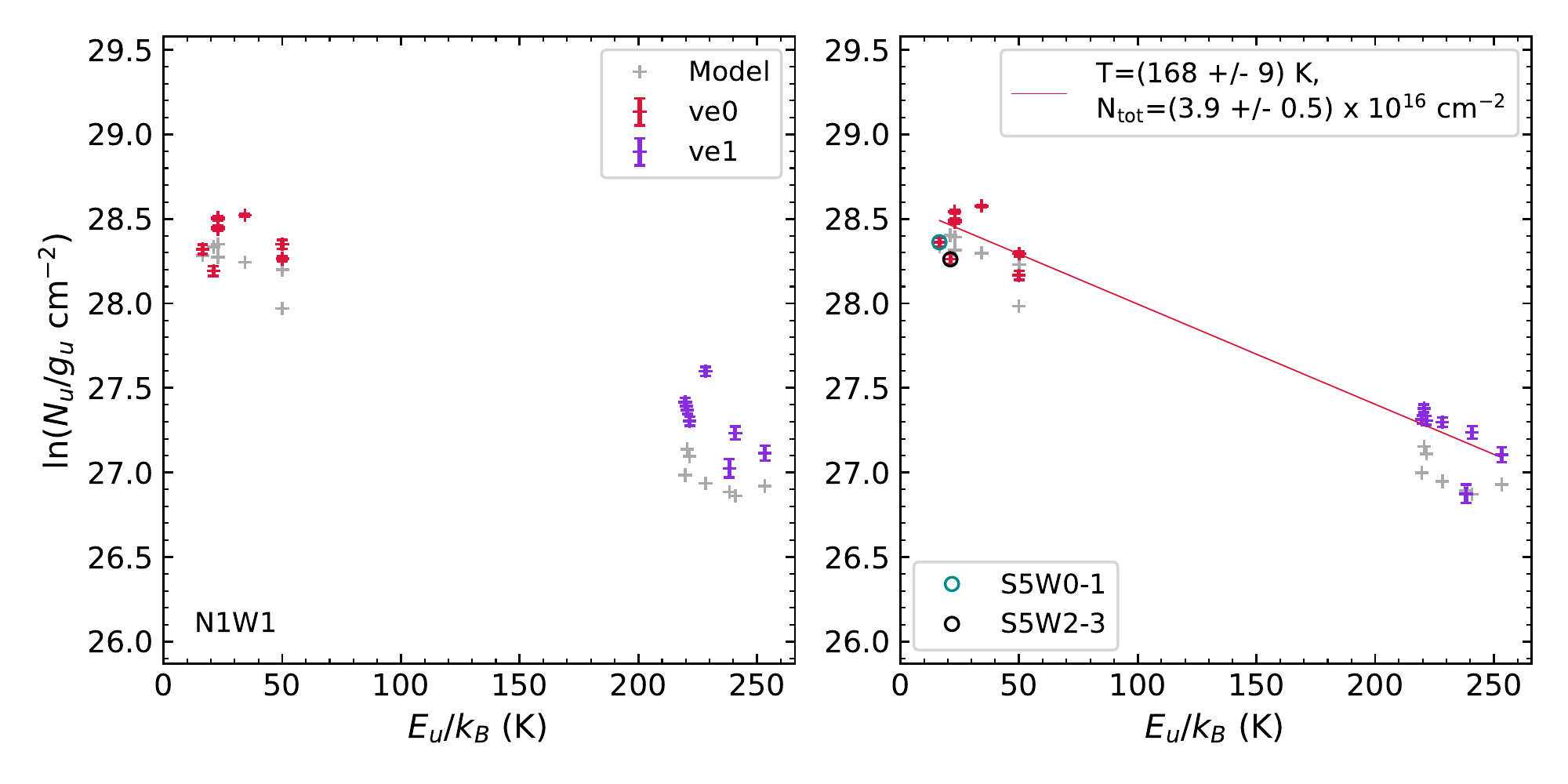}
    \includegraphics[width=0.49\textwidth]{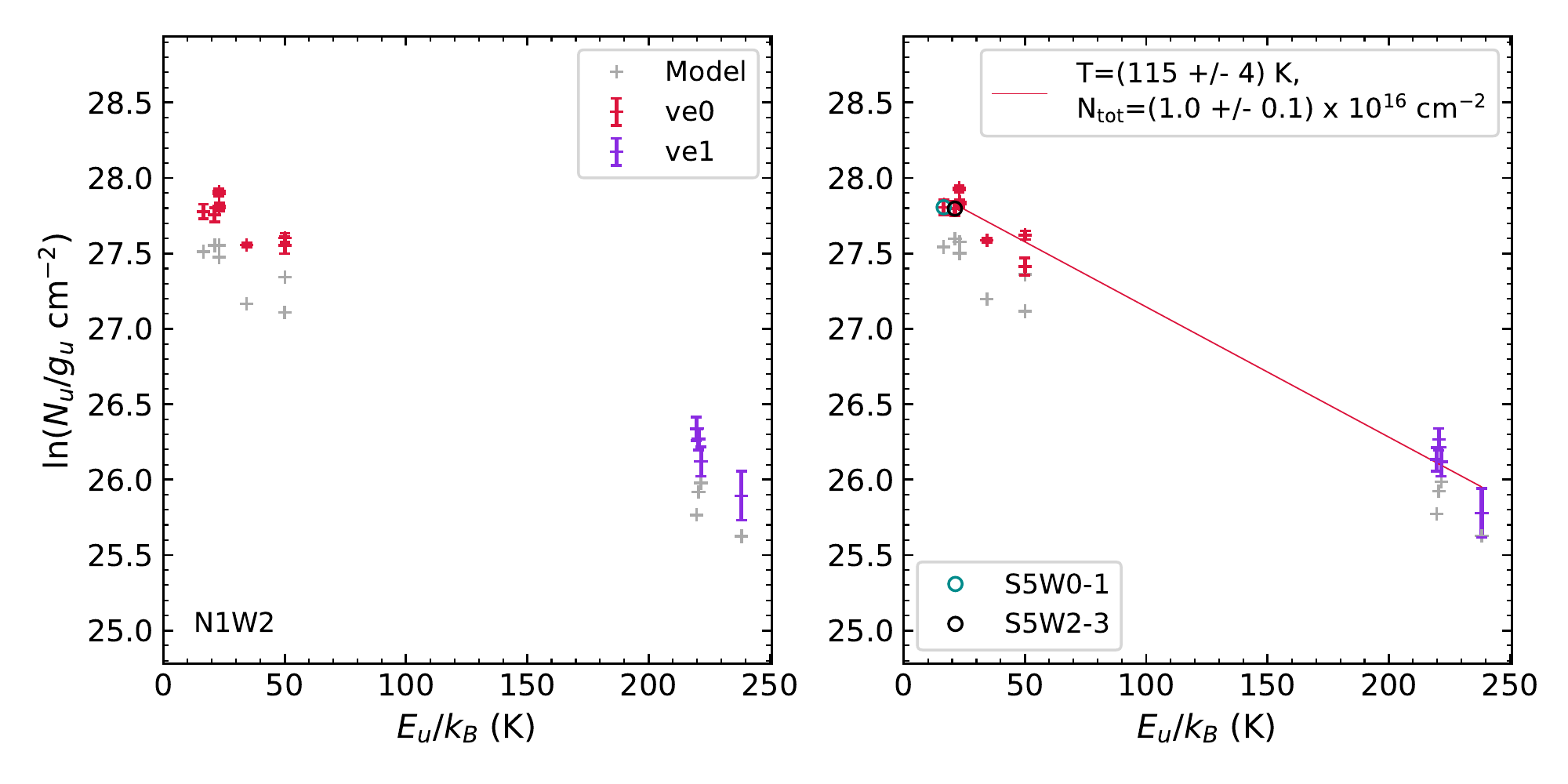}
    \includegraphics[width=0.49\textwidth]{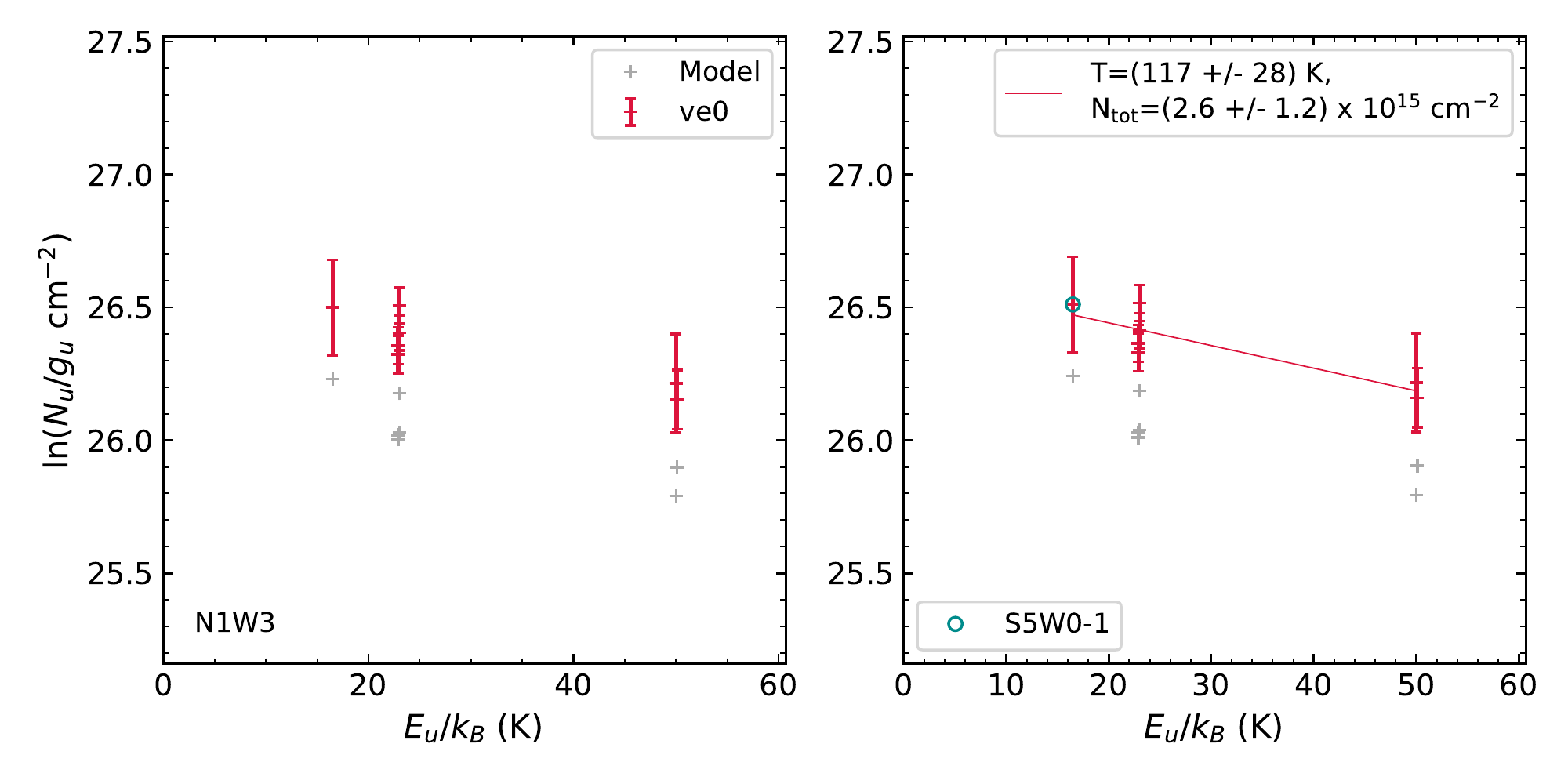}
    \caption{Same as Fig.\,\ref{fig:PD_met}, but for \ad and for all positions to the west where the molecule is detected.}
    \label{fig:wPD_ad}
\end{figure*}

\begin{figure*}[h]
    \includegraphics[width=0.49\textwidth]{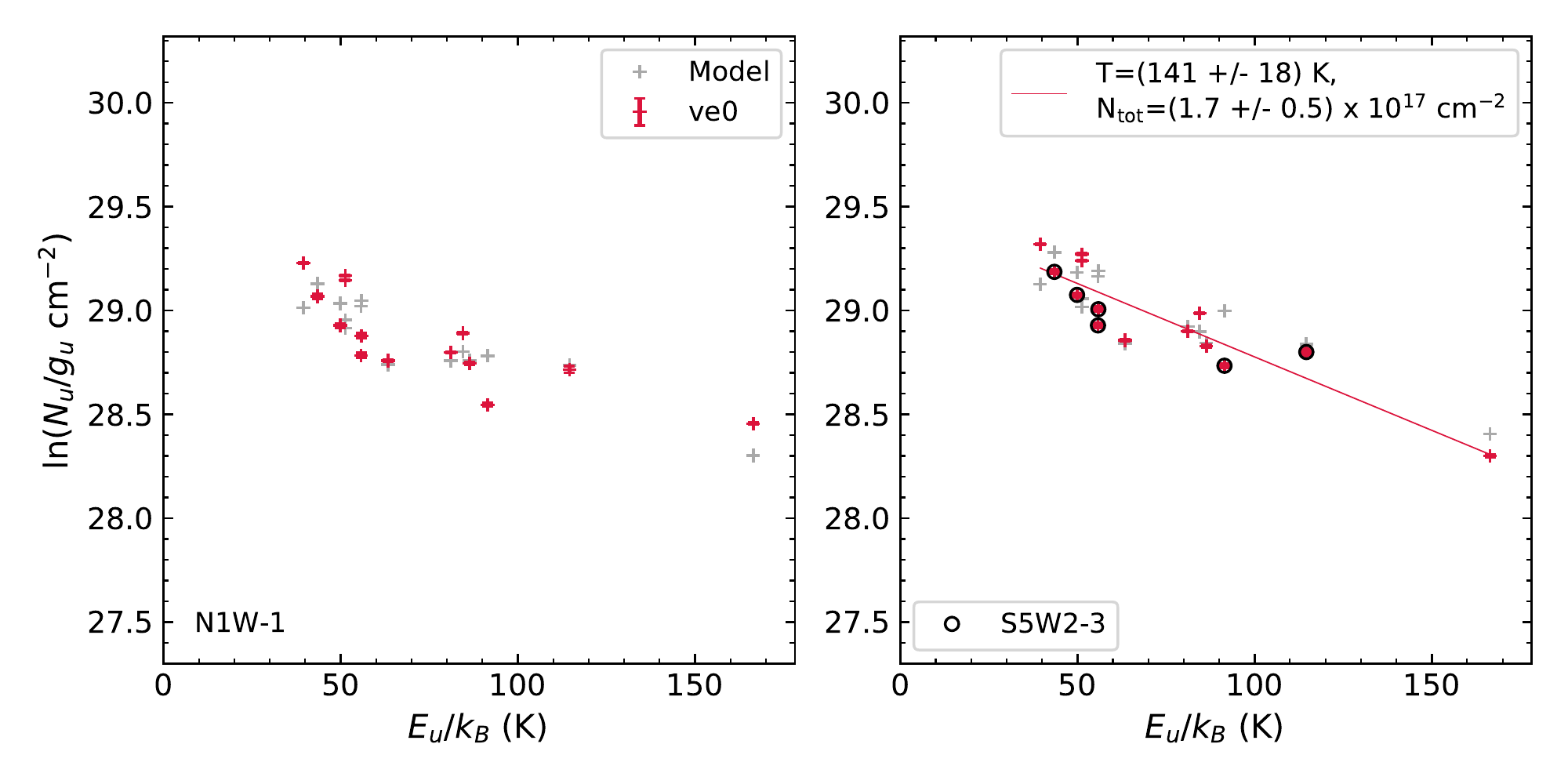}
    \includegraphics[width=0.49\textwidth]{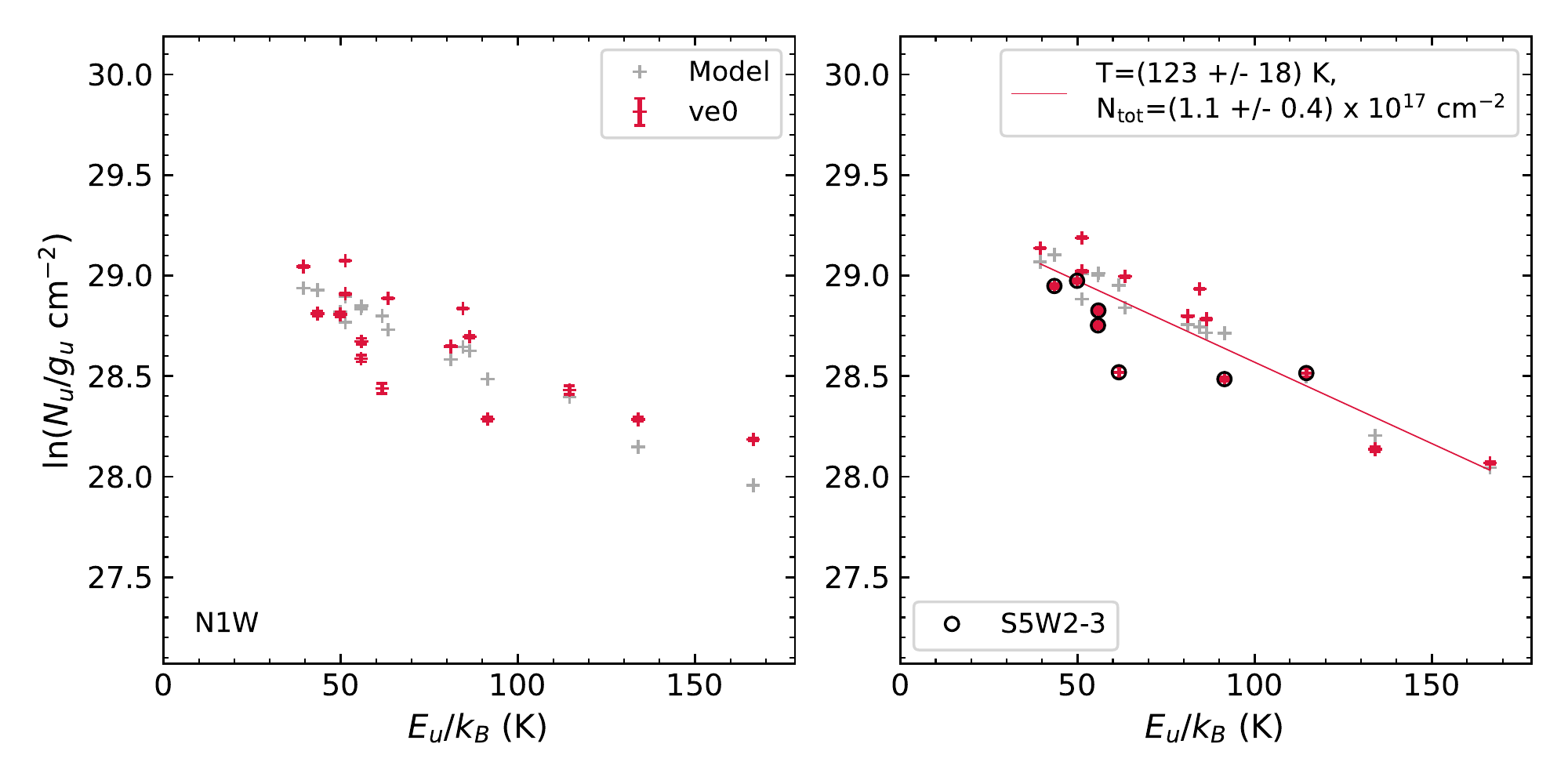}
    \includegraphics[width=0.49\textwidth]{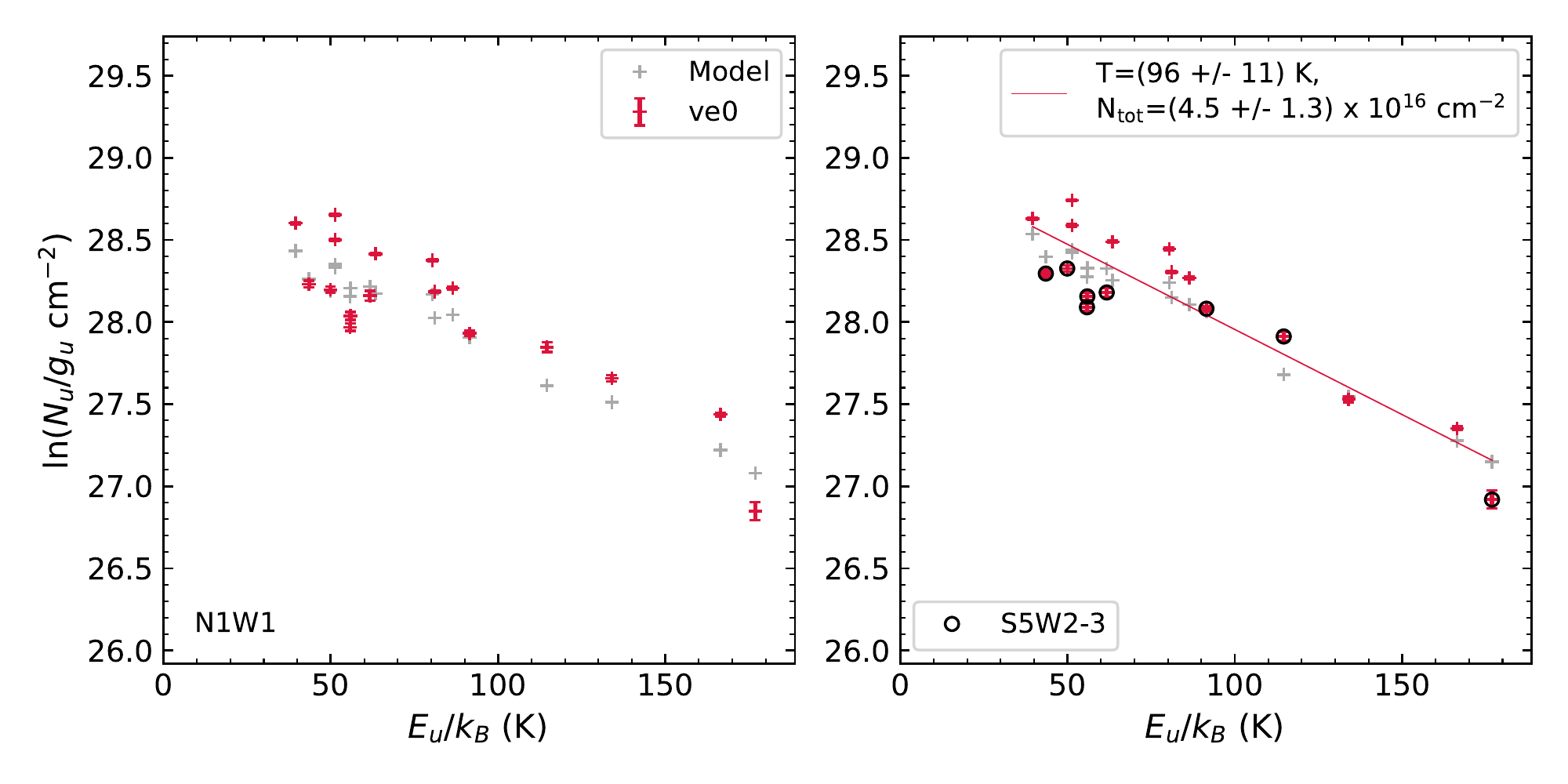}
    \includegraphics[width=0.49\textwidth]{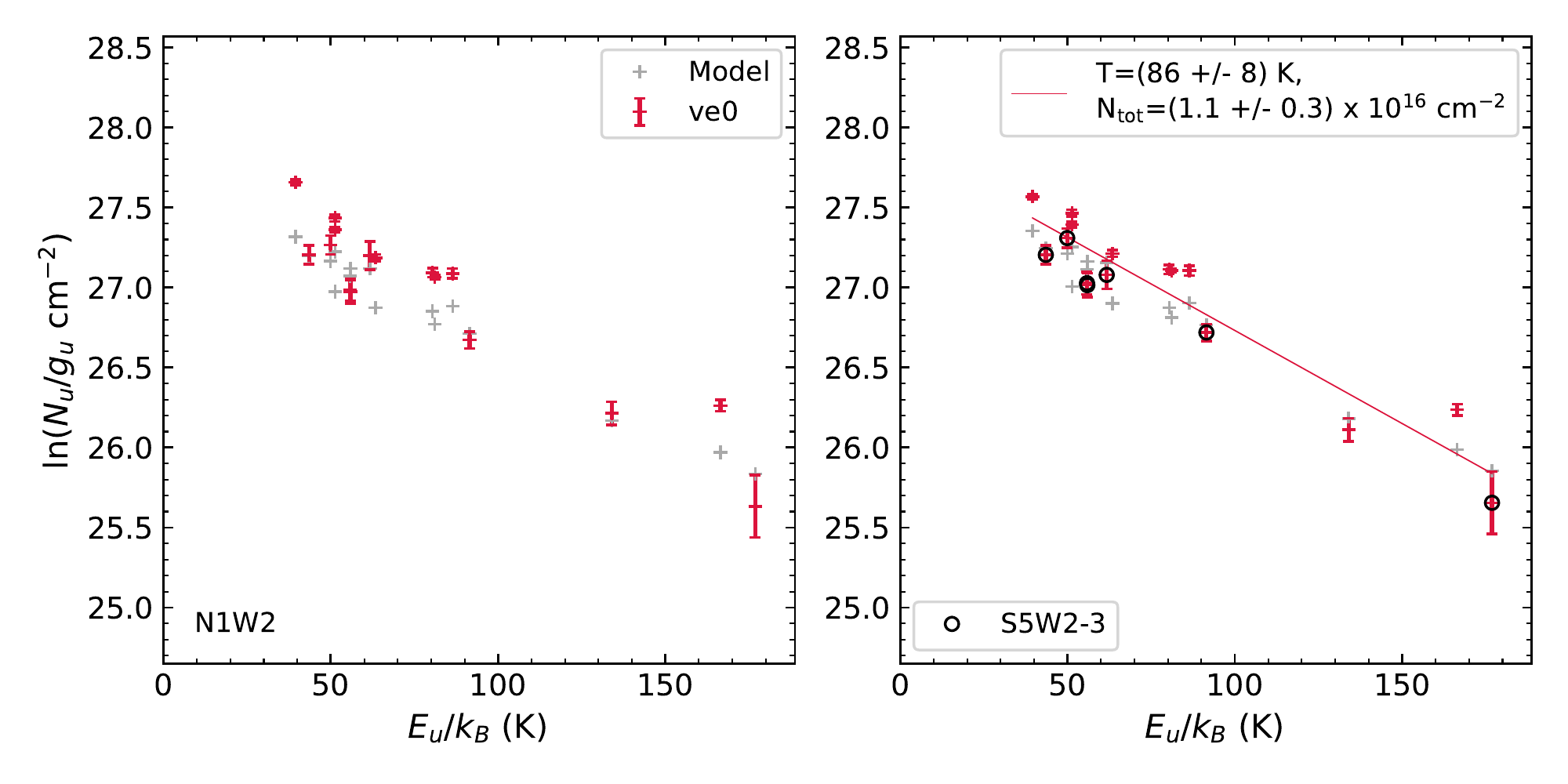}
    \includegraphics[width=0.49\textwidth]{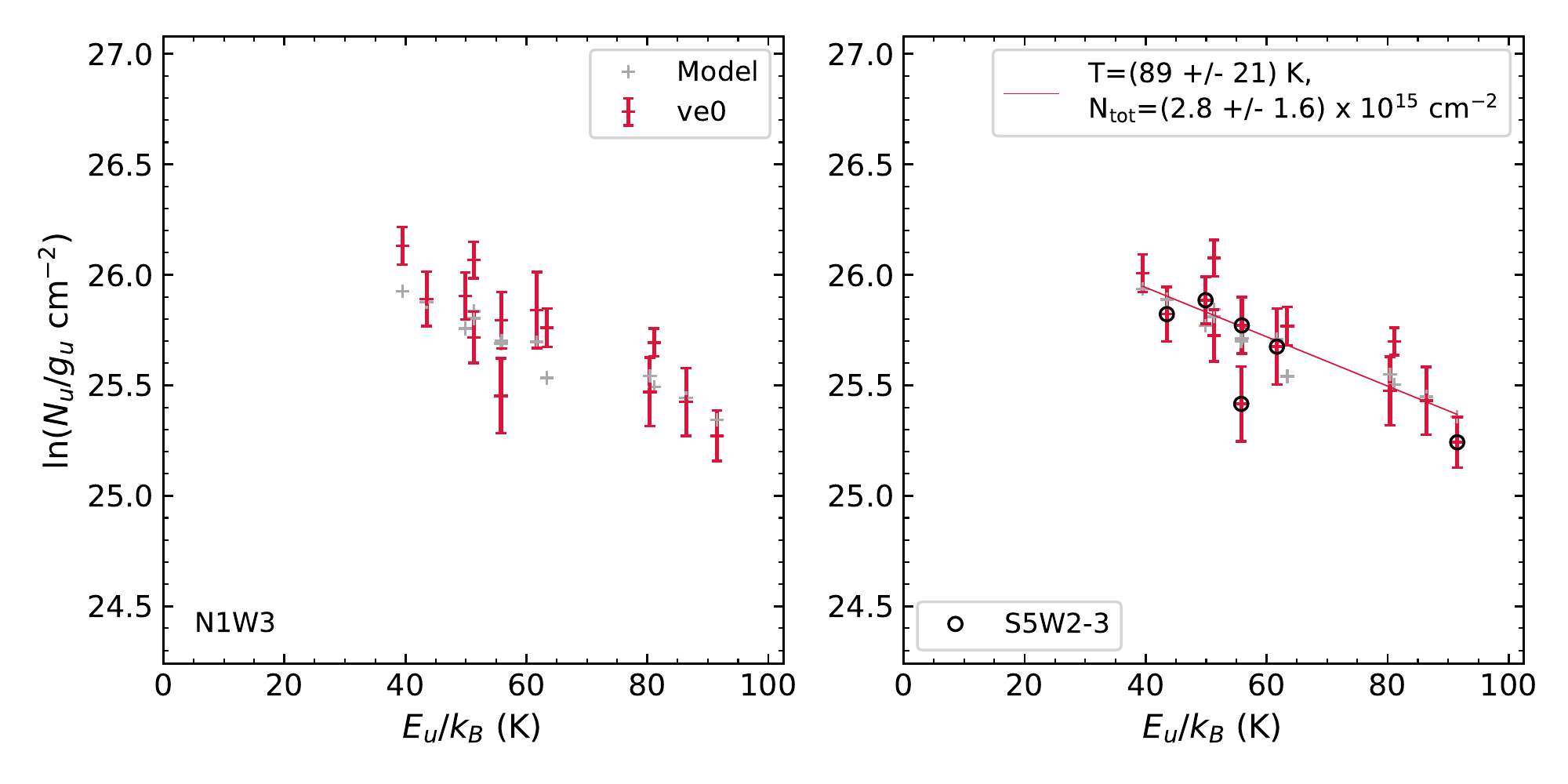}
    \caption{Same as Fig.\,\ref{fig:PD_met}, but for \mic and for all positions to the west where the molecule is detected.}
    \label{fig:wPD_mic}
\end{figure*}

\begin{figure*}[h]
    \includegraphics[width=0.49\textwidth]{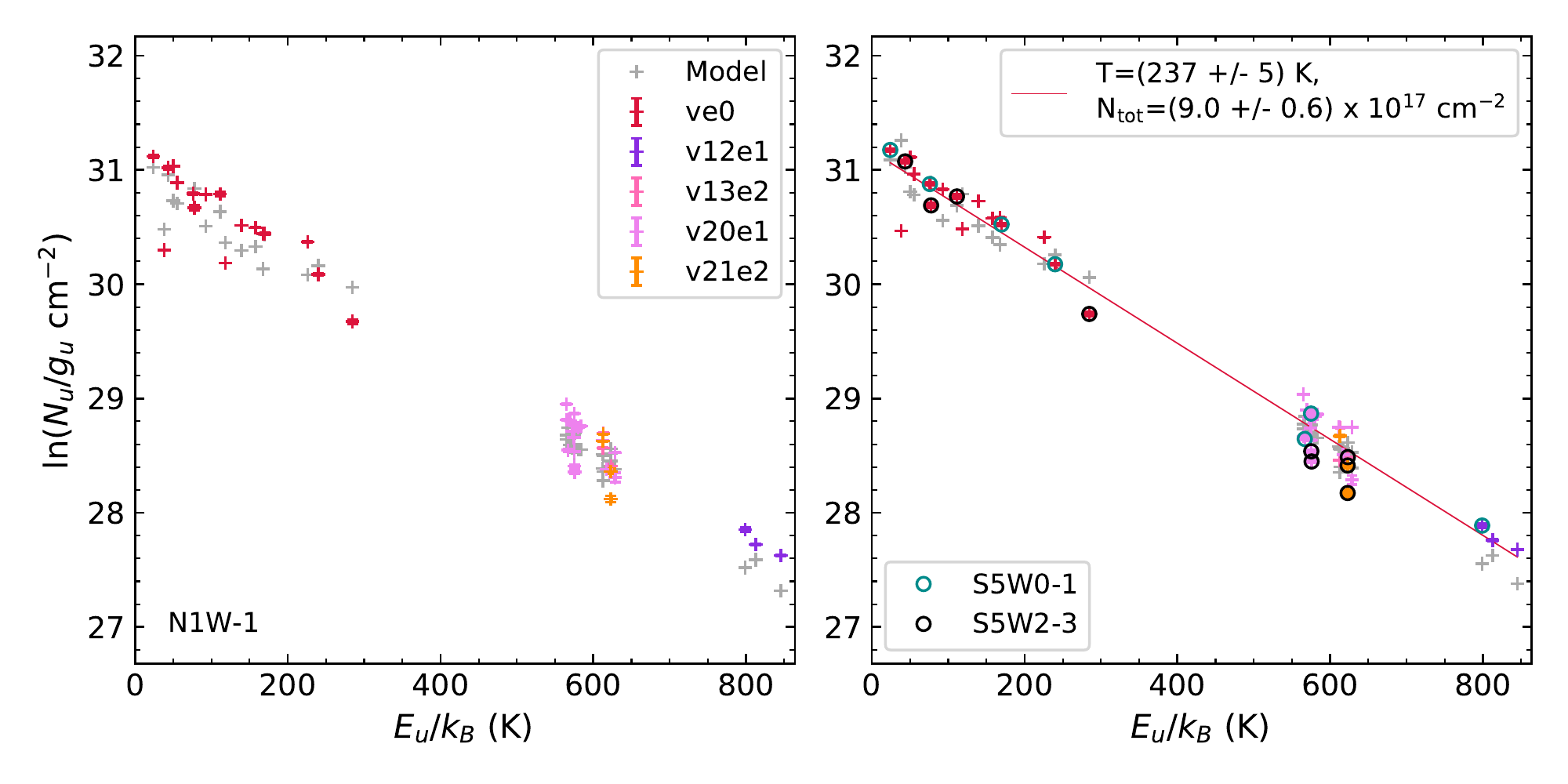}
    \includegraphics[width=0.49\textwidth]{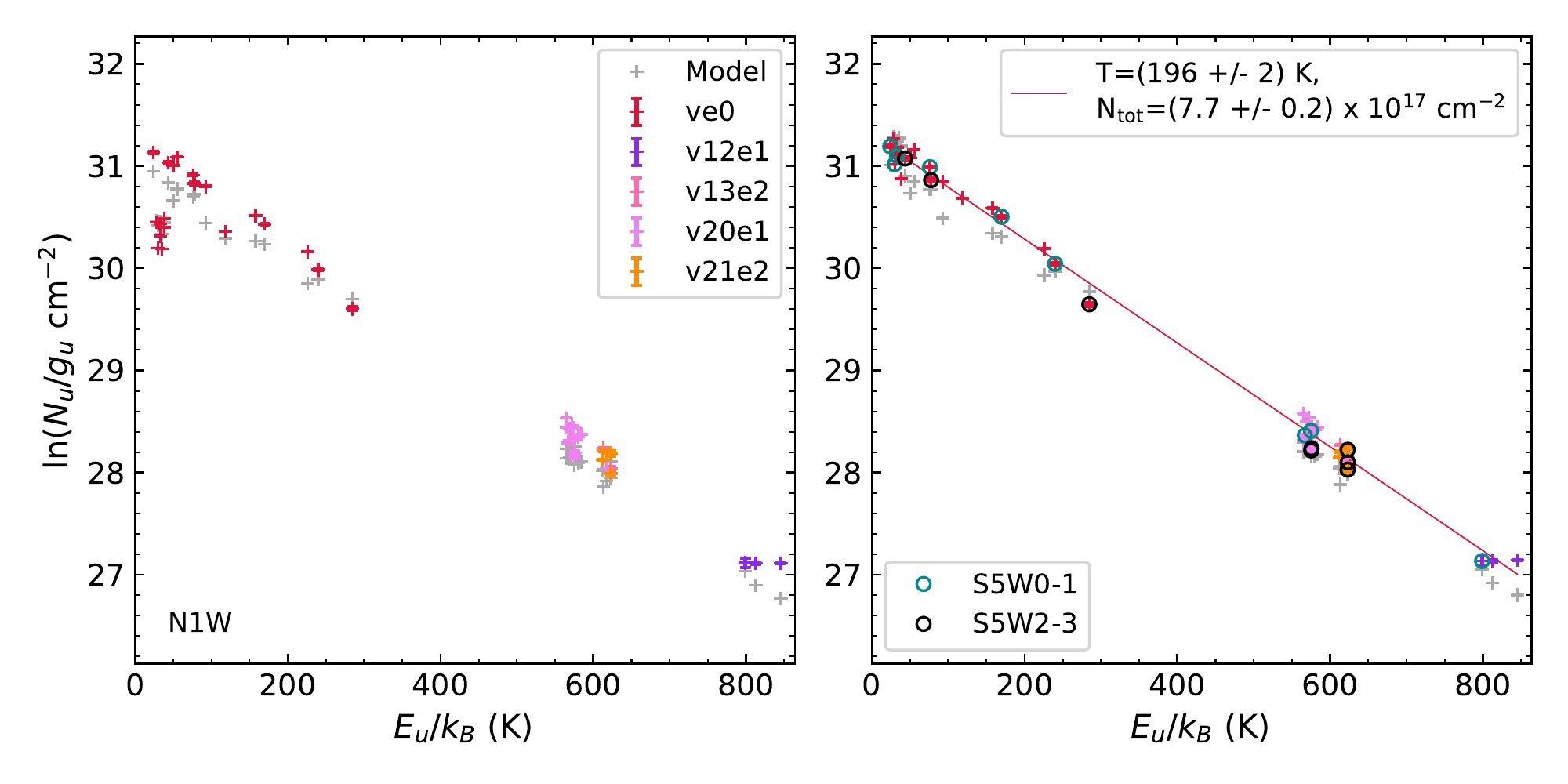}
    \includegraphics[width=0.49\textwidth]{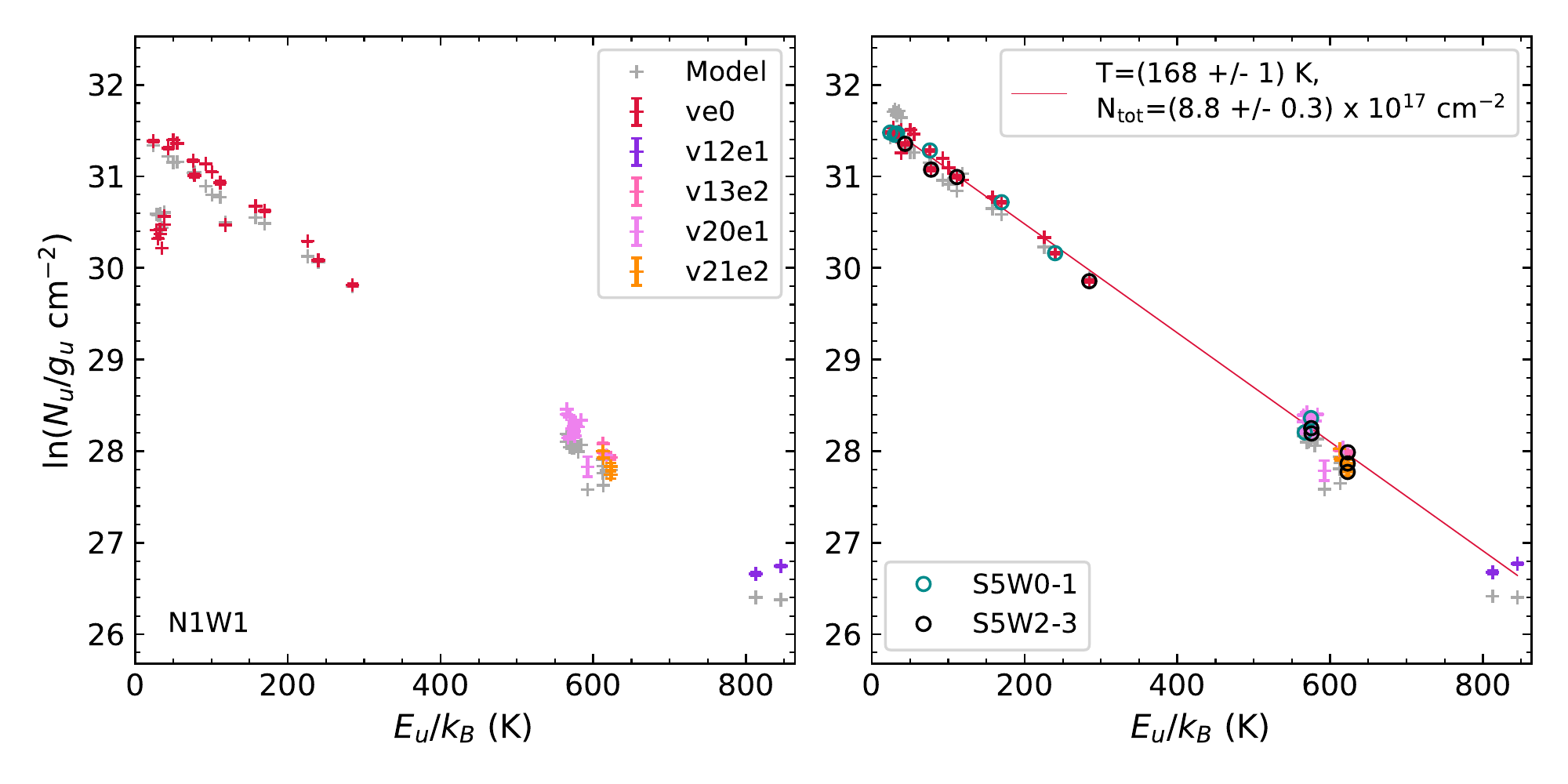}
    \includegraphics[width=0.49\textwidth]{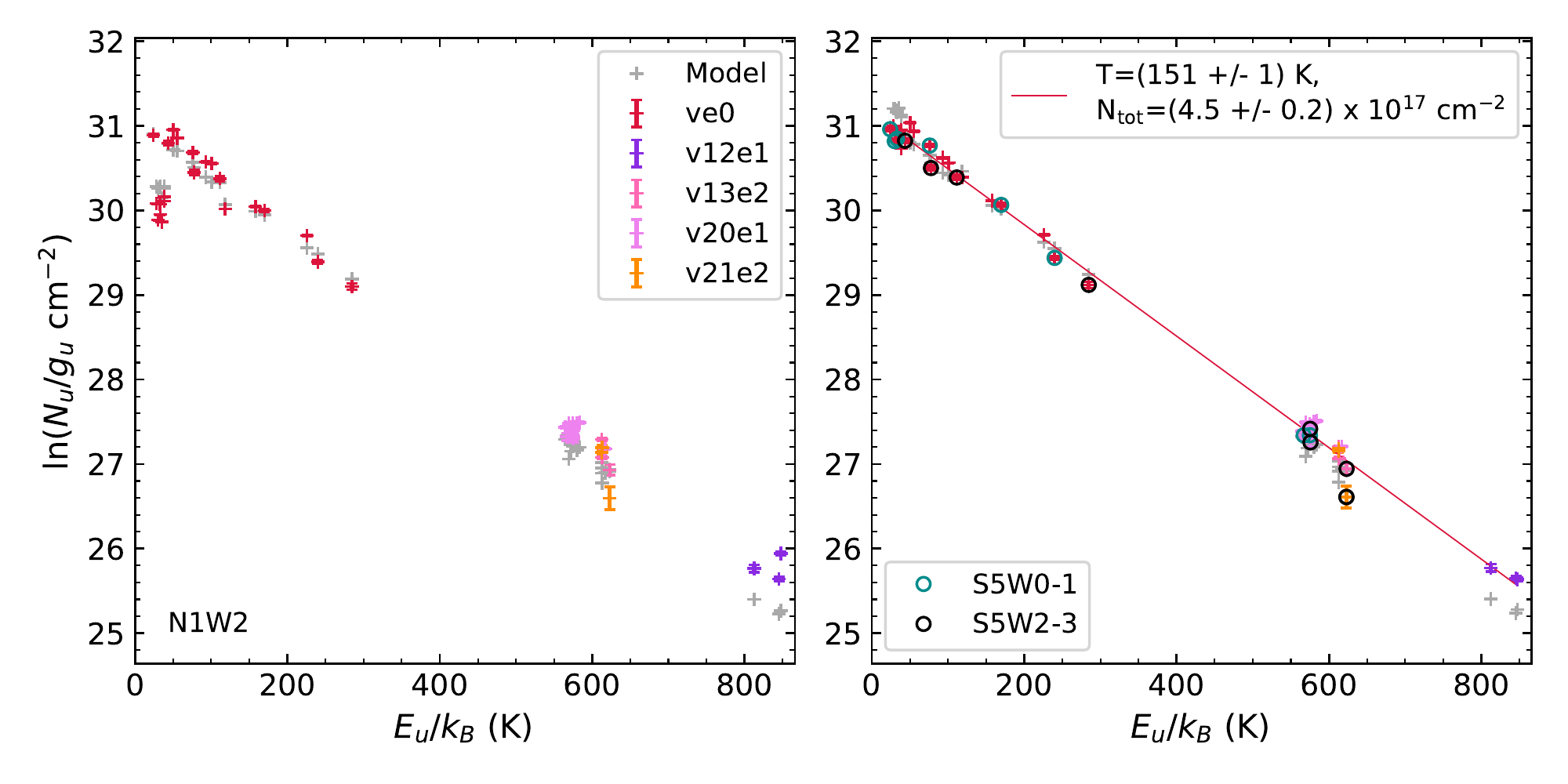}
    \includegraphics[width=0.49\textwidth]{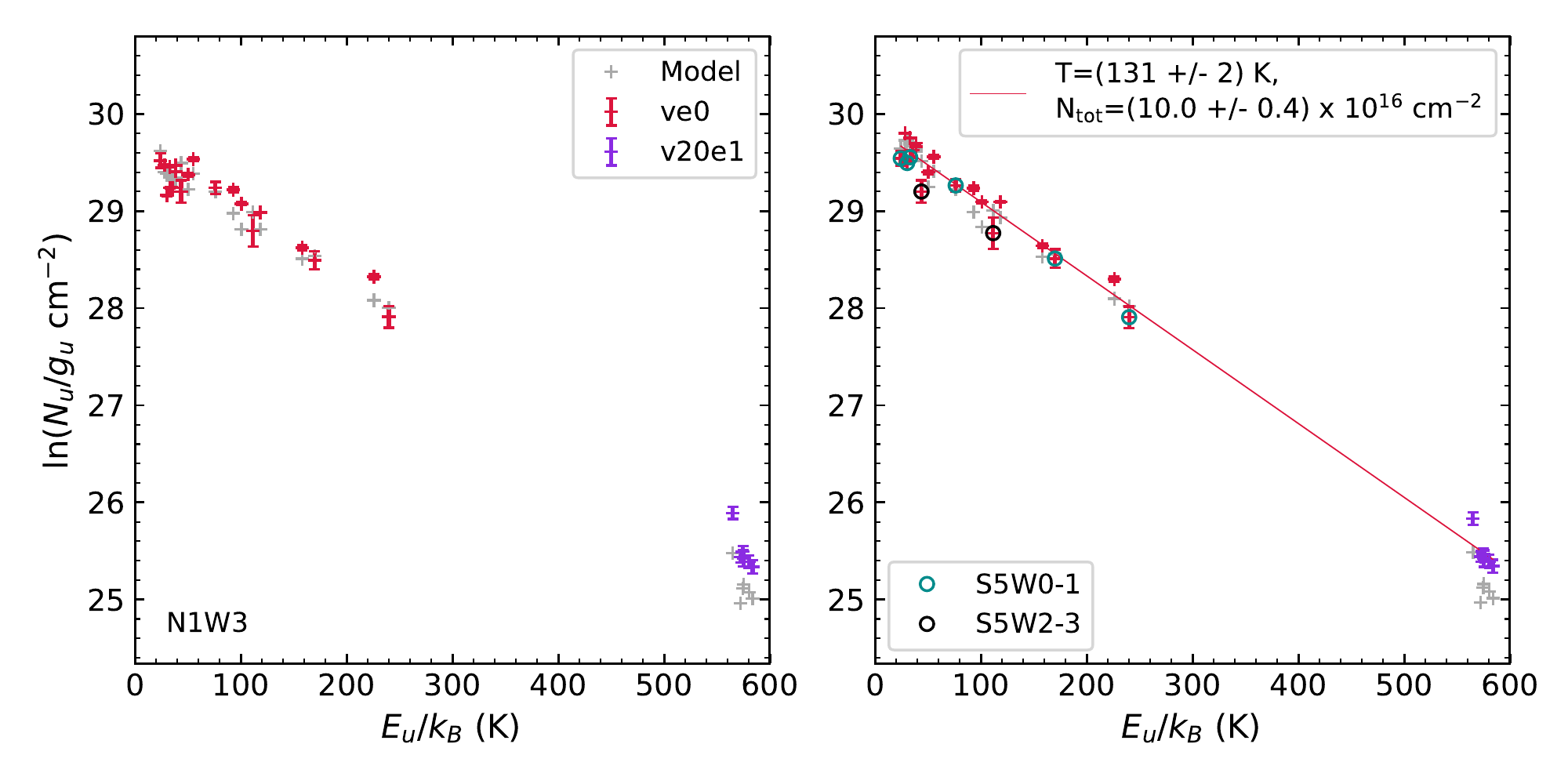}
    \includegraphics[width=0.49\textwidth]{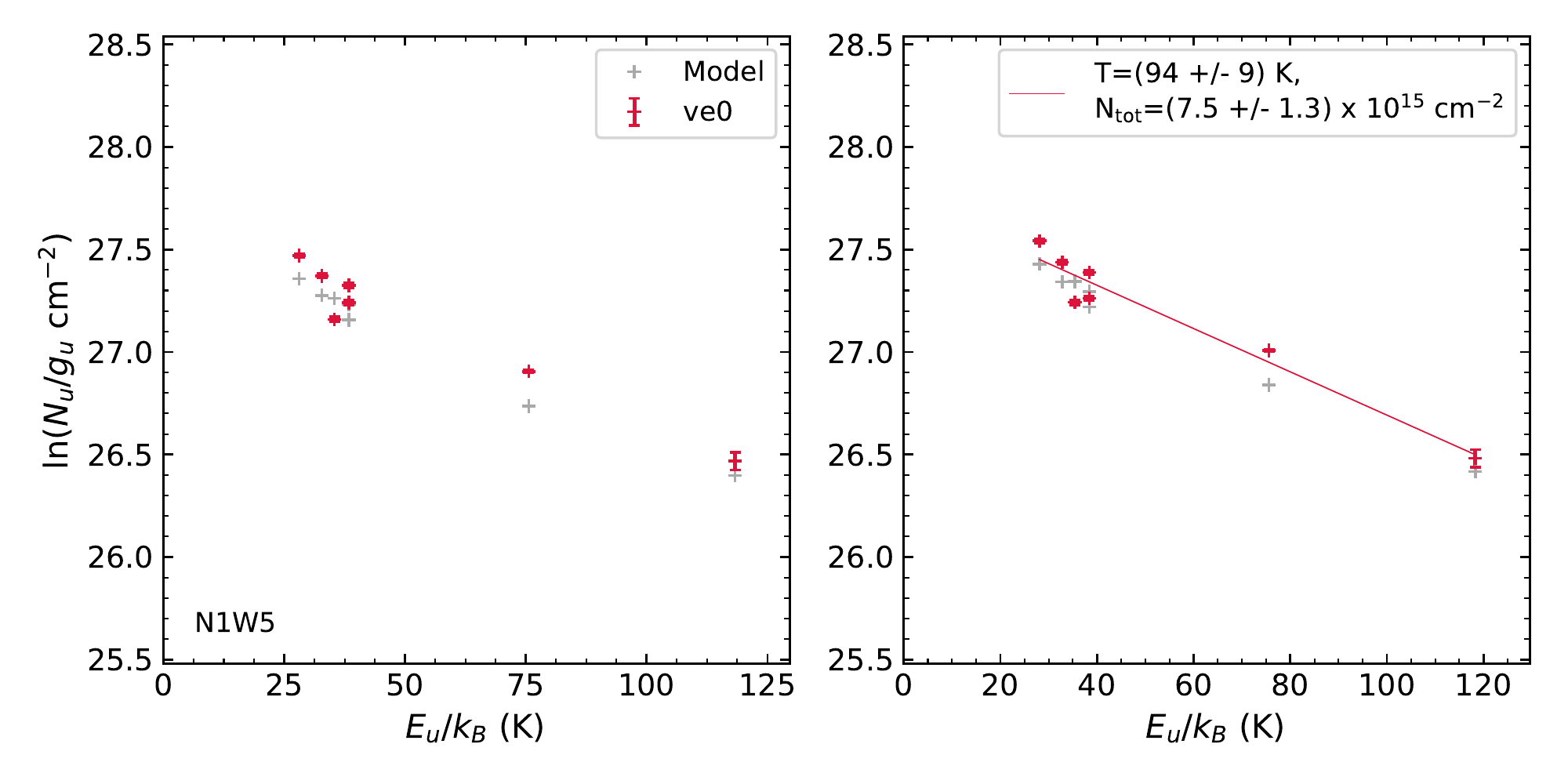}
    \includegraphics[width=0.49\textwidth]{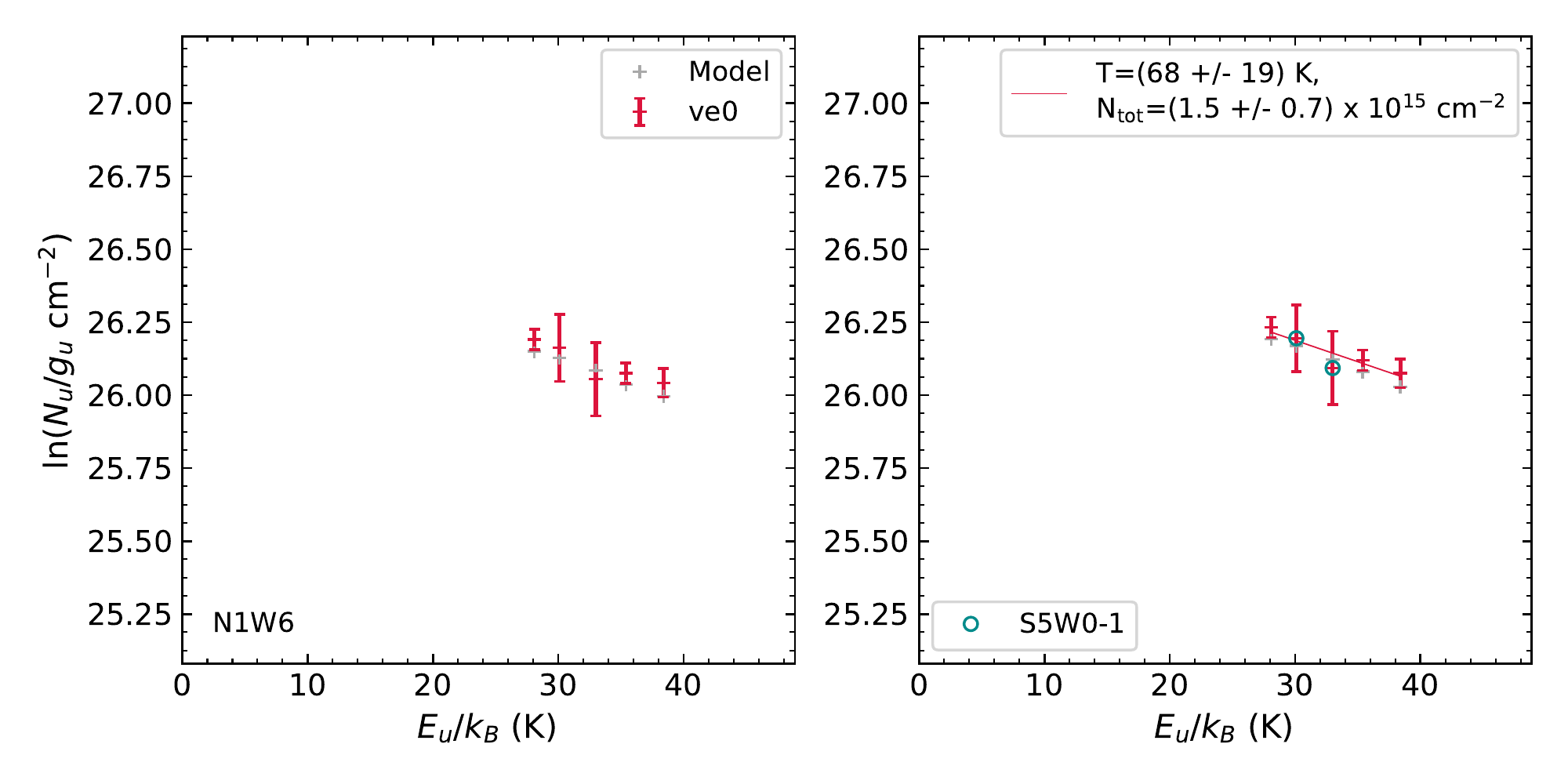}
    \includegraphics[width=0.49\textwidth]{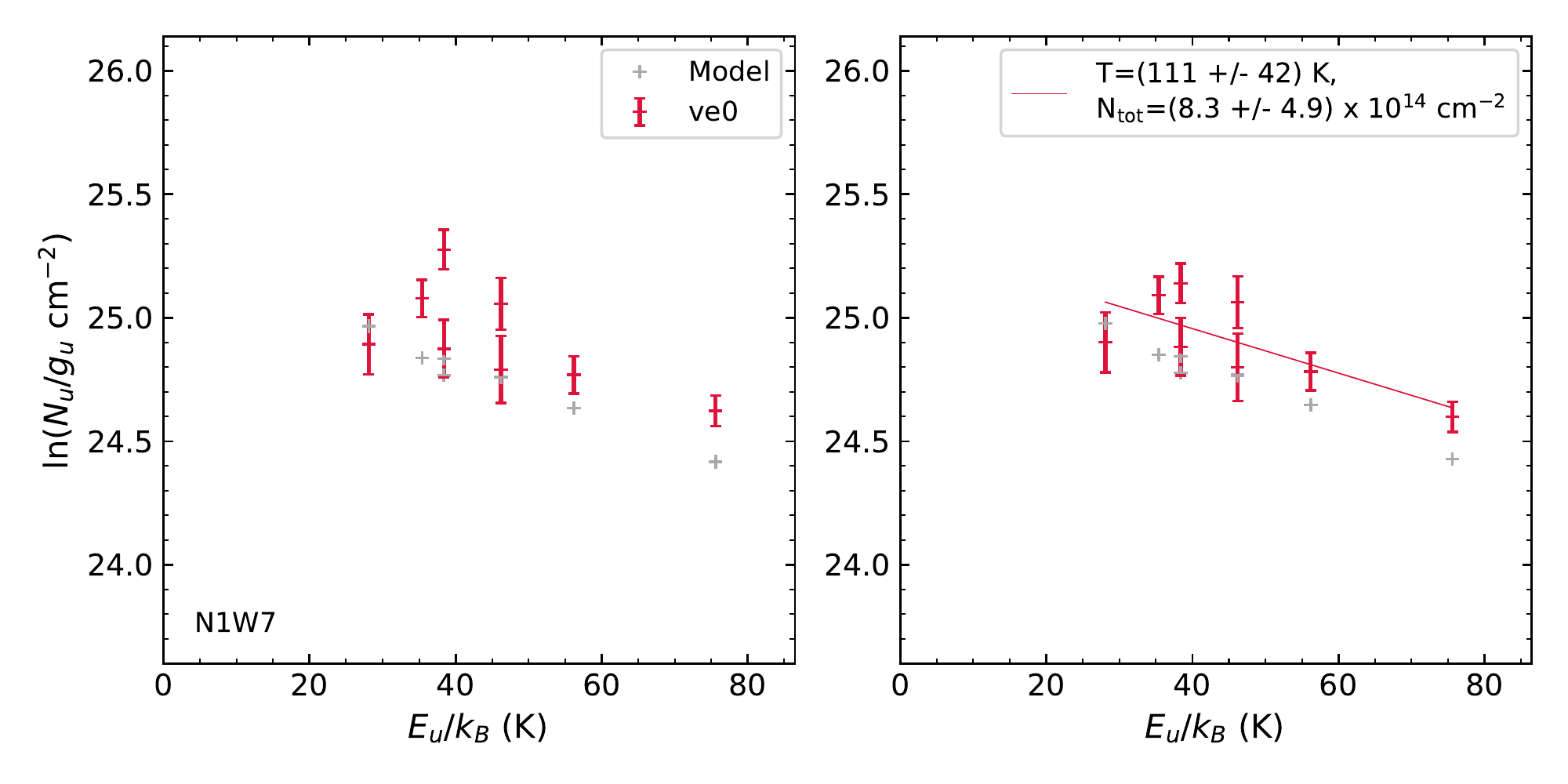}
    \includegraphics[width=0.49\textwidth]{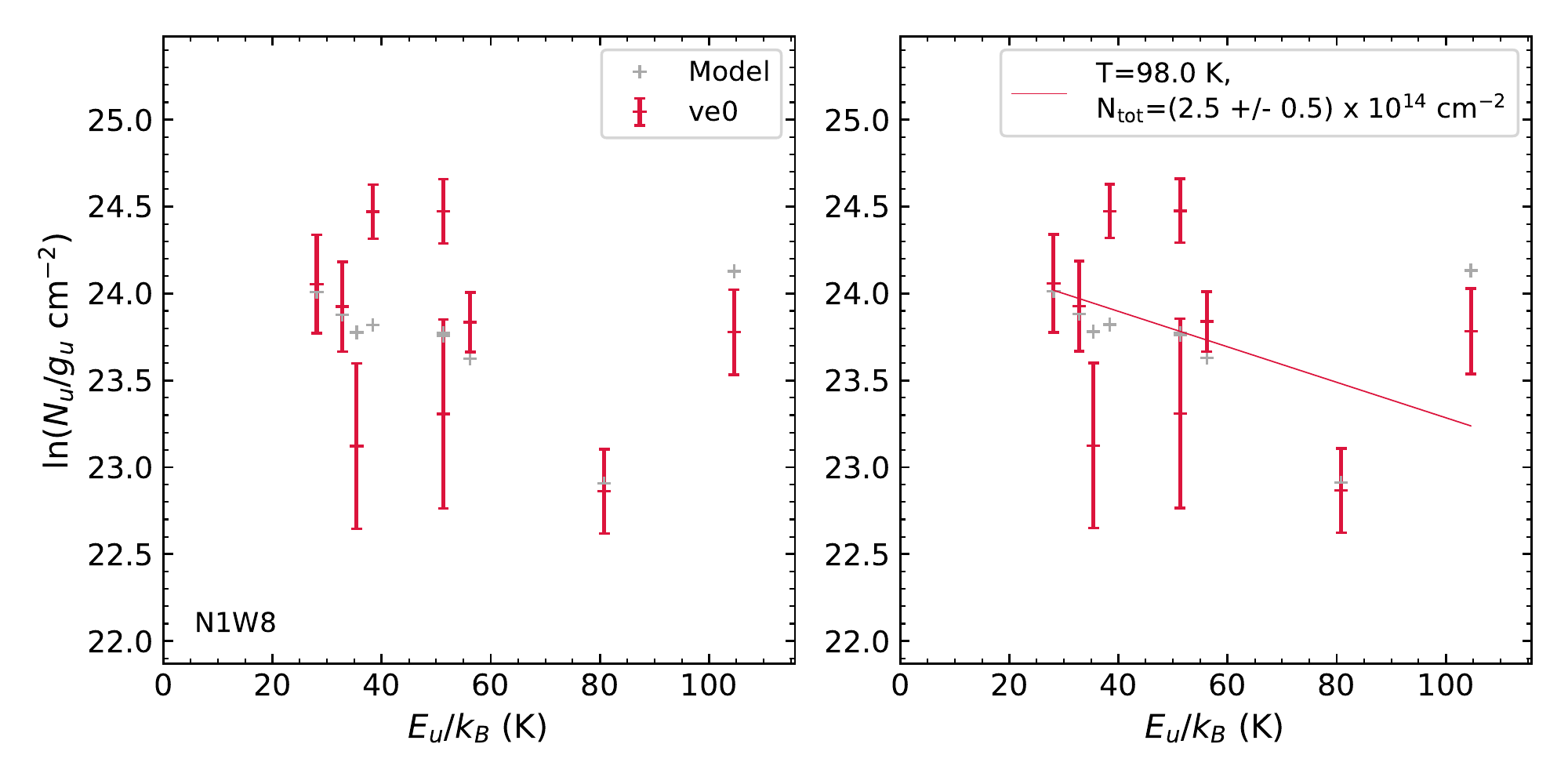}
    \caption{Same as Fig.\,\ref{fig:PD_met}, but for \etc and for all positions to the west where the molecule is detected.}
    \label{fig:wPD_etc}
\end{figure*}

\begin{figure*}[h]
    \includegraphics[width=0.49\textwidth]{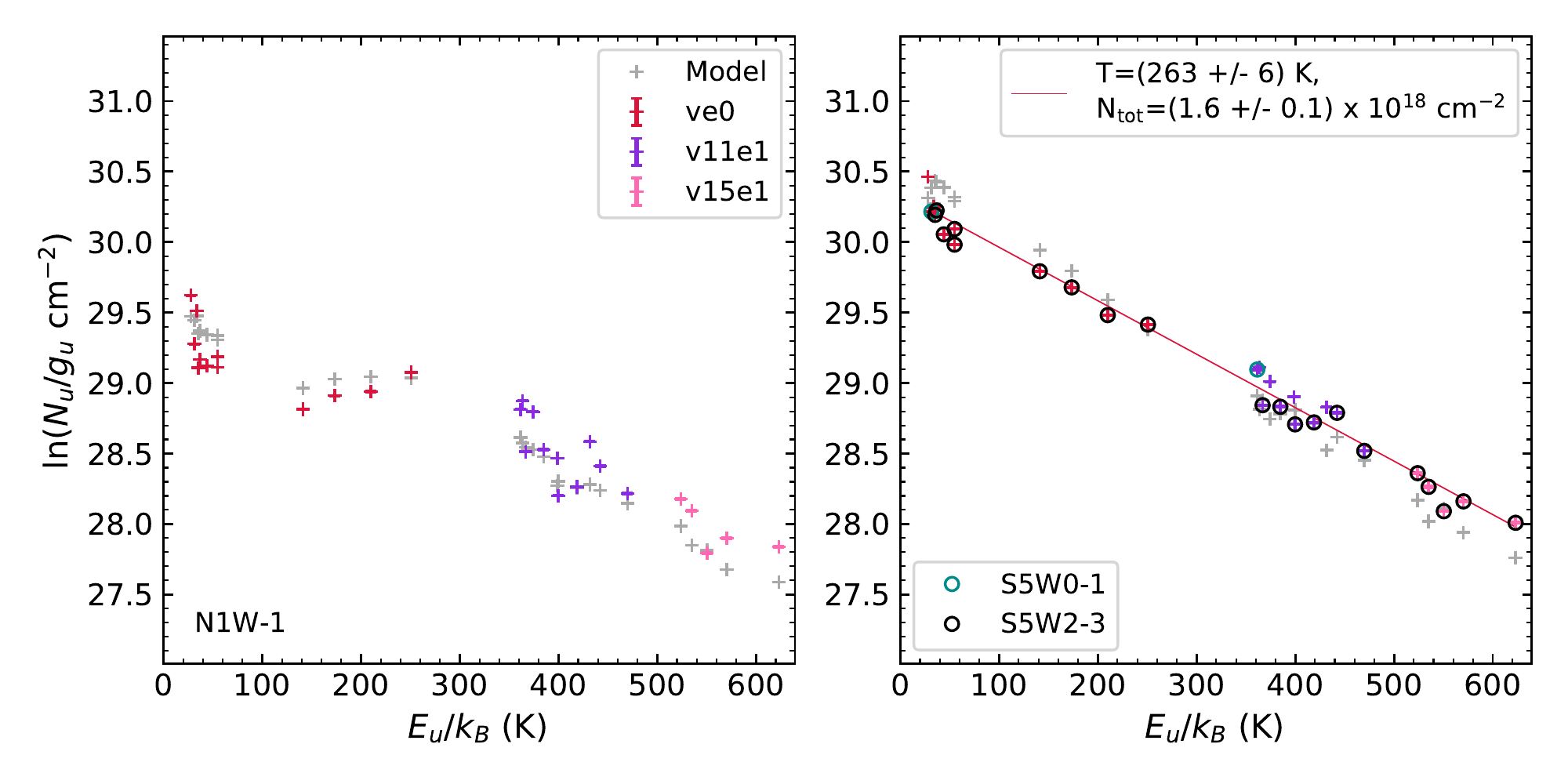}
    \includegraphics[width=0.49\textwidth]{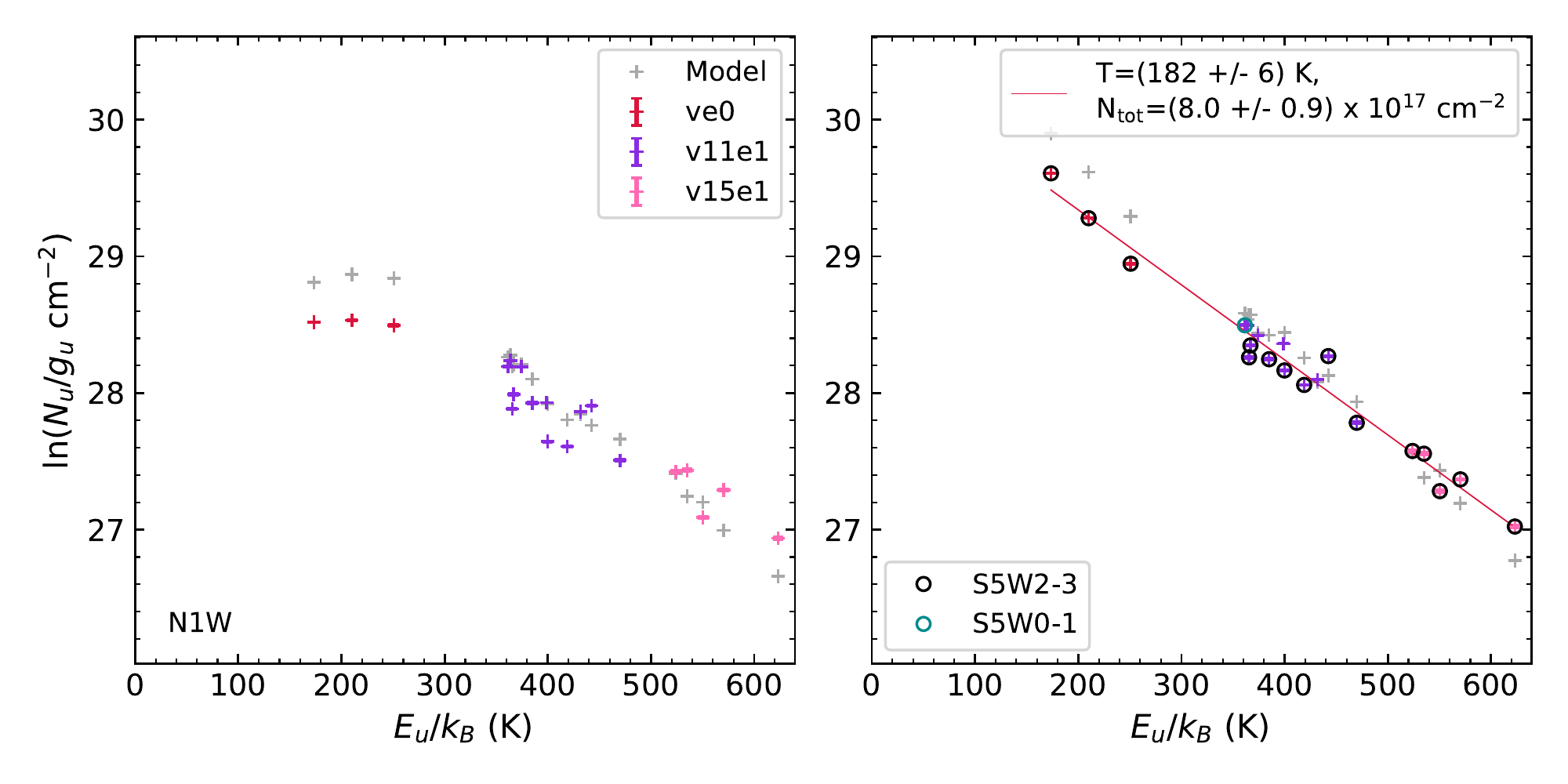}
    \includegraphics[width=0.49\textwidth]{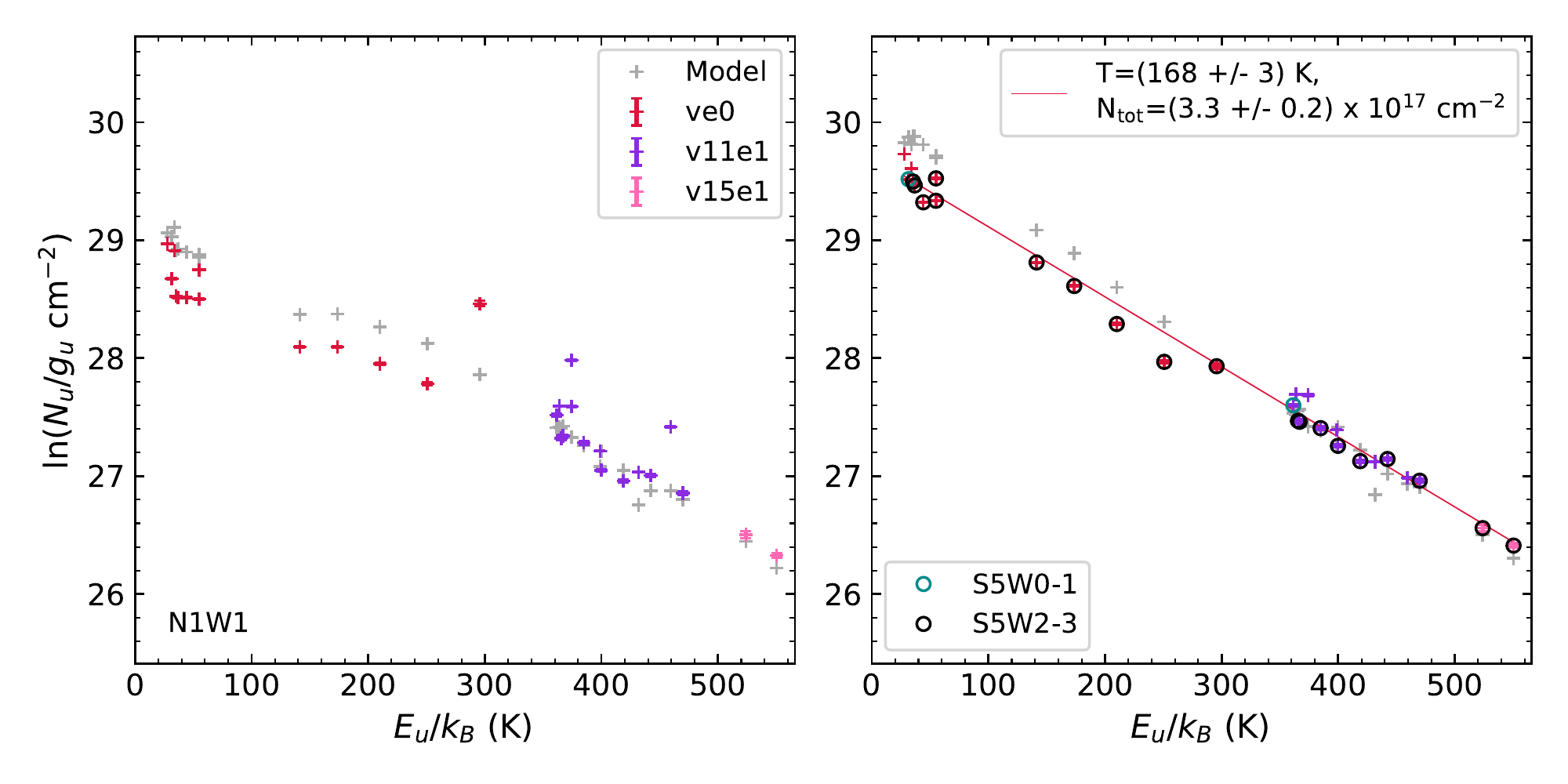}
    \includegraphics[width=0.49\textwidth]{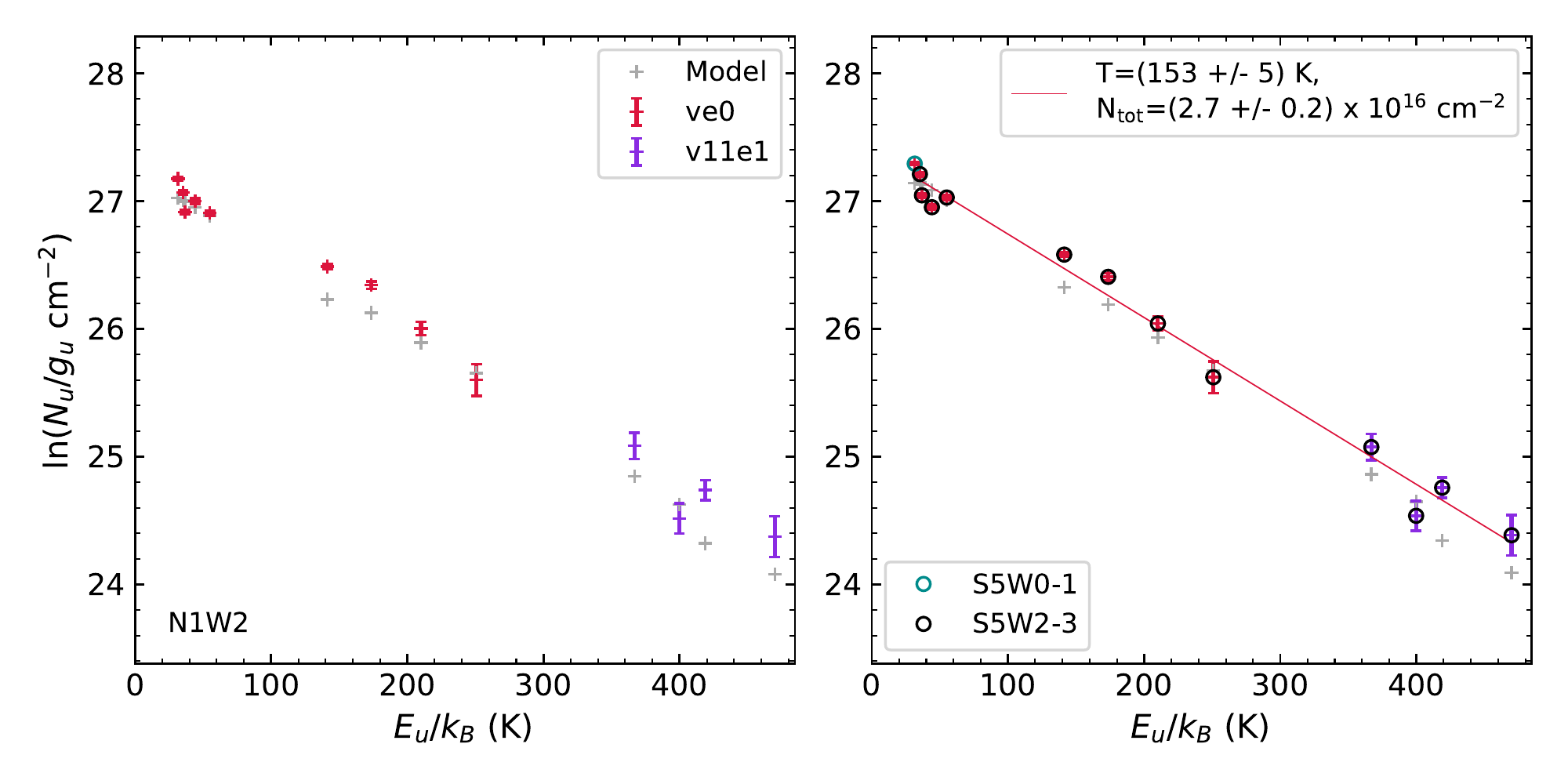}
    \includegraphics[width=0.49\textwidth]{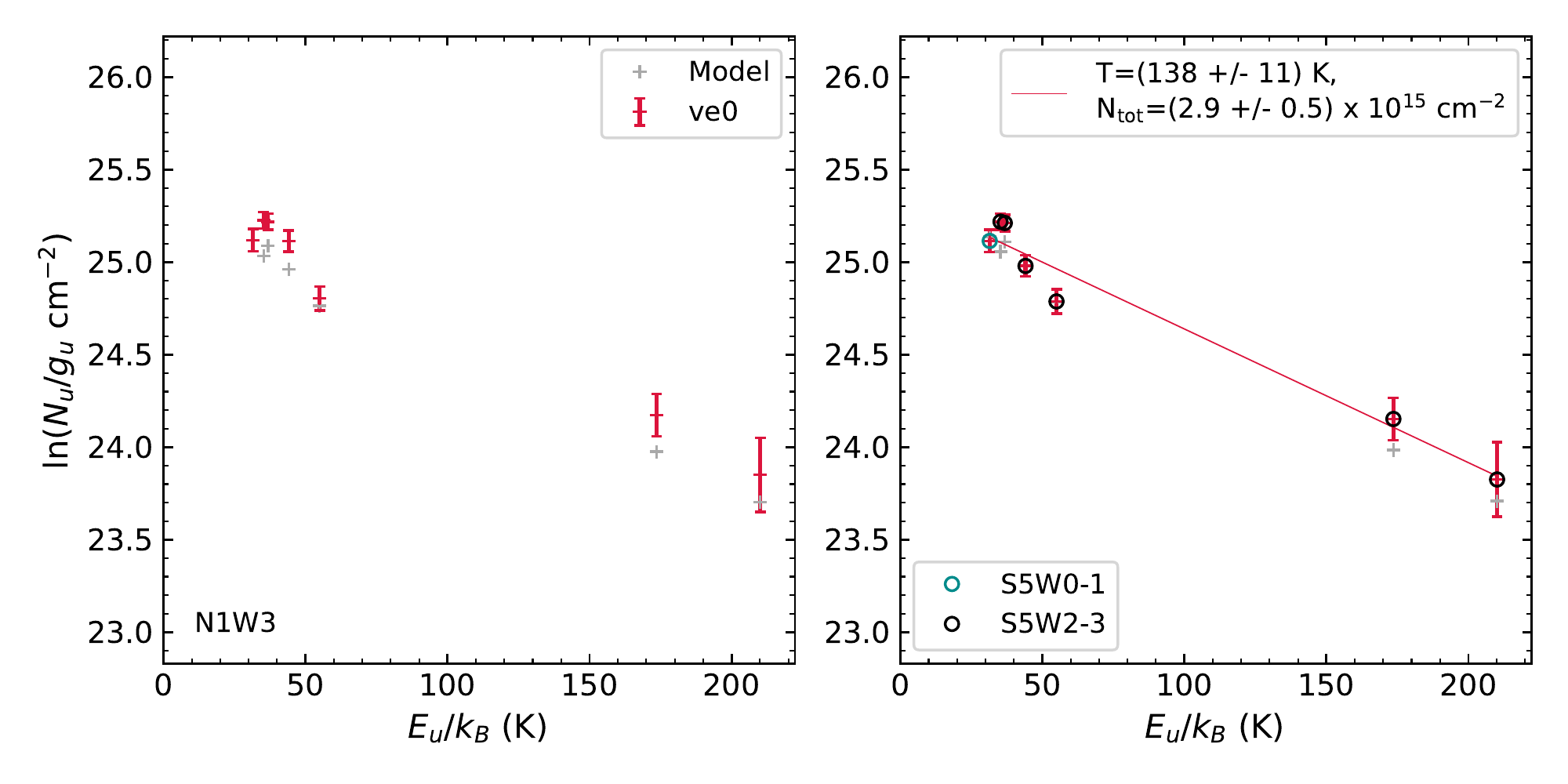}
    \includegraphics[width=0.49\textwidth]{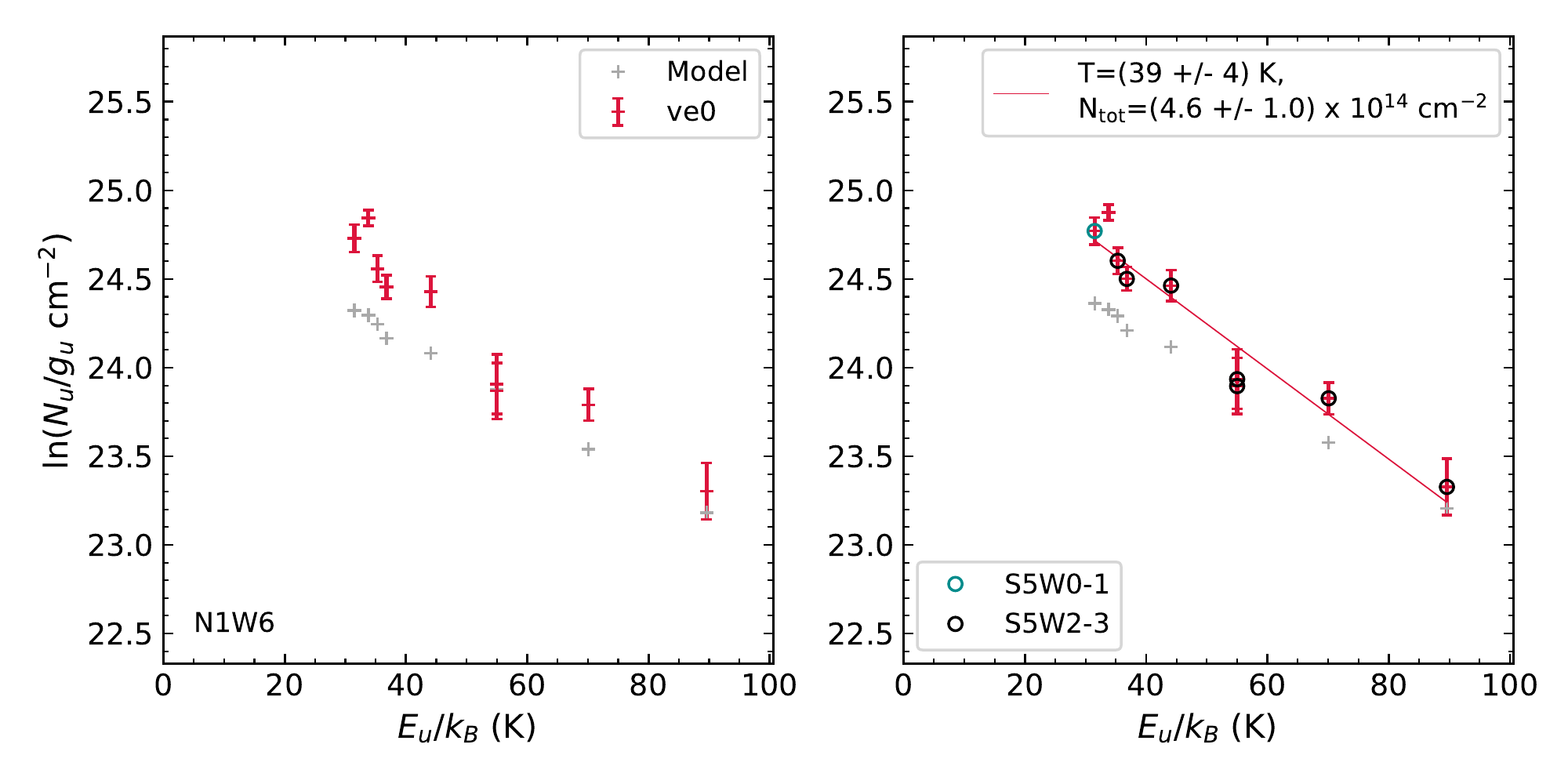}
    \caption{Same as Fig.\,\ref{fig:PD_met}, but for \vc and for all positions to the west where the molecule is detected.}
    \label{fig:wPD_vc}
\end{figure*}

\begin{figure*}[h]
    \includegraphics[width=0.49\textwidth]{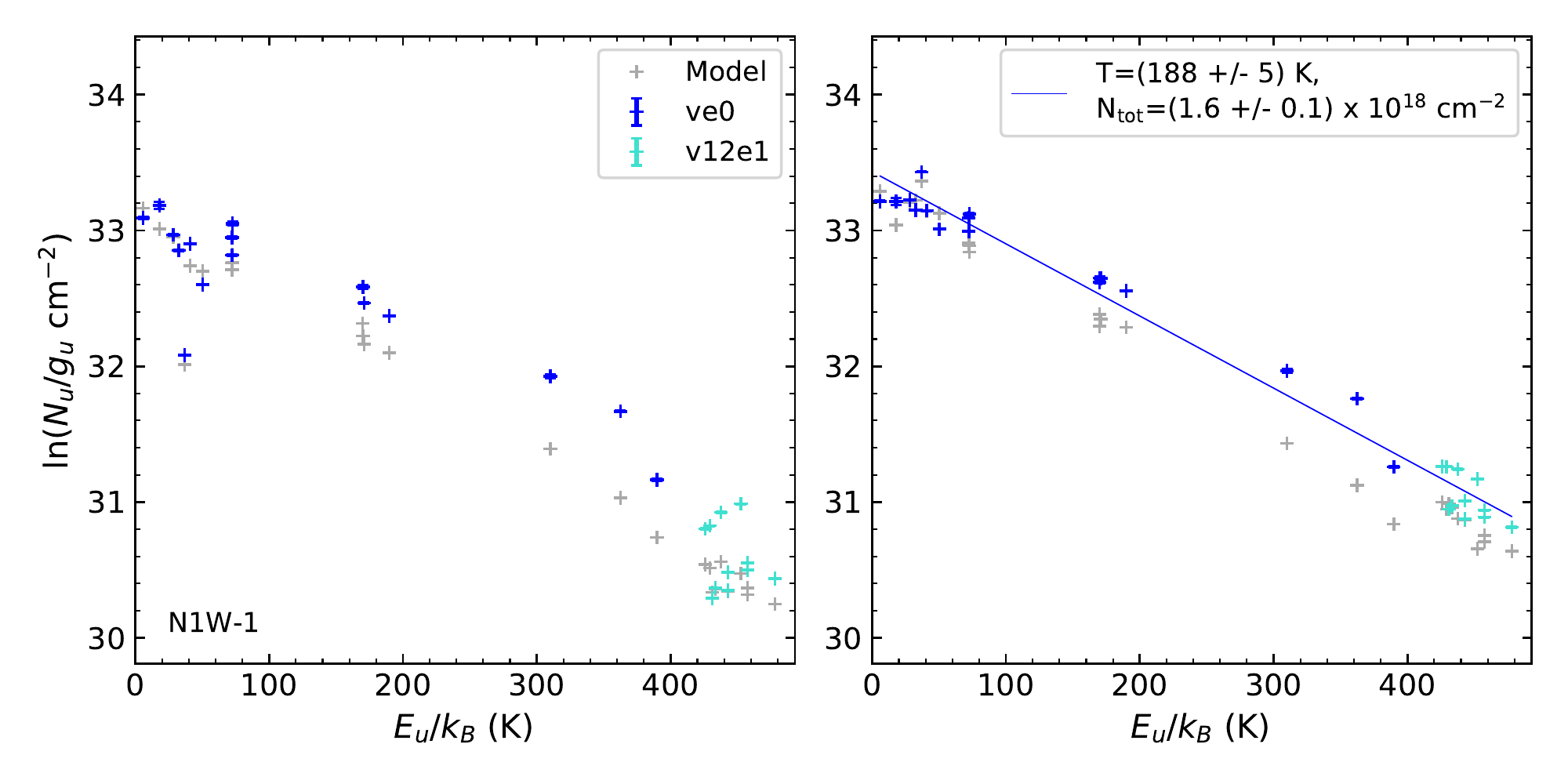}
    \includegraphics[width=0.49\textwidth]{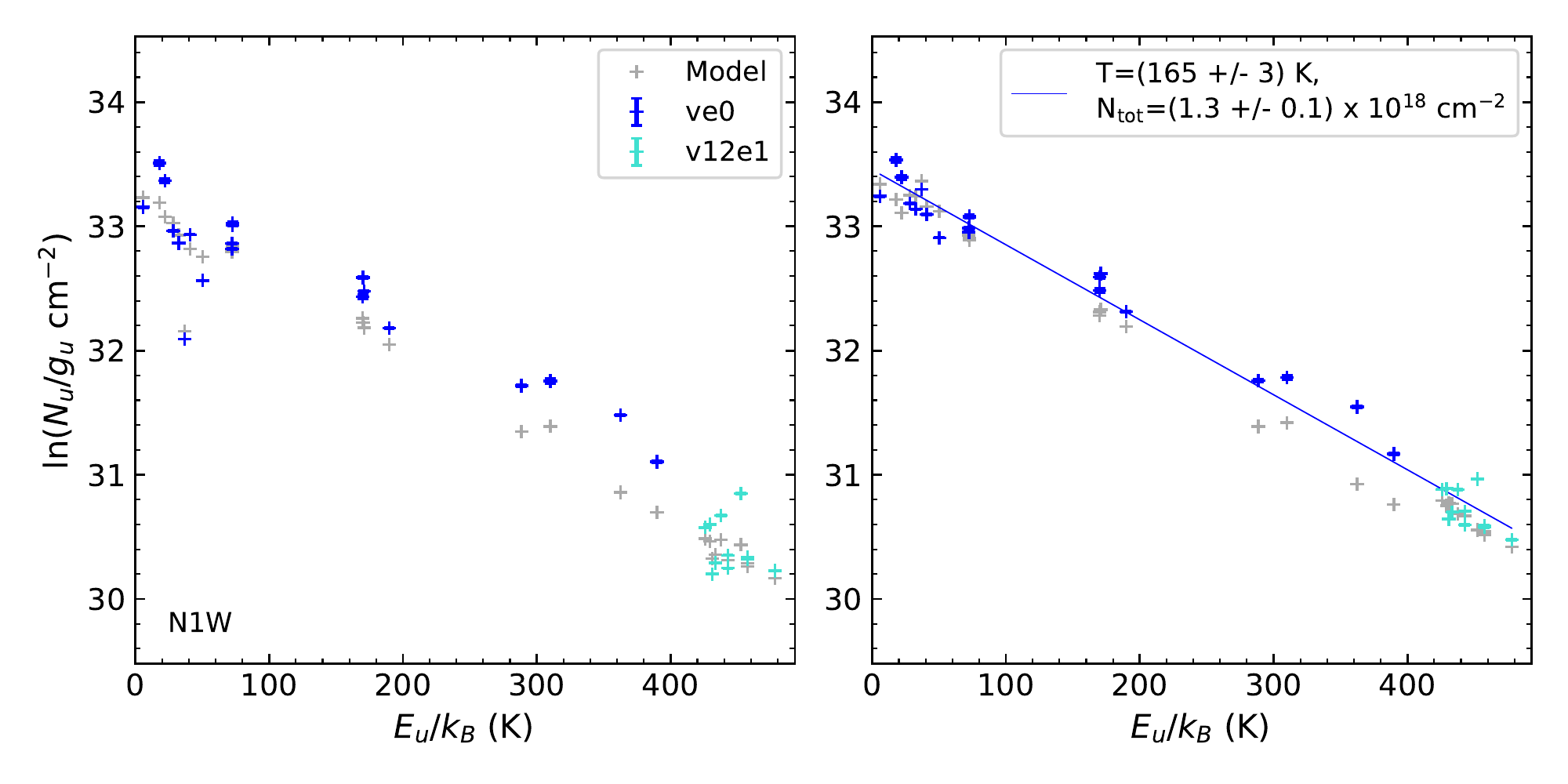}
    \includegraphics[width=0.49\textwidth]{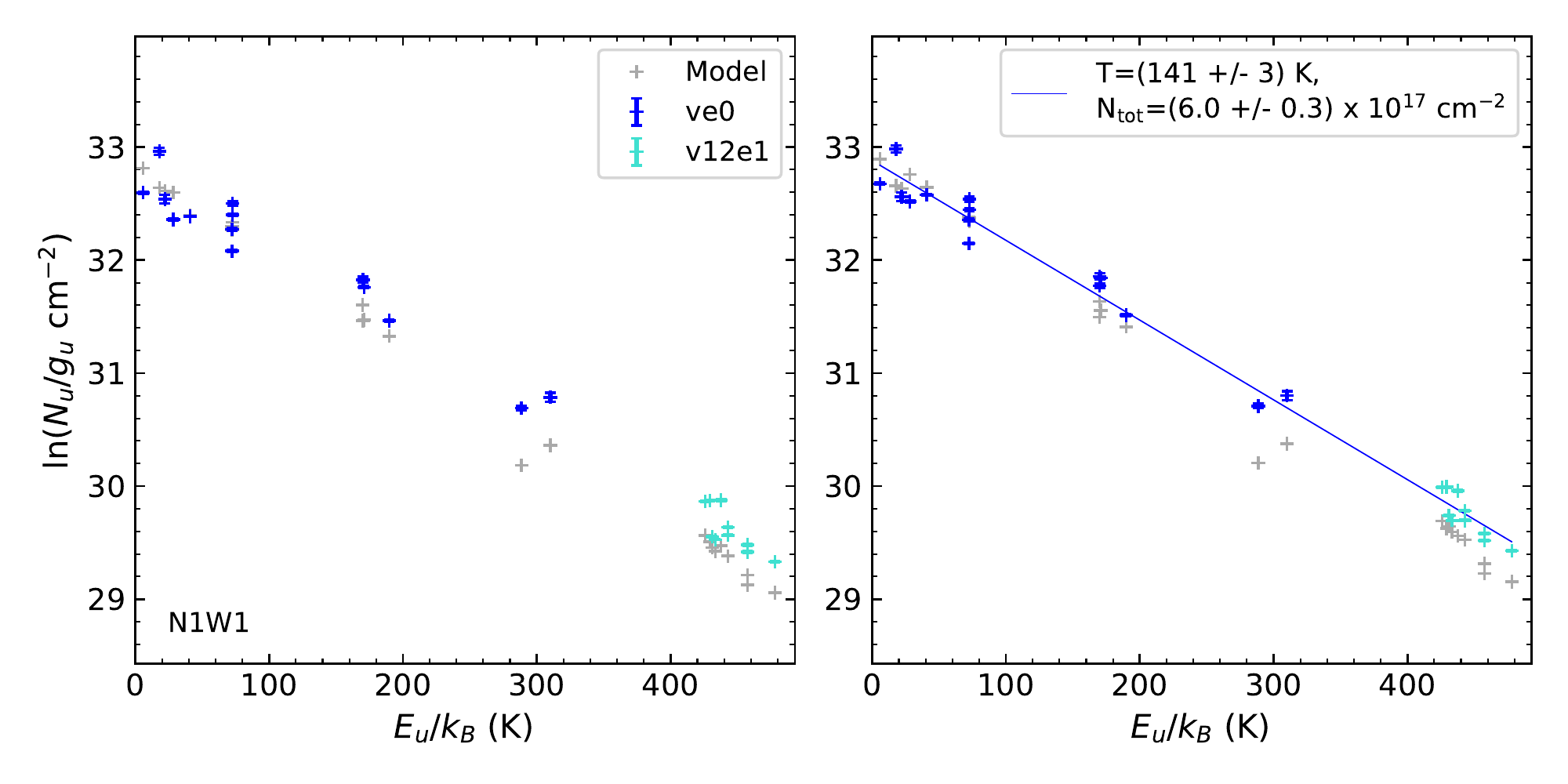}
    \includegraphics[width=0.49\textwidth]{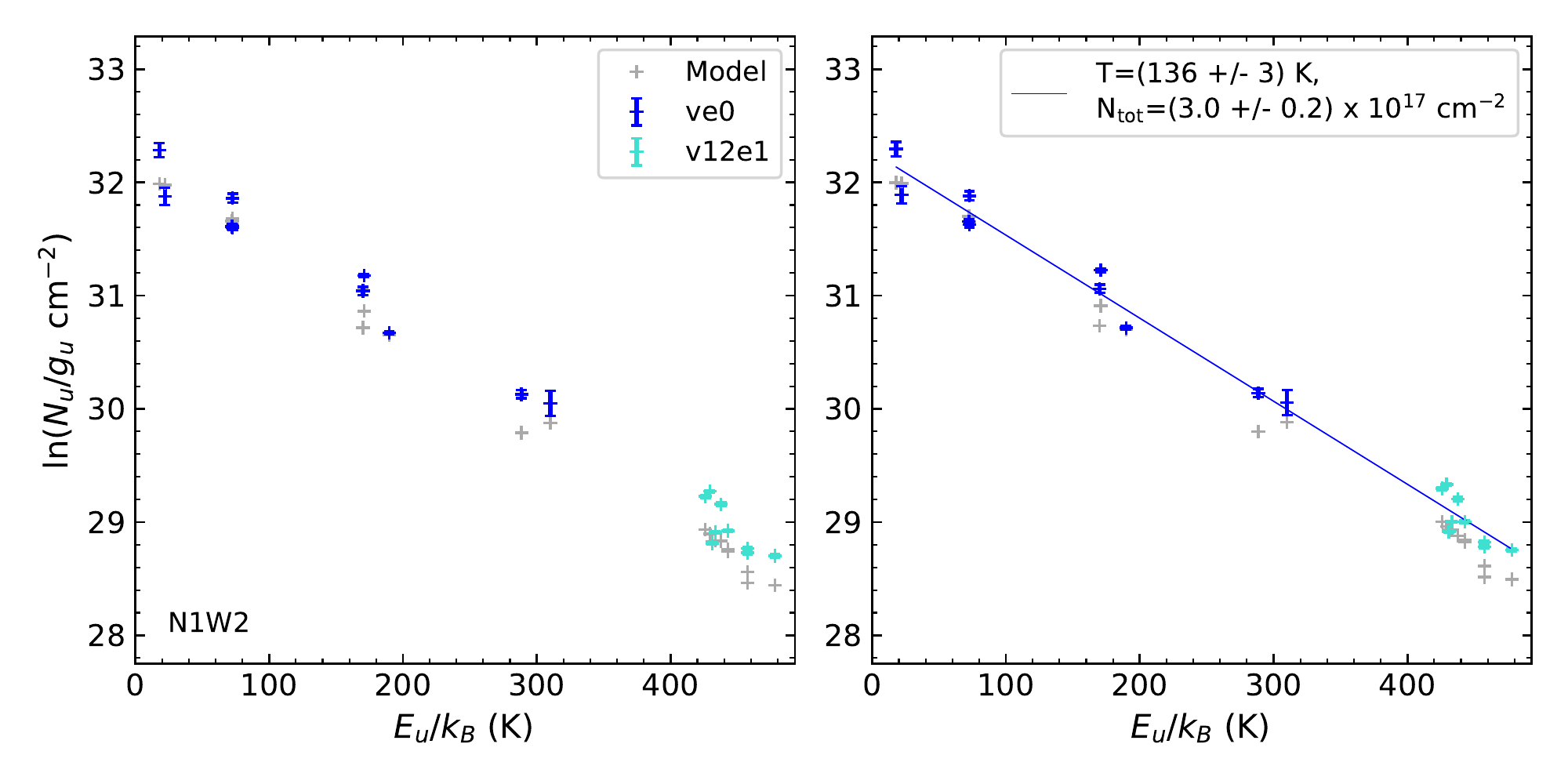}
    \includegraphics[width=0.49\textwidth]{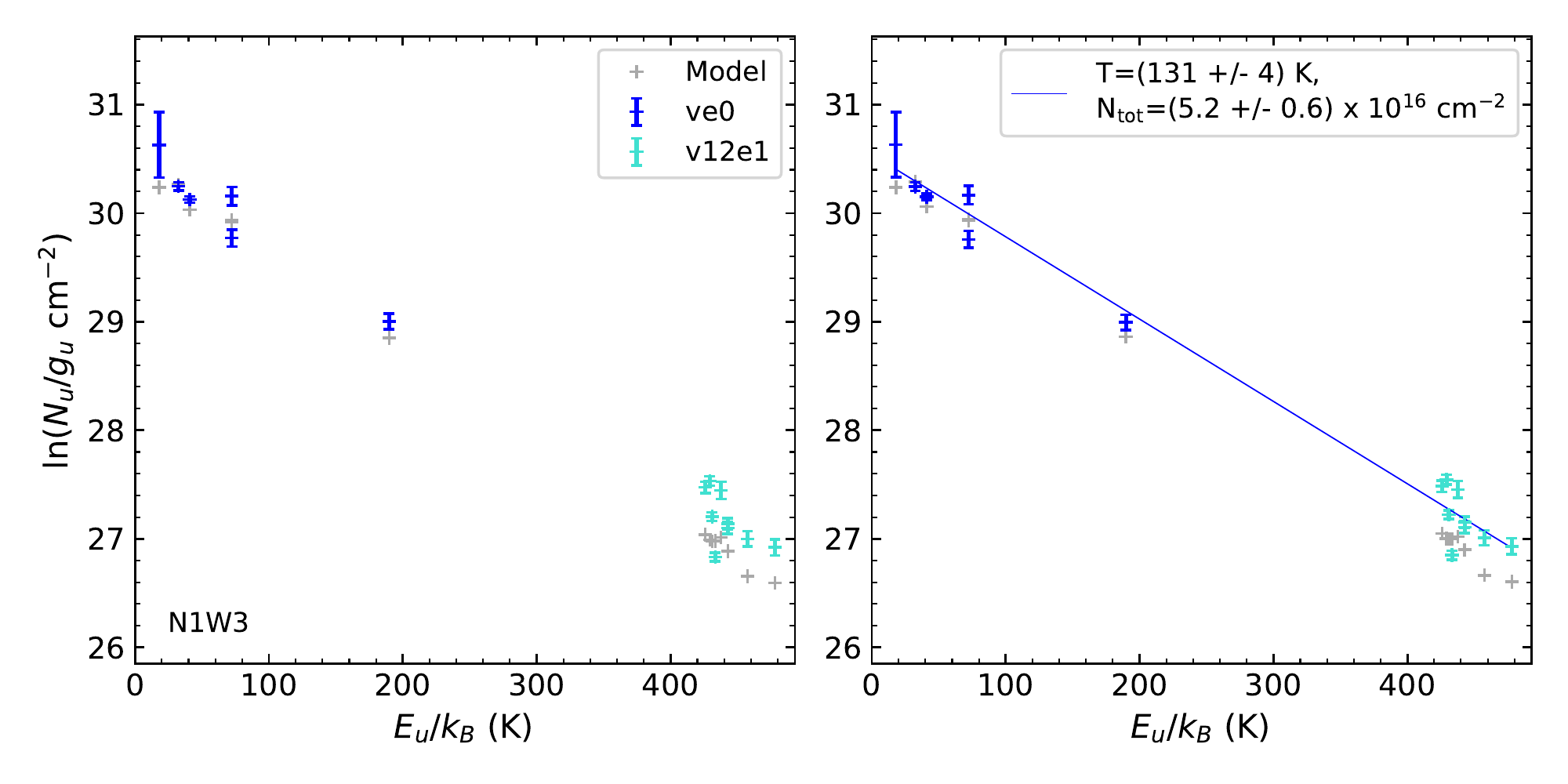}
    \caption{Same as Fig.\,\ref{fig:PD_met}, but for \fmm.}
    \label{fig:wPD_fmm}
\end{figure*}

Figures\,\ref{fig:PD_met}--\ref{fig:PD_fmm} show population diagrams of \met, its $^{13}$C isotopologue, \et, \dme, \mf, \ad, \mic, \etc, \vc, and \fmm for all positions to the south where they are detected. The vibrational states used for each COM are listed in Table\,\ref{tab:vibstates}. Figures\,\ref{fig:wPD_met}--\ref{fig:wPD_fmm} show the same for positions to the west. 
\begin{table}[]
    \caption{Vibrational states of COMs used during the analysis.}
    \centering
    \begin{tabular}{ll}
       \hline\hline\\[-0.2cm]
        COM &  vib. states\\[0.1cm]\hline\\[-0.3cm]
        \met & $v=0$, $v=1$, $v=2$ \\ 
        $^{13}$\met & $v=0$ \\
        \et & $v=0$ \\
        \dme & $v=0$, $v_{11}=1$, $v_{15}=1$ \\
        \mf & $v=0$, $v=1$ \\
        \ad & $v=0$, $v=1$, $v=2$ \\
        \mic & $v=0$ \\
        \et & $v=0$, $v_{12}=1$, $v_{20}=1$, $v_{13}=2$, $v_{21}=2$ \\
        \vc & $v=0$, $v_{11}=1$, $v_{15}=1$ \\
        \fmm & $v=0$, $v_{12}=1$ \\
        \hline\hline
    \end{tabular}
    \label{tab:vibstates}
\end{table}
\clearpage


\end{appendix}

\end{document}